%% file: main.tex
\documentclass[11pt,twoside]{book}

\DeclareFixedFont{\titlefont}{T1}{ppl}{b}{it}{0.4in}
\DeclareFixedFont{\subtitlefont}{T1}{ppl}{b}{it}{0.25in}

\title{%
	\titlefont{Concerto Grosso} \brk 
	\subtitlefont{for} \brk
	\titlefont{Sessions} \\~\\
	\subtitlefont{Fair Termination of Sessions}
	}

\usepackage{stmaryrd}
\usepackage{etoolbox}
\usepackage{mathpartir}
\usepackage{mathtools}
\usepackage{macros/prooftree}
\usepackage{amsmath , amsfonts , amsthm , amssymb}
\usepackage{xspace} 
\usepackage{xcolor}

\usepackage{listings}
\let\union\cup
\usepackage{wrapfig}
\usepackage{tikz}

\usepackage{hyperref}
\usepackage{cleveref}
\usepackage{anyfontsize}
\usepackage{natbib}
\usepackage{wrapfig}
\usepackage{lettrine}
\usepackage{pdfpages}
\includepdfset{pages=-}

\input{macros/env}

\begin{document}

\input{macros/processes}
\input{macros/macros}

\input{macros/is-macros}
\input{macros/is-examples}
\input{macros/sessiontypes}
\input{macros/pi-calc}
\input{macros/macros-lib}
\input{macros/macros-agda}
\input{macros/macros-app}

\author{Luca Ciccone}

\input{first}

\chapter*{\textit{Gratias Vobis Ago}}
\input{acks}

\newpage
\clearpage

\chapter*{Overture}

\begintreble
\emph{Sessions} are a fundamental notion in message-passing systems. A session is an abstract notion of 
communication between parties where each one owns an endpoint.
\emph{Session types} are types that are assigned to the endpoints and that are used to statically and dynamically 
enforce some desired properties of the communications, such as the absence of deadlocks.

Properties of concurrent systems are usually
divided in \emph{safety} and \emph{liveness} ones and depending on the class it belongs to, a property is defined
using different (dual) techniques. However, there exist some properties that require to mix both techniques
and the challenges in defining them
are exacerbated in \emph{proof assistants} (\eg Agda, Coq, $\dots$), 
that is, tools that allow users to formally characterize and prove theorems. In particular, we mechanize 
the meta-theory of \textit{inference systems} in Agda.

Among the interesting properties that can be studied in the session-based context, we study
\emph{fair termination} which is the property of those sessions that can always eventually reach
\emph{successful} termination under a \emph{fairness} assumption. Fair termination 
implies many desirable and well known properties, such as lock freedom. Moreover, a lock free session
does not imply that other sessions are lock free as well. On the other hand, if we consider a session and 
we assume that all the other ones are fairly terminating, we can conclude that the one under analysis is
fairly terminating as well.


\addcontentsline{toc}{chapter}{Contents}
\tableofcontents

\chapter*{Praeludium}
\addcontentsline{toc}{chapter}{Praeludium}
\input{prelude}

\part[Key Notions]{Key Notions\\\textit{(Vivace)}}


\chapter{Introductory Notions}
\input{keynotions/intro}

	\section{Safety \& Liveness}
	\label{sec:sft_lvn}
	\input{keynotions/sft-lvn}
	\section{(Generalized) Inference Systems}
	\label{sec:gis}
	\input{keynotions/gis}
	\section{Session Types}
	\label{sec:st}
	\input{keynotions/sessiontypes}
	\section{Related Work}
	\label{sec:related}
	\input{keynotions/related}

\chapter{Fair Termination}
\label{ch:ft}
\input{fairtermination/intro}

	\section{Introduction}
	\label{sec:ft_intro}
		\input{fairtermination/secintro}
		\subsection{What}
		\input{fairtermination/what}
		\subsection{Why}
		\input{fairtermination/why}

	\section{Fair Termination - Formally}
	\label{sec:ft_formally}
	\input{fairtermination/formally}

	\section{Fair Termination and Session Types}
	\label{sec:ft_st}
	\input{fairtermination/sessiontypes}

\chapter{Fair Subtyping}
\label{ch:fs}
\input{fairsub/chintro}

	\section{Original Subtyping}
	\label{sec:original_sub}
	\input{fairsub/original}
		\subsection{Unfair Subtyping}
		\label{ssec:unfair_sub}
		\input{fairsub/unfair}		
	\section{Fair Subtyping}
	\label{sec:fair_sub}
	\input{fairsub/fair}
		\subsection{Correctness - Soundness}
		\label{ssec:fsub_sound}
		\input{fairsub/fsub-sound}
		\subsection{Correctness - Completeness}
		\label{ssec:fsub_complete}
		\input{fairsub/fsub-complete}
		\subsection{Generalized Inference System}
		\label{ssec:fsub_gis}
		\input{fairsub/fsub-gis}

\part[Enforcing Fair Termination]{Enforcing\\Fair Termination\\\textit{(Largo)}}
\label{pt:type_systems}


\chapter{Fair Termination Of Binary Sessions}
\label{ch:ft_bin}
\input{ts-bin/chintro}

	\section{Calculus}
	\label{sec:ts_bin_proc}
	\input{ts-bin/calculus}
	\section{Type System}
	\label{sec:ts_bin_ts}
	\input{ts-bin/ts}
	\section{Proofs}
	\label{sec:ts_bin_corr}
	\input{ts-bin/soundness}

\chapter{Fair Termination Of Multiparty Sessions}
\label{ch:ft_multi}
\input{ts-multi/chintro}

		\section{Calculus}
		\label{sec:ts_multi_proc}
		\input{ts-multi/calculus}
		\section{Type System}
		\label{sec:ts_multi_ts}
		\input{ts-multi/ts}
		\section{Correctness}
		\label{sec:ts_multi_corr}
		\input{ts-multi/soundness} 
		\section{Related Work}
		\label{sec:ts_multi_related}
		\input{ts-multi/related}
		

\chapter{Linear Logic Based Approach}
\label{ch:ft_ll}
\input{ts-ll/chintro}

		\section{Types and Formulas}
		\label{sec:ts_ll_types}
		\input{ts-ll/types}
		\section{Calculus}
		\label{sec:ts_ll_proc}
		\input{ts-ll/calculus}
		\section{Type System}
		\label{sec:ts_ll_ts}
		\input{ts-ll/ts}
		\section{Correctness}
		\label{sec:ts_ll_corr}
		\input{ts-ll/soundness}
		\section{Comparing the Type Systems}
		\label{sec:comparison}
		\input{ts-ll/comparison}
		\section{Related Work}
		\label{sec:ts_ll_related}
		\input{ts-ll/related}
		
\part[Agda Mechanizations]{Agda Mechanizations\\ \textit{(Allegro Moderato)}}
\label{pt:agda}


\chapter[A Library For GIS]{A Library For (Generalized) Inference Systems}
\label{ch:gis_lib}
\input{gis-lib/chintro}

	\section{Meta Theory}
	\label{sec:agda_gis_meta}
	\input{gis-lib/meta}
	\section{Divergence In The Lambda-Calculus}
	\label{sec:agda_gis_lambda}
	\input{gis-lib/lambda}

	\section{Indexed (Endo) Containers}
	\label{sec:agda_gis_container}
	\input{gis-lib/container}


\chapter{Properties Of Session Types}
\label{ch:agda_prop}
\input{agda-prop/chintro}

	\section{Fair Termination}
	\label{sec:agda_fairt}
	\input{agda-prop/fair-termination}
	\section{Session Types}
	\label{sec:agda_st}
	\input{agda-prop/st}

	\section{Fair Termination}
	\label{sec:agda_ft}
	\input{agda-prop/ft}
	\section{Fair Compliance}
	\label{sec:agda_fc}
	\input{agda-prop/fc}
	\section{Fair Subtyping}
	\label{sec:agda_fs}
	\input{agda-prop/fs}
	\section{Correctness of Fair Compliance}
	\label{sec:agda_fc_corr}
	\input{agda-prop/fc-corr}

	\section{Related Work}
	\label{sec:agda_related}
	\input{agda-prop/related}

\chapter*{Postludium}
\addcontentsline{toc}{chapter}{Postludium}
\input{finale}


\bibliography{main}


\clearpage
\appendix
\input{appendix}

\clearpage
\newpage
\includepdf{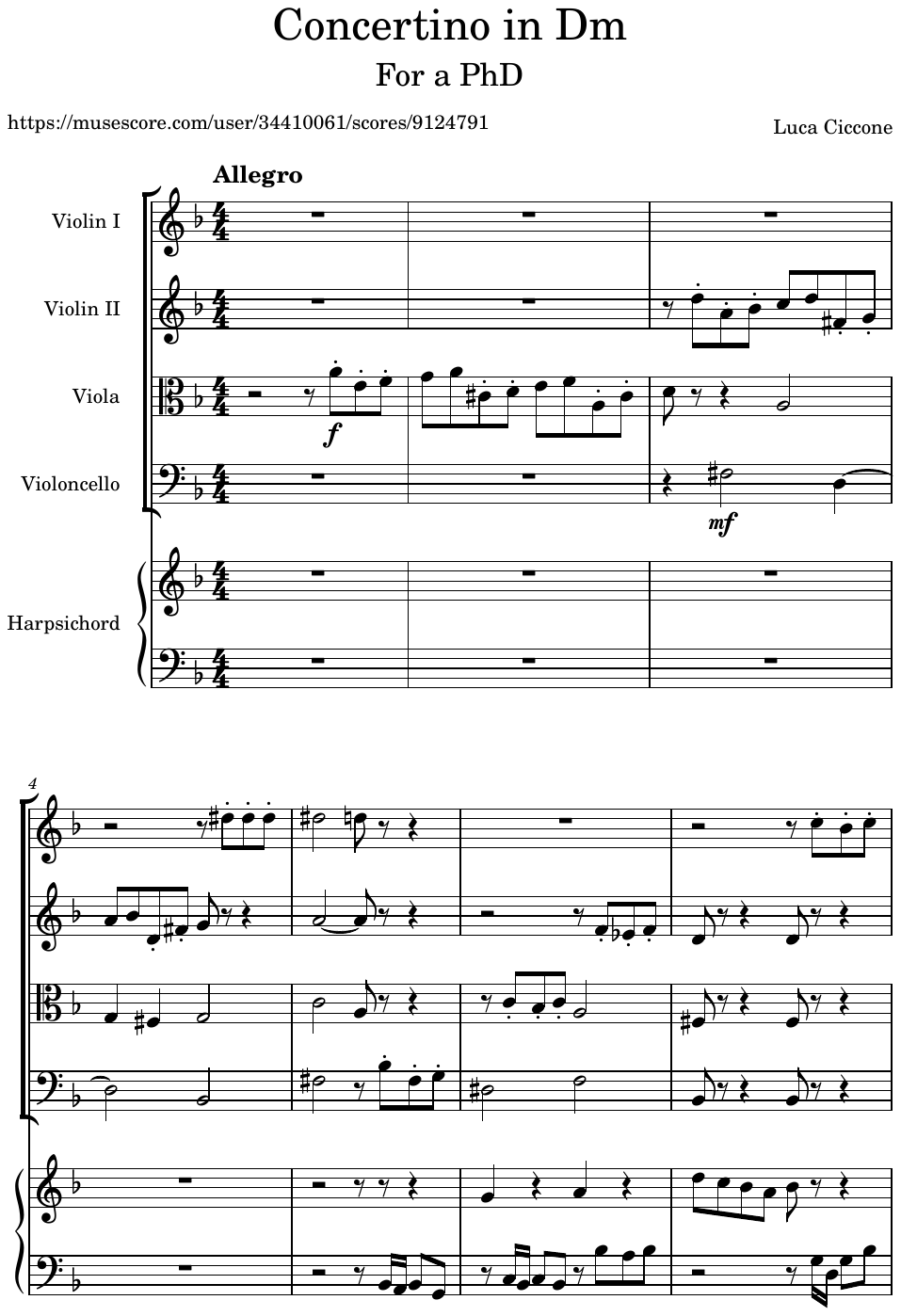}

\end{document}

%% file: macros/env.tex
\newtheorem{theorem}{Theorem}[section]
\newtheorem{example}{Example}[section]
\newtheorem{corollary}{Corollary}[section]
\newtheorem{proposition}{Proposition}[section]

\newtheorem{notation}{Notation}

\theoremstyle{definition}
\newtheorem{definition}{Definition}[section]

\newtheorem{lemma}{Lemma}[section]

\theoremstyle{remark}
\newtheorem{remark}{Remark}[section]

\newcommand{\eoe}{\hfill$\lrcorner$}
\newcommand{\eor}{\eoe}

%% file: macros/processes.tex
\newcommand{\fn}[1]{\mathsf{fn}(#1)}
\newcommand{\bn}[1]{\mathsf{bn}(#1)}
\newcommand{\pcong}{\preccurlyeq}
\newcommand{\subst}[2]{\{#1/#2\}}

\newcommand{\ple}[1]{\ensuremath{\langle #1 \rangle}} 
\newcommand{\N}{\mathbb{N}} 
\newcommand{\kw}[1]{\mathsf{#1}} 
\newcommand{\seqof}[1]{\overline{#1}}

\newcommand{\PP}{P}
\newcommand{\aP}{Q}
\newcommand{\bP}{R} 
\newcommand{\var}{x}
\newcommand{\avar}{y}
\newcommand{\bvar}{z}
\newcommand{\sn}{s}
\newcommand{\asn}{t}
\newcommand{\role}[1][p]{\kw{\color{purple}#1}}
\newcommand{\rolep}{\role[p]}
\newcommand{\roleq}{\role[q]}
\newcommand{\roler}{\role[r]}
\newcommand{\arole}{\kw{q}} 
\newcommand{\brole}{\kw{r}} 
\newcommand{\ep}[2]{#1[#2]} 
\newcommand{\chvar}{u}
\newcommand{\achvar}{v} 
\newcommand{\la}{l}
\newcommand{\ala}{m} 
\newcommand{\pcn}{c}
\newcommand{\pdn}{A}
\newcommand{\apdn}{B}
\newcommand{\bpdn}{C} 

\newcommand{\pspace}{\,} 
\newcommand{\psep}{.} 
\newcommand{\oact}{\kw{!}}
\newcommand{\iact}{\kw{?}} 
\newcommand{\pdone}{\mkkeyword{done}} 
\newcommand{\pclose}[1]{\mkkeyword{close}\pspace #1} 
\newcommand{\pwait}[2][\psep]{\mkkeyword{wait}\pspace #2#1} 
\newcommand{\act}[3] {#1[#2]#3} 
\newcommand{\pact}[4]{\act{#1}{#2}{#3}#4} 
\newcommand{\pbranch}[6][]{\pact{#2}{#3}{#4}{\ifblank{#1}{#5\psep #6}{\{#5\psep #6\}_{#1}}}}
\newcommand{\pobranch}[5][]{\pbranch[#1]{#2}{#3}{\oact}{#4}{#5}}
\newcommand{\pibranch}[5][]{\pbranch[#1]{#2}{#3}{\iact}{#4}{#5}}
\newcommand{\poch}[4]{\pact{#1}{#2}{\oact}{#3\psep #4}} 
\newcommand{\pich}[4]{\pact{#1}{#2}{\iact}{(#3)\psep #4}} 
\newcommand{\pcast}[1]{\lceil #1 \rceil}
\newcommand{\pres}[2]{(#1)(#2)}
\newcommand{\ppar}{\mathbin{\mid}} 
\newcommand{\pinvk}[2]{#1\langle #2 \rangle}
\newcommand{\pdef}[4][]{\Definition[#1]{#2}{#3}{#4}} 
\newcommand{\pchoice}[1][]{\oplus\ifblank{#1}{}{_{#1}}}


\newcommand{\angles}[1]{\langle#1\rangle}
\newcommand{\parens}[1]{(#1)}
\newcommand{\braces}[1]{\{#1\}}

\newcommand{\Done}{\pdone}
\newcommand{\Call}[2]{#1\ifblank{#2}{}{\angles{#2}}}
\newcommand{\POutput}[2]{#1 {\Out} #2}
\newcommand{\PInput}[2]{#1 {\In} #2}
\newcommand{\PSend}[2]{\POutput{#1}{\set{#2}}}
\newcommand{\PRecv}[2]{\PInput{#1}{\set{#2}}}
\newcommand{\PBranch}[5]{{#1}{#2}\{{#3}\psep{#4}\}_{#5}}
\newcommand{\Close}[1]{\pclose{#1}}
\newcommand{\Wait}[2]{\pwait{#1}{#2}}
\newcommand{\Cast}[1]{\pcast{#1}}
\newcommand{\NewPar}[3]{(#1)(#2 \parop #3)}
\newcommand{\Program}{\mathcal{P}}


\newcommand{\procs}[1]{\seqof{#1}}
\newcommand{\hasrole}[2]{#1[#2]} 

\newcommand{\peq}{\mathrel{\smash{\stackrel\vartriangle=}}} 
\newcommand{\Definition}[4][]{#2\ifblank{#3}{}{(#3)} #1 \peq #1 #4}

\newcommand{\tass}[3]{#1 : [\ifblank{#2}{()}{#2}; #3]}

\newenvironment{lines}[1][t]{
  \begin{array}[#1]{@{}l@{}}
}{
  \end{array}
}


\newcommand{\Main}{\textit{Main}}
\newcommand{\Buyer}{\textit{Buyer}}
\newcommand{\Seller}{\textit{Seller}}
\newcommand{\Carrier}{\textit{Carrier}}

\newcommand{\rbuyer}{\role[b]}
\newcommand{\rseller}{\role[s]}
\newcommand{\rcarrier}{\role[c]}

\newcommand{\tquery}{\Tag[query]}
\newcommand{\tprice}{\Tag[price]}
\newcommand{\tok}{\Tag[ok]}
\newcommand{\tcancel}{\Tag[cancel]}
\newcommand{\tbox}{\Tag[box]}
\newcommand{\tsplit}{\Tag[split]}
\newcommand{\tgiveup}{\Tag[giveup]}
\newcommand{\tyes}{\Tag[yes]}
\newcommand{\tno}{\Tag[no]}

\newcommand{\player}{\role[player]}
\newcommand{\rplayer}{\role[p]}
\newcommand{\tplay}{\Tag[play]}
\newcommand{\tquit}{\Tag[quit]}
\newcommand{\twin}{\Tag[win]}
\newcommand{\tlose}{\Tag[lose]}

\newcommand{\Sort}{\textit{Sort}}
\newcommand{\Merge}{\textit{Merge}}

\newcommand{\rworker}{\role[w]}
\newcommand{\rmaster}{\role[m]}

\newcommand{\treq}{\Tag[req]}
\newcommand{\tres}{\Tag[res]}

\newcommand{\SlotMachine}{\mathit{Slot}}


\newcommand{\rk}{\mathsf{wg}} 


\newcommand{\EmptyCtx}{\emptyset}
\newcommand{\Ctx}{\Gamma}
\newcommand{\CtxC}{\Gamma}
\newcommand{\CtxD}{\Delta}

\newcommand{\wtpx}[4]{#1 \vdash\ifblank{#2}{}{_{#2}}\ifblank{#4}{}{^{#4}} #3}
\newcommand{\wtp}[3][]{\wtpx{#2}{}{#3}{#1}}
\newcommand{\wtpc}[3][]{\wtpx{#2}\CoInd{#3}{#1}}
\newcommand{\wtpi}[3][]{\wtpx{#2}\Ind{#3}{#1}}
\newcommand{\wtpn}[4][]{#3 \vDash^{#2} #4}

\newcommand{\Measure}{\MeasureM}
\newcommand{\MeasureM}{\mu}
\newcommand{\MeasureN}{\nu}


\newcommand{\PCtx}{\mathcal{C}}
\newcommand{\PCtxC}{\mathcal{C}}
\newcommand{\PCtxD}{\mathcal{D}}
\newcommand{\Hole}{[~]}


\newcommand{\Xnf}[1]{#1^{\mathit{nf}}}
\newcommand{\Pnf}{\Xnf{P}}
\newcommand{\Qnf}{\Xnf{Q}}
\newcommand{\Rnf}{\Xnf{R}}
\newcommand{\Xpar}[1]{#1^{\mathit{par}}}
\newcommand{\Ppar}{\Xpar P}
\newcommand{\Qpar}{\Xpar Q}
\newcommand{\Rpar}{\Xpar R}
\newcommand{\Pth}{P^{\mathit{th}}}
\newcommand{\Qth}{Q^{\mathit{th}}}

%% file: macros/macros.tex

\newcommand{\eqdef}{\stackrel{\text{\tiny\sf def}}=}

\newcommand{\set}[1]{\{#1\}}

\newcommand{\brk}{~\\}

\setlength{\intextsep}{0.5pt}%
\newenvironment{musickey}{%
  \setlength\intextsep{0.5pt}
  \wrapfigure[4]{l}{0pt}
}{\endwrapfigure}

\newcommand{\beginnote}[1]{
		\begin{music}
			#1
		\end{music}}

\newcommand{\begintreble}{
	\lettrine{\makebox[1cm]{\includegraphics[height=1.7cm]{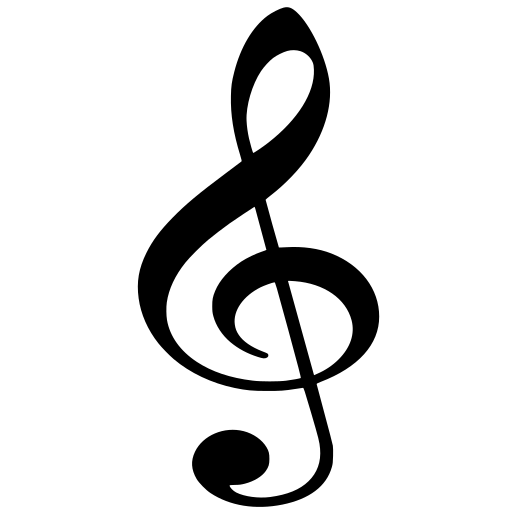}}}{}
	}

\newcommand{\beginalto}{
	\lettrine{\makebox[1cm]{\includegraphics[height=1cm]{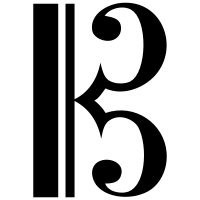}}}{}
	}
	
\newcommand{\beginbass}{
	\lettrine{\makebox[1cm]{\includegraphics[height=0.8cm]{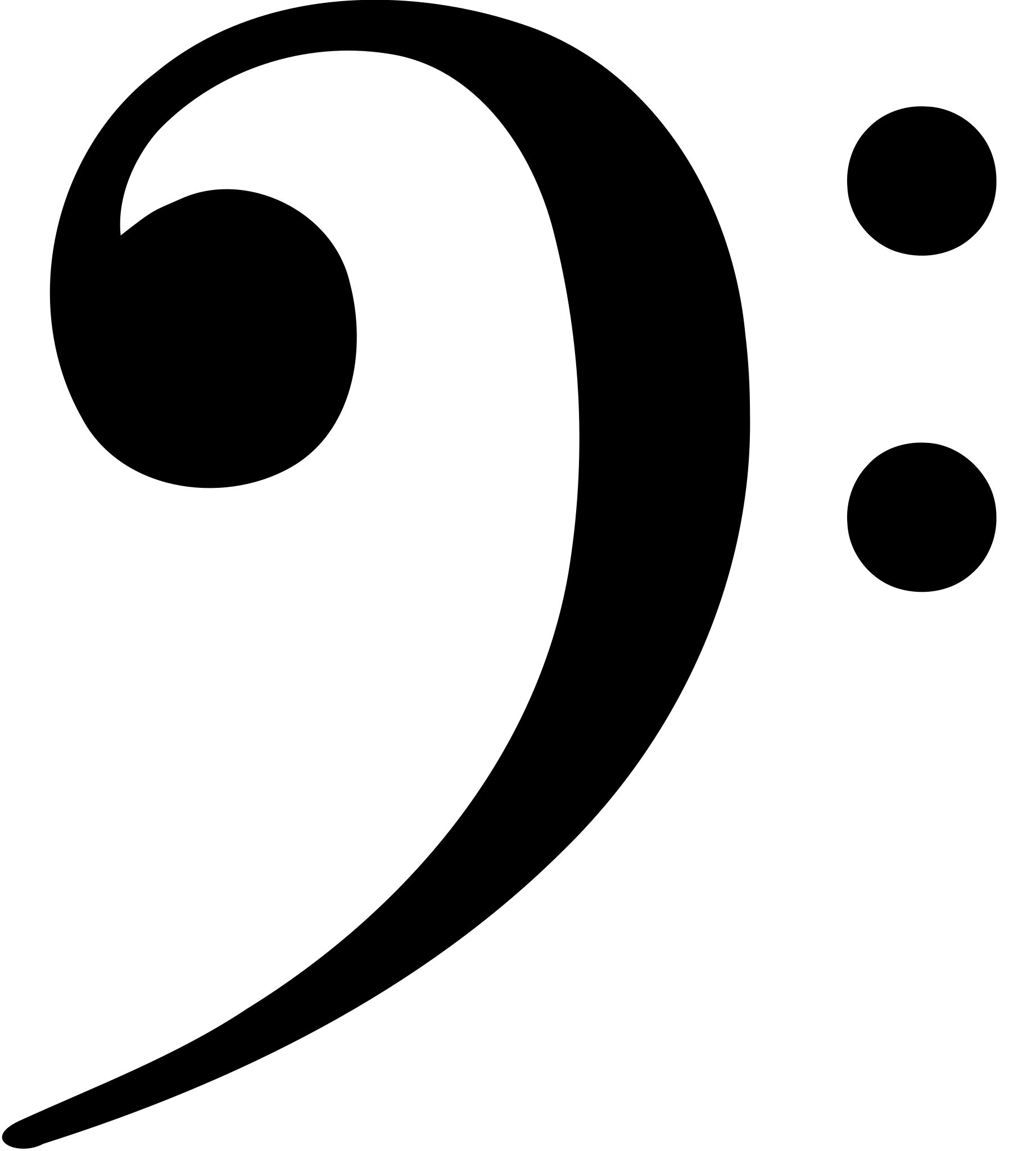}}}{}
	}


\newif\ifcomments
\commentstrue

\newcommand{\marginnote}[2]{%
  \ifcomments%
  $^{\color{magenta}\mathclap\star}$\marginpar{\sf\textbf{#1}: #2}%
  \fi%
}

\newcommand{\LC}[1]{\marginnote{\color{red}{LC}}{\color{red}#1}}
\newcommand{\LP}[1]{\marginnote{LP}{\color{blue}#1}}


\newcommand{\eg}{\emph{e.g.}\xspace}
\newcommand{\ie}{\emph{i.e.}\xspace}


\newcommand{\x}{x}
\newcommand{\y}{y}
\newcommand{\z}{z}


\newcommand{\GIS}{GIS\xspace}
\newcommand{\GISs}{GISs\xspace}
\newcommand{\Pair}[2]{\langle#1,#2\rangle}
\newcommand{\fun}[2]{#1 \to #2}
\newcommand{\annotation}[1]{\mathsf{#1}}
\newcommand{\Ind}{\annotation{ind}}
\newcommand{\CoInd}{\annotation{coind}}


\newcommand{\terminates}[2][]{#2{\Downarrow_{#1}}}
\newcommand{\compliance}[3][]{#2 \dashv_{#1} #3}
\newcommand{\subt}[1][]{\leqslant\ifblank{#1}{}{_{#1}}}
\newcommand{\usubt}{\subt[*]}
\newcommand{\ssubt}{\sqsubseteq}
\newcommand{\srel}{\mathcal{S}}

\newcommand{\isubt}{\subt[\Ind]}
\newcommand{\csubt}{\subt[\CoInd]}


\newcommand{\States}{\mathcal{S}}
\newcommand{\StateSet}{\mathcal{C}}
\newcommand{\FinalStates}{\mathcal{F}}
\newcommand{\run}{\rho}
\newcommand{\FairA}{\Phi}
\newcommand{\TF}{\mathbb{T}}


\newcommand{\rulename}[1]{\textsc{\small[#1]}}
\newcommand{\defrule}[2][]{\hypertarget{rule:\ifblank{#1}{#2}{#1}}{\ifblank{#2}{}{\rulename{#2}}}}
\newcommand{\refrule}[2][]{\hyperlink{rule:\ifblank{#1}{#2}{#1}}{\rulename{#2}}}
\newcommand{\proofcase}[1]{\textit{#1}.}
\newcommand{\proofrule}[1]{\proofcase{Case \refrule{#1}}}

\newcommand{\hl}[1]{{\color{lucared}#1}}

%% file: macros/is-macros.tex

\newcommand{\universe}{\mathcal{U}}
\newcommand{\is}[1][I]{\mathcal{#1}}
\newcommand{\cois}[1][I]{\is[#1]_{\mathsf{co}}}
\newcommand{\judg}{\mathit{j}} 
\newcommand{\InfOp}[1]{\mathit{F}_{#1}} 
\newcommand{\gis}[2]{\langle#1,#2\rangle}

\newcommand{\RulePair}[2]{\Pair{#1}{#2}} 
\newcommand{\Rule}[2]{\inferrule{#1}{#2}} 
\newcommand{\MetaRule}[4]{\rn{#1}\Rule{#2}{#3}\, \begin{array}{l} #4 \end{array} } 
\newcommand{\MetaCoRule}[4]{\rn{#1}\CoRule{#2}{#3}\, \begin{array}{l} #4 \end{array} } 
\newcommand{\rn}[1]{\textsf{\scriptsize{(#1)}}}
\newcommand{\prem}{\mathit{pr}}
\newcommand{\Restrictis}[2]{{#1}_{\mid #2}} 
\newcommand{\inst}[2]{#1(#2)}

\newcommand{\mycofr}[2]{\hbox{$\genfrac{}{}{1.5pt}{0}{#1}{#2}$}}

\newcommand{\infercorule}[3][]{%
  {\mprset{myfraction=\mycofr}\inferrule[#1]{#2}{#3}}%
}

\newcommand{\sem}[1]{\llbracket #1 \rrbracket} 
\newcommand{\Inductive}[1]{\textup{\textsf{Ind}}\sem{#1}}
\newcommand{\CoInductive}[1]{\textup{\textsf{CoInd}}\sem{#1}}
\newcommand{\FlexCo}[2]{\textup{\textsf{Gen}}\sem{#1,#2}}
\newcommand{\valid}[3][{}]{ #2 \vdash_{#1} #3 }
\newcommand{\validInd}[2]{ \valid[\mu]{#1}{#2} } 
\newcommand{\validCo}[2]{ \valid[\nu]{#1}{#2} }
\newcommand{\validFCo}[3]{\validCo{\ple{#1,#2}}{#3}} 
\newcommand{\Trees}[1]{\mathcal{T}} 
\newcommand{\TInfOp}[1]{\mathit{T}_{#1}} 
\newcommand{\tr}{\tau}
\newcommand{\rt}[1]{\mathsf{r}(#1)}
\newcommand{\chld}[1]{\mathsf{chl}(#1)} 
\newcommand{\subtr}[2]{{#1}_{\mid_{#2}}}
\newcommand{\dsubtrn}{\mathsf{dst}} 
\newcommand{\dsubtr}[1]{\dsubtrn (#1)} 
\newcommand{\rtdir}[1]{\img{\mathsf{r}}(#1)}

\newcommand{\Spec}{\mathcal{S}}
\newcommand{\SpecInfty}{\Spec^\infty}
%

\newcommand{\MCtx}{\mathcal{C}}
\newcommand{\mc}{c} 
\newcommand{\mr}{\rho}
\newcommand{\mj}{\mathit{J}} 
\newcommand{\mpr}{\mathit{Pr}} 
\newcommand{\sd}{\Sigma} 
\newcommand{\mis}{\mathcal{I}} 
\newcommand{\mcois}{\mis_{\mathsf{co}}}
\newcommand{\Ground}[1]{\|#1\|} 

%% file: macros/is-examples.tex
\newcommand{\listl}{l}
\newcommand{\nil}{[]}
\newcommand{\cons}[2]{#1 \mathbin{::} #2}
\newcommand{\maxelem}{\mathit{maxElem}}
\newcommand{\member}{\mathit{member}}
\newcommand{\allpos}{\mathit{allPos}\thinspace}

\newcommand{\Nat}{\mathbb{N}}

\newcommand{\listin}[2]{#1[#2]}

%% file: macros/sessiontypes.tex
\definecolor{lucared}{rgb}{0.5,0,0}
\definecolor{lucagreen}{rgb}{0,0.5,0}
\definecolor{lucablue}{rgb}{0,0,0.5}

\newcommand{\mkkeyword}[1]{\textsf{\upshape\color{lucablue}#1}}
\newcommand{\mktag}[1]{\textsf{\upshape\color{lucagreen}#1}}

\newcommand{\Tag}[1][m]{\mktag{#1}}

\newcommand{\tadd}{\Tag[add]}
\newcommand{\tpay}{\Tag[pay]}
\newcommand{\tship}{\Tag[ship]}
\newcommand{\tsearch}{\Tag[search]}

\newcommand{\ladd}{\Tag[add]}
\newcommand{\lpay}{\Tag[pay]}
\newcommand{\lship}{\Tag[ship]}
\newcommand{\lsearch}{\Tag[search]}
\newcommand{\lmore}{\Tag[more]}
\newcommand{\lstop}{\Tag[stop]}
\newcommand{\lzero}{\Tag[zero]}
\newcommand{\lone}{\Tag[one]}

\renewcommand{\l}{\Tag[m]}
\renewcommand{\la}{\Tag[a]}
\newcommand{\lb}{\Tag[b]}
\newcommand{\lc}{\Tag[c]}
\newcommand{\ld}{\Tag[d]}

\newcommand{\T}{T}
\newcommand{\R}{R}
\renewcommand{\S}{S}
\newcommand{\U}{U}
\newcommand{\V}{V}


\newcommand{\actor}[1]{\textbf{#1}}
\newcommand{\traces}[1]{\mathtt{tr}(#1)}
\newcommand{\prefix}{\leq}
\newcommand{\paths}[1]{\mathsf{paths}(#1)}

\newcommand{\co}[1]{\overline{#1}}
\newcommand{\SessionType}{\mathbb{S}}
\newcommand{\Message}{\mathbb{V}}
\newcommand{\WeaklyTerminating}{\mathbb{W}}
\newcommand{\FairlyTerminating}{\mathbb{F}}

\newcommand{\session}[2]{#1 \mathrel{\#} #2}

\newcommand{\vtrue}{\mathsf{true}}
\newcommand{\vfalse}{\mathsf{false}}
\newcommand{\Bool}{\mathbb{B}}

\newcommand{\f}{f}
\newcommand{\g}{g}

\newcommand{\End}[1][\Pol]{#1\mkkeyword{end}}
\newcommand{\Win}{\End[\Out]}
\newcommand{\TNil}{\mathsf{\color{red}nil}}
\newcommand{\Out}{{!}}
\newcommand{\In}{{?}}

\newcommand{\red}{\to}
\newcommand{\nred}{\arrownot\red}
\newcommand{\wred}{\Rightarrow}
\newcommand{\wlred}[1]{\stackrel{#1}\Longrightarrow}
\newcommand{\lred}[1]{\stackrel{#1}\longrightarrow}

\newcommand{\action}{\actionA}
\newcommand{\actionA}{\alpha}
\newcommand{\actionB}{\beta}
\newcommand{\actionC}{\gamma}

\newcommand{\actions}{\actionsA}
\newcommand{\actionsA}{\varphi}
\newcommand{\actionsB}{\psi}
\newcommand{\actionsC}{\chi}

\newcommand{\es}{\varepsilon}

\newcommand{\parop}{\mathbin{|}}
\newcommand{\Sys}[2]{{#1 \parop #2}}

\newcommand{\rank}[1]{\|#1\|}
\newcommand{\rankb}[2]{\|#1,#2\|}


\newcommand{\Branch}[1]{\In\set{#1}}
\newcommand{\Choice}[1]{\Out\set{#1}}
\newcommand{\branch}{+}
\newcommand{\choice}{+}

\newcommand{\Labels}{\mathcal{L}}

\newcommand{\compatible}[1][]{\sim\ifblank{#1}{}{_{\!#1}}}


\newcommand{\ft}{~\text{coherent}}
\newcommand{\coherent}{{\#}}

\newcommand{\buyer}{\role[buyer]}
\newcommand{\seller}{\role[seller]}
\newcommand{\carrier}{\role[carrier]}

\newcommand{\dom}[1]{\mathsf{dom}(#1)}
\newcommand{\RoleSet}{\mathsf{Roles}}
\newcommand{\targets}[1]{\mathsf{targets}(#1)}

\renewcommand{\rolep}{\role[p]}
\renewcommand{\roleq}{\role[q]}

\newcommand{\Tags}[1][i\in I]{\sum_{#1}}
\newcommand{\JTags}[1][j\in J]{\sum_{#1}}
\newcommand{\Pol}{\pi}

\newcommand{\Map}[1]{#1 \triangleright}

\newcommand{\terminated}{\checkmark}

\newcommand{\xlred}[1]{\xrightarrow{#1}}
\newcommand{\xwlred}[1]{\xRightarrow{#1}}

\newcommand{\converge}{\mathrel\downarrow}
\newcommand{\diverge}{\mathrel\uparrow}

%% file: macros/pi-calc.tex
\newcommand{\muMALL}{$\mu\textsf{\upshape MALL}^\infty$\xspace}
\newcommand{\piLIN}{\texorpdfstring{$\pi\textsf{LIN}$}{piLIN}\xspace}
\newcommand{\piDILL}{$\pi\textsf{DILL}$\xspace}
\newcommand{\CP}{$\textsf{CP}$\xspace}
\newcommand{\muCP}{$\mu\textsf{CP}$\xspace}

\newcommand{\mkfunction}[1]{\mathsf{#1}}
\newcommand{\rankof}[1]{|#1|}
\newcommand{\fc}[2]{\mathsf{fc}(#1,#2)}
\newcommand{\depth}[1]{\mathsf{depth}\parens{#1}}
\newcommand{\Resolve}[1]{\lfloor#1\rfloor}

\newcommand{\InfOften}[1]{\mkfunction{inf}\parens{#1}}


\newcommand{\Work}{\textit{Work}}
\newcommand{\Gather}{\textit{Gather}}
\newcommand{\ComplexTag}{\mktag{complex}}
\newcommand{\SimpleTag}{\mktag{simple}}
\newcommand{\Forwarder}{\textit{Fwd}}
\newcommand{\Player}{\textit{Player}}
\newcommand{\Machine}{\textit{Machine}}

\newcommand{\RankSet}{\Nat^{\infty}}
\newcommand{\rankR}{r}
\newcommand{\rankS}{s}

\newcommand\Run{\rho}

\newcommand{\InTag}{\mktag{in}}
\newcommand{\LeftTag}{\InTag_1}
\newcommand{\RightTag}{\InTag_2}
\newcommand{\AddTag}{\mktag{add}}
\newcommand{\PayTag}{\mktag{pay}}
\newcommand{\PlayTag}{\mktag{play}}
\newcommand{\QuitTag}{\mktag{quit}}
\newcommand{\WinTag}{\mktag{win}}
\newcommand{\LoseTag}{\mktag{lose}}

\newcommand{\out}[1]{\overline{#1}}
\newcommand{\inp}[1]{#1}

\newcommand{\Let}[3]{#1\ifblank{#2}{}{\parens{#2}}=#3}
\newcommand{\Link}[2]{#1\leftrightarrow#2}
\newcommand{\PiClose}[1]{\out{#1}\parens{}}
\newcommand{\PiWait}[1]{\inp{#1}\parens{}}
\newcommand{\Fail}[1]{\mkkeyword{case}\,\inp{#1}\braces{}}
\newcommand{\Select}[3][]{#2\,\out{#3}\ifblank{#1}{}{\parens{#1}}}
\newcommand{\Left}[2][]{\Select[#1]{\InTag_1}{#2}}
\newcommand{\Right}[2][]{\Select[#1]{\InTag_2}{#2}}
\newcommand{\CaseX}[4][]{\mkkeyword{case}\,\inp{#2}\ifblank{#1}{}{\parens{#1}}\braces{#3,#4}}
\newcommand{\Case}[4][]{\CaseX[#1]{#2}{#3}{#4}}
\newcommand{\Fork}[5][]{\out{#2}\parens{#3\ifblank{#1}{}{,#1}}\parens{#4\parop#5}}
\newcommand{\FreeFork}[3]{\out{#1}\angles{#2,#3}}
\renewcommand{\Join}[3][]{\inp{#2}\parens{#3\ifblank{#1}{}{,#1}}}
\newcommand{\Cut}[3]{\parens{#1}\parens{#2\parop#3}}
\newcommand{\PiChoice}[3][]{#2 \oplus\ifblank{#1}{}{_{#1}} #3}
\newcommand{\Rec}[2][]{\mkkeyword{rec}\,\out{#2}\ifblank{#1}{}{\parens{#1}}}
\newcommand{\Corec}[2][]{\mkkeyword{corec}\,\inp{#2}\ifblank{#1}{}{\parens{#1}}}


\newcommand{\AddressSet}{\mathcal{A}}
\newcommand{\address}{\addressA}
\newcommand{\addressA}{\alpha}
\newcommand{\addressB}{\beta}

\newcommand{\Formulas}{\Phi}
\newcommand{\Formula}{\FormulaF}
\newcommand{\FormulaF}{\varphi}
\newcommand{\FormulaG}{\psi}

\newcommand{\Type}{\TypeT}
\newcommand{\TypeT}{T}
\newcommand{\TypeS}{S}

\newcommand{\X}{X}
\newcommand{\Y}{Y}

\newcommand{\mkformula}[1]{#1}

\newcommand{\Bot}{\mkformula\bot}
\newcommand{\Top}{\mkformula\top}
\newcommand{\One}{\mkformula{\mathbf{1}}}
\newcommand{\Zero}{\mkformula{\mathbf{0}}}
\newcommand{\plinchoice}{\mathbin{\mkformula\oplus}}
\newcommand{\plinbranch}{\mathbin{\mkformula\binampersand}}
\newcommand{\tfork}{\mathbin{\mkformula\otimes}}
\newcommand{\tjoin}{\mathbin{\mkformula\bindnasrepma}}
\newcommand{\tmu}{\mkformula\mu}
\newcommand{\tnu}{\mkformula\nu}

\newcommand{\dual}[1]{#1^{\bot}}
\newcommand{\subf}{\preceq}
\newcommand{\minf}{\mkfunction{min}\,}
\newcommand{\strip}[1]{\overline{#1}}

\newcommand{\tred}{\leadsto}

\newcommand{\ffun}{\mathcal{F}}


\newcommand{\qtp}[3][]{#3 \vdash\ifblank{#1}{}{_{\color{red}XXX#1}} #2}
\newcommand{\piwtp}[3][]{#3 \Vdash\ifblank{#1}{}{_{\color{red}XXX#1}} #2}

\newcommand{\LinkRule}{ax}
\newcommand{\CutRule}{cut}
\newcommand{\FailRule}{$\Top$}
\newcommand{\PiCloseRule}{$\One$}
\newcommand{\PiWaitRule}{$\Bot$}
\newcommand{\ForkRule}{$\tfork$}
\newcommand{\JoinRule}{$\tjoin$}
\newcommand{\SelectRule}{$\plinchoice$}
\newcommand{\CaseRule}{$\plinbranch$}
\newcommand{\RecRule}{$\tmu$}
\newcommand{\CorecRule}{$\tnu$}
\newcommand{\PiChoiceRule}{choice}

%% file: macros/macros-lib.tex
\newcommand{\xs}{\mathit{xs}}
\newcommand{\FIList}[1]{#1^\infty}

\newcommand{\produzione}[3]{#1&::=&#2&\mbox{#3}}

\newcommand{\te}{\mathit{t}}
\newcommand{\Val}{\mathsf{Val}} 
\newcommand{\val}{\mathit{v}}
\newcommand{\infval}{\val^\infty} 

\newcommand{\Lam}[1]{\lambda #1.}
\newcommand{\appop}{\, }
\newcommand{\LambdaExp}[2]{\Lam{#1}#2}
\newcommand{\AppExp}[2]{#1\appop #2}

\newcommand{\ev}{\Rightarrow}
\newcommand{\evalsc}[3]{ {#1}\, {\Downarrow_{\scriptstyle #2}}\, {#3} }
\newcommand{\eval}[2]{ \evalsc{#1}{}{#2} }  
\newcommand{\lamsubst}[3]{{#1}[{#2}/{#3}]}

\newcommand{\SmallStep}[2]{#1\ev#2}

%% file: macros/macros-agda.tex
\newcommand{\NatPlus}{\Nat^{\texttt{+}}}

%% file: macros/macros-app.tex
\newif\ifextension
\extensiontrue

\newcommand{\p}{\Pol}

\newcommand{\aset}{\mathcal{A}}
\newcommand{\bset}{\mathcal{B}}

\newcommand{\Rank}[2][\aset]{\|#2\|\ifblank{#1}{}{_{#1}}}
\newcommand{\RankE}[1]{\|{#1}\|}

\newcommand{\psup}{\sqcup}

\newcommand{\IAlg}{\mathsf{alg}}

\newcommand{\wtpa}[3][]{\wtpx{#2}\IAlg{#3}{{\color{red}#1}}}

\newcommand{\pbounded}[2][\aset]{#1 \Vdash #2}

%% file: first.tex
\thispagestyle{empty}

\centerline {\Large{\textsc{ UNIVERSIT\`A DEGLI STUDI DI TORINO}}}
\vskip 10 pt

\centerline {\Large{\textsc DIPARTIMENTO DI INFORMATICA}}

\vskip 10 pt

\centerline {{\textsc SCUOLA DI SCIENZE DELLA NATURA}}

\vskip 10 pt
	
\centerline{\Large{\textsc Doctoral School in Computer Science - Cycle XXXV}}

\vskip 20 pt

\centerline{\titlefont{Concerto Grosso}}
\vskip 6 pt
\centerline{\subtitlefont{for}}
\centerline{\titlefont{Sessions}}
\vskip 6 pt
\centerline{\subtitlefont{Fair Termination of Sessions}}

\vskip 60 pt


\begin{center}
	\includegraphics[width=\linewidth , height=2.5cm]{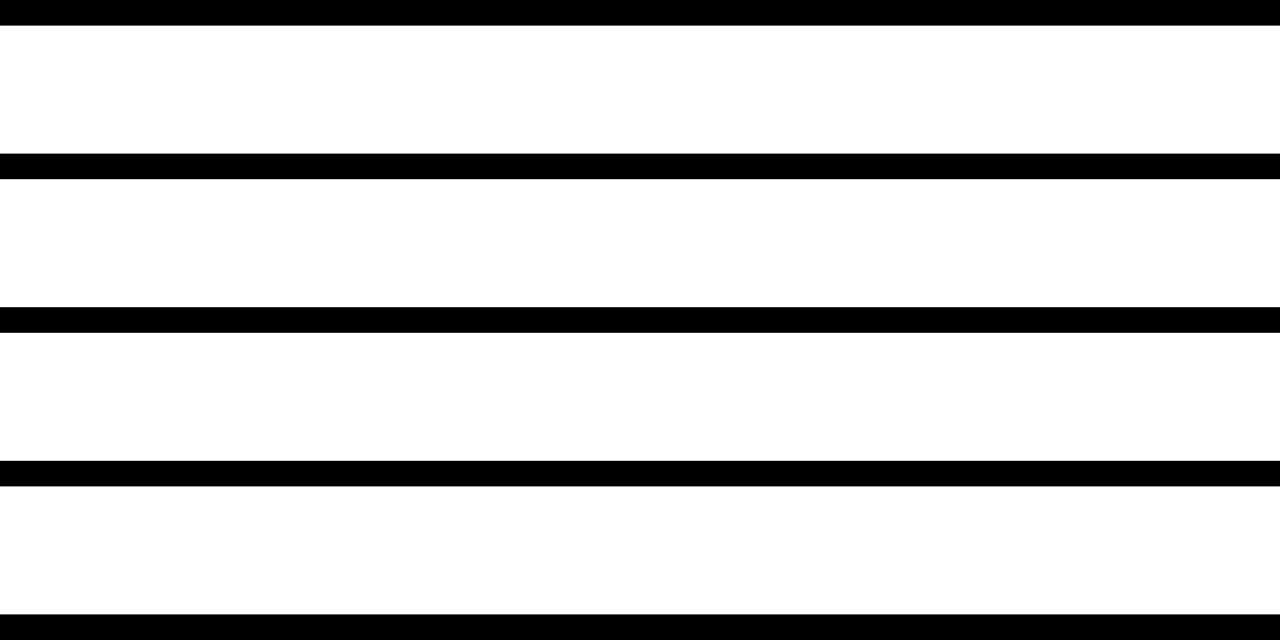}

	\raisebox{-20pt}[0pt][0pt]{
		\hspace{-3.3cm}
		\raisebox{2pt}[0pt][0pt]{\includegraphics[height=4.5cm]{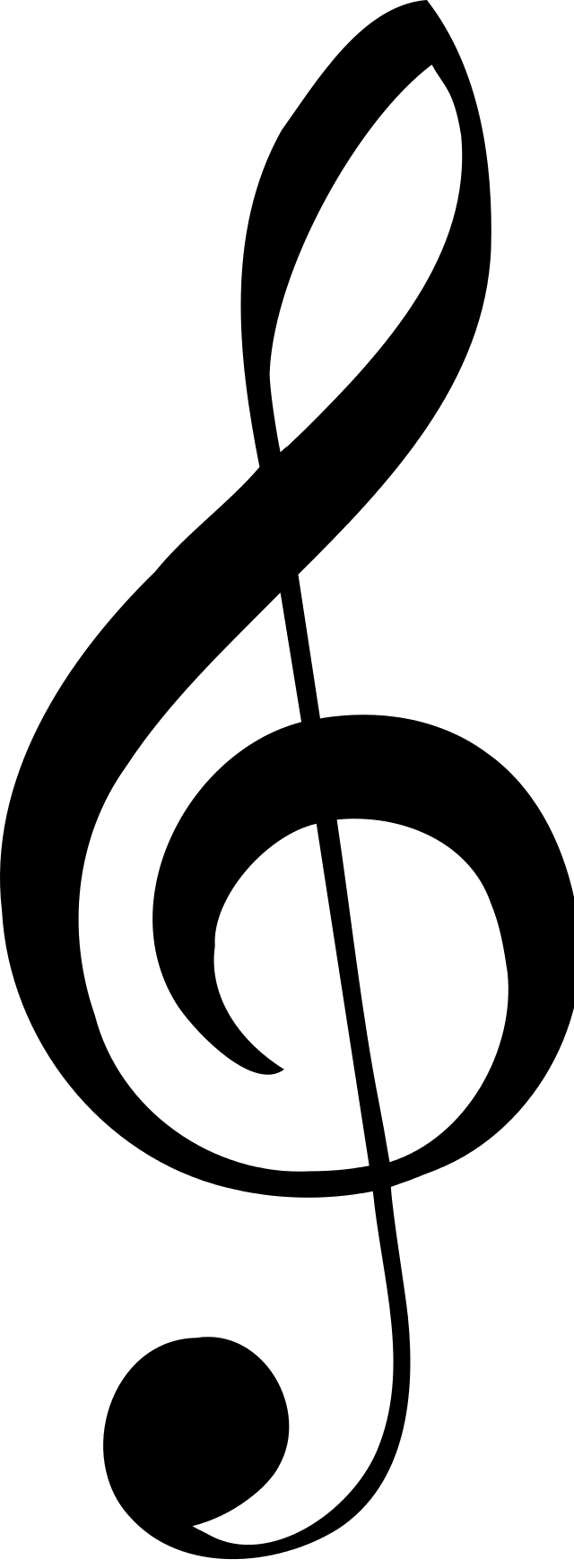}}
		\hspace{2.7cm}
		\includegraphics[width=5cm]{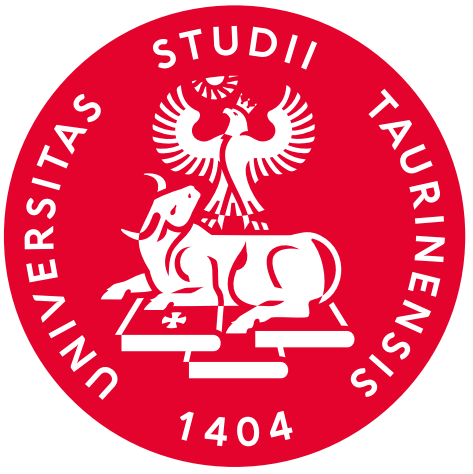}
		}
\end{center}

\vskip 20 pt

\par
\noindent
\begin{minipage}[t]{0.47\textwidth}\raggedright
	{\normalsize{\textbf{Candidate}\\ Luca Ciccone}}
\end{minipage}
\hfill
\begin{minipage}[t]{0.47\textwidth}\raggedleft
	{\normalsize{\textbf{Supervisors}\\ Prof. Luca Padovani\\ Prof. Ugo de'Liguoro}}
\end{minipage}

\vskip 0.50cm

\begin{center}
	\textbf{Reviewers}\\ Prof. Eugenio Moggi\\ Prof. Ornela Dardha
\end{center}

\vskip 0.50cm

\begin{center}
	\[
	\begin{array}{rl}
		\text{\textbf{\textit{PhD Coordinator}}:} & \text{Prof. Viviana Patti}
		\\
		\text{\textbf{\textit{Submitted on}}:} & \text{December 2022}
	\end{array}
	\]
\end{center}

%% file: acks.tex

\section*{\textit{Institutional}}

\paragraph{To University of Torino and the entire PhD board}
2020 and 2021 have been hard years for everyone. However, UniTo successfully managed to provide
courses and side activities in the best way possible. At the beginning of 2020 TYPES conference should have been
held in Torino but COVID was spreading in Italy. The conference moved to a virtual space. 
No one knew it was the first of a long sequence of remote events. 
PhD related events have been obviously affected as well. 
Nonetheless, courses have been well organized fast in a remote fashion.

\paragraph{To my supervisor Luca Padovani}
I skip the COVID part as it would be obvious and I move to mere professional considerations.
As soon as my PhD started Luca and I discussed about the ways in which we could have taken advantage of
the work I did during the master thesis. This way we immediately found the right intersection between by knowledge
and my PhD project and we started working together. I never worked \emph{for} but \emph{with} him. 
Such an approach led to fast and satisfactory research outcomes. Moreover Luca always shared his connections with other 
research groups, \eg at University of Glasgow.
I have been his first PhD student so I hope that these lines will give him a feedback about his supervision.

\paragraph{To University of Genova and Elena Zucca}
My master thesis supervisor Elena Zucca agreed with Luca for giving me an office at DIBRIS, UniGe. 
This way we could start a fruitful research collaboration. 
During last year we managed to publish together a paper at ECOOP that earned the \emph{distinguished paper} award.
Hence, I'm grateful to both the University of Genova and Elena for allowing me to use such office.

\paragraph{To EuroProofNet, University of Glasgow and Ornela Dardha}
After two years of remote human interactions, in-person events started again. During the pandemic I have been in touch with 
Ornela Dardha whose research interests were very close to mine. Unfortunately I could not have met her research group.
In the meantime EuroProofNet, a european project aimed at connecting researchers on formal proofs, was born.  
Such project gave me the funds for making two visits to Ornela Dardha at University of Glasgow. 
I have been welcomed in the best way. 

\paragraph{To the reviewers}
I'm grateful to Eugenio Moggi and Ornela Dardha for having spent a lot of time 
reading this thesis. They provided useful comments and important references.
I hope this work will be useful for their research groups.


\section*{\textit{...a little flattery}}

\paragraph{To my super-visor Luca Padovani}
Coming to personal considerations, I am grateful to Luca for many reasons.
In general, as an empathetic person I found his approach of sharing not only his ideas
but also his feelings very inspiring. I learned how to face a variety of situations 
thanks to him, from the anxiety in meeting deadlines to the tricks in providing 
successful presentations. These skills will be helpful for my career.
Last but not least, I'd like to thank Luca for his support during the pandemic when I had 
serious COVID problems in my family; we were writing the ICALP paper and the result has been 
successful.

\paragraph{To my unofficial supervisor Elena Zucca}
I'd like to express my gratitude to Elena for her precious suggestions during the PhD and before
starting it. She put me in touch with Luca Padovani at the end of my master thesis.
Although she was not officially reported as my co-supervisor, she demonstrated a high 
involvement in working together as well as with my colleague and friend Francesco Dagnino (UniGe).
As for Luca, I'd like to thank her for her support when I had COVID problems in my family. 
When I was working on the ICALP paper with Luca, in the meantime I was writing the ITP one
with Elena and Francesco. Such paper has been accepted and together with the PPDP one it 
introduced me to the formal proofs community.

\paragraph{To Ornela Dardha}
I had the opportunity of meeting Ornela only last year. She welcomed me in Glasgow 
as a friend more than as a professor. 
I immediately discovered that it was not just 
a special treatment for visitors as she has a very special way of dealing with her research group.
\begin{wrapfigure}{r}{0.45\textwidth}
\centering
    \includegraphics[scale=0.23]{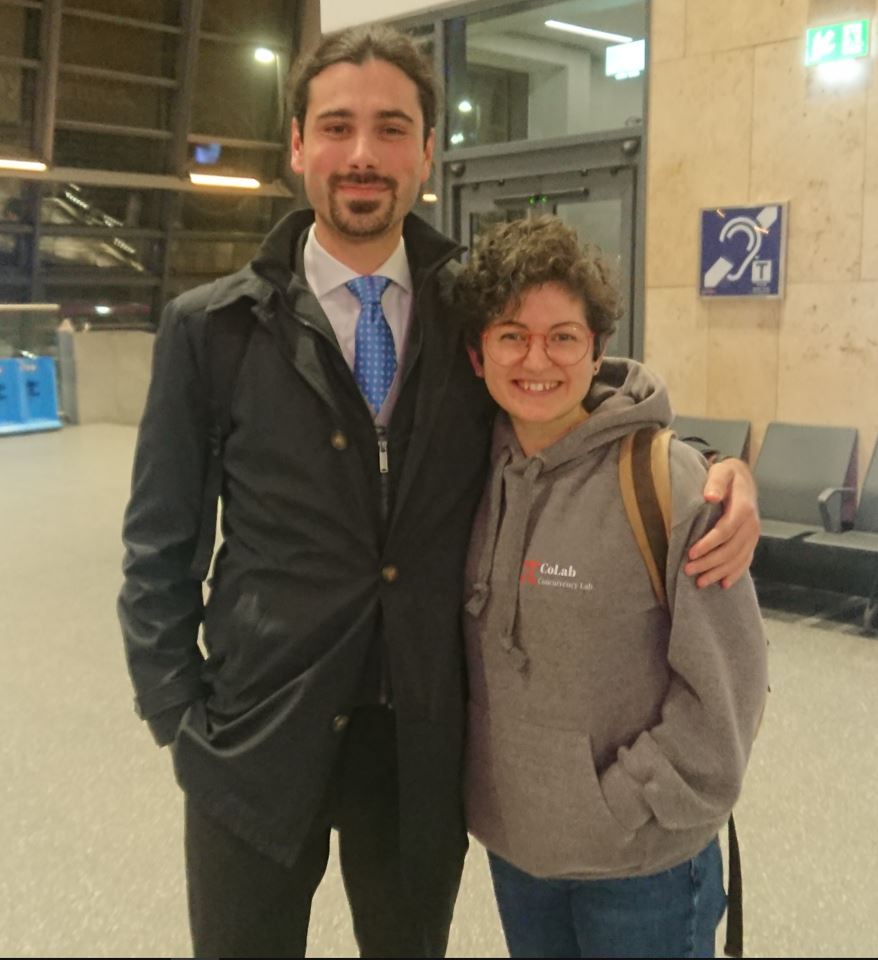}
    \par\vspace{2pt}
    \textit{SPLS, 26th October 2022.}
\end{wrapfigure}
Although I'm usually not so keen on those situations in which 
\textit{colleagues} and \textit{friends} automatically overlap, I'm sure that Ornela's approach 
is fantastic both for her students and visitors. The reason behind my opinion is that
Ornela is a very empathetic person and she cares about the mood of others a lot.
For this reason I decided to visit her two more times the same year and I really hope to meet her again
even if I will not be involved in the research world anymore.\\
\textit{That's okay for today. Now you should go to the pub! (cit.)}


\section*{\textit{Personal}}

\paragraph{To my family}
This is a special paragraph. I'm usually in trouble when I have to write something
to express my gratitude to my family as putting on paper deep feelings is a very hard task 
(and I also have to deal with my eyes filled with tears). What lies behind the usual sentence
\textit{for their support and blah blah}? I would like to completely change the point of view
and try something new. Obviously, the focus is on my parents Renato Ciccone and Marirosa Gaggero.
Let me start with a consideration: I trust stereotypes and in Genova we are famous as people that 
do not want to spend money. My studies were not free of course, so I'm very grateful to my parents 
for the sacrifices they made for paying a lot for so many years. 
Although I managed to win scolarships different times,
I always used such money for whatever but paying university subscriptions; I was younger and naive.
Furthermore, when I decided to start a PhD, I always asked myself whether such a title would have
been useful for my career outside the academy. 
I honestly shared my doubts with my parents. In the end I found 
a satisfactory number of job opportunities positively influenced by these three years of research.
At last, the years of PhD, and in particular the last one, made me grow up a lot as a person.
Hence, I hope to have made my parents proud of the results.

\paragraph{To my grandmother Anna (Ciccone) Peruzzi}
Things become harder. She is a dresskmaker for children, she loves classical music, 
she attends theatres and music associations...more in general, she loves beauty in its essence.
\textit{What is a person without any cultural interest?}, she always asked.
From her I learned the things to appreciate and study in my free time and that I deepened every day.
I'm grateful to her for her ambitions towards me that I hope to have satisfied.

\paragraph{To my ggf Ruggero Peruzzi}
\begin{wrapfigure}{r}{0.35\textwidth}
\centering
    \includegraphics[scale=0.3, angle=180]{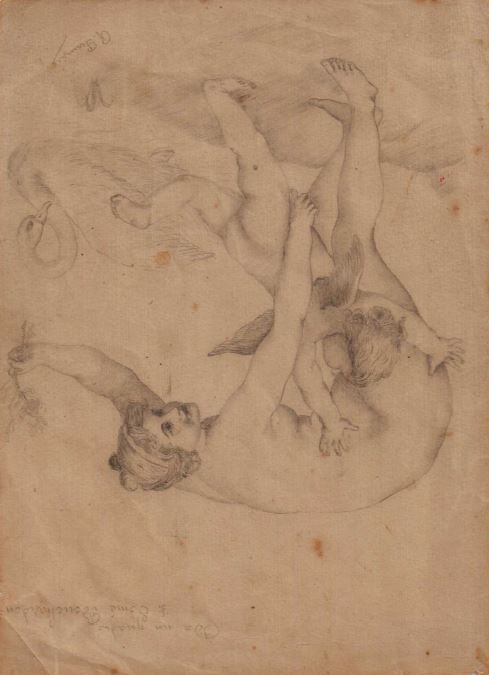}
    \par\vspace{2pt}
    \textit{Ruggero Peruzzi after Edmé Bouchardon.}
\end{wrapfigure}
He was born on 19 Novemeber, 1891. 
He was an antiquarian in Genova. 
My grandmother learned from him the love of beauty. 
During a journey he met the painter C. Tafuri and he paid for him an office in
Palazzo della Borsa in Genova. 
During WW2 he has been captured by fascists as they wanted the names of 
some partisans. He never told such information and fascists started torturing him. 
He had only one opportunity to say goodbye to her daughter.
He has been killed by fascists and his body has been found on 18 February 1945
in front of Brignole train station in Genova where you can find his tombstone.

\vfill

\paragraph{Concerto Grosso?}
This last acknowledgment goes to A. Schnittke, A. Schönberg, A. Berg, J. S. Bach, B. Galuppi, A. Pärt
as well as to many other great composers. In particular, Schnittke's music has been the soundtrack of
the last three years. His \emph{Concerti Grossi} obviously recall baroque music but at the same time they
are progressively cracked and filled with XXth century influences.

%% file: prelude.tex
\begintreble
Nowadays concurrent and distributed systems are ubiquitous in the computer science world.
It is not surprising that there exists a wide research about how to enforce some desired 
properties of communications in such systems.

\section*{A Gentle Introduction}

\paragraph{Sessions}
The $\pi$-calculus is a process calculus used for modeling concurrent systems and it is based
on \emph{channels} that are owned by the entities involved in the communication and that are
used to exchange messages.
Session types \citep{Honda93} are a formalism to describe communication protocols. 
A session is represented by a channel that connects processes and session types are assigned to them.
They describe how the channels will be used by the processes that own them.
Session types have been originally introduced to model the interaction between \emph{two} entities
and they have been later generalized to multiparty \citep{HondaYoshidaCarbone08}.
There exists a broad family of desired properties that one would like to enforce in a communication.
Among them, the most common ones are that messages are exchanged in the right order (\emph{protocol fidelity}) 
or that there are no communication errors (\emph{communication safety}).

\paragraph{Safety vs Liveness - Fair Termination}
In general, properties can be divided in \emph{safety} (nothing bad ever happens) and 
\emph{liveness} (something good will eventually happen) ones \citep{OwickiLamport82}.
According to the class to which it belongs, a property is defined using different and dual
techniques: \emph{induction} for safety and \emph{coinduction} for liveness.
Actually, there exist some properties that fall in between as they combine safety and
liveness aspects. We can informally describe them using the motto ``something good always eventually
happens'' and their dual. Hence, such properties are hard to deal with as they mix induction
and coinduction.
In a communication-based scenario, protocol fidelity is a safety property and the eventual termination
of a session is a liveness one.
In this thesis we study \emph{fair termination} which is the property of those sessions that can always
eventually (successfully) terminate under a fairness assumption. As the reader might guess, fair termination
joins a safety and a liveness part. We investigate such property as it entails many well knows properties
that are studied in the literature.

\paragraph{Fair Subtyping}
Subtyping between session types is a fundamental relation in this thesis. It induces
a substitution principle \citep{LiskovWing94} which states that a process that uses an
endpoint according to some type can be substituted with another process that uses the
endpoint according to a supertype. Although the principle seems in the wrong direction
it is proved to be equivalent to the right-to-left variant by taking into account endpoints
instead of processes \citep{Gay16}. 
The original subtyping relation \citep{GayHole05} is studied to preserve the safety properties of
a session but not the livess ones. Indeed an application of the original subtyping can 
break the termination property of the involved session.
In this thesis we rely on \emph{fair subtyping} \citep{Padovani14,Padovani16} which is
a liveness-preserving variant of the subtyping in \cite{GayHole05}.

\paragraph{(Generalized) Inference Systems}
Inference systems are a widely used framework in computer science for defining predicates by means of (meta) rules.
They support both inductive and coinductive definitions and they have been recently generalized
to deal with those predcates that mix the two approaches \citep{AnconaDagninoZucca17}.
Roughly, an inference system, that is, a set of (meta)rules, is interpreted either
inductively or coinductively. In a generalized inference system the key ingredient
is the addition of (meta)corules and the interpretation is obtained by properly mixing the inductive and
the coinductive one. 
Then, inference systems also provide \emph{proof principles} for proving the correctness of
a definition with respect to a semantic one called \emph{specification} in a canonical way.
Correctness is usually expressed in terms of \emph{soundness} and \emph{completeness}.
The reference principles are \emph{induction}, \emph{coinduction} which involve
inductive and coinductive predicates, respectively.
\emph{Bounded coinduction} is a generalization of coinduction and is used when dealing
with definitions through generalized inference systems.
We will rely on (generalized) inference systems for defining most
of the notions we will introduce.

\paragraph{Proof Assistants}
These days there is an increasing interest in \emph{theorem provers}, that is, tools
that allow users to write proofs by relying on a language that resembles a programming language.
The advantage of using proof assistants is that the proofs are certified (provided that the tool is
itself consistent). 
Due to the novelty of the research field, there is a focus in the literature in providing big libraries
with the aim of making the community agree on some design techniques. For example,
when formalizing calculi, De Bruijn indices are usually preferred for dealing with variables.
Many research communities are starting to rely on theorem provers to mechanize existing theories.
The community working on behavioral types is affected by this phenomenon as well.
Moreover, apart from traditional conferences on formal proofs (\eg CPP, ITP), a very first
workshop on \emph{Verification of Session Types} has been organized in 2020 and 2021. 
The main output of such event was the lack of a reference methodology for mechanizing
session-based theories. In this thesis we will refer to the Agda proof assistant \citep{Agda}.
Notably, we refer to theorem provers and proof assistant interchangeably.

\paragraph{Agda}
Agda \citep{Agda} is an interactive proof assistant based on intuitionistic logic. The interactivity aspect
is fundamental as the user is guided step-by-step while mechanizing a proof.
The tool can be download for different platforms and comes with only the very basic
definitions, \eg natural numbers. The \emph{standard library} can be added separately
and it is constantly updated in order to keep the modules consistent with the latest
version of Agda. 
Moreover, Agda has built-in support for both inductive and coinductive predicates.
In \cite{Ciccone20} we made an in-depth analysis of all the techniques that users can
follow when mechanizing predicates and we found out that pure (co)inductive
inference system have a natural and intuitive encoding in Agda.
However, for what concerns those predicates defined by generalized inference systems,
there is no support. Indeed, although a generalized inference system can still be 
defined, the resulting code becomes hard to understand and it contains many duplicated
notions.

\section*{Contributions}

\begin{center}
\Large{Can well-typed programs go well?}
\end{center}

\textbf{Developing type systems for enforcing \emph{liveness} properties of processes is not 
an easy task}. For example, in the context of this thesis the type systems 
of \cite{Kobayashi02,Padovani14} adopt a sophisticated technique for guaranteeing 
that all pending actions will be eventually executed. The question we want to answer
is in sharp contrast with the famous Robin Milner's sentence ``Well typed programs cannot
go wrong'' \citep{Milner78} which points out the fact that type systems are usually made for enforcing 
\emph{safety} aspects.

\paragraph{Fair Termination}
We present type systems for \textbf{enforcing fair termination} in three different scenarios.
First, we deal with \emph{binary} sessions, that is, sessions involving only two
participants. This base case is fundamental to highlight all the technical aspects
and additional properties that the type system must ensure in order to guarantee
fair termination.
Then, we generalize such type system to the \emph{multiparty} case. Although
many notions can be easily adapted from the binary case, there are some technical
details that are hard to deal with. For what concerns the additional properties that
the type system has to enforce, we can still rely on the binary type system since binary
sessions are the simplest multiparty ones. 
At last, we show a type for enforcing fair termination in the linear $\pi$-calculus.
This last part falls in between the previous type systems and those type systems
interpreting session-based calculi in the context of linear logic.
The interesting difference between the type systems for sessions and the one for the $\pi$-calculus
is that in the second there is no notion of fair subtyping.

\begin{center}
\Large{
	Can Agda support definitions mixing induction and coinduction?\\
	Can Agda proofs match the pen-and-paper ones?}
\end{center}

As we mentioned before, Agda supports purely (co)inductive definitions. Moreover,
a predicate defined using an inference system can be naturally encoded.
On the other hand, when we deal with a predicate that involves both induction and coinduction, 
there is no built-in support \citep{Ciccone20}. 
As a result, we can still mechanize such predicates but
the resulting code contains many duplicated notions and it differs from the definition
given through a generalized inference system.
Another drawback of mechanizing notions in Agda is that the proofs become very hard to
be understood as they differ from the hand written ones since they require many additional
lemmas related to the technicalities of the proof assistant, \eg equalities.
Hence, we first ask whether it is possible to \textbf{encode generalized inference systems in Agda
in a natural way}. Then we ask if we can ``hide'' all the Agda related technicalities
so that the \textbf{proofs are shown to the user in a more friendly flavor}.

\paragraph{An Agda Library for Inference Systems}
(Generalized) inference systems are used throughout the entire thesis. 
Since we aimed at mechanizing some results in Agda, we thought about a way for
supporting the framework in the proof assistant. For this sake we present a
\textbf{library} that allows users to formalize inference systems using a syntax that
resembles the one used on paper. Moreover, the library supports the usage of
corules that are not built-in supported (see \cite{Ciccone20} for more details).
Clearly, all the proof principles are encoded as well.
As an important and useful side effect, the codes of the proofs based on the library
resemble the proofs written on paper. We think that this contribution 
can be very helpful for the community since closed proofs are usually very hard
to read and understand.

\paragraph{Mechanizing Properties of Session Types}
We take advantage of the library we developed to \textbf{mechanize some notions in the
session types context}.
We first mechanize an alternative definition of session types that aims at
simplifying the codes as much as possible. Such characterization has also the 
advantage of supporting \emph{dependency} with respect to previously exchanged messages.
Then we focus on three interesting properties of session types and we characterize them
using generalized inference systems: fair termination of a session type, fair compliance
(successful fair termination) of a binary session and fair subtyping.
We mechanize the three inference system as well as their correctness proofs with
respect to the semantic definitions.
As mentioned before, the community working on session types is increasingly interested
in proof assistants but so far there is a lack of a reference methodology.
We hope that our results and our experience will be used for finding 
a guideline on which all the community agrees.

\section*{Structure of the Thesis}

\paragraph{Concerto Grosso}
As the title might suggest, the thesis is structured in three parts as the original italian
\emph{Concerto Grosso} form of baroque music.
Moreover, since binary sessions are the simplest multiparty ones, we will jump between the two
scenarios according to our needs. In general, we will show paradigmatic examples in the binary setting
and we will prove the main notions in the more general multiparty one. 
The continuous alternation of communications between two and more participants 
resembles again the \emph{Concerto Grosso} which is based on a dialogue between a small ensemble
and the full orchestra.
The tempo indications at the beginning of each part are consistent with the fast-slow-fast schema
of the concerto grosso and we use them to jokingly indicate that the part that should be read more
carefully is the central one. At last, in order to help the reader we use treble, soprano and bass clefs to
denote texts inside chapters, sections and subsections, respectively.

\paragraph{Part I: Key Notions}
There are different notions that we will use throughout \Cref{pt:type_systems,pt:agda}.
Thus, we decided to dedicate the first part to introduce all the main ingredients of
the thesis. 
At the beginning we discuss the usual classification of properties in safety and liveness ones in \Cref{sec:sft_lvn}.
Then, in \Cref{sec:gis} we introduce (Generalized) inference systems as a reference framework
and in \Cref{sec:st} we introduce syntax and semantics of binary and multiparty session types.
We dedicate \Cref{ch:ft} to explain what is \emph{fair termination} and to express it in a general
setting, that is, considering an arbitrary transition system. Then, we instantiate the property
in the session types context.
At last, in \Cref{ch:fs} we focus on the subtyping relation between session types.
We first introduce the original subtyping relation \citep{GayHole05} and we show why 
it can break the liveness properties on the session on which it is applied.
We then introduce \emph{fair subtyping} as a liveness-preserving refinement of the original relation.
We present fair subtyping in two different flavours; first using a purely coinductive
inference system and then using a generalized inference system.

\paragraph{Part II: Type Systems}
This part is dedicated to the study of the three type systems for fair termination
and it is split according to the calculus under analysis.
In \Cref{ch:ft_bin} we present a type system for a calculus with binary sessions.
\Cref{ch:ft_multi} generalizes such type system to the multiparty case.
At last, in \Cref{ch:ft_ll} we show a type system for a $\pi$-calculus that is
based on linear logic.
The three chapters are stuctured similarly.
We first introduce the calculus (and the types) and then we show the typing rules.
We dedicate the rest of each chapter to detail the soundness proofs.
As we mentioned before, in \Cref{ch:ft_bin} we focus on those paradigmatic 
examples that point out the additional properties that the type system must enforce.
Since such examples hold in the multiparty setting as well, in \Cref{ch:ft_multi}
we present some more involved processes.

\paragraph{Part III: Agda Mechanizations} 
This third and last part focuses on the mechanization aspects.
In \Cref{ch:gis_lib} we present the library for supporting in Agda generalized
inference systems. We first mechanize the meta theory of inference systems as well
as the proof principles and then we show some basic examples of usage.
Then in \Cref{ch:agda_prop} we present the definitions through generalized inference
systems of three properties of session types: fair termination of a session type,
fair compliance of a binary session and fair subtyping.
Notably, we mechanize the general notion of \emph{fair termination} and we 
relate it to the semantic definitions of the first two properties.
For each property we show the mechanization of the inference system and
we show how proof principles are used in the correctness proofs.
At last, we detail the proofs of fair compliance.

\section*{Published Works}

We conclude the introduction by relating the content of the thesis with
the published works of the author. 

\paragraph{Fair Termination}
\Cref{ch:ft_bin} is a refined version of \cite{CicconePadovani22}.
Differently from \cite{CicconePadovani22}, here we adopt the new characterization
of fair subtyping that we introduced later in \cite{CicconeDP22}.
\Cref{ch:ft_multi}, \Cref{ch:ft_ll} show the type systems that have been presented in \cite{CicconeDP22}
and in \cite{CicconeP22b}, respectively.

\paragraph{Agda}
The Agda library for (Generalized) Inference Systems in \Cref{ch:gis_lib} has been presented
in \cite{CicconeDagninoZucca21} which is based on \cite{Ciccone20}. 
At last, the Agda formalizations of the properties of session types in \Cref{ch:agda_prop}
have been presented in \cite{CicconeP22@lmcs,CicconePadovani21}.

\paragraph{Not in this thesis}
In \cite{CicconePadovani20} we provide an Agda mechanization of a linear $\pi$-calculus
with dependent types. One of the interesting features of such formalization is that we 
can inject Agda terms in the calculus by using a proper constructor 
for the terms of the calculus.
Although that work covers some topics that are related to the present thesis
(\eg $\pi$-calculus, Agda mechanization, induction/coinduction), we do not give details
about it as it falls outside the core topic the thesis.
Moreover, we are working on an Agda unification library and on 
co-contextual typing inference algorithms for the linear and shared $pi$-calculus
proved to be correct in the proof assistant. Such work is based on \cite{ZalakainD21}.

%% file: keynotions/intro.tex
\begintreble
This chapter is dedicated to the analysis of the main notions that are treated throughout the
thesis.
First, we recall the usual distinction of properties of concurrent systems in 
\emph{safety} and \emph{liveness} ones \citep{OwickiLamport82}.
According to the class to which it belongs, a property must be defined using a different (dual) technique.
The same happens when we want to prove the correctness of such definitions.
Then, we introduce \emph{Generalized Inference Systems} \citep{AnconaDagninoZucca17}. 
Inference systems are a well established framework in the literature to define purely (co)inductive predicates 
as they provide canonical techniques for proving the correctness of a definition.
In this thesis we mainly focus on their generalization with \emph{corules} in order to treat those predicates
that require a mix of induction and coinduction.
After that we move to the main topics of the thesis.
We give an overview of \emph{session types} by describing what such types are and what they are used for.
In a few words, they are used to model the communication in message passing systems and to statically
enforce some desired properties. They have been originally introduced by \cite{Honda93} in a binary context
and later generalized by \cite{HondaYoshidaCarbone08} to \emph{multiparty} to deal with those scenarios
in which more entities participate to the communication.
We then review some interesting desirable properties that fall in the intersection between safety and liveness ones.
%

The chapter is organized as follows.
In \Cref{sec:sft_lvn} we review the usual classification of properties into safety and liveness ones and we
show that some fall in the intersection of the two classes.
In \Cref{sec:gis} we introduce generalized inference systems as a reference framework that can be used to deal with
such kind of properties.
At last, in \Cref{sec:st} we introduce session types in both their binary and multiparty variants.

%% file: keynotions/sft-lvn.tex
\beginalto
It is well known that properties can be
distinguished between \emph{safety} and \emph{liveness} one.
Informally, the two classes are identified by the mottos
``nothing bad ever happens'' and
``something good eventually happens'' \citep{OwickiLamport82}.
For example, in a network of communicating processes, the absence of
communication errors and of deadlocks are safety properties, whereas the fact
that a protocol or a process can always successfully terminate is a liveness
property.
Because of their different nature, characterizations and proofs of safety and
liveness properties rely on fundamentally different (dual) techniques: safety
properties are usually based on invariance (coinductive) arguments, whereas
liveness properties are usually based on well foundedness (inductive)
arguments \citep{AlpernSchneider85,AlpernSchneider87}.

The correspondence and duality between safety/coinduction and liveness/induction
is particularly apparent when properties are specified as formulas in the modal
\emph{$\mu$-calculus} \citep{Kozen83,Stirling01,BradfieldStirling07}, a modal
logic equipped with least and greatest fixed points: safety properties are
expressed in terms of greatest fixed points, so that the ``bad things'' are
ruled out along all (possibly infinite) program executions; liveness properties
are expressed in terms of least fixed points, so that the ``good things'' are
always within reach along all program executions. Since the $\mu$-calculus
allows least and greatest fixed points to be interleaved arbitrarily, it makes
it possible to express properties that combine safety and liveness aspects,
although the resulting formulas are sometimes difficult to understand.

In the next examples we informally state some (co)inductive properties that 
we will characterize in details in \Cref{sec:gis} and that we will characterize in
Agda in \Cref{sec:agda_gis_meta}.

\begin{example}
	\label{ex:memberof}
	\emph{``$\x$ belongs to the possibly infinite list $\listl$''}
	
	This is an example of a liveness property. Indeed, even if the list $\listl$ is infinite,
	if $\x$ belongs to $\listl$, then we need to inspect only a finite portion of
	the list. Hence, induction is required.
	\eoe
\end{example}

\begin{example}
	\label{ex:allpos}
	\emph{``All the numbers in the possibly infinite list $\listl$ are positive''}
	
	This is an example of a safety property. Indeed, in order to actually state that
	a list is made of only positive numbers, we have to inspect it entirely. It is
	clear that coinduction is required since the list can be infinite.
	\eoe
\end{example}

\begin{example}
	\label{ex:maxelem}
	\emph{`` $\x$ is the maximum element of the possibly infinite list $\listl$''}
	
	The predicate is composed by a safety and a liveness part.
	\begin{itemize}
	\item \emph{Safety}: $\x$ is greater that all the elements in $\listl$
	\item \emph{Liveness}: $\x$ belongs to $\listl$
	\end{itemize}
	Hence, this property requires a mix of induction and coinduction.
	\eoe
\end{example}

%% file: keynotions/gis.tex
\beginalto
A different way of specifying (and enforcing) properties is by means of
\emph{inference systems} \citep{Aczel77}. Inference systems admit two natural
interpretations, an inductive and a coinductive one respectively corresponding
to the least and the greatest fixed points of their associated inference
operator. Unlike the $\mu$-calculus, however, they lack the flexibility of
mixing their different interpretations since the inference rules are interpreted
either all inductively or all coinductively. For this reason, it is generally
difficult to specify properties that combine safety and liveness aspects by
means of a single inference system.
\emph{Generalized Inference Systems} (\GISs)
\citep{AnconaDagninoZucca17,Dagnino19} admit a wider range of interpretations,
including intermediate fixed points of the inference operator associated with
the inference system different from the least or the greatest one.  This is made
possible by the presence of \emph{corules}, whose purpose is to provide an
inductive definition of a space within which a coinductive definition is used.
Although \GISs do not achieve the same flexibility of the modal $\mu$-calculus
in combining different fixed points, they allow for the specification of
properties that can be expressed as the intersection of a least and a greatest
fixed point. This feature of \GISs resonates well with one of the fundamental
results in model checking stating that every property can be decomposed into a
conjunction of a safety property and a liveness
one \citep{AlpernSchneider85,AlpernSchneider87,BaierKatoen08}.

\subsection{Definitions and interpretations}  

\beginbass
We first summarize the main notions and definitions of inference systems.
Then we generalize them by introducing (meta)corules. Given an inference
system, one has to take an \emph{interpretation} that identifies
the set of defined judgments.
When dealing with a generalized inference system, such interpretation is obtained
by combining both the inductive and the coinductive ones and by 
taking into account corules in the proper way.

\begin{definition}[Inference Systems \& Rules]
	\label{def:infsys}
	An \emph{inference system} \citep{Aczel77} $\mis$ over a \emph{universe}
	$\universe$ of \emph{judgments} is a set of \textit{rules}, which are pairs
	$\RulePair\prem\judg$ where $\prem\subseteq\universe$ is the set of
	\emph{premises} of the rule and $\judg\in\universe$ is the \emph{conclusion} of
	the rule. A rule without premises is called \emph{axiom}. Rules are typically
	presented using the syntax
	\[
  	\Rule\prem\judg
	\]
	where the line separates the premises (above the line) from the
	conclusion (below the line).
\end{definition}

Since rules may be infinite, inference systems are usually described 
by using \emph{meta-rules} written in a meta-language. 
Informally, meta-rules group rules according to their shape.

\begin{definition}[Metarule]
	A \emph{meta-rule} $\mr$  is a tuple \ple{\MCtx,\mpr,\mj,\sd} where 
 	\begin{itemize}
 	\item $\MCtx$ is a set called \emph{context}
 	\item $\mpr = \ple{\mj_1,\ldots,\mj_n}$ is a finite sequence of functions 
 	of type $\MCtx\rightarrow\universe$ called \emph{(meta-)premises}
 	\item $\mj$ is a function of type $\MCtx\rightarrow\universe$ called \emph{(meta-)conclusion} 
 	\item $\sd$ is a subset of $\MCtx$  called \emph{side condition}
 	\end{itemize}
\end{definition}

Given $\mr=\ple{\MCtx,\ple{\mj_1,\ldots,\mj_n},\mj,\sd}$,
the inference system $\Ground{\mr}$ \emph{denoted by $\mr$} is defined by:\\
\[\Ground{\mr}=\{\RulePair{\{\mj_1(\mc),\ldots,\mj_n(\mc)\}}{\mj(\mc)}\mid c\in\sd\}\]

\begin{definition}[Meta System]
	An \emph{inference meta-system} $\mis$ is a set of meta-rules. 
	The inference system $\Ground{\mis}$ is the union of $\Ground{\mr}$ for all $\mr\in\mis$.  
\end{definition}

A \emph{predicate} on $\universe$ is any subset of $\universe$. An
\emph{interpretation} of an inference system $\mis$ identifies a
\emph{predicate} on $\universe$ whose elements are called \emph{derivable
judgments}.

To define the interpretation of an inference system $\mis$, consider
the \emph{inference operator} associated with $\mis$, which is the function
$\InfOp\mis : \wp(\universe) \to \wp(\universe)$ such that
\[
  \InfOp\mis(X)=\set{ \judg\in\universe \mid \exists\prem\subseteq X: \RulePair\prem\judg\in\mis }
\]
for every $X \subseteq \universe$.
Intuitively, $\InfOp\mis(X)$ is the set of judgments that can be
derived in one step from those in $X$ by applying a rule of $\mis$.
Note that $\InfOp\mis$ is a monotone endofunction on the complete
lattice $\wp(\universe)$, hence it has least and greatest fixed
points.
It can also be defined for a meta-rule
$\mr$ and an inference meta-system $\mis$ as
\[
\begin{array}{cc}
	\InfOp{\mr}(X) = \set{ \mj(\mc) \mid \mc\in\sd, \mj_i(\mc) \in X \text{ for all } i \in 1..n }
	&
	\InfOp{\mis}(X) = \bigcup_{\mr\in\is} \InfOp{\mr}(X)
\end{array}
\]

\begin{definition}[(Co)Inductive Interpretations]
  The \emph{inductive interpretation} $\Inductive\mis$ of an
  inference system $\mis$ is the least fixed point of $\InfOp\mis$
  and the \emph{coinductive interpretation} $\CoInductive\mis$ is
  the greatest one.
\end{definition}

From a proof theoretical point of view, $\Inductive\mis$ and $\CoInductive\mis$
are the sets of judgments derivable with well-founded and non-well-founded proof
trees, respectively.

Generalized Inference Systems enable the definition of (some)
predicates for which neither the inductive interpretation nor the
coinductive one give the expected meaning.

\begin{definition}[Generalized Inference System]
  \label{def:gis}
  A \emph{generalized inference system} is a pair $\Pair\mis\mcois$ where $\mis$
  and $\mcois$ are inference systems (over the same $\universe$) whose elements
  are called \emph{rules} and \emph{corules}, respectively.
  The interpretation of a generalized inference system $\Pair\mis\mcois$,
  denoted by $\FlexCo\mis\mcois$, is the greatest post-fixed point of
  $\InfOp\mis$ that is included in $\Inductive{\mis\cup\mcois}$.
  
  Note that meta-corules are usually represented with a thicker line.
\end{definition}

From a proof theoretical point of view, a \GIS $\ple{\mis,\mcois}$ identifies
those judgments derivable with an arbitrary (not necessarily well-founded) proof
tree in $\mis$ and whose nodes (the judgments occurring in the proof tree) are
derivable with a well-founded proof tree in $\mis\cup\mcois$.
For more details we refer to \cite{Dagnino19}.

\subsection{Proving correctness - Proof principles}

\beginbass
The advantage of using inference systems is that they provide canonical
techniques for proving the correctness of a given definition.
Consider now a \emph{specification} $\Spec\subseteq\universe$, that is an
arbitrary subset of $\universe$. We can relate $\Spec$ to the interpretation of
a (generalized) inference system using one of the following proof principles.
The \emph{induction principle} \citep[Corollary 2.4.3]{Sangiorgi11} allows us to
prove the \emph{soundness} of an inductively defined predicate by showing that
$\Spec$ is \emph{closed} with respect to $\mis$. That is, whenever the premises
of a rule of $\mis$ are all in $\Spec$, then the conclusion of the rule is also
in $\Spec$.

\begin{proposition}[Induction]
	\label{prop:indp}
	$\InfOp\mis(\Spec) \subseteq \Spec$
	implies $\Inductive\mis \subseteq \Spec$.
\end{proposition}

The \emph{coinduction principle} \citep[Corollary 2.4.3]{Sangiorgi11} allows us
to prove the \emph{completeness} of a coinductively defined predicate by showing
that $\Spec$ is \emph{consistent} with respect to $\mis$.  That is, every
judgment of $\Spec$ is the conclusion of a rule whose premises are also in
$\Spec$.

\begin{proposition}[Coinduction]
	\label{prop:coindp}
	$\Spec \subseteq \InfOp\mis(\Spec)$ implies $\Spec \subseteq
  \CoInductive\mis$.
\end{proposition}

The \emph{bounded coinduction principle} \citep{AnconaDagninoZucca17} allows us
to prove the \emph{completeness} of a predicate defined by a generalized
inference system $\ple{\mis,\mcois}$. In this case, one needs to show not only
that $\Spec$ is consistent with respect to $\mis$, but also that $\Spec$ is
\emph{bounded} by the inductive interpretation of the inference system $\mis
\cup \mcois$. Formally:

\begin{proposition}[Bounded Coinduction]
	\label{prop:bcp}
  $\Spec \subseteq \Inductive{\mis\cup\mcois}$ and
  $\Spec \subseteq \InfOp\mis(\Spec)$ imply
  $\Spec\subseteq\FlexCo\mis\mcois$.
\end{proposition}

Proving the boundedness of $\Spec$ amounts to proving the completeness of
$\mis\cup\mcois$ (inductively interpreted) with respect to $\Spec$.

\subsection{Examples}

\beginbass
We now characterize the examples from \Cref{sec:sft_lvn}. For the sake of simplicity
we only present the definitions and we informally state where the proof principles
are used to prove the correctness (see \cite{Ciccone20} for the detailed proofs).
In \Cref{sec:agda_gis_meta} we will show Agda mechanizations of all the definitions.
In the following we write $\listin\xs{i}$ for the i-th element of list $\xs$.

\begin{example}
	Recall the inductive property from \Cref{ex:memberof}. Consider the inference system $\is$
	
	\begin{mathpar}
		\inferrule{\mathstrut}{\member(\x , \cons\x\xs)}
		\and
		\inferrule{\member(\x ,\xs)}{\member(\x , \cons\y\xs)}
	\end{mathpar}
	where the axiom states that $\x$ is inside the list containing only $\x$ while the rule tells that,
	if $\x$ is inside $\xs$, then it is in the same list with an additional element $\y$.
	Such inference system $\is$ must be \emph{inductively interpreted}. Hence, we consider only those finite
	proof trees whose leaves are applications of the axiom.
	
	Let $\Spec = \set{(\x , \xs) \mid \exists i \in \Nat. \listin\xs{i} = \x}$. 
	We use the \emph{induction principle} to prove the \emph{soundness} of the definition
	(\ie $\Inductive\is \subseteq \Spec$).
	\eoe
\end{example}

\begin{example}
	Recall the coinductive property from \Cref{ex:allpos}. Consider the inference system $\is$
	
	\begin{mathpar}
	\inferrule{\mathstrut}{\allpos\nil}
	\and
	\inferrule{\allpos\xs}{\allpos \cons\x\xs} ~ \x > 0
	\end{mathpar}
	where the axiom tells that the predicate trivially holds on the empty list and the rule
	states that, if a list $\xs$ is made of strictly positive numbers, then the predicate holds
	on such list with an additional element $\x$ provided that $\x > 0$.
	Such inference system $\is$ must be \emph{coinductively interpreted}. Hence, we consider either those finite
	proof trees whose leaves are the axiom or the infinite ones that are obtained by applying infinitely
	many times the rule. Note that is not compulsory to interpret the rule coinductively. On the other hand,
	if we interpret it inductively we obtain the expected predicate if and only if the lists under analysis are
	finite.
	
	Let $\Spec = \set{\xs \mid \forall i \in \Nat. \listin\xs{i} > 0}$. 
	We use the \emph{coinduction principle} to prove the \emph{completeness} of the definition
	(\ie $\Spec \subseteq \CoInductive\is$).
	\eoe
\end{example}

\begin{example}
	Recall the property from \Cref{ex:maxelem}. We first consider the following inference system $\is$
	
	\begin{mathpar}
		\inferrule{\mathstrut}{\maxelem(\x , \cons\x\nil)}
		\and
		\inferrule{\maxelem(\x , \xs)}{\maxelem(max(\x , \y) , \cons\y\xs)}
	\end{mathpar}
	where the axiom states that $\x$ is the maximum of the list that contains only $\x$, while the rule
	states that, if $\x$ is the maximum of $\xs$, then, when we add a new number $\y$ to $\xs$, the maximum
	becomes the greatest between $\x$ and $\y$.
	Such inference system cannot be inductively interpreted since we have to entirely inspect a possibly infinite
	list. On the other hand, the coinductive interpretation does not give us the expected meaning. Indeed, for example 
	we can derive $\maxelem(2 , \xs)$ where $\xs = \cons{1}\xs$ and $2$ actually does not belong to $\xs$.
	The infinite proof tree is obtained by applying the rule infinitely many times. 
	\[
	\begin{prooftree}
		\[
			\vdots
			\justifies
			\maxelem(2 , \xs)
		\]
		\justifies
		\maxelem(2 , \cons{1}{\xs})
	\end{prooftree}
	\]	
	Hence, $\is$ characterize only the \emph{safety} component of $\maxelem$ which states that the maximum must be
	greater than all the elements in the list.
	
	Now we consider the following $\cois$ consisting of a single \emph{coaxiom}
	\begin{mathpar}
		\infercorule{\mathstrut}{\maxelem(\x , \cons\x\xs)}
	\end{mathpar}  
	Such coaxiom forces the membership of the maximum into the list.
	The interpretation $\FlexCo\is\cois$ characterize the expected predicate. If we manage to find a finite proof tree
	in the inductive interpretation of the inference system with the additional coaxiom it means that 
	the maximum belongs to the list.
	
	Let $\Spec = \set{(\x , \xs) \mid \forall i \in \Nat. \x \geq \listin\xs{i}$ and $\member(\x , \xs)}$. 
	We use the \emph{bounded coinduction principle} to prove the \emph{completeness} of the definition
	(\ie $\Spec \subseteq \FlexCo\is\cois$).
	\eoe
\end{example}

As noted at the beginning of the section, all the proofs of the examples are detailed in \citep{Ciccone20}.

%% file: keynotions/sessiontypes.tex
\beginalto
Session types have been originally introduced by \cite{Honda93}.
A session type describes the communication protocol that takes place over a
channel, namely the allowed sequences of input/output actions performed by a
process on that channel. This way it is possible to statically guarantee
some desirable properties of the communication such as 
\emph{protocol fidelity}, \emph{communication safety} 
and \emph{deadlock freedom}. Original session types concerned with
the communication between two entities (\emph{binary}) where each
one owns an endpoint of the channel. 
They have been generalized to \emph{multiparty} 
by \cite{HondaYoshidaCarbone08} to model the interaction between 
multiple participants.


\subsection{Binary session types}
\label{ssec:bin}
\input{keynotions/binary}


\subsection{Multiparty session types}
\label{ssec:multi}
\input{keynotions/multiparty}

%% file: keynotions/binary.tex
\beginbass
We use \emph{polarities} $\Pol \in \set{\In,\Out}$ to
distinguish \emph{input actions} ($\In$) from \emph{output actions} ($\Out$) and
we write $\co\Pol$ for the \emph{opposite} or \emph{dual} polarity of $\Pol$ so that
$\co\In = \Out$ and $\co\Out = \In$.
We use $\Tag$ to denote an element of a given set of \emph{message tags} 
which may include values with a specific
interpretation such as booleans, natural numbers, and so forth.
The different terminology for labels 
is needed to avoid confusion with the labels used in the operational semantics.

\begin{definition}[Binary Session Types]
	Session types \citep{Honda93} are the possibly infinite, regular trees \citep{Courcelle83}
	coinductively generated by the grammar
	\[
    S, T, U, V ::= \End[\Pol] \mid \Tags\Pol\Tag_i.\S_i \mid \Pol \S.\T
	\]
\end{definition}

Session types of the form $\End[\Pol]$ describe channels used for exchanging a
session termination signal and on which no further communication takes place.
Session types of the form $\Pol\U.\S$ describe channels used for exchanging another
channel of type $U$ and then behaving according to $S$.
Finally, session types of the form $\Tags\Pol\Tag_i.S_i$ describe channels
used for exchanging a tag $\Tag_k$ and then behaving according to $S_k$. Session types of
the form $\Tags\In\Tag_i.S_i$ and $\Tags\Out\Tag_i.S_i$ are
sometimes referred to as \emph{external} and \emph{internal} choices
respectively, to emphasize that the label/tag being received or sent is always
chosen by the sender process. In a session type $\Tags\Pol\Tag_i.S_i$ we
assume that $I$ is not empty and that $i \ne j$ implies $\Tag_i \ne \Tag_j$ for every
$i,j\in I$. Note that $I$ is not necessarily finite, although regularity implies
that there must be finitely many \emph{distinct} $S_i$.

To improve readability we write $\Pol\Tag.\S$ when $I$ is the singleton $\set{\Tag}$
(when the choice is trivial) and we define the partial $\choice$ such
that

\[
  \begin{array}{lcl}
  \Tags\Pol\Tag_i.\S_i \branch \Tags[i \in J]\Pol\Tag_i.S_i & = &
  \Tags[i \in I \union J]\Pol\Tag_i.S_i
  \end{array}
\]

when $I, J \ne \emptyset$ and $I \cap J = \emptyset$.
Hereafter we specify possibly infinite session types by means of equations $S = \dots$ where
the right hand side of the equation may contain guarded occurrences of the metavariable $S$. 
Guardedness guarantees that a session type $S$ satisfying the equation exists and is unique \citep{Courcelle83}.

We equip session types with a \emph{labeled transition system} (LTS) that allows
us to describe, at the type level, the sequences of actions performed by a
process on a channel. We distinguish two kinds of transitions:
\emph{unobservable transitions} $S \lred\tau T$ are made autonomously by the process;
\emph{observable transitions} $S \lred{\smash\action} T$ are made by the process
in cooperation with the one it is interacting with through the channel. The
label $\action$ describes the kind of interaction and has either the form $\Pol U$
(indicating the exchange of a channel of type $U$) or the form $\Pol\Tag$ (indicating
the exchange of tag $\Tag$). The polarity $\Pol$ indicates whether the message is
received ($\In$) or sent ($\Out$).
If we use a term of the form $\Sys\S\T$ to describe the binary session as a whole, with
the two interacting processes behaving as $S$ and $T$, then we can formalize
their interaction (at the type level) using the LTS for session types and the
reduction rules in \Cref{fig:lts_bin}.

\begin{figure}[t]
\framebox[\textwidth]{
	\begin{mathpar}
		\inferrule[lb-channel]{\mathstrut}{\Pol\U.\S \lred{\mathstrut\Pol\U} \S} \defrule[lb-channel]{}
  	\\
  	\inferrule[lb-pick]{\mathstrut}
  		{\textstyle
  		\Tags\Out\Tag_i.\S_i \lred\tau \Out\Tag_k.\S_k}
  	~ k \in I \defrule[lb-pick]{}
  	\and
  	\inferrule[lb-tag]{\mathstrut}
  		{\textstyle
  		\Tags\Pol\Tag_i.\S_i \lred{\Pol\Tag_k} \S_k} 
  		~ k \in I \defrule[lb-tag]{}
  	\\
  	\inferrule[lb-tau-l]{
    	\S \lred\tau  \S'
  	}{
    	\Sys\S\T \lred\tau \Sys{\S'}\T
  	} \defrule[lb-tau-l]{}
  	\and
  	\inferrule[lb-tau-r]{
    	\T \lred\tau  \T'
  	}{
    	\Sys\S\T \lred\tau \Sys\S{\T'}
  	} \defrule[lb-tau-r]{}
  	\and
  	\inferrule[lb-sync]{
    	\S \lred{\co\action} \S'
    	\\
    	\T \lred\action  \T'
  	}{
    	\Sys\S\T \lred\tau \Sys{\S'}{\T'}
  	} \defrule[lb-sync]{}
	\end{mathpar}
}
\caption{Labeled Transition System for Binary Session Types}
\label{fig:lts_bin}
\end{figure}

We write $\co\action$ for the dual of $\action$, obtained by
changing the polarity of $\action$ with the opposite one.
A reduction occurs whenever one of the connected processes performs an
unobservable transition (\refrule{lb-tau-l} and \refrule{lb-tau-r}) 
or when the two processes exchange a message by
proposing complementary actions (\refrule{lb-sync}).
As usual, we let $\wred$ stand for the reflexive, transitive closure of $\lred\tau$ and 
we write $\Sys\S\T \nred$ if there are no $\S'$ and $\T'$ such that $\Sys\S\T \lred\tau \Sys{\S'}{\T'}$.
Note the different behaviors described by session types of the form
$\Tags\Pol\Tag_i.\S_i$ depending on the polarity $\Pol$.
According to \refrule{lb-tag}, a process using a channel of type $\Tags\In\Tag_i.\S_i$ performs an
observable transition for each of the tags $\Tag_i$ it is willing to receive.
On the contrary, a process using a channel of type $\Tags\Out\Tag_i.\S_i$
can first \emph{choose} a particular tag $\Tag = \Tag_k$ for some $k\in I$ by \refrule{lb-pick} (this choice
is internal to the process and is therefore unobservable) and then \emph{send}
the tag $\Tag$. As an example, the chain of transitions
\[
  \Out\Tag.\S \choice \T \lred\tau \Out\Tag.\S \lred{\Out\Tag} \S
\]
models a process that first chooses and then sends the tag $\Tag$. The choice of
the tag is irrevocable and not negotiable with the receiver process. Note
that, according to the definition of $\choice$, $\T$ must
be an internal choice of tags different from $\Tag$, hence $\Out\Tag.\S \choice \T$
is a non-trivial choice among two or more tags.

%% file: keynotions/multiparty.tex
\beginbass
A \emph{multiparty} session type describes the communication protocol between
at least two participants. For this reason, the types associated to each
endpoint are exactly those presented in \Cref{ssec:bin} with the addition of
\emph{roles} ($\rolep$,$\roleq\dots)$. Indeed, a binary session can be considered
as the simplest multiparty one with only two participants; hence, the roles
can be omitted in the types.

\begin{definition}[Local Types]
	A \emph{local session type} is a regular tree \cite{Courcelle83} coinductively
	generated by the productions
	\[
  	S, T, U, V ::= \End \mid \Tags\rolep\Pol\Tag_i.S_i \mid \rolep\Pol{S}.T
	\]
\end{definition}

The session type $\End$ describes the behavior of a process that sends or receives
a termination signal.
The session type $\Tags\rolep\Pol\Tag_i.S_i$ describes the behavior of a process
that sends to or receives from the participant $\rolep$ one of the tags $\Tag_i$
and then behaves according to $S_i$. Note that the source or destination role
$\rolep$ and the polarity $\Pol$ are the same in every branch. We require that
$I$ is not empty and $i, j \in I$ with $i \ne j$ implies $\Tag_i \ne \Tag_j$.
Occasionally we write $\rolep\Pol\Tag_1.S_1 + \cdots + \rolep\Pol\Tag_n.S_n$
instead of $\Tags[i=1]^n \rolep\Pol\Tag_i.S_i$.
Finally, a session type $\rolep\Pol{S}.T$ describes the behavior of a process
that sends to or receives from the participant $\rolep$ an endpoint of type $S$
and then behaves according to $T$.
We often specify infinite session types as solutions of equations of the form $S
= \cdots$ where the metavariable $S$ may occur on the right hand side of $=$
guarded by at least one prefix. A regular tree satisfying such equation is
guaranteed to exist and to be unique~\cite{Courcelle83}.

In order to describe a whole multiparty session at the level of types we
introduce the notion of \emph{session map}.

\begin{definition}[session map]
  \label{def:session_map}
  A \emph{session map} is a finite, partial map from roles to session types
  written $\set{\Map{\role_i} S_i}_{i\in I}$.  We let $M$ and $N$ range over
  session maps, we write $\dom{M}$ for the domain of $M$, we write $M \parop N$
  for the union of $M$ and $N$ when $\dom{M} \cap \dom{N} = \emptyset$, and we
  abbreviate the singleton map $\set{\Map\rolep{S}}$ as $\Map\rolep{S}$.
\end{definition}

Again, we describe the evolution of a session at the level of types by means of a
\emph{labeled transition system} for session maps. Labels are generated by the
grammar below:
\[
  \textbf{Label}
  \qquad
  \ell ::= \tau \mid \action
  \qquad\qquad
  \textbf{Action}
  \qquad
  \actionA, \actionB ::= \Pol\terminated \mid \Map\rolep{\roleq\Pol\Tag} \mid \Map\rolep{\roleq\Pol S}
\]

The label $\tau$ represents either an internal action performed by a participant
independently of the others or a synchronization between two participants.
The labels of the form $\Pol\terminated$ describe the input/output of
termination signals, whereas the labels of the form $\Map\rolep\roleq\Pol\Tag$
and $\Map\rolep\roleq\Pol S$ represent the input/output of a tag $\Tag$ or of an
endpoint of type $S$. 

\begin{figure}[t]
\framebox[\textwidth]{
  \begin{mathpar}
    \inferrule[lm-end]{ }{
      \Map\role\End \xlred{\Pol\terminated} \Map\role\End
    } \defrule[lm-end]{}
    \and
    \inferrule[lm-channel]{ }{
      \Map\rolep\roleq\Pol U.S \xlred{\Map\rolep\roleq\Pol U} \Map\rolep{S}
    } \defrule[lm-channel]{}
    \and
    \inferrule[lm-pick]{ }{
      \textstyle
      \Map\rolep{\Tags\roleq\Out\Tag_i.S_i}
      \lred\tau
      \Map\rolep{\roleq\Out\Tag_k.S_k}
    }
    ~k\in I \defrule[lm-pick]{}
    \and
    \inferrule[lm-tag]{ }{
      \textstyle
      \Map\rolep{\Tags\roleq\Pol\Tag_i.S_i}
      \xlred{\Map\rolep{\roleq\Pol\Tag_k}}
      \Map\rolep{S_k}
    }
    ~ k \in I \defrule[lm-tag]{}
    \and
    \inferrule[lm-tau]{
      M \lred\tau M'
    }{
      M \parop N \lred\tau M' \parop N
    } \defrule[lm-tau]{}
    \and
    \inferrule[lm-terminate]{
      M \lred{\In\terminated} M'
      \\
      N \lred{\Out\terminated} N'
    }{
      M \parop N \lred{\In\terminated} M' \parop N'
    } \defrule[lm-terminate]{}
    \and
    \inferrule[lm-sync]{
      M \lred{\co\action} M'
      \\
      N \lred{\action} N'
    }{
      M \parop N \lred\tau M' \parop N'
    } \defrule[lm-sync]{}
  \end{mathpar}
  }
  \caption{Labeled Transition System for Session Maps}
  \label{fig:lts_multi} 
\end{figure}

The labeled transition system is defined by the rules in \Cref{fig:lts_multi}, most of
which are straightforward. Rule \refrule{lm-pick} models the fact that the
participant $\rolep$ may internally choose one particular tag $\Tag_k$ before
sending it to $\roleq$. The chosen tag is not negotiable with the receiver.
Rule \refrule{lm-terminate} models termination of a session. A session terminates
when there is exactly one participant waiting for the termination signal and all
the others are sending it. This property follows from a straightforward
induction on the derivation of $M \lred{\In\terminated} N$ using
\refrule{lm-terminate} and \refrule{lm-end}.

The existence of a single participant waiting for the termination signal ensures
that there is a uniquely determined continuation process after the session has
been closed.
Finally, rule \refrule{lm-sync} models the synchronization between two
participants performing complementary actions. The complement of an action
$\action$, denoted by $\co\action$, is the partial operation defined by the
equations
\[
  \co{\Map\rolep\roleq\Pol\Tag} \eqdef \Map\roleq\rolep{\co\Pol}\Tag
  \qquad
  \co{\Map\rolep\roleq\Pol S} \eqdef \Map\roleq\rolep{\co\Pol} S
\]
where $\co\Pol$ denotes the complement of the polarity $\Pol$. The complement of
actions of the form $\Pol\terminated$ is undefined, so rule \refrule{lm-sync}
cannot be applied to terminated sessions.
Hereafter we write $\wred$ for the reflexive, transitive closure of $\lred\tau$
and $\wlred\action$ for the composition ${\wred}{\lred\action}$.

%% file: keynotions/related.tex
\beginalto
Now we have all the ingredients for discussing about some well known properties
that can be found in the literature. 
Notably, such properties have been studied both on the pure $\pi$-calculus and
on session-based calculi. We split the discussion according the property and
the scenario under analysis.

\paragraph{Termination of Binary Sessions.}
Termination is a liveness property that can be guaranteed when finite session
types are considered \citep{PerezCairesPfenningToninho12}. As soon as infinite
session types are considered, many session type systems weaken the guaranteed
property to deadlock freedom.
\cite{LindleyMorris16} define a type system for a functional
language with session primitives and recursive session types that is strongly
normalizing.

\paragraph{Liveness Properties in the $\pi$-Calculus.}
\cite{Kobayashi02} defines a behavioral type system that guarantees
lock freedom in the $\pi$-calculus. Lock freedom is a liveness property akin to
progress for sessions, except that it applies to \emph{any} communication
channel (shared or private). 
As a paradigmatic example of lock we can consider a variant of \Cref{ex:bsc}
in which the \actor{buyer} never pays the products. Thus, the \actor{carrier}
is locked.
\cite{Padovani14} adapts and extends the type system of
\cite{Kobayashi02} to enforce lock freedom in the \emph{linear
$\pi$-calculus} \citep{KobayashiPierceTurner99}, into which binary sessions can
be encoded \citep{DardhaGiachinoSangiorgi17}.
All of these works annotate types with numbers representing finite upper bounds
to the number of interactions needed to unblock a particular input/output
action.

\paragraph{Deadlock Freedom.}
Deadlock freedom is a safety properties according to which a communication
cannot get stuck. A paradigmatic example of deadlocked process is (we adopt binary session
type syntax)
\[
\x\iact\Tag[true].\y\oact\Tag[false]\dots \parop \y\iact\Tag[false].\x\oact\Tag[true]\dots
\]
where we have a couple of processes running in parallel.
The former is waiting for $\Tag[true]$ on $\x$ which is sent by the second 
as soon as it receives $\Tag[false]$ by the former.

\cite{Kobayashi02,Padovani14} guarantee deadlock freedom using the same technique for
lock freedom.
There exist a number of session based type systems inspired by linear logic
\citep{Wadler14,CairesPfenningToninho16,LindleyMorris16,CarboneLMSW16,CarboneMontesiSchurmannYoshida17}
in which deadlock freedom is dealt with
by using a \emph{process composition} rule that resembles the \emph{cut} rule.

\paragraph{Liveness Properties of Multiparty Sessions.}
The enforcement of liveness properties has always been a key aspect of session
type systems, although previous works have almost exclusively focused on
progress rather than on (fair) termination.
\cite{ScalasYoshida19} define a general framework for
ensuring safety and liveness properties of multiparty sessions. In particular,
they define a hierarchy of three liveness predicates to characterize ``live''
sessions that enjoy progress. They also point out that the coarsest liveness
property in this hierarchy, which is the one more closely related to fair
termination, cannot be enforced by their type system.
The work of \cite{GlabbeekHofnerHorne21} presents a type
system for multiparty sessions that ensures progress and is not only sound but
also complete.
\cite{CarboneDM14} characterize progress in terms 
of the standard notion of lock-freedom by using catalysers.

%% file: fairtermination/intro.tex
\begintreble
Among the properties mixing safety and liveness aspects in the context of session types, we decided
to develop type systems enforcing \emph{fair termination}. Such property is desirable for many reasons. 
Indeed, as we briefly mentioned at the beginning of the thesis, a lock free session
does not imply that other sessions are lock free as well. On the other hand, if we consider a session and 
we assume that all the other ones are fairly terminating, we can conclude that the one under analysis is
fairly terminating as well.

In \Cref{sec:ft_intro} we describe the property and we relate it with other properties that are
frequently studied in the literature. 
In \Cref{sec:ft_formally} we formally define \emph{fair termination}
in its general form on \emph{reduction systems} such that it 
will be instantiated in the different session type based scenarios
in \Cref{sec:ft_st}.
%
Notably, in \Cref{pt:agda} we will provide a characterization of fair termination based on generalized
inference systems (\Cref{sec:gis}) and we will provide a mechanization of both its definition
and correctness proofs in Agda.

%% file: fairtermination/secintro.tex
\beginalto
The decomposition of a distributed program into sessions enables its
modular static analysis and the enforcement of useful properties through a type
system. Examples of such properties are \emph{communication safety} (no message
of the wrong type is ever exchanged), \emph{protocol fidelity} (messages are
exchanged in the order prescribed by session types) and \emph{deadlock freedom}
(the program keeps running unless all sessions have terminated). These are all
instances of \emph{safety properties}, implying that ``nothing bad'' happens. In
general, one is also interested in reasoning and possibly enforcing
\emph{liveness properties}, those implying that ``something good'' happens
\cite{OwickiLamport82}. Examples of liveness properties are \emph{junk freedom}
(every message is eventually received), \emph{progress} (every non-terminated
participant of multiparty a session eventually performs an action) and \emph{termination}
(every session eventually comes to an end).

%% file: fairtermination/what.tex
\beginbass
Fair termination is \emph{termination} under a \emph{fairness assumption}. We decided to
investigate such property since current type systems cannot deal with those scenarios 
in which the communication between two entities \emph{depends} on the communication between 
others.
To explain why,
we show a paradigmatic scenario that we will instantiate in different ways in the next chapters.

\begin{example}[Buyer - Seller - Carrier]
	\label{ex:bsc}
	\brk
	Consider the interaction between the following three entities:
	\begin{itemize}
	\item \actor{buyer}: he can tell a \actor{seller} either that he adds an item to the cart or that he pays the total amount.
	We assume that the messages being sent are $\tadd$ and $\tpay$, respectively
	\item \actor{seller}: if the \actor{buyer} decides to pay the amount (\ie $\tpay$ is received) he contacts the \actor{carrier}
	to ship the items. The message being sent is $\tship$
	\item \actor{carrier}: he sends the items as soon as he is contacted by the \actor{seller} (\ie $\tship$ is received)
	\end{itemize}
\end{example}

At the moment we do not care about the technicalities on how the three actors interact.
What makes this scenario somewhat difficult to reason
about is that \emph{the progress of the carrier is not unconditional but depends
on the choices performed by the buyer}: the carrier can make progress only if
the buyer eventually pays the seller.

The \emph{fairness assumption} that we used can be informally stated as
\begin{center}
	\emph{If termination is always possible, then it is inevitable}
\end{center}

Note that \Cref{ex:bsc} admits an infinite execution in which the \actor{buyer} only
adds items to the cart and never pays the amount. Such execution is \emph{unfair}
according to our assumption since the \actor{buyer} can always send $\tpay$ (and terminate)
but he always avoid to do so.

%% file: fairtermination/why.tex
\beginbass
The reader might wonder why we focus on fair termination instead of considering
some fair version of progress. There are three reasons why we think that fair
termination is overall more appropriate than just progress.
First of all, ensuring that sessions (fairly) terminate is consistent with the
usual interpretation of the word ``session'' as an activity that lasts for a
\emph{finite amount of time}, even when the maximum duration of the activity is
not known \emph{a priori}.
Second, \emph{fair termination implies progress} when it is guaranteed along
with the usual safety properties of sessions. Indeed, if the session eventually
terminates, it must be the case that any non-terminated participant (think of
the carrier waiting for a \textit{"ship"} message) is guaranteed to eventually make
progress, even when such progress \emph{depends} on choices made by others
(like the buyer sending \textit{"pay"} to the seller).
Last but not least, \emph{fair session termination enables compositional
reasoning} in the presence of multiple sessions. This is not true for progress:
if an action on a session $s$ is blocked by actions on a different session $t$,
then knowing that the session $t$ enjoys progress does not necessarily guarantee
that the action on $s$ will eventually be performed (the interaction on $t$
might continue forever). On the contrary, knowing that $t$ fairly terminates
guarantees that the action on $s$ will eventually be scheduled and performed, so
that $s$ may in turn progress towards termination.

%% file: fairtermination/formally.tex
\beginalto
Since the notion of fair termination will apply to several different entities in this chapter
(session types, binary and multiparty sessions, processes) here we formally define it for a generic
reduction system. Later on we will show various instantiations of this
definition.

\begin{definition}[Reduction System]
	A \emph{reduction system} is a pair $(\States, {\red})$ where 
	\begin{itemize}
	\item $\States$ is a set of \emph{states}
	\item ${\red} \subseteq \States \times \States$ is a \emph{reduction relation}
	\end{itemize}
	We adopt the following notation:
	we let $C$ and $D$ range over states;
	we write $C \red$ if there exists $D \in \States$ such that $C \red D$; we write
	$C \nred$ if not $C \red$; we write $\wred$ for the reflexive, transitive
	closure of $\red$.
	We say that $D$ is \emph{reachable} from $C$ if $C \wred D$.
\end{definition}

As an example, the reduction system $(\set{A,B}, \set{(A,A),(A,B)})$
models an entity that can be in two states, $A$ or $B$, and such that the entity
may perform a reduction to remain in state $A$ or a reduction to move from state
$A$ to state $B$. To formalize the evolution of an entity from a particular
state we define \emph{runs}.

\begin{definition}[runs and maximal runs]
  \label{def:run}
  A \emph{run} of $C$ is a (finite or infinite) sequence
  $C_0C_1\dots C_i\dots$ of states such that $C_0 = C$ and
  $C_i \red C_{i+1}$ for every valid $i$. A run is \emph{maximal} if
  either it is infinite or if its last state $C_n$ is such that
  $C_n \nred$.
\end{definition}

Hereafter we let $\run$ range over runs. Each run in the previously defined
reduction system is either of the form $A^n$ -- a finite sequence of $A$ -- or
of the form $A^nB$ -- a finite sequence of $A$ followed by one $B$ -- or
$A^\omega$ -- an infinite sequence of $A$. Among these, the runs of the form
$A^nB$ and $A^\omega$ are maximal, whereas no run of the form $A^n$ is maximal.

We now use runs to define different termination properties of states:
we say that $C$ is \emph{weakly terminating} if there exists a maximal run of
$C$ that is finite;
we say that $C$ is \emph{terminating} if every maximal run of $C$ is finite;
we say that $C$ is \emph{diverging} if every maximal run of $C$ is infinite.
\emph{Fair termination} \citep{Francez86} is a termination property that only
considers a subset of all (maximal) runs of a state, those that are considered
to be ``realistic'' or ``fair'' according to some fairness assumption.
The assumption that we make in this work, and that we stated in words in
\Cref{sec:ft_intro}, is formalized thus:

\begin{definition}[Fair run]
  \label{def:fair_run}
  A run is \emph{fair} if it contains finitely many weakly terminating states.
  Conversely, a run is \emph{unfair} if it contains infinitely many weakly
  terminating states.
\end{definition}

Continuing with the previous example, the runs of the form $A^n$ and $A^nB$ are
fair, whereas the run $A^\omega$ is unfair. In general, an unfair run is an
execution in which termination is always within reach, but is never reached.

A key requirement of any fairness assumption is that it must be possible to
extend every finite run to a maximal fair one. This property is called
\emph{feasibility} \citep{AptFrancezKatz87,GlabbeekHofner19} or \emph{machine
closure} \citep{Lamport00}.
It is easy to see that our fairness assumption is feasible:

\begin{lemma}
  \label{lem:feasibility}
  If $\run$ is a finite run, then there exists $\run'$ such that $\run\run'$ is
  a maximal fair run.
\end{lemma}
\begin{proof}
    Let $D$ be the last state of $\run$. We distinguish two possibilities: if $D$
    is weakly terminating, then there exists a finite maximal run $D\run'$ of $D$;
    if $D$ is diverging, then there exists an infinite run $D\run'$ of $D$ such
    that no state in $\run'$ is weakly terminating. In both cases we conclude by
    noting that $\run\run'$ is a maximal fair run. 
\end{proof}

Fair termination is finiteness of all maximal fair runs:

\begin{definition}[Fair Termination]
  \label{def:fair_termination}
  We say that $C$ is \emph{fairly terminating} if every maximal fair run of $C$
  is finite.
\end{definition}

In the reduction system given above, $A$ is fairly terminating. Indeed, all the
maximal runs of the form $A^nB$ are finite whereas $A^\omega$, which is the only
infinite run of $A$, is unfair.

For the particular fairness assumption that we make, it is possible to provide a
sound and complete characterization of fair termination that does not mention
fair runs. This characterization will be useful to relate fair termination with
the notion of correct session (\Cref{def:compatibility,def:coherence}) 
and the soundness property of the type systems we are going 
to introduce in \Cref{pt:type_systems}. 

\begin{theorem}
  \label{thm:fair_termination}
  Let $(\States, {\red})$ be a reduction system and $C\in\States$. Then $C$ is
  fairly terminating if and only if every state reachable from $C$ is weakly
  terminating.
\end{theorem}
\begin{proof}
	
  \proofcase{$\Rightarrow$} Let $D$ be a state reachable from $C$. That is, there exists a
  finite run $\run$ of $C$ ending with $D$. By \Cref{lem:feasibility} we deduce
  that this run can be extended to a maximal fair one $\run\run'$. From the
  hypothesis that $C$ is fairly terminating we deduce that $\run\run'$ is
  finite. Hence, $D$ is weakly terminating.

  \proofcase{$\Leftarrow$} Let $C_0C_1\dots$ be an infinite fair run of $C$.  Using the
  hypothesis we deduce that each $C_i$ is weakly terminating, which is absurd.
  Hence, either there are no maximal fair runs or 
  every maximal fair run of $C$ is finite, but the first case is not possible 
  by \Cref{lem:feasibility}, thus $C$ is fairly terminating. 
\end{proof}

\begin{remark}[Fair Reachability of Predicates]
  Most fairness assumptions have the form ``if \emph{something} is infinitely
  often possible then \emph{something} happens infinitely often'' and, in this
  respect, our formulation of fair run (\cref{def:fair_run}) looks slightly
  unconventional. However, it is not difficult to realize that
  \cref{def:fair_run} is an instance of the notion of fair reachability of
  predicates as defined by \cite[Definition 3]{QueilleSifakis83}. 
  According to Queille and Sifakis, a run $\run$ is fair
  with respect to some predicate $\StateSet \subseteq \States$ if, whenever in
  $\run$ there are infinitely many states from which a state in $\StateSet$ is
  reachable, then in $\run$ there are infinitely many occurrences of states in
  $\StateSet$. When we take $\StateSet$ to be $\nred$, that is the set of
  terminated states that do not reduce, pretending that irreducible states
  should occur infinitely often in the run is nonsensical. So, the fairness
  assumption boils down to assuming that such states should \emph{not} be
  reachable infinitely often, which is precisely the formulation of
  \cref{def:fair_run}.
  \eor
\end{remark}

%% file: fairtermination/sessiontypes.tex
\beginalto
In this section we instantiate \emph{fair termination} in session type based scenarios.
First, we consider binary sessions and then we generalize to the multiparty case.
We instantiate \Cref{ex:bsc} as well since it will be the running example
in \Cref{ch:ft_bin} and \Cref{ch:ft_multi}.
For what concerns the used syntaxes and labeled transition systems, we refer to \Cref{sec:st}.

\begin{definition}[Compatibility of a binary session]
	\label{def:compatibility}
	We say that $S$ and $T$ are \emph{compatible}, notation
  $S \compatible T$, if $\Sys\S\T \wred \Sys{S'}{T'}$ implies
  $\Sys{S'}{T'} \wred \Sys{\End[\co\Pol]}{\End[\Pol]}$ for some $\Pol$.
\end{definition}

\begin{definition}[Coherence of a session map]
	\label{def:coherence}
  We say that a session map $M$ is \emph{coherent}, notation $\coherent M$, if $M \wred N$
  implies $N \wlred{\In\terminated}$.
\end{definition}

The term ``coherence'' is borrowed from \cite{CarboneLMSW16,CarboneMontesiSchurmannYoshida17}, 
although the property is actually stronger than the one of 
\cite{CarboneLMSW16,CarboneMontesiSchurmannYoshida17}
as it entails fair termination of multiparty sessions 
through \cref{thm:fair_termination}.
In particular, if we consider the reduction system whose states are session maps
and whose reduction relation is $\lred\tau$, then $\coherent M$ implies $M$
fairly terminating. The same applies to a $\S \compatible \T$ binary session.

\begin{remark}[Successful fair termination]
	The properties stated in \Cref{def:compatibility,def:coherence}
	are stronger than fair termination. Indeed, a \emph{deadlocked} session
	is \emph{fairly terminating} as well as it cannot reduce. 
	\Cref{def:compatibility,def:coherence} are equivalent to a
	\emph{successful} form of fair termination since we ask that the
	involved sessions correctly terminate. 
\end{remark}

Now we can revise \Cref{ex:bsc} in both scenarios.

\begin{example}[Binary Buyer - Seller - Carrier]
	\label{ex:bin_bsc}
	Consider the types $\S_b,\S_s$ and $\T_s,\T_c$ that model 
	the two binary sessions connecting \actor{buyer} - \actor{seller}
	and \actor{seller} - \actor{carrier}, respectively, according to 
	\Cref{ex:bsc}.
	\[
	\begin{array}{lcllcl}
		\S_b & = & \Out\tadd.\S_b \choice \Out\tpay.\End[\Out] \qquad & \qquad \T_s & = & \Out\tship.\End[\Out] \\
		\S_s & = & \In\tadd.\S_s \choice \In\tpay.\End[\In] \qquad & \qquad \T_c & = & \In\tship.\End[\In]
	\end{array}
	\]
	We focus on the session $\Sys{\S_b}{\S_s}$. According to
	\Cref{def:compatibility}, $\S_b$ and $\S_s$ are \emph{compatible}
	because, no matter how many times the \actor{buyer} adds an item to the cart,
	he always has the possibility to $\tpay$ the amount. The infinite
	run in which the \actor{buyer} only adds items is \emph{unfair} according to
	\Cref{def:fair_run}.
	
	Note that $\S_b \compatible \S_s$ implies \emph{progress}
	of the session $\Sys{\T_s}{\T_c}$. Indeed, the communication between 
	the \actor{seller} and the \actor{carrier} takes place after the \actor{buyer}
	sends $\tpay$ which is guaranteed to happen by $\S_b \compatible \S_s$. 
	Furthermore, although this example can be easily adapted to the multiparty case, 
	it shows a very simple and realistic scenario in which more sessions are
	involved, no matter if they are binary or multiparty.
	\eoe
\end{example}

\Cref{ex:bin_bsc} clearly holds in the multiparty context as well,
now we can model the same communication protocol using a single
multiparty session.

\begin{example}[Multiparty Buyer - Seller - Carrier]
\label{ex:bsc_ty_multi}
  Consider the session types
  \[
  \begin{array}{l@{~}c@{~}l}
    S_b & = & \seller\Out\tadd.S_b + \seller\Out\tpay.\End[\Out]\\
    S_s & = & \buyer\In\tadd.S_s + \buyer\In\tpay.\carrier\Out\tship.\End[\Out]\\
    S_c & = & \seller\In\tship.\End[\In]
  \end{array} 
  \]
  which describe the behavior of the three participants
  in \cref{ex:bsc}. The session map $\Map{\buyer}{S_b} \parop
  \Map{\seller}{S_s} \parop  \Map{\carrier}{S_c}$ is \emph{coherent}. 
  The explanation is just the same as the one given in \Cref{ex:bin_bsc}.
  \eoe
\end{example}

Notably, in the multiparty context, fair termination implies \emph{progress}
of all non terminated participants (see $\carrier$ in \Cref{ex:bsc_ty_multi}).

%% file: fairsub/chintro.tex
\begintreble
The mere \emph{assumption} of fairness (see \Cref{def:fair_run}) does not
turn an ordinary session type system into one that ensures fair session
termination because the correspondence imposed by the type system between the
structure of processes and that of the protocols they implement is generally
(often necessarily) a loose one.
Indeed, processes may be ``more accommodating'' than the protocols they
implement by handling more messages than those mentioned in the protocols. For
example, the \actor{seller} in \Cref{ex:bsc} could handle a $\tsearch$ message in
addition to $\tadd$ and $\tpay$, even if the session type associated with $x$
does not mention $\tsearch$.
At the same time, processes may also be ``less demanding'' than the protocols they 
implement by sending fewer messages than those allowed by the protocols. For example, 
the \actor{buyer} in \Cref{ex:bsc} could always purchase an even/odd number of items, 
or at least $n$ items, or no more than $n$ items, even if the session type 
associated with the channel allows sending an arbitrary number of $\tadd$ messages.
These mismatches between processes and protocols are usually reconciled by a
\emph{subtyping relation} for session types \citep{GayHole05,BernardiHennessy16}. 
The problem is that this subtyping relation is \emph{too coarse} because 
it has been conceived to preserve the
\emph{safety} properties of sessions but not termination, which is a
\emph{liveness} property: if session types are not sufficiently precise
descriptions of the actual behavior of processes, a session that appears to be
fairly terminating at the level of types may not terminate at all at the level
of processes.
To solve this problem we adopt \emph{fair subtyping}
\citep{Padovani13,Padovani16,BravettiLangeZavattaro21}, a
\emph{liveness-preserving} refinement of the subtyping relation defined by
\cite{GayHole05}.

The chapter is organized as follows. First, in \Cref{sec:original_sub} we present the original
subtyping relation for session types \citep{GayHole05}.
We conclude such section showing why such relation is \emph{unfair} by applying it
to a variant of \Cref{ex:bsc} (see \Cref{ssec:unfair_sub}).
Then, in \Cref{sec:fair_sub} we introduce \emph{fair subtyping}
\citep{Padovani13,Padovani16,BravettiLangeZavattaro21} .
We show an inference system for characterizing it and we prove its correctness
with respect to a semantic definition. At last, we
show a characterization of fair subtyping by using a generalized inference
system (see \Cref{ssec:fsub_gis}). We will refer to this definition
in \Cref{sec:agda_fs}.

%% file: fairsub/original.tex
\beginalto
\begin{figure}[t]
  \framebox[\textwidth]{
    \begin{mathpar}
        \inferrule[us-end]{\mathstrut}{
          \End \usubt \End
        } \defrule[us-end]{}
        \and
        \inferrule[us-channel-in]{
          \S \usubt \T \\ \U \usubt \V
        }{
          \In\U.\S \usubt \In\V.\T
        } \defrule[us-channel-in]{}
        \and
        \inferrule[us-channel-out]{
          \S \usubt \T \\ \V \usubt \U
        }{
          \Out\U.\S \usubt \Out\V.\T
        } \defrule[us-channel-out]{}
        \and
        \inferrule[us-tag-in]{
           \forall i\in I: \S_i \usubt \T_i
        }{
        	\textstyle
          \Tags\In\Tag_i.S_i \usubt 
          \Tags[i \in I \union J]\In\Tag_i.T_i
        } \defrule[us-tag-in]{}
        \and
        \inferrule[us-tag-out]{
           \forall i\in I: \S_i \usubt \T_i
        }{
        	\textstyle
          \Tags[i \in I \union J]\Out\Tag_i.S_i \usubt 
          \Tags[i \in I]\Out\Tag_i.T_i
        } \defrule[us-tag-out]{}
    \end{mathpar}
  }
	\caption{Rules for subtyping \citep{GayHole05}}
	\label{fig:usub}
\end{figure}
The original subtyping relation for session types has 
been introduced by \cite{GayHole05} and it is obtained
by coinductively interpreting the rules in \Cref{fig:usub}.
Notably, we only show the relation for binary session types 
since in the multiparty case it is defined in the same way and it
can be obtained by simply adapting the syntax.
The inference system in \Cref{fig:usub} derives judgments of
the form $\S \usubt \T$ meaning that $\S$ is a \emph{subtype}
of $\T$. Let us analyze the rules in details.

Rule \refrule{us-end} is used to relate terminated sessions.
\refrule{us-channel-in} and \refrule{us-channel-out} relate 
input and output of channels, respectively. Note that they
show a different behavior. While the former is \emph{covariant}
with respect to both the channel being received and the continuation,
the second one is \emph{contravariant} in the exchanged channel.
Rules \refrule{us-tag-in} and \refrule{us-tag-out} relate input
and output of message tags, respectively. Similarly to the 
exchange of a channel, the former is covariant while the second
contravariant in the set of messages being exchanged.

The rules are consistent with the informal example that we gave
at the beginning of \Cref{ch:fs} by referring to \Cref{ex:bsc}.
Indeed, the \actor{seller} in could handle a $\tsearch$ message in
addition to $\tadd$ and $\tpay$ while the \actor{buyer} 
could always purchase an even/odd number of items.

\begin{example}
	\label{ex:original_sub}
	Consider the session types $\S_b$ and $\S_s$ from 
	\Cref{ex:bin_bsc} describing the communication protocol between
	the \actor{buyer} and the \actor{seller}.
	We can derive
	\[
	\begin{array}{lcl}
		\S_b = \Out\tadd.\S_b \choice \Out\tpay.\End[\Out]
				 & \usubt & 
				 \S_b' = \Out\tadd\Out\tadd.\S'_b \choice \Out\tpay.\End[\Out]
		\\
		\S_s = \In\tadd.\S_s \choice \In\tpay.\End[\In]
				 & \usubt & 
				 \S_s' = \In\tadd.\S_s' \choice \In\tpay.\End[\In]
				 						\choice \In\tsearch.\S_s'
	\end{array}
	\]
	We can focus on the infinite derivation trees.
	\[
	\begin{prooftree}
		\[
			\[
				\vdots
				\justifies
				\S_b \usubt \S_b'
			\]
			\justifies
			\Out\tadd.\S_b \choice \Out\tpay.\End[\Out] \usubt
			\Out\tadd.\S_b'
			\using\refrule{us-tag-out}
		\]
		\[
			\justifies
			\End[\Out] \usubt \End[\Out]
			\using\refrule{us-end}
		\]
		\justifies
		\Out\tadd.\S_b \choice \Out\tpay.\End[\Out] \usubt
		\Out\tadd\Out\tadd.\S'_b \choice \Out\tpay.\End[\Out]
		\using\refrule{us-tag-out}
	\end{prooftree}
	\]\[
	\begin{prooftree}
		\[
			\vdots
			\justifies
			\S_s \usubt \S_s'
		\]
		\[
			\justifies
			\End[\In] \usubt \End[\In]
			\using\refrule{us-end}
		\]
		\justifies
		\In\tadd.\S_s \choice \In\tpay.\End[\In] \usubt
		\In\tadd.\S_s' \choice \In\tpay.\End[\In]
				 						\choice \In\tsearch.\S_s'
		\using\refrule{us-tag-in}
	\end{prooftree}
	\]
	
	As previously noted, the same judgments can be derived
	for the multiparty session types in \Cref{ex:bsc_ty_multi}.
	\eoe
\end{example}

The subtyping relation we presented induces a substitution
principle.

\begin{proposition}[\emph{Safe} substitution principle]
	\label{prop:safe_sub}
	If $\S \usubt \T$, then a process that uses an endpoint according
	to $\S$ can be \emph{safely} substituted with a process that
	uses the endpoint according to $\T$.
\end{proposition}

Note that we highlight the word \emph{safe} as we want to
point out that the subtyping relation under analysis is
studied to preserve the safety of the communication.
We will give more details in \Cref{ssec:unfair_sub}.

\begin{remark}
	The reader might be confused by the formulation of \Cref{prop:safe_sub}
	as it seems to treat the substitution in the wrong direction 
	(left-to-right). However, it is important to think about the
	subject of the principle, \ie processes in \Cref{prop:safe_sub}. 
	Indeed, the same principle
	can be formulated in a right-to-left fashion as in \cite{LiskovWing94} 
	by taking into account the endpoints and not the processes. 
	The two formulations turn out to be equivalent (see \cite{Gay16} for 
	more details).
	\eor
\end{remark}

%% file: fairsub/unfair.tex
\beginbass
In \Cref{sec:original_sub} we pointed out the \emph{safety}-preserving
property of \Cref{prop:safe_sub}.
Indeed, the two substitutions proposed in \Cref{ex:original_sub}
do not break the correctness of the communication between the entities.
However, the subtyping relation introduced by \cite{GayHole05}
can break the \emph{liveness} of the session on which it is applied.

\begin{example}
	\label{ex:unfair_sub}
	Consider the session types $\S_b$ and $\S_s$ from \Cref{ex:bin_bsc}. 
	The following judgment can be derived
	\[
	\begin{array}{lclclcl}
		\S_b & = 			& \Out\tadd.\S_b \choice \Out\tpay.\End[\Out]
				 & \usubt & \S_b^\infty
				 & = 			& \Out\tadd.\S_b^\infty
	\end{array}
	\]
	Let us look at the infinite derivation tree
	\[
	\begin{prooftree}
		\[
			\[
				\vdots
				\justifies
				\S_b \usubt \S_b^\infty
				\using\refrule{us-tag-out}
			\]
			\justifies
			\S_b \usubt \S_b^\infty
			\using\refrule{us-tag-out}
		\]
		\justifies
		\Out\tadd.\S_b \choice \Out\tpay.\End[\Out] \usubt
		\Out\tadd.\S_b^\infty
		\using\refrule{us-tag-out}
	\end{prooftree}
	\]
	As a consequence, \actor{buyer} can be replaced with
	\actor{buyer$_\infty$} behaving as $\S_b^\infty$
	by using \Cref{prop:safe_sub} (it only $\tadd$s items). 
	No matter how it reduces,
	the \actor{buyer$_\infty$} will never $\tpay$ the amount and the
	session will never terminate. Moreover, the only run of the
	session is infinite and \emph{fair} since it does not contain
	any weakly terminating state.
	This example applies to \Cref{ex:bsc_multi} as well.
	\eoe
\end{example}

Intuitively, the problem of the subtyping we investigated so far lies
in the contravariance of the output rule \refrule{us-tag-out}.
When such rule is applied, some branches are cut and it might happen
that all those branches leading to termination are removed. 
This is what happens in \Cref{ex:unfair_sub}.
Hence, the subtyping of \cite{GayHole05} must be refined
to preserve the termination property of the session.

%% file: fairsub/fair.tex
\beginalto
Fair subtyping is a \emph{liveness}-preserving variant
of the subtyping proposed by \cite{GayHole05}.
In this section we present such relation as a purely
coinductive predicate. However, in \Cref{ssec:fsub_gis}
we will show an equivalent and more involved formulation
obtained by equipping the rules in \Cref{fig:usub} with
a \emph{corule} (see \Cref{sec:gis}). While the generalized 
inference system is intriguing since it points out the safety 
and liveness aspects of the predicate, the formulation we give in
this section significantly simplifies the correctness proofs and
it will be easily integrated in the type systems in \Cref{pt:type_systems}.
Since binary sessions are the simplest multiparty ones, we refer to the binary
case when it allows to simplify the presentation of some features of the property
(\eg examples, description of the rules) while we present all the results
in the more general multiparty one.

\begin{figure}[t]
	\framebox[\textwidth]{
  \begin{mathpar}
    \inferrule[fsb-end]{\mathstrut}{
      \End \subt[n] \End
    } \defrule[fsb-end]{}
    \and
    \inferrule[fsb-tag-in]{
      \forall i\in I: S_i \subt[n_i] T_i
      \\
      \forall i\in I: n_i \leq n
    }{
      \textstyle
      \Tags\In\Tag_i.S_i \subt[n] \Tags[i\in I \union J]\In\Tag_i.T_i
    } \defrule[fsb-tag-in]{}
    \\
    \inferrule[fsb-channel]{
      S \subt[n] T
    }{
      \Pol U.S \subt[n] \Pol U.T
    } \defrule[fsb-channel]{}
    \and
    \inferrule[fsb-tag-out-1]{
      \forall i\in I: S_i \subt[n_i] T_i
      \\
      \forall i\in I: n_i \leq n
    }{
      \textstyle
      \Tags[i\in I]\Out\Tag_i.S_i \subt[n] \Tags\Out\Tag_i.T_i
    } \defrule[fsb-tag-out-1]{}
    \and
    \inferrule[fsb-tag-out-2]{
      \forall i\in I: S_i \subt[n_i] T_i
      \\
      \exists i\in I: n_i < n
    }{
      \textstyle
      \Tags[i \in I \union J]\Out\Tag_i.S_i \subt[n] \Tags\Out\Tag_i.T_i
    } \defrule[fsb-tag-out-2]{}
  \end{mathpar}
  }
  \caption{Inference system for fair subtyping - Binary}
  \label{fig:fsub_bin}
\end{figure}

\begin{figure}[t]
	\framebox[\textwidth]{
  \begin{mathpar}
    \inferrule[fsm-end]{\mathstrut}{
      \End \subt[n] \End
    } \defrule[fsm-end]{}
    \and
    \inferrule[fsm-tag-in]{
      \forall i\in I: S_i \subt[n_i] T_i
      \\
      \forall i\in I: n_i \leq n
    }{
      \textstyle
      \Tags\rolep\In\Tag_i.S_i \subt[n] \Tags[i\in I \union J]\rolep\In\Tag_i.T_i
    } \defrule[fsm-tag-in]{}
    \and
    \inferrule[fsm-channel]{
      S \subt[n] T
    }{
      \role\Pol U.S \subt[n] \role\Pol U.T
    } \defrule[fsm-channel]{}
    \and
    \inferrule[fsm-tag-out-1]{
      \forall i\in I: S_i \subt[n_i] T_i
      \\
      \forall i\in I: n_i \leq n
    }{
      \textstyle
      \Tags[i\in I]\role\Out\Tag_i.S_i \subt[n] \Tags\role\Out\Tag_i.T_i
    } \defrule[fsm-tag-out-1]{}
    \and
    \inferrule[fsm-tag-out-2]{
      \forall i\in I: S_i \subt[n_i] T_i
      \\
      \exists i\in I: n_i < n
    }{
      \textstyle
      \Tags[i \in I \union J]\role\Out\Tag_i.S_i \subt[n] \Tags\role\Out\Tag_i.T_i
    } \defrule[fsm-tag-out-2]{}
  \end{mathpar}
  }
  \caption{Inference system for fair subtyping - Multiparty}
  \label{fig:fsub_multi}
\end{figure}

Fair subtyping for binary session types is defined as the
relation $\subt[n]$ coinductively defined by the inference system
in \Cref{fig:fsub_bin}, where $n$ ranges over natural numbers. The characterization
of fair subtyping that we consider is the relation ${\subt} \eqdef
\bigcup_{n\in\Nat} {\subt[n]}$.
Fair subtyping for multiparty session types is analogously defined
(see \Cref{fig:fsub_multi}).
The rules for deriving $S \subt[n] T$ are quite similar to those of the standard
subtyping relation for session types \citep{GayHole05}: \refrule{fsb-end} states
reflexivity of subtyping on terminated session types; \refrule{fsb-channel}
relates higher-order session types with the same polarity and payload type;
\refrule{fsb-tag-in} is the usual covariant rule for the input of tags (the set
of tags in the larger session type includes those in the smaller one);
\refrule{fsb-tag-out-2} is the usual contravariant rule for the output of tags
(the set of tags in the smaller session type includes those in the larger one).
Overall, these rules entail a ``simulation'' between the behaviors described by
$\S_b$ and $\S_b'$ whereby all inputs offered by $\S_b$ are also offered by $\S_b'$ and all
outputs performed by $\S_b'$ are also performed by $\S_b$.
The main differences between $\subt$ and the subtyping relation of \cite{GayHole05}
are the presence of an invariant rule for outputs
\refrule{fsb-tag-out-1} and the natural number $n$ annotating each subtyping
judgment $S \subt[n] T$. Intuitively, this number estimates how much $\S_b$ and $\S_b'$
differ in terms of performed outputs. In all rules but \refrule{fsb-tag-out-2},
the annotation in the conclusion of the rule is just an upper bound of the
annotations found in the premises. In \refrule{fsb-tag-out-2}, where the sets of
output tags in related session types may differ, the annotation $n$ is required
to be a \emph{strict} upper bound for at least one of the premises. That is,
there must be at least one premise in which the annotation strictly decreases,
while no restriction is imposed on the others. Intuitively, this ensures the
existence of a tag shared by the two related session types whose corresponding
continuations are slightly less different. So, the annotation $n$ provides an
upper bound to the number of applications of \refrule{fsb-tag-out-2} along any
path (\ie any sequence of actions) shared by $\S_b$ and $\S_b'$ that leads to
termination.  In the particular case when $n=0$, the rule \refrule{fsb-tag-out-2}
cannot be applied, so that $\S_b'$ may perform all the outputs also performed by
$\S_b$.

\begin{remark}[Invariant delegation]
	\label{rmk:fsub_invariant}
	As it can be noted from \refrule{fsb-channel} and \refrule{fsm-channel},
	we require that the type of the channel being exchanged in the supertype
	matches that in the subtype. We made such a decision
	since we found out that allowing co/contravariance of input/output of channels
  may break the liveness of the session. 
	We will give more details in \Cref{ssec:invariant_ch}.
	\eor
\end{remark}

Fair subtyping allows us to reject the subtyping instance in \Cref{ex:unfair_sub}
that breaks the liveness of the involved session.

\begin{example}
	\label{ex:bsc_fair_sub}
  Consider the session types 
  \[
  \begin{array}{rclrcl}
  	S_b & = & \Out\tadd.S_b + \Out\tpay.\End[\Out]
  	\qquad &
  	\S_b' & = & \Out\tadd.\Out\tadd.\S_b' + \Out\tpay.\End[\Out]
  \end{array}
  \]
  from \Cref{ex:bin_bsc,ex:original_sub} 
  describing the \actor{buyer} purchasing an arbitrary number of items
  and the behavior of the \actor{buyer} always purchasing an even
  number of items, respectively. Consider also
  \[
  	\S_b^\infty = \Out\tadd.\S_b^\infty
  \]
  from \Cref{ex:unfair_sub}, 
  which describes the behavior of a \actor{buyer} attempting to $\tadd$ an 
  infinite number of items without ever $\tpay$ing the \actor{seller}.
  We have $\S_b \subt \S_b'$ and $\S_b \not\subt \S_b^\infty$. Indeed, we can derive
  \[
    \begin{prooftree}
      \[
        \[
          \mathstrut\smash\vdots
          \justifies
          \S_b \subt[1] \S_b' 
        \]
        \justifies
        \S_b \subt[2] \Out\tadd.\S_b'
        \using\refrule{fsb-tag-out-2}
      \]
      \[
        \justifies
        \End[\Out] \subt[0] \End[\Out]
        \using\refrule{fsb-end}
      \]
      \justifies
      \S_b \subt[1] \S_b'
      \using\refrule{fsb-tag-out-2}
    \end{prooftree}
  \]
  but there is no derivation for $\S_b \subt[n] \S_b^\infty$ no matter how large $n$ is
  chosen.
  Note that there are infinitely many sequences of actions of $\S_b$ that cannot be
  performed by both $\S_b'$ and $\S_b^\infty$. In particular, $\S_b'$ cannot perform any sequence
  of actions consisting of an odd number of $\Tag[add]$ outputs followed by a
  $\Tag[pay]$ output, whereas $\S_b^\infty$ cannot perform any sequence of $\Tag[add]$
  outputs followed by a $\Tag[pay]$ output. Nonetheless, there is a path shared
  by $\S_b$ and $\S_b'$ that leads into a region of $\S_b$ and $\S_b'$ in which no more
  differences are detectable. The annotations in the derivation tree measures
  the distance of each judgment from such region. In the case of $\S_b$ and $\S_b^\infty$,
  there is no shared path that leads to a region where no differences are
  detectable.
  \eoe
\end{example}

\begin{example}
  \label{ex:slot_fair_sub}
  Consider the session types 
	\[
	\begin{array}{rcl}
		S & = & \In\tplay.(\Out\twin.S + \Out\tlose.S) + \In\tquit.\End[\Out]
		\\
		T & = & \In\tplay.\Out\tlose.T + \In\tquit.\End[\Out]
	\end{array}
	\]  
  describing
  the behavior of two slot machines, an unbiased one in which the player may win
  at every play and a biased one in which the player never wins. If we try to
  build a derivation for $S \subt[n] T$ we obtain
  \[
    \begin{prooftree}
      \[
        \[
          \mathstrut\smash\vdots
          \justifies
          S \subt[n-1] T
        \]
        \justifies
        \Out\twin.S + \Out\tlose.S \subt[n] \Out\tlose.T
        \using\refrule{f-tag-out-1}
      \]
      \[
        \justifies
        \End[\Out] \subt[n] \End[\Out]
        \using\refrule{f-end}
      \]
      \justifies
      S \subt[n] T
      \using\refrule{f-tag-in}
    \end{prooftree}
  \]
  which would contain an infinite branch with strictly decreasing annotations.
  Therefore, we have $S \not\subt T$.
  In this case there exists a shared path leading into a region of $S$ and $T$
  in which no more differences are detectable between the two protocols, but
  this path starts from an input. The fact that $S$ is \emph{not} a fair subtype
  of $T$ has a semantic justification. Think of a player that deliberately
  insists on playing until it wins. This is possible when the player interacts
  with the unbiased slot machine $S$ but not with the biased one $T$.
  \eoe
\end{example}

Now we can provide a semantic characterization of fair subtyping for binary and multiparty
session types as the relation preserving \emph{compliance} (\Cref{def:compatibility}) and
\emph{coherence} (\Cref{def:coherence}) of the involved session, respectively.
Since the notion of \emph{coherence} boils down to \emph{compatibility} when there 
are exactly two participants, we only state the property in the multiparty scenario. 

\begin{definition}[Semantic fair subtyping - Multiparty]
  \label{def:ssubt}
  We say that $\S$ is a \emph{fair subtype} of\/ $\T$, notation
  $\S \ssubt \T$, if\/ $M \parop \Map\role\S$ \emph{coherent} implies
  $M \parop \Map\role\T$ \emph{coherent} for every $M$ and $\role$.
\end{definition}

We conclude this section by stating and proving a fundamental property of
fair subtyping.

\begin{theorem}
	\label{thm:fsub_preorder}
  	$\subt$ is a preorder.
\end{theorem}

While reflexivity of $\subt$ is trivial to prove (since \refrule{fsm-tag-out-2} is
never necessary, it suffices to only consider judgments with a $0$ annotation),
transitivity is surprisingly complex. The challenging part of proving that from
$S \subt[m] U$ and $U \subt[n] T$ we can derive $S \subt[k] T$ is to come up
with a feasible annotation $k$. As it turns out, such $k$ depends not only on
$m$ and $n$, but also on annotations found in different regions of the
derivation trees that prove $S \subt[m] U$ and $U \subt[n] T$. In particular,
the ``difference'' of $S$ and $T$ is not simply the ``maximum difference'' or
``the sum of the differences'' of $S$ and $U$ and of $U$ and $T$.
More in detail, we first show that we can always find a derivation of $S\subt[m]
U$ where the rank annotations of all judgements occurring in it are below some
$h \geq m$; then, the judgement $S\subt[k] T$ is provable for $k = m + (1+h)n$. 

\begin{lemma}
  \label{lem:bounded_derivation}
  Let $S \subt[n]^m T$ if and only if $S \subt[n] T$ is the
  conclusion of a derivation in which every rank annotation is at
  most $m$. Then $S \subt[n] T$ if and only $S \subt[n]^m T$ for
  some $m$.
\end{lemma}
\begin{proof}
  \newcommand{\msubt}[1]{\subt[#1{\leq}]}
  The ``if'' part is obvious. Concerning the ``only if'' part, it
  suffices to show that each judgment in the set
  \[
    \srel \eqdef \set{U \subt[m] V \mid U \subt[m] V \wedge \nexists
      n < m: U \subt[n] V}
  \]
  is derivable from premises that are also in $\srel$. This is
  enough to prove $S \subt[n]^m T$ from $S \subt[n] T$, because in
  $\srel$ there is at most one judgment $U \subt[n] V$ for each pair
  of session types $U$ and $V$ and, by regularity of $U$ and $V$,
  the derivation of $U \subt[n] V$ obtained using judgments in
  $\srel$ contains finitely many annotations, which must have a
  maximum.

  Suppose $U \subt[m] V \in \srel$. Then $U \subt[m] V$ is
  derivable. We reason by cases on the last rule applied to derive
  this judgment.

  \proofrule{fsm-end}
  Then $U = V = \End$ and there is nothing left to prove since
  \refrule{fsm-end} has no premises.

  \proofrule{fsm-tag-in}
  Then $U = \Tags\role\In\Tag_i.U_i$ and
  $V = \Tags[i\in J]\role\In\Tag_i.V_i$ and $I \subseteq J$ and
  $U_i \subt[n_i] V_i$ and $n_i \leq m$ for every $i\in I$.
  By definition of $\srel$ we have that, for every $i\in I$, there
  exists $m_i \leq n_i$ such that $U \subt[m_i] V_i \in \srel$.
  Then $U \subt[m] V$ is derivable by \refrule{fsm-tag-in} using
  premises in $\srel$.

  \proofrule{fsm-tag-out-1}
  Analogous to the previous case.

  \proofrule{fsm-tag-out-2}
  Then $U = \Tags\role\Out\Tag_i.U_i$ and
  $V = \Tags[i\in J]\role\Out\Tag_i.V_i$ and $J \subseteq I$ and
  $U_i \subt[n_i] V_i$ for every $i\in J$ and $n_k < m$ for some
  $k\in I$.
  By definition of $\srel$ we have that, for every $i\in J$, there
  exists $m_i \leq n_i$ such that $U \subt[m_i] V_i \in \srel$.
  In particular, $m_k \leq n_k < m$.
  Then $U \subt[m] V$ is derivable by \refrule{fsm-tag-out-2} using
  premises in $\srel$.
\end{proof}

\begin{proof}[Proof of \Cref{thm:fsub_preorder}]
The proof that $\subt$ is reflexive is trivial, since
  $S \subt[n] S$ is derivable for every $n$. Concerning
  transitivity, by \cref{lem:bounded_derivation} it suffices to show
  that each judgment in the set
  \[
    \srel \eqdef \set{
      S \subt[n_1 + (1 + m)n_2] T \mid S \subt[n_1]^m U \wedge U \subt[n_2] T
    }
  \]
  is derivable using the rules in \Cref{fig:fsub_multi} from premises that
  are also in $\srel$.
  Suppose $S \subt[n] T \in \srel$. Then there exist $U$, $n_1$, $m$
  and $n_2$ such that $S \subt[n_1]^m U$ and $U \subt[n_2] T$ and
  $n = n_1 + (1 + m)n_2$.
  We reason by cases on the last rules applied to derive
  $S \subt[n_1] U$ and $U \subt[n_2] T$.

  \proofrule{fsm-end}
  Then $S = U = T = \End$ hence $S \subt[n] T$ is derivable by \refrule{fsm-end}.

  \proofrule{fsm-tag-in}
  Then $S = \Tags\role\In\Tag_i.S_i$ and $U = \Tags[i\in J]\role\In\Tag_i.U_i$
  and $T = \Tags[i\in K]\role\In\Tag_i.T_i$ and $I \subseteq J \subseteq K$ and
  $S_i \subt[n_{1i}]^m U_i$ and $n_{1i} \leq n_1$ for every $i\in I$ and $U_i
  \subt[n_{2i}] T_i$ and $n_{2i} \leq n_2$ for every $i\in J$.
  By definition of $\srel$ we have that $S_i \subt[n_{1i} + (1 + m)n_{2i}] T_i
  \in \srel$ for every $i\in I$.
  Observe that $n_{1i} + (1 + m)n_{2i} \leq n_1 + (1 + m)n_2 = n$ for every
  $i\in I$ hence $S \subt[n] T$ is derivable by \refrule{fsm-tag-in}.

  \proofrule{fsm-tag-out-1}
  Then $S = \Tags\role\Out\Tag_i.S_i$ and
  $U = \Tags\role\Out\Tag_i.U_i$ and $T = \Tags\role\Out\Tag_i.T_i$
  and $S_i \subt[n_{1i}]^m U_i$ and $n_{1i} \leq n_1$ and
  $U_i \subt[n_{2i}] T_i$ and $n_{2i} \leq n_2$ for every $i\in I$.
  By definition of $\srel$ we have that
  $S_i \subt[n_{1i} + (1 + m)n_{2i}] T_i \in \srel$ for every
  $i\in I$.
  Observe that $n_{1i} + (1 + m)n_{2i} \leq n_1 + (1 + m)n_2 = n$
  for every $i\in I$ hence $S \subt[n] T$ is derivable by
  \refrule{fsm-tag-out-1}.

  \proofcase{Case \refrule{fsm-tag-out-1} and \refrule{fsm-tag-out-2}}
  Then $S = \Tags\role\Out\Tag_i.S_i$ and
  $U = \Tags\role\Out\Tag_i.U_i$ and
  $T = \Tags[i\in J]\role\Out\Tag_i.T_i$ and $J \subseteq I$ and
  $S_i \subt[n_{1i}]^m U_i$ and $n_{1i} \leq n_1$ for every $i\in I$
  and $U_i \subt[n_{2i}] T_i$ for every $i\in J$ and $n_{2k} < n_2$
  for some $k\in J$.
  By definition of $\srel$ we have that
  $S_i \subt[n_{1i} + (1 + m)n_{2i}] T_i \in \srel$ for every
  $i\in J$.
  Observe that $n_{1k} + (1 + m)n_{2k} < n_1 + (1 + m)n_2 = n$ hence
  $S \subt[n] T$ is derivable by \refrule{f-tag-out-2}.

  \proofcase{Case \refrule{fsm-tag-out-2} and \refrule{fsm-tag-out-1}}
  Then $S = \Tags\role\Out\Tag_i.S_i$ and
  $U = \Tags[i\in J]\role\Out\Tag_i.U_i$ and
  $T = \Tags[i\in J]\role\Out\Tag_i.T_i$ and $J \subseteq I$ and
  $S_i \subt[n_{1i}]^m U_i$ for every $i\in J$ and $n_{1k} < n_1$
  for some $k\in J$ and $U_i \subt[n_{2i}] T_i$ for every $i\in J$
  and $n_{2i} \leq n_2$ for every $i\in J$.
  By definition of $\srel$ we have that
  $S_i \subt[n_{1i} + (1 + m)n_{2i}] T_i \in \srel$ for every
  $i\in J$.
  Observe that $n_{1k} + (1 + m)n_{2k} < n_1 + (1 + m)n_2 = n$ hence
  $S \subt[n] T$ is derivable by \refrule{fsm-tag-out-2}.

  \proofrule{fsm-tag-out-2}
  Then $S = \Tags\role\Out\Tag_i.S_i$,
  $U = \Tags[i\in J]\role\Out\Tag_i.U_i$ and
  $T = \Tags[i\in K]\role\Out\Tag_i.T_i$ and
  $K \subseteq J \subseteq I$ and $S_i \subt[n_{1i}]^m U_i$ for
  every $i\in J$ and $n_{1j} < n_1$ for some $j\in J$ and
  $U_i \subt[n_{2i}] T_i$ for every $i\in K$ and $n_{2k} < n_2$ for
  some $k\in K$.
  By definition of $\srel$ we have that
  $S_i \subt[n_{1i} + (1 + m)n_{2i}] T_i \in \srel$ for every
  $i\in K$.
  Observe that
  \[
    \begin{array}{rcll}
      n_{1k} + (1 + m)n_{2k} & \leq & m + (1 + m)n_{2k} & \text{since $n_{1k} \leq m$}
      \\
      & < & 1 + m + (1 + m)n_{2k}
      \\
      & = & (1 + m)(1 + n_{2k})
      \\
      & \leq & (1 + m)n_2 & \text{since $n_{2k} < n_2$}
      \\
      & < & n_1 + (1 + m)n_2 & \text{since $n_{1j} < n_1$}
    \end{array}
  \]
  hence $S \subt[n] T$ is derivable by \refrule{fsm-tag-out-2}.
\end{proof}

%% file: fairsub/fsub-sound.tex
\beginbass
Now we prove the \emph{soundness} of $\subt$ with respect
to $\ssubt$ (\Cref{def:ssubt}). That is, we prove that $\subt$ 
is \emph{coherence}-preserving just like $\ssubt$ is.
The proof of this result relies on a key property of $\subt$ not enjoyed by the
usual subtyping relation on session types \citep{GayHole05}: when $S\subt T$ and
$M \parop \Map\rolep S$ is coherent, the session map $M \parop \Map\rolep T$ can
successfully terminate (\Cref{lem:fsub_term}).
The rank annotation on subtyping judgements is used to
set up an appropriate inductive argument for proving this property.

\begin{theorem}[Soundness]
	\label{thm:fsub_sound}
 	If $S \subt T$ then $S \ssubt T$.
\end{theorem}

We start with an auxiliary result formalizing the simulation entailed by the
relation $S \subt T$.

\begin{lemma}
	\label{lem:subt_sim}
  	If\/ $S \subt T$ and $M \parop \Map\rolep{S}$ is coherent and $M \parop
  	\Map\rolep{T} \wred N \parop \Map\rolep{T'}$, then $M \parop \Map\rolep{S}
  	\wred N \parop \Map\rolep{S'}$ for some $S' \subt T'$.
\end{lemma}
\begin{proof}
  We prove the result for a single reduction
  $M \parop \Map\rolep{T} \lred\tau N \parop \Map\rolep{T'}$. The
  general statement then follows by a straightforward induction on
  the length of the reduction
  $M \parop \Map\rolep{T} \wred N \parop \Map\rolep{T'}$ using the
  fact that coherence is preserved by reductions.

  \proofcase{Case $M \lred\tau N$}
  Then $T' = T$ and we conclude by taking $S' \eqdef S$.

  \proofcase{Case $\Map\rolep{T} \lred\tau \Map\rolep{T''}$}
  Then
  $\Map\rolep{T} = \Map\rolep{\Tags\roleq\Out\Tag_i.T_i} \lred\tau
  \Map\rolep{\roleq\Out{\Tag_k}.T_k} = \Map\rolep{T'}$ for
  some $k\in I$.
  From the hypothesis $S \subt T$ we deduce
  $S = \Tags[i\in J]\roleq\Out\Tag_i.S_i$ where $I \subseteq J$ and
  $S_i \subt T_i$ for every $i\in I$.
  Now we have
  $M \parop \Map\rolep{S} \lred\tau M \parop
  \Map\rolep{\roleq\Out\Tag_k.S_k}$ and also
  $\roleq\Out\Tag_k.S_k \subt T'$.
  We conclude by taking $S' \eqdef \roleq\Out\Tag_k.S_k$.

  \proofcase{Case $M \xlred{\Map\roleq{\rolep\Out\Tag}} N$ and
    $\Map\rolep{T} \xlred{\Map\rolep{\roleq\In\Tag}}
    \Map\rolep{T'}$}
  Then $T = \Tags\roleq\In\Tag_i.T_i$ and $\Tag = \Tag_k$ and
  $T' = T_k$ for some $k\in I$.
  From the hypothesis $S \subt T$ we deduce
  $S = \Tags[i\in J]\roleq\In\Tag_i.S_i$ and $J \subseteq I$ and
  $S_i \subt T_i$ for every $i\in J$.
  From the hypothesis $M \parop \Map\rolep{S}$ coherent we deduce
  $k\in J$ or else the participant $\rolep$ would not be able to
  receive the $\Tag$ tag.
  We conclude by taking $S' \eqdef S_k$.

  \proofcase{Case $M \xlred{\Map\roleq{\rolep\In\Tag}} N$ and
    $\Map\rolep{T} \xlred{\Map\rolep{\roleq\Out\Tag}}
    \Map\rolep{T'}$}
  Then $T = \Tags\roleq\Out\Tag_i.T_i$ and $\Tag = \Tag_k$ and
  $T' = T_k$ for some $k\in I$.
  From the hypothesis $S \subt T$ we deduce
  $S = \Tags[i\in J]\roleq\Out\Tag_i.S_i$ and $I \subseteq J$ and
  $S_i \subt T_i$ for every $i\in I$.
  We conclude by taking $S' \eqdef S_k$.

  \proofcase{Case $M \xlred{\Map\roleq\rolep\Out U} N$ and
    $\Map\rolep{T} \xlred{\Map\rolep\roleq\In U} \Map\rolep{T'}$}
  Then $T = \roleq\In U.T'$.
  From the hypothesis $S \subt T$ we deduce $S = \roleq\In U.S'$
  and $S' \subt T'$.
  We conclude by observing that
  $M \parop \Map\rolep{S} \lred\tau N \parop \Map\rolep{S'}$.

  \proofcase{Case $M \xlred{\Map\roleq\rolep\In U} N$ and
    $\Map\rolep{T} \xlred{\Map\rolep\roleq\Out U} \Map\rolep{T'}$}
  Then $T = \roleq\Out U.T'$.
  From the hypothesis $S \subt T$ we deduce $S = \roleq\Out U.S'$
  and $S' \subt T'$.
  We conclude by observing that
  $M \parop \Map\rolep{S} \lred\tau N \parop \Map\rolep{S'}$.
\end{proof}

Next we show that $S \subt T$ preserves the termination of any session map that
completes $S$ into a coherent one.

\begin{lemma}
	\label{lem:fsub_term}
  	If\/ $S \subt[n] T$ and $M \parop \Map\rolep{S}$ is coherent, then $M \parop
  	\Map\rolep{T} \wlred{\In\terminated}$.
\end{lemma}
\begin{proof}
  By induction on the lexicographically ordered tuple
  $(n, |\actions|)$ where $\actions$ is any string of actions such
  that $M \xwlred{\co\actions\co\Pol\terminated}$ and
  $\Map\role{S} \xwlred{\actions\Pol\terminated}$.  We know that at
  least one such $\actions$ least does exist from the hypothesis
  $M \parop \Map\rolep{S}$ is coherent.
  We now reason by cases on the shape of $\actions$.

  \proofcase{Case $\actions = \varepsilon$}
  Then $S = \End$.
  From the hypothesis $S \subt[n] T$ and \refrule{fsm-end} we deduce
  $T = \End$ and we conclude
  $M \parop \Map\rolep{T} \wlred{\In\terminated}$.

  \proofcase{Case $\actions = \Map\rolep\roleq\In\Tag\actionsB$}
  Then $S = \Tags\roleq\In\Tag_i.S_i$ and $\Tag = \Tag_k$ for some
  $k\in I$.
  From the hypothesis $S \subt[n] T$ and \refrule{fsm-tag-in} we
  deduce $T = \Tags[i\in J]\roleq\In\Tag_i.T_i$ and $I \subseteq J$
  and $S_i \subt[n_i] T_i$ and $n_i \leq n$ for every $i\in I$.
  We conclude using the induction hypothesis.

  \proofcase{Case $\actions = \Map\rolep\roleq\Out\Tag\actionsB$}
  Then $S = \Tags\roleq\Out\Tag_i.S_i$ and $\Tag = \Tag_k$ for some
  $k\in I$. We distinguish two sub-cases, according to the last rule
  used in the derivation of $S \subt[n] T$.
  If the last rule was \refrule{fsm-tag-out-1}, then
  $T = \Tags[i\in I]\roleq\Out\Tag_i.T_i$ and $S_i \subt[n_i] T_i$
  and $n_i \leq n$ for every $i\in I$.
  In particular, $S_k \subt[n_k] T_k$ and $n_k \leq n$ and we
  conclude using the induction hypothesis.
  If the last rule was \refrule{fsm-tag-out-2}, then
  $T = \Tags[i\in J]\roleq\Out\Tag_i.T_i$ with $J \subseteq I$ and
  $S_i \subt[n_i] T_i$ for every $i\in J$ and $n_j < n$ for some
  $j\in J$.
  In particular, we have $S_j \subt[n_j] T_j$ and we conclude using
  the induction hypothesis.
\end{proof}

\begin{proof}[Proof of \Cref{thm:fsub_sound}]
  Consider a run
  $M \parop \Map\rolep{T} \wred N \parop \Map\rolep{T'}$.
  From \Cref{lem:subt_sim} we deduce that there exists $S' \subt T'$
  such that $M \parop \Map\rolep{S} \wred N \parop \Map\rolep{S'}$.
  From the hypothesis that $M \parop \Map\rolep{S}$ is coherent we
  deduce $N \parop \Map\rolep{S'}$ is also coherent.
  From \Cref{lem:fsub_term} we conclude
  $N \parop \Map\rolep{T'} \wlred{\In\terminated}$.
\end{proof}

%% file: fairsub/fsub-complete.tex
\beginbass
\Cref{thm:fsub_sound} alone suffices to justify the adoption of $\subt$ as
fair subtyping relation, but we are interested in understanding to which extent
$\subt$ covers $\ssubt$. In this respect, it is quite easy to see that there
exist session types that are related by $\ssubt$ but not by $\subt$. For
example, consider $S = \role\Out\Tag[a].S$ and $T = \role\In\Tag[b].T$ and
observe that these two session types describe completely different protocols
(the output of infinitely many $\Tag[a]$'s in the case of $S$ and the input of
infinitely many $\Tag[b]$'s in the case of $T$). In particular, we have $S
\not\subt T$ and $T \not\subt S$ but also $S \ssubt T$ and $T \ssubt S$. That
is, $S$ and $T$ are \emph{unrelated} according to $\subt$ but they are
\emph{equivalent} according to $\ssubt$. This equivalence is justified by the
fact that there exists no coherent session map in which $S$ and $T$ could play
any role, because none of them can ever terminate.

This discussion hints at the possibility that, if we restrict the attention to
those session types that \emph{can} terminate, which are the interesting ones as
far as this work is concerned, then we can establish a tighter correspondence
between $\subt$ and $\ssubt$. We call such session types \emph{bounded}, because
they describe protocols for which termination is always within reach.

\begin{definition}[Bounded session type]
	\label{def:bounded_type}
	We say that a session type is \emph{bounded} if all of its subtrees contain a
	$\End$ leaf.
\end{definition}

Note that a \emph{finite} session type is always bounded but not every bounded
session type is finite. If we consider the reduction system in which states are
session types and we have $S \red T$ if $T$ is an immediate subtree of $S$, then
$S$ is bounded if and only if $S$ is fairly terminating.
Now, for the family of bounded session types we can prove a \emph{relative
completeness} result for $\subt$ with respect to $\ssubt$.

\begin{theorem}[Relative completeness]
	\label{thm:fsub_complete}
 	$S$ bounded, $S \ssubt T$ imply $S \subt T$.
\end{theorem}

The proof of \Cref{thm:fsub_complete} is done by contradiction. We show
that, for any bounded $S$, if $S\subt T$ does not hold then we can build a
session map $M$ called \emph{discriminator} such that $M\parop \Map\rolep S$ is
coherent and $M\parop \Map\rolep T$ is not, which contraddicts the hypothesis $S
\ssubt T$.
The boundedness of $S$ is necessary to make sure that it is always possible to
find a session map $N$ such that $N\parop \Map\rolep S$ is coherent.

For this proof we need some auxiliary notions and notation.
First of all, we consider \emph{unfair subtyping} $\usubt$ from 
\Cref{fig:usub} and we fix the invariance of the type of the
channel being sent by substituting rules \refrule{us-channel-in}
and \refrule{us-channel-out} with
\[
	\inferrule[us-channel]{
      S \usubt T
    }{
      \role\Pol U.S \usubt \role\Pol U.T
    }
\]
It is straightforward to see that ${\subt} \subseteq {\usubt}$.
Then, we introduce some convenient notation for building session maps. To do
this, we assume the existence of an arbitrary total order $<$ on the set of
roles.
Now, given a finite set of roles $\set{\role_1,\dots,\role_n}$ where $\role_1 <
\cdots < \role_n$, we write $\set{\role_1,\dots,\role_n}\Out\Tag.S$ for the
session type $\role_1\Out\Tag\cdots\role_n\Out\Tag.S$.
Given a finite family $\set{M_i}_{i\in I}$ of session maps all having the same
domain $\set\roleq \subseteq D \subseteq \RoleSet\setminus\set\rolep$, we write
$\Map\roleq{\Tags{\rolep\Pol\Tag_i.M_i}}$ for the session map $M$ having domain
$D$ and such that
\[
  M(\roler) \eqdef
  \begin{cases}
    \Tags\rolep\Pol\Tag_i.  D\setminus\set{\roleq}\Out\Tag_i.
    M_i(\roleq) & \text{if $\roler = \roleq$}
  \\
  \Tags\roleq\In\Tag_i.M_i(\roler) & \text{if
    $\roler \neq \roleq$}
  \end{cases}
\]
for every $\roler \in D$.
As suggested by the notation, this session map realizes a conversation in which
$\roleq$ first interacts with $\rolep$ by exchanging a tag $\Tag_i$ and then it
informs all the other participants about the tag that has been exchanged. This
session map has the property
\[
  M \xlred{\roleq:\rolep\Pol\Tag_k}\wred M_i
\]
for every $k\in I$.

Similarly, given a session map $N$ with domain $D \supseteq\set\roleq$, we write
$\Map{\roleq}{\rolep\Pol{U}.N}$ for the session map $M$ with domain $D$ such
that
\[
  M(\roler) \eqdef
  \begin{cases}
    \rolep\Pol{U}.N(\roleq) & \text{if $\roler = \roleq$}
    \\
    N(\roler) & \text{if $\roler \in D\setminus\set{\rolep,\roleq}$}
  \end{cases}
\]
for every $\roler \in D$. Note that $M$ has the property
\[
  M \xlred{\roleq:\rolep\Pol{U}} N
\]

The first key step is showing that $S \ssubt T$ implies $S \usubt T$ when $S$ is
a bounded session type. That is, unfair subtyping is a \emph{necessary
condition} for fair subtyping to hold.

\begin{lemma}
  \label{lem:usub_complete}
  If\/ $S$ is bounded and $S \ssubt T$ then $S \usubt T$.
\end{lemma}
\begin{proof}
  Using the coinduction principle (\Cref{prop:coindp}) 
  it suffices to show that each judgment in the set
  \[
    \srel \eqdef \set{ S \usubt T \mid \text{$S$ is bounded and $S \ssubt T$} }
  \]
  is derivable by the rules in \Cref{fig:usub} from premises that
  satisfy the same property. Let $S \usubt T \in \srel$. Then $S$ is
  bounded and $S \ssubt T$.
  We reason by cases on the shape of $S$.

  \proofcase{Case $S = \End$}
  Consider $M \eqdef \Map{\roleq}{\End[\co\Pol]}$ and observe that
  $M \parop \Map{\rolep} S$ is coherent. Then $M \parop \Map{\rolep} T$ is
  coherent as well, which implies $T = \End$.
  We conclude by observing that $\End \usubt \End$ is derivable with
  \refrule{us-end}.

  \proofcase{Case $S = \Tags\roleq\Out\Tag_i.S_i$}
  Let $\set{M_i}_{i\in I}$ be a family of session maps such that
  $M_i \parop \Map{\rolep} S_i$ is coherent for every $i\in I$. Such
  family is guaranteed to exist from the hypothesis that $S$ is
  bounded.
  Without loss of generality we may assume that the $M_i$ all have
  the same domain $D \supseteq \set\roleq$.
  Let $M \eqdef \Map{\roleq}{\Tags\rolep\In\Tag_i.M_i}$ and observe that
  $M \parop \Map{\rolep}{S}$ is coherent by definition of $M$.  Then
  $M \parop \Map{\rolep}{T}$ is coherent as well.
  We deduce that $T = \Tags[i\in J]\roleq\Out\Tag_i.T_i$ and
  $J \subseteq I$ and also that $M_i \parop \Map{\rolep}{T_i}$ is coherent
  for every $i\in J$. Hence $S_i \ssubt T_i$ for every $i\in J$,
  namely $S_i \usubt T_i \in \srel$ for every $i\in J$ by definition
  of $\srel$.
  We conclude by observing that $S \usubt T$ is derivable by
  \refrule{us-tag-out}.

  \proofcase{Case $S = \Tags\roleq\In\Tag_i.S_i$}
  Let $\set{M_i}_{i\in I}$ be a family of session maps such that
  $M_i \parop \Map{\rolep}{S_i}$ is coherent for every $i\in I$. Such
  family is guaranteed to exist from the hypothesis that $S$ is
  bounded. Without loss of generality we may assume that the $M_i$
  all have the same domain $D \supseteq \set\roleq$.
  Let $M \eqdef \Map{\roleq}{\Tags\rolep\Out\Tag_i.M_i}$ and observe that
  $M \parop \Map{\rolep}{S}$ is coherent by definition of $M$.
  We deduce that $T = \Tags[i\in J]\roleq\In\Tag_i.T_i$ and
  $I \subseteq J$ and also that $M_i \parop \Map{\rolep}{T_i}$ is coherent
  for every $i\in I$. Hence $S_i \ssubt T_i$ for every $i\in I$,
  namely $S_i \usubt T_i \in \srel$ for every $i\in I$ by definition
  of $\srel$.
  We conclude by observing that $S \usubt T$ is derivable by
  \refrule{us-tag-in}.

  \proofcase{Case $S = \roleq\Pol{U}.S'$}
  Let $N$ be a session map such that $N \parop \Map{\rolep}{S'}$ is
  coherent. Such $N$ is guaranteed to exist from the hypothesis that
  $S$ is bounded.
  Let $M \eqdef \Map{\roleq}{\rolep\co\Pol{U}.N}$ and observe that
  $M \parop \Map{\rolep}{S}$ is coherent by definition of $M$.
  We deduce that $T = \roleq\Pol{U}.T'$ and also that
  $N \parop \Map{\rolep}{T'}$ is coherent. Hence $S' \ssubt T'$, namely
  $S' \usubt T' \in \srel$ by definition of $\srel$.
  We conclude by observing that $S \usubt T$ is derivable by
  \refrule{us-channel}.
\end{proof}

\newcommand{\dualof}[3][D]{\mathsf{dual}_{#1}(\Map{#2}#3)}

Next we show that every bounded session type may be part of a coherent session
map. This result is somewhat related to the notion of \emph{duality} in binary
session type theories \citep{Honda93,HondaVasconcelosKubo98}, showing that every
behavior can be completed by a matching -- dual -- one.

\begin{definition}[Duality]
  Let $\targets{\cdot}$ be the function that yields the set of roles occurring
  in a session type, let $S$ be a bounded session type and $D$ be a non-empty
  set of roles that includes $\targets{S}$ but not $\rolep$.
  Let $\dualof\rolep{S}$ be the session map corecursively defined by
  the following equations:
  \[
    \begin{array}{r@{~}c@{~}ll}
      \dualof\rolep{\End[\In]} & = & \set{\Map\roleq \End[\Out]}_{\roleq\in D}
      \\
      \dualof\rolep{\End[\Out]} & = & \Map{\min D} \End[\In] \parop \set{\Map\roleq \End[\Out]}_{\roleq\in D \setminus \set{\min D}}
      \\
      \dualof\rolep{\Tags\roleq\Pol\Tag_i.S_i} & = &
      \Map\roleq \Tags\rolep\co\Pol\Tag_i.\dualof\rolep{S_i}
      \\
      \dualof\rolep{\roleq\Pol{U}.S} & = &
      \Map\roleq \rolep\co\Pol{U}.\dualof\rolep{S}
    \end{array}
  \]
\end{definition}

\begin{lemma}[Duality]
	\label{lem:duality}
	$\dualof\rolep{S} \parop \Map\rolep S$ is coherent.
\end{lemma}
\begin{proof}
  Follows from the definition of $\dualof\rolep{S}$.
\end{proof}

Now we provide an algorithmic way of computing the ``difference'' between two
session types related by unfair subtyping.

\begin{definition}[Subtyping weight]
	\label{def:fsub_wg}
  Under the hypothesis $S\usubt T$, let $\rk(S,T)\in\N\union\set\infty$ 
  be the least solution of the system of equations below:
\[
  \begin{array}{@{}r@{~}c@{~}ll@{}}
    \rk(\End, \End) & = & 0
    \\
    \rk(\Tags\role\In\Tag_i.S_i, \Tags[i\in J]\role\In\Tag_i.T_i)
    & = & \max_{i\in I} \rk(S_i, T_i)
    & I \subseteq J
    \\
    \rk(\Tags\role\Out\Tag_i.S_i, \Tags[i\in J]\role\Out\Tag_i.T_i)
    & = & 1 + \min_{i\in J} \rk(S_i, T_i)
    & J \subsetneq I
    \\
    \rk(\Tags\role\Out\Tag_i.S_i, \Tags\role\Out\Tag_i.T_i)
    & = & \min\set{ \\
    	& & ~~ 1 + \min_{i\in I} \rk(S_i,T_i), \\
    	& & ~~ \max_{i\in I} \rk(S_i,T_i)}
    \\
    \rk(\role\Pol{U}.S', \role\Pol{U}.T') & = & \rk(S', T')
  \end{array}
\]
\end{definition}

To see that $\rk(S,T)$ is well defined, observe that the system of equations
defining $\rk(S,T)$ under the hypothesis $S \usubt T$ contains finitely many
equations, say $n$, by regularity of $S$ and $T$. The system is representable as
a monotone endofunction $F$ on the complete lattice $(\Nat\union\set\infty)^n$.
Thus, $F$ has a least solution of which $\rk(S,T)$ is a component.
We call two session types $S$ and $T$ divergent if they are related by unfair
subtyping and have infinite rank.

\begin{definition}[Divergence]
	\label{def:diverge}
	\brk
 	 We write $S \diverge T$ if $S \usubt T$ and $\rk(S, T) = \infty$.
\end{definition}

\begin{lemma}
	\label{lem:diverge}
  	If $S \diverge T$ then the derivation of $S \usubt T$ contains at least one application of \refrule{u-tag-out} with $J \subsetneq I$ and one of the following holds:
  	\begin{enumerate}
  	\item $S = \Tags \role\In\Tag_i.S_i$ and
    	$T = \Tags[i \in J] \role\In\Tag_i.T_i$ with $I \subseteq J$ and
    	$S_k \diverge T_k$ for some $k\in I$, or
  	\item $S = \Tags \role\Out\Tag_i.S_i$ and
    	$T = \Tags[i\in J] \role\Out\Tag_j.T_j$ with $J \subseteq I$ and
    	$S_i \diverge T_i$ for every $i\in J$, or
  	\item $S = \role\Pol{U}.S'$ and $T = \role\Pol{U}.T'$ and
    	$S' \diverge T'$.
  	\end{enumerate}
\end{lemma}
\begin{proof}
  If the derivation of $S \usubt T$ contained no application of
  \refrule{u-tag-out} with $J \subsetneq I$ we would have
  $\rk(S, T) = 0$. Now we reason by cases on the last rule used to
  derive $S \usubt T$.

  \proofrule{us-end}
  Then $S = T = \End$.  This case is impossible because
  $\rk(\End, \End) = 0$ by definition.

  \proofrule{us-tag-in}
  Then $S = \Tags\role\In\Tag_i.S_i$ and
  $T = \Tags[i\in J] \role\In\Tag_i.T_i$ with $I \subseteq J$ and
  $S_i \usubt T_i$ for every $i\in I$ and
  $\infty = \rk(S, T) = \max_{i\in I} \rk(S_i, T_i)$.
  That is, $\rk(S_k,T_k) = \infty$ for some $k \in I$, hence we
  conclude $S_k \diverge T_k$.

  \proofrule{us-tag-out}
  Then $S = \Tags\role\Out\Tag_i.S_i$ and
  $T = \Tags[i\in J]\role\Out\Tag_i.T_i$ with $J \subseteq I$ and
  $S_i \usubt T_i$ for every $i\in J$.
  We distinguish two sub-cases.
  If $J \subsetneq I$ then
  $\infty = \rk(S, T) = 1 + \min_{i\in J} \rk(S_i, T_i)$, that is
  $\rk(S_i, T_i) = \infty$ for every $i\in J$.
  If $J = I$ then
  $\infty = \rk(S, T) = \min\set{1 + \min_{i\in I} \rk(S_i, T_i),
    \max_{i\in I} \rk(S_i, T_i)} \leq 1 + \min_{i\in I} \rk(S_i,
  T_i)$ and we have $\rk(S_i, T_i) = \infty$ for every $i\in I$.
  Therefore, in both cases, we conclude $S_i \diverge T_i$ for every
  $i \in J$.

  \proofrule{us-channel}
  Then $S = \role\Pol{U}.S'$ and $T = \role\Pol{U}.T'$ and
  $S' \usubt T'$ and $\infty = \rk(S, T) = \rk(S', T')$ hence we
  conclude $S' \diverge T'$.
\end{proof}

\newcommand{\discriminator}[3]{\mathsf{disc}(#1,#2,#3)}

Finally, the key aspect of the proof of \Cref{lem:divergence} is 
how we build the session map $M$ such that $M \parop \Map{\rolep}{S}$ 
is coherent while $M \parop \Map{\rolep}{T}$ is not.

\begin{definition}[Discriminator]
	\label{def:discriminator}
	Let $\discriminator\rolep{S}{T}$ be the session map
  	corecursively defined by the following equations:
  	\[
    	\begin{array}{r@{~}c@{~}l}
      	\discriminator\rolep{
        	\Tags\roleq\In\Tag_i.S_i
      	}{
        	\Tags[i\in J]\roleq\In\Tag_i.T_i
      	}
      	& = &
      	\Map\roleq \Tags[i\in I, S_i \diverge T_i]\rolep\Out\Tag_i.\discriminator\rolep{S_i}{T_i}
      	\\ & & \hfill \text{if $I \subseteq J$}
      	\\
      	\discriminator\rolep{
        	\Tags\roleq\Out\Tag_i.S_i
      	}{
        	\Tags[i\in J]\roleq\Out\Tag_i.T_i
      	}
      	& = &
      	\Map\roleq \Tags[i\in J]\roleq\In\Tag_i.\discriminator\rolep{S_i}{T_i} 
        \\
        & + & \Tags[i\in I\setminus J] \roleq\In\Tag_i.\dualof\rolep{S_i}
      	\\ & & \hfill \text{if $J \subseteq I$}
      	\\
      	\discriminator\rolep{
        	\roleq\Pol U.S 
      	}{
        	\roleq\Pol U.T
      	}
      	& = &
      	\Map\roleq \rolep\co\Pol U.\discriminator\rolep{S}{T}
    	\end{array}
  	\]
\end{definition}

\begin{lemma}
	\label{lem:divergence}
  If $S$ is bounded and $S \diverge T$ then $S \not\ssubt T$.
\end{lemma}
\begin{proof}
  From the hypothesis that
  $S \diverge T$ and \Cref{lem:diverge} we deduce that the
  derivation of $S \usubt T$ contains at least one application of
  \refrule{us-tag-out} with $J \subsetneq I$.
  Consider $\discriminator\rolep{S}{T}$ from \Cref{def:discriminator}.
  Note that $\discriminator\rolep{S}{T}$ always sends a subset of
  the labels accepted by $S$, it is willing to receive any label
  sent by $S$, and it can always terminate successfully when
  interacting with $S$. Note also that it terminates successfully
  only after receiving a label from $S$ that $T$ cannot sent.
  Therefore, we have
  $\discriminator\rolep{S}{T} \parop \Map\rolep{S}$ coherent and
  $\discriminator\rolep{S}{T} \parop \Map\rolep{T}$ incoherent,
  which proves $S \not\ssubt T$.
\end{proof}

\begin{proof}[Proof of \Cref{thm:fsub_complete}]
  \newcommand{\foosubt}{\subt[@]}
  Let $S \foosubt T$ if $S \usubt T$ and $\rk(U, T) < \infty$ for
  every judgment $U \usubt V$ in the derivation of $S \usubt T$.
  From \Cref{lem:usub_complete,lem:divergence} we have that
  $S \ssubt T$ implies $S \foosubt T$. Indeed, if there is a
  judgment $U \usubt V$ in the derivation of $S \usubt T$ such that
  $\rk(U,V) = \infty$, then it is possible to build a session $M$
  such that $M \parop \Map\role{S}$ is coherent and
  $M \parop \Map\role{T}$ is not by induction on the minimum depth
  of the judgment $U \usubt V$ in the derivation using the
  hypothesis that $S$ is bounded and \Cref{lem:divergence}.

  Now, using the principle of coinduction, it suffices to show that
  each judgment in the set
  \[
    \srel \eqdef \set{ S\subt[\rk(S,T)] T \mid S \foosubt T}
  \]
  is derivable using one of the rules in \Cref{fig:fsub_multi} whose
  premises all belong to $\srel$.
\end{proof}

%% file: fairsub/fsub-gis.tex
\beginbass
The purely \emph{coinductive} characterization of fair subtyping 
presented in \Cref{fig:fsub_bin,fig:fsub_multi} has been first used in \cite{CicconeDP22}.
Previous works 
\citep{Padovani13,Padovani16,CicconePadovani21,CicconePadovani22}
relied on a different characterization based on a generalized
inference system (see \Cref{sec:gis}) that consists of the same rules
of the original subtyping relation (\Cref{fig:usub}) equipped with
a corule. The generalized inference system for binary session types
is presented in \Cref{fig:fsub_gis}. Note the invariance in the type
of the channel being exchanged according to \Cref{rmk:fsub_invariant}.
In addition to $\subt$, we write $\isubt$ for the
relation defined by the \emph{inductive} interpretation of the same inference
system and $\csubt$ for the relation defined by the \emph{coinductive}
interpretation of the inference system in \cref{fig:fsub_gis} 
excluding the corule \refrule{fs-converge}. We introduce the
notions of \emph{paths} and \emph{residual} of a session type.

\begin{figure}[t]
  \framebox[\textwidth]{
    \begin{mathpar}
        \inferrule{\mathstrut}{
          \End \usubt \End
        }
        \and
        \inferrule{
          \S \usubt \T
        }{
          \Out\U.\S \usubt \Out\U.\T
        }
        \\
    		\inferrule{
           \forall i\in I: \S_i \usubt \T_i
        }{
        	\textstyle
          \Tags\In\Tag_i.S_i \usubt 
          \Tags[i \in I \union J]\In\Tag_i.T_i
        }
        \and
        \inferrule{
           \forall i\in I: \S_i \usubt \T_i
        }{
        	\textstyle
          \Tags[i \in I \union J]\Out\Tag_i.S_i \usubt 
          \Tags[i \in I]\Out\Tag_i.T_i
        }
        \and
        \infercorule[fs-converge]{
          \forall\actionsA\in\paths\S\setminus\paths\T:
          \exists\actionsB \prefix \actionsA, \Tag:
          S(\actionsB\Out\Tag) \subt T(\actionsB\Out\Tag)
        }{
          S \subt T
        } \defrule[fs-converge]{}
    \end{mathpar}
  }
  \caption{
    Generalized inference system for fair subtyping}
  \label{fig:fsub_gis}
\end{figure}

\begin{definition}[Paths of a session type]
	\label{def:path}
  We say that $\actions$ is a \emph{path} of $S$ if $S \wlred\actions$. We write
  $\paths\S$ for the (prefix-closed) set of paths of $S$, that is $\paths\S
  \eqdef \set{\actions \mid S \wlred\actions }$.
\end{definition}

\begin{definition}[Residual of a session type]
  \label{def:residual}
  The \emph{residual} of a session type $S$ with respect to a path $\actions \in
  \paths\S$, denoted by $S(\actions)$, is the unique session type $T$ such that
  $S \wlred\actions T$.
\end{definition}

The subtle difference between fair and unfair subtyping is due to the corule
\refrule{fs-converge}. Since this corule is somewhat obscure, 
we explain it gradually starting with the following observations:
\begin{enumerate}
\item Recall from \Cref{sec:gis} that $S \subt T$ implies $S \csubt T$ and $S
  \isubt T$. Hence, $\subt$ is a refinement of $\csubt$ such that, for each pair
  of related session types $S$ and $T$, there exists a \emph{finite-depth}
  derivation tree for the judgment $S \subt T$ using the rules and possibly the
  corule \refrule{fs-converge}.
\item When $S \csubt T$ holds, it is not possible to establish a general
  correlation between $\paths\S$ and $\paths\T$. Indeed, \refrule{us-tag-in}
  entails that some paths of $T$ may not be present in $S$ and
  \refrule{us-tag-out} entails that some paths of $S$ may not be present in $T$.
\item The judgment $S \subt T$ is trivially derivable using \refrule{fs-converge}
  if the path inclusion relation $\paths\S \subseteq \paths\T$ holds. Since
  \refrule{us-tag-out} is the only rule that allows $T$ to have fewer paths
  than $S$, we deduce that \refrule{fs-converge} limits (but does not always
  forbid) applications of \refrule{us-tag-out}.
\item In general \refrule{fs-converge} requires that, whenever a path $\actions$
  of $S$ is no longer present in $T$, it must be possible to find a prefix
  $\actionsB$ of $\actions$ and an output $\Out\Tag$ shared by both $S$ and $T$
  such that $S(\actionsB\Out\Tag)$ and $T(\actionsB\Out\Tag)$ are one step closer to
  the region of $S$ and $T$ where path inclusion holds.
\end{enumerate}

The reason why path inclusion plays such an important role in the definition of
fair subtyping is that a process using a channel of type $T$ keeps using it
according to $T$ even if it is replaced by another channel of type $S \subt T$,
without even realizing that the replacement has taken place. After all, this is
what the ``safe substitution principle'' (\Cref{prop:safe_sub}) is based on. 
As a consequence, none of
the paths in $S$ that have disappeared in $T$ will be offered to the process at
the other end of the session. If there are ``too few'' paths in $T$ compared
to $S$, then the replacement might compromise the termination of the process at
the other end of the session, should it crucially rely on those paths to
terminate.
When $S \subt T$ (and therefore $S \isubt T$) holds, the corule
\refrule{fs-converge} makes sure that the process using the channel of type $S$ 
believing
that it has type $T$ is always at \emph{finite distance} from the region where
path inclusion between (some subtrees of) $S$ and (the corresponding subtrees
of) $T$ holds. Moreover, this region is always reachable by means of
\emph{output actions} (those $\Out\l$ mentioned in \refrule{fs-converge}) which
are performed actively by the process using the channel. In other words, the process
using such channel is always able, in a finite amount of time and relying on choices and
actions it can perform autonomously, to steer the interaction towards a region
of the protocol where path inclusion holds, hence where a common path to session
termination is guaranteed to exist.

To conclude, the generalized inference system in \Cref{fig:fsub_gis}
has the advantage of highlighting the \emph{liveness}-preserving 
feature of fair subtyping by using \refrule{fs-converge}. However,
the purely coinductive characterization that we gave in \Cref{sec:fair_sub}
allowed us to to provide a direct proof of \Cref{thm:fsub_preorder}.
Indeed, in previous works
\citep{Padovani13,Padovani16,CicconePadovani21,CicconePadovani22},
transitivity has been established indirectly by relating the 
generalized inference system of
fair subtyping with its semantic definition (\Cref{def:ssubt}).

In \Cref{pt:agda} we will characterize properties of session types
mixing safety and liveness aspects. Starting from the definitions
of the safety (coinductive) parts, we will show how to use corules
to obtain the desired predicates. Hence, for what concerns
fair subtyping, we will refer to \Cref{fig:fsub_gis} to provide
a sound and complete Agda mechanization of such predicate (see \Cref{sec:agda_fs}).

%% file: ts-bin/chintro.tex
\begintreble
In this chapter we present the first type system for enforcing fair
termination of \emph{binary sessions}. Although in \Cref{ch:ft_multi} we will
present the same approach applied to multiparty ones, hence to a more general context,
the binary case succeeds in highlighting the main requirements, properties and
challenges of such a type system.
For this reason, in this chapter we focus on paradigmatic examples in order
to guide the reader across the main developed features. 
In \Cref{ssec:proc_ex_multi,ssec:ts_multi_ex} we will show some more involved scenarios.
It is worth noting that the proof technique that we adopt to prove the soundness
of the type system is shared by both the binary and the multiparty case. However,
the two proofs differ in some technicalities (\eg deadlock freedom).
The type system that we present in this chapter is a refined version of the one
used by \cite{CicconePadovani22}. Indeed, \cite{CicconePadovani22} rely on
the characterization of fair subtyping presented in \Cref{ssec:fsub_gis}.
For the sake of uniformity with respect to \cite{CicconeDP22} on which
\Cref{ch:ft_multi} will be based on,
we refer to the purely coinductive characterization presented in \Cref{sec:fair_sub}.

The chapter is organized as follows.
In \Cref{sec:ts_bin_proc} we present the syntax and the semantics of the
session based calculus on which we apply static analysis.
\Cref{sec:ts_bin_ts} shows the type system for such calculus and introduce
some additional properties that are required to enforce fair termination.
Finally, in \Cref{sec:ts_bin_corr} we detail the soundness proof the type system. 

Concerning a comparison of the type system we show in this chapter with respect
to existing works, we delay the discussion to \Cref{ch:ft_multi}.

%% file: ts-bin/calculus.tex
\beginalto
In this section we introduce the calculus for binary sessions.
We recall some basic notions.
We use an infinite set of \emph{channel names} ranged over by $x$, $y$, $z$,
a set of \emph{message tags} ranged over by $\Tag$, and a set of \emph{process names} ranged over by
$A$, $B$, $C$.
We write $\seqof x$ for a possibly empty sequences of
channels, extending this notation to other entities and 
we use $\Pol$ to range over the elements of the set $\set{\iact,\oact}$ of
\emph{polarities} to distinguish input actions ($\iact$) from output actions
($\oact$).


\subsection{Syntax of Processes}
\input{ts-bin/proc-syntax}


\subsection{Operational Semantics}
\input{ts-bin/proc-sem}

%% file: ts-bin/proc-syntax.tex
\beginbass
A \emph{program} is a finite set of definitions of the form
$\pdef\pdn{\seqof\var}{P}$, at most one for each process name, where
$P$ is a \emph{process} generated by the grammar in \Cref{fig:proc_syntax_bin}.
\begin{figure}[t]
	\framebox[\textwidth]{
	\begin{math}
		\begin{array}[t]{@{}rcll@{}}
    	P, Q & ::= & \Done & \text{termination}
    	\\
    	& | & \Wait\x{P} & \text{signal in}
    	\\
    	& | & \PInput\x{(\y)}.{P} & \text{channel in}
    	\\
    	& | & \PBranch\x\Pol{\l_i}{P_i}{i \in I} & \text{label in/out}
    	\\
    	& | & \NewPar\x{P}{Q} & \text{session}
  	\end{array}
  	~
  	\begin{array}[t]{@{}rcll@{}}
    	& | & \Call{A}{\seqof\x} & \text{invocation}
    	\\
    	& | & \Close\x & \text{signal out}
    	\\
    	& | & \POutput\x{(\y)}.{P} & \text{channel out}
    	\\
    	& | & P \pchoice Q & \text{choice}
    	\\
    	& | & \Cast\x{P} & \text{cast}
  	\end{array}
	\end{math}
	}
	\caption{Syntax of processes}
	\label{fig:proc_syntax_bin}
\end{figure}
The process $\Done$ is terminated and performs no action.
The invocation $\Call A {\seqof\x}$ behaves as $P$ if $\Definition A {\seqof\x} P$ is the definition of $A$. 
When $\seqof\x$ is empty, we write $\Call A {}$ and $\Definition A {} P$ 
instead of $\Call A {}$ and $\Definition A {{}} P$.
The process $\Wait\x{P}$ waits for a signal from channel $x$ indicating that the
session $x$ is being closed and then continues as $P$. The process $\Close\x$
sends the termination signal on $x$.
The process $\PInput\x{(\y)}.{P}$ receives a channel $y$ from channel $x$ and then continues as $P$. 
Dually, $\POutput\x{(\y)}.{P}$ sends $y$ on $x$ and then continues as $P$.
The process $x\Pol\set{\l_i:P_i}_{i\in I}$ exchanges a tag $\l_i$ on channel $x$ and then continues as $P_i$. 
As for session types, we assume that the set $I$ in these forms is always non-empty and 
that $i\ne j$ implies $\l_i \ne \l_j$ for every $i,j\in I$. 
Also, we write $x \Pol \l_i.P_i$ instead of $x \Pol\set{\l_i : P_i}_{i\in I}$ when $I$ is a singleton $\set{i}$.
A non-deterministic choice $P_1 \pchoice P_2$ reduces to either $P_1$ or $P_2$.
The annotation $k\in\set{1,2}$ has no operational meaning, it is only used to
record that $P_k$ leads to the termination of the process and is omitted when
irrelevant.
A session $\NewPar\x P Q$ is the parallel composition of $P$ and $Q$ connected by $x$.
Note that composition and restriction is atomic. This approach is inspired by linear-logic
based calculi; we provide more details later on.
Finally, a \emph{cast} $\Cast \x P$ behaves exactly as $P$. This form simply
records the fact that the type of $x$ is subject to an application of fair
subtyping in the typing derivation for $P$. As we will see in
\Cref{sec:ts_bin_ts}, we use this form to precisely account for
all places in (the typing derivation of) a process where fair subtyping is used.
Occasionally we write $\Cast{x_1\cdots x_n}P$ for $\Cast{x_1}\cdots\Cast{x_n}P$.

The only binders are channel inputs $\PInput\x{(y)}.P$ and sessions
$\NewPar\x{P}{Q}$. We write $\fn{P}$ and $\bn{P}$ for the sets of free and bound
channel names occurring in $P$ and we identify processes modulo renaming of
bound names.
The program $\Program$ that provides the meaning to the process names occurring
in processes is often left implicit. Sometimes we write a process definition
$\Definition{A}{\seqof\x}{P}$ as a proposition or side condition, intending that
such definition is part of the implicit program $\Program$.
Note that this approach differs from the one used by \cite{ScalasYoshida19} as they 
introduce process definition in the syntax. We adopt a different approach in order
to keep the calculus as light as possible.
 
\begin{example}
  \label{ex:bsc_bin_proc}
  Let us revisit and complete the example we sketched in \Cref{ex:bsc}. 
  We let the \actor{buyer} $\tadd$ an \emph{odd} number of items.
  We can model the whole system as the following set of process definitions:
  \[
      \begin{array}{@{}r@{~}l@{}}
        \Definition\Main{}{&
          \NewPar\y{\NewPar\x{\Cast\x \Call \Buyer x}{\Call \Seller {x,y}}}{\Call \Carrier y}
        }
        \\
        \Definition{\Buyer}{x}{&
          \POutput\x\ladd.{
          	\PSend\x{
            	\ladd : \Call \Buyer x,
            	\lpay : \Close\x
          	}
        }}
				\\
				\Definition{\Seller}{x,y}{&
              \PRecv\x{
                  \ladd : \Call \Seller {x,y},
                  \lpay : \Wait\x{\POutput\y\lship.{\Close\y}
              }}
        }
				\\
        \Definition{\Carrier}{y}{&
            \PInput\y\lship.{\Wait\y\Done}
        }
      \end{array}
  \]

  Note that $\Buyer$ deterministically sends $\tadd$ to $\Seller$ as the
  first message, whereas he chooses among $\tadd$ and $\tpay$ every other
  interaction. After $\Buyer$ has sent $\tpay$, it closes the session $x$
  with $\Seller$. At this point, $\Seller$ sends $\tship$ to
  $\Carrier$ and closes the session $y$.
  The cast $\Cast\x$ before the invocation of $\Call \Buyer x$ in $\Main$ is meant to
  account for the mismatch between the behavior of the buyer, which always
  adds an odd number of items to the cart, and that of the business, which
  accepts any number of items added to the shopping cart.
  \eoe
\end{example}

%% file: ts-bin/proc-sem.tex
\beginbass
\begin{figure}[t]
  \framebox[\textwidth]{
    \begin{math}
      \displaystyle
      \begin{array}{@{}lr@{~}c@{~}ll@{}}
        \defrule{sb-par-comm} & \NewPar\x{P}{Q} & \pcong & \NewPar\x{Q}{P}
        \\
        \defrule{sb-par-assoc} & \NewPar\x{P}{\NewPar\y{Q}{R}} & \pcong & \NewPar\y{\NewPar\x{P}{Q}}{R}
        & \text{if $x \in \fn{Q}$}
        \\
        & & & & \text{and $y \not\in \fn{P}$}
        \\
        & & & & \text{and $x \not\in \fn{R}$}
        \\
        \defrule{sb-cast-comm} & \Cast\x\Cast\y P & \pcong & \Cast\y\Cast\x P
        \\
        \defrule{sb-cast-new} & \NewPar\x{\Cast\x{P}}{Q} & \pcong & \NewPar\x{P}{Q} &
        \\
        \defrule{sb-cast-swap} & \NewPar\x{\Cast\y{P}}{Q} & \pcong & \Cast\y\NewPar\x{P}{Q}
        & \text{if $x \ne y$}
        \\
        \defrule{sb-call} & \Call{A}{\seqof\x} & \pcong & P
        & \text{if $\Definition{A}{\seqof\x}{P}$}
      \end{array}
    \end{math}
  }
  \caption{Structural pre-congruence of processes}
  \label{fig:pcong_bin} 
\end{figure}
\begin{figure}[t]
  \framebox[\textwidth]{
    \begin{mathpar}
      \displaystyle
      	\inferrule[rb-choice]{\mathstrut}{P_1 \pchoice P_2 \red P_k}
        ~ k\in\set{1,2}
        \defrule[rb-choice]{}
        \and
        \inferrule[rb-signal]{\mathstrut}{\NewPar\x{\Close\x}{\Wait\x{P}} \red P}
        \defrule[rb-signal]{}
        \and
        \inferrule[rb-channel]
        	{\mathstrut}
        	{\NewPar\x{\POutput\x\y.P}{\PInput\x{(y)}.Q} \red \NewPar\x{P}{Q}}
        	\defrule[rb-channel]{}
        \and
        \inferrule[rb-pick]
        	{\mathstrut}
        	{\NewPar\x{\PSend\x{\l_i : P_i}_{i\in I}}{Q} \red \NewPar\x{\POutput\x{\l_k}.P_k}{Q}} 
        	~ k\in I, |I| > 1
        	\defrule[rb-pick]{}
        \and
        \inferrule[rb-tag]
        	{\mathstrut}
        	{\NewPar\x{\POutput\x{\l_k}.P}{\PRecv\x{\l_i : Q_i}_{i\in I}} \red \NewPar\x{P}{Q_k}}
        	~ k\in I
        	\defrule[rb-tag]{}
        \and
        \inferrule[rb-par]
        	{P \red Q}
        	{\NewPar\x{P}{R} \red \NewPar\x{Q}{R}}
        	\defrule[rb-tag]{}
        \and
        \inferrule[rb-cast]
        	{P \red Q}
        	{\Cast\x P \red \Cast\x Q}
        	\defrule[rb-cast]{}
        \and
        \inferrule[rb-struct]
        	{P \pcong P' \\ P' \red Q' \\ Q' \pcong Q}
        	{P \red Q}
        	\defrule[rb-struct]{}
    \end{mathpar}
  }
  \caption{Reduction of processes}
  \label{fig:red_bin} 
\end{figure}
The operational semantics of processes is defined using a structural
pre-congruence relation $\pcong$ and a reduction relation $\red$, both of which
are defined in \Cref{fig:pcong_bin,fig:red_bin} and described hereafter.
Rules \refrule{sb-par-comm} and \refrule{sb-par-assoc} express the usual
commutativity and associativity of parallel composition. In the case of
\refrule{sb-par-assoc}, the side condition $x\in\fn{Q}$ makes sure that the
session $\NewPar\x P Q$ we obtain on the right hand side does indeed connect $P$
and $Q$ through $x$.
The other side conditions $y \not\in \fn{P}$ and $x \not\in \fn{R}$ will
always hold when dealing with well-typed processes.
Also note that \refrule{sb-par-assoc} only describes a right-to-left
associativity of parallel composition and that left-to-right associativity is
derivable by the chain of relations
 \[
   \begin{array}{r@{~}l}
     \NewPar\x{\NewPar\y{P}{Q}}{R}
     &
     \pcong
     \NewPar\x{R}{\NewPar\y{P}{Q}}
     \\
     & \pcong
     \NewPar\x{R}{\NewPar\y{Q}{P}}
     \\
     & \pcong
     \NewPar\y{\NewPar\x{R}{Q}}{P}
     \\
     & \pcong
     \NewPar\y{\NewPar\x{Q}{R}}{P}
     \pcong
     \NewPar\y{P}{\NewPar\x{Q}{R}}
   \end{array}
 \]
 when $x\in\fn{Q}$.
Axiom \refrule{sb-cast-new} annihilates a cast on $x$ nearby
the binder for $x$, making sure that casts can only be removed and never added.
Axioms \refrule{sb-cast-comm} and \refrule{sb-cast-swap} are used to move casts
closer to their binder so that they can be annihilated with
\refrule{sb-cast-new}. Rule \refrule{sb-call} unfolds process invocations to their
definition.

\begin{remark}[On structural pre-congruence]
	\label{rmk:pcong}
	The reader might be wonder why we adopted a pre-congruence relation instead of a more common
	congruence one. The reason behind this choice is that in general it reduces the number of
	cases to be analyzed in some proofs (see \Cref{lem:subj_cong_bin}) without compromising 
	the properties of the calculus. Moreover, the adoption of a congruence relation would
	lead to some unrealistic cases from the computation point of view. Indeed,
	\refrule{sb-cast-new} would introduce a reflexive application of fair subtyping from
	a session restriction and \refrule{sb-call} would fold a process call.
	\eor
\end{remark}

The reduction rules are mostly unremarkable: \refrule{rb-choice} models the
non-deterministic choice between alternative behaviors; \refrule{rb-pick} models
a non-trivial choice among a set of labels to send; \refrule{rb-signal},
\refrule{rb-label} and \refrule{rb-channel} model synchronizations between a
sender (on the left hand side of the parallel composition) and a receiver (on
the right hand side of the parallel composition) with \refrule{rb-signal}
removing the binder of a closed session; \refrule{rb-par}, \refrule{rb-cast} and
\refrule{rb-struct} close reductions under parallel compositions, under casts and
by structural pre-congruence.
In the following we write $\wred$ for the reflexive, transitive closure of
$\red$ and $\wred^+$ for $\wred\red$.

\begin{example}
  \label{ex:bsc_bin_reduction}
  With the definitions given in \Cref{ex:bsc_bin_proc}, it is easy to see that
  there is an infinite reduction sequence starting from $\Main$ in which the
  acquirer keeps adding items to the cart:
  \[
    \begin{array}{@{}c@{~}l@{}}
    	&
      \NewPar\y{
        \NewPar\x{
          \Cast x
          \Call \Buyer x
        }{
          \Call \Seller {x,y}
        }
      }{
        \Call \Carrier y
      }
      \\
      \wred &
      \NewPar\y{
        \NewPar\x{
          \PSend\x{
            \ladd : \Call \Buyer x,
            \lpay : \Close x
          }
        }{
          \Call \Seller {x,y}
        }
      }{
        \Call \Carrier y
      }
      \\
      \red &
      \NewPar\y{
        \NewPar\x{
          \POutput\x\ladd.
          \Call \Buyer x
        }{
          \Call \Seller {x,y}
        }
      }{
        \Call \Carrier y
      }
      \\
      \wred &
      \NewPar\y{
        \NewPar\x{
          \Call \Buyer x
        }{
          \Call \Seller {x,y}
        }
      }{
        \Call \Carrier y
      } 
      \\
      \red & \cdots
    \end{array}
  \]
  
  Nonetheless, $\Main$ is fairly terminating. For example, we have:
  \[
    \begin{array}{c@{~}l}
    	&
      \NewPar\y{
      \NewPar\x{
         \PSend\x{
           \ladd : \Call \Buyer x,
           \lpay : \Close x
         }
       }{
         \Call \Seller {x,y}
       }
     }{
       \Call \Carrier y
     }
      \\
      \red &
      \NewPar\y{
        \NewPar\x{
          \POutput\x\lpay.
          \Close x
        }{
          \Call \Seller {x,y}
        }
      }{
        \Call \Carrier y
      }
      \\
      \wred &
      \NewPar\y{
        \NewPar\x{
          \Close x
        }{
          \Wait\x.
          \POutput\y\lship.
          \Close\y
        }
      }{
        \Call \Carrier y
      }
      \\
      \red &
      \NewPar\y{
        \POutput\y\lship.
        \Close\y
      }{
        \Call \Carrier y
      }
      \\
      \wred &
      \NewPar\y{
        \Close\y
      }{
        \Wait\y
        \Done
      }
      \\
      \red &
      \Done
    \end{array}
  \]

  Note that in general it might be necessary for the acquirer to add one more item to the cart before it can send the payment to the business and the carrier receives a $\lship$ message.
  \eoe
\end{example}

%% file: ts-bin/ts.tex
\beginalto
In this section we present the type system for binary sessions that we introduced in \Cref{sec:ts_bin_proc}.
Before looking at the typing rules we motivate, through a series of examples, the key properties enforced by the type system that, 
taken together, guarantee fair termination. There are two families of problems that can compromise fair termination.
First of all, the process (or part thereof) may be unable to reduce further but is not $\Done$. 
In our model, this can happen for many reasons, for example: a process attempts at sending a 
tag on a session that the receiver is not willing to accept; a process attempts at sending a 
termination signal when the receiver expects a channel; the processes at the two ends of the same 
session are both waiting for a message from that session. These are all examples of \emph{safety violations}, 
which are prevented by any ordinary session type system. 
In \Cref{ssec:boundedness} we focus instead on \emph{liveness violations}. Roughly
speaking, liveness is violated when a process (or part thereof) engages an
infinite computation that cannot possibly terminate.


\subsection{Boundedness Properties}
\label{ssec:boundedness}
\input{ts-bin/ts-boundedness}


\subsection{Typing Rules}
\label{ssec:ts_bin_rules}
\input{ts-bin/ts-rules}


\subsection{Examples}
\input{ts-bin/ts-examples}


\subsection{On Higher-Order Session Types}
\label{ssec:invariant_ch}
\input{ts-bin/ts-higherorder}

%% file: ts-bin/ts-boundedness.tex
\beginbass
In \Cref{ch:fs}
we have introduced a fair subtyping relation that is liveness preserving but,
as we will see in a moment, the adoption of fair subtyping alone is not enough
to rule out all potential liveness violations. The type system must also enforce
three properties that we call \emph{action boundedness}, \emph{session
boundedness} and \emph{cast boundedness} guaranteeing that the overall effort
required to terminate the process is finite. In the rest of the section we
describe informally these properties and we show that violating even just one of
them may compromise fair process termination.


\begin{definition}[Action Boundedness]
	We say that a process is \emph{action bounded} if there is a finite upper bound
	to the number of actions it has to perform in order to terminate. An
	action-unbounded process cannot terminate.
\end{definition}

\begin{example}
	\label{ex:action_boundedness}
	Compare the following processes
	\[
  \begin{array}{ll}
    \Definition{A}{}{A \pchoice \Done}
  	\qquad & \qquad
  	\Definition{B}{}{B \pchoice B}
  \end{array}
  \]
  and observe that $A$ may always reduce to $\Done$, whereas $B$ can only reduce
	forever into itself. So $A$ is action bounded whereas $B$ is not. 
	\eoe
\end{example}

We consider a
parallel composition action bounded if so are \emph{both} processes composed in
parallel.
Action boundedness is a necessary condition for (fair) process termination,
hence the type system must guarantee that well-typed processes are action
bounded. As we will see in \Cref{ssec:ts_bin_rules}, this can be easily achieved by means
of \emph{typing corules} (see \Cref{sec:gis}). 
Besides, action boundedness carries along two welcome side effects.
The first one is that degenerate process definitions such as
$\Definition{A}{}{A}$ are not action bounded and therefore are flagged as ill
typed by the type system. This guarantees that finitely many unfoldings of
recursive process invocations always suffice to expose some observable process
behavior.
The second is that action boundedness allows us to detect recursive processes
that claim to use a channel in a certain way when in fact they never do so. 
\begin{example}
	Consider the following processes
	\[
	\begin{array}{ll}
		\Definition{A}{x,y}{
    	\POutput\x\la.\Call{A}{x,y} \pchoice \POutput\x\lb.\Close\x
  	}
  	\qquad & \qquad
  	\Definition{B}{x,y}{\POutput\x\la.\Call{B}{x,y}}
	\end{array}
	\]
	where $\Call{A}{x,y}$ is action bounded and $\Call{B}{x,y}$ is not. An ordinary
	session type system with coinductively interpreted typing rules would accept
	$\Call{B}{x,y}$ regardless of $y$'s type on the grounds that $y$ occurs once in
	the body of $B$, hence it is ``used'' linearly. This is unfortunate, since $y$
	is not used in any meaningful way other than being passed as an argument of $B$.
	In $A$, the same linearity check promptly detects that $y$ is not used along the
	path to $\Close\x$ that proves the boundedness of $\Call{A}{x,y}$.
	\eoe
\end{example}


\begin{definition}[Session Boundedness]
	We say that a process is \emph{session bounded} if there is a finite upper bound
	to the number of sessions it has to create in order to terminate.
\end{definition}

 It is easy
to construct non-terminating processes by chaining together an infinite number
of finite (or fairly terminating) sessions.

\begin{example}
  \label{ex:session_bounded}
  Compare the following processes
  \[
  \begin{array}{rl}
  	\Definition{A}{}{& \NewPar\x{\Close\x}{\Wait\x{\Call{A}{}}} \pchoice \Done}
  	\\
  	\Definition{B_1}{}{& \NewPar\x{\Close\x}{\Wait\x{\Call{B_1}{}}}}
  \end{array}
  \]
	where $A$ always has a possibility to terminate without creating new sessions
	(it is session bounded) while $B_1$ does not (it is session unbounded). It could
	be argued that $B_1$ is already ruled out because it is not action bounded.
	Indeed, while the left-hand side of the
	parallel composition in $B_1$ is finite, the right hand side is not (recall that
	we require \emph{both} sides of a parallel composition to admit a finite path to
	either $\Done$ or $\Close\x$). 
	\eoe
\end{example}

\begin{example}
	Below is a slightly more complex variation of
	$B_1$ that is action bounded and session unbounded. The trick is to have a
	finite branch on one side of the parallel composition matched by an infinite one
	on the other side:
	\[
	  \Definition{B_2}{}{
    \NewPar\x{
      \PSend\x{
        \la : \Close\x,
        \lb : \Wait\x{\Call{B_2}{}}
      }
    }{
      \PRecv\x{
        \la : \Wait\x{\Call{B_2}{}},
        \lb : \Close\x
      }
    }
  }
  \]
  \eoe
\end{example}

\Cref{ex:session_bounded} shows that a session bounded process like $A$ may
still create an unbounded number of sessions. Below is another example of
session bounded process that creates unboundedly many \emph{nested} sessions,
such that the first session being created is also the last one being completed.
\begin{example}
  \label{ex:infinite_sessions}
  \[
  \begin{array}{ll}
    \NewPar\x{\Call{C}{x}}{\Wait\x\Done}
  	&
  	\Definition{C}{x}{
    	\NewPar\y{\Call{C}{y}}{\Wait\y{\Close\x}} \pchoice \Close\x
  	}
  \end{array}
  \]
	While both $A$ and $C$ \emph{may} create an arbitrary number of sessions, they
	do not \emph{have to} do so in order to terminate. This is what sets them apart
	from $B_1$ and $B_2$.
	\eoe
\end{example}

To ensure session boundedness, we check that no new sessions are found in loops that 
occur along inevitable paths leading to the termination of a process. For example, 
the creation of a new session $x$ is inevitable in both $B_1$ and $B_2$ but it is not in $A$ and $C$.


\begin{definition}[Cast Boundedness]
	We say that a process is \emph{cast bounded} if there is a finite upper bound to
	the number of casts it has to perform in order to terminate.
\end{definition}

Performing a cast
means applying \refrule{sm-cast-new}, which corresponds to a usage of fair
subtyping. The reason why cast boundedness is fundamental is that the
liveness-preserving property of fair subtyping holds as long as fair
subtyping is used finitely many times. Conversely, infinitely many usages of
fair subtyping may have the overall effect of a single usage of unfair subtyping
(see \Cref{ssec:unfair_sub}). By ``infinitely many usages'' we mean usages of
fair subtyping that occur within a loop in a recursive process. 

\begin{remark}[0-weight casts]
	Notably, cast boundedness refers to those situations in which a cast with a strictly
	positive weight is applied inside a loop. It might be the case that a 0-weighted cast
	is performed (\eg no output contravariance). 
	In this case we allow such application of fair subtyping as the 
	liveness of the session is preserved. This is a substantial difference with respect
	to \cite{CicconePadovani22}.
	\eor
\end{remark}

\begin{example}
  \label{ex:infinite_fair}
  Consider the following variants of \actor{buyer} and \actor{seller} from \Cref{ex:bsc} 
  where we can assume that the \actor{seller} closes the session as soon as he receives 
  a tag $\lpay$.
  \[
  \begin{array}{ll}
    \NewPar\x{\Call \Buyer x}{\Call \Seller x}
  	&
  	\begin{array}{r@{~}l}
    	\Definition{\Buyer}{x}{& \Cast\x \POutput\x\ladd.\Call \Buyer x} \\
    	\Definition{\Seller}{x}{&
      	\PRecv\x{
        	\ladd : \Call \Seller x,
        	\lpay : \dots
      	}
    	}
  	\end{array}
  \end{array}
  \]
  and the session type $S_b = \Out\ladd.S_b \choice \Out\lpay.\End[\Out]$.
	It can be argued that the channel $x$ is used according to $S_b$ in $\Buyer(x)$ and
	according to $\co{\S_b}$ in $\Seller(x)$. Indeed, the structure of $\Seller(x)$ matches
	perfectly that of $\co{\S_b}$ and note that $\POutput\x\ladd.\Buyer(x)$ uses $x$ according to
	$\Out\ladd.S_b$, which is a fair supertype of $S_b$ accounted for by the cast
	$\Cast\x$ in $\Buyer$. With this cast it is as if $\Buyer(x)$ promises to make a choice
	between sending $\ladd$ and sending $\lpay$ at each iteration, but
	systematically favors $\ladd$ over $\lpay$. The overall effect of these
	unfulfilled promises is that the actual behavior of $\Buyer(x)$ over $x$ is better
	described by the session type $S_b^\infty = \Out\ladd.S_b^\infty$, which is
	\emph{not} a fair supertype of $S_b$ as we have seen in
	\Cref{ex:unfair_sub}.
	\eoe
\end{example}

Although $\Buyer(x)$ could be rejected on the grounds that it is not action bounded,
it is possible to find an action-bounded (but slightly more involved) variation
of \Cref{ex:infinite_fair} in which the same phenomenon
occurs.

\begin{example}
  \label{ex:infinite_fair_bounded}
  \[
  \begin{array}{r@{~}l}
    \Definition{A}{x}{&
      \Cast\x
      \POutput\x\lmore.
      \PRecv\x{
        \lmore : \Call A x,
        \lstop : \Wait\x\Done
      }
    }
    \\
    \Definition{B}{x}{&
      \PRecv\x{
        \lmore : \Cast\x \POutput\x\lmore. \Call B x,
        \lstop : \Wait\x\Done
      }
    }
  \end{array}
  \]
  both $A(x)$ and $B(x)$ have a chance to continue or to terminate the session by 
  sending either $\lmore$ or $\lstop$, except that they systematically favor $\lmore$ over $\lstop$.
	Now, if we consider the session types
	\[
	\begin{array}{lcl}
		S & = & \Out\lmore.(\In\lmore.S \branch \In\lstop.\End[\In]) \choice \Out\lstop.\End[\Out]
		\\
		S_A & = & \Out\lmore.(\In\lmore.S \branch \In\lstop.\End[\In])
		\\
		S_B & = & \In\lmore.\Out\lmore.\co\S \branch \In\lstop.\End[\In]
	\end{array}
	\]
	it can be argued that $A(x)$
	uses $x$ according to $S_A$, which is a fair supertype of $S$, and that $B(x)$ uses $x$
	according to $S_B$, which
	is a fair supertype of $\co\S$. The casts
	account for the differences between $S$ and $S_A$ in $A(x)$ and between $\co\S$
	and $S_B$ in $B(x)$, but they occur within loops along paths that lead to
	process termination, hence $A$ and $B$ are not cast bounded.
	\eoe
\end{example}

It is worth discussing one last attempt to work around the problem, by moving
the casts outward from within $A(x)$ and $B(x)$.

\begin{example}
  \label{ex:finite_unfair}
  \[
  	\begin{array}{@{}r@{~}l@{}}
  		& \NewPar\x{\Cast\x \Call A x}{\Cast\x \Call B x}
  		\\
    	\Definition{A}{x}{&
      	\POutput\x\lmore.
      	\PRecv\x{
        	\lmore : \Call A x,
        	\lstop : \Wait\x\Done
      	}
    	}
    	\\
    	\Definition{B}{x}{&
      	\PRecv\x{
        	\lmore : \POutput\x\lmore. \Call B x,
        	\lstop : \Wait\x\Done
      	}
    	}
  	\end{array}
  \]
  Now $A(x)$ uses $x$  according to $T_A = \Out\lmore.(\In\lmore.T_A
	\branch \In\lstop.\End[\In])$ and $B(x)$ uses $x$ according to $T_B =
	\In\lmore.\Out\lmore.T_B \branch \In\lstop.\End[\In]$, but while $S \usubt T_A$
	and $\co\S \usubt T_B$ both hold neither $S \subt T_A$ nor $\co\S \subt T_B$
	does.
	\eoe
\end{example}

In summary, the non-terminating process in \Cref{ex:finite_unfair} is
action bounded, session bounded and cast bounded, but it is typeable only using
\emph{unfair} subtyping.
As we will see in \Cref{ssec:ts_bin_rules}, we enforce cast boundedness using the same 
technique already introduced for session boundedness. 
That is, we check that casts do not occur along ``inevitable'' paths leading to recursive process invocations.

%% file: ts-bin/ts-rules.tex
\beginbass
The typing rules resemble those of a traditional session type system but differ
in a few key aspects. First of all, they establish a tighter-than-usual
correspondence between types and processes so that any discrepancy between
actual and expected types is accounted for by explicit casts. This way, we make
sure that actions leading to the termination of a session \emph{at the type
level} are matched by corresponding actions \emph{at the process level}, a key
property used in the soundness proof of the type system.
In addition, the typing rules enforce the boundedness properties informally
described in the previous section.
Action boundedness is enforced by specifying the typing rules as a generalized
inference system and using two corules to make sure that every well-typed
process is at finite distance from $\Done$ or a $\Close\x$.
Concerning session and cast boundedness, we annotate typing judgments with a
\emph{rank}, that is an upper bound to the \emph{weights} of casts that must be
performed and of sessions that must be created in order to terminate the process
in the judgment.

\begin{figure}[t]
  \framebox[\textwidth]{
      \begin{mathpar}
        \displaystyle
          \inferrule[tb-done]{\mathstrut}{
            \wtp[n]\EmptyCtx\Done
          } \defrule[tb-done]{}
          \and
          \inferrule[tb-wait]
          {
            \wtp[n]\Ctx{P}
          }{
            \wtp[n]{\Ctx, x :  \End[\In]}{\Wait\x{P}}
          } \defrule[tb-wait]{}
          \and
          \inferrule[tb-close]{\mathstrut}
          {
            \wtp[n]{x : \End[\Out]}{\Close\x}
          } \defrule[tb-close]{}
          \and
          \inferrule[tb-channel-in]{
            \wtp[n]{\Ctx, x : S, y : T}{P}
          }{
            \wtp[n]{\Ctx, x :  \In\T.S}{\PInput\x{(y)}.P}
          } \defrule[tb-channel-in]{}
          \and
          \inferrule[tb-channel-out]{
            \wtp[n]{\Ctx, x : S}{P}
          }{
            \wtp[n]{\Ctx, x : \Out\T.S, y : T}{\POutput\x\y.P}
          } \defrule[tb-channel-out]{}
          \and
          \inferrule[tb-tag]
          {
          	\forall i\in I:
            \wtp[n]{\Ctx, x : S_i}{P_i}
          }{
            \textstyle
            \wtp[n]{
              \Ctx, x : \Pol\set{\l_i : S_i}_{i \in I}
            }{
              x\Pol\set{\l_i : P_i}_{i \in I}
            }
          } \defrule[tb-tag]{}
          \and
          \inferrule[tb-choice]{
            \wtp[n_1]\Ctx{P}
            \\
            \wtp[n_2]\Ctx{Q}
          }{
            \wtp[n_k]\Ctx{P \pchoice_k Q}
          }
          ~
          k \in \set{1,2}
          \defrule[tb-choice]{}
          \and
          \inferrule[tb-cast]{
            \wtp[n]{\Ctx, x : T}{P}
          }{
            \wtp[n+m]{\Ctx, x : S}{\Cast\x P}
          }
          ~ S \subt[m] T \defrule[tb-cast]{}
          \and
          \inferrule[tb-par]{
            \wtp[m]{\Ctx, x : S}{P}
            \\
            \wtp[n]{\CtxD, x : T}{Q}
          }{
            \wtp[1+m+n]{
              \Ctx, \CtxD
            }{
              \NewPar\x{P}{Q}
            }
          }
          ~
          S \compatible T \defrule[tb-par]{}
          \and
          \inferrule[tb-call]{
            \wtp[n]{\seqof{x:S}}{P}
          }{
            \wtp[m+n]{\seqof{x:S}}{\Call{A}{\seqof\x}}
          }
          ~
          \tass{A}{\seqof{S}}{n},
          \Definition{A}{\seqof\x}{P} \defrule[tb-call]{}
          \and
          \infercorule[cob-tag]{
            \wtp[n]{\Ctx, x : S_k}{P_k}
          }{
            \wtp[n]{
              \Ctx, x : \Pol\set{\l_i : S_i}_{i \in I}
            }{
              x\Pol\set{\l_i : P_i}_{i \in I}
            }
          }
          ~
          k \in  I \defrule[cob-tag]{}
          \and
          \infercorule[cob-choice]{
            \wtp[n]\Ctx{P_k}
          }{
            \wtp[n]\Ctx{P_1 \pchoice_k P_2}
          } \defrule[cob-choice]{}
      \end{mathpar}
    }
    \caption{Typing rules}
    \label{fig:ts_bin}
\end{figure}

The typing rules are defined by the generalized inference system in
\Cref{fig:ts_bin} and derive judgements of the form $\wtp[n]\Ctx{P}$, meaning
that $P$ is well typed in the \emph{typing context} $\Ctx$ and has rank $n$.
A typing context is a finite map from channels to session types written $x_1 :
S_1, \dots, x_n : S_n$ or $\seqof{x : S}$. We use $\Ctx$ and $\CtxD$ to
range over typing contexts, we write $\EmptyCtx$ for the empty context and
$\Ctx,\CtxD$ for the union of $\Ctx$ and $\CtxD$ when they
have disjoint domains.
We type check a program $\set{\Definition{A_i}{\seqof{x_i}}{P_i}}_{i\in I}$
under a global set of type assignments $\set{\tass{A_i}{\seqof{S_i}}{n_i}}_{i\in
I}$ associating each process name $A_i$ with a tuple of session types
$\seqof{S_i}$ and a rank $n_i$. The program is well typed if
$\wtp[n_i]{\seqof{x_i : S_i}}{P_i}$ for every $i\in I$, establishing that the
tuple $\seqof{S_i}$ corresponds to the way the channels $\seqof{x_i}$ are used
by $P_i$ and that $n_i$ is a feasible rank annotation for $P_i$. Hereafter, we
omit the rank from judgments when it is not important. 

Let us look at the typing (co)rules in detail.
\refrule{tb-done} is the usual axiom requiring that the terminated process leaves no unused channels behind. 
Since $\Done$ performs no casts and creates no sessions, it can have any rank.
Rules \refrule{tb-wait} and \refrule{tb-close} concern the exchange of session
termination signals. There is nothing remarkable here except noting once again
that the rank of $\Close\x$ can be arbitrary.
Rules \refrule{tb-channel-in} and \refrule{tb-channel-out} are similar, but they
concern the exchange of channels. Note that, in \refrule{tb-channel-out}, the
type $T$ of the message $y$ is required to match \emph{exactly} that in the type
of the channel $x$ used for the communication, whereas \citep{GayHole05} allow
the type of $y$ to be a subtype of $T$. This is one instance of the ``tight
correspondence'' that we mentioned earlier (see \Cref{ex:invariant_ch}).
The rule \refrule{tb-label} deals with the input/output of labels. As usual, any
channel other than the one affected by the communication must be used in exactly
the same way in every branch. However, the rule is stricter than that of
\citet{GayHole05} because it requires an exact correspondence between the labels
that can be exchanged on $x$ by the process and those in the type of $x$. The
fact that a conclusion and premises are all annotated with the same rank $n$
means that $n$ is an upper bound for the rank of all branches of a label
input/output.
The corule \refrule{cob-label} does not impose additional constraints compared to
\refrule{tb-label} and has \emph{exactly one premise}, corresponding to one
branch of the process in the conclusion. The effect of \refrule{cob-label}, when
interpreted inductively together with the other rules, is to ensure the
existence of a finite typing derivation whose leaves are applications of
\refrule{tb-done} or \refrule{tb-close}, hence action boundedness.

Rule \refrule{tb-choice} is a standard typing rule for non-deterministic choices,
requiring that both branches are well typed in exactly the same typing context.
Notice that the rank of a choice $P_1 \pchoice_k P_2$ is determined by the branch
indexed by the $k$ annotation, which is elected as the branch that leads to
termination. Like \refrule{cob-label}, the associated corule
\refrule{cob-choice} ensures that the same branch gets closer to $\Done$ or a
$\Close\x$ to enforce action boundedness. Without this corule, it would not be
possible to find a \emph{finite-depth} derivation tree for an action-bounded
process such as $A$ in \Cref{ex:action_boundedness}. Coherently with
\refrule{tb-choice}, the same branch that leads to termination is also the one
that determines the rank of the choice as a whole.

Rule \refrule{tb-cast} is Liskov's substitution principle formulated as an
inference rule. It states that a channel $x$ of type $S$ can be safely used
where a channel of type $T$ is expected, provided that $S \subt T$. The most
important detail to notice here is that the rank of a cast is the \emph{weight}
of the subtyping judgment plus that of
the process in which the cast has effect. This way we account for this cast in
the rank of the process so as to guarantee cast boundedness.
Rule \refrule{tb-par} concerns parallel composition and session creation. The
rule is shaped after the cut rule of linear logic also adopted in other session
type systems based on linear logic
\citep{CairesPfenningToninho16,Wadler14,LindleyMorris16}. In particular, the
parallel processes $P$ and $Q$ share no channel other than the session $x$ that
connects them, so as to prevent mutual dependencies between sessions and
guarantee deadlock freedom. The side condition $S \compatible T$ requires that
the way in which $P$ and $Q$ use channel $x$ is such that the session $x$ can
fairly terminate (see \Cref{def:compatibility}). We \emph{do not} require that $S$
and $T$ are dual to each other because reductions (see \refrule{rb-pick}) and
structural pre-congruence (see \refrule{sb-cast-new}) do not necessarily preserve
session type duality. Also, duality does not always imply compatibility.
The rank of a parallel composition is one plus that of the composed processes.
By accounting for each occurrence of parallel compositions in the rank, we
guarantee that well-typed processes are session bounded.

Finally, rule \refrule{tb-call} states that a process invocation
$\Call{A}{\seqof\x}$ is well typed provided that the types associated with
$\seqof\x$ match those of the global assignment $\tass{A}{\seqof{S}}{n}$. Note
that \refrule{tb-call} is \emph{not} an axiom: its premise (re)checks that the
body $P$ in the definition of $A$ is coherent with the global type assignment
$\tass{A}{\seqof{S}}{n}$. With this formulation of \refrule{tb-call}, the only
axioms are \refrule{tb-done} and \refrule{tb-close} so that the inductive
interpretation of the typing (co)rules ensures action boundedness. Note also
that the rank of the conclusion may be greater than the rank $n$ associated with
$A$. This overapproximation grants more flexibility when typing different
branches in \refrule{tb-label}.

\begin{remark}[On structural pre-congruence...continuation]
	Now we have all the ingredients to understand why the choice of a pre-congruence
	over a congruence relation is just a design one (see \Cref{rmk:pcong}).
	Indeed, the such choice was compulsory in \cite{CicconePadovani22}.
	As mentioned before, in such work we relied on the characterization of fair subtyping based on
	a generalized inference system (see \Cref{ssec:fsub_gis}) and the subsumption rule
	\refrule{tb-cast} always increased the \emph{rank} by one. This way, \refrule{sb-cast-new} interpreted
	in a congruence way would increase the rank of the process due to the introduced cast.
	Using the actual notions, a reflexive application of fair subtyping has weight zero.
	\eor
\end{remark}

Well-typed processes enjoy the expected properties, including typing
preservation under structural pre-congruence and reduction. Most importantly,
they fairly terminate:

\begin{theorem}{Soundness}
  \label{thm:ts_bin_sound}
  If $\wtp[n]\EmptyCtx P$ and $P \wred Q$, then $Q \wred\pcong \pdone$.
\end{theorem}

The proof of \Cref{thm:ts_bin_sound} follows \Cref{thm:fair_termination}. 
Moreover the proof that all the reducts of a process are \emph{weakly terminating}
(see \Cref{lem:weak_termination_bin})
is loosely based on the method of helpful directions
\citep{Francez86}, namely on the property that a (well-typed) process \emph{may}
reduce in such a way that its measure strictly decreases
(see \Cref{lem:helpful_direction_bin}). Recall that this
property is not true for every reduction.

There are several valuable implications of \Cref{thm:ts_multi_sound} on a well-typed,
closed process $P$:
\begin{description}
  \item[Deadlock freedom.] If $Q$ cannot reduce any further, then it must be
  $\pdone$ (structurally precongruent to), namely there are no residual
  input or output actions.
  \item[Fair termination.] Under the fairness assumption,
  \Cref{thm:fair_termination} assures that $P$ eventually reduces to $\pdone$.
  This also implies that every session created by $P$ eventually terminates.
  \item[Junk freedom.] Each message produced as $P$ executes is eventually
  consumed. Indeed, if $Q$ contains a pending message, the fact that $Q$ may
  reduce to $\pdone$ means that some process is able to consume the message and
  will eventually do so under the fairness assumption.
  \item[Progress.] If $Q$ contains a sub-process with pending input/output
  actions, the fact that $Q$ may reduce to $\pdone$ means that these actions are
  eventually performed.
\end{description}

\begin{remark}
  \label{rem:internal_choice_rank}
  The rank of a non-deterministic choice $P \pchoice Q$ can usually be chosen to
  be the minimum among those of the branches $P$ and $Q$, so that the type
  system can handle processes like those in \cref{ex:infinite_sessions}, which
  \emph{may} create new sessions or perform casts but they need not do so in
  order to terminate.
  On the contrary, the rank of a label output $\PSend\x{\l_i:P_i}_{i\in I}$ has
  to be an upper bound of that of all branches $P_i$.
  The motivation for such different ways of determining the rank of these
  process forms, despite both represent an \emph{internal choice}, lies in the
  proof of \Cref{lem:helpful_direction_bin}.
  In $P \pchoice Q$, both branches are typed in \emph{exactly the same} typing
  context, meaning that the choice of one branch or the other has no substantial
  impact on the shortest paths that terminate the sessions used by $P$ and $Q$.
  Thus, the ``helpful'' reduction can be solely driven by the rank of the chosen
  branch.
  In a label output $\PSend\x{\l_i:P_i}_{i\in I}$ it could happen that all
  branches with minimum rank increase the length of the shortest path that leads
  to the termination of $x$. In this case, the choice of the ``helpful''
  reduction must prioritize the termination of $x$, but then the rank of the
  whole process has to be an upper bound of that of the branches to be sure that
  the measure of the reduct decreases.
  \eor
\end{remark}

%% file: ts-bin/ts-examples.tex
\beginbass
In this section we present some examples of well-typed processes using the 
type system introduced in the previous section.
In particular, we first complete the variant (see \Cref{ex:bsc_bin_proc}) of the
running example (see \Cref{ex:bsc}). 
Then, we show some examples where some of them are 
related to the boundedness properties that we 
presented in \Cref{ssec:boundedness}.

\begin{example}
    \label{ex:bsc_ts}
    To show that the program defined in \Cref{ex:bsc_bin_proc} is well typed,
    consider the session types $S_b'$ and $S_s$, $\T_s$, $\T_c$ (see \Cref{ex:bin_bsc}).
		\[
		\begin{array}{ll}
			S_b' = \In\ladd.S_b' \branch \In\lpay.\End[\In]
			& \qquad
			S_s = \Out\ladd.(\Out\ladd.S_s \choice \Out\lpay.\End[\Out])
			\\
			T_s = \Out\lship.\End[\Out]
			& \qquad
			T_c = \In\lship.\End[\In]
		\end{array}
		\]    
    and the global type assignments
    \[
    \begin{array}{ll}
    	\tass{\Buyer}{S_b'}{0}
    	& \qquad
    	\tass{\Seller}{S_s, T_s}{0}
    	\\
    	\tass{\Carrier}{T_c}{0}
    	& \qquad
    	\tass\Main{()}{3}
    \end{array}
    \]
    We can obtain typing derivations for $\Buyer$, $\Seller$ and $\Carrier$ using a null rank.
    In particular, we derive
    \[
        \begin{prooftree}
            \[
                \[
                    \smash\vdots\mathstrut
                    \justifies
                    \wtp[0]{
                        x : \S_b'
                    }{
                        \Call{\Buyer}{x}
                    }
                    \using\refrule{tb-call}
                \]
                \[
                    \justifies
                    \wtp[0]{
                        x : \End[\Out]
                    }{
                        \Close\x
                    }
                    \using\refrule{tb-close}
                \]
                \justifies
                \wtp[0]{
                    x : \Out\ladd.\S_b' \choice \Out\lpay.\End[\Out]
                }{
                    \PSend\x{
                        \ladd : \Call \Buyer x,
                        \lpay : \Close\x
                    }
                }
                \using\refrule{tb-label}
            \]
            \justifies
            \wtp[0]{
                x : \S_b'
            }{
                \POutput\x\ladd.
                \PSend\x{
                  \ladd : \Call \Buyer x,
                  \lpay : \Close\x
                }    
            }
            \using\refrule{tb-label}
        \end{prooftree}
    \]
    for the definition of $\Buyer$ and
    \[
        \begin{prooftree}
            \[
                \smash\vdots
                \justifies
                \wtp[0]{
                    x : \S_s,
                    y : \T_s
                }{
                    \Call \Seller {x,y}
                }
            \]
            \[
                \[
                    \[
                        \justifies
                        \wtp[0]{
                            y : \End[\Out]
                        }{
                            \Close\y
                        }
                        \using\refrule{tb-close}
                    \]
                    \justifies
                    \wtp[0]{
                        y : \T_s
                    }{
                        \POutput\y\lship.\Close\y
                    }
                    \using\refrule{tb-label}
                \]
                \justifies
                \wtp[0]{
                    x : \End[\Out],
                    y : \T_s
                }{
                    \Wait\x{\dots}
                }
                \using\refrule{tb-wait}
            \]
            \justifies
            \wtp[0]{
                x : \S_s,
                y : \T_s
            }{
                \PRecv\x{
                    \ladd : \Call \Seller {x,y},
                    \lpay : \Wait\x{\POutput\y\lship.\Close\y}
                }  
            }
            \using\refrule{tb-label}
        \end{prooftree}
    \]
    for the definition of $\Seller$. Note that the left branch starts with an application of \refrule{tb-call}.
    An analogous (but finite) derivation can be easily obtained for the body of process $\Carrier$ and is omitted here for space limitations.
    Now we have
    \[
        \begin{prooftree}
            \[
                \[
                    \[
                        \smash\vdots\mathstrut
                        \justifies
                        \wtp[0]{
                            x : \S_b'
                        }{
                            \Call \Buyer x
                        }
                        \using\refrule{tb-call}
                    \]
                    \justifies
                    \wtp[1]{
                        x : \S_b
                    }{
                        \Cast\x \Call \Buyer x
                    }
                    \using\refrule{tb-cast}
                \]
                \vdots
                \justifies
                \wtp[2]{
                    y : \T_s
                }{
                    \NewPar\x{\Cast\x \Call \Buyer x}{\Call \Seller {x,y}}
                }
                \using\refrule{tb-par}
            \]
            \[
                \vdots
                \justifies
                \wtp[0]{
                    y : \T_c
                }{
                    \Call \Carrier y
                }
            \]
            \justifies
            \wtp[3]\emptyset{
                \NewPar\y{\NewPar\x{\Cast\x \Call \Buyer x}{\Call \Seller {x,y}}}{\Call \Carrier y}
            }
        \end{prooftree}
    \]
    showing that $\Main$ too is well typed. Note that $\S_b$ from \Cref{ex:bin_bsc} is a fair subtype of
    $\S_b'$ with weight $1$; indeed \refrule{tb-cast} increases the rank by such amount.
    In all cases, we have truncated the proof trees above the applications of \refrule{tb-call}.
    Of course, for each judgment occurring in these proof trees, 
    we also have to exhibit a finite proof tree possibly using \refrule{cob-label} proving action boundedness. 
    This can be easily achieved for the given process definitions, observing that none of $\Buyer$, $\Seller$ and $\Carrier$ 
    creates new sessions and that all of their typing derivations have a finite branch.
    \eoe
\end{example}

\begin{example}[Infinite sessions/casts]
  In this example we demonstrate that well-typed processes may still create an unbounded number of (nested) sessions. 
  To this aim, let us consider again the process $C$ defined in \Cref{ex:infinite_sessions}. 
  Notice that $C$ is a choice whose left branch creates a new session and whose right branch does not. 
  For this reason, we elect the right choice as the one that leads to termination, 
  and therefore that determines the rank of the process.
  We derive
  \[
    \begin{prooftree}
      \[
        \[
          \mathstrut\smash\vdots
          \justifies
          \wtp[0]{
            y : \End[\Out]
          }{
            \Call{C}{y}
          }
        \]
        \[
          \[
            \justifies
            \wtp[0]{
              x : \End[\Out]
            }{
              \Close\x
            }
          \]
          \justifies
          \wtp[0]{
            x : \End[\Out], y : \End[\In]
          }{
            \Wait\y\dots
          }
        \]
        \justifies
        \wtp[1]{
          x : \End[\Out]
        }{
          \NewPar\y{\Call{C}{y}}{\Wait\y\dots}
        }
      \]
      \[
        \justifies
        \wtp[0]{
          x : \End[\Out]
        }{
          \Close\x
        }
      \]
      \justifies
      \wtp[0]{
        x : \End[\Out]
      }{
        \NewPar\y{\Call{C}{y}}{\Wait\y\dots}
        \pchoice_2
        \Close\x
      }
    \end{prooftree}
  \]

  In a similar way, there exist well-typed processes that perform an unbounded number 
  of casts but whose rank is finite. For example, it is easy to obtain a typing derivation 
  for the following alternative definition of the process $\Buyer$ discussed in \Cref{ex:bsc_bin_proc}:
  \[
    \Definition{\Buyer}{x}{\POutput\x\ladd.(\Cast\x\POutput\x\ladd.\Call{\Buyer}{x} \pchoice_2 \Cast\x\POutput\x\lpay.\Close\x)}
  \]
  
  Even though this process uses fair subtyping an unbounded number of times, the right branch of the choice has rank $1$, which is all we need to conclude that the process has rank $1$ overall.
  \eoe
\end{example}



%% file: ts-bin/ts-higherorder.tex
\beginbass
We have defined fair subtyping in such a way that higher-order session types are \emph{invariant} 
with respect to the type of the channel being exchanged (see \refrule{fs-channel}). 
This is a limitation compared to traditional presentations of unfair subtyping 
\citep{GayHole05,CastagnaDezaniGiachinoPadovani09,BernardiHennessy16}, 
where the covariant/contravariant rules shown below are adopted (\refrule{us-channel-in}, \refrule{us-channel-out}):
\[
        \inferrule{
          \S \usubt \T \\ \U \usubt \V
        }{
          \In\U.\S \usubt \In\V.\T
        }
        \qquad\qquad
        \inferrule{
          \S \usubt \T \\ \V \usubt \U
        }{
          \Out\U.\S \usubt \Out\V.\T
        }
\]

The problem of these rules is that a single application of fair subtyping allowing for 
co-/contra-variance of higher-order session types may have the same overall effect of 
infinitely many applications of fair subtyping on first-order session types and, 
as we have seen in \Cref{ssec:boundedness}, unbounded applications of fair subtyping may compromise fair termination.
Below is an example that illustrates the problem. The example is not large \emph{per se}, 
but it is a bit contrived because it has to involve two sessions 
(or else there would be no need for higher-order session types), 
it must be bounded (or else it could be ruled out by the 
action/session/cast boundedness requirements) and non-terminating.

\begin{example}
    \label{ex:invariant_ch}
    \[
    	\begin{array}{@{}r@{~}l@{}}
    	& \NewPar\y{\NewPar\x{\Call A {x,y}}{\Call B x}}{\Call B y}
    	\\
    	\Definition{A}{x,y}{& \POutput\x\lmore.\POutput\x\y.\Call B x}
    	\\
    	\Definition{B}{x}{&
        \PRecv\x{
        	\lmore : \PInput\x{(y)}.\Call A {y,x},
        	\lstop : \Wait\x\Done
        }
    	}	
    	\end{array}
    \]
    The process models a \emph{master} $\Call A {x,y}$ connected with a
	\emph{primary slave} $\Call B x$ and a \emph{secondary slave} $\Call B y$
	through the sessions $x$ and $y$. The interaction among the three processes
	proceeds in rounds. At each round, the master may decide whether to continue or
	stop the interaction by sending either $\lmore$ or $\lstop$ on the session $x$
	to the primary slave. If the master decides to continue the interaction (which
	it does deterministically), it also delegates $y$ to the primary slave so that,
	at the next round, the roles of the three processes rotate: the master is
	downgraded to secondary slave, the primary slave is promoted to master, and the
	secondary slave becomes the primary one.
	\eoe
\end{example}

Below is a graphical representation of
the network topology modeled by the process in \Cref{ex:invariant_ch} and of its evolution:
\[
    \tikzset{baseline=1.25em}
    \tikz[thick]{
    \node (M) at ( 0   ,1) {$\Call A {x,y}$};
    \node (P) at (-0.75,0) {$\Call B x$};
    \node (S) at ( 0.75,0) {$\Call B y$};
    \draw (P) -- (M) -- (S);
    }
    \wred
    \tikz[thick]{
    \node (S) at ( 0   ,1) {$\Call B x$};
    \node (M) at (-0.75,0) {$\Call A {y,x}$};
    \node (P) at ( 0.75,0) {$\Call B y$};
    \draw (P) -- (M) -- (S);
    }
    \wred
    \tikz[thick]{
    \node (P) at ( 0   ,1) {$\Call B x$};
    \node (S) at (-0.75,0) {$\Call B y$};
    \node (M) at ( 0.75,0) {$\Call A {x,y}$};
    \draw (P) -- (M) -- (S);
    }
    \wred
    \tikz[thick]{
    \node (M) at ( 0   ,1) {$\Call A {y,x}$};
    \node (P) at (-0.75,0) {$\Call B y$};
    \node (S) at ( 0.75,0) {$\Call B x$};
    \draw (P) -- (M) -- (S);
    }
\] 

It is clear that the process in \Cref{ex:invariant_ch} does not terminate since 
there is no $\Close\x$ to match the $\Wait\x\Done$.
It is also relatively easy to infer the types of $x$ and $y$ from the structure of 
$A(x,y)$ and $B(x)$. In particular, if we call $S_A$ and $T_A$ the types of $x$ and 
$y$ in $A(x,y)$ and $S_B$ the type of $x$ in $B(x)$ we see that these types must satisfy the equations
\[
\begin{array}{rcl}
	 	S_A & = & \Out\lmore.\Out\T_A.S_B
    \\
    S_B & = & \In\lmore.\In\S_A.T_A \branch \In\lstop.\End[\In]
    \\
    T_A & = & \Out\lmore.\Out\T_A.S_B \choice \Out\lstop.\End[\Out]
\end{array}
\]

Note that $T_A \subt[1] S_A$ holds because $T_A$ and $S_A$ differ only for the topmost output. 
The validity of this relation is unquestionable as it relies on the definition of fair subtyping 
that we have given in \Cref{sec:fair_sub}, which is invariant with respect to higher-order session types.
\begin{remark}
	In the following we assume to have two rules for fair subtyping allowing for 
	co-/contra-variance of channel input and output, respectively.
	\[
        \inferrule[ch-in]{
          \S \subt[m] \T \\ \U \subt[n] \V
        }{
          \In\U.\S \subt[k] \In\V.\T
        }\defrule[ch-in]{}
        \qquad\qquad
        \inferrule[ch-out]{
          \S \subt[m] \T \\ \V \subt[n] \U
        }{
          \Out\U.\S \subt[k] \Out\V.\T
        }\defrule[ch-out]{}
	\]
	where $k \in F(m,n)$ concerning the input and $k \in G(m,n)$ concerning the output.
	Note that $F,G : \Nat \times \Nat \rightarrow \mathcal{P_*}(\Nat)$.
	However, \refrule{tb-channel-in} and \refrule{tb-channel-out} are still invariant.
	\eor
\end{remark}
If fair subtyping allowed for covariance of higher-order inputs (see \refrule{us-channel-in}, \refrule{us-channel-out}),
then $\co{T_A} \subt S_B$, $\co{S_B} \subt T_A$ (along with $\co{S_B} \subt S_A$ by transitivity of $\subt$)  
would also hold and we would be able to establish that the process 
in \Cref{ex:invariant_ch} is well typed, provided that casts are placed appropriately. 

\begin{example}
	We show the derivation for the judgment $\co{T_A} \subt[k] S_B$ for some $k$.
	\[
	\begin{prooftree}
		\[
			\[
				\vdots
				\justifies
				T_A \subt[1] S_A
			\]
			\[
				\vdots
				\justifies
				\co{T_A} \subt[k] S_B
			\]
			\justifies
			\Out{S_A}.\co{T_A} \subt[m] \Out{T_A}.S_b
			\using\refrule{ch-out}
		\]
		\[
			\vdots
			\justifies
			\Out\lstop.\End[\Out] \subt[0] \Out\lstop.\End[\Out] 
		\]
		\justifies
		\co{S_B} \subt[1] T_A
		\using\refrule{fsb-tag-out-2}
	\end{prooftree}
	\]
	\[
	\begin{prooftree}
		\[
			\[
				\vdots
				\justifies
				T_A \subt[1] S_A
			\]
			\[
				\vdots
				\justifies
				\co{S_B} \subt[1] T_A
			\]
			\justifies
			\In{T_A}.\co{S_B} \subt[k] \In{S_A}.T_A
			\using\refrule{ch-in}
		\]
		\[
			\vdots
			\justifies
			\In\lstop.\End[\In] \subt[0] \In\lstop.\End[\In]
		\]
		\justifies
		\co{T_A} \subt[k] S_B
		\using\refrule{fsb-tag-in}
	\end{prooftree}
	\]
	Note that $m \in G(1,k)$ and $k \in F(1,1)$. Hence, a solution is guaranteed to exist.
	For this reason, in the following we omit the weight of the subtyping judgments involving
	\refrule{ch-in} and \refrule{ch-out}. It suffices to know that such judgments are derivable.
	\eoe
\end{example}

\begin{example}
    \label{ex:annotated_delegation}
    We show a version of the process in \Cref{ex:invariant_ch} in which 
    we have annotated restrictions with the types $\Sys\S\T$ 
		of the two endpoints and casts with the target type of the channel affected by subtyping. 
    \[
    \NewPar{y : \hl{\Sys{T_A}{\co{T_A}}}}{
        \NewPar{x : \hl{\Sys{T_A}{\co{T_A}}}}{
            \Cast{x : \hl{S_A}}
            \Call A {x,y}
        }{
            \Cast{x : \hl{S_B}}
            \Call B x
        }
    }{
        \Cast{y : \hl{S_B}}
        \Call B y
    }
    \]
    We provide another formulation of such process which is well-typed as well.
    \[
    \NewPar{y : \hl{\Sys{\co{S_B}}{S_B}}}{
        \NewPar{x : \hl{\Sys{\co{S_B}}{S_B}}}{
            \Cast{x : \hl{S_A}}
            \Cast{y : \hl{T_A}}
            \Call A {x,y}
        }{
            \Call B x
        }
    }{
        \Call B y
    }
		\]
    \eoe
\end{example}

\begin{example}
	We show that the process in \Cref{ex:annotated_delegation} is well typed.
	Below is the partial proof tree showing that $\Call{A}{x,y}$ is well typed. 
	Each judgment is implicitly annotated with the rank $0$:
	\[
    \begin{prooftree}
    \[
        \[
            \[
                \smash\vdots\mathstrut
                \justifies
                \wtp{x : S_B}{\Call{B}{x}}
                \using\refrule{tb-call}
            \]
            \justifies
            \wtp{x : \Out\T_A.S_B, y : T_A}{
                \POutput\x\y.\Call{B}{x}
            }
            \using\refrule{tb-channel-out}
        \]
        \justifies
        \wtp{x : S_A, y : T_A}{
            \POutput\x\lmore.\POutput\x\y.\Call{B}{x}
        }
        \using\refrule{tb-tag}
    \]
    \justifies
    \wtp{x : S_A, y : T_A}{\Call{A}{x,y}}
    \using\refrule{tb-call}
    \end{prooftree}
	\]
	Below is the partial proof tree showing that $\Call{B}{x}$ is well typed. Each judgment is implicitly annotated with the rank $0$:
	\[
    \begin{prooftree}
    \[
        \[
            \[
                \smash\vdots\mathstrut
                \justifies
                \wtp{x : T_A, y : S_A}{
                \Call{A}{y,x}
                }
                \using\refrule{tb-call}
            \]
            \justifies
            \wtp{x : \In\S_A.T_A}{
                \PInput\x{(y)}.\Call{A}{y,x}
            }
            \using\refrule{tb-channel-in}
        \]
        \[
            \[
                \justifies
                \wtp\emptyset\Done
            \]
            \justifies
            \wtp{x : \End[\In]}{
                \Wait\x\Done
            }
        \]
        \justifies
        \wtp{x : S_B}{
        \PRecv\x{
            \lmore : \PInput\x{(y)}.\Call{A}{y,x},
            \lstop : \Wait\x\Done
        }
        }
    \]
    \justifies
    \wtp{x : S_B}{
        \Call{B}{x}
    }
    \end{prooftree}
	\]
	Finally, here is the partial proof tree showing that the process shown in \Cref{ex:annotated_delegation} is well typed
	(again, we omit rank information since \refrule{ch-in} and \refrule{ch-out} are used):
	\[
    \begin{prooftree}
    \[
        \[
            \[
                \smash\vdots\mathstrut
                \justifies
                \wtp{x : T_A, y : S_A}{\Call{A}{x,y}}
            \]
            \justifies
            \wtp{
                x : S_A,
                y : S_A
            }{
                \Cast\x \Call{A}{x,y}
            }
        \]
        \[
            \[
                \smash\vdots\mathstrut
                \justifies
                \wtp{x : T_B}{\Call{B}{x}}
            \]
            \justifies
            \wtp{
                x : \co{S_A}
            }{
                \Cast\x \Call{B}{x}
            }
        \]
        \justifies
        \wtp{y : S_A}{
            \NewPar\x{
                \Cast\x \Call{A}{x,y}
            }{
                \Cast\y \Call{B}{x}
            }
        }
    \]
    \[
        \[
            \smash\vdots\mathstrut
            \justifies
            \wtp{
                y : T_B
            }{
                \Call{B}{y}
            }
        \]
        \justifies
        \wtp{
            y : \co{S_A}
        }{
            \Cast\y \Call{B}{y}
        }
    \]
    \justifies
    \wtp\emptyset{
        \NewPar\y{
            \NewPar\x{
                \Cast\x \Call{A}{x,y}
            }{
                \Cast\x \Call{B}{x}
            }
        }{
            \Cast\y \Call{B}{y}
        }
    }
    \end{prooftree}
	\]
	\eoe
\end{example}

By restricting fair subtyping of higher-order session types to invariant inputs and outputs, 
the only chance we have to build a typing derivation for the process in \Cref{ex:invariant_ch} 
is by casting $y$ each time it is delegated, either before it is sent or after it is received.

\begin{example}
	Consider the following variant of \Cref{ex:invariant_ch}:
	\[
    \begin{array}{r@{~}l}
    \Definition{A}{x : \hl{U_A},y : \hl{V_A}}{&
        \POutput\x\lmore.
        \Cast{y : \hl{U_A}}
        \POutput\x\y.
        \Call B x
    } \\
    \Definition{B}{x : \hl{U_B}}{&
        \PRecv\x{
            \lmore : \PInput\x{(y : \hl{U_A})}.\Call A {y,x},
            \lstop : \Wait\x\Done
        }
    }
    \end{array}
	\]
	where
	\[
	\begin{array}{rcl}
		U_A & = & \Out\lmore.\Out\U_A.U_B \\
		U_B & = & \In\lmore.\In\U_A.V_A \branch \In\lstop.\End[\In] \\
		V_A & = & \Out\lmore.\Out\U_A.U_B \choice \Out\lstop.\End[\Out]
	\end{array}
	\]
	Note that $\Cast{y : U_A}$ is a ``first-order'' cast, in the sense that the relation $V_A \subt[1] U_A$ 
	holds for fair subtyping as defined in \Cref{sec:fair_sub} without using \refrule{us-channel-in} or \refrule{us-channel-out}, 
	but the cast is now placed in a region within the definition of $A$ that prevents finding a finite rank for $A$.
	\eoe
\end{example}

%% file: ts-bin/soundness.tex
\beginalto
In this section we detail the proof of \Cref{thm:ts_bin_sound}.
The most intriguing aspect in such proof is that a
closed, well-typed process admits a reduction sequence to $\Done$.
The proof technique is related to the \emph{method of helpful directions} \citep{Francez86}:
we define a well-founded \emph{measure} for (well-typed) processes and we prove
that this measure decreases strictly as the result of ``helpful'' reductions.
In our case, the measure of a (well-typed) process $P$ is a lexicographically
ordered pair $(m, n)$ of natural numbers such that $m$ is an upper bound to the
number of sessions that $P$ may need to create and of weights of casts that $P$ may need to
perform \emph{in the future} in order to terminate, whereas $n$ is the
cumulative efforts to terminate the sessions that $P$ has created \emph{in the past} and that
are not terminated yet. 
We account for this effort by measuring the shortest reduction that terminates a
compatible session (\Cref{def:compatibility}).
Notably, a session terminates by \refrule{rb-close}; a cast is
performed when it is absorbed by the corresponding restriction, namely by
\refrule{sb-cast-new}.

We first show two standard results, that is, typing is preserved by structural
precongruence and reduction.
Then, we formally introduce the measure and we characterize some \emph{normal forms} that we
need in order to achieve the proof that such measure reduces by following the right
reductions.


\subsection{Subject Reduction}
\input{ts-bin/proof-sr}


\subsection{Measure}
\input{ts-bin/proof-measure}


\subsection{Normal Forms}
\label{ssec:nf_bin}
\input{ts-bin/proof-nf}


\subsection{Soundness}
\input{ts-bin/proof-soundness}

%% file: ts-bin/proof-sr.tex
\beginbass
We first prove that typing is preserved by structural pre-congruence.
This lemma is required to prove subject reduction when dealing with
the reduction under structural pre-congruence (\refrule{rb-struct}).

\begin{lemma}
  \label{lem:subj_cong_bin}
  If\/ $\wtp[n]\Ctx{P}$ and $P \pcong Q$, then $\wtp[m]\Ctx{Q}$ for some $m \leq n$.
\end{lemma}
\begin{proof}
  The proof is by induction on the derivation of $P \pcong Q$ and by cases on
  the last rule applied. We only discuss a few representative cases, the
  remaining ones are analogous.

  \proofrule{sb-par-comm}
  Then $P = \NewPar\x{P_1}{P_2} \pcong \NewPar\x{P_2}{P_1} = Q$.
  From \refrule{tb-par} we deduce that there exist $\Ctx_1$,
  $\Ctx_2$, $x$, $S_1$, $S_2$, $n_1$ and $n_2$ such that:
  \begin{itemize}
  \item $\Ctx = \Ctx_1, \Ctx_2$
  \item $\wtp[n_i]{\Ctx_i, x : S_i}{P_i}$ for $i=1,2$
  \item $S_1 \compatible S_2$
  \item $n = 1 + n_1 + n_2$  
  \end{itemize}

  We conclude $\wtp\Ctx{Q}$ with one application of \refrule{tb-par} by taking $m \eqdef n$.

  \proofrule{sb-par-assoc}
  Then $P = \NewPar\x{P_1}{\NewPar\y{P_2}{P_3}} \pcong
  \NewPar\y{\NewPar\x{P_1}{P_2}}{P_3} = Q$ and $x \in \fn{P_2}$.
  From \refrule{tb-par} we deduce that there exist $\Ctx_1$, $\Ctx_{23}$,
  $T_1$, $S_1$, $n_1$ and $n_{23}$ such that:
  \begin{itemize}
  \item $\Ctx = \Ctx_1, \Ctx_{23}$
  \item $\wtp[n_1]{\Ctx_1, x : T_1}{P_1}$
  \item $\wtp[n_{23}]{\Ctx_{23}, x : S_1}{\NewPar\y{P_2}{P_3}}$
  \item $T_1 \compatible S_1$
  \item $n = 1 + n_1 + n_{23}$
  \end{itemize}

  From \refrule{tb-par} we deduce that there exist $\Ctx_2$, $\Ctx_3$,
  $T_2$, $S_2$, $n_2$ and $n_3$ such that:
  \begin{itemize}
  \item $\Ctx_{23} = \Ctx_2, \Ctx_3$
  \item $\wtp[n_2]{\Ctx_2, x : S_1, y : T_2}{P_2}$
  \item $\wtp[n_3]{\Ctx_3, y : S_2}{P_3}$
  \item $T_2 \compatible S_2$
  \item $n_{23} = 1 + n_2 + n_3$
  \end{itemize}

  Using \refrule{tb-par} we derive $\wtp[1 + n_1 + n_2]{\Ctx_1, \Ctx_2, y
  : T_2}{\NewPar\x{P_1}{P_2}}$.
  We conclude $\wtp[m]\Ctx{\NewPar\y{\NewPar\x{P_1}{P_2}}{P_3}}$ with another application of \refrule{tb-par} by taking $m \eqdef n$.

  \proofrule{sb-cast-new}
  Then $P = \NewPar\x{\Cast\x P_1}{P_2} \pcong \NewPar\x{P_1}{P_2}$.
  From \refrule{tb-par} we deduce that there exist $\Ctx_1$, $\Ctx_2$, $S_1$, $T$, $n_1$ and $n_2$ such that:
  \begin{itemize}
  \item $\Ctx = \Ctx_1, \Ctx_2$
  \item $\wtp[n_1]{\Ctx_1, x : S_1}{\Cast\x P_1}$
  \item $\wtp[n_2]{\Ctx_2, x : T}{P_2}$
  \item $S_1 \compatible T$
  \item $n = 1 + n_1 + n_2$
  \end{itemize}

  From \refrule{tb-cast} we deduce that there exist $S_2$, $n_3$, $m_x$ such that
  \begin{itemize}
  \item $\wtp[n_3]{\Ctx_1, x : S_2}{P_1}$
  \item $S_1 \subt[m_x] S_2$
  \item $n_1 = m_x + n_3$
  \end{itemize}

  From $S_1 \compatible T$ and $S_1 \subt S_2$ and \Cref{thm:fsub_sound} we
  deduce $S_2 \compatible T$. We conclude with one application of
  \refrule{tb-par} by taking $m \eqdef 1 + n_3 + n_2$ and observing that
  \[
  	m \eqdef 1 + n_3 + n_2 \le 1 + m_x + n_3 + n_2 = 1 + n_1 + n_2 = n
  \]

  \proofrule{sb-cast-swap}
  Then $P = \NewPar\x{\Cast\y P_1}{P_2} \pcong \Cast\y\NewPar\x{P_1}{P_2}$ and $\x \ne \y$.
  From \refrule{tb-par} we deduce that there exist $\Ctx_1$, $S_1$, $S_2$,
  $T_1$, $n_1$ and $n_2$ such that:
  \begin{itemize}
  \item $\Ctx = \Ctx_1, \Ctx_2, y : T_1$
  \item $\wtp[n_1]{\Ctx_1, x : S_1, y : T_1}{\Cast\y P_1}$
  \item $\wtp[n_2]{\Ctx_2, x : S_2}{P_2}$
  \item $S_1 \compatible S_2$
  \item $n = 1 + n_1 + n_2$
  \end{itemize}

  From \refrule{tb-cast} we deduce that there exist $T_2$, $n_3$ and $m_y$ such that
  \begin{itemize}
  \item $T_1 \subt[m_y] T_2$
  \item $\wtp[n_3]{\Ctx_1, x : S_1, y : T_2}{P_1}$
  \item $n_1 = m_y + n_3$
  \end{itemize}

  We derive $\wtp[1 + n_3 + n_2]{\Ctx_1, \Ctx_2, y :
  T_2}{\NewPar\x{P_1}{P_2}}$ with one application of \refrule{tb-par} and we
  conclude with one application of \refrule{tb-cast} by taking $m \eqdef n$.

  \proofrule{sb-cast-comm}
  Then $P = \Cast\x\Cast\y P' \pcong \Cast\y\Cast\x P' = Q$. We can assume $x \ne y$ or else $P = Q$ and there is nothing to prove.
  From \refrule{tb-cast} we deduce that there exist $\Ctx_1$, $S_1$, $S_2$, $n_1$ and $m_x$ such that
  \begin{itemize}
  \item $\Ctx = \Ctx_1, x : S_1$
  \item $\wtp[n_1]{\Ctx_1, x : S_2}{\Cast\y P'}$
  \item $S_1 \subt[m_x] S_2$
  \item $n = m_x + n_1$
  \end{itemize}
  
  From \refrule{tb-cast} and the hypothesis $x \neq y$ we deduce that there exist $\Ctx_2$, $T_1$, $T_2$, $n_2$ and $m_y$ such that
  \begin{itemize}
  \item $\Ctx_1 = \Ctx_2, y : T_1$
  \item $\wtp[n_2]{\Ctx_2, x : S_2, y : T_2}{P'}$
  \item $T_1 \subt[m_y] T_2$
  \item $n_1 = m_y + n_2$
  \end{itemize}

  We derive $\wtp[n_1]{\Ctx_2, x : S_1, y : T_2}{\Cast\x P'}$ with one application of 
  \refrule{tb-cast} and we conclude with another application of \refrule{tb-cast} by taking $m \eqdef n$.

  \proofrule{sb-call}
  Then $P = \Call{A}{\seqof\x} \pcong Q$ and $\Definition{A}{\seqof\x}Q$.
  From \refrule{tb-call} we conclude that there exist $\seqof\S$ and $m$ such
  that $\tass{A}{\seqof{S}}{m}$ and $\Ctx = \seqof{x : S}$ and
  $\wtp[m]{\seqof{x : S}}{Q}$ and $m \leq n$.
\end{proof}

Then we have subject reduction, stating that typing is preserved also by reductions. 
Note that in this case we are not able to establish a general relation between the rank 
of the reducible process and that of the reduct. In particular, the rank may increase.

\begin{lemma}[subject reduction]
  \label{lem:subj_red_bin}
  If\/ $\wtp[n]\Ctx{P}$ and $P \red Q$, then $\wtp[m]\Ctx{Q}$ for some $m$.
\end{lemma}
\begin{proof}
  By induction on the derivation of $P \red Q$ and by cases on the last rule
  applied.

  \proofrule{rb-choice}
  Then $P = P_1 \pchoice P_2 \red P_k = Q$ where $k\in\set{1,2}$.
  From \refrule{tb-choice} we deduce that $\wtp[m]\Ctx{Q}$ for some $m$, which is all we need to conclude.

  \proofrule{rb-pick}
  Then $P = \NewPar\x{\PSend\x{\l_i : P_i}_{i\in I}}{R} \red
  \NewPar\x{\POutput\x\l_k.P_k}{R} = Q$ where $k\in I$ and $|I| > 1$.
  From \refrule{tb-par} we deduce that there exist $\Ctx_1$, $\Ctx_2$, $S$, $T$, $n_1$ and $n_2$ such that
  \begin{itemize}
  \item $\Ctx = \Ctx_1, \Ctx_2$
  \item $\wtp[n_1]{\Ctx_1, x : S}{\PSend\x{\l_i : P_i}_{i\in I}}$
  \item $\wtp[n_2]{\Ctx_2, x : T}{R}$
  \item $S \compatible T$
  \item $n = 1 + n_1 + n_2$
  \end{itemize}

  From \refrule{tb-tag} we deduce that there exists a family $S_{i\in I}$ such
  that:
  \begin{itemize}
  \item $S = \Choice{\l_i : S_i}_{i\in I}$
  \item $\wtp[n_1]{\Ctx_1, x : S_i}{P_i}$ for every $i\in I$
  \end{itemize}

  From $S \compatible T$ and $S \red \Out\l_k.S_k$ and
  \Cref{def:compatibility} we deduce $\Out\l_k.S_k \compatible T$.
  We conclude with one application of \refrule{tb-tag} and one application of \refrule{tb-par} by taking $m \eqdef n$.
 	
  \proofrule{rb-signal}
  Then $P = \NewPar\x{\Close\x}{\Wait\x{Q}} \red Q$.
  From \refrule{tb-par}, \refrule{tb-close} and \refrule{tb-wait} we deduce that there exist $n'$ and $m$ such that:
  \begin{itemize}
  \item $\wtp[n']{x : \End[\Out]}{\Close\x}$
  \item $\wtp[m]{\Ctx, x : \End[\In]}{\Wait\x{Q}}$
  \item $\wtp[m]\Ctx{Q}$
  \item $n = 1 + n' + m$
  \end{itemize}

  There is nothing left to prove.

  \proofrule{rb-label}
  Then $P = \NewPar\x{\POutput\x\l_k.R}{\PRecv\x{\l_i : Q_i}_{i \in I}} \red
  \NewPar\x{R}{Q_k} = Q$ with $k \in I$.
  From \refrule{tb-par} we deduce that there exist $\Ctx_1$, $\Ctx_2$,
  $S$, $T$, $n_1$ and $n_2$ such that:
  \begin{itemize}
  \item $\Ctx = \Ctx_1, \Ctx_2$
  \item $\wtp[n_1]{\Ctx_1, x : S}{\POutput\x\l_k.R}$
  \item $\wtp[n_2]{\Ctx_2, x : T}{\PRecv\x{\l_i : Q_i}_{i \in I}}$
  \item $S \compatible T$
  \item $n = 1 + n_1 + n_2$
  \end{itemize}

  From \refrule{tb-tag} we deduce that there exists $S_1$ such that
  $S = \Out\l_k. S_1$ and $\wtp[n_1]{\Ctx_1, x : S_1}{R}$.
  From \refrule{tb-tag} we deduce that there exists a family $T_{i\in I}$ such
  that:
  \begin{itemize}
  \item $T = \Branch{\l_i : T_i}_{i \in I}$
  \item $\wtp[n_2]{\Ctx_2, x : T_i}{Q_i}$ for every $i\in I$
  \end{itemize}

  From $S \compatible T$ we deduce $S_1 \compatible T_k$.
  We conclude with one application of \refrule{tb-par} by taking $m \eqdef n$.

  \proofrule{rb-channel}
  Then $P = \NewPar\x{\POutput\x\y.P'}{\PInput\x{(\y)}.Q'} \red
  \NewPar\x{P'}{Q'} = Q$.
  From \refrule{tb-par} we deduce that there exist $\Ctx_1$, $\Ctx_2$, $S$, $T$, $n_1$ and $n_2$ such that
  \begin{itemize}
  \item $\Ctx = \Ctx_1, \Ctx_2$
  \item $\wtp[n_1]{\Ctx_1, x : S}{\POutput\x\y.P'}$
  \item $\wtp[n_2]{\Ctx_2, x : T}{\PInput\x{(\y)}.Q'}$
  \item $S \compatible T$
  \item $n = 1 + n_1 + n_2$
  \end{itemize}

  From \refrule{tb-channel-out} we deduce that there exist $\Ctx_1'$, $S_1$ and $S_2$ such that
  \begin{itemize}
  \item $\Ctx_1 = \Ctx_1', y : S_1$
  \item $S = \Out\S_1.S_2$
  \item $\wtp[n_1]{\Ctx_1', x : S_2}{P'}$
  \end{itemize}

  From \refrule{tb-channel-in} we deduce that there exist $T_1$ and $T_2$ such that
  \begin{itemize}
  \item $T = \In\T_1.T_2$
  \item $\wtp[n_2]{\Ctx_2, x : T_2, y : T_1}{Q'}$
  \end{itemize}

  From $S \compatible T$ we deduce $S_1 = T_1$ and $S_2 \compatible T_2$.
  We conclude with one application of \refrule{tb-par} by taking $m \eqdef n$.

  \proofrule{rb-cast}
  Then $P = \Cast\x P' \red \Cast\x Q' = Q$ and $P' \red Q'$.
  From \refrule{tb-cast} we deduce that there exist $\Ctx'$, $S$, $T$, $n'$ 
  and $m_x$ such that
  \begin{itemize}
  \item $\Ctx = \Ctx', x : S$
  \item $\wtp[n']{\Ctx', x : T}{P'}$
  \item $S \subt[m_x] T$
  \item $n = m_x + n'$
  \end{itemize}

  Using the induction hypothesis on $\wtp[n']{\Ctx', x : T}{P'}$ and $P'
  \red Q'$ we derive $\wtp[m']{\Ctx', x : T}{Q'}$ for some $m'$. We conclude
  with one application of \refrule{tb-cast} by taking $m \eqdef m_x + m'$.
  
  \proofrule{rb-par}
  Then $P = \NewPar\x{P'}{R} \red \NewPar\x{Q'}{R} = Q$ and $P' \red Q'$.
  From \refrule{tb-par} we deduce that there exist $\Ctx_1$, $\Ctx_2$,
  $S$, $T$, $n_1$ and $n_2$ such that:
  \begin{itemize}
  \item $\Ctx = \Ctx_1, \Ctx_2$
  \item $\wtp[n_1]{\Ctx_1, x : S}{P'}$
  \item $\wtp[n_2]{\Ctx_2, x : T}{R}$
  \item $S \compatible T$
  \item $n = 1 + n_1 + n_2$
  \end{itemize}

  Using the induction hypothesis we deduce that $\wtp[m_1]{\Ctx_1, x :
  S}{Q'}$ for some $m_1$. We conclude with one application of \refrule{tb-par} by
  taking $m \eqdef m_1 + n_2$.

  \proofrule{rb-struct}
  Then $P \pcong P' \red Q' \pcong Q$.
  From \Cref{lem:subj_cong_bin} we deduce $\wtp[n']\Ctx{P'}$ for some $n' \leq n$.
  Using the induction hypothesis we deduce $\wtp[m']\Ctx{Q'}$ for some $m'$.
  We conclude using \Cref{lem:subj_cong_bin} once more.
\end{proof}

%% file: ts-bin/proof-measure.tex
\beginbass
At the beginning on \Cref{sec:ts_bin_corr} we informally introduced the
\emph{measure} of a process as a lexicographically ordered pair $(m, n)$ 
where the two component refer to \emph{past} and \emph{future}, respectively.
To distinguish between past and future ofa process $P$ we look at its structure: all
sessions that occur unguarded in $P$ have been created and are not terminated;
all casts that occur in $P$ are yet to be performed; all sessions that occur
guarded in $P$ have not been created yet.
First, we need to formalize the notion of \emph{rank} of a session which represents
the minimum effort for reaching termination.

\begin{definition}[rank]
  \label{def:rank_bin}
  The \emph{rank} of $S$ and $T$, written $\rankb\S\T$, is the element of $\Nat
  \cup \set\infty$ defined as
  \[
    \rankb\S\T \eqdef \min\set{ 1 + |\actions| \mid \exists\actions, \Pol: S \wlred{\co\actions} \End[\co\Pol], T \wlred\actions \End[\Pol] }
  \]
  where $|\actions|$ denotes the length of $\actions$, $\co\actions$ is the
  string of actions obtained by dualizing all the actions in $\actions$, and we
  postulate that $\min\emptyset = \infty$.
\end{definition}

Note that the rank $\rankb\S\T$ is generally unrelated to the lengths of the
shortest paths of $S$ and $T$ that lead to termination. For example, if we take
$S = \In\la.\Out\lc.\In\la.\End[\In] \branch \In\lb.\End[\In]$ and $T =
\Out\la.(\In\lc.\Out\la.\End[\Out] \branch \In\ld.\End[\Out])$ we see that the
shortest path $\actionsA$ such that $S(\actionsA) = \End[\In]$ is $\In\lb$ of
length 1 and the shortest path $\actionsB$ such that $T(\actionsB) = \End[\Out]$
is $\Out\la\In\ld$ of length 2, but $\rankb\S\T = 4$. Indeed, both $S$ and $T$ must
reach termination; so we have to consider the exchange of labels $\la,\lc,\la$ and 
increment by one according to \Cref{def:rank_bin}.

\begin{theorem}
    \label{thm:rank_bin}
    If $U \compatible S$, then
    \begin{enumerate}
    	\item\label{item:rank_finite} $\rankb\U\S \in \Nat$ and
    	\item\label{item:rank_subt} $S \csubt T$ implies $\rankb\U\S \leq \rankb\U\T$.
    \end{enumerate}
\end{theorem}
\begin{proof}
    \Cref{item:rank_finite} follows from the observation that $U \compatible S$
    implies $\Sys\U\S \wred \Sys{\End[\co\Pol]}{\End[\Pol]}$ for some $\Pol$, hence there exists $\actions$
    such that $U \wlred{\co\actions} \End[\co\Pol]$ and $S \wlred\actions
   	\End[\Pol]$.
    \Cref{item:rank_subt} is trivial to prove if we establish that
    \[
        \set{ \actions \mid U \wlred{\co\actions}, T \wlred\actions }
        \subseteq
        \set{ \actions \mid U \wlred{\co\actions}, S \wlred\actions }
    \]
    under the hypotheses $U \compatible S$ and $S \csubt T$.
    We prove that $U \wlred{\co\actions}$ and $T \wlred\actions$ implies $S
    \wlred\actions$ by induction on $\actions$ and by cases on its first action.
    
    The base case $\actions = \es$ is trivial.
    If $\actions = \Pol\V\actionsB$, then $T = \Pol\V.T'$ for some $T'$ and by definition of 
    $\csubt$ we deduce $S = \Pol\V.S'$ and $S' \csubt T'$ for some $S'$. 
    From $U \compatible S$ we deduce $U = \co\\Pol\V.U'$ for some $U' \compatible S'$.
    Using the induction hypothesis we obtain $S' \wlred\actionsB$, which is enough to conclude $S \wlred\actions$.
    
    If $\actions = \Out\l\actionsB$, then $T = \Choice{l_j:T_j}_{j\in J}$ and $l
    = l_k$ for some $k\in J$. By definition of $\csubt$ we deduce $S = \Choice{l_i:S_i}_{i\in I}$ for some $I \supseteq J$, 
    hence $S \wlred{\Out\l}$.
    From $U \compatible S$ we deduce $U = \Branch{l_k:U_k}_{k\in K}$ for some $K \supseteq I$ and $U_k \compatible S_k$.
    Using the induction hypothesis we obtain $S_k \wlred\actionsB$, from which we conclude $S \wlred\actions$.

    If $\actions = \In\l\actionsB$, then $T = \Branch{l_j:T_j}_{j\in J}$ and $l = l_k$ for some $k\in J$. 
    By definition of $\csubt$ we deduce $S = \Branch{l_i:S_i}_{i\in I}$ for some $I \subseteq J$.
    From $U \compatible S$ we deduce $U = \Choice{l_k:U_k}_{k\in K}$ for some $K \subseteq I$. 
    Also, it must be the case that $k\in K$ and $U_k \compatible S_k$.
    Using the induction hypothesis we obtain $S_k \wlred\actionsB$, from which we conclude $S \wlred\actions$.
\end{proof}

\Cref{thm:rank_bin} shows that every usage of (fair) subtyping may increase the
amount of work that is necessary to terminate a session (\Cref{item:rank_subt}),
although such amount is guaranteed to remain finite as long as compatibility is
preserved (\Cref{item:rank_finite}). This property justifies the adoption of
fair subtyping over unfair subtyping, since fair subtyping preserves
compatibility whereas unfair subtyping in general does not (\Cref{ssec:unfair_sub}).

To compute the measure of a process, we introduce three refined typing rules to
derive judgments of the form $\wtpn\Measure\Ctx{P}$, stating that $P$ is
well typed in $\Ctx$ and has measure $\Measure$.
Such rules are shown in \Cref{fig:measure_bin}.

\begin{figure}[t]
	\framebox[\linewidth]{
	\begin{mathpar}
	\inferrule[mtb-thread]{
        \wtp[n]\Ctx{P}
    }{
        \wtpn{(n, 0)}\Ctx{P}
    } \defrule[mtb-thread]{}
    \and
    \inferrule[mtb-cast]{
        \wtpn\Measure{\Ctx, x : T}{P}
    }{
        \wtpn{\Measure + (m,0)}{\Ctx, x : S}{\Cast\x P}
    }
    ~
    S \subt[m] T \defrule[mtb-cast]{}
    \and
    \inferrule[mtb-par]{
        \wtpn{\MeasureM}{\Ctx, x : S}{P}
        \\
        \wtpn{\MeasureN}{\CtxD, x : T}{P}
    }{
        \wtpn{\MeasureM + \MeasureN + (0, \rankb{S}{T})}{
            \Ctx, \CtxD
        }{
            \NewPar\x{P}{Q}
        }
    }
    ~
    S \compatible T \defrule[mtb-par]{}
	\end{mathpar}
	}
	\caption{Typing rules with measure}
	\label{fig:measure_bin}
\end{figure}

Rule \refrule{mtb-thread} has \emph{lower priority} than \refrule{mtb-par} and
\refrule{mtb-cast}, in the sense that it applies only to processes that are not a
cast or a parallel composition. We call such processes \emph{threads} and their
measure is solely determined by their rank: every cast occurring in a thread is
yet to be performed and every session occurring in a thread is yet to be created.
Rule \refrule{mtb-par} states that the measure of a parallel composition is the
(pointwise) sum of the measures of the composed processes, taking into account
the rank of the session $x$ by which they are connected.
Finally, \refrule{mtb-cast} states that the measure of a cast is the same measure
of the process in which the cast has effect, but with the first component
increased by one to account for the fact that the cast is yet to be performed.

Note that, as a well-typed process reduces, its measure may vary arbitrarily. In
particular, its measure \emph{may increase} if the process chooses to create new
sessions (see the left choice of process $C$ in \Cref{ex:infinite_sessions}) or
if it picks a label that lengthens the shortest path leading to session
termination (see \Cref{ex:bsc_bin_reduction}).

\begin{lemma}
    \label{lem:measure_rank_bin}
    The following properties hold:
    \begin{enumerate}
        \item $\wtp[n]\Ctx{P}$ implies $\wtpn\Measure\Ctx{P}$ for some
        $\Measure \leq (n, 0)$;
        \item $\wtpn\Measure\Ctx{P}$ implies $\wtp[n]\Ctx{P}$ for some $n$ such that $\Measure \leq (n, 0)$.
    \end{enumerate}
\end{lemma}
\begin{proof}
    We prove item 1 by induction on the structure of $P$ and
    by cases on the last rule applied. The proof of item 2 is by induction over
    $\wtpn\Measure\Ctx{P}$.

    \proofcase{Case $P = \NewPar\x{P_1}{P_2}$}
    Then from \refrule{tb-par} we deduce that there exist $\Ctx_1$, $\Ctx_2$,
    $S_1$, $S_2$, $n_1$, $n_2$ such that:
    \begin{itemize}
    \item $\Ctx = \Ctx_1, \Ctx_2$
    \item $\wtp[n_i]{\Ctx_i, x : S_i}{P_i}$ for $i=1,2$
    \item $S_1 \compatible S_2$
    \item $n = 1 + n_1 + n_2$
    \end{itemize}
    Using the induction hypothesis we deduce that there exist $\Measure_1$ and
    $\Measure_2$ such that $\wtpn{\Measure_i}{\Ctx_i, x : S_i}{P_i}$ and
    $\Measure_i \leq (n_i, 0)$ for $i=1,2$.
    We conclude with one application of \refrule{mtb-par} by taking $\Measure
    \eqdef \Measure_1 + \Measure_2 + (0, \rankb{S_1}{S_2})$ and observing that
    $\Measure < (n_1, 0) + (n_2, 0) + (1, 0) = (n, 0)$.

    \proofcase{Case $P = \Cast\x Q$}
    Then from \refrule{tb-cast} we deduce that there exist 
    $\CtxD$, $S$, $T$, $n$, $m$, $m_x$ such that:
    \begin{itemize}
    \item $\Ctx = \CtxD, x : S$
    \item $\wtp[m]{\CtxD, x : T}{Q}$
    \item $S \subt[m_x] T$
    \item $n = m_x + m$
    \end{itemize}
    Using the induction hypothesis we deduce $\wtpn\MeasureN{\CtxD, x :
    T}{Q}$ for some $\MeasureN \leq (m, 0)$.
    We conclude with one application of \refrule{mtb-cast} by taking $\Measure
    \eqdef \MeasureN + (m_x, 0)$ and observing that $\Measure \leq (m, 0) + (m_x, 0)
    = (n, 0)$.

    \proofcase{In all the other cases}
    We conclude with one application of \refrule{mtb-thread} by taking $\Measure
    \eqdef (n, 0)$.
\end{proof}

The next lemma states that structural pre-congruence does not increase the measure
of a process.

\begin{lemma}
\label{lem:m_pcong_bin}
If\/ $\wtpn\MeasureM\Ctx P$ and $P \pcong Q$, then there exists $\MeasureN
\le \MeasureM$ such that $\wtpn\MeasureN\Ctx Q$.
\end{lemma}
\begin{proof}
By induction on the derivation of $P \pcong Q$ and by cases on the last rule
applied. We only consider the base cases.

\proofrule{sb-par-comm}
Then $P = \NewPar\x{P_1}{P_2} \pcong \NewPar\x{P_2}{P_1} = Q$.
From \refrule{mtb-par} we deduce that there exist $\Ctx_1$, $\Ctx_2$,
$S$, $T$, $\MeasureM_1$ and $\MeasureM_2$ such that
\begin{itemize}
\item $\Ctx = \Ctx_1, \Ctx_2$
\item $\MeasureM = \MeasureM_1 + \MeasureM_2 + (0, \rankb{S}{T})$
\item $\wtpn{\MeasureM_1}{\Ctx_1, x : S}{P_1}$
\item $\wtpn{\MeasureM_2}{\Ctx_2, x : T}{P_2}$
\end{itemize}
We conclude with one application of \refrule{mtb-par} by taking $\MeasureN \eqdef
\MeasureM$.

\proofrule{sb-par-assoc}
Then $P = \NewPar\x{P_1}{\NewPar\y{Q_1}{Q_2}} \pcong
\NewPar\y{\NewPar\x{P_1}{Q_1}}{Q_2} = Q$.
From \refrule{mtb-par} we deduce that there exist $\Ctx_1$, $\CtxD_1$,
$\CtxD_2$, $\MeasureM_1$, $\MeasureN_1$, $\MeasureN_2$, $S_1$, $S_2$, $T_1$
and $T_2$ such that
\begin{itemize}
\item $\Ctx = \Ctx_1, \CtxD_1, \CtxD_2$
\item $\MeasureM = \MeasureM_1 + (\MeasureN_1 + \MeasureN_2 + (0,
\rankb{T_1}{T_2})) + (0, \rankb{S_1}{S_2})$
\item $\wtpn{\MeasureM_1}{\Ctx_1, x : S_1}{P_1}$
\item $\wtpn{\MeasureN_1}{\CtxD_1, y : T_1, x : S_2}{Q_1}$
\item $\wtpn{\MeasureN_2}{\CtxD_2, y : T_2}{Q_2}$
\end{itemize}
We conclude with two applications of \refrule{mtb-par} by taking $\MeasureN
\eqdef \MeasureM$.

\proofrule{sb-cast-comm}
Then $P = \Cast\x \Cast\y P' \pcong \Cast\x \Cast\y P' = Q$.
We only consider the case $x \ne y$ or else there is nothing interesting to
prove. From \refrule{mtb-cast} we deduce that there exist $\Ctx'$,
$\MeasureM'$, $S_1$, $T_1$, $S_2$, $T_2$, $m_x$ and $m_y$ such that
\begin{itemize}
\item $S_1 \subt[m_x] S_2$
\item $T_1 \subt[m_y] T_2$
\item $\Ctx = \Ctx', x : S_1, y : T_1$
\item $\MeasureM = \MeasureM' + (m_x, 0) + (m_y, 0)$
\item $\wtpn{\MeasureM'}{\Ctx', x : S_2, y : T_2} P'$
\end{itemize}
We conclude with two applications of \refrule{mtb-cast} by taking $\MeasureN
\eqdef \MeasureM$.

\proofrule{sb-cast-new}
Then $P = \NewPar\x{\Cast\x P_1}{P_2} \pcong \NewPar\x{P_1}{P_2} = Q$.
From \refrule{mtb-par} we deduce that there exist $\Ctx_1$, $\Ctx_2$,
$\MeasureM_1$, $\MeasureM_2$, $S_1$ and $T$ such that:
\begin{itemize}
\item $\Ctx = \Ctx_1, \Ctx_2$
\item $\MeasureM = \MeasureM_1 + \MeasureM_2 + (0, \rankb{S_1}{T})$
\item $\wtpn{\MeasureM_1}{\Ctx_1, x : S_1}{\Cast\x P_1}$ and
$\wtpn{\MeasureM_2}{\Ctx_2, x : T}{P_2}$
\item $S_1 \compatible T$
\end{itemize}

From \refrule{mtb-cast} we deduce that there exist $\MeasureM_1'$, $S_2$ and $m_x$ such
that
\begin{itemize}
\item $S_1 \subt[m_x] S_2$
\item $\MeasureM_1 = \MeasureM_1' + (m_x , 0)$
\item $\wtpn{\MeasureM_1'}{\Ctx_1, x : S_2}{P_1}$
\end{itemize}

From $S_1 \compatible T$ and $S_1 \subt[m_x] S_2$ and \Cref{def:compatibility} we
deduce $S_2 \compatible T$.
We conclude with one application of \refrule{mtb-par} taking $\MeasureN \eqdef
\MeasureM_1' + \MeasureM_2 + (0, \rankb{S_2}{T}) < \MeasureM$.

\proofrule{sb-cast-swap}
Then $P = \NewPar\x{\Cast\y P_1}{P_2} \pcong \Cast\y{\NewPar\x{P_1}{P_2}} = Q$
and $x \ne y$.
From \refrule{mtb-par} we deduce that there exist $\Ctx_1$, $\MeasureM_1$,
$\MeasureM_2$, $S_1$, $S_2$ and $T_1$ such that:
\begin{itemize}
\item $\Ctx = \Ctx_1, \Ctx_2, y : T_1$
\item $\MeasureM = \MeasureM_1 + \MeasureM_2 + (0, \rankb{S_1}{S_2})$
\item $\wtpn{\MeasureM_1}{\Ctx_1, x : S_1, y : T_1}{\Cast\y P_1}$ and
$\wtpn{\MeasureM_2}{\Ctx_2, x : S_2}{P_2}$
\item $S_1 \compatible S_2$
\end{itemize}

From \refrule{mtb-cast} we deduce that there exist $\MeasureM_1'$, $T_2$ and $m_y$ such
that
\begin{itemize}
\item $T_1 \subt[m_y] T_2$
\item $\MeasureM_1 = \MeasureM_1' + (m_y, 0)$
\item $\wtp{\Ctx_1, x : S_1, y : T_2}{P_1}$
\end{itemize}

We derive $\wtpn{\MeasureM_1' + \MeasureM_2 + (0, \rankb{S_1}{S_2})}{\Ctx_1,
\Ctx_2, y : T_2}{\NewPar\x{P_1}{P_2}}$ with one application of
\refrule{mtb-par} and we conclude with one application of \refrule{mtb-cast} by
taking $\MeasureN \eqdef \MeasureM$.

\proofrule{sb-call}
Then $P = \Call{A}{\seqof\x} \pcong Q$ where $\Definition{A}{\seqof\x} Q$.
From \refrule{mtb-thread} we deduce that $\wtp[n]\Ctx{\Call A {\seqof\x}}$
for some $n$ such that $\MeasureM = (n, 0)$.
Using \Cref{lem:subj_cong_bin} we deduce that $\wtp[m]\Ctx Q$ for some $m \leq n$.
Using \Cref{lem:measure_rank_bin} we deduce that $\wtpn\MeasureN\Ctx Q$ for some
$\MeasureN \le (m, 0)$.
We conclude by observing that $\MeasureN \le (m, 0) \leq (n, 0) = \MeasureM$.
\end{proof}

%% file: ts-bin/proof-nf.tex
\beginbass
In this section we introduce some normal forms that are instrumental to the
soundness proof of the type system. In particular, they allow us to prove
a \emph{quasi} deadlock freedom result which states that a well types process
can be rearranged in such a way that one of the reduction rule is ready to be 
applied. Deadlock freedom is achieved by considering the process in such
shape and the fact that the session under analysis is \emph{compatible}.

To this aim, it is useful to also introduce
\emph{process contexts} as a convenient way of referring to sub-processes. A
process context $\PCtxC$ is essentially a process with a \emph{hole} denoted by
$\Hole$:
\[
	\textbf{Process context}
	\quad
	\PCtxC, \PCtxD ~~::=~~ \Hole \mid \NewPar\x\PCtxC{P} \mid \NewPar\x{P}\PCtxC \mid \Cast\x\PCtxC
\]

As usual, we write $\PCtxC[P]$ for the process obtained by replacing the hole in $\PCtxC$ with $P$. 
Note that this operation may capture channel names that occur free in $P$ and that are bound by $\PCtxC$.


\begin{definition}[Choice Normal Form]
	\label{def:cnf_bin}
	We say that $P_1 \pchoice P_2$ is an \emph{unguarded choice} of $P$ if there exists 
	$\PCtxC$ such that $P \pcong \PCtxC[P_1 \pchoice P_2]$. 
	We say that $P$ is in \emph{choice normal form} if it has no unguarded choices.
\end{definition}

We introduce a normal form that makes it easier to locate the components 
of a process that may interact with each other. 
Intuitively, a process is in \emph{thread normal form} if it consists of 
an initial prefix of casts followed by a parallel composition of threads, 
where a thread is either $\Done$ or a process waiting to perform an input/output 
action on some channel $x$. In this latter case, we say that the thread is an $x$-thread. 
Note that a process invocation $\Call A {\seqof\x}$ is \emph{not} a thread. Formally:

\begin{definition}[Thread Normal Form]
	\label{def:tnf_bin}
	A process is in \emph{thread normal form} if it is generated by the grammar below:
	\[
		\begin{array}{@{}rcl@{}}
			\Pnf, \Qnf & ::= & \Cast\x\Pnf \mid \Ppar
			\\
			\Ppar, \Qpar & ::= & \NewPar\x\Ppar\Qpar \mid \Pth
			\\
			\Pth & ::= &
			\Done \mid
			\Close\x \mid \Wait\x{P} \mid 
			\x\Pol\set{\l_i : P_i}_{i \in I} \mid
			\POutput\x\y.P \mid \PInput\x{(\y)}.P
		\end{array}
	\] 
\end{definition}

At last, a process in proximity normal form is such that there exist at least two 
$x$-threads that are next to each other. Since each thread is waiting to perform 
an operation on the same session $x$, the two thread may potentially 
reduce if the operations are complementary ones.

\begin{definition}[Proximity Normal Form]
	\label{def:pnf}
	We say that $\Pnf$ is in \emph{proximity normal form} if $\Pnf =
	\PCtxC[\NewPar\x\Pth\Qth]$ for some $\PCtxC$, $x$, $\Pth$ and $\Qth$ where
	$\Pth$ and $\Qth$ are $x$-threads.
\end{definition}


We can reduce any well-typed process into a process that is in choice normal form. 
The fact that the original process is well typed guarantees that this reduction 
eventually terminates when all the unguarded choices have been resolved.

\begin{lemma}
	\label{lem:cnf2_bin}
	If $\wtp[n]\Ctx{P}$ and $\wtpi\Ctx{P}$, then there exists $Q$ in choice
	normal form such that $P \wred Q$ and $\wtp[m]\Ctx{Q}$ for some $m \le n$.
\end{lemma}
\begin{proof}
By induction on $\wtpi\Ctx{P}$ and by cases on the last rule applied.

\proofcase{Case $P$ is already in choice normal form} 
We conclude taking $Q \eqdef P$ and $m \eqdef n$.

\proofrule{tb-call}
Then $P = \pinvk{A}{\seqof{u}}$ and $\Definition{A}{\seqof{x}}{R}$.
We deduce $\Ctx = \seqof{u : S}$, $\tass{A}{\seqof{S}}{n'}$ and $\wtpi\Ctx{R\subst{\seqof u}{\seqof x}}$. Moreover, it must be the case that
$\wtp[n']\Ctx{R\subst{\seqof u}{\seqof x}}$ and $n' \leq n$ since \refrule{tb-call} is used in the coinductive judgment as well.
Using the induction hypothesis we deduce that there exist $Q$ in choice normal form and $m \le n'$ such that $R\subst{\seqof u}{\seqof x} \wred Q$ 
and $\wtp[m]\Ctx{Q}$.
We conclude by observing that $P \wred Q$ using \refrule{rb-struct} and that $m \leq n' \leq n$.

\proofrule{cob-choice}
Then $P = P_1 \pchoice P_2$.
We deduce $\wtpi\Ctx{P_k}$ with $k \in \set{1,2}$.
Moreover, it must be the case that $\wtp[n]\Ctx{P_k}$ since \refrule{tb-choice} is used in the coinductive judgment.
Using the induction hypothesis we deduce that there exist $Q$ in choice normal form and $m \leq n$ such that $P_k \wred Q$ and $\wtp[m]\Ctx{Q}$.
We conclude by observing that $P \red P_k$ by \refrule{rb-choice}.

\proofrule{tb-choice}
Analogous to the previous case but we consider the premise in which the rank is the same of the conclusion to keep sure that it does not increase.

\proofrule{tb-par}
Then $P = \pres{x}{P_1 \parop P_2}$. 
We deduce that there exist $\Ctx_1, \Ctx_2, S_1, S_2$ such that
\begin{itemize}
\item $\Ctx = \Ctx_1, \Ctx_2$
\item $\wtpi{\Ctx_1, x : S_1}{P_1}$
\item $\wtpi{\Ctx_2, x : S_2}{P_2}$
\item $S_1 \compatible S_2$
\end{itemize}
Furthermore, it must be the case that there exist $n_1$ and $n_2$ such that
$\wtp[n_i]{\Ctx_i, x : S_i}{P_i}$ for $i=1,2$ 
and $n = 1 + n_1 + n_2$ since \refrule{tb-par} is used in the coinductive judgment as well.
Using the induction hypothesis we deduce that there exist $Q_i$ in choice normal form and $m_i \leq n_i$ such that 
$P_i \wred Q_i$ and $\wtp[m_i]{\Ctx_i, x : S_i}{Q_i}$ for $i=1,2$.
We conclude by taking $m \eqdef 1 + m_1 + m_2$ and $Q \eqdef \pres{s}{Q_1 \parop Q_2}$ 
with one application of \refrule{tb-par}, observing that 
$m = 1 + + m_1 + m_2 \leq 1 + n_1 + n_2 = n$ and that $P \wred Q$ by \refrule{rb-par}.

\proofrule{tb-cast}
Then $P = \pcast{x} P'$.
Analogous to the previous case, just simpler.
\end{proof}

\begin{lemma}
	\label{lem:cnf1_bin}
	If $\wtp[n]\Ctx{P}$, then there exists $Q$ in choice normal form such that $P \wred Q$ and $\wtp[m]\Ctx{Q}$ for some $m \le n$.
\end{lemma}
\begin{proof}
	\brk
	Consequence of \Cref{lem:cnf2_bin} noting that $\wtp[n]\Ctx{P}$ implies $\wtpi\Ctx{P}$.
\end{proof}

\begin{lemma}
	\label{lem:cnf_exists_bin}
	If\/ $\wtpn\MeasureM\Ctx{P}$, then there exist $Q$ in choice normal form and 
	$\MeasureN \leq \MeasureM$ such that $P \wred Q$ and $\wtpn\MeasureN\Ctx{Q}$.
\end{lemma}
\begin{proof}
By induction on $\wtpn\Measure\Ctx{P}$ and by cases on the last rule applied.

\proofrule{mtb-thread}
Then $P$ is a thread. We deduce that
\begin{itemize}
\item $\Measure = (n , 0)$ for some $n$
\item $\wtp[n]\Ctx{P}$
\end{itemize}
From \Cref{lem:cnf1_bin} we deduce that there exist $Q$ and $m \le n$ such that $P \wred Q$ and $\wtp[m]\Ctx{Q}$. 
From \Cref{lem:measure_rank_bin} we deduce $\wtpn\MeasureN\Ctx{Q}$ for some $\MeasureN \le (m , 0)$. 
We conclude observing that $\MeasureN \le (m , 0) \le (n , 0) = \Measure$.

\proofrule{mtb-cast}
Then $P = \pcast{x}{P'}$. We deduce that
\begin{itemize}
\item $\Ctx = \CtxD, x : S$
\item $S \subt[n] T$
\item $\Measure = \Measure' + (n,0)$
\item $\wtpn{\Measure'}{\Ctx', x : T}{P'}$
\end{itemize}
Using the induction hypothesis we deduce that there exist $Q'$ and $\MeasureN' \le \Measure'$ such that $P' \wred Q'$ 
and $\wtpn{\MeasureN'}{\Ctx', x : T}{Q'}$. We conclude with an application of \refrule{mtb-cast} taking 
$Q \eqdef \pcast{x}{Q'}$, $\MeasureN \eqdef \MeasureN' + (n,0)$ and observing that $P \wred Q$ using \refrule{rb-cast}. 

\proofrule{mtb-par}
Then $P = \pres{s}{P_1 \parop P_2}$. 
We deduce that there exist $\Ctx_1, \Ctx_2, S_1, S_2, \Measure_1, \Measure_2$ such that
\begin{itemize}
\item $\Ctx = \Ctx_1, \Ctx_2$
\item $\wtpn{\Measure_1}{\Ctx_1, x : S_1}{P_1}$
\item $\wtpn{\Measure_2}{\Ctx_2, x : S_2}{P_2}$
\item $\Measure = \Measure_1 + \Measure_2 + (0, \rankb{S_1}{S_2})$
\item $S_1 \compatible S_2$
\end{itemize}
Using the induction hypothesis we deduce that there exist 
$Q_i$ in choice normal form and $\MeasureN_i \leq \Measure_i$ such that $P_i \wred Q_i$ and 
$\wtpn{\MeasureN_i}{\Ctx_i, x : S_i}{Q_i}$ for $i=1,2$.
We conclude by taking $\MeasureN \eqdef \MeasureN_i + \MeasureN_2 + (0, \rankb{S_1}{S_2})$ and 
$Q \eqdef \pres{s}{Q_1 \parop Q_2}$ with one application of \refrule{mtb-par}, 
observing that 
\[
\MeasureN = \MeasureN_1 + \MeasureN_2 + (0, \rankb{S_1}{S_2}) \leq 
						\Measure_1 + \Measure_2 + (0, \rankb{S_1}{S_2}) = \Measure
\]
and that $P \wred Q$ by \refrule{rb-par}.
\end{proof}

It is easy to rewrite any \emph{well-typed} process that is in choice normal 
form into thread normal form using structural pre-congruence. 
The hypothesis that the process is well typed, at least according to the 
inductive interpretation of the typing rules with the corule \refrule{cob-tag},
is necessary to guarantee that a process invocation may eventually be expanded 
to a term other than another process invocation. 
For example, the process $A$ defined by $\Definition{A}{}{A}$ has 
no thread normal form and is ill typed. By combining this result with 
\Cref{lem:subj_cong_bin} we can deduce that the obtained thread normal form is also well typed.

\begin{lemma}
\label{lem:tnf_exists_bin}
	If\/ $\wtpi\Ctx P$ and $P$ is in choice normal form, then there exists
	$\Pnf$ such that $P \pcong \Pnf$. 
\end{lemma}
\begin{proof}
	By induction on $\wtpi\Ctx P$ and by cases on the last rule applied.

	\proofcase{Cases \refrule{tb-choice} and \refrule{co-choice}} 
	These cases are impossible from the hypothesis that $P$ is in choice normal form.

	\proofcase{Cases \refrule{tb-done}, \refrule{tb-wait}, \refrule{tb-close}, 
	\refrule{tb-channel-in}, \refrule{tb-channel-out}, \refrule{tb-label}, \refrule{co-label}}
	Then $P$ is a thread and is already in thread normal form and we conclude by
	reflexivity of $\pcong$.

	\proofrule{tb-call}
	Then there exist $A$, $Q$, $\seqof\x$ and $\seqof\S$ such that
	\begin{itemize}
	\item $P = \Call{A}{\seqof\x}$
	\item $\Definition{A}{\seqof\x}Q$
	\item $\Ctx = \seqof{x : S}$
	\item $\wtpi{\seqof{x : S}}{Q}$
	\end{itemize}

	Using the induction hypothesis on $\wtpi{\seqof{x : S}}{Q}$ 
	we deduce that there exists $\Pnf$ such that $Q \pcong \Pnf$.
	We conclude $P \pcong \Pnf$ using \refrule{sb-call} and the transitivity of $\pcong$.

	\proofrule{tb-par}
	Then there exist $x$, $P_1$, $P_2$, $\Ctx_1$, $\Ctx_2$, $S_1$ and $S_2$ such that
	\begin{itemize}
	\item $P = \NewPar\x{P_1}{P_2}$
	\item $\Ctx = \Ctx_1, \Ctx_2$
	\item $\wtpi{\Ctx_i, x : S_i}{P_i}$ for $i=1,2$
	\end{itemize}

	Using the induction hypothesis on $\wtpi{\Ctx_i, x : S_i}{P_i}$ we deduce 
	that there exists $\Pnf_i$ such that $P_i \pcong \Pnf_i$ for $i=1,2$.
	By definition of thread normal form, it must be the case that 
	$\Pnf_i = \Cast{\seqof{x_i}} \Ppar_i$ for some $\seqof{x_i}$ and $\Ppar_i$.
	Let $\seqof{y_i}$ be the same sequence as $\seqof{x_i}$ except that occurrences of $x$ have been removed.
	We conclude by taking 
	$\Pnf \eqdef \Cast{\seqof{y_1}\seqof{y_2}}\NewPar\x{\Ppar_1}{\Ppar_2}$ and observing that
	\[
		\begin{array}{rcll}
			P & = & \NewPar\x{P_1}{P_2} & \text{by definition of $P$}
			\\
			& \pcong & \NewPar\x{\Pnf_1}{\Pnf_2} & \text{using the induction hypothesis}
			\\
			& = & \NewPar\x{\Cast{\seqof{x_1}}\Ppar_1}{\Cast{\seqof{x_2}}\Ppar_2}
			& \text{by definition of thread normal form}
			\\
			& \pcong & \Cast{\seqof{y_1}\seqof{y_2}}\NewPar\x{\Ppar_1}{\Ppar_2}
			& \text{by \refrule{sb-cast-new},} \\
			& & & \text{\refrule{sb-cast-swap} and \refrule{sb-par-comm}}
			\\
			& = & \Pnf & \text{by definition of $\Pnf$}
		\end{array}
	\]

	\proofrule{tb-cast}
	Then there exist $x$, $Q$, $\Ctx'$, $S$ and $T$ such that
	\begin{itemize}
	\item $P = \Cast\x Q$
	\item $\Ctx = \Ctx', x : S$
	\item $\wtpi{\Ctx', x : T}{Q}$
	\item $S \subt T$
	\end{itemize}

	Using the induction hypothesis on $\wtpi{\Ctx', x : T}{Q}$ we 
	deduce that there exists $\Qnf$ such that $Q \pcong \Qnf$.
	We conclude by taking $\Pnf \eqdef \Cast\x\Qnf$ using the fact that $\pcong$ is a pre-congruence.
\end{proof}

In order to show that every well-typed, closed process in thread normal form can 
also be rewritten in proximity normal form we prove \Cref{lem:proximity_bin}, 
which pushes a restriction $(x)$ next to a process in which $x$ occurs free, 
which might as well be an $x$-thread.

\begin{lemma}[Proximity]
  \label{lem:proximity_bin}
  If $x\in\fn{P} \setminus \bn\PCtxC$, then $\NewPar\x{\PCtxC[P]}{Q}
  \pcong \PCtxD[\NewPar\x{P}{Q}]$ for some $\PCtxD$.
\end{lemma}
\begin{proof}
  By induction on the structure of $\PCtxC$ and by cases on its shape.

  \proofcase{Case $\PCtxC = \Hole$}
  We conclude by taking $\PCtxD \eqdef \Hole$ using the reflexivity
  of $\pcong$.

  \proofcase{Case $\PCtxC = \NewPar\y{\PCtxC'}{R}$}
  From the hypothesis $x \in \fn{P} \setminus \bn\PCtxC$ 
  we deduce $x \ne y$ and $x \in \fn{P} \setminus \bn{\PCtxC'}$.
  Using the induction hypothesis we deduce that there exists $\PCtxD'$ 
  such that $\NewPar\x{\PCtxC'[P]}{Q} \pcong \PCtxD'[\NewPar\x{P}{Q}]$.
  Take $\PCtxD \eqdef \NewPar\y{\PCtxD'}{R}$. We conclude
  \[
    \begin{array}{rcll}
      \NewPar\x{\PCtxC[P]}{Q}
      & = & \NewPar\x{\NewPar\y{\PCtxC'[P]}{R}}{Q}
      & \text{by definition of $\PCtxC$}
      \\
      & \pcong & \NewPar\x{Q}{\NewPar\y{\PCtxC'[P]}{R}}
      & \text{by \refrule{sb-par-comm}}
      \\
      & \pcong & \NewPar\y{\NewPar\x{Q}{\PCtxC'[P]}}{R}
      & \text{by \refrule{sb-par-assoc}} \\
      & & & \text{and $x \in \fn{\PCtxC'[P]}$}
      \\
      & \pcong & \NewPar\y{\NewPar\x{\PCtxC'[P]}{Q}}{R}
      & \text{by \refrule{sb-par-comm}}
      \\
      & \pcong & \NewPar\y{\PCtxD'[\NewPar\x{P}{Q}]}{R}
      & \text{by induction hypothesis}
      \\
      & = & \PCtxD[\NewPar\x{P}{Q}]
      & \text{by definition of $\PCtxD$}
    \end{array}
  \]
  where, in using \refrule{sb-par-assoc}, we note that
  $x \in \fn{\PCtxC'[P]}$ since $x \in \fn{P} \setminus \bn\PCtxC$.

  \proofcase{Case $\PCtxC = \NewPar\y{R}{\PCtxC'}$}
  Symmetric of the previous case.

  \proofcase{Case $\PCtxC = \Cast\y\PCtxC'$ and $x \ne y$}
  Using the induction hypothesis we deduce that there exists $\PCtxD'$ such that
  $\NewPar\x{\PCtxC'[P]}{Q} \pcong \PCtxD'[\NewPar\x{P}{Q}]$.
  Take $\PCtxD \eqdef \Cast\y\PCtxD'$. We conclude
  \[
    \begin{array}{rcll}
      \NewPar\x{\PCtxC[P]}{Q}
      & = & \NewPar\x{\Cast\y\PCtxC'[P]}{Q}
      & \text{by definition of $\PCtxC$}
      \\
      & \pcong & \Cast\y\NewPar\x{\PCtxC'[P]}{Q}
      & \text{by \refrule{sb-cast-swap} and $x \ne y$}
      \\
      & \pcong & \Cast\y\PCtxD'[\NewPar\x{P}{Q}]
      & \text{using the induction hypothesis}
      \\
      & = & \PCtxD[\NewPar\x{P}{Q}]
      & \text{by definition of $\PCtxD$}
    \end{array}
  \]

  \proofcase{Case $\PCtxC = \Cast\x\PCtxC'$}
  Using the induction hypothesis we deduce that there exists $\PCtxD$ such that
  $\NewPar\x{\PCtxC'[P]}{Q} \pcong \PCtxD[\NewPar\x{P}{Q}]$.
  We conclude
  \[
    \begin{array}[b]{rcll}
      \NewPar\x{\PCtxC[P]}{Q}
      & = & \NewPar\x{\Cast\x\PCtxC'[P]}{Q}
      & \text{by definition of $\PCtxC$}
      \\
      & \pcong & \NewPar\x{\PCtxC'[P]}{Q}
      & \text{by \refrule{sb-cast-new}}
      \\
      & \pcong & \PCtxD[\NewPar\x{P}{Q}]
      & \text{using the induction hypothesis}
    \end{array}
    \qedhere
  \]
\end{proof}

We can now prove the fact that every well-typed, closed process in thread normal form can 
be rewritten using structural pre-congruence either to $\Done$ or to a process in proximity normal form. 
Note that this property the one that, combined with the \emph{compatibility} of the involved session,
guarantees deadlock freedom.

\begin{lemma}[Quasi Deadlock Freedom]
	\label{lem:dl_freedom_bin}
	If\/ $\wtpn\Measure\EmptyCtx\Pnf$, then $\Pnf = \Done$ or $\Pnf
	\pcong \Qnf$ for some $\Qnf$ in proximity normal form.
\end{lemma}
\begin{proof}
	A simple induction on the derivation of $\wtpn\MeasureM\EmptyCtx\Pnf$
	allows us to deduce that $\Pnf$ consists of $k$ sessions and $k+1$ threads.
	If $k = 0$, then we conclude $\Pnf = \Done$.
  	If $k > 0$, then each of the $k+1$ threads is an $x_i$-thread for some
	$x_i$. Since there are $k+1$ threads but only $k$ distinct sessions, it must
	be the case that $x_i = x_j$ for some $1 \leq i < j \leq k+1$. In other
	words, there exist $\PCtxC$, $\PCtxC_1$, $\PCtxC_2$, $\Pth_1$ and $\Pth_2$ such
	that $\Pth_1$ and $\Pth_2$ are $x$-threads and $\Pnf =
	\PCtxC[\NewPar\x{\PCtxC_1[\Pth_1]}{\PCtxC_2[\Pth_2]}]$.
	We conclude
	\[
		\begin{array}{rcll}
			\Pnf & = & \PCtxC[\NewPar\x{\PCtxC_1[\Pth_1]}{\PCtxC_2[\Pth_2]}]
			& \text{by definition of $\Pnf$}
			\\
			& \pcong & \PCtxC[\PCtxD_1[\NewPar\x{\Pth_1}{\PCtxC_2[\Pth_2]}]]
			& \text{for some $\PCtxD_1$ by \Cref{lem:proximity_bin}}
			\\
			& \pcong & \PCtxC[\PCtxD_1[\NewPar\x{\PCtxC_2[\Pth_2]}{\Pth_1}]]
			& \text{by \refrule{s-par-comm}}
			\\
			& \pcong & \PCtxC[\PCtxD_1[\PCtxD_2[\NewPar\x{\Pth_2}{\Pth_1}]]]
			& \text{for some $\PCtxD_2$ by \Cref{lem:proximity_bin}}
			\\
			& \eqdef & \Qnf
		\end{array}
	\]
	where $x = x_i = x_j$. The fact that $\Qnf$ is in thread normal form follows
	 from the observation that $\Pnf$ does not have unguarded casts (it is a closed process in thread normal form) 
	 so the pre-congruence rules applied here and in \Cref{lem:proximity_bin} do not move casts around. 
	 We conclude that $\Qnf$ is in proximity normal form by its shape.
\end{proof}

%% file: ts-bin/proof-soundness.tex
\beginbass
Here we prove the soundness of the type system (see \Cref{thm:ts_bin_sound}). As already hinted at in
\Cref{ssec:ts_bin_rules}, the proof is loosely based on the method of helpful directions
\Citep{Francez86}, namely on the property that a (well-typed) process \emph{may}
reduce in such a way that its measure strictly decreases. Recall that this
property is not true for every reduction. Here we assume that the reducing
process is in proximity normal form. The same result will be generalized later
on.

\begin{lemma}
	\label{lem:pnf_helpful_direction_bin}
	If\/ $\wtpn\MeasureM\Ctx\Pnf$ where $\Pnf$ is in proximity normal
	form, then there exist $Q$ and $\MeasureN < \MeasureM$ such that $\Pnf
	\wred^+ Q$ and $\wtpn\MeasureN\Ctx{Q}$.
\end{lemma}
\begin{proof}
From the hypothesis that $\Pnf$ is in proximity normal form we know that 
$\Pnf = \PCtxC[\pres{x}{\Pth_1 \parop \Pth_2}]$ for some $\PCtxC$, $x$ and $\Pth_1, \Pth_2$ $x$-threads.
We reason by induction on $\PCtxC$ and by cases on its shape.
	
\proofcase{Case $\PCtxC = \Hole$}
From \refrule{mtb-thread} and \refrule{mtb-par} we deduce that there exist $\Ctx_i, S_i, n_i$ for $i=1,2$ such that
\begin{itemize}
\item $\Ctx = \Ctx_1,\Ctx_2$
\item $\Measure = (n_1 + n_2 , \rankb{S_1}{S_2})$
\item $\wtp[n_1]{\Ctx_1, x : S_1}{\Pth_1}$
\item $\wtp[n_2]{\Ctx_2, x : S_2}{\Pth_2}$
\item $S_1 \compatible S_2$
\end{itemize}
From the hypothesis that $S_1 \compatible S_2$ we deduce 
$S_1 \wlred{\co\actions} \End[\co\Pol], S_2 \wlred\actions \End[\Pol]$ for some $\actions$. 
We now reason on the rank of the session and on the shape of $S_i$. 
For the sake of simplicity, we implicitly apply \refrule{sb-par-comm} at process level.\\\\
If $\rankb{S_1}{S_2} = 0$, then $S_1 = \End[\co\Pol]$ and $S_2 = \End[\Pol]$ for some $\Pol$.
\begin{itemize}
\item \proofcase{Case $S_1 = \End[\In]$ and $S_2 = \End[\Out]$}
Then 
	\begin{itemize}
	\item $\Ctx_2 = \EmptyCtx$ and $\Pth_2 = \pclose{x}$
	\item $\Pth_1 = \pwait{x}{Q}$
	\item $\wtp[n_1]{\Ctx_1}{Q}$
	\end{itemize}
From \Cref{lem:measure_rank_bin} we deduce that $\wtpn{\MeasureN}{\Ctx_1}{Q}$ for some $\MeasureN \le (n_1 , 0)$. 
We conclude observing that $\Pnf \red Q$ by \refrule{rb-signal} and that 
$\MeasureN \le (n_1 , 0) < (n_1 + n_2 , \rankb{S_1}{S_2}) = \Measure$.
\end{itemize}
If $\rankb{S_1}{S_2} > 1$, then $\Sys{S_1}{S_2} \lred{\tau} \dots \lred\tau \Sys{\End[\co\Pol]}{\End[\Pol]}$ 
first using \refrule{lb-tau-l}/\refrule{lb-tau-r} and \refrule{lb-sync}. 
Observe that \refrule{lb-pick} is never used since we are considering the minimum reduction sequence; 
a synchronization through \refrule{lb-pick} and \refrule{lb-sync} would lead to a longer reduction. T
hen $S_1 \xlred{\Out{\Tag_k}}$ and $S_2 \xlred{\In{\Tag_k}}$ for some 
$\Tag_k$ or $S_1 \xlred{\Out U}$ and $S_2 \xlred{\In U}$ for some $U$.
\begin{itemize}
\item \proofcase{Case $S_1 = \Tags\Out \Tag_i.S'_i$ and $S_2 = \JTags\In \Tag_j.T_j$ with $k \in I$}
From the hypothesis that $S_1 \compatible \S_2$ we deduce $I \subseteq J$. 
From \refrule{tb-tag} we deduce that 
	\begin{itemize}
	\item $\Pth_1 = \x\oact\set{{\Tag_i}.{P'_i}}_{i \in I}$
	\item $\Pth_2 = \x\iact\set{{\Tag_j}.{Q_j}}_{j \in J}$
	\item $\wtp[n_1]{\Ctx_1, x : S'_i}{P'_i}$ for all $i \in I$
	\item $\wtp[n_2]{\Ctx_2, x : T_j}{Q_j}$ for all $j \in J$
	\end{itemize}
Let $Q \eqdef \pres{x}{P'_k \parop Q_k}$ and observe that $\Pnf \wred^+ Q$ by \refrule{rb-pick} and \refrule{rb-tag}. 
From \Cref{lem:measure_rank_multi} we deduce that there exist $\Measure_1 \le (n_1 , 0), \Measure_2 \le (n_2 , 0)$ such that
	\begin{itemize}
	\item $\wtpn{\Measure_1}{x : S'_k}{P'_k}$
	\item $\wtpn{\Measure_2}{x : T_k}{Q_k}$
	\end{itemize}
Let 
$\MeasureN \eqdef \Measure_1 
	+ \Measure_2 
	+ (0 , \rankb{S'_k}{T_k})$. 
We conclude with one application of \refrule{mtb-par} observing that
\[
\begin{array}{rcll}
	\MeasureN & = & \Measure_1 + \Measure_2 + (0 , \rankb{S'_k}{T_k}) & \text{by def. of $\MeasureN$}
	\\
	& \le & (n_1 + n_2 , \rankb{S'_k}{T_k}) & \text{by \Cref{lem:measure_rank_bin}}
	\\
	& < & (n_1 + n_2 , \rankb{S_1}{S_2}) & \text{before $\red$}
	\\
	& = & \Measure
\end{array}
\]
\end{itemize}
\begin{itemize}
\item \proofcase{Case $S_1 = \Out{S}.T_1$ and $S_2 = \In{S}.T_2$}
\end{itemize}
From the hypothesis that $S_1 \compatible S_2$ and \Cref{def:compatibility} we deduce 
$T_1 \compatible T_2$ and from 
\refrule{tb-channel-out} and \refrule{tb-channel-in} we deduce that
	\begin{itemize}
	\item $\Pth_1 = \POutput{\x}{\y}.{P'_1}$
	\item $\Pth_2 = \PInput{\x}{\y}.{P'_2}$
	\item $\wtp[n_1]{\Ctx_1, x : T_1}{P'_1}$
	\item $\wtp[n_2]{\Ctx_2, x : T_2, y : S}{P'_2}$
	\end{itemize}
Let $Q \eqdef \pres{x}{P'_1 \parop P'_2}$ and observe that $\Pnf \red Q$ by \refrule{rb-channel}. 
From \Cref{lem:measure_rank_bin} we deduce that there exist $\Measure_1 \le (n_1 , 0), \Measure_2 \le (n_2 , 0)$ such that
	\begin{itemize}
	\item $\wtpn{\Measure_1}{\Ctx_1, x : T_1}{P'_1}$
	\item $\wtpn{\Measure_2}{\Ctx_2, x, y : S}{P'_2}$
	\end{itemize}
Let $\MeasureN \eqdef \Measure_1 
	+ \Measure_2 
	+ (0 , \rankb{T_1}{T_2})$. 
We conclude with one application of \refrule{mtb-par} observing that
\[
\begin{array}{rcll}
	\MeasureN & = & \Measure_1 + \Measure_2 + (0 , \rankb{T_1}{T_2}) & \text{by definition of $\MeasureN$}
	\\
	& \le & (n_1 + n_2 , \rankb{T_1}{T_2}) & \text{by \Cref{lem:measure_rank_multi}}
	\\
	& < & (n_1 + n_2 , \rankb{S_1}{S_2}) & \text{before reductions}
	\\
	& = & \Measure
\end{array}
\]

\proofcase{Case $\PCtxC = \pres{y}{\PCtxD \parop \Qpar}$}
Let $\Rnf \eqdef \PCtxD[\pres{x}{\Pth_1 \parop \Pth_2}]$ and observe that $\Rnf$ is in proximity normal form. 
From \refrule{mtb-par} we deduce that there exist $\Ctx_i, S_i, \Measure_i$ for $i=1,2$ such that
\begin{itemize}
\item $\Ctx = \Ctx_1,\Ctx_2$
\item $\wtpn{\Measure_1}{\Ctx_1, y : S_1}{\Rnf}$
\item $\wtpn{\Measure_2}{\Ctx_2, y : S_2}{\Qpar}$
\item $\Measure = \Measure_1 + \Measure_2 + (0 , \rankb{S_1}{S_2})$
\end{itemize}
Using the induction hypothesis on $\wtpn{\Measure_1}{\Ctx_1, y : S_1}{\Rnf}$ we deduce that there 
exists $Q'$ and $\MeasureN' < \Measure_1$ such that
\begin{itemize}
\item $\Rnf \wred^+ Q'$
\item $\wtpn{\MeasureN'}{\Ctx_1, y : S_1}{Q'}$
\end{itemize}
We conclude taking $Q \eqdef \pres{t}{Q' \parop \Qpar}$ and
\[
	\MeasureN \eqdef \MeasureN' + \Measure_2 + (0 , \rankb{S_1}{S_2})
\] 
and observing that $\MeasureN < \Measure$ and $\Pnf \wred^+ Q$ by \refrule{rb-par}.

\proofcase{Case $\PCtxC = \pres{x}{\Qpar \parop \PCtxD}$}
Symmetric to the previous case.

\proofcase{Case $\PCtxC = \pcast{y}{\PCtxD}$}
Observe that $x \ne y$. Let $\Rnf \eqdef \PCtxD[\pres{x}{\Pth_1 \parop \Pth_2}]$ 
and note that $\Rnf$ is in proximity normal form. 
From \refrule{mtb-cast} we deduce that there exists $\CtxD, \Measure', S, T, m_y$ such that
\begin{itemize}
\item $\Ctx = \CtxD, y : S$
\item $S \subt[m_y] T$
\item $\Measure = \Measure' + (m_y , 0)$
\item $\wtpn{\Measure'}{\CtxD, y : T}{\Rnf}$
\end{itemize}
Using the induction hypothesis on $\wtpn{\Measure'}{\CtxD, y : T}{\Rnf}$ we deduce that 
there exist $Q'$ and $\MeasureN' < \Measure'$ such that $\Rnf \wred^+ Q'$ and $\wtpn{\MeasureN'}{\CtxD, y : T}{Q'}$. 
We conclude taking $Q \eqdef \pcast{y}{Q'}$ and $\MeasureN \eqdef \MeasureN' + (m_y , 0)$ 
and observing that $\MeasureN < \Measure$ and $\Pnf \wred^+ Q$ by \refrule{rb-cast}.
\end{proof}

Now we prove that any well-typed, closed process can be either rewritten to $\Done$ 
using structural pre-congruence or reduced so as to obtain a strictly smaller measure.

\begin{lemma}
  \label{lem:helpful_direction_bin}
  If\/ $\wtpn\MeasureM\EmptyCtx{P}$, then either $P \pcong \Done$ or $P
  \wred^+ Q$ and $\wtpn\MeasureN\EmptyCtx{Q}$ for some $Q$ and $\MeasureN
  < \MeasureM$.
\end{lemma}
\begin{proof}
    Using \Cref{lem:cnf_exists_bin} we deduce that there exist $P'$ in choice normal form such that 
    $P \wred P'$ and $\wtpn{\MeasureM'}\EmptyCtx{P'}$ and $\MeasureM' \leq \MeasureM$.
    By \Cref{lem:measure_rank_bin} we deduce $\wtp\EmptyCtx{P'}$.
    Using \Cref{lem:tnf_exists_bin} we deduce that there exist $\Pnf$ such that $P' \pcong
    \Pnf$.
    If $\Pnf = \Done$ there is nothing left to prove.
    If $\Pnf \neq \Done$, by \Cref{lem:dl_freedom_bin} we deduce $\Pnf \pcong \Qnf$ for
    some $\Qnf$ in proximity normal form.
    From \Cref{lem:m_pcong_bin} we deduce $\wtpn{\MeasureM''}\EmptyCtx\Qnf$ for some $\MeasureM'' \leq \MeasureM'$.
    Using \Cref{lem:pnf_helpful_direction_bin} we conclude that $\Qnf \wred^+ Q$ and
    $\wtpn\MeasureN\EmptyCtx Q$ for some $Q$ and $\MeasureN < \MeasureM'' \leq \MeasureM' \leq \MeasureM$.
\end{proof}

Using \Cref{lem:helpful_direction_bin} we can prove that a well-typed, closed process weakly terminates. 
Namely, that there exists a finite reduction sequence to $\Done$.

\begin{lemma}[Weak Termination]
	\label{lem:weak_termination_bin}
	If\/ $\wtp[n]\EmptyCtx{P}$, then either $P \pcong \Done$ or $P \wred^+
	\Done$.
\end{lemma}
\begin{proof}
	From \Cref{lem:measure_rank_bin} we deduce that there exists $\MeasureM \le (n,
	0)$ such that $\wtpn\MeasureM\EmptyCtx P$.
	We proceed doing an induction on the lexicographically ordered pair $\MeasureM$.
    From \Cref{lem:helpful_direction_bin} we deduce either $P \pcong \Done$ or $P
    \wred^+ Q$ and $\wtpn\MeasureN\EmptyCtx{Q}$ for some $\MeasureN <
    \MeasureM$.
    In the first case there is nothing left to prove.
    In the second case we use the induction hypothesis to deduce that either $Q
    \pcong \Done$ or $Q \wred^+ \Done$.
    We conclude using either \refrule{rb-struct} or the transitivity of
    $\wred^+$, respectively.
\end{proof}

The soundness of the type system is now a simple corollary.

\begin{proof}[Proof of \Cref{thm:ts_bin_sound}]
    Consequence of \Cref{lem:subj_red_bin} and \Cref{lem:weak_termination_bin}.
\end{proof}

%% file: ts-multi/chintro.tex
\begintreble
In this chapter we generalize the type system in \Cref{ch:ft_bin} to multiparty session types.
In particular, we provide a type system for enforcing (successful) fair termination
of multiparty sessions. In this case we refer to the notion of \emph{coherence} (\Cref{def:coherence}).
Again, we embed \emph{fair subtyping} as the coherence-preserving relation.
Notably, the \emph{boundedness} properties mentioned in \Cref{ssec:boundedness} are still required.
For the sake of simplicity we do not analyze them step by step as the motivating examples can be
straightforwardly adapted. Instead, we directly show how to enforce them.

\begin{remark}
	The need of the type system that we present in this chapter comes from the fact the there 
	exist multiparty sessions that cannot be decomposed in binary ones.
	Consider the following local types
	\[
	\begin{array}{lcl}
		\S_{\rolep} & : & \roleq\Out\set{\Tag[a].\roler\Out\Tag[c].\End[\Out],\, \Tag[b].\End[\Out]}
		\\
		\S_{\roleq} & : & \rolep\In\set{\Tag[a].\roler\Out\Tag[ok].\End[\Out],\, \Tag[b].\roler\Out\Tag[no].\End[\Out]}
		\\
		\S_{\roler} & : & \roleq\In\set{\Tag[ok].\rolep\In\Tag[c].\End[\In],\, \Tag[no].\End[\In]}
	\end{array}
	\]
	This example is paradigmatic because of the dependencies in the communications.
	Indeed, $\roler$ receives $\Tag[c]$ from $\role$ only if $\roleq$ receives $\Tag[a]$ from $\role$.
	Furthermore, the class of multiparty sessions that can be decomposed into linear
	ones consists of those sessions whose local types can be \emph{projected} to all the participants
	appearing in them.
	(\ie the \emph{projections} exist).
	This is not true for the example above because the projections of $\S_{\role}$ on $\roler$ and
	of $\S_{\roler}$ on $\role$ are not defined as is such types the communication with $\roler$ and $\role$
	does not take place in all the branches. 
	\eor
\end{remark}

The chapter is organized as follows.
In \Cref{sec:ts_multi_proc} we present the syntax and the semantics of the calculus based on multiparty sessions. 
\Cref{sec:ts_multi_ts} shows the typing rules for such calculus. Differently from \Cref{sec:ts_bin_ts}, we focus
on some involved examples of processes instead of analyzing additional required properties (see \Cref{ssec:boundedness}).
Then, in \Cref{sec:ts_multi_corr} we detail the soundness proof of the type system.
At last, in \Cref{sec:ts_multi_related} we compare the present type system, and consequently
the one in \Cref{ch:ft_ll}, to existing works.

%% file: ts-multi/calculus.tex
\beginalto
In this section we define the calculus for multiparty sessions on which we apply
our static analysis technique. The calculus is an extension of the one presented
by \cite{CicconePadovani22} to multiparty sessions in the
style of \cite{ScalasYoshida19} and that has been presented in \cite{CicconeDP22}.
We recall some basic notions.
We use an infinite set of \emph{variables} ranged over by $x$, $y$, $z$, an
infinite set of \emph{session names} ranged over by $s$ and $t$, a set of
\emph{roles} ranged over by $\rolep$, $\roleq$, $\roler$, a set of \emph{message
tags} ranged over by $\Tag$, and a set of \emph{process names} ranged over by
$A$, $B$, $C$. As in the binary case, the different terminology for labels 
is needed to avoid confusion with the labels in \Cref{sec:st}.
We use roles to distinguish the participants of a session. In particular, an
\emph{endpoint} $\ep\sn\role$ consists of a session name $\sn$ and a role
$\role$ and is used by the participant with role $\role$ to interact with the
other participants of the session $s$.
We use $u$ and $v$ to range over \emph{channels}, which are either variables or
session endpoints.
We write $\seqof x$ and $\seqof u$ to denote possibly empty sequences of
variables and channels, extending this notation to other entities.
We use $\Pol$ to range over the elements of the set $\set{\iact,\oact}$ of
\emph{polarities}, distinguishing input actions ($\iact$) from output actions
($\oact$).


\subsection{Syntax of Processes}
\input{ts-multi/proc-syntax}


\subsection{Operational Semantics}
\input{ts-multi/proc-sem}


\subsection{Examples}
\label{ssec:proc_ex_multi}
\input{ts-multi/proc-ex}

%% file: ts-multi/proc-syntax.tex
\beginbass
\begin{figure}[t]
	\framebox[\textwidth]{
  \begin{math}
    \begin{array}[t]{@{}rcll@{}}
      P, Q, R & ::= & & \textbf{Process} \\
      &   & \pdone & \text{termination} \\
      & | & \pwait\chvar{P} & \text{signal in} \\
      & | & \pich\chvar\role{x}{P} & \text{channel in} \\
      & | & \pbranch[i\in I]\chvar\role\Pol{\Tag_i}{P_i} & \text{tag in/out} \\
      & | & \pres\sn{P_1\ppar\cdots\ppar P_n} & \text{session} \\
    \end{array}
    ~
    \begin{array}[t]{@{}rcll@{}}
      \\
      & | & \pinvk\pdn{\seqof\chvar} & \text{invocation} \\
      & | & \pclose\chvar & \text{signal out} \\
      & | & \poch\chvar\role\achvar{P} & \text{channel out} \\
      & | & P \pchoice Q &\text{choice} \\
      & | & \pcast\chvar P & \text{cast} \\
    \end{array}
  \end{math}
  }
  \caption{Syntax of processes}
  \label{fig:proc_syntax_multi}
\end{figure}
A \emph{program} is a finite set of \emph{definitions} of the form
$\pdef\pdn{\seqof\var}{P}$, at most one for each process name, where $P$ is a
term generated by the syntax shown in \Cref{fig:proc_syntax_multi}.
The term $\pdone$ denotes the terminated process that performs no action.
The term $\pinvk\pdn{\seqof\chvar}$ denotes the invocation of the process with
name $\pdn$ passing the channels $\seqof\chvar$ as arguments. When
$\seqof\chvar$ is empty we just write $A$ instead of $\pinvk{A}{}$.
The term $\pclose\chvar$ denotes the process that sends a termination signal on
the channel $\chvar$, whereas $\pwait\chvar P$ denotes the process that waits
for a termination signal from channel $\chvar$ and then continues as $P$.
The term $\poch\chvar\role\achvar{P}$ denotes the process that sends the channel
$\achvar$ on the channel $\chvar$ to the role $\rolep$ and then continues as
$P$. Dually, $\pich\chvar\role\var{P}$ denotes the process that receives a
channel from the role $\rolep$ on the channel $\chvar$ and then continues as $P$
where $\var$ is replaced with the received channel.
The term $\pbranch[i\in I]\chvar\role\Pol{\Tag_i}{P_i}$ denotes a process that
exchanges one of the tags $\Tag_i$ on the channel $\chvar$ with the role $\role$
and then continues as $P_i$. Whether the tag is sent or received depends on the
polarity $\Pol$ and, as it will be clear from the operational semantics, the
polarity $\Pol$ also determines whether the process behaves as an internal
choice (when $\Pol$ is $\oact$) or an external choice (when $\Pol$ is $\iact$).
In the first case the process chooses \emph{actively} the tag being sent,
whereas in the second case the process reacts \emph{passively} to the tag being
received.
We assume that $I$ is finite and non-empty and also that the tags $\Tag_i$ are
pairwise distinct. For brevity, we write $\pbranch\chvar\role\Pol{\Tag_k}{P_k}$
instead of $\pbranch[i\in I]\chvar\role\Pol{\Tag_i}{P_i}$ when $I$ is the
singleton set $\set{k}$.
The term $P \pchoice Q$ denotes a process that non-deterministically behaves
either as $P$ or as $Q$.

A term $\pres{s}{P_1\ppar\cdots\ppar P_n}$ with $n\geq 1$ denotes the parallel
composition of $n$ processes, each of them being a participant of the session
$s$. Each process is associated with a distinct a role $\role_i$ and
communicates in $s$ through the endpoint $\ep{s}{\role_i}$. Combining session
creation and parallel composition in a single form is common in session type
systems based on linear
logic \citep{CairesPfenningToninho16,Wadler14,LindleyMorris16} and helps
guaranteeing deadlock freedom. 
Finally, a \emph{cast} $\pcast\chvar P$ denotes a process that behaves exactly
as $P$. This form is only relevant for the type system and
denotes the fact that the type of $\chvar$ is subject to an application of
subtyping.

The free and bound names of a process are defined as usual, the latter ones
being easily recognizable as they occur within round parenteses. We write
$\fn{P}$ for the set of free names of $P$ and we identify processes modulo
renaming of bound names. Note that $\fn{P}$ may contain variables and session
names, but not endpoints.
Occasionally we write $\pdef{A}{\seqof{x}}{P}$ as a predicate or side condition,
meaning that $P$ is the process associated with the process name $A$. For each
of such definitions we assume that $\fn{P} \subseteq \set{\seqof{x}}$.

%% file: ts-multi/proc-sem.tex
\beginbass
\begin{figure}[t]
	\framebox[\textwidth]{
  \begin{math}
    \begin{array}{@{}lr@{~}c@{~}ll@{}}
      \defrule{sm-par-comm} & \pres\sn{\procs{P} \ppar P \ppar Q \ppar \procs{Q}} 
      & \pcong & 
      \pres\sn{\procs{P} \ppar Q \ppar P \ppar \procs{Q}}
      \\
      \defrule{sm-par-assoc} & \pres\sn{\procs{P} \ppar \pres\asn{R \ppar \procs{Q}}} 
      & \pcong & 
      \pres\asn{\pres\sn{\procs{P} \ppar R} \ppar \procs{Q}}
      & \text{if $s \in \fn{R}$}
      \\
      & & & & \text{and $t \not\in \fn{P}$,}
      \\
      & & & & \text{$\forall Q. s \not\in \fn{Q}$}
      \\
      \defrule{sm-cast-comm} & \pcast{u}{\pcast{v}{P}} & \pcong & \pcast{v}{\pcast{u}{P}}
      \\
      \defrule{sm-cast-new} & \pres{s}{\pcast{\ep{s}\role} P \ppar \procs{Q}}
      & \pcong & 
      \pres{s}{P \ppar \procs{Q}}
      \\
      \defrule{sm-cast-swap} & \pres{s}{\pcast{\ep{t}\role}{P} \ppar \procs{Q}} 
      & \pcong & 
      \pcast{\ep{t}\role}{\pres{s}{P \ppar \procs{Q}}}
      & \text{if $s \ne t$} 
      \\
      \defrule{sm-call} & \pinvk{A}{\seqof{u}} & \pcong & P\subst{\seqof{u}}{\seqof{x}}
      & \text{if $\pdef{A}{\seqof{x}}{P}$}
    \end{array}
  \end{math}
  }
  \caption{Structural precongruence of processes}
  \label{fig:pcong_multi}
\end{figure}
\begin{figure}[t]
	\framebox[\textwidth]{
  \begin{mathpar}
    \inferrule[rm-choice]{ }{
      P_1 \pchoice P_2 \red P_k
    }
    ~ k\in\set{1,2} \defrule[rm-choice]{}
    \and
    \inferrule[rm-signal]{ }{
      \pres{s}{\pwait{\ep{s}\role}{P} \parop \pclose{\ep{s}{\roleq_1}} \parop \cdots \parop \pclose{\ep{s}{\roleq_n}}} 
      \red
      P
    } \defrule[rm-signal]{}
    \and
    \inferrule[rm-channel]{ }{
      \pres{s}{\poch{\ep{s}{\rolep}}{\roleq}{v}{P} \parop \pich{\ep{s}{\roleq}}{\rolep}{x}{Q} \parop \procs{R}}
      \red
      \pres{s}{P \parop Q\subst{v}{x}	\parop \procs{R}}
    } \defrule[rm-channel]{}
    \and
    \inferrule[rm-pick]{ }{
      \pres{s}{\pobranch[i\in I]{\ep{s}\role}\roleq{\Tag_i}{P_i} \parop \procs{Q}} 
      \red
      \pres{s}{\pobranch{\ep{s}\role}\roleq{\Tag_k}{P_k}\parop \procs{Q}}
    }
    ~ k\in I \defrule[rm-pick]{}
    \and
    \inferrule[rm-tag]{ }{
      \pres{s}{\pobranch{\ep{s}\role}{\roleq}{\Tag_k}{P} \parop \pibranch[i\in I]{\ep{s}\roleq}\rolep{\Tag_i}{Q_i} \parop \procs{R}} 
      \red
      \pres{s}{P \parop Q_k \parop \procs{R}}
    }
    ~ k\in I \defrule[rm-tag]{}
    \and
    \inferrule[rm-par]{
      P \red Q
    }{
      \pres{s}{P \parop \procs{R}} \red \pres{s}{Q \parop \procs{R}}
    } \defrule[rm-par]{}
    \and
    \inferrule[rm-cast]{
      P \red Q
    }{
      \pcast{u}{P} \red \pcast{u}{Q}
    } \defrule[rm-cast]{}
    \and
    \inferrule[rm-struct]{
      P \pcong P'
      \\
      P' \red Q'
      \\
      Q' \pcong Q
    }{
      P \red Q
    } \defrule[rm-struct]{}
  \end{mathpar}
  }
  \caption{Reduction of processes}
  \label{fig:red_multi}
\end{figure}
The operational semantics of processes is given by the structural precongruence
relation $\pcong$ defined in \Cref{fig:pcong_multi} and the reduction relation $\red$
defined in \Cref{fig:red_multi}. As usual, structural precongruence allows us to
rearrange the structure of processes without altering their meaning, whereas
reduction expresses an actual computation or interaction step.
The adoption of a structural \emph{pre}congruence (as opposed to a more common
congruence relation) is not strictly necessary, but it simplifies the technical
development by reducing the number of cases we have to consider in proofs
without affecting the properties of the calculus in any way (see \Cref{rmk:pcong}).

Rules \refrule{sm-par-comm} and \refrule{sm-par-assoc} state commutativity and
associativity of parallel composition of processes (we write $\seqof{P}$ to
denote possibly empty parallel compositions of processes). In
\refrule{sm-par-assoc}, the side condition $s \in \fn{R}$ makes sure that $R$ is
indeed a participant of the session $s$. 
We write $\forall Q. s \not\in \fn{Q}$
to state that $s$ is not free in each of the processes $\procs{Q}$.
Moreover, note that this rule only states
right-to-left associativity. Left-to-right associativity is derivable from this
rule and repeated uses of \refrule{sm-par-comm}.
Rule \refrule{sm-cast-comm} allows us to swap two consecutive casts.
Rule \refrule{sm-cast-new} removes an unguarded cast on an endpoint of the
restricted session (we refer to this operation as ``performing the cast'').
Rule \refrule{sm-cast-swap} swaps a cast and a restricted session as long as the
endpoint in the cast refers to a different session.
Finally, rule \refrule{sm-call} unfolds a process invocation to its definition.
Hereafter, we write $\subst{u}{x}$ for the capture-avoiding substitution of each
free occurrence of $x$ with $u$ and $\subst{\seqof{u}}{\seqof{x}}$ for its
natural extension to equal-length tuples of variables and names.
The rules \refrule{sm-cast-new}, \refrule{sm-cast-swap} and \refrule{sm-call} are
not invertible: by \refrule{sm-cast-new} casts can only be removed but never
added; by \refrule{sm-cast-swap} casts can only be moved closer to their
restriction, so that they can be eventually performed by \refrule{sm-cast-new};
by \refrule{sm-call} process invocations can only be unfolded.

The reduction relation is quite standard.
Rule \refrule{rm-choice} reduces $P_1\pchoice P_2$ to either $P_1$ or $P_2$, non
deterministically.
Rule \refrule{rm-signal} terminates a session in which all participants
($\roleq_1,\ldots,\roleq_n$) but one ($\rolep$) are sending a termination signal
and $\rolep$ is waiting for it; the resulting process is the continuation of the
participant $\rolep$.
Rule \refrule{rm-channel} models the exchange of a channel among two participants
of a session.
Rule \refrule{rm-pick} models an internal choice whereby a process picks one
particular tag $\Tag_k$ to send on a session.
Rule \refrule{rm-tag} synchronizes two participants $\rolep$ and $\roleq$ on
the tag chosen by $\rolep$.
Finally, rules \refrule{rm-par}, \refrule{rm-cast} and \refrule{rm-struct} close
reductions under parallel compositions and casts and by structural precongruence. 

%% file: ts-multi/proc-ex.tex
\beginbass
In the rest of this section we illustrate the main features of the calculus with
some examples. For none of them the existing multiparty session type systems are
able to guarantee progress.
First of all, we formally define in our calculus the processes corresponding to 
(a slightly different) \actor{buyer}, \actor{seller} and \actor{carrier} from \Cref{ex:bsc} that was only informally 
presented. 

\begin{example}[Buyer - Seller - Carrier]
  \label{ex:bsc_multi}
  Consider the following definitions:
  \begin{align*}
    \Main & \peq \pres\sn{ \pinvk\Buyer{\ep\sn\buyer} \ppar \pinvk\Seller{\ep\sn\seller} \ppar \pinvk\Carrier{\ep\sn\carrier}} \\ 
    \Buyer(x) & \peq \act{x}\seller\oact\set{
    			  \tadd.\act{x}\seller\oact\tadd.\pinvk\Buyer{x},
                  \tpay.\pclose{x}
                } \\ 
    \Seller(x) & \peq \act{x}\buyer\iact\set{
                  \tadd.\pinvk\Seller{x},
                  \tpay.\act{x}\carrier\oact\tship.\pclose{x} 
                } \\ 
    \Carrier(x) & \peq \act{x}\seller\iact\tship.\pwait{x}\pdone
  \end{align*}
  Note that the buyer either sends $\tpay$ or it sends two $\tadd$ messages in a
  row before repeating this behavior. That is, this particular buyer always adds
  an even number of items to the shopping cart.
  Nonetheless, the buyer periodically has a chance to send a $\tpay$ message and
  terminate. Therefore, the execution of the program in which the buyer only
  sends $\tadd$ is unfair according to \Cref{def:fair_run} hence this program is
  fairly terminating. 
  \eoe
\end{example}

\begin{example}[Purchase with negotiation]
  \label{ex:2bsc_multi}
  Consider a variation of \Cref{ex:bsc} in which the buyer, before making the
  payment, negotiates with a secondary buyer for an arbitrarily long time. The
  interaction happens in two nested sessions, an outer one involving the primary
  buyer, the seller and the carrier, and an inner one involving only the two
  buyers. We model the interaction as the program below, in which we collapse
  role names to their initials.
    \begin{align*}
    \Main & \peq \pres\sn{ \pinvk\Buyer{\ep\sn\rbuyer} \ppar \pinvk\Seller{\ep\sn\rseller} \ppar \pinvk\Carrier{\ep\sn\rcarrier} }
    \\ 
    \Buyer(x) & \peq \act{x}\rseller\oact\tquery.
                \act{x}\rseller\iact\tprice.
                \pres{t}{ \pinvk{\Buyer_1}{x,\ep\asn{\rbuyer_1}} \ppar \pinvk{\Buyer_2}{\ep\asn{\rbuyer_2}}} 
    \\ 
    \Seller(x) & \peq \act{x}\rbuyer\iact\tquery.
                \act{x}\rbuyer\oact\tprice.
                \act{x}\rbuyer\iact\{
                	\begin{lines}
                		\tpay.\act{x}\rcarrier\oact\tship.\pclose{x}, \\
                		\tcancel.\act{x}\rcarrier\oact\tcancel.\pclose{x} \}
                	\end{lines}
    \\     
    \Carrier(x) & \peq \act{x}\rseller\iact\set{
                  \tship.\act{x}\rbuyer\oact\tbox.\pclose{x},
                  \tcancel.\pclose{x} 
                }
    \\
    \Buyer_1(x,y) & \peq \act{y}{\rbuyer_2}\oact\{
                      \begin{lines}
                        \tsplit.\act{y}{\rbuyer_2}\iact\{
                          \begin{lines}
                            \tyes.\pcast{x}
                                  \act{x}\rseller\oact\tok.
                                  \act{x}\rcarrier\iact\tbox.
                                  \pwait{x}
                                  \pwait{y}
                                  \pdone,
                                  \\
                            \tno.\pinvk{\Buyer_1}{x,y} \},
                          \end{lines}
                        \\
                        \tgiveup.
                          \pwait{y}
                          \pcast{x}
                          \act{x}\rseller\oact\tcancel.
                          \pwait{x}
                          \pdone \}
                      \end{lines}
    \\
    \Buyer_2(y) & \peq \act{y}{\rbuyer_1}\iact\set{
                    \tsplit.\act{y}{\rbuyer_1}\oact\set{
                      \tyes.\pclose{y},
                      \tno.\pinvk{\Buyer_2}{y} 
                    },
                    \tgiveup.\pclose{y} 
                  }
  \end{align*} 

  The buyer queries the seller which replies with a price. At this point,
  $\Buyer$ creates a new session $t$ and forks as a primary buyer $\Buyer_1$ and
  a secondary buyer $\Buyer_2$. The interaction between the two sub-buyers goes
  on until either $\Buyer_1$ gives up or $\Buyer_2$ accepts its share of the
  price. In the former case, the primary buyer waits for the internal session to
  terminate and $\tcancel$s the order with the seller which, in turn, aborts the
  transaction with the carrier. In the latter case, the buyer confirms the order
  to the seller, which then instructs the carrier to $\tship$ a $\tbox$ to the
  buyer.
  
  Note that the outermost session $s$, taken in isolation, terminates in a
  bounded number of interactions, but its progress cannot be established without
  assuming that the innermost session $t$ terminates. In particular, if the two
  buyers keep negotiating forever, the seller and the carrier starve.
  However, the innermost session can terminate if $\Buyer_1$ sends $\tgiveup$ to
  $\Buyer_2$ or if $\Buyer_2$ sends $\tyes$ to $\Buyer_1$. Thus, the run in
  which the two buyers negotiate forever is unfair, the session $t$ fairly
  terminates and the session $s$ terminates as well.
  
  On the technical side, note that the definition of $\Buyer_1$ contains two
  casts on the variable $x$. As we will see in \cref{ex:2bsc-ts}, these casts
  are necessary for the typeability of $\Buyer_1$ to account for the fact that
  $x$ is used \emph{differently} in two distinct branches of the process.
  \eoe
\end{example}

\begin{example}[Parallel merge sort]
  \label{ex:pms_multi}
  To illustrate an example of program that creates an unbounded number of
  sessions we model a parallel version of the merge sort algorithm.
  \begin{align*}
    \Main & \peq \pres{s}{
                    \act{\ep{s}\rmaster}\rworker\oact\treq.
                    \act{\ep{s}\rmaster}\rworker\iact\tres.
                    \pwait{s}
                    \pdone \parop
                    \pinvk\Sort{\ep{s}\rworker}
                  } \\ 
    \Sort(x) & \peq \act{x}\rmaster\iact\treq.
    \\
    & (
                  \pres{t}{
                    \pinvk\Merge{x,\ep{t}\rmaster} \parop
                    \pinvk\Sort{\ep{t}{\rworker_1}} \parop
                    \pinvk\Sort{\ep{t}{\rworker_2}}
                  }
                  \pchoice
                  \act{x}\rmaster\oact\tres.\pclose{x}
                )
    \\ 
    \Merge(x,y) & \peq \act{y}{\rworker_1}\oact\treq.
                  \act{y}{\rworker_2}\oact\treq.
                  \act{y}{\rworker_1}\iact\tres.
                  \act{y}{\rworker_2}\iact\tres.
                  \pwait{y}
                  \act{x}\rmaster\oact\tres.
                  \pclose{x}
  \end{align*}

  The program starts as a single session $s$ in which a master $\rmaster$ sends
  the initial collection of data to the worker $\rworker$ as a $\treq$ message
  and waits for the $\tres$ult. The worker is modeled as a process $\Sort$ that
  decides whether to sort the data by itself (right branch of the choice in
  $\Sort$), in which case it sends the $\tres$ult directly to the master, or to
  partition the collection (left branch of the choice in $\Sort$). In the latter
  case, it creates a new session $t$ in which it sends $\treq$uests to two
  sub-workers $\rworker_1$ and $\rworker_2$, it gathers the partial $\tres$ults
  from them and gets back to the master with the complete $\tres$ult.

  Since a worker may always choose to start two sub-workers in a new session,
  the number of sessions that may be created by this program is unbounded. At
  the same time, each worker may also choose to complete its task without
  creating new sessions. So, while in principle there exists a run of this
  program that keeps creating new sessions forever, this run is unfair according
  to \Cref{def:fair_run}.
  \eoe
\end{example}

%% file: ts-multi/ts.tex
\beginalto
In this section we describe the type system for the calculus of multiparty
sessions of \Cref{sec:ts_multi_proc}. 
The typing judgments have the form $\wtp[n]\Ctx{P}$, meaning that the process
$P$ is well typed in the typing context $\Ctx$ and has rank $n$. As usual, the
\emph{typing context} is a map associating channels with session types (\Cref{ssec:multi}) and is
meant to contain an association for each name in $\fn{P}$. We write $u_1 : S_1,
\dots, u_n : S_n$ for the map with domain $\set{u_1,\dots,u_n}$ that associates
$u_i$ with $S_i$. Occasionally we write $\seqof{u : S}$ for the same context,
when the number and the specific associations are unimportant. We also assume
that endpoints occurring in a typing context have different session names. That
is, $\ep{s}\rolep, \ep{s}\roleq \in \dom\Ctx$ implies $\rolep = \roleq$. 
This constraint makes sure that each well-typed process plays exactly one role
in each of the sessions in which it participates. It is also a common assumption
made in all multiparty session calculi.
We use $\Ctx$ and $\CtxD$ to range over typing contexts, we write $\EmptyCtx$
for the empty context and $\Ctx,\CtxD$ for the union of $\Ctx$ and $\CtxD$ when
they have disjoint domains and disjoint sets of session names.
The \emph{rank} $n$ in a typing judgment estimates the number of sessions that
$P$ has to create and the number of casts that $P$ has to perform in order to
terminate. The fact that the rank is finite suggests that so is the effort
required by $P$ to terminate.


\subsection{Typing Rules}
\input{ts-multi/ts-rules}


\subsection{Examples}
\label{ssec:ts_multi_ex}
\input{ts-multi/ts-examples}

%% file: ts-multi/ts-rules.tex
\beginbass
\begin{figure}[t]
	\framebox[\textwidth]{
  \begin{mathpar}
    \inferrule[tm-done]{\mathstrut}{
          \wtp[n]\EmptyCtx\pdone
        } \defrule[tm-done]{}
        \and
        \inferrule[tm-call]{
          \wtp[n]{\seqof{u : S}}{P\subst{\seqof{u}}{\seqof{x}}}
        }{
          \wtp[n+m]{\seqof{u : S}}{\pinvk{A}{\seqof u}}
        }~ \tass{A}{\seqof{S}}{n}, \Definition{A}{\seqof x}{P} \defrule[tm-call]{}
        \and
        \inferrule[tm-wait]
        {
          \wtp[n]{\Ctx}{P}
        }{
          \wtp[n]{\Ctx, u : \End[\In]}{\pwait{u}{P}}
        } \defrule[tm-wait]{}
        \and
        \inferrule[tm-close]{\mathstrut}
        {
          \wtp[n]{u : \End[\Out]}{\pclose{u}}
        } \defrule[tm-close]{}
        \and
        \inferrule[tm-channel-in]{
        \wtp[n]{\Ctx, u : T, x : S}{P}
        }{
        \wtp[n]{\Ctx, u : \rolep\In{S}.T}{\pich{u}{\rolep}{x}{P}}
        } \defrule[tm-channel-in]{}
        \and
        \inferrule[tm-channel-out]{
        \wtp[n]{\Ctx, u : T}{P}
        }{
        \wtp[n]{\Ctx, u : \role\Out{S}.T , v : S}{\poch{u}{\rolep}{v}{P}}
        } \defrule[tm-channel-out]{}
        \\
        \inferrule[tm-tag]
        {
          \forall i\in I:
          \wtp[n]{\Ctx, u : S_i}{P_i}
        }{
          \textstyle
          \wtp[n]{
            \Ctx, u : \Tags\rolep\Pol \Tag_i.S_i
          }{
            \pbranch[i \in I]{u}{\role}{\Pol}{\Tag_i}{P_i}
          }
        } \defrule[tm-tag]{}
        \and
        \inferrule[tm-choice]{
          \wtp[n_1]\Ctx{P_1}
          \\
          \wtp[n_2]\Ctx{P_2}
        }{
          \wtp[n_k]\Ctx{P_1 \choice P_2}
        } ~ k \in \set{1,2}
        \and
        \inferrule[tm-cast]{
          \wtp[n]{\Ctx, u : T}{P}
        }{
          \wtp[m+n]{\Ctx, u : S}{\pcast{u}{P}}
        }
        ~ S \subt[m] T \defrule[tm-cast]{}
        \and
        \inferrule[tm-par]{
          \forall i\in\set{1,\dots,h}:
          \wtp[n_i]{\Ctx_i, \ep{s}{\role_i} : S_i}{P_i}
        }{
          \wtp[1+n_1+\cdots+n_h]{
            \Ctx_1, \dots, \Ctx_h
          }{
            \pres{s}{P_1 \parop \cdots \parop P_h}
          }
        }
        ~ \coherent\set{\Map{\role_i} S_i}_{i=1..h} \defrule[tm-par]{}
        \and
        \infercorule[com-tag]{
          \wtp[n]{\Ctx, u : S_k}{P_k}
        }{
          \textstyle
          \wtp[n]{\Ctx, u : \Tags\rolep\Pol \Tag_i.S_i}{\pbranch[i \in I]{u}{\role}{\Pol}{\Tag_i}{P_i}}
        }~ k \in I \defrule[com-tag]{}
        \and
        \infercorule[com-choice]{
          \wtp[n]{\Ctx}{P_k}
        }{
          \wtp[n]{\Ctx}{P_1 \pchoice P_2}
        } ~ k \in \set{1,2}  \defrule[com-choice]{}
  \end{mathpar}
  }
  \caption{Typing rules}
  \label{fig:ts_multi}
\end{figure}
The typing rules are shown in \Cref{fig:ts_multi} as a \emph{generalized inference
system} (see \Cref{sec:gis})
in which, we recall, the singly-lined rules are interpreted coinductively
and the called \emph{corules} are interpreted inductively. 
For more details about the interpretation, we refer to \Cref{sec:gis}.
We type check a program $\set{\Definition{A_i}{\seqof{x_i}}{P_i}}_{i\in I}$
under a global set of assignments $\set{\tass{A_i}{\seqof{S_i}}{n_i}}_{i\in I}$
associating each process name $A_i$ with a tuple of session types $\seqof{S_i}$,
one for each of the variables in $\seqof{x_i}$, and a rank $n_i$. The program is
well typed if $\wtp[n_i]{\seqof{x_i : S_i}}{P_i}$ is derivable for every $i\in
I$, establishing that the tuple $\seqof{S_i}$ corresponds to the way the
variables $\seqof{x_i}$ are used by $P_i$ and that $n_i$ is a feasible rank
annotation for $P_i$. We now describe the typing rules in detail.


The rule \refrule{tm-done} states that the terminated process is well typed in
the empty context, to make sure that no unused channels are left behind. Note
that $\pdone$ can be given any rank, since it performs no casts and it creates
no new sessions. 
The rule \refrule{tm-call} checks that a process invocation
$\pinvk{A}{\seqof{u}}$ is well typed by unfolding $A$ into the process
associated with $A$. The types associated with $\seqof{u}$ must match those of
the global assignment $\tass{A}{\seqof{S}}{n}$ and the rank of the process must
be no greater than that of the invocation. The potential mismatch between the
two ranks improves typeability in some corner cases.
The rules \refrule{tm-wait} and \refrule{tm-close} concern processes that exchange
termination signals. The channel being closed is consumed and, in the case of
\refrule{tm-wait}, no longer available in the continuation $P$. Again,
$\pclose{u}$ can be typed with any rank whereas the rank of $\pwait{u}P$
coincides with that of $P$.
The rules \refrule{tm-channel-in} and \refrule{tm-channel-out} deal with the
exchange of channels in a quite standard way. As in \Cref{ssec:ts_bin_rules}, 
note that the actual type of the
exchanged channel is required to coincide with the expected one. In particular,
no covariance or contravariance of input and output respectively is allowed.
Relaxing the typing rule in this way would introduce implicit applications of
subtyping that may compromise fair termination (see \Cref{ex:invariant_ch}). In our
type system, each application of subtyping must be explicitly accounted for as
we will see when discussing \refrule{tm-cast}.
Rule \refrule{tm-tag} deals with the exchange of tags. Channels that are not used for such
communication must be used in the same way in all branches, whereas the type of
the channel on which the message is exchanged changes accordingly. All branches
are required to have the same rank, which also corresponds to the rank of the
process. Unlike other presentations of this typing rule \citep{GayHole05}, we
require the branches in the process to be matched exactly by those in the type.
Again, this is to avoid implicit application of subtyping, which might
jeopardize fair termination. 
The rule \refrule{tm-choice} deals with non-deterministic choices and requires
both continuations to be well typed in the same typing context. The judgment in
the conclusion inherits the rank of one of the processes, typically the one with
minimum rank. As we will see in \Cref{ex:pms_ts}, this makes it possible to
model finite-rank processes that may create an unbounded number of sessions or
that perform an unbounded number of casts.

The rule \refrule{tm-cast} models the substitution principle induced by fair
subtyping: when $S \subt[m] T$ (\Cref{fig:fsub_multi}), a channel of type $S$ can be used where a
channel of type $T$ is expected or, in dual fashion \citep{Gay16}, a process
using $u$ according to $T$ can be used in place of a process using $u$ according
to $S$. To keep track of this cast, the rank in the conclusion is augmented by
the weight $m$ of the subtyping relation between $S$ and $T$.
Note that the typing rule guesses the target type of the cast.

Finally, the rule \refrule{tm-par} deals with session creation and parallel
composition. This rule is inspired to the \emph{multiparty cut} rule found in
linear logic interpretations of multiparty session
types \citep{CarboneLMSW16,CarboneMontesiSchurmannYoshida17} and provides a
straightforward way for enforcing deadlock freedom. Each process in the
composition must be well typed in a slice of the typing context augmented with
the endpoint corresponding to its role. The session map of the new session must
be coherent (\Cref{def:coherence}), implying that it fairly terminates. 
The rank of the composition is
one plus the aggregated rank of the composed processes, to account for the fact
that one more session has been created. Recall that coherence is a property
expressed on the LTS of session maps (\Cref{def:coherence}) in line with the
approach of \cite{ScalasYoshida19}.

The typing rules described so far are interpreted \emph{coinductively}. That is,
in order for a rank $n$ process $P$ to be well typed in $\Ctx$ there must be a
\emph{possibly infinite} derivation tree built with these rules and whose
conclusion is the judgment $\wtp[n]\Ctx{P}$. But in a generalized inference
system like the one we are defining, this is not enough to establish that $P$ is
well typed. In addition, it must be possible to find \emph{finite} derivation
trees for all of the judgments occurring in this possibly infinite derivation
tree using the discussed rules \emph{and possibly} the corules, which we are
about to describe. 
Since the additional derivation trees must be finite, all of their branches must
end up with an application of \refrule{tm-done} or \refrule{tm-close}, which are
the only axioms in \Cref{fig:ts_multi} corresponding to the only terminated processes
in \Cref{fig:proc_syntax_multi}. So, the purpose of these finite typing derivations is
to make sure that in every well-typed (sub-)process there exists a path that
leads to termination. On the one hand, this is a sensible condition to require
as our type system is meant to enforce fair process termination. On the other
hand, insisting that these finite derivations can be built using only the typing
rules discusses thus far is overly restrictive, for a process might have
\emph{one} path that leads to termination, but also alternative paths that lead
to (recursive) process invocations. In fact, all of the processes we have
discussed in \Cref{ex:bsc_multi,ex:2bsc_multi,ex:pms_multi} are structured like this. The two
corules \refrule{com-choice} and \refrule{com-tag} in \Cref{fig:ts_multi} establish
that, whenever a multi-branch process is dealt with, it suffices for \emph{one}
of the branches to lead to termination. A key detail to note in the case of
\refrule{com-choice} is that the rank of the non-deterministic choice coincides
with that of the branch that leads to termination. This makes sense recalling
that the rank associated with a process represents the overall effort required
for that process to terminate.

Let us recap the notion of well-typed process resulting from the typing rules of
\Cref{fig:ts_multi}.

\begin{definition}[Well-typed process]
  \label{def:wtp_multi}
  We say that $P$ is \emph{well typed} in the context $\Ctx$ and has rank $n$ if
  \begin{enumerate}
  \item There exists an arbitrary (possibly infinite) derivation tree obtained
  using the rules in \Cref{fig:ts_multi} and whose conclusion is
  $\wtp[n]\Ctx{P}$
  \item For each judgment in such tree there is a finite
  derivation obtained using the rules and the corules
  \end{enumerate}
\end{definition}

Now we can prove a strong soundness result for our type system, stating that 
well-typed, closed processes can always successfully terminate no matter how they reduce.
We analyze the proof in details in \Cref{sec:ts_multi_corr}.

\begin{theorem}[Soundness]
	\label{thm:ts_multi_sound}
	If $\wtp[n]\EmptyCtx P$ and $P \wred Q$, then $Q \wred\pcong \pdone$.
\end{theorem}

The implications of \Cref{thm:ts_bin_sound} that we presented in \Cref{sec:ts_bin_ts}
still hold. Notably, the proof schema of \Cref{thm:ts_multi_sound} follows that of \Cref{thm:ts_bin_sound}.

%% file: ts-multi/ts-examples.tex
\beginbass
We dedicate the rest of \Cref{sec:ts_multi_ts} to the analysis of some examples
that integrate all the features of the presented type system. We start from some
basic examples and then we move to more involved ones. First, in \Cref{ex:bsc_ts_multi} we take into
account our slightly different variant of the running example (\Cref{ex:bsc_multi}). 
For what concerns the problematic processes in \Cref{ssec:boundedness}, they are still
valid in the multiparty context (see \Cref{rm:boundedness_multi}). 
We use the rest of the examples to deal with the processes introduced in \Cref{ssec:proc_ex_multi}.


\begin{remark}[Boundedness]
\label{rm:boundedness_multi}
	All the problematic processes that we presented in \Cref{ssec:boundedness} are still valid
	in the multiparty scenario and can be dealt with using the techniques that we mentioned for
	the binary case. In particular
	\begin{description}
	\item[Action-boundedness.] Type system with corules.
	\end{description}
	\[
    A \peq A
    \qquad
    B \peq {B \pchoice B}
    \qquad
    C \peq {C \pchoice \pdone}
  \]
	\begin{description}
	\item[Session-boundedness.] Rule \refrule{tm-par} increases the rank by one.
	\end{description}
	\[
    A \peq
      \pres{s}{
        \act{\ep{s}{\rolep}}{\roleq}\oact\set{
          \Tag[a].\pclose{\ep{s}{\rolep}},
          \Tag[b].\pwait{\ep{s}{\rolep}}{A}
        }
        \parop
        \act{\ep{s}{\roleq}}{\rolep}\iact\set{
          \Tag[a].\pwait{\ep{s}{\roleq}}{A},
          \Tag[b].\pclose{\ep{s}{\roleq}}
        }
      }
  \]
	\begin{description}
	\item[Cast-boundedness.] Rule \refrule{tm-cast} increases the rank by the weight of the subtyping
		being applied.
	\end{description}
	\[
  	B(x) \peq \pcast{x}\act{x}\seller\oact\tadd.\pinvk{B}{x}
  \]
	\eor
\end{remark}


\begin{example}
  \label{ex:bsc_ts_multi} 
  Let us show some typing derivations for fragments of \Cref{ex:bsc_multi}. 
  Let $\S_b$, $S_s$ and $S_c$ be the types from \Cref{ex:bsc_ty_multi}.
  We collapse roles to their initials.
  Let $\S'_b = \rseller\Out\tadd\rseller\Out\tadd.\S'_b + \rseller\Out\tpay.\End[\Out]$.
  Concerning $\Buyer$, we obtain the infinite derivation
  \[
    \begin{prooftree}
      \[
        \[
          \mathstrut\smash\vdots
          \justifies
          \wtp[0]{
            x : \S'_b
          }{
            \pinvk\Buyer{x}
          }
          \using\refrule{tm-call}
        \]
        \justifies
        \wtp[0]{
          x : \rseller\Out\tadd.\S'_b
        }{
          \act{x}\rseller\oact\tadd.\pinvk\Buyer{x}
        }
        \using\refrule{tm-tag}
      \]
      \[
        \justifies
        \wtp[0]{
          x : \End[\Out]
        }{
          \pclose{x}
        }
        \using\refrule{tm-close}
      \]
      \justifies
      \wtp[0]{
        x : \S'_b
      }{
        \act{x}\rseller\oact\set{
          \tadd.\act{x}\rseller\oact\tadd.\pinvk\Buyer{x},
          \tpay.\pclose{x}
        }
      }
      \using\refrule{tm-tag}
    \end{prooftree}
  \]
  and, for each judgment in it, it is easy to find a finite derivation possibly
  using \refrule{com-tag}. Concerning $\Main$ we obtain
  \[
    \begin{prooftree}
      \[
        \mathstrut\smash\vdots
        \justifies
        \wtp[0]{
          \ep{s}\rbuyer : \S'_b
        }{
          \pinvk\Buyer{\ep{s}\rbuyer}
        }
        \using\refrule{tm-call}
      \]
      \[
        \smash\vdots
        \justifies
        \wtp[0]{
          \ep{s}\rseller : S_s
        }{
          \pinvk\Seller{\ep{s}\rseller}
        }
      \]
      \vdots
      \justifies
      \wtp[1]{
        \EmptyCtx
      }{
        \pres\sn{
          \pinvk\Buyer{\ep{s}\rbuyer} \ppar \pinvk\Seller{\ep{s}\rseller} \ppar \pinvk\Carrier{\ep{s}\rcarrier}
        }
      }
      \using\refrule{tm-par}
    \end{prooftree}
  \]
  where the application of \refrule{tm-par} is justified by the fact that
  $\Map\rbuyer{\S'_b} \parop \Map\rseller{S_s} \parop \Map\rcarrier{S_c}$ is coherent.
  We recall that $\S_b \subt[1] \S'_b$ (\Cref{ex:bsc_fair_sub}).
  No participant creates new sessions or performs casts, so they all have zero
  rank. The rank of $\Main$ is 1 since it creates the session $s$.
  \eoe
\end{example}


\begin{example}
\label{ex:2bsc-ts}
In this example we show that the process $\Buyer_1$ playing the role $\rbuyer_1$
in the inner session of \Cref{ex:2bsc_multi} is well typed. For clarity, we recall its
definition here: 
\[
Buyer_1(x,y) \peq \act{y}{\rbuyer_2}\oact\{
	\begin{lines}
	  \tsplit.\act{y}{\rbuyer_2}\iact\{
		\begin{lines}
			\tyes.\pcast{x}
				\act{x}\rseller\oact\tok.
				\act{x}\rcarrier\iact\tbox.
				\pwait{x}
				\pwait{y}
				\pdone,
				\\
			\tno.\pinvk{Buyer_1}{x,y} \},
		\end{lines}
	  \\
	  \tgiveup.
		\pwait{y}
		\pcast{x}
		\act{x}\rseller\oact\tcancel.
		\pwait{x}
		\pdone \}
	\end{lines}
\]

We wish to build a typing derivation showing that $Buyer_1$ has rank $1$ and
uses $x$ and $y$ respectively according to $S$ and $T$, where $S =
\rseller\Out\tok.\rcarrier\In\tbox.\End[\In] + \rseller\Out\tcancel.\End[\In]$
and $T = \rbuyer_2\Out\tsplit.(\rbuyer_2\In\tyes.\End[\In] + \rbuyer_2\In\tno.T)
+ \rbuyer_2\Out\tgiveup.\End[\In]$.
As it has been noted previously, what makes this process interesting is that it
uses the endpoint $x$ differently depending on the messages it exchanges with
$\rbuyer_2$ on $y$. Since rule \refrule{tm-tag} requires any endpoint other
than the one on which messages are exchanged to have the same type, the only way
$\Buyer_2$ can be declared well typed is by means of the casts that occur in its
body.
For the branch in which $\Buyer_1$ proposes to $\tsplit$ the payment we obtain
the following derivation tree (we show only the $\tyes$ branch, the $\tno$ one is trivial):
\[
	\begin{prooftree}
		\[
			\[
				\[
					\[
						\[
							\[
								\justifies
								\wtp[0]\EmptyCtx\pdone
								\using\refrule{tm-done}
							\]
							\justifies
							\wtp[0]{
								y : \End[\In]
							}{
								\pwait{y}\pdone
							}
							\using\refrule{tm-wait}
						\]
						\justifies
						\wtp[0]{
							x : \End[\In],
							y : \End[\In]
						}{
							\pwait[\dots]{x}
						}
						\using\refrule{tm-wait}
					\]
					\justifies
					\wtp[0]{
						x : \rcarrier\In\tbox.\End[\In],
						y : \End[\In]
					}{
						\act{x}\rcarrier\iact\tbox\dots
					}
					\using\refrule{tm-tag}
				\]
				\justifies
				\wtp[0]{
					x : \rseller\Out\tok.\rcarrier\In\tbox.\End[\In],
					y : \End[\In]
				}{
					\act{x}\rseller\oact\tok\dots
				}
				\using\refrule{tm-tag}
			\]
			\justifies
			\wtp[1]{
				x : S,
				y : \End[\In]
			}{
				\pcast{x}\dots
			}
			\using\refrule{tm-cast}
		\]
		\vdots
		\justifies
		\wtp[1]{
			x : S,
			y : \rbuyer_2\In\tyes.\End[\In] + \rbuyer_2\In\tno.T
		}{
			\act{y}{\rbuyer_2}\iact\set{\tyes\dots, \tno\dots}
		}
		\using\refrule{tm-tag}
	\end{prooftree}
\]

Note how the application of \refrule{tm-cast} is key to change the type of $x$ in
the branch where the proposed split is accepted by $\rbuyer_2$. In that branch,
$x$ is deterministically used to send an $\tok$ message and we leverage on the
fair subtyping relation $S \subt[1] \rseller\Out\tok.\rcarrier\In\tbox.\End[\In]$.

For the branch in which $\Buyer_1$ sends $\tgiveup$ we obtain the following
derivation tree:
\[
	\begin{prooftree}
		\[
			\[
				\[
					\[
						\justifies
						\wtp[0]\EmptyCtx{
							\pdone
						}
						\using\refrule{tm-done}
					\]
					\justifies
					\wtp[0]{
						x : \End[\In]
					}{
						\pwait{x}\pdone
					}
					\using\refrule{tm-wait}
				\]
				\justifies
				\wtp[0]{
					x : \rseller\Out\tcancel.\End[\In]
				}{
					\act{x}\rseller\oact\tcancel.
					\pwait{x}
					\pdone
				}
			\]
			\justifies
			\wtp[1]{
				x : S
			}{
				\pcast{x}
				\act{x}\rseller\oact\tcancel.
				\pwait{x}
				\pdone
			}
			\using\refrule{tm-cast}
		\]
		\justifies
		\wtp[1]{
			x : S,
			y : \End[\In]
		}{
			\pwait{y}
			\pcast{x}
			\act{x}\rseller\oact\tcancel.
			\pwait{x}
			\pdone
		}
		\using\refrule{tm-wait}
	\end{prooftree}
\]

Once again the cast is necessary to change the type of $x$, but this time
leveraging on the fair subtyping relation $S \subt[1]
\rseller\Out\tcancel.\End[\In]$.
These two derivations can then be combined to complete the proof that the body
of $\Buyer_1$ is well typed:
\[
	\begin{prooftree}
		\qquad
		\mathstrut\smash\vdots
		\qquad
		\qquad
		\qquad
		\smash\vdots
		\qquad
		\justifies
		\wtp[1]{
			x : S,
			y : T
		}{
			\act{y}{\rbuyer_2}\oact\set{\tsplit\dots, \tgiveup\dots}
		}
		\using\refrule{tm-tag}
	\end{prooftree}
\]

Clearly, it is also necessary to find finite derivation trees for all of the
judgments shown above. This can be easily achieved using the corule
\refrule{com-tag}.
\eoe
\end{example}

\begin{example}
	\label{ex:non-det}
	Casts can be useful to reconcile the types of a channel that is used
	differently in different branches of a non-deterministic choice. For
	example, below is an alternative modeling of $\Buyer$ from \Cref{ex:bsc_multi}
	where we abbreviate $\role[\seller]$ to $\rseller$ for convenience:
	\[
		\Definition{B}{x}{
			\pcast{x}
			\act{x}\rseller\oact\tadd.
			\act{x}\rseller\oact\tadd.\pinvk{B}{x}
			\pchoice
			\pcast{x}
			\act{x}\rseller\oact\tpay.
			\pclose{x}
		}
	\]

	Note that $x$ is used for sending two $\tadd$ messages in the left branch of
	the non-deterministic choice and for sending a single $\tpay$ message in the
	right branch. Given the session type $S = \rseller\Out\tadd.S +
	\rseller\Out\tpay.\End[\Out]$ and using the fair subtyping relations $S
	\subt[2] \rseller\Out\tadd.\rseller\Out\tadd.S$ and $S \subt[1]
	\rseller\Out\tpay.\End[\Out]$ we can obtain the following typing derivation
	for the body of $B$ (we show only the left branch as the right one contains a
	straightforward application of $S \subt[1] \rseller\Out\tpay.\End[\Out]$):
	\[
		\begin{prooftree}
			\[
				\[
					\[
						\[
							\mathstrut\smash\vdots
							\justifies
							\wtp[1]{
								x : S
							}{
								\pinvk{B}{x}
							}
							\using\refrule{tm-call}
						\]
						\justifies
						\wtp[1]{
							x : \rseller\Out\tadd.S
						}{
							\act{x}\rseller\oact\tadd.\pinvk{B}{x}
						}
						\using\refrule{tm-tag}
					\]
					\justifies
					\wtp[1]{
						x : \rseller\Out\tadd.\rseller\Out\tadd.S
					}{
						\act{x}\rseller\oact\tadd.
						\act{x}\rseller\oact\tadd.\pinvk{B}{x}
					}
					\using\refrule{tm-tag}
				\]
				\justifies
				\wtp[3]{
					x : S
				}{
					\pcast{x}
					\act{x}\rseller\oact\tadd.
					\act{x}\rseller\oact\tadd.\pinvk{B}{x}	
				}
				\using\refrule{tm-cast}
			\]
			\vdots
			\justifies
			\wtp[1]{
				x : S
			}{
				\pcast{x}
				\act{x}\rseller\oact\tadd.
				\act{x}\rseller\oact\tadd.\pinvk{B}{x}
				\pchoice
				\pcast{x}
				\act{x}\rseller\oact\tpay.
				\pclose{x}	
			}
			\using\refrule{tm-choice}
		\end{prooftree}
	\]
	In general, the transformation $\pbranch[i=1..n]{u}{\role}\oact{\Tag_i}{P_i}
	\leadsto \pcast{u}\act{u}{\role}\oact{\Tag_1}.P_1 \pchoice \cdots \pchoice
	\pcast{u}\act{u}\role\oact{\Tag_n}.P_n$ does not always preserve typing, so
	it is not always possible to encode the output of tags using casts and
	non-deterministic choices. As an example, the definition
	\[
		\Definition\SlotMachine{x}{
			\act{x}\rplayer\iact\set{
				\tplay.\act{x}\rplayer\oact\set{
					\twin.\pinvk\SlotMachine{x},
					\tlose.\pinvk\SlotMachine{x}
				},
				\tquit.\pclose{x}
			}
		}
	\]
	implements the unbiased slot machine of \Cref{ex:slot_fair_sub} that waits
	for a message indicating whether a $\rplayer$ wants to $\tplay$ another game or
	to $\tquit$ (we assign role $\rplayer$ to the player). 
	In the former case, the slot machine notifies $\rplayer$ of the
	outcome (either $\twin$ or $\tlose$).
	It is easy to see that $\SlotMachine$ is well typed under the global type
	assignment $\tass\SlotMachine{T}{0}$ where $T =
	\rplayer\In\tplay.(\rplayer\Out\twin.T + \rplayer\Out\tlose.T) +
	\rplayer\In\tquit.\End[\Out]$. In particular, $\SlotMachine$ has rank $0$
	since it performs no casts and it creates no sessions. If we encode the tag
	output in $\SlotMachine$ using casts and non-deterministic choices we end up
	with the following process definition, which is ill typed because it cannot
	be given a finite rank:
	\[
		\Definition{\SlotMachine}{x}{
			\act{x}\rplayer\iact\set{
				\tplay.(
					\pcast{x}
					\act{x}\rplayer\oact\twin.
					\pinvk{\SlotMachine}{x}
					\pchoice
					\pcast{x}
					\act{x}\rplayer\oact\tlose.
					\pinvk{\SlotMachine}{x}
				),
				\tquit.\pclose{x}
			}
		}
	\]

	The difference between this version of $\SlotMachine$ and the above
	definition of $B$ is that $\SlotMachine$ always recurs after a cast, so it
	is not obvious that finitely many casts suffice in order for $\SlotMachine$
	to terminate. 
	\eoe
\end{example}


\begin{example} 
	\Cref{ex:2bsc-ts}
	shows that casts are essential in the type derivation.
	However, the process would be well typed if we considered a subtyping relation that does not preserve coherence
	\citep{GayHole05} for the involved types are finite.
	Now we refine the buyer from \Cref{ex:bsc_multi} in order to consider more involved sessions. Again, we
	collapse role names to their initials.
	\begin{align*}
		B(x) & \peq
			{\pcast{x}\pinvk{B_1}{x} \pchoice \pcast{x}\pinvk{B_2}{x}} 
		\\
		B_1(x) & \peq
			{\act{x}{\rseller}\oact\Tag[add].\act{x}{\rseller}\oact\set{
  			\Tag[add].\pinvk{B_1}{x},\, 
  			\Tag[pay].\pwait{x}{\pdone} 
			}} 
	  \\  
		B_2(x) & \peq
			{\act{x}{\rseller}\oact\set{
				\Tag[add].\act{x}{\rseller}\oact\Tag[add]. \pinvk{B_2}{x},\, 
  				\Tag[pay].\pwait{x}{\pdone} 
			}}
	\end{align*}
	$B_2$ corresponds to the buyer in \Cref{ex:bsc_multi} while $B_1$ is the acquirer that adds an odd number of items to the cart. 
	$B$ non deterministically chooses to behave according to $B_1$ or $B_2$.
	Let $S_{b_1}$ and $S_{b_2}$ be the session types such that 
	$x : S_{b_1}$ in $B_1$ and $x : S_{b_2}$ in $B_2$ respectively:
	\[
	\begin{array}{ll}
		S_{b_1} = \rseller\Out\Tag[add].(\rseller\Out\Tag[add].S_{b_1} + \rseller\Out\Tag[pay].\End[\Out])
		&
		S_{b_2} = \rseller\Out\Tag[add].\rseller\Out\Tag[add].S_{b_2} + \rseller\Out\Tag[pay].\End[\Out]
	\end{array}
	\]
	In \cref{ex:bsc_fair_sub} we showed that $S \subt[1] S_{b_2}$ where $S =\rseller\Out\tadd.S + \rseller\Out\tpay.\End[\Out]$
	models the acquirer that adds arbitrarily many items to the cart. Analogously, we can prove that $S \subt[2] S_{b_1}$.
	Hence we derive
	\[
	\begin{prooftree}
		\[
			\[
				\vdots
				\justifies
				\wtp[0]{x : S_{b_1}}{\pinvk{B_1}{x}}
				\using\refrule{tm-call}
			\]
			\justifies
			\wtp[2]{x : S}{\pcast{x}\pinvk{B_1}{x}}
			\using\refrule{tm-cast}
		\]
		\[
			\[
				\vdots
				\justifies
				\wtp[0]{x : S_{b_2}}{\pinvk{B_2}{x}}
				\using\refrule{tm-call}
			\]
			\justifies
			\wtp[1]{x : S}{\pcast{x}\pinvk{B_2}{x}}
			\using\refrule{tm-cast}
		\]
		\justifies
		\wtp[1]
			{x : S}
			{\pcast{x}\pinvk{B_1}{x} \pchoice \pcast{x}\pinvk{B_2}{x}}
		\using\refrule{tm-choice}
	\end{prooftree}
	\]
	Again, the casts are crucial to obtain the type derivation of process $B$ because rule 
	\refrule{tm-choice} requires that $B_1$ and $B_2$ are typed in the same context.
	Note that $B_1$ and $B_2$ are typed with rank 0 since no sessions are created and no casts are performed by the processes.
\end{example}


\begin{example}
	\label{ex:pms_ts}
	Here we provide evidence that the process definitions in \Cref{ex:pms_multi} are
	well typed, even if they model processes that can open arbitrarily many
	sessions. In that example, the most interesting process definition is that
	of the worker $\Sort$, which is recursive and may create a new session. In
	contrast, $\Merge$ is finite and $\Main$ only refers to $\Sort$. We claim
	that these process definitions are well typed under the global type
	assignments
	\[
		\tass\Main{}{1}
		\qquad
		\tass\Sort{U}{0}
		\qquad
		\tass\Merge{T, V}{0}
	\]
	where 
	\[
	\begin{array}{lll}
		T = \rmaster\Out\tres.\End[\Out]
		&
		U = \rmaster\In\treq.T
		&
		V = \rworker_1\Out\treq.\rworker_2\Out\treq.\rworker_1\In\tres.\rworker_2\In\tres.\End[\In]
	\end{array}
	\]
	For the branch of $\Sort$ that creates a new session we obtain the
	derivation tree
	\[
		\begin{prooftree}
			\[
				\mathstrut\smash\vdots
				\justifies
				\wtp[0]{
					x : T,
					\ep{t}\rmaster : V
				}{
					\pinvk\Merge{x,\ep{t}\rmaster}
				}
				\using\refrule{tm-call}
			\]
			\[
				\smash\vdots
				\justifies
				\wtp[0]{
					\ep{t}{\rworker_i} : U
				}{
					\pinvk\Sort{\ep{t}{\rworker_i}}
				}
			\]
			\justifies
			\wtp[1]{
				x : T
			}{
				\pres{t}{
					\pinvk\Merge{x,\ep{t}\rmaster} \parop
					\pinvk\Sort{\ep{t}{\rworker_1}} \parop
					\pinvk\Sort{\ep{t}{\rworker_2}}
				}
			}
			\using\refrule{tm-par}
		\end{prooftree}
	\]
	where $i=1,2$. The rank $1$ derives from the fact that the created session involves
	three zero-ranked participants.
	For the body of $\Sort$ we obtain the following derivation tree:
	\[
		\begin{prooftree}
			\[
				\[
					\smash\vdots
					\justifies
					\wtp[1]{
						x : T
					}{
						\pres{t}{
							\pinvk\Merge{x,\ep{t}\rmaster} \parop
							\cdots
						}
					}
					\using\refrule{tm-par}
				\]
				\[
					\[
						\justifies
						\wtp[0]{
							x : \End[\Out]
						}{
							\pclose{x}
						}
						\using\refrule{tm-close}
					\]
					\justifies
					\wtp[0]{
						x : T
					}{
						\act{x}\rmaster\oact\tres\dots
					}
					\using\refrule{tm-tag}
				\]
				\justifies
				\wtp[0]{
					x : T
				}{
					\pres{t}{
						\pinvk\Merge{x,\ep{t}\rmaster} \parop
						\cdots
					}
					\pchoice
					\act{x}\rmaster\oact\tres\dots
				}
				\using\refrule{tm-choice}
			\]
			\justifies
			\wtp[0]{
				x : U
			}{
				\act{x}\rmaster\iact\treq.(
					\pres{t}{
						\pinvk\Merge{x,\ep{t}\rmaster} \parop
						\cdots
					}
					\pchoice
					\act{x}\rmaster\oact\tres\dots
				)
			}
			\using\refrule{tm-tag}
		\end{prooftree}
	\]
	
	In the application of the rule \refrule{tm-choice}, the rank of the whole
	choice coincides with that of the branch in which no new sessions are
	created. This way we account for the fact that, even though $\Sort$
	\emph{may} create a new session, it does not \emph{have to} do so in order
	to terminate.
	\eoe
\end{example}

%% file: ts-multi/soundness.tex
\beginalto
In this section we discuss the proof of \Cref{thm:ts_multi_sound}.
The proof technique follows that of \Cref{thm:ts_bin_sound}.
Such proof is essentially composed of a standard subject
reduction result showing that typing is preserved by reductions and a proof that
every well-typed process other than $\pdone$ may always reduce in such a way
that a suitably defined \emph{well-founded measure} strictly decreases. The
measure is a lexicographically ordered pair of natural numbers with the
following meaning: the first component measures the number of sessions that must
be created and the total weight of casts that must be performed in order for the
process to terminate (this information is essentially the rank we associate with
typing judgments); the second component measures the overall effort required to
terminate every session that has already been created (these sessions are
identified by the fact that their restriction occurs unguarded in the process).
We account for this effort by measuring the shortest reduction that terminates a
coherent session map (\Cref{def:coherence}). The reason why we need two
quantities in the measure is that in general every application of fair subtyping
may \emph{increase} the length of the shortest reduction that terminates a
coherent session map. So, when casts are performed the second component of the
measure may increase, but the first component reduces.
As a final remark, it should be noted that the overall measure associated with a
well-typed process \emph{may also increase}, for example if new sessions are
created (\Cref{ex:pms_multi}). However, one particular reduction that decreases the
measure is always guaranteed to exist.
For the sake of clarity, in the following we often write $M \ft$ instead of $\coherent M$.


\subsection{Subject Reduction}
\input{ts-multi/proof-sr}


\subsection{Measure}
\input{ts-multi/proof-measure}


\subsection{Normal Forms}
\input{ts-multi/proof-nf}


\subsection{Soundness}
\input{ts-multi/proof-soundness}

%% file: ts-multi/proof-sr.tex
\beginbass
We show the proof of a standard subject reduction theorem (see \Cref{lem:subj_red_multi}). For this sake,
we need an additional result, that we dub subject congruence,
stating that well-typedness is preserved by the structural precongruence
relation for processes in \Cref{fig:pcong_multi} (\Cref{lem:subj_cong_multi}).
Notably, \Cref{lem:subj_cong_multi} tells that the rank does not incrase.
Such result is not provable in \Cref{lem:subj_red_multi} as the non deterministic choice
can reduce to a branch that increases the rank.

\begin{lemma}
\label{lem:substitution_multi}
	If $\wtp[n]{\Ctx, x : S}{P}$ and $\Ctx, u : S$ is defined, then $\wtp[n]{\Ctx, u : S}{P \subst{u}{x}}$.
	A typing context is \emph{defined} if the endpoints occurring in it all have different session names.
\end{lemma}
\begin{proof}
By bounded coinduction (see \Cref{prop:bcp}).
\end{proof}

\begin{lemma}[Subject Congruence]
\label{lem:subj_cong_multi}
	If\/ $\wtp[n] \Ctx {P}$ and $P \pcong Q$, then $\wtp[m] \Ctx {Q}$ for some $m \leq n$.
\end{lemma}
\begin{proof}
By induction on the derivation of $P \pcong Q$ and by cases on the last rule applied.

\proofrule{sm-par-comm} 
Then $P = \pres{s}{\procs{P} \ppar P' \ppar Q' \ppar \procs{Q}} \pcong \pres{s}{\procs{P} \ppar Q' \ppar P' \ppar \procs{Q}} = Q$.
From rule \refrule{tm-par} we deduce that there exist $\Ctx_i, \role_i, S_i, n_i$ for $i = 1,\dots,h$ such that
\begin{itemize}
\item $\Ctx = \Ctx_1,\dots,\Ctx_h$
\item $n = 1 + \sum_{i=1}^h n_i$
\item $\prod_{i=1}^h \Map{\role_i}{S_i} \ft$
\item $\wtp[n_i]{\Ctx_i, \ep{s}{\role_i} : S_i}{P_i}$ for $i = 1,\dots,k$
\item $\wtp[n_{k+1}]{\Ctx_{k+1}, \ep{s}{\role_{k+1}} : S_{k+1}}{P'}$
\item $\wtp[n_{k+2}]{\Ctx_{k+2}, \ep{s}{\role_{k+2}} : S_{k+2}}{Q'}$
\item $\wtp[n_i]{\Ctx_i, \ep{s}{\role_i} : S_i}{Q_i}$ for $i = k+3,\dots,h$
\end{itemize}

We conclude $\wtp[m]\Ctx{Q}$ with one application of \refrule{tm-par} by taking $m \eqdef n$.

\proofrule{sm-par-assoc}
Then $P = \pres{s}{\procs{P} \ppar \pres{t}{R \ppar \procs{Q}}} \pcong \pres{t}{\pres{s}{\procs{P} \ppar R} \ppar \procs{Q}} = Q$ and $s \in \fn{R}$.
From rule \refrule{tm-par} we deduce that there exist $\Ctx_i, \role_i, S_i, n_i$ for $i = 1,\dots,h$ such that
\begin{itemize}
\item $\Ctx = \Ctx_1,\dots,\Ctx_h$
\item $n = 1 + \sum_{i=1}^h n_i$
\item $\prod_{i=1}^h \Map{\role_i}{S_i} \ft$
\item $\wtp[n_i]{\Ctx_i, \ep{s}{\role_i} : S_i}{P_i}$ for $i = 1,\dots,h - 1$
\item $\wtp[n_h]{\Ctx_h, \ep{s}{\role_h} : S_h}{\pres{t}{R \ppar \procs{Q}}}$
\end{itemize}
From rule \refrule{tm-par} and the hypothesis that $s \in \fn{R}$ we deduce that there exist $\CtxD_i, \roleq_i, T_i, m_i$ for $i = 1,\dots,k$ such that
\begin{itemize}
\item $\Ctx_h = \CtxD_1,\dots,\CtxD_k$
\item $n_h = 1 + \sum_1^k m_i$
\item $\prod_1^k \Map{\roleq_i}{T_i} \ft$
\item $\wtp[m_1]{\CtxD_1, \ep{s}{\role_h} : S_h, \ep{t}{\roleq_1} : T_1}{R}$
\item $\wtp[m_{i+1}]{\CtxD_{i+1}, \ep{t}{\roleq_{i+1}} : T_{i+1}}{Q_i}$ for $i = 1,\dots,k-1$
\end{itemize}
Using \refrule{tm-par} we deduce 
$\wtp[1 + \sum_{i=1}^{h-1}{n_i} + m_1]{\Ctx_1,\dots,\Ctx_{h-1},\CtxD_1, \ep{t}{\roleq_1} : T_1}{\pres{s}{\procs{P} \ppar R}}$.
We conclude $\wtp[m]{\Ctx}{\pres{t}{\pres{s}{\procs{P} \ppar R} \ppar \procs{Q}}}$ with another application of \refrule{tm-par} by taking $m \eqdef n$.

\proofrule{sm-cast-comm} 
Then $P = \pcast{u}{\pcast{v}{R}} \pcong \pcast{v}{\pcast{u}{R}} = Q$. We can assume $u \ne v$ or else $P = Q$.
From rule \refrule{tm-cast} we deduce that there exist $\Ctx_1, S, T, n_1, n_u$ such that
\begin{itemize}
\item $\Ctx = \Ctx_1, u : S$
\item $S \subt[n_u] T$
\item $n = n_u + n_1$
\item $\wtp[n_1]{\Ctx_1, u : T}{\pcast{v}{R}}$
\end{itemize} 
From rule \refrule{tm-cast} we deduce that there exist $\Ctx_2, S', T', n_2, n_v$ such that
\begin{itemize}
\item $\Ctx_1 = \Ctx_2, v : S'$
\item $S' \subt[n_v] T'$
\item $n_1 = n_v + n_2$
\item $\wtp[n_2]{\Ctx_2, u : T, v : T'}{R}$
\end{itemize}
We derive $\wtp[n_u + n_2]{\Ctx_2, u : S, v : T'}{\pcast{u}{R}}$ with one application of \refrule{tm-cast} 
and we conclude with another application of \refrule{tm-cast} by taking $m \eqdef n$.

\proofrule{sm-cast-new}
Then $P = \pres{s}{\pcast{\ep{s}{\role}}{R} \ppar \procs{P}} \pcong \pres{s}{R \ppar \procs{P}} = Q$.
From rule \refrule{tm-par} we deduce that there exist $\CtxD, n'$ and $\Ctx_i, \roleq_i, n_i$ for $i = 1,\dots,h$ such that
\begin{itemize}
\item $\Ctx = \CtxD, \Ctx_1, \dots, \Ctx_h$ for $i = 1,\dots,h$
\item $n = 1 + n' + \sum_{i=1}^h n_i$
\item $\Map{\role}{S} \ppar \prod_{i = 1}^h \Map{\roleq_i}{S_i} \ft$
\item $\wtp[n']{\CtxD, \ep{s}{\role} : S}{\pcast{\ep{s}{\role}}{R}}$
\item $\wtp[n_i]{\Ctx_i, \ep{s}{\roleq_i} : S_i}{P_i}$ for $i = 1,\dots,h$
\end{itemize} 
From rule \refrule{tm-cast} we deduce that there exist $T, m', m_s$ such that
\begin{itemize}
\item $S \subt[m_s] T$
\item $n' = m_s + m'$
\item $\wtp[m']{\CtxD, \ep{s}{\role} : T}{R}$
\end{itemize}
From $\Map{\role}{S} \ppar \prod_{i = 1}^h \Map{\roleq_i}{S_i} \ft$, $S \subt[m_s] T$ and \cref{def:ssubt} we deduce 
$\Map{\role}{T} \ppar \prod_{i = 1}^h \Map{\roleq_i}{S_i} \ft$.
We conclude with an application of \refrule{tm-par} by taking $m \eqdef 1 + m' + \sum_{i=1}^h n_i \le n$.

\proofrule{sm-cast-swap}
Then $P = \pres{s}{\pcast{\ep{t}{\role}}{R} \ppar \procs{P}} \pcong \pcast{\ep{t}{\role}}{\pres{s}{R \ppar \procs{P}}} = Q$ and $t \ne s$.
From rule \refrule{tm-par} we deduce that there exist $\Ctx_i, \roleq_i, n_i$ for $i = 1,\dots,h$ such that
\begin{itemize}
\item $\Ctx = \Ctx_1, \dots, \Ctx_h$ for $i = 1,\dots,h$
\item $n = 1 + \sum_{i=1}^h n_i$
\item $\prod_{i = 1}^h \Map{\roleq_i}{S_i} \ft$
\item $\wtp[n_1]{\Ctx_1, \ep{s}{\roleq_1} : S_1}{\pcast{\ep{t}{\role}}{R}}$
\item $\wtp[n_i]{\Ctx_i, \ep{s}{\roleq_i} : S_i}{P_i}$ for $i = 2,\dots,h$
\end{itemize} 
From rule \refrule{tm-cast} we deduce that there exist $\CtxD, T, n', m_t$ such that
\begin{itemize}
\item $\Ctx_1 = \CtxD, \ep{t}{\role} : S$
\item $S \subt[m_t] T$
\item $n_1 = m_t + n'$
\item $\wtp[n']{\CtxD, \ep{t}{\role} : T, \ep{s}{\roleq_1} : S_1}{R}$
\end{itemize}
We derive $\wtp[1 + n' + \sum_{i=2}^h n_i]{\CtxD, \ep{t}{\role} : T,\Ctx_2,\dots,\Ctx_h}{\pres{s}{R \ppar \procs{P}}}$ 
with an application of \refrule{tm-par}. We conclude with an application of \refrule{tm-cast} by taking $m \eqdef n$.

\proofrule{sm-call} 
Then $P = \pinvk{A}{\seqof u} \pcong R\subst{\seqof u}{\seqof x} = Q$ and
$\Definition{A}{\seqof x}{R}$.
From \refrule{tm-call} we conclude that there exist $\seqof S$ and $m$ such that
$\tass{A}{\seqof{S}}{m}$ and $\Ctx = \seqof{u : S}$ and $\wtp[m]{\seqof{u :
S}}{Q}$ and $m \leq n$.
\end{proof}

\begin{lemma}[Subject Reduction]
	\label{lem:subj_red_multi}
	If\/ $\wtp[n] \Ctx {P}$ and $P \red Q$, then $\wtp[m] \Ctx {Q}$ for some $m$.
\end{lemma}
\begin{proof}
By induction on the derivation of $P \red Q$ and by cases on the last rule applied.

\proofrule{rm-choice}
Then $P = P_1 \pchoice P_2 \red P_k = Q$ and $k \in \set{1,2}$.
From \refrule{tm-choice} we deduce that $\wtp[m]{\Ctx}{Q}$ for some $m$.

\proofrule{rm-signal}
Then $ P = \pres{s}{\pwait{\ep{s}{\role}}{Q} \parop \pclose{\ep{s}{\roleq_1}} \parop \cdots \parop \pclose{\ep{s}{\roleq_h}}} \red Q$.
From \refrule{tm-par}, \refrule{tm-wait} and \refrule{tm-close} we deduce that there exist $m$ and $n_i$ for $i=1,\dots,h$ such that
\begin{itemize}
\item $n = 1 + m + \sum_{i=1}^h n_i$
\item $\wtp[m]{\Ctx, \ep{s}{\role} : \End[\In]}{\pwait{\ep{s}{\role}}{Q}}$
\item $\wtp[m]{\Ctx}{Q}$
\item $\wtp[n_i]{\ep{s}{\roleq_i} : \End[\Out]}{\pclose{\ep{s}{\roleq_i}}}$ for
$i=1,\dots,h$
\end{itemize}
There is nothing left to prove.

\proofrule{rm-channel}
Then $P = \pres{s}{\poch{\ep{s}{\rolep}}{\roleq}{v}{P'} \parop
\pich{\ep{s}{\roleq}}{\rolep}{x}{Q'} \parop \procs{R}} \red \pres{s}{P' \parop
Q'\subst{v}{x} \parop \procs{R}} = Q$.
From \refrule{tm-par} we deduce that there exist $\Ctx_i, S_i, \role_i, n_i$ for $i=1,\dots,h$ such that
\begin{itemize}
\item $\Ctx = \Ctx_1,\dots,\Ctx_h$
\item $n = 1 + \sum_{i=1}^h n_i$
\item $\prod_{i=1}^h \Map{\role_i}{S_i} \ft$
\item $\role = \role_1$ and $\roleq = \role_2$
\item $\wtp[n_1]{\Ctx_1, \ep{s}{\role} : S_1}{\poch{\ep{s}{\rolep}}{\roleq}{v}{P'}}$
\item $\wtp[n_2]{\Ctx_2, \ep{s}{\roleq} : S_2}{\pich{\ep{s}{\roleq}}{\rolep}{x}{Q'}}$
\item $\wtp[n_i]{\Ctx_i, \ep{s}{\role_i} : S_i}{R_i}$ for $i = 3,\dots,h$
\end{itemize}
From \refrule{tm-channel-out} and \refrule{tm-channel-in} we deduce that there exist $S_v, T_1, T_2, \CtxD_1$ such that
\begin{itemize}
\item $S_1 = \roleq\Out{S_v}.T_1$
\item $\Ctx_1 = \CtxD_1, v : S_v$
\item $\wtp[n_1]{\CtxD_1, \ep{s}{\rolep} : T_1}{P'}$
\item $S_2 = \rolep\In{S_v}.T_2$
\item $\wtp[n_2]{\Ctx_2, \ep{s}{\roleq} : T_2, x : S_v}{Q'}$
\end{itemize} 
Using \cref{lem:substitution_multi} we deduce $\wtp[n_2]{\Ctx_2, \ep{s}{\roleq} : T_2,
v : S_v}{Q'\subst{v}{x}}$. Using \cref{def:coherence} we deduce $\Map{\rolep}{T_1}
\parop \Map{\roleq}{T_2} \parop \prod_{i=3}^h \Map{\role_i}{S_i} \ft$. We conclude with
one application of \refrule{tm-par} taking $m \eqdef n$.

\proofrule{rm-pick}
Then $P = \pres{s}{\pobranch[i\in I]{\ep{s}{\role}}{\roleq}{\Tag_i}{\PP_i} \parop \procs{Q}} \red 
\pres{s}{\pobranch{\ep{s}{\role}}{\roleq}{\Tag_k}{\PP_k}\parop \procs{Q}} = Q$ and $k \in I$. 
From \refrule{tm-par} we deduce that there exist $\Ctx_i, \role_i, n_i, S_i$ for $i=1,\dots,h$ such that
\begin{itemize}
\item $\Ctx = \Ctx_1, \dots, \Ctx_h$
\item $n = 1 + \sum_{i=1}^h n_i$
\item $\prod_{i=1}^h \Map{\role_i}{S_i} \ft$
\item $\role = \role_1$ and $\roleq = \role_i$ for some $i \in \set{2,\dots,h}$
\item $\wtp[n_1]{\Ctx_1, \ep{s}{\role} : S_1}{\pobranch[i\in I]{\ep{s}{\role}}{\roleq}{\Tag_i}{\PP_i}}$
\item $\wtp[n_i]{\Ctx_i, \ep{s}{\role_i} : S_i}{Q_i}$ for $i=2,\dots,h$
\end{itemize}
From \refrule{tm-tag} we deduce that there exist $T_i$ for all $i \in I$ such that
\begin{itemize}
\item $S_1 = \Tags\roleq\Out \Tag_i.T_i$
\item $\wtp[n_1]{\Ctx_1, \ep{s}{\role} : T_i}{P_i}~{}^{(i\in I)}$
\end{itemize}
From the hypothesis that $k \in I$ we deduce that $\wtp[n_1]{\Ctx_1, \ep{s}{\role} : T_k}{P_k}$ and from \refrule{tm-tag} we deduce 
$\wtp[n_1]{\Ctx_1, \ep{s}{\role} : \roleq\Out\Tag_k.T_k}{\pobranch{\ep{s}{\role}}{\roleq}{\Tag_k}{\PP_k}}$. 
From \cref{def:coherence} we deduce that $\roleq\Out\Tag_k.T_k \parop \prod_{i=2}^h \Map{\role_i}{S_i} \ft$. 
We conclude with an application of \refrule{tm-par} taking $m \eqdef n$.

\proofrule{rm-tag}
Then $P = \pres{s}{\pobranch{\ep{s}{\role}}{\roleq}{\Tag_k}{\PP'} \parop \pibranch[i\in I]{\ep{s}{\roleq}}{\role}{\Tag_i}{Q_i} \parop \procs{R}} 
        \red 
        \pres{s}{P' \parop Q_k \parop \procs{R}} = Q$ and $k \in I$.
From \refrule{tm-par} we deduce that there exist $\Ctx_i, S_i, \role_i, n_i$ for $i=1,\dots,h$ such that
\begin{itemize}
\item $\Ctx = \Ctx_1,\dots,\Ctx_h$
\item $n = 1 + \sum_{i=1}^h n_i$
\item $\prod_{i=1}^h \Map{\role_i}{S_i} \ft$
\item $\role = \role_1$ and $\roleq = \role_2$
\item $\wtp[n_1]{\Ctx_1, \ep{s}{\role} : S_1}{\pobranch{\ep{s}{\role}}{\roleq}{\Tag_k}{\PP'}}$
\item $\wtp[n_2]{\Ctx_2, \ep{s}{\roleq} : S_2}{\pibranch[i\in I]{\ep{s}{\roleq}}{\role}{\Tag_i}{Q_i}}$
\item $\wtp[n_i]{\Ctx_i, \ep{s}{\role_i} : S_i}{R_i}$ for $i = 3,\dots,h$
\end{itemize}
From \refrule{tm-tag} we deduce that there exist $S'_1$ and $T_i$ for every $i \in I$ such that
\begin{itemize}
\item $S_1 = \roleq\Out \Tag_k.S'_1$
\item $\wtp[n_1]{\Ctx_1, \ep{s}{\role} : S'_1}{P'}$
\item $S_2 = \Tags\rolep\In \Tag_i.T_i$
\item $\wtp[n_2]{\Ctx_2, \ep{s}{\roleq} : T_i}{Q_i}~{}^{(i\in I)}$
\end{itemize}
From \cref{def:coherence} we deduce that $\Map{\role}{S'_1} \parop \Map{\roleq}{T_k} \parop \prod_{i=2}^h \Map{\role_i}{S_i} \ft$. 
We conclude with an application of \refrule{tm-par} by taking $m \eqdef n$.

\proofrule{rm-par}
Then $P = \pres{s}{P' \parop \procs{R}} \red \pres{s}{Q' \parop \procs{R}} = Q$ and $P' \red Q'$.
From \refrule{tm-par} we deduce that there exist $\Ctx_i, \role_i, S_i, n_i$ for $i=1,\dots,h$ such that
\begin{itemize}
\item $\Ctx = \Ctx_1,\dots,\Ctx_h$
\item $n = 1 + \sum_{i=1}^h n_i$
\item $\prod_{i=1}^h \Map{\role_i}{S_i} \ft$
\item $\wtp[n_1]{\Ctx_1, \ep{s}{\role_1} : S_1}{P'}$
\item $\wtp[n_i]{\Ctx_i, \ep{s}{\role_i} : S_i}{R_i}$ for $i = 2,\dots, h$
\end{itemize}
Using the induction hypothesis on $\wtp[n_1]{\Ctx_1, \ep{s}{\role_1} : S_1}{P'}$ and $P' \red Q'$
we deduce $\wtp[n'_1]{\Ctx_1, \ep{s}{\role_1} : S_1}{Q'}$ for some $n'_1$. We conclude with an application 
of \refrule{tm-par} taking $m \eqdef 1 + n'_1 + \sum_{i=2}^h n_i$.

\proofrule{rm-cast}
Then $P = \pcast{u}{P'} \red \pcast{u}{Q' } = Q$ and $P' \red Q'$.
From \refrule{tm-cast} we deduce that there exist $S, T, \Ctx', n', m_u$ such that
\begin{itemize}
\item $\Ctx = \Ctx', u : S$
\item $S \subt[m_u] T$
\item $n = m_u + n'$
\item $\wtp[n']{\Ctx', u : T}{P'}$
\end{itemize}
Using the induction hypothesis on $\wtp[n']{\Ctx', u : T}{P'}$ and $P' \red Q'$ we deduce $\wtp[m']{\Ctx', u : T}{Q'}$ for some $m'$.
We conclude with an application of \refrule{tm-cast} taking $m \eqdef m_u + m'$.

\proofrule{rm-struct}
Then $P \pcong P' \red Q' \pcong Q$.
From \Cref{lem:subj_cong_multi} we deduce that $\wtp[n']{\Ctx}{P'}$ for some $n' \le n$.
Using the induction hypothesis on $\wtp[n']{\Ctx}{P'}$ and $P' \red Q'$ we deduce $\wtp[m']{\Ctx}{Q'}$ for some $m'$. 
We conclude using \cref{lem:subj_cong_multi} once more.
\end{proof}

%% file: ts-multi/proof-measure.tex
\beginbass
We introduce two fundamental notions for the soundness proof of the type system.
First, we introduce the \emph{rank} of a session map $M$ as the minimum length to reach
successful termination of session $M$. 
Then, we introduce the \emph{measure} of a process which takes into account the rank in the typing judgment
as well as the ranks of the session that have been already opened.
We embed such measure in the typing derivations by using a refined set of rules.
At last, we compare the typing judgments labeled with the usual rank with those including the measure (see \Cref{lem:measure_rank_multi})
and we prove that structural precongruence of processes does not increase the measure (see \Cref{lem:measure_pcong_multi}).

\begin{figure}[t]
\framebox[\textwidth]{
\begin{mathpar}
    \inferrule[mtm-thread]{
        \mathstrut
    }{
        \wtpn{(n, 0)}\Ctx{P}
    }
    \wtp[n]\Ctx{P}
    \defrule[mtm-thread]{}
    \and
    \inferrule[mtm-cast]{
        \wtpn\Measure{\Ctx, u : T}{P}
    }{
        \wtpn{\Measure + (n,0)}{\Ctx, u : S}{\pcast{u} P}
    }
    ~
    S \subt[n] T
    \defrule[mtm-cast]{}
    \and
    \inferrule[mtm-par]{
        \wtpn{\Measure_i}{\Ctx_i, \ep{s}{\role_i} : S_i}{P_i}~{}^{(i=1,\dots,h)}
    }{
        \wtpn{\sum_{i=1}^h \Measure_i + (0, \rank{\set{\Map{\role_i}{S_i}}_{i=1,\dots,h}})}{
            \Ctx_1,\dots,\Ctx_h
        }{
            \pres{s}{P_1 \parop \dots \parop P_h}
        }
    }
    ~ \coherent{\set{\Map{\role_i}{S_i}}_{i=1..h}}
    \defrule[mtm-par]{}
\end{mathpar}
}
\caption{Typing rules with measure}
\label{fig:measure_multi}
\end{figure}

\begin{definition}[rank]
  \label{def:rank_multi}
  The \emph{rank} of a session map $M = \prod_{i=1}^h \Map{\role_i}{S_i}$, written $\rank{M}$, is the element of $\Nat
  \union \set\infty$ defined as
  \begin{center}
  	\begin{math}
  		\rank{M} \eqdef \min | M \wlred{\In\terminated}|
  	\end{math}
  \end{center}
  where $|M \wlred{\action} N|$ denotes the length of the sequence $\tau,\dots,\tau,\action$ and we postulate that $\min\emptyset = \infty$.
\end{definition}

\begin{definition}[Measure]
\label{def:measure_multi}
The measure of a process is a lexicographically ordered pair of natural numbers
$(m , n)$ where:
\begin{itemize}
\item $m$ is an upper bound to the number of sessions that the process may open
and of weights of casts that the process may perform \emph{in the future} before
it terminates;
\item $n$ is the overall effort for terminating the sessions that have been
already opened \emph{in the past}, \ie the sum of their rank (\Cref{def:rank_multi}).
\end{itemize}
\end{definition}

In \Cref{fig:measure_multi} we introduce a refined set of typing rules for processes that allow us to
associate them with their measure, not just with their rank.
The idea behind these rules (similarly to \Cref{fig:measure_bin}) is that they distinguish between \emph{past} and
\emph{future} of a process by looking at its structure. Indeed, unguarded
sessions have been created, casts have not been performed yet and sessions that
occur guarded have not been created yet.
\refrule{mtm-thread} adopts the rank of the process inside the usual typing
judgment (\Cref{fig:ts_multi}) as first component of the measure. This rule has lower
priority with respect to the other rules so that it is applied to processes that
are not casts or restrictions.
In \refrule{mtm-cast} the first component of the measure is increased by the
weight of the cast.
\refrule{mtm-par} increases the second component of the measure by the rank of
the involved session.

\begin{lemma}
    \label{lem:measure_rank_multi}
    The following properties hold:
    \begin{enumerate}
        \item $\wtp[n]\Ctx{P}$ implies $\wtpn\Measure\Ctx{P}$ for some
        $\Measure \leq (n, 0)$;
        \item $\wtpn\Measure\Ctx{P}$ implies $\wtp[n]\Ctx{P}$ for some $n$ such that $\Measure \leq (n, 0)$.
    \end{enumerate}
\end{lemma}
\begin{proof}
    We prove item 1 by induction on the structure of $P$. 
    The proof of item 2 is by a straightforward induction over $\wtpn\Measure\Ctx{P}$.
    
\proofcase{Case $P = \pres{s}{\procs{P}}$}
From \refrule{tm-par} we deduce that there exist $\Ctx_i, \role_i, S_i, n_i$ for $i = 1,\dots,h$ such that
\begin{itemize}
\item $\Ctx = \Ctx_1,\dots,\Ctx_h$
\item $n = 1 + \sum_{i=1}^h n_i$
\item $\prod_{i=1}^h \Map{\role_i}{S_i} \ft$
\item $\wtp[n_i]{\Ctx_i, \ep{s}{\role_i} : S_i}{P_i}~{}^{(i=1,\dots,h)}$
\end{itemize} 
Using the induction hypothesis on $\wtp[n_i]{\Ctx_i, \ep{s}{\role_i} : S_i}{P_i}~{}^{(i=1,\dots,h)}$ we deduce 
that there exist $\Measure_i$ for $i=1,\dots,h$ such that 
\begin{itemize}
\item $\wtpn{\Measure_i}{\Ctx_i, \ep{s}{\role_i} : S_i}{P_i}~{}^{(i=1,\dots,h)}$
\item $\Measure_i \le (n_i,0)$ for $i=1,\dots,h$
\end{itemize}
We conclude with one application of \refrule{mtm-par} by taking 
$\Measure \eqdef \sum_{i=1}^h \Measure_i + (0, \rank{\prod_{i=1}^h \Map{\role_i}{S_i}})$ and observing that 
$\Measure < (n_1,0) + (n_2,0) + \dots + (n_h,0) + (1,0) = (n,0)$.

\proofcase{Case $P = \pcast{u}{Q}$}
From \refrule{tm-cast} we deduce that there exist $\CtxD, S, T, m$ and $m_u$ such that
\begin{itemize}
\item $\Ctx = \CtxD, u : S$
\item $S \subt[m_u] T$
\item $n = m_u + m$
\item $\wtp[m]{\CtxD, u : T}{Q}$
\end{itemize}
Using the induction hypothesis on $\wtp[m]{\CtxD, u : T}{Q}$ we deduce $\wtpn{\MeasureN}{\CtxD, u : T}{Q}$ for some $\MeasureN \le (m,0)$.
We conclude with an application of \refrule{mtm-cast} by taking $\Measure \eqdef \MeasureN + (m_u,0)$ 
and observing that $\Measure \le (m,0) + (m_u,0) = (n,0)$.

\proofcase{In all the other cases} We conclude with an application of \refrule{mtm-thread} by taking $\Measure \eqdef (n,0)$.
\end{proof}

\begin{lemma}
	\label{lem:measure_pcong_multi}
	If $\wtpn\MeasureM\Ctx P$ and $P \pcong Q$, then there exists $\MeasureN \le \MeasureM$ such that $\wtpn\MeasureN\Ctx Q$.
\end{lemma}
\begin{proof}
By induction on the derivation of $P \pcong Q$ and by cases on the last rule applied. We only consider the base cases.

\proofrule{sm-par-comm} 
Then $P = \pres{s}{\procs{P} \ppar P' \ppar Q' \ppar \procs{Q}} \pcong \pres{s}{\procs{P} \ppar Q' \ppar P' \ppar \procs{Q}} = Q$.
From rule \refrule{mtm-par} we deduce that there exist $\Ctx_i, \role_i, S_i, \Measure_i$ for $i = 1,\dots,h$ such that
\begin{itemize}
\item $\Ctx = \Ctx_1,\dots,\Ctx_h$
\item $\Measure = \sum_{i=1}^h \Measure_i + (0 , \rank{\prod_{i=1}^h \Map{\role_i}{S_i}})$
\item $\prod_{i=1}^h \Map{\role_i}{S_i} \ft$
\item $\wtpn{\Measure_i}{\Ctx_i, \ep{s}{\role_i} : S_i}{P_i}$ for $i = 1,\dots,k$
\item $\wtpn{\Measure_{k+1}}{\Ctx_{k+1}, \ep{s}{\role_{k+1}} : S_{k+1}}{P'}$
\item $\wtpn{\Measure_{k+2}}{\Ctx_{k+2}, \ep{s}{\role_{k+2}} : S_{k+2}}{Q'}$
\item $\wtpn{\Measure_i}{\Ctx_i, \ep{s}{\role_i} : S_i}{Q_i}$ for $i = k+3,\dots,h$
\end{itemize}
We conclude $\wtpn{\MeasureN}\Ctx{Q}$ with one application of \refrule{mtm-par} by taking $\MeasureN \eqdef \Measure$.

\proofrule{sm-par-assoc}
Then $P = \pres{s}{\procs{P} \ppar \pres{t}{R \ppar \procs{Q}}} \pcong \pres{t}{\pres{s}{\procs{P} \ppar R} \ppar \procs{Q}} = Q$ and $s \in \fn{R}$.
From rule \refrule{mtm-par} we deduce that there exist $\Ctx_i, \role_i, S_i, \Measure_i$ for $i = 1,\dots,h$ such that
\begin{itemize}
\item $\Ctx = \Ctx_1,\dots,\Ctx_h$
\item $\Measure = \sum_{i=1}^h \Measure_i + (0 , \rank{\prod_{i=1}^h \Map{\role_i}{S_i}})$
\item $\prod_{i=1}^h \Map{\role_i}{S_i} \ft$
\item $\wtpn{\Measure_i}{\Ctx_i, \ep{s}{\role_i} : S_i}{P_i}$ for $i = 1,\dots,h - 1$
\item $\wtpn{\Measure_h}{\Ctx_h, \ep{s}{\role_h} : S_h}{\pres{t}{R \ppar \procs{Q}}}$
\end{itemize}
From rule \refrule{mtm-par} and the hypothesis that $s \in \fn{R}$ we deduce that there exist 
$\CtxD_i, \roleq_i, T_i, \MeasureN_i$ for $i = 1,\dots,k$ such that
\begin{itemize}
\item $\Ctx_h = \CtxD_1,\dots,\CtxD_k$
\item $\Measure_h = \sum_1^k \MeasureN_i + (0 , \rank{\prod_{i=1}^k \Map{\roleq_i}{T_i}})$
\item $\prod_{i=1}^k \Map{\roleq_i}{T_i} \ft$
\item $\wtpn{\MeasureN_1}{\CtxD_1, \ep{s}{\role_h} : S_h, \ep{t}{\roleq_1} : T_1}{R}$
\item $\wtpn{\MeasureN_{i+1}}{\CtxD_{i+1}, \ep{t}{\roleq_{i+1}} : T_{i+1}}{Q_i}$ for $i = 1,\dots,k-1$
\end{itemize}
Using \refrule{tm-par} we deduce
\begin{itemize}
\item $\wtpn{\sum_{i=1}^{h-1}{\Measure_i} 
	+ \MeasureN_1 
	+ \rank{\prod_{i=1}^h \Map{\role_i}{S_i}}}{\Ctx_1,\dots,\Ctx_{h-1},\CtxD_1, \ep{t}{\roleq_1} : T_1}{\pres{s}{\procs{P} \ppar R}}$
\end{itemize}   
We conclude $\wtpn{\MeasureN}{\Ctx}{\pres{t}{\pres{s}{\procs{P} \ppar R} \ppar \procs{Q}}}$ with another 
application of \refrule{mtm-par} by taking $\MeasureN \eqdef \Measure$.

\proofrule{sm-cast-comm} 
Then $P = \pcast{u}{\pcast{v}{R}} \pcong \pcast{v}{\pcast{u}{R}} = Q$. We can assume $u \ne v$ or else $P = Q$.
From rule \refrule{mtm-cast} we deduce that there exist $\Ctx_1, S, T, \Measure_1, m_u$ such that
\begin{itemize}
\item $\Ctx = \Ctx_1, u : S$
\item $S \subt[m_u] T$
\item $\Measure = \Measure_1 + (m_u , 0)$
\item $\wtpn{\Measure_1}{\Ctx_1, u : T}{\pcast{v}{R}}$
\end{itemize} 
From rule \refrule{tm-cast} we deduce that there exist $\Ctx_2, S', T', \Measure_2, m_v$ such that
\begin{itemize}
\item $\Ctx_1 = \Ctx_2, v : S'$
\item $S' \subt[m_v] T'$
\item $\Measure_1 = \Measure_2 + (m_v , 0)$
\item $\wtpn{\Measure_2}{\Ctx_2, u : T, v : T'}{R}$
\end{itemize}
We derive $\wtpn{\Measure_2 + (m_u,0)}{\Ctx_2, u : S, v : T'}{\pcast{u}{R}}$ with one application of \refrule{mtm-cast} 
and we conclude with another application of \refrule{mtm-cast} by taking $\MeasureN \eqdef \Measure$.

\proofrule{sm-cast-new}
Then $P = \pres{s}{\pcast{\ep{s}{\role}}{R} \ppar \procs{P}} \pcong \pres{s}{R \ppar \procs{P}} = Q$.
From rule \refrule{mtm-par} we deduce that there exist $\CtxD, \Measure', S$ and $\Ctx_i, \roleq_i, S_i, \Measure_i$ for $i = 1,\dots,h$ such that
\begin{itemize}
\item $\Ctx = \CtxD, \Ctx_1, \dots, \Ctx_h$
\item $\Measure = \Measure' + \sum_{i=1}^h \Measure_i + (0 , \rank{\Map{\role}{S} \ppar \prod_{i = 1}^h \Map{\roleq_i}{S_i}})$
\item $\Map{\role}{S} \parop \prod_{i = 1}^h \Map{\roleq_i}{S_i} \ft$
\item $\wtpn{\Measure'}{\CtxD, \ep{s}{\role} : S}{\pcast{\ep{s}{\role}}{R}}$
\item $\wtpn{\Measure_i}{\Ctx_i, \ep{s}{\roleq_i} : S_i}{P_i}$ for $i = 1,\dots,h$
\end{itemize} 
From rule \refrule{mtm-cast} we deduce that there exist $T, \MeasureN', m_s$ such that
\begin{itemize}
\item $S \subt[m_s] T$
\item $\MeasureN' = \Measure' + (m_s , 0)$
\item $\wtpn{\MeasureN'}{\CtxD, \ep{s}{\role} : T}{R}$
\end{itemize}
From $\Map{\role}{S} \ppar \prod_{i = 1}^h \Map{\roleq_i}{S_i} \ft$, $S \subt[m_s] T$ and \cref{def:ssubt} 
we deduce $\Map{\role}{T} \ppar \prod_{i = 1}^h \Map{\roleq_i}{S_i} \ft$. 
We conclude with an application of \refrule{mtm-par} by taking 
$\MeasureN = \MeasureN' + \sum_{i=1}^h \Measure_i + (0 , \rank{\Map{\role}{S} \ppar \prod_{i = 1}^h \Map{\roleq_i}{S_i}}) \le n$.

\proofrule{sm-cast-swap}
Then $P = \pres{s}{\pcast{\ep{t}{\role}}{R} \ppar \procs{P}} \pcong \pcast{\ep{t}{\role}}{\pres{s}{R \ppar \procs{P}}} = Q$ and $t \ne s$.
From rule \refrule{mtm-par} we deduce that there exist $\Ctx_i, \roleq_i, \Measure_i$ for $i = 1,\dots,h$ such that
\begin{itemize}
\item $\Ctx = \Ctx_1, \dots, \Ctx_h$
\item $\Measure = \sum_{i=1}^h \Measure_i + (0 , \rank{\prod_{i = 1}^h \Map{\roleq_i}{S_i}})$
\item $\prod_{i = 1}^h \Map{\roleq_i}{S_i} \ft$
\item $\wtpn{\Measure_1}{\Ctx_1, \ep{s}{\roleq_1} : S_1}{\pcast{\ep{t}{\role}}{R}}$
\item $\wtpn{\Measure_i}{\Ctx_i, \ep{s}{\roleq_i} : S_i}{P_i}$ for $i = 2,\dots,h$
\end{itemize} 
From rule \refrule{mtm-cast} we deduce that there exist $\CtxD, T, \Measure', m_t$ such that
\begin{itemize}
\item $\Ctx_1 = \CtxD, \ep{t}{\role} : S$
\item $S \subt[m_t] T$
\item $\Measure_1 = \Measure' + (m_t , 0)$
\item $\wtpn{\Measure'}{\CtxD, \ep{t}{\role} : T, \ep{s}{\roleq_1} : S_1}{R}$
\end{itemize}
We derive 
$\wtpn{\Measure' 
	+ \sum_{i=2}^h \Measure_i 
	+ (0 , \rank{\prod_{i = 1}^h \Map{\roleq_i}{S_i}})}{\CtxD, \ep{t}{\role} : T,\Ctx_2,\dots,\Ctx_h}{\pres{s}{R \ppar \procs{P}}}$ 
with an application of \refrule{mtm-par}. We conclude with an application of \refrule{mtm-cast} by taking $m \eqdef n$.

\proofrule{sm-call} 
Then $P = \pinvk{A}{\seqof u} \pcong R\subst{\seqof x}{\seqof u} = Q$ and $\Definition{A}{\seqof x}{R}$.
From \refrule{mtm-thread} we deduce that $\wtpn{n}{\Ctx}{\pinvk{A}{\seqof u}}$ for some $n$ such that $\Measure = (n , 0)$. 
Using \Cref{lem:subj_cong_multi} we deduce $\wtp[m]{\Ctx}{Q}$ for some $m \le n$.
Using \Cref{lem:measure_rank_multi} we deduce that $\wtpn{\MeasureN}{\Ctx}{Q}$ for some $\MeasureN \le (m , 0)$. 
We conclude observing that $\MeasureN \le (m , 0) \le (n , 0) = \Measure$.
\end{proof}

%% file: ts-multi/proof-nf.tex
\beginbass
As in \Cref{ssec:nf_bin} we need to prove that a well typed process is deadlock free. The key lemma of this is \Cref{lem:dl_freedom_multi}
that we dub \emph{quasi deadlock freedom}. Informally, we want to rearrange the process under analysis 
in order to put together the creation of some session $s$ with all the subprocesses that start with
some communication on $s$. This way we can try to apply a reduction rule (see \Cref{fig:red_multi}). 
This is what \Cref{lem:dl_freedom_multi} does.
However, a process organized as described is not guaranteed to make progress.
Indeed, for example all the subprocesses can be waiting for a message leading to a stuck process.
Hence, we called the lemma \emph{quasi} deadlock freedom. 
Deadlock freedom is achieved by taking into account the \emph{coherence} of the session as well.
We called the shape of the process mentioned before \emph{proximity normal form} (\Cref{def:pnf_multi}).
In order to obtain a process in such form from a well typed one we require several steps.
We introduce additional normal forms (\Cref{def:cnf_multi,def:tnf_multi}) to describe those processes in the middle of the procedure.
We introduce \emph{process contexts} to easily refer to unguarded sub-processes:
\[
	\textbf{Process context}
	\quad
	\PCtxC, \PCtxD ~~::=~~ \Hole \mid \pres{s}{\procs{P} \parop \PCtxC \parop \procs{Q}} \mid \pcast{u}\PCtxC
\]


\begin{definition}[Choice Normal Form]
	\label{def:cnf_multi}
	We say that $P_1 \pchoice P_2$ is an \emph{unguarded choice} of $P$ if there
	exists $\PCtxC$ such that $P \pcong \PCtxC[P_1 \pchoice P_2]$. We say that
	$P$ is in \emph{choice normal form} if it has no unguarded choices.
\end{definition}

\begin{definition}[Thread Normal Form]
	\label{def:tnf_multi}
	A process is in \emph{thread normal form} if it is generated by the grammar below:
	\[
		\begin{array}{@{}rcl@{}}
			\Pnf, \Qnf & ::= & \pcast{u}\Pnf \mid \Ppar
			\\
			\Ppar, \Qpar & ::= & \pres{s}{\procs{\Ppar}} \mid \Pth
			\\
			\Pth & ::= &
				\pdone \mid
				\pclose{u} \mid \pwait{u}{P} \mid 
				\pbranch[i\in I]{\chvar}{\rolep}\Pol{\la_i}{\PP_i} \\
				 
			& &	\mid \poch{u}{\rolep}{v}{P} \mid
					\pich{u}{\rolep}{x}{P}
		\end{array}
	\] 
\end{definition}

Intuitively, a process is in \emph{thread normal form} if it consists of an initial prefix of casts followed 
by a parallel composition of threads, where a thread is either $\pdone$ or a process waiting to perform an input/output 
action on some channel $u = \ep{s}{\role}$ for some $\role$. In this latter case, we say that the thread is an $s$-thread.

\begin{definition}[Proximity Normal Form]
	\label{def:pnf_multi}
	We say that $\Pnf$ is in \emph{proximity normal form} if $\Pnf = \PCtxC[\pres{s}{\procs{\Pth}}]$ for some $\PCtxC$, $s$, $\procs{\Pth}$ where
	each $\Pth_i$ for $i=1,\dots,h$ is a $s$-thread.
\end{definition}


\begin{lemma}
	\label{lem:cnf2_multi}
	If $\wtp[n]\Ctx{P}$ and $\wtpi\Ctx{P}$, then there exists $Q$ in choice
	normal form such that $P \wred Q$ and $\wtp[m]\Ctx{Q}$ for some $m \le n$.
\end{lemma}
\begin{proof}
By induction on $\wtpi\Ctx{P}$ and by cases on the last rule applied.

\proofcase{Case $P$ is already in choice normal form} 
We conclude taking $Q \eqdef P$ and $m \eqdef n$.

\proofrule{tm-call}
Then $P = \pinvk{A}{\seqof{u}}$ and $\Definition{A}{\seqof{x}}{R}$.
We deduce $\Ctx = \seqof{u : S}$, $\tass{A}{\seqof{S}}{n'}$ and $\wtpi\Ctx{R\subst{\seqof u}{\seqof x}}$. Moreover, it must be the case that
$\wtp[n']\Ctx{R\subst{\seqof u}{\seqof x}}$ and $n' \leq n$ since \refrule{tm-call} is used in the coinductive judgment as well.
Using the induction hypothesis we deduce that there exist $Q$ in choice normal form and $m \le n'$ such that $R\subst{\seqof u}{\seqof x} \wred Q$ 
and $\wtp[m]\Ctx{Q}$.
We conclude by observing that $P \wred Q$ using \refrule{rm-struct} and that $m \leq n' \leq n$.

\proofrule{com-choice}
Then $P = P_1 \pchoice P_2$.
We deduce $\wtpi\Ctx{P_k}$ with $k \in \set{1,2}$.
Moreover, it must be the case that $\wtp[n]\Ctx{P_k}$ since \refrule{tm-choice} is used in the coinductive judgment.
Using the induction hypothesis we deduce that there exist $Q$ in choice normal form and $m \leq n$ such that $P_k \wred Q$ and $\wtp[m]\Ctx{Q}$.
We conclude by observing that $P \red P_k$ by \refrule{rm-choice}.

\proofrule{tm-choice}
Analogous to the previous case but we consider the premise in which the rank is the same of the conclusion to keep sure that it does not increase.

\proofrule{tm-par}
Then $P = \pres{s}{P_1 \parop \dots \parop P_h}$. 
We deduce 
\begin{itemize}
\item $\Ctx = \Ctx_1, \dots, \Ctx_h$
\item $\wtpi{\Ctx_i, \ep{s}{\role_i} : S_i}{P_i}$ for $i=1,\dots,h$
\item $\prod_{i=1}^h \Map{\role_i}{S_i} \ft$
\end{itemize}
Furthermore, it must be the case that $\wtp[n_i]{\Ctx_i, \ep{s}{\role_i} : S_i}{P_i}$ for $i=1,\dots,h$ 
and $n = 1 + \sum_{i=1}^h n_i$ since \refrule{tm-par} is used in the coinductive judgment as well.
Using the induction hypothesis we deduce that there exist $Q_i$ in choice normal form and $m_i \leq n_i$ such that 
$P_i \wred Q_i$ and $\wtp[m_i]{\Ctx_i, \ep{s}{\role_i} : S_i}{Q_i}$ for $i=1,\dots,h$.
We conclude by taking $m \eqdef 1 + \sum_{i=1}^h m_i$ and $Q \eqdef \pres{s}{Q_1 \parop \cdots \parop Q_h}$ 
with one application of \refrule{tm-par}, observing that 
$m = 1 + \sum_{i=1}^h m_i \leq 1 + \sum_{i=1}^h n_i = n$ and that $P \wred Q$ by \refrule{rm-par}.

\proofrule{tm-cast}
Then $P = \pcast{u} P'$.
Analogous to the previous case, just simpler.
\end{proof}

\begin{lemma}
	\label{lem:cnf1_multi}
	If $\wtp[n]\Ctx{P}$, then there exists $Q$ in choice normal form such that $P \wred Q$ and $\wtp[m]\Ctx{Q}$ for some $m \le n$.
\end{lemma}
\begin{proof}
	\brk
	Consequence of \Cref{lem:cnf2_multi} noting that $\wtp[n]\Ctx{P}$ implies $\wtpi\Ctx{P}$.
\end{proof}

\begin{lemma}
	\label{lem:cnf_exists_multi}
	If $\wtpn\Measure\Ctx{P}$, then there exist $Q$ in choice normal form and $\MeasureN \leq \Measure$ such that $P \wred Q$ and $\wtpn\MeasureN\Ctx{Q}$.
\end{lemma}
\begin{proof}
By induction on $\wtpn\Measure\Ctx{P}$ and by cases on the last rule applied.

\proofrule{mtm-thread}
Then $P$ is a thread. We deduce that
\begin{itemize}
\item $\Measure = (n , 0)$ for some $n$
\item $\wtp[n]\Ctx{P}$
\end{itemize}
From \Cref{lem:cnf1_multi} we deduce that there exist $Q$ and $m \le n$ such that $P \wred Q$ and $\wtp[m]\Ctx{Q}$. 
From \Cref{lem:measure_rank_multi} we deduce $\wtpn\MeasureN\Ctx{Q}$ for some $\MeasureN \le (m , 0)$. 
We conclude observing that $\MeasureN \le (m , 0) \le (n , 0) = \Measure$.

\proofrule{mtm-cast}
Then $P = \pcast{u}{P'}$. We deduce that
\begin{itemize}
\item $\Ctx = \CtxD, u : S$
\item $S \subt[n] T$
\item $\Measure = \Measure' + (n,0)$
\item $\wtpn{\Measure'}{\Ctx', u : T}{P'}$
\end{itemize}
Using the induction hypothesis we deduce that there exist $Q'$ and $\MeasureN' \le \Measure'$ such that $P' \wred Q'$ 
and $\wtpn{\MeasureN'}{\Ctx', u : T}{Q'}$. We conclude with an application of \refrule{mtm-cast} taking 
$Q \eqdef \pcast{u}{Q'}$, $\MeasureN \eqdef \MeasureN' + (n,0)$ and observing that $P \wred Q$ using \refrule{rm-cast}. 

\proofrule{mtm-par}
Then $P = \pres{s}{P_1 \parop \dots \parop P_h}$. 
We deduce 
\begin{itemize}
\item $\Ctx = \Ctx_1, \dots, \Ctx_h$
\item $\Measure = \sum_{i=1}^h \Measure_i + (0, \rank{\prod_{i=1}^h \Map{\role_i}{S_i}})$
\item $\wtpn{\Measure_i}{\Ctx_i, \ep{s}{\role_i} : S_i}{P_i}$ for $i=1,\dots,h$
\item $\prod_{i=1}^h \Map{\role_i}{S_i} \ft$
\end{itemize}
Using the induction hypothesis we deduce that there exist $Q_i$ in choice normal form and $\MeasureN_i \leq \Measure_i$ such that $P_i \wred Q_i$ and 
$\wtpn{\MeasureN_i}{\Ctx_i, \ep{s}{\role_i} : S_i}{Q_i}$ for $i=1,\dots,h$.
We conclude by taking $\MeasureN \eqdef \sum_{i=1}^h \MeasureN_i + (0, \rank{\prod_{i=1}^h \Map{\role_i}{S_i}})$ and 
$Q \eqdef \pres{s}{Q_1 \parop \cdots \parop Q_h}$ with one application of \refrule{mtm-par}, 
observing that 
$\MeasureN = \sum_{i=1}^h \MeasureN_i + (0, \rank{\prod_{i=1}^h \Map{\role_i}{S_i}}) \leq 
						 \sum_{i=1}^h \Measure_i + (0, \rank{\prod_{i=1}^h \Map{\role_i}{S_i}}) = \Measure$ 
and that $P \wred Q$ by \refrule{rm-par}.
\end{proof}

\begin{lemma}
	\label{lem:tnf_exists_multi}
	If $\wtpi\Ctx P$ and $P$ is in choice normal form, then there exists $\Pnf$ such that $P \pcong \Pnf$. 
\end{lemma}
\begin{proof}
By induction on $\wtpi\Ctx P$ and by cases on the last rule applied.

\proofcase{Cases \refrule{tm-choice} and \refrule{com-choice}} These cases are impossible from the hypothesis that $P$ is in choice normal form.

\proofcase{Cases \refrule{tm-done}, \refrule{tm-wait} and\refrule{tm-close}}
Then $P$ is a thread and is already in thread normal form and we conclude by reflexivity of $\pcong$.

\proofcase{\refrule{tm-channel-in}, \refrule{tm-channel-out}, \refrule{tm-tag} and \refrule{com-tag}}
Then $P$ is a thread and is already in thread normal form and we conclude by reflexivity of $\pcong$.

\proofrule{tm-call}
Then there exist $A$, $Q$, $\seqof u$ and $\seqof S$ such that
\begin{itemize}
\item $P = \pinvk{A}{\seqof u}$
\item $\Definition{A}{\seqof x}Q$
\item $\Ctx = \seqof{u : S}$
\item $\wtpi{\seqof{u : S}}{Q\subst{\seqof u}{\seqof x}}$
\end{itemize}
Using the induction hypothesis on $\wtpi{\seqof{u : S}}{Q\subst{\seqof u}{\seqof x}}$ we deduce that there exists 
$\Pnf$ such that $Q\subst{\seqof u}{\seqof x} \pcong \Pnf$.
We conclude $P \pcong \Pnf$ using \refrule{sm-call} and the transitivity of $\pcong$.

\proofrule{tm-par}
Then there exist $s$ and $P_i, \Ctx_i, S_i, \role_i$ for $i=1,\dots,h$ such that
\begin{itemize}
\item $P = \pres{s}{P_1 \parop \cdots \parop P_h}$
\item $\Ctx = \Ctx_1, \dots,\Ctx_h$
\item $\wtpi{\Ctx_i, \ep{s}{\role_i} : S_i}{P_i}$ for $i=1,\dots,h$
\end{itemize}
Using the induction hypothesis on $\wtpi{\Ctx_i, \ep{s}{\role_i} : S_i}{P_i}$ we deduce that there exist 
$\Pnf_i$ such that $P_i \pcong \Pnf_i$ for $i=1,\dots,h$.
By definition of thread normal form, it must be the case that $\Pnf_i = \pcast{\seqof{u_i}} \Ppar_i$ 
for some $\seqof{u_i}$ and $\Ppar_i$. Let $\seqof{v_i}$ be the same sequence as $\seqof{u_i}$
 except that occurrences of $\ep{s}{\role_i}$ have been removed.
We conclude by taking $\Pnf \eqdef \pcast{\seqof{v_1}\dots\seqof{v_h}}\pres{s}{\Ppar_1 \parop \dots \parop \Ppar_h}$ 
and using the fact that $\pcong$ is a pre-congruence and observing that
	\[
		\begin{array}{rcll}
			P & = & \pres{s}{P_1 \parop \cdots \parop P_h} & \text{by definition of $P$}
			\\
			& \pcong & \pres{s}{\Pnf_1 \parop \cdots \parop \Pnf_h} & \text{using the induction hypothesis}
			\\
			& = & \pres{s}{\pcast{\seqof{u_1}}\Ppar_1 \parop \cdots \parop \pcast{\seqof{u_h}}\Ppar_h}
			& \text{by \Cref{def:tnf_multi}}
			\\
			& \pcong & \pcast{\seqof{v_1} \dots \seqof{v_h}}\pres{s}{\Ppar_1 \parop \cdots \parop \Ppar_h}
			& \text{by \refrule{sm-cast-new},} \\
			& & & \text{\refrule{sm-cast-swap}, \refrule{sm-par-comm}}
			\\
			& = & \Pnf & \text{by definition of $\Pnf$}
		\end{array}
	\]

\proofrule{tm-cast}
Then there exist $u$, $Q$, $\Ctx'$, $S$ and $T$ such that
\begin{itemize}
\item $P = \pcast{u} Q$
\item $\Ctx = \Ctx', u : S$
\item $\wtpi{\Ctx', u : T}{Q}$
\item $S \subt T$
\end{itemize}
Using the induction hypothesis on $\wtpi{\Ctx', u : T}{Q}$ we deduce that there exists $\Qnf$ such that $Q \pcong \Qnf$.
We conclude by taking $\Pnf \eqdef \pcast{u}\Qnf$ using the fact that $\pcong$ is a pre-congruence.
\end{proof}

\begin{lemma}[Proximity]
  \label{lem:proximity_multi}
  If $s\in\fn{P} \setminus \bn\PCtxC$, then $\pres{s}{\PCtxC[P] \parop \procs{Q}} \pcong \PCtxD[\pres{s}{P \parop \procs{Q}}]$ for some $\PCtxD$.
\end{lemma}
\begin{proof}
By induction on the structure of $\PCtxC$ and by cases on its shape.

\proofcase{Case $\PCtxC = \Hole$}
We conclude by taking $\PCtxD \eqdef \Hole$ using the reflexivity of $\pcong$. 

\proofcase{Case $\PCtxC = \pres{t}{\procs{P'} \parop \PCtxC' \parop \procs{Q'}}$}
From the hypothesis $s \in \fn{P} \setminus \bn\PCtxC$ we deduce $s \ne t$ and $s \in \fn{P} \setminus \bn{\PCtxC'}$.
Using the induction hypothesis and \refrule{sm-par-comm} we deduce that there exists $\PCtxD'$ such that 
$\pres{s}{\PCtxC'[P] \parop \procs{Q}} \pcong \PCtxD'[\pres{s}{P \parop \procs{Q}}]$.
Take $\PCtxD \eqdef \pres{t}{\PCtxD' \parop \procs{P'} \parop \procs{Q'}}$. We conclude
  \[
    \begin{array}{rcll}
      \pres{s}{\PCtxC[P] \parop \procs{Q}}
      & = & \pres{s}{\pres{t}{\procs{P'} \parop \PCtxC'[P] \parop \procs{Q'}} \parop \procs{Q}}
      & \text{by definition of $\PCtxC$}
      \\
      & \pcong & \pres{s}{\procs{Q} \parop \pres{t}{\PCtxC'[P] \parop \procs{P'} \parop \procs{Q'}}}
      & \text{by \refrule{sm-par-comm}}
      \\
      & \pcong & \pres{t}{\pres{s}{\procs{Q} \parop \PCtxC'[P]} \parop \procs{P'} \parop \procs{Q'}}
      & \text{by \refrule{sm-par-assoc}} \\
      & & & \text{and $s \in \fn{\PCtxC'[P]}$}
      \\
      & \pcong & \pres{t}{\pres{s}{\PCtxC'[P] \parop \procs{Q}} \parop \procs{P'} \parop \procs{Q'}}
      & \text{by \refrule{sm-par-comm}}
      \\
      & \pcong & \pres{t}{\PCtxD'[\pres{s}{P \parop \procs{Q}}] \parop \procs{P'} \parop \procs{Q'}}
      & \text{by induction hypothesis}
      \\
      & = & \PCtxD[\pres{s}{P \parop \procs{Q}}]
      & \text{by definition of $\PCtxD$}
    \end{array}
  \]
  
\proofcase{Case $\PCtxC = \pcast{\ep{t}{\role}}\PCtxC'$ and $s \ne t$}
Using the induction hypothesis we deduce that there exists $\PCtxD'$ such that 
$\pres{s}{\PCtxC'[P] \parop \procs{Q}} \pcong \PCtxD'[\pres{s}{P \parop \procs{Q}}]$.
Take $\PCtxD \eqdef \pcast{\ep{t}{\role}}\PCtxD'$. We conclude
  \[
    \begin{array}{rcll}
      \pres{s}{\PCtxC[P] \parop \procs{Q}}
      & = & \pres{s}{\pcast{\ep{t}{\role}}\PCtxC'[P] \parop \procs{Q}}
      & \text{by definition of $\PCtxC$}
      \\
      & \pcong & \pcast{\ep{t}{\role}}\pres{s}{\PCtxC'[P] \parop \procs{Q}}
      & \text{by \refrule{sm-cast-swap} and $t \ne s$}
      \\
      & \pcong & \pcast{\ep{t}{\role}}\PCtxD'[\pres{s}{P \parop \procs{Q}}]
      & \text{using the induction hypothesis}
      \\
      & = & \PCtxD[\pres{s}{P \parop \procs{Q}}]
      & \text{by definition of $\PCtxD$}
    \end{array}
  \]

\proofcase{Case $\PCtxC = \pcast{\ep{s}{\role}}\PCtxC'$}
Using the induction hypothesis we deduce that there exists $\PCtxD$ such that 
$\pres{s}{\PCtxC'[P] \parop \procs{Q}} \pcong \PCtxD[\pres{s}{P \parop \procs{Q}}]$.
We conclude
  \[
    \begin{array}[b]{rcll}
      \pres{s}{\PCtxC[P] \parop \procs{Q}}
      & = & \pres{s}{\pcast{\ep{s}{\role}}\PCtxC'[P] \parop \procs{Q}}
      & \text{by definition of $\PCtxC$}
      \\
      & \pcong & \pres{s}{\PCtxC'[P] \parop \procs{Q}}
      & \text{by \refrule{sm-cast-new}}
      \\
      & \pcong & \PCtxD[\pres{s}{P \parop \procs{Q}}]
      & \text{using the induction hypothesis}
    \end{array}
    \qedhere
  \]
\end{proof}

\begin{lemma}[Quasi - Deadlock Freedom]
	\label{lem:dl_freedom_multi}
	If $\wtpn\Measure\EmptyCtx\Pnf$, then $\Pnf = \pdone$ or $\Pnf \pcong \Qnf$ for some $\Qnf$ in proximity normal form.
\end{lemma}
\begin{proof}
By induction on the derivation of $\wtpn\MeasureM\EmptyCtx\Pnf$ we deduce that 
$\Pnf$ consists of $s_1,\dots,s_h$ sessions and $\sum_{i=1}^h k_i - h + 1$ threads where 
$k_i$ is the number of roles in $s_i$. The scenarios in which no communication is possible are those 
in which for each session $s_i$ there are less than $k_i$ $s_i$-threads. If we assume that 
for each $s_i$ there are $k_i - 1$ threads, then we obtain 
\[
\sum_{i=1}^h k_i - h + 1 - \sum_{i=1}^h(k_i - 1) = \sum_{i=1}^h k_i - h + 1 - \sum_{i=1}^hk_i + h = 1
\]
$s_i$-thread for some $s_i$; hence, there exist $k_i$ $s_i$-threads. 
In other words, there exist $\PCtxD,\PCtxC_1,\dots,\PCtxC_{k_i}$ and $\Pth_1,\dots,\Pth_{k_i}$ $s_i$-threads 
such that 
\[
	\Pnf = \PCtxD[\pres{s_i}{\PCtxC_1[\Pth_1] \parop \cdots \parop \PCtxC_{k_i}[\Pth_{k_i}]}]
\] 
We conclude
	\[
		\begin{array}{rcl}
			\Pnf & = & \PCtxD[\pres{s_i}{\PCtxC_1[\Pth_1] \parop \cdots \parop \PCtxC_{k_i}[\Pth_{k_i}]}] \\
			& & \hfill\text{by definition of $\Pnf$}
			\\
			& \pcong & \PCtxD[\PCtxD_1[\pres{s_i}{\Pth_1 \parop \PCtxC_2[\Pth_2] \parop \cdots \parop \PCtxC_{k_i}[\Pth_{k_i}]}]] \\
			& & \hfill\text{by \Cref{lem:proximity_multi}}
			\\
			& \pcong & \PCtxD[\PCtxD_1[\pres{s_i}{\PCtxC_2[\Pth_2] \parop \Pth_1 \parop \cdots \parop \PCtxC_{k_i}[\Pth_{k_i}]}]] \\
			& & \hfill\text{by \refrule{sm-par-comm}}
			\\
			& \dots \\
			& \pcong & \PCtxD[\PCtxD_1[\PCtxD_2[ \dots \PCtxD_{k_i}[\pres{s_i}{\Pth_{k_i} \parop \cdots \parop \Pth_2 \parop \Pth_1}] \dots ]]] \\
			& & \hfill\text{for some $\PCtxD_2,\dots,\PCtxD_{k_i}$ by \Cref{lem:proximity_multi}}
			\\
			& \eqdef & \Qnf
		\end{array}
	\]
	The fact that $\Qnf$ is in thread normal form follows from the observation
	that $\Pnf$ does not have unguarded casts (it is a closed process in thread
	normal form) so the pre-congruence rules applied here and in
	\Cref{lem:proximity_multi} do not move casts around. We conclude that $\Qnf$ is in
	proximity normal form by its shape.
\end{proof}

As mentioned at the beginning, 
\Cref{lem:dl_freedom_multi} is dubbed ``quasi-deadlock freedom'' because it does not
say that $\Qnf$ reduces. Indeed, a process in proximity normal form is only
\emph{ready to communicate} thanks to its shape (see reduction rules). We can
prove that a well typed process of such kind actually reduces by observing that
\refrule{tm-par} requires that the involved session is coherent. This result is
the key ingredient for proving \Cref{lem:pnf_helpful_direction_multi}.

%% file: ts-multi/proof-soundness.tex
\beginbass
The next key result we prove is inspired by the \emph{helpful direction}. Given a well typed process in
proximity normal form (see \Cref{def:pnf_multi}), we prove that there exists a reduct that is well typed
with a \emph{strictly smaller} measure (see \Cref{lem:pnf_helpful_direction_multi}). 
Notably, this result implies deadlock freedom.
Then, it is easy to prove that a well typed process is either $\pdone$ or it can reach $\pdone$ in finitely
many steps by induction over the measure.

\begin{lemma}
	\label{lem:pnf_helpful_direction_multi}
	If $\wtpn\Measure\Ctx\Pnf$ where $\Pnf$ is in proximity normal form, then there exist $Q$ and $\MeasureN < \Measure$ 
	such that $\Pnf \wred^+ Q$ and $\wtpn\MeasureN\Ctx{Q}$.
\end{lemma}
\begin{proof}
From the hypothesis that $\Pnf$ is in proximity normal form we know that 
$\Pnf = \PCtxC[\pres{s}{\Pth_1 \parop \cdots \parop \Pth_h}]$ for some $\PCtxC$, $s$ and $\Pth_1, \dots, \Pth_h$ $s$-threads.
We reason by induction on $\PCtxC$ and by cases on its shape.
	
\proofcase{Case $\PCtxC = \Hole$}
From \refrule{mtm-thread} and \refrule{mtm-par} we deduce that there exist $\Ctx_i, S_i, \role_i, n_i$ for $i=1,\dots,h$ such that
\begin{itemize}
\item $\Ctx = \Ctx_1,\dots,\Ctx_h$
\item $\prod_{i=1}^h \Map{\role_i}{S_i} \ft$
\item $\Measure = (\sum_{i=1}^h n_i , \rank{\prod_{i=1}^h \Map{\role_i}{S_i}})$
\item $\wtp[n_i]{\Ctx_i, \ep{s}{\role_i} : S_i}{\Pth_i}$ for $i=1,\dots,h$
\end{itemize}
From the hypothesis that $\prod_{i=1}^h \Map{\role_i}{S_i} \ft$ we deduce $\prod_{i=1}^h \Map{\role_i}{S_i} \wlred{\In\terminated}$. 
We now reason on the rank of the session and on the shape of $S_i$. 
For the sake of simplicity, we implicitly apply \refrule{s-par-comm} at process level.\\\\
If $\rank{\prod_{i=1}^h \Map{\role_i}{S_i}} = 1$, then $\prod_{i=1}^h \Map{\role_i}{S_i} \lred{\In\terminated}$ using \refrule{lm-terminate}.
\begin{itemize}
\item \proofcase{Case $S_1 = \End[\In]$ and $S_j = \End[\Out]$ for $j=2,\dots,h$}
Then 
	\begin{itemize}
	\item $\Ctx_j = \EmptyCtx$ and $\Pth_j = \pclose{\ep{s}{\role_j}}$ for $j=2,\dots,h$
	\item $\Pth_1 = \pwait{\ep{s}{\role_1}}{Q}$
	\item $\wtp[n_1]{\Ctx_1}{Q}$
	\end{itemize}
From \Cref{lem:measure_rank_multi} we deduce that $\wtpn{\MeasureN}{\Ctx_1}{Q}$ for some $\MeasureN \le (n_1 , 0)$. 
We conclude observing that $\Pnf \red Q$ by \refrule{rm-signal} and that 
$\MeasureN \le (n_1 , 0) < (\sum_{i=1}^h n_i , \rank{\prod_{i=1}^h \Map{\role_i}{S_i}}) = \Measure$.
\end{itemize}
If $\rank{\prod_{i=1}^h \Map{\role_i}{S_i}} > 1$, then $\prod_{i=1}^h \Map{\role_i}{S_i} \lred{\tau} \dots \lred{\In\terminated}$ 
first using \refrule{lm-tau} and \refrule{lm-sync}. 
Observe that \refrule{lm-pick} is never used since we are considering the minimum reduction sequence; 
a synchronization through \refrule{lm-pick} and \refrule{lm-sync} would lead to a longer reduction. T
hen $S_1 \xlred{\Map{\role_1}{\role_2\Out\Tag_k}}$ and $S_2 \xlred{\Map{\role_2}{\role_1\In\Tag_k}}$ for some 
$\Tag_k$ or $S_1 \xlred{\Map{\role_1}{\role_2}\Out S}$ and $S_2 \xlred{\Map{\role_1}{\role_2}\In S}$.
\begin{itemize}
\item \proofcase{Case $S_1 = \Tags\role_2\Out \Tag_i.S'_i$ and $S_2 = \JTags\role_1\In \Tag_j.T_j$ with $k \in I$}
From the hypothesis that $\prod_{i=1}^h \Map{\role_i}{S_i} \ft$ we deduce $I \subseteq J$. 
From \Cref{def:coherence} we deduce 
$\prod_{i = 3}^h \Map{\role_i}{S_i} \parop \Map{\role_1}{S'_{k}} \parop \Map{\role_2}{T_{k}} \ft$ 
and from \refrule{tm-tag} we deduce that 
	\begin{itemize}
	\item $\Pth_1 = \pbranch[i \in I]{\ep{s}{\role_1}}{\role_2}{\Out}{\Tag_i}{P'_i}$
	\item $\Pth_2 = \pbranch[j \in J]{\ep{s}{\role_2}}{\role_1}{\In}{\Tag_j}{Q_j}$
	\item $\wtp[n_1]{\Ctx_1, \ep{s}{\role_1} : S'_i}{P'_i}$ for all $i \in I$
	\item $\wtp[n_2]{\Ctx_2, \ep{s}{\role_2} : T_j}{Q_j}$ for all $j \in J$
	\end{itemize}
Let $Q \eqdef \pres{s}{P'_k \parop Q_k \parop P_3 \parop \dots \parop P_h}$ and observe that $\Pnf \wred^+ Q$ by \refrule{rm-pick} and \refrule{rm-tag}. 
From \Cref{lem:measure_rank_multi} we deduce that there exist $\Measure_1 \le (n_1 , 0), \Measure_2 \le (n_2 , 0)$ such that
	\begin{itemize}
	\item $\wtpn{\Measure_1}{\Ctx_1, \ep{s}{\role_1} : S'_k}{P'_k}$
	\item $\wtpn{\Measure_2}{\Ctx_2, \ep{s}{\role_2} : T_k}{Q_k}$
	\end{itemize}
Let 
$\MeasureN \eqdef \Measure_1 
	+ \Measure_2 
	+ (\sum_{i=3}^h n_i, \rank{\Map{\role_1}{S'_k} \parop \Map{\role_2}{T_k} \parop \prod_{i=3}^h \Map{\role_i}{S_i}})$. 
We conclude with one application of \refrule{mtm-par} observing that
\[
\begin{array}{rcll}
	\MeasureN & = & \Measure_1 + \Measure_2 + \\
	& & (\sum_{i=3}^h n_i, \rank{\Map{\role_1}{S'_k} \parop \Map{\role_2}{T_k} \parop \prod_{i=3}^h \Map{\role_i}{S_i}})
	& \text{by def. of $\MeasureN$}
	\\
	& \le & (\sum_{i=1}^h n_i, \rank{\Map{\role_1}{S'_k} \parop \Map{\role_2}{T_k} \parop \prod_{i=3}^h \Map{\role_i}{S_i}})
	& \text{by \Cref{lem:measure_rank_multi}}
	\\
	& < & (\sum_{i=1}^h n_i, \rank{\prod_{i=1}^h \Map{\role_i}{S_i}})
	& \text{before $\red$}
	\\
	& = & \Measure
\end{array}
\]
\end{itemize}
\begin{itemize}
\item \proofcase{Case $S_1 = \role_2\Out{S}.T_1$ and $S_2 = \role_1\In{S}.T_2$}
\end{itemize}
From the hypothesis that $\prod_{i=1}^h \Map{\role_i}{S_i} \ft$ and \Cref{def:coherence} we deduce 
$\Map{\role_1}{T_1} \parop \Map{\role_2}{T_2} \parop \prod_{i=3}^h \Map{\role_i}{S_i} \ft$ and from 
\refrule{tm-channel-out} and \refrule{tm-channel-in} we deduce that
	\begin{itemize}
	\item $\Pth_1 = \poch{\ep{s}{\role_1}}{\role_2}{u}{P'_1}$
	\item $\Pth_2 = \pich{\ep{s}{\role_2}}{\role_1}{x}{P'_2}$
	\item $\wtp[n_1]{\Ctx_1, \ep{s}{\role_1} : T_1}{P'_1}$
	\item $\wtp[n_2]{\Ctx_2, \ep{s}{\role_2} : T_2, x : S}{P'_2}$
	\end{itemize}
Let $Q \eqdef \pres{s}{P'_1 \parop P'_2\subst{u}{x} \parop \Pth_3 \parop \cdots \parop \Pth_h}$ and observe that $\Pnf \red Q$ by \refrule{rm-channel}. 
Using \Cref{lem:substitution_multi} we deduce $\wtp[n_2]{\Ctx_2, \ep{s}{\role_2} : T_2, u : S}{P'_2\subst{u}{x}}$ and 
from \Cref{lem:measure_rank_multi} we deduce that there exist $\Measure_1 \le (n_1 , 0), \Measure_2 \le (n_2 , 0)$ such that
	\begin{itemize}
	\item $\wtpn{\Measure_1}{\Ctx_1, \ep{s}{\role_1} : T_1}{P'_1}$
	\item $\wtpn{\Measure_2}{\Ctx_2, \ep{s}{\role_2} : T_2, u : S}{P'_2\subst{u}{x}}$
	\end{itemize}
Let $\MeasureN \eqdef \Measure_1 
	+ \Measure_2 
	+ (\sum_{i=3}^h n_i, \rank{\Map{\role_1}{T_1} \parop \Map{\role_2}{T_2} \parop \prod_{i=3}^h \Map{\role_i}{S_i}})$. 
We conclude with one application of \refrule{mtm-par} observing that
\[
\begin{array}{rcll}
	\MeasureN & = & \Measure_1 + \Measure_2 + \\
	& & (\sum_{i=3}^h n_i, \rank{\Map{\role_1}{T_1} \parop \Map{\role_2}{T_2} \parop \prod_{i=3}^h \Map{\role_i}{S_i}})
	& \text{by definition of $\MeasureN$}
	\\
	& \le & (\sum_{i=1}^h n_i, \rank{\Map{\role_1}{T_1} \parop \Map{\role_2}{T_2} \parop \prod_{i=3}^h \Map{\role_i}{S_i}})
	& \text{by \Cref{lem:measure_rank_multi}}
	\\
	& < & (\sum_{i=1}^h n_i, \rank{\prod_{i=1}^h \Map{\role_i}{S_i}}) 
	& \text{before reductions}
	\\
	& = & \Measure
\end{array}
\]

\proofcase{Case $\PCtxC = \pres{t}{\procs{\Ppar} \parop \PCtxD \parop \procs{\Qpar}}$}
Let $\Rnf \eqdef \PCtxD[\pres{s}{\Pth_1 \parop \dots \parop \Pth_h}]$ and observe that $\Rnf$ is in proximity normal form. 
From \refrule{mtm-par} we deduce that there exist $\Ctx_i, S_i, \Measure_i, \role_i$ for $i=1,\dots,h$ and $k \le h$ such that
\begin{itemize}
\item $\Ctx = \Ctx_1,\dots,\Ctx_h$
\item $\wtpn{\Measure_1}{\Ctx_1, \ep{t}{\role_1} : S_1}{\Rnf}$
\item $\wtpn{\Measure_i}{\Ctx_i, \ep{t}{\role_i} : S_i}{\Ppar_i}$ for $i=1,\dots,k$
\item $\wtpn{\Measure_i}{\Ctx_i, \ep{t}{\role_i} : S_i}{\Qpar_i}$ for $i=k+1,\dots,h$
\item $\Measure = \sum_{i=1}^h \Measure_i + (0 , \rank{\prod_{i=1}^h \Map{\role_i}{S_i}})$
\end{itemize}
Using the induction hypothesis on $\wtpn{\Measure_1}{\Ctx_1, \ep{t}{\role_1} : S_1}{\Rnf}$ we deduce that there 
exists $Q'$ and $\MeasureN' < \Measure_1$ such that
\begin{itemize}
\item $\Rnf \wred^+ Q'$
\item $\wtpn{\MeasureN'}{\Ctx_1, \ep{t}{\role_1} : S_1}{Q'}$
\end{itemize}
We conclude taking $Q \eqdef \pres{t}{Q' \parop \procs\Ppar \parop \procs\Qpar}$ and
\[
	\MeasureN \eqdef \MeasureN' + \sum_{i=2}^h \Measure_i + (0 , \rank{\prod_{i=1}^h \Map{\role_i}{S_i}})
\] 
and observing that $\MeasureN < \Measure$ and $\Pnf \wred^+ Q$ by \refrule{rm-par}.

\proofcase{Case $\PCtxC = \pcast{\ep{t}{\roleq}}{\PCtxD}$}
Observe that $t \ne s$. Let $\Rnf \eqdef \PCtxD[\pres{s}{\Pth_1 \parop \cdots \parop \Pth_h}]$ 
and note that $\Rnf$ is in proximity normal form. 
From \refrule{mtm-cast} we deduce that there exists $\CtxD, \Measure', S, T, m_t$ such that
\begin{itemize}
\item $\Ctx = \CtxD, \ep{t}{\roleq} : S$
\item $S \subt[m_t] T$
\item $\Measure = \Measure' + (m_t,0)$
\item $\wtpn{\Measure'}{\CtxD, \ep{t}{\roleq} : T}{\Rnf}$
\end{itemize}
Using the induction hypothesis on $\wtpn{\Measure'}{\CtxD, \ep{t}{\roleq} : T}{\Rnf}$ we deduce that 
there exist $Q'$ and $\MeasureN' < \Measure'$ such that $\Rnf \wred^+ Q'$ and $\wtpn{\MeasureN'}{\CtxD, \ep{t}{\roleq} : T}{Q'}$. 
We conclude taking $Q \eqdef \pcast{\ep{t}{\roleq}}{Q'}$ and $\MeasureN \eqdef \MeasureN' + (m_t , 0)$ 
and observing that $\MeasureN < \Measure$ and $\Pnf \wred^+ Q$ by \refrule{rm-cast}.
\end{proof}

\begin{lemma}
	\label{lem:helpful_direction_multi}
	If $\wtpn{\Measure}{\EmptyCtx}{P}$, then either $P \pcong \pdone$ or $P \wred^+ Q$ 
	and $\wtpn{\MeasureN}{\EmptyCtx}{Q}$ for some $Q$ and $\MeasureN < \Measure$.
\end{lemma}
\begin{proof}
Using \Cref{lem:cnf_exists_multi} we deduce that there exist $P'$ in choice normal form such that
$P \wred P'$ and $\wtpn{\MeasureM'}\EmptyCtx{P'}$ and $\MeasureM' \leq \MeasureM$.
By \Cref{lem:measure_rank_multi} we deduce $\wtp\EmptyCtx{P'}$. 
Using \Cref{lem:tnf_exists_multi} we deduce that there exist $\Pnf$ such that $P' \pcong \Pnf$.

If $\Pnf = \pdone$ there is nothing left to prove.

If $\Pnf \neq \pdone$, by \Cref{lem:dl_freedom_multi} we deduce $\Pnf \pcong \Qnf$ for some $\Qnf$ in proximity normal form.
From \Cref{lem:measure_pcong_multi} we deduce $\wtpn{\MeasureM''}\EmptyCtx\Qnf$ for some $\MeasureM'' \leq \MeasureM'$.
Using \Cref{lem:pnf_helpful_direction_multi} we conclude that $\Qnf \wred^+ Q$ and $\wtpn\MeasureN\EmptyCtx Q$ 
for some $Q$ and $\MeasureN < \MeasureM'' \leq \MeasureM' \leq \MeasureM$.
\end{proof}

\begin{lemma}
	\label{lem:weak_termination_multi}
	If $\wtp[n]\EmptyCtx{P}$, then either $P \pcong \pdone$ or $P \wred^+ \pdone$.
\end{lemma}
\begin{proof}
From \Cref{lem:measure_rank_multi} we deduce that there exists $\MeasureM \le (n,0)$ such that $\wtpn\MeasureM\EmptyCtx P$.
We proceed doing an induction on the lexicographically ordered pair $\MeasureM$.
From \Cref{lem:helpful_direction_multi} we deduce either $P \pcong \pdone$ or $P \wred^+ Q$ 
and $\wtpn\MeasureN\EmptyCtx{Q}$ for some $\MeasureN < \MeasureM$.
In the first case there is nothing left to prove.
In the second case we use the induction hypothesis to deduce that either $Q \pcong \pdone$ or $Q \wred^+ \pdone$.
We conclude using either \refrule{rm-struct} or the transitivity of $\wred^+$, respectively.
\end{proof}

\begin{proof}[Proof of \Cref{thm:ts_multi_sound}]
	\brk
	Immediate consequence of \Cref{lem:subj_red_multi,lem:weak_termination_multi}.
\end{proof}

%% file: ts-multi/related.tex
\beginalto
We conclude the chapter by relating the type systems in 
\Cref{ch:ft_bin,ch:ft_multi} to others that can be found
in the literature. In particular, we mainly recall
the references we pointed out in \Cref{sec:related}.

\paragraph{Termination of Binary Sessions.}
\cite{LindleyMorris16} define a type system for a functional
language with session primitives and recursive session types that is strongly
normalizing. That is, a well-typed program along with all the sessions it
creates is guaranteed to terminate. This strong result is due to the fact that
the type language is equipped with least and greatest fixed point operators that
are required to match each other by duality.
Notably, strong normalization is stronger than fair termination. 
As an example, \Cref{ex:bsc_bin_proc,ex:bsc_multi} are fairly terminating but not
strongly terminating as the buyer can add arbitrarily many, possibly infinitely
many, items to the cart.

\paragraph{Liveness Properties in the $\pi$-Calculus.}
\cite{Kobayashi02,Padovani14} define a behavioral type system that guarantees
lock freedom in the $\pi$-calculus. .
These works annotate types with numbers representing finite upper bounds
to the number of interactions needed to unblock a particular input/output
action. For this reason, none of our key examples (\Cref{ex:bsc_multi,ex:2bsc_multi,ex:pms_multi})
is in the scope of these analysis techniques.
\cite{KobayashiSangiorgi10} show how to enforce lock
freedom by combining deadlock freedom and termination. Our work can be seen as a
generalization of this approach whereby we enforce lock freedom by combining
deadlock freedom (through a mostly conventional session type system) and
\emph{fair} termination. Since fair termination is coarser than termination, the
family of programs for which lock freedom can be proved is larger as well.

\paragraph{Deadlock Freedom.}
Our type system enforces deadlock freedom essentially thanks to the shape of the
rule \refrule{tm-par} (\refrule{tb-par} in \Cref{ch:ft_bin}) which is inspired 
to the cut rule of linear logic. This
rule has been applied to session type systems for binary
sessions \citep{Wadler14,CairesPfenningToninho16,LindleyMorris16} and
subsequently extended to multiparty
sessions \citep{CarboneLMSW16,CarboneMontesiSchurmannYoshida17}. In the latter
case, the rule -- dubbed \emph{multiparty cut} -- requires a coherence condition
among cut types establishing that the session types followed by the single
participants adhere to a so-called global type describing the multiparty session
as a whole. The rule \refrule{tm-par} adopts the same schema, except that the
coherence condition is stronger to entail fair session termination. 
The key principle of these formulations of the cut rule as a typing rule for
parallel processes is to impose a tree-like network topology, whereby two
parallel processes can share at most one channel. In the multiparty case, cyclic
network topologies can be modeled within each session (\Cref{ex:pms_multi}) since
coherence implies deadlock freedom.

Having a single construct that merges session restriction and parallel
composition allows for a simple formulation of the typing rules so that dealock
freedom is easily guaranteed. However, many session calculi separate these two
forms in line with the original presentation of the $\pi$-calculus. We think
that our type system can be easily reformulated to support distinct session
restriction and parallel composition by means of
hypersequents \citep{KokkeMontesiPeressotti18,KokkeMontesiPeressotti19}.

A more liberal version of the cut rule, named multi-cut and inspired to
Gentzen's ``mix'' rule, is considered by \citep{AbramskyGN96}
enabling processes to share more than one channel. In this setting,
deadlock freedom is lost but can be recovered by means of a richer type structure
that keeps track of the dependencies between different channels. This approach
has been pioneered by \cite{Kobayashi02, Kobayashi06} for the
$\pi$-calculus and later on refined by \cite{Padovani14}.
Other approaches to ensure deadlock freedom based on
\emph{dependency/connectivity graphs} that capture the network topology
implemented by processes have been studied by 
\cite{CarboneDebois10,KobayashiLaneve17,deLiguoroP18,JacobsBalzerKrebbers22}.

\paragraph{Liveness Properties of Multiparty Sessions.}
In \cite{ScalasYoshida19} the authors point out that the coarsest liveness
property in the hierarchy of liveness properties that they take into account, which is the one more closely related to fair
termination, cannot be enforced by their type system. In part, this is due to
the fact that their type system relies on a standard subtyping relation for
session types \citep{GayHole05} instead of fair subtyping
\citep{Padovani13,Padovani16}. As we have seen in \Cref{ch:ft_bin}, even for
single-session programs the mere adoption of fair subtyping is not enough and it
is necessary to meet additional requirements.
\cite{GlabbeekHofnerHorne21} propose a type
system for multiparty sessions that ensures progress and is not only sound but
also complete. The fairness assumption they make -- called \emph{justness} -- is
substantially weaker than our own (\Cref{def:fair_run}) and such that the unfair
runs are those in which some interactions between participants are
systematically discriminated in favor of other interactions involving a disjoint
set of independent participants. For this reason, their progress property is in
between the two more restrictive liveness predicates of \cite{ScalasYoshida19} 
and can only be guaranteed when it is independent
of the behavior of the other participants of the same session.

%% file: ts-ll/chintro.tex
\begintreble
%
%
In this chapter we propose a type system for \piLIN, a linear $\pi$-calculus with
(co)recursive types, such that well-typed processes are \emph{fairly
terminating}.
Our type system is a conservative extension of
\muMALL \citep{BaeldeDoumaneSaurin16,Doumane17,BaeldeEtAl22}, the infinitary
proof system for the multiplicative additive fragment of linear logic with least
and greatest fixed points. In fact, the modifications we make to \muMALL are
remarkably small: we add one (standard) rule to deal with
\emph{non-deterministic choices}, those performed autonomously by a process, and
we relax the validity condition on \muMALL proofs so that it only considers the
``fair behaviors'' of the program it represents.
The fact that there is such a close correspondence between the typing rules of
\piLIN and the inference rules of \muMALL is not entirely surprising. After all,
there have been plenty of works investigating the relationship between
$\pi$-calculus terms and linear logic proofs, from those of \cite{Abramsky94}, 
\cite{BellinScott94} to those on
the interpretation of linear logic formulas as session
types \citep{DeYoungCairesPfenningToninho12,Wadler14,CairesPfenningToninho16,LindleyMorris16,RochaCaires21,QianKavvosBirkedal21}.

Nonetheless, we think that the connection between \piLIN and \muMALL stands out
for two reasons.
First, \piLIN is conceptually simpler and more general than the session-based
calculi that can be encoded in it. In particular, all the session calculi based
on linear logic rely on an asymmetric interpretation of the multiplicative
connectives $\tfork$ and $\tjoin$ so that $\FormulaF \tfork \FormulaG$
(respectively, $\FormulaF \tjoin \FormulaG$) is the type of a session endpoint
used for sending (respectively, receiving) a message of type $\FormulaF$ and
then used according to $\FormulaG$. In our setting, the connectives $\tfork$ and
$\tjoin$ retain their symmetry since we interpret $\FormulaF \tfork \FormulaG$
and $\FormulaF \tjoin \FormulaG$ formulas as the output/input of pairs, in the
same spirit of the original encoding of linear logic proofs proposed by 
\cite{BellinScott94}. This interpretation gives \piLIN the ability of
modeling \emph{bifurcating protocols} of which binary sessions are just a
special case.
The second reason why \piLIN and \muMALL get along has to do with the cut
elimination result for \muMALL. In finitary proof systems for linear logic, cut
elimination may proceed by removing \emph{topmost cuts}. In \muMALL there is no
such notion as a topmost cut since \muMALL proofs may be infinite. As a
consequence, the cut elimination result for \muMALL is proved by eliminating
\emph{bottom-most cuts} \citep{BaeldeEtAl22}. This strategy fits perfectly with
the reduction semantics of {\piLIN} -- and that of any other conventional
process calculus, for that matter -- whereby reduction rules act only on the
exposed (\ie unguarded) part of processes but not behind prefixes. As a result,
the reduction semantics of \piLIN is completely ordinary, unlike other
logically-inspired process calculi that incorporate commuting conversions
\citep{Wadler14,LindleyMorris16}, perform reductions behind
prefixes \citep{QianKavvosBirkedal21} or swap prefixes \citep{BellinScott94}.

In \Cref{ch:ft_bin,ch:ft_multi} we have proposed a type system
ensuring the fair termination of binary/multiparty sessions. 
In the present chapter we achieve the
same objective using a more basic process calculus and exploiting its strong
logical connection with \muMALL. In fact, the soundness proof of our type system
piggybacks on the cut elimination property of \muMALL.
Other session typed calculi based on linear logic with fixed points have been
studied by \cite{LindleyMorris16} and \cite{DerakhshanPfenning19,Derakhshan21}.
The type systems described in these works respectively guarantee termination and
strong progress, whereas our type system guarantees fair termination which is
somewhat in between these properties. Overall, our type system seems to hit a
sweet spot: on the one hand, it is deeply rooted in linear logic and yet it can
deal with common communication patterns (like the buyer/seller interaction
described above) that admit potentially infinite executions and therefore are
out of scope of other logic-inspired type systems; on the other hand, it
guarantees lock freedom \citep{Kobayashi02,Padovani14}, strong
progress \citep[Theorem 12.3]{DerakhshanPfenning19} and also termination, under a
suitable fairness assumption.

The chapter is organized as follows.
In \Cref{sec:ts_ll_types} we show the definition of the types bu relying on
the formulas of linear logic. Notably, these contain least/greatest fixed points.
In \Cref{sec:ts_ll_proc} we present syntax and semantics of \piLIN.
\Cref{sec:ts_ll_ts} introduces the type system for \piLIN as well as the 
validity conditions.
As usual, we dedicate \Cref{sec:ts_ll_corr} to detail the soundness proof 
of the type system and \Cref{sec:ts_ll_related} to discuss about related works.
Finally, we are now able to make a comparison between the present type system and
those presented in \Cref{ch:ft_bin,ch:ft_multi}.
We make such comparison by examples in \Cref{sec:comparison}.

%% file: ts-ll/types.tex
\beginalto
The types of \piLIN are built using the multiplicative additive fragment of
linear logic enriched with least and greatest fixed points. In this section we
specify the syntax of types along with all the auxiliary notions that are needed
to present the type system and prove its soundness.

\begin{definition}[Pre-formula]
	The syntax of pre-formulas relies on an infinite set of \emph{propositional
	variables} ranged over by $X$ and $Y$ and is defined by the grammar below:
\[
\begin{array}{rcl}
    \FormulaF, \FormulaG &
    ::= &
    \Zero \mid
    \Top \mid
    \One \mid
    \Bot \mid
    \FormulaF \plinchoice \FormulaG \mid
    \FormulaF \plinbranch \FormulaG \mid
    \FormulaF \tfork \FormulaG \mid
    \FormulaF \tjoin \FormulaG \mid
    \tmu\X.\Formula \mid
    \tnu\X.\Formula \mid
    X
\end{array}
\]
\end{definition}

As usual, $\tmu$ and $\tnu$ are the binders of propositional variables and the
notions of free and bound variables are defined accordingly. We assume that the
body of fixed points extends as much as possible to the right of a pre-formula,
so $\tmu\X.X \plinchoice \One$ means $\tmu\X.(X \plinchoice \One)$ and not $(\tmu\X.X)
\plinchoice \One$. We write $\subst\Formula\X$ for the capture-avoiding substitution
of all free occurrences of $X$ with $\Formula$. 

\begin{definition}[Dual of a formula]
	We write $\dual\Formula$ for the
	\emph{dual} of $\Formula$, which is the involution defined by the equations
\[
    \begin{array}{*{6}{@{}r@{~}c@{~}l@{\quad}}@{}}
        \dual\Zero & = & \Top & \qquad
        \dual{(\FormulaF \plinchoice \FormulaG)} & = & \dual\FormulaF \plinbranch \dual\FormulaG & \qquad
        \dual{(\tmu\X.\Formula)} & = & \tnu\X.\dual\Formula &
        \\
        \dual\One & = & \Bot &
        \dual{(\FormulaF \tfork \FormulaG)} & = & \dual\FormulaF \tjoin \dual\FormulaG &
        \dual\X & = & X
    \end{array}
\]
\end{definition}

\begin{definition}[Formula]
	A \emph{formula} is a closed pre-formula.
\end{definition}

In the context of \piLIN, formulas
describe how linear channels are used. Positive formulas (those built with the
constants $\Zero$ and $\One$, the connectives $\plinchoice$ and $\tfork$ and the
least fixed point) indicate output operations whereas negative formulas (the
remaining forms) indicate input operations. The formulas $\FormulaF \plinchoice
\FormulaG$ and $\FormulaF \plinbranch \FormulaG$ describe a linear channel used for
sending/receiving a tagged channel of type $\FormulaF$ or $\FormulaG$. The tag
(either $\LeftTag$ or $\RightTag$) distinguishes between the two possibilities.
The formulas $\FormulaF \tfork \FormulaG$ and $\FormulaF \tjoin \FormulaG$
describe a linear channel used for sending/receiving a pair of channels of type
$\FormulaF$ and $\FormulaG$; $\tmu\X.\Formula$ and $\tnu\X.\Formula$ describe a
linear channel used for sending/receiving a channel of type
$\Formula\subst{\tmu\X.\Formula}\X$ or $\Formula\subst{\tnu\X.\Formula}\X$
respectively. The constants $\One$ and $\Bot$ describe a linear channel used for
sending/receiving the unit. Finally, the constants $\Zero$ and $\Top$
respectively describe channels on which nothing can be sent and from which
nothing can be received.

\begin{example}
    \label{ex:bsc_ll_formulas}
    Looking at the structure of \actor{buyer} and \actor{seller} in
    \Cref{ex:bsc}, we can make an educated guess on the type of the
    channel they use. Indeed, we see that it is used according to
    $\FormulaF \eqdef \tmu\X.X \plinchoice \One$ by \actor{buyer} and according to
    $\FormulaG \eqdef \tnu\X.X \plinbranch \Bot$ in \actor{seller}. 
    Note that $\FormulaF = \dual\FormulaG$, 
    suggesting that \actor{seller} and \actor{seller} may interact correctly when connected.
    \eoe
\end{example}

\begin{definition}[Subformula ordering]
	We write $\subf$ for the \emph{subformula ordering}, that is the least partial
	order such that $\FormulaF \subf \FormulaG$ if $\FormulaF$ is a subformula of
	$\FormulaG$.
\end{definition}

For example, consider $\FormulaF \eqdef \tmu\X.\tnu\Y.X \plinchoice Y$
and $\FormulaG \eqdef \tnu\Y.\FormulaF \plinchoice Y$. Then we have $\FormulaF \subf
\FormulaG$ and $\FormulaG \not\subf \FormulaF$. When $\Formulas$ is a set of
formulas, we write $\minf\Formulas$ for its $\subf$-minimum formula if it
is defined.
Occasionally we let $\star$ stand for an arbitrary binary connective $\plinchoice$,
$\tfork$, $\plinbranch$, or $\tjoin$ and $\sigma$ stand for an arbitrary fixed point
operator $\tmu$ or $\tnu$.

When two \piLIN processes interact on some channel $x$, they may exchange other
channels on which their interaction continues. We can think of these subsequent
interactions stemming from a shared channel $x$ as being part of the same
conversation (the literature on \emph{sessions} \citep{Honda93,HuttelEtAl16}
builds on this idea \citep{Kobayashi02b,DardhaGiachinoSangiorgi17}). The
soundness proof of the type system is heavily based on the proof of the cut
elimination property of \muMALL, which relies on the ability to uniquely
identify the types of the channels that belong to the same conversation and to
trace conversations within typing derivations. Following the literature on
\muMALL \citep{BaeldeDoumaneSaurin16,Doumane17,BaeldeEtAl22}, we annotate
formulas with addresses.
We assume an infinite set $\AddressSet$ of \emph{atomic addresses},
$\dual\AddressSet$ being the set of their duals such that $\AddressSet \cap
\dual\AddressSet = \emptyset$ and $\dual{\dual\AddressSet{}} = \AddressSet$. We
use $a$ and $b$ to range over elements of $\AddressSet \cup \dual\AddressSet$.

\begin{definition}[Address]
	An \emph{address} is a string $aw$ where $w \in \set{i,l,r}^*$. The dual of an
	address is defined as $\dual{(aw)} = \dual{a}w$.
\end{definition}

We use $\addressA$ and $\addressB$ to range over addresses, we write $\prefix$
for the prefix relation on addresses and we say that $\addressA$ and $\addressB$
are \emph{disjoint} if $\addressA \not\prefix \addressB$ and $\addressB
\not\prefix \addressA$.

\begin{definition}[Type]
	A \emph{type} is a formula $\Formula$ paired with an address $\address$ written
	$\Formula_\address$.
\end{definition}

We use $S$ and $T$ to range over types and we extend to
types several operations defined on formulas: we use logical connectives to
compose types so that $\FormulaF_{\address l} \mathbin\star \FormulaG_{\address
r} \eqdef (\FormulaF \mathbin\star \FormulaG)_\address$ and
$\sigma\X.\Formula_{\address i} \eqdef (\sigma\X.\Formula)_\address$; the dual
of a type is obtained by dualizing both its formula and its address, that is
$\dual{(\Formula_\address)} \eqdef \dual\Formula_{\dual\address}$; type
substitution preserves the address in the type within which the substitution
occurs, but forgets the address of the type being substituted, that is
$\FormulaF_\addressA\subst{\FormulaG_\addressB}\X \eqdef
\FormulaF\subst\FormulaG\X_\addressA$.

We often omit the address of constants (which represent terminated
conversations) and we write $\strip{S}$ for the formula obtained by forgetting
the address of $S$. Finally, we write $\tred$ for the least reflexive relation
on types such that $S_1 \star S_2 \tred S_i$ and $\sigma\X.S \tred
S\subst{\sigma\X.S}\X$.

\begin{example}
    \label{ex:bsc_ll_types}
    Consider once again the formula $\FormulaF \eqdef \tmu\X.X \plinchoice \One$ that
    describes the behavior of \actor{buyer} (\Cref{ex:bsc_ll_formulas}) and let
    $a$ be an arbitrary atomic address. We have
    \[
        \Formula_a
        \tred (\Formula \plinchoice \One)_{ai}
        \tred \Formula_{ail}
        \tred (\Formula \plinchoice \One)_{aili}
        \tred \One_{ailir}
    \]
    where the fact that the types in this sequence all share a common non-empty
    prefix `$a$' indicates that they belong to the same conversation.
    Note how the symbols $i$, $l$ and $r$ composing an address indicate the step
    taken in the syntax tree of types for making a move in this sequence: $i$
    means ``inside'', when a fixed point operator is unfolded, whereas $l$ and
    $r$ mean ``left'' and ``right'', when the corresponding branch of a
    connective is selected.
    \eoe
\end{example}

%% file: ts-ll/calculus.tex
\beginalto
In this section we define syntax and reduction semantics of \piLIN, a variant of
the linear $\pi$-calculus \citep{KobayashiPierceTurner99} in which all channels
are meant to be used for \emph{exactly} one communication. The calculus supports
(co)recursive data types built using units, pairs and disjoint sums. These data
types are known to be the essential ingredients for the encoding of sessions in
the linear
$\pi$-calculus \citep{Kobayashi02b,DardhaGiachinoSangiorgi17,ScalasDardhaHuYoshida17}.


\subsection{Syntax of Processes}
\input{ts-ll/proc-syntax}


\subsection{Operational Semantics}
\input{ts-ll/proc-sem}

%% file: ts-ll/proc-syntax.tex
\beginbass
\begin{figure}[t]
	\framebox[\textwidth]{
    \begin{math}
        \displaystyle
        \begin{array}[t]{@{}r@{~}c@{~}ll@{}}
            P, Q
            & ::= & \Link\x\y & \text{link}
            \\
            & | & \Fail\x & \text{empty in}
            \\
            & | & \PiWait\x.P & \text{unit in}
            \\
            & | & \Join[y]\x\z.P & \text{pair in}
            \\
            & | & \Case[y]\x{P}{Q} & \text{sum in}
            \\
            & | & \Corec[y]\x.P & \text{corec}
        \end{array}
        ~
        \begin{array}[t]{@{}r@{~}c@{~}lll@{}}
            & | & \Cut\x{P}{Q} & \text{comp}
            \\
            & | & \PiChoice{P}{Q} & \text{choice}
            \\
            & | & \PiClose\x & \text{unit out}
            \\
            & | & \Fork[y]\x\z{P}{Q} & \text{pair out}
            \\
            & | & \Select[y]{\InTag_i}\x.P & \text{sum out} & i\in\set{1,2}
            \\
            & | & \Rec[y]\x.P & \text{rec}
        \end{array}
    \end{math}
  	}
    \caption{Syntax of \piLIN}
    \label{fig:proc_syntax_ll}
\end{figure}
We assume an infinite set of \emph{channels} ranged over by $x$, $y$ and
$z$. \piLIN processes are coinductively generated by the productions of the
grammar shown in \Cref{fig:proc_syntax_ll}. 
A \emph{link} $\Link\x\y$ acts as a \emph{linear
forwarder} \cite{GardnerLaneveWischik07} that forwards a single message either
from $x$ to $y$ or from $y$ to $x$. The uncertainty in the direction of the
message is resolved once the term is typed and the polarity of the types of $x$
and $y$ is fixed (as we will show in \Cref{sec:ts_ll_ts}).
The term $\Fail\x$ represents a process that receives an empty message from $x$
and then fails. This form is only useful in the metatheory: the type system
guarantees that well-typed processes never fail, since it is not possible to
send empty messages.
The term $\PiClose\x$ models a process that sends the unit on $x$, effectively
indicating that the interaction is terminated, whereas $\PiWait\x.P$ models a
process that receives the unit from $x$ and then continues as $P$.
The term $\Fork[z]\x\y{P}{Q}$ models a process that creates two new channels $y$
and $z$, sends them in a pair on channel $x$ and then forks into two parallel
processes $P$ and $Q$. Dually, $\Join[z]\x\y.P$ models a process that receives a
pair containing two channels $y$ and $z$ from channel $x$ and then continues as
$P$.
The term $\Select[y]{\InTag_i}\x.P$ models a process that creates a new channel
$y$ and sends $\InTag_i\parens\y$ (that is, the $i$-th injection of $y$ in a
disjoint sum) on~$x$. Dually, $\Case[y]\x{P_1}{P_2}$ receives a disjoint sum
from channel $x$ and continues as either $P_1$ or $P_2$ depending on the tag
$\InTag_i$ it has been built with. For clarity, in some examples we will use
more descriptive labels such as $\AddTag$ and $\PayTag$ instead of $\LeftTag$
and $\RightTag$.
The terms $\Rec[y]\x.P$ and $\Corec[y]\x.P$ model processes that respectively
send and receive a new channel $y$ and then continue as $P$. They do not
contribute operationally to the interaction being modeled, but they indicate the
points in a program where (co)recursive types are unfolded.
A term $\Cut\x{P}{Q}$ denotes the parallel composition of two processes $P$ and
$Q$ that interact through the fresh channel $x$.
Finally, the term $\PiChoice{P}{Q}$ models a non-deterministic choice between two
behaviors $P$ and $Q$.

\piLIN binders are easily recognizable because they enclose channel names in
round parentheses. Note that all outputs are in fact \emph{bound outputs}. The
output of free channels can be modeled by combining bound outputs with
links \citep{LindleyMorris16}. For example, the output $\FreeFork\x\y\z$ of a
pair of free channels $y$ and $z$ can be modeled as the term
$\Fork[z']\x{y'}{\Link\y{y'}}{\Link\z{z'}}$.
We identify processes modulo renaming of bound names, we write $\fn{P}$ for the
set of channel names occurring free in $P$ and we write $\subst\y\x$ for the
capture-avoiding substitution of $y$ for the free occurrences of $x$.
We impose a well-formedness condition on processes so that, in every sub-term of
the form $\Fork[z]\x\y{P}{Q}$, we have $y \not\in\fn{Q}$ and $z\not\in\fn{P}$.

We omit any concrete syntax for representing infinite processes. Instead, we
work directly with infinite trees obtained by corecursively unfolding
contractive equations of the form $A(x_1,\dots,x_n) = P$. For each such
equation, we assume that $\fn{P} \subseteq \set{x_1,\dots,x_n}$ and we write
$\Call{A}{y_1,\dots,y_n}$ for its unfolding $P\subst{y_i}{x_i}_{1\leq i\leq n}$.

\begin{notation}
    \label{not:sessions}
    To reduce clutter due to the systematic use of bound outputs, by convention
    we omit the continuation called $y$ in \Cref{fig:proc_syntax_ll} when its name is
    chosen to coincide with that of the channel $x$ on which $y$ is
    sent/received.
    For example, with this notation we have $\Join\x\z.P = \Join[x]\x\z.P$ and
    $\Select{\InTag_i}\x.P = \Select[x]{\InTag_i}\x.P$ and $\Case\x{P}{Q} =
    \Case[x]\x{P}{Q}$.
    \eoe
\end{notation}

A welcome side effect of adopting \Cref{not:sessions} is that it gives the
illusion of working with a session calculus in which the same channel $x$ may be
used repeatedly for multiple input/output operations, while in fact $x$ is a
linear channel used for exchanging a single message along with a fresh
continuation that turns out to have the same name. If one takes this notation as
native syntax for a session calculus, its linear $\pi$-calculus
encoding \citep{DardhaGiachinoSangiorgi17} turns out to be precisely the \piLIN
term it denotes.
Besides, the idea of rebinding the same name over and over is widespread in
session-based functional languages \citep{GayVasconcelos10,Padovani17} as it
provides a simple way of ``updating the type'' of a session endpoint after each
use.

\begin{example}
    \label{ex:bsc_ll_proc}
    Below we model the interaction informally described in
    \Cref{ex:bsc} between \actor{buyer} and \actor{seller} using  the syntactic sugar
    defined in \Cref{not:sessions}:
    \[
    \begin{array}{@{}r@{~}l@{}}
    				& \Cut\x{\Call\Buyer\x}{\Call\Seller{x,y}} \\
            \Buyer(x) & =
            \Rec\x.\parens{
                \PiChoice{
                    \Select\AddTag\x.\Call\Buyer\x
                }{
                    \Select\PayTag\x.\PiClose\x
                }
            }
            \\
            \Seller(x,y) & =
            \Corec\x.
            \Case\x{
                \Call\Seller{x,y}
            }{
                \PiWait\x.\PiClose\y
            }
        \end{array}
    \]

    At each round of the interaction, the buyer decides whether to $\AddTag$ an
    item to the shopping cart and repeat the same behavior (left branch of the
    choice) or to $\PayTag$ the seller and terminate (right branch of the
    choice). The seller reacts dually and signals its termination by sending a
    unit on the channel $y$. As we will see in \Cref{sec:ts_ll_ts}, $\Rec\x$
    and $\Corec\x$ identify the points within processes where (co)recursive
    types are unfolded.

    If we were to define $\Buyer$ using distinct bound names we would write an
    equation like
    \[
        \Buyer(x) =
        \Rec[y]\x.
        \parens{
            \PiChoice{
                \Select[z]\AddTag\y.\Call\Buyer\z
            }{
                \Select[z]\PayTag\y.\PiClose\z
            }
        }
    \]
    and similarly for $\Seller$. 
    \eoe
\end{example}

%% file: ts-ll/proc-sem.tex
\beginbass
\begin{figure}[t]
		\framebox[\textwidth]{
    \begin{math}
        \displaystyle
        \begin{array}{@{}lr@{~}c@{~}ll@{}}
            \defrule{sp-link} &
            \Link\x\y & \pcong & \Link\y\x
            \\
            \defrule{sp-comm} &
            \Cut\x{P}{Q} & \pcong & \Cut\x{Q}{P}
            \\
            \defrule{sp-assoc} &
            \Cut\x{P}{\Cut\y{Q}{R}} & \pcong & \Cut\y{\Cut\x{P}{Q}}{R} &
            \text{if $x\in\fn{Q}$,}
            \\
            & & & & \text{$y\not\in\fn{P}$, $x\not\in\fn{R}$}
        \end{array}
    \end{math}
    }
    \caption{Structural pre-congruence of \piLIN}
    \label{fig:pcong_ll}
\end{figure}
\begin{figure}[t]
    \framebox[\textwidth]{
    \begin{mathpar}
        \displaystyle
        \inferrule[rp-link]{\mathstrut}{\Cut\x{\Link\x\y}{P} \red P\subst\y\x}
        \defrule[rp-link]{}
        \and
        \inferrule[rp-unit]{\mathstrut}{\Cut\x{\PiClose\x}{\PiWait\x.P} \red P}
        \defrule[rp-unit]{}
        \and
        \inferrule[rp-pair]
        	{\mathstrut}
        	{\Cut\x{\Fork[y]\x\z{P_1}{P_2}}{\Join[y]\x\z.Q} \red \Cut\z{P_1}{\Cut\y{P_2}{Q}}}
        	\defrule[rp-pair]{}
        \and
        \inferrule[rp-sum]
        	{\mathstrut}
        	{\Cut\x{\Select[y]{\InTag_i}\x.P}{\Case[y]\x{P_1}{P_2}} \red \Cut\y{P}{P_i}}
          ~ i\in\set{1,2}
          \defrule[rp-sum]{}
        \and
        \inferrule[rp-rec]
        	{\mathstrut}
        	{\Cut\x{\Rec[y]\x.P}{\Corec[y]\x.Q} \red \Cut\y{P}{Q}}
        	\defrule[rp-rec]{}
        \and
        \inferrule[rp-choice]
        	{\mathstrut}
        	{\PiChoice{P_1}{P_2} \red P_i}
        	~ i\in\set{1,2}
        	\defrule[rp-choice]{}
				\and
				\inferrule[rp-cut]
					{P \red Q}
					{\Cut\x{P}{R} \red \Cut\x{Q}{R}}
					\defrule[rp-cut]{}
        \and
        \inferrule[rp-struct]
        	{P \pcong P' \\ P' \red Q' \\ Q' \pcong Q}
        	{P \red Q}
        	\defrule[rp-struct]{}
    \end{mathpar}
    }
    \caption{Reduction of \piLIN}
    \label{fig:red_ll}
\end{figure}
The operational semantics of the calculus is given in terms of the
\emph{structural precongruence} relation $\pcong$ and the \emph{reduction
relation} $\red$ defined in \Cref{fig:pcong_ll,fig:red_ll}.
As usual, structural precongruence relates processes that are syntactically
different but semantically equivalent. In particular, \refrule{sp-link} states
that linking $x$ with $y$ is the same as linking $y$ with $x$, whereas
\refrule{sp-comm} and \refrule{sp-assoc} state the expected  commutativity and
associativity laws for parallel composition. Concerning the latter, the side
condition $x\in\fn{Q}$ makes sure that $Q$ (the process brought closer to $P$
when the relation is read from left to right) is indeed connected with $P$ by
means of the channel $x$.
Note that \refrule{sp-assoc} only states the right-to-left associativity of
parallel composition and that the left-to-right associativity law
$\Cut\x{\Cut\y{P}{Q}}{R} \pcong \Cut\y{P}{\Cut\x{Q}{R}}$ is derivable when
$x\in\fn{Q}$.
The reduction relation is mostly unremarkable. Links are reduced with
\refrule{rp-link} by effectively merging the linked channels. All the reductions
that involve the interaction between processes except \refrule{rp-unit} create
new continuations channels that connect the reducts. The rule \refrule{rp-choice}
models the non-deterministic choice between two behaviors. Finally,
\refrule{rp-cut} and \refrule{rp-struct} close reductions by cuts and structural
precongruence.
In the following we write $\wred$ for the reflexive, transitive closure of
$\red$ and we say that $P$ is \emph{stuck} if there is no $Q$ such that $P \red
Q$.

%% file: ts-ll/ts.tex
\beginalto
In this section we present the typing rules for \piLIN.
As usual we introduce typing contexts to track the type of the names occurring
free in a process. A \emph{typing context} is a finite map from names to types
written $x_1 : S_1, \dots, x_n : S_n$. We use $\CtxC$ and $\CtxD$ to
range over contexts, we write $\dom\Ctx$ for the domain of $\Ctx$ and
$\Ctx,\CtxD$ for the union of $\Ctx$ and $\CtxD$ when
$\dom\Ctx \cap \dom\CtxD = \emptyset$.
Typing judgments have the form $\qtp\Ctx{P}$ and we say that $P$ is \emph{quasi
typed} in $\Ctx$.
For the time being we say ``quasi typed'' and not ``well typed'' because some
infinite derivations using the rules in \Cref{ssec:ts_ll} are invalid.
Well-typed processes are quasi-typed processes whose typing derivation satisfies
some additional validity conditions that we detail in \Cref{ssec:ts_ll_wtp}.


\subsection{Typing Rules}
\label{ssec:ts_ll}
\input{ts-ll/ts-rules}


\subsection{From Quasi Typed to Well Typed Processes}
\label{ssec:ts_ll_wtp}
\input{ts-ll/ts-wtp}


\subsection{Examples}
\label{ssec:ts_ll_ex}
\input{ts-ll/ts-examples}

%% file: ts-ll/ts-rules.tex
\beginbass
\begin{figure}[t]
  	\framebox[\textwidth]{
    \begin{mathpar}
        \inferrule{\mathstrut}{
            \qtp{x : \Formula_\addressA, y : \dual\Formula_\addressB}{\Link\x\y}
        }
        ~\defrule\LinkRule
        \and
        \inferrule{
            \qtp{\CtxC, x : S}{P}
            \\
            \qtp{\CtxD, x : \dual{S}}{Q}
        }{
            \qtp{\CtxC, \CtxD}{\Cut\x{P}{Q}}
        }
        ~\defrule\CutRule
        \and
        \inferrule{\mathstrut}{
            \qtp{\Ctx, x : \Top}{\Fail\x}
        }
        ~\defrule[fail]\FailRule
        \and
        \inferrule{
            \qtp\Ctx{P}
        }{
            \qtp{\Ctx, x : \Bot}{\PiWait\x.P}
        }
        ~\defrule[wait]\PiWaitRule
        \and
        \inferrule{\mathstrut}{
            \qtp{x : \One}{\PiClose\x}
        }
        ~\defrule[close]\PiCloseRule
        \and
        \inferrule{
            \qtp{\Ctx, y : S, z : T}{P}
        }{
            \qtp{\Ctx, x : S \tjoin T}{\Join[z]\x\y.P}
        }
        ~\defrule[join]\JoinRule
        \and
        \inferrule{
            \qtp{\CtxC, y : S}{P}
            \\
            \qtp{\CtxD, z : T}{Q}
        }{
            \qtp{\CtxC, \CtxD, x : S \tfork T}{\Fork[z]\x\y{P}{Q}}
        }
        ~\defrule[fork]\ForkRule
        \and
        \inferrule{
            \qtp{\Ctx, y : S}{P}
            \\
            \qtp{\Ctx, y : T}{Q}
        }{
            \qtp{\Ctx, x : S \plinbranch T}{\Case[y]\x{P}{Q}}
        }
        ~\defrule[case]\CaseRule
        \and
        \inferrule{
            \qtp{\Ctx, y : S_i}{P}
        }{
            \qtp{\Ctx, x : S_1 \plinchoice S_2}{\Select[y]{\InTag_i}\x.P}
        }
        ~\defrule[select]\SelectRule
        \and
        \inferrule{
            \qtp{\Ctx, y : S\subst{\tnu\X.S}X}{P}
        }{
            \qtp{\Ctx, x : \tnu\X.S}{\Corec[y]\x.P}
        }
        ~\defrule[corec]\CorecRule
        \and
        \inferrule{
            \qtp{\Ctx, y : S\subst{\tmu\X.S}X}{P}
        }{
            \qtp{\Ctx, x : \tmu\X.S}{\Rec[y]\x.P}
        }
        ~\defrule[rec]\RecRule
        \and
        \inferrule{
            \qtp\Ctx{P}
            \\
            \qtp\Ctx{Q}
        }{
            \qtp\Ctx{\PiChoice{P}{Q}}
        }
        ~\defrule\PiChoiceRule
    \end{mathpar}
    }
    \caption{Typing rules for \piLIN}
    \label{fig:ts_ll}
\end{figure}
We say that $P$ is \emph{quasi
typed} in $\Ctx$ if the judgment $\qtp\Ctx{P}$ is coinductively
derivable using the rules shown in \Cref{fig:ts_ll}
Rule \refrule{\LinkRule} states that a link $\Link\x\y$ is quasi typed provided
that $x$ and $y$ have dual types, but not necessarily dual addresses.
Rule \refrule{\CutRule} states that a process composition $\Cut\x{P}{Q}$ is
quasi typed provided that $P$ and $Q$ use the linear channel $x$ in
complementary ways, one according to some type $S$ and the other according to
the dual type $\dual{S}$. Note that the context $\CtxC,\CtxD$ in the
conclusion of the rule is defined provided that $\CtxC$ and $\CtxD$ have
disjoint domains. This condition entails that $P$ and $Q$ do not share any
channel other than $x$ ensuring that the interation between $P$ and $Q$ may
proceed without deadlocks.
Rule \refrule[fail]{\FailRule} deals with a process that receives an empty
message from channel $x$. Since this cannot happen, we allow the process to be
quasi typed in any context.
Rules \refrule[close]{\PiCloseRule} and \refrule[wait]{\PiWaitRule} concern the
exchange of units. The former rule states that $\PiClose\x$ is quasi typed in a
context that contains a single association for the $x$ channel with type $\One$,
whereas the latter rule removes $x$ from the context (hence from the set of
usable channels), requiring the continuation process to be quasi typed in the
remaining context.
Rules \refrule[fork]{\ForkRule} and \refrule[join]{\JoinRule} concern the
exchange of pairs. The former rule requires the two forked processes $P$ and $Q$
to be quasi typed in the respective contexts enriched with associations for the
continuation channels $y$ and $z$ being created. The latter rule requires the
continuation process to be quasi typed in a context enriched with the channels
extracted from the received pair.
Rules \refrule[select]{\SelectRule} and \refrule[case]{\CaseRule} deal with the
exchange of disjoint sums in the expected way.
Rules \refrule[rec]{\RecRule} and \refrule[corec]{\CorecRule} deal with fixed
point operators by unfolding the (co)recursive type of the channel $x$. As in
\muMALL, the two rules have exactly the same structure despite the fact that the
two fixed point operators being used are dual to each other. Clearly, the
behavior of least and greatest fixed points must be distinguished by some other
means, as we will see in \Cref{ssec:ts_ll_wtp} when discussing the validity of a
typing derivation.
Finally, \refrule{\PiChoiceRule} deals with non-deterministic choices by requiring
that each branch of a choice must be quasi typed in exactly the same typing
context as the conclusion.

Besides the structural constraints imposed by the typing rules, we implicitly
require that the types in the range of all typing contexts have pairwise
disjoint addresses. This condition ensures that it is possible to uniquely trace
a communication protocol in a typing derivation: if we have two channels $x$ and
$y$ associated with two types $\FormulaF_\addressA$ and $\FormulaG_\addressB$
such that $\addressA \prefix \addressB$, then we know that $y$ is a continuation
resulting from a communication that started from $x$. In a sense, $x$ and $y$
represent different moments in the same conversation.

\begin{remark}
	The typing rules in \Cref{fig:ts_ll} except \refrule{\PiChoiceRule} are in
	one-to-one correspondence with those of the \muMALL proof
	system \citep{BaeldeDoumaneSaurin16,Doumane17}. Concerning \refrule{\PiChoiceRule},
  it does not alter in any way the context of the process being typed.
  This implies that the type system is a conservative extension
  of \muMALL. That is, if $\qtp{x_1:S_1,\dots,x_n:S_n}{P}$ is coinductively
  derivable using the rules in \Cref{fig:ts_ll}, then $\vdash
  S_1,\dots,S_n$ is coinductively derivable in \muMALL. The converse is also true,
  although the proof term $P$ is not uniquely determined.
\end{remark}

\begin{example}[Buyer - Seller]
    \label{ex:bsc_ll_ts}
    Let us show that the system described in \Cref{ex:bsc_ll_proc} is quasi
    typed. To this aim, let $\FormulaF \eqdef \tmu\X.X \plinchoice \One$ and
    $\FormulaG \eqdef \tnu\X.X \plinbranch \Bot$ respectively be the formulas
    describing the behavior of $\Buyer$ and $\Seller$ on the channel $x$. Note
    that $\FormulaG = \dual\FormulaF$ and let $a$ be an arbitrary atomic
    address. We derive
    \[
        \begin{prooftree}
            \[
                \[
                    \[
                        \mathstrut\smash\vdots
                        \justifies
                        \qtp{
                            x : \FormulaF_{ail}
                        }{
                            \Call\Buyer\x
                        }
                    \]
                    \justifies
                    \qtp{
                        x : (\FormulaF \plinchoice \One)_{ai}                        
                    }{
                        \Select\AddTag\x.\Call\Buyer\x
                    }
                    \using\refrule[select]\SelectRule
                \]
                \qquad
                \[
                    \[
                        \justifies
                        \qtp{
                            x : \One
                        }{
                            \PiClose\x
                        }
                        \using\refrule[close]{\PiCloseRule}
                    \]
                    \justifies
                    \qtp{
                        x : (\FormulaF \plinchoice \One)_{ai}
                    }{
                        \Select\PayTag\x.\PiClose\x
                    }
                    \using\refrule[select]{\SelectRule}
                \]
                \justifies
                \qtp{
                    x : (\FormulaF \plinchoice \One)_{ai}
                }{
                    \PiChoice{
                        \Select\AddTag\x.\Call\Buyer\x
                    }{
                        \Select\PayTag\x.\PiClose\x
                    }
                }
                \using\refrule{\PiChoiceRule}
            \]
            \justifies
            \qtp{
                x : \FormulaF_a
            }{
                \Call\Buyer\x
            }
            \using\refrule[rec]{\RecRule}
        \end{prooftree}
    \]
    and also
    \[
        \begin{prooftree}
            \[
                \[
                    \mathstrut\smash\vdots
                    \justifies
                    \qtp{
                        x : \FormulaG_{\dual{a}il},
                        y : \One
                    }{
                        \Call\Seller{x,y}
                    }
                \]
                \qquad
                \[
                    \[
                        \justifies
                        \qtp{
                            y : \One
                        }{
                            \PiClose\y
                        }
                        \using\refrule[close]{\PiCloseRule}
                    \]
                    \justifies
                    \qtp{
                        x : \Bot,
                        y : \One
                    }{
                        \PiWait\x.\PiClose\y
                    }
                    \using\refrule[wait]{\PiWaitRule}
                \]
                \justifies
                \qtp{
                    x : (\FormulaG \plinbranch \Bot)_{\dual{a}i},
                    y : \One
                }{
                    \CaseX\x{
                        \Call\Seller{x,y}
                    }{
                        \PiWait\x.\PiClose\y
                    }            
                }
                \using\refrule[case]{\CaseRule}
            \]
            \justifies
            \qtp{
                x : \FormulaG_{\dual{a}},
                y : \One
            }{
                \Call\Seller{x,y}
            }
            \using\refrule[corec]{\CorecRule}
        \end{prooftree}
    \]
    showing that $\Buyer$ and $\Seller$ are quasi typed. Note that both
    derivations are infinite, but for dual reasons. In $\Buyer$ the infinite
    branch corresponds to the behavior in which $\Buyer$ chooses to add one more
    item to the shopping cart. This choice is made independently of the behavior
    of other processes in the system. In $\Seller$, the infinite branch
    corresponds to the behavior in which $\Seller$ receives one more $\AddTag$
    message from $\Buyer$.
    By combining these derivations we obtain
    \[
        \begin{prooftree}
            \[
                \mathstrut\smash\vdots
                \justifies
                \qtp{
                    x : \FormulaF_a
                }{
                    \Call\Buyer\x
                }
            \]
            \qquad
            \[
                \mathstrut\smash\vdots
                \justifies
                \qtp{
                    x : \FormulaG_{\dual{a}},
                    y : \One
                }{
                    \Call\Seller{x,y}
                }
            \]
            \justifies
            \qtp{
                y : \One
            }{
                \Cut\x{\Call\Buyer\x}{\Call\Seller{x,y}}
            }
            \using\refrule{\CutRule}
        \end{prooftree}
    \]
    showing that the system as a whole is quasi typed.
    \eoe
\end{example}

%% file: ts-ll/ts-wtp.tex
\beginbass
As we have anticipated, there exist infinite typing derivations that are unsound
from a logical standpoint, because they allow us to prove $\Zero$ or the empty
sequent. Hence, the typing rules presented in \Cref{fig:ts_ll} must be combined
with additional \emph{validity conditions}.

\begin{example}
	\label{ex:omega}
	Consider the non-terminating process $\Omega(x) = \PiChoice{\Call\Omega\x}{\Call\Omega\x}$.
	We obtain the following infinite derivation showing that $\Call\Omega\x$ is quasi typed.
	\[
    \begin{prooftree}
        \[
            \mathstrut\smash\vdots
            \justifies
            \qtp{x : \Zero}{\Call\Omega\x}
        \]
        \qquad
        \[
            \mathstrut\smash\vdots
            \justifies
            \qtp{x : \Zero}{\Call\Omega\x}
        \]
        \justifies
        \qtp{x : \Zero}{\Call\Omega\x}
        \using\refrule\PiChoiceRule
    \end{prooftree}
    \]
\end{example}

As illustrated by the next example,
there exist non-terminating processes that are quasi typed also in logically
sound contexts.

\begin{example}[Compulsive Buyer]
    \label{ex:compulsive_buyer_ll}
    Consider the following variant of the $\Buyer$ process
    \[
        \Buyer(x) = \Rec\x.\Select\AddTag\x.\Call\Buyer{x}
    \] 
    that models a ``compulsive buyer'', namely a buyer that adds infinitely many
    items to the shopping cart but never pays. Using $\FormulaF \eqdef \tmu\X.X
    \plinchoice \One$ and an arbitrary atomic address $a$ we can build the following
    infinite derivation
    \[
        \begin{prooftree}
            \[
                \[
                    \mathstrut\smash\vdots
                    \justifies
                    \qtp{
                        x : \Formula_{ail}
                    }{
                        \Call\Buyer{x}
                    }
                \]
                \justifies
                \qtp{
                    x : (\Formula \plinchoice \One)_{ai}
                }{
                    \Select\AddTag\x.\Call\Buyer{x}
                }
                \using\refrule[select]{\SelectRule}
            \]
            \justifies
            \qtp{
                x : \Formula_a
            }{
                \Call\Buyer{x}
            }
            \using\refrule[rec]{\RecRule}
        \end{prooftree}
    \]
    showing that this process is quasi typed. By combining this derivation with
    the one for $\Seller$ in \Cref{ex:bsc_ll_ts} we obtain a
    derivation establishing that $\Cut\x{\Call\Buyer\x}{\Call\Seller{x,y}}$ is
    quasi typed in the context $y : \One$, although this composition cannot
    terminate.
    \eoe
\end{example}

To rule out unsound derivations like those in
\Cref{ex:omega,ex:compulsive_buyer_ll} it is necessary to impose a validity
condition on derivations \citep{BaeldeDoumaneSaurin16,Doumane17}. Roughly
speaking, \muMALL's validity condition requires every infinite branch of a
derivation to be supported by the continuous unfolding of a greatest fixed
point. In order to formalize this condition, we start by defining
\emph{threads}, which are sequences of types describing sequential interactions
at the type level.

\begin{definition}[Thread]
    \label{def:thread}
    A \emph{thread} of $S$ is a sequence of types $(S_i)_{i\in o}$ for some
    $o\in\omega + 1$ such that $S_0 = S$ and $S_i \tred S_{i+1}$ whenever
    $i+1\in o$.
\end{definition}

Hereafter we use $t$ to range over threads. 

\begin{example}
	\label{ex:bsc_ll_thread}
	Consider
	$\Formula \eqdef \tmu\X.X \plinchoice \One$ from \Cref{ex:bsc_ll_formulas} we have that
	$t \eqdef (\Formula_a,(\Formula\plinchoice\One)_{ai},\Formula_{ail},\dots)$ is an
	infinite thread of $\Formula_a$.
\end{example}

A thread is \emph{stationary} if it has an
infinite suffix of equal types. The thread $t$ from \Cref{ex:bsc_ll_thread} is not stationary.
Among all threads, we are interested in finding those in which a $\tnu$-formula
is unfolded infinitely often. These threads, called $\tnu$-threads, are precisely
defined thus:

\begin{definition}[$\tnu$-thread]
    \label{def:nu-thread}
    Let $t = (S_i)_{i\in\omega}$ be an infinite thread, let $\strip{t}$ be the
    corresponding sequence $(\strip{S_i})_{i\in\omega}$ of formulas and let
    $\InfOften{t}$ be the set of elements of $\strip{t}$ that occur infinitely
    often in $\strip{t}$. We say that $t$ is a \emph{$\tnu$-thread} if
    $\minf\InfOften{t}$ is defined and is a $\tnu$-formula.
\end{definition}

\begin{example}
	Consider the infinite thread $t$ from \Cref{ex:bsc_ll_thread}.
	We have $\InfOften{t} = \set{\Formula, \Formula \choice \One}$ 
	and $\minf\InfOften{t} = \Formula$, so $t$
	is \emph{not} a $\tnu$-thread because $\Formula$ is not a $\tnu$-formula.
\end{example}

\begin{example}
	Consider the following formulas 
	\[
	\begin{array}{rclrcl}
		\FormulaF & \eqdef & \tnu\X.\tmu\Y.X \choice Y 
		& \qquad
		\FormulaG & \eqdef & \tmu\Y.\FormulaF \choice Y
	\end{array}
	\]	
	Observe that $\FormulaG$ is the ``unfolding'' of $\FormulaF$. 
	Now 
	\[
	t_1 \eqdef (\FormulaF_a, \FormulaG_{ai},
	(\FormulaF \choice \FormulaG)_{aii}, \FormulaF_{aiil}, \dots)
	\] is a thread of
	$\FormulaF_a$ such that $\InfOften{t_1} = \set{\FormulaF, \FormulaG, \FormulaF
	\choice \FormulaG}$ and we have $\minf\InfOften{t_1} = \FormulaF$ because
	$\FormulaF \subf \FormulaG$, so $t_1$ is a $\tnu$-thread.
	If, on the other hand, we consider the thread 
	\[
	t_2 \eqdef (\FormulaF_a,
	\FormulaG_{ai}, (\FormulaF \choice \FormulaG)_{aii}, \FormulaG_{aiir},
	(\FormulaF \choice \FormulaG)_{aiiri}, \dots)
	\]
	such that $\InfOften{t_2} =
	\set{\FormulaG, \FormulaF \choice \FormulaG}$ we have $\minf\InfOften{t_2} =
	\FormulaG$ because $\FormulaG \subf \FormulaF \choice \FormulaG$, so $t_2$ is
	not a $\tnu$-thread.
\end{example}

Intuitively, the $\subf$-minimum formula among those that occur infinitely often
in a thread is the outermost fixed point operator that is being unfolded
infinitely often. It is possible to show that this minimum formula is always
well defined \citep{Doumane17}. If such minimum formula is a greatest fixed point
operator, then the thread is a $\tnu$-thread.

Now we proceed by identifying threads along branches of typing derivations. To
this aim, we provide a precise definition of \emph{branch}.

\begin{definition}[Branch]
    \label{def:branch}
    A \emph{branch} of a typing derivation is a sequence
    $(\qtp{\Ctx_i}{P_i})_{i\in o}$ of judgments for some $o\in\omega+1$ such
    that $\qtp{\Ctx_0}{P_0}$ occurs somewhere in the derivation and
    $\qtp{\Ctx_{i+1}}{P_{i+1}}$ is a premise of the rule application that
    derives $\qtp{\Ctx_i}{P_i}$ whenever $i+1\in o$.
\end{definition}

An infinite branch is valid if supported by a $\tnu$-thread that originates
somewhere therein.

\begin{definition}[Valid Branch]
    \label{def:valid_branch}
    Let $\gamma = (\qtp{\Ctx_i}{P_i})_{i\in\omega}$ be an infinite branch in
    a derivation. We say that $\gamma$ is \emph{valid} if there exists
    $j\in\omega$ such that $(S_k)_{k\geq j}$ is a non-stationary $\tnu$-thread
    and $S_k$ is in the range of $\Ctx_k$ for every $k \geq j$.
\end{definition}

\begin{example}
	The infinite branch in the typing derivation for $\Seller$ of
	\Cref{ex:bsc_ll_formulas} is valid since it is supported by the $\tnu$-thread
	$(\FormulaG_{\dual{a}}, (\FormulaG \plinbranch \Bot)_{\dual{a}i},
	\FormulaG_{\dual{a}il},\dots)$ where $\FormulaG \eqdef \tnu\X.X \plinbranch \Bot$
	happens to be the $\subf$-minimum formula that is unfolded infinitely often.
\end{example}

\begin{example}
	The infinite branch in the typing derivation for $\Buyer$ of
	\Cref{ex:compulsive_buyer_ll} is invalid, because the only infinite thread
	in it is $(\FormulaF_a, (\FormulaF \plinchoice \One)_{ai}, \FormulaF_{ail}, \dots)$
	which is not a $\tnu$-thread.
\end{example}

A \muMALL derivation is valid if so is every infinite branch in
it \citep{BaeldeDoumaneSaurin16,Doumane17}. For the purpose of ensuring fair
termination, this condition is too strong because some infinite branches in a
typing derivation may correspond to unfair executions that, by definition, we
neglect insofar its termination is concerned. For example, the infinite branch
in the derivation for $\Buyer$ of \Cref{ex:bsc_ll_formulas} corresponds to
an unfair run in which the buyer insists on adding items to the shopping cart,
despite it periodically has a chance of paying the seller and terminate the
interaction. That typing derivation for $\Buyer$ would be considered an invalid
proof in \muMALL because the infinite branch is not supported by a $\tnu$-thread
(in fact, there is a $\tmu$-formula that is unfolded infinitely many times along
that branch, as in \Cref{ex:compulsive_buyer_ll}).

It is generally difficult to understand if a branch corresponds to a fair or
unfair run because the branch describes the evolution of an incomplete process
whose behavior is affected by the interactions it has with processes found in
other branches of the derivation.
However, we can detect (some) unfair branches by looking at the
non-deterministic choices they traverse, since choices are made autonomously by
processes. To this aim, we introduce the notion of \emph{rank} to estimate the
least number of choices a process can possibly make during its lifetime.

\begin{definition}[Rank]
    \label{def:rank}
    Let $\rankR$ and $\rankS$ range over the elements of $\RankSet \eqdef
    \Nat\cup\set\infty$ equipped with the expected total order $\leq$ and
    operation $+$ such that $\rankR + \infty = \infty + \rankR = \infty$.
    The \emph{rank} of a process $P$, written $\rankof{P}$, is the least element
    of $\RankSet$ such that
    \[
    	\begin{array}{lr}
        \begin{array}{@{}r@{~}c@{~}l@{}}
            \rankof{\Link\x\y} & = & 0 \\
            \rankof{\Fail\x} & = & 0 \\
            \rankof{\PiClose\x} & = & 0 \\
            \rankof{\PiWait\x.P} & = & \rankof{P} \\
            \rankof{\Join[z]\x\y.P} & = & \rankof{P} \\
            \rankof{\Select[y]{\InTag_i}\x.P} & = & \rankof{P}
        \end{array}
				& \qquad
        \begin{array}{@{}r@{~}c@{~}l@{}}
        		\rankof{\Rec[y]\x.P} & = & \rankof{P} \\
            \rankof{\Corec[y]\x.P} & = & \rankof{P} \\
            \rankof{\Case[y]\x{P}{Q}} & = & \max\set{\rankof{P},\rankof{Q}} \\
            \rankof{\PiChoice{P}{Q}} & = & 1 + \min\set{\rankof{P},\rankof{Q}} \\
            \rankof{\Cut\x{P}{Q}} & = & \rankof{P} + \rankof{Q} \\
            \rankof{\Fork[z]\x\y{P}{Q}} & = & \rankof{P} + \rankof{Q}
        \end{array}
    	\end{array}
    \]
\end{definition}

Roughly, the rank of terminated processes is $0$, that of processes with a
single continuation $P$ coincides with the rank of $P$, and that of processes
spawning two continuations $P$ and $Q$ is the sum of the ranks of $P$ and $Q$.
Then, the rank of a sum input with continuations $P$ and $Q$ is conservatively
estimated as the maximum of the ranks of $P$ and $Q$, since we do not know which
one will be taken, whereas the rank of a choice with continuations $P$ and $Q$
is 1 plus the minimum of the ranks of $P$ and $Q$.

\begin{example}
	Consider $\Buyer$ and $\Seller$ from \Cref{ex:bsc_ll_proc} and $\Omega$ from \Cref{ex:omega}.
 	Then have $\rankof{\Call\Buyer\x} = 1$, $\rankof{\Call\Seller{x,y}} = 0$
 	and $\rankof{\Call\Omega\x} = \infty$.
\end{example}

Note that $\rankof{P}$ only depends on the structure of $P$ but not on the
actual names occurring in $P$. As a consequence, when $P$ is defined by means of a
\emph{finite} system of equations, the value of $\rankof{P}$ too can be
determined by a \emph{finite} system of equations.

\begin{example}
	\label{ex:eq_system}
	Consider the definition of $\Buyer$ found in
	\Cref{ex:bsc_ll_proc}. In order to compute $\rankof{\Call\Buyer\x}$ we consider
	the system of equations
	{
    \renewcommand{\x}{\bullet}
    \renewcommand{\y}{\bullet}
    \renewcommand{\z}{\bullet}
    \[
    \begin{array}{rcl}
    	\rankof{\Call\Buyer\x} & = & \rankof{
            \PiChoice{
                \Select[\z]\AddTag\y.\Call\Buyer\z
            }{
                \Select[\z]\PayTag\y.\PiClose\z
            }
        }
      \\
      \rankof{
            \PiChoice{
                \Select[\z]\AddTag\y.\Call\Buyer\z
            }{
                \Select[\z]\PayTag\y.\PiClose\z
            }
        } & = & 1 + \min\set{ 
            \\ & & \qquad\quad \rankof{\Select[\z]\AddTag\y.\Call\Buyer\z},
            \\ & & \qquad\quad \rankof{\Select[\z]\PayTag\y.\PiClose\z}
        }
        \\
        \rankof{\Select[\z]\AddTag\y.\Call\Buyer\z} & = & \rankof{\Call\Buyer\z}
        \\
        \rankof{\Select[\z]\PayTag\y.\PiClose\z} & = & \rankof{\PiClose\z}
        \\
        \rankof{\PiClose\z} & = & 0
    \end{array}
    \]
	}
	where we have used a placeholder $\bullet$ in place of every channel name
	occurring in these terms.
	\eoe
\end{example}

Every such system of equations can be thought of as a function $\ffun :
\parens\RankSet^n \to \parens\RankSet^n$ on the complete lattice
$\parens\RankSet^n$ ordered by the pointwise extension of $\leq$ in $\RankSet$.
Note that $\ffun$ is monotone, because all the operators occurring in the
definition of rank (see \Cref{def:rank}) are monotone. So, $\ffun$ has a least fixed
point by the Knaster-Tarski theorem and the rank of $P$ is the component of this
fixed point that corresponds to $\rankof{P}$.

\begin{example}
	For system of equations in \Cref{ex:eq_system} we have $n = 5$ and we have
	\[
    \ffun(x_1,x_2,x_3,x_4,x_5) = (x_2,1+\min\set{x_3,x_4},x_1,x_5,0)
	\]
	whose least solution is $(1,1,1,0,0)$. Now $\rankof{\Call\Buyer\x}$ corresponds
	to the first component of this solution, that is $1$.
	\eoe
\end{example}

\begin{definition}
    \label{def:fair_branch}
    A branch is \emph{fair} if it traverses finitely many, finitely-ranked
    choices.
\end{definition}

A finitely-ranked choice is at finite distance from a region of the process in
which there are no more choices. An \emph{unfair} branch gets close to such
region infinitely often, but systematically avoids entering it.
Note that every finite branch is also fair, but there are fair branches that are
infinite. 

\begin{example}
	All the infinite branches inside the derivation of
	\Cref{ex:omega} and the only infinite branch in the derivation for
	$\Call\Seller{x,y}$ of \Cref{ex:bsc_ll_ts} are fair since they do not
	traverse any finitely-ranked choice. On the contrary, the only infinite branch
	in the derivation for $\Call\Buyer\x$ of the \Cref{ex:bsc_ll_ts} is
	unfair since it traverses infinitely many finitely-ranked choices. All fair
	branches in the same derivation for $\Buyer$ are finite.
\end{example}

At last we can define our notion of well-typed process.

\begin{definition}[Well-Typed Process]
    \label{def:wtp}
    We say that $P$ is \emph{well typed} in $\Ctx$, written
    $\piwtp\Ctx{P}$, if the judgment $\qtp\Ctx{P}$ is derivable and each
    fair, infinite branch in its derivation is valid.
\end{definition}

\begin{example}
	 $\Omega$ is ill typed since the fair, infinite branches in
	\Cref{ex:omega} are all invalid.
\end{example}

\begin{theorem}[Soundness]
	\label{thm:ts_ll_sound}
    If $\piwtp{x : \One}{P}$ and $P \wred Q$ then $Q \wred \PiClose\x$.
\end{theorem}

As for the session-based calculi (\Cref{ch:ft_bin,ch:ft_multi}) 
\Cref{thm:ts_ll_sound} entails all the good properties we expect from well-typed
processes: \emph{failure freedom} (no unguarded sub-process $\Fail\y$ ever
appears), \emph{deadlock freedom} (if the process stops it is terminated),
\emph{lock freedom} \citep{Kobayashi02,Padovani14} (every pending action can be
completed in finite time) and \emph{junk freedom} (every channel can be
depleted).

%% file: ts-ll/ts-examples.tex
\beginbass
We dedicate the rest of the section to show some more involved
examples. In \Cref{ex:parallel_programming} we show the implementation of 
a \emph{parallel programming pattern}. 
\Cref{ex:forwarder} models a simple $\Forwarder$ that has the peculiarity 
of unfolds a leas fixed point infinitely many times.
At last, in \Cref{ex:slot_machine} we model a slot machine.

\begin{example}[Parallel Programming]
    \label{ex:parallel_programming}
    In this example we see a \piLIN modeling of a \emph{parallel programming
    pattern} whereby a $\Work$ process creates an unbounded number of workers
    each one dedicated to an independent task and a $\Gather$ process collects
    and combines the partial results from the workers. The processes $\Work$ and
    $\Gather$ are defined as follows:
    \begin{align*}
        \Work(x) & =
        \Rec\x.\parens{
            \PiChoice{
                \Select\ComplexTag\x.\Fork\x\y{\PiClose\y}{\Call\Work\x}
            }{
                \Select\SimpleTag\x.\PiClose\x
            }
        }
        \\
        \Gather(x,z) & =
        \Corec\x.
        \Case\x{
            \Join\x\y.\PiWait\y.\Call\Gather{x,z}
        }{
            \PiWait\x.\PiClose\z
        }
    \end{align*}

    At each iteration the $\Work$ process non-deterministically decides whether
    the task is $\ComplexTag$ (left hand side of the choice) or $\SimpleTag$
    (right hand side of the choice). In the first case, it bifurcates into a new
    worker, which in the example simply sends a unit on $y$, and another
    instance of itself. In the second case it terminates.
    The $\Gather$ process joins the results from all the workers before
    signalling its own termination by sending a unit on $z$. Note that the
    number of actions $\Gather$ has to perform before terminating is unbounded,
    as it depends on the non-deterministic choices made by $\Work$.

    Below is a typing derivation for $\Work$ where $\FormulaF \eqdef
    \tmu\X.(\One \tfork X) \plinchoice \One$ and $a$ is an arbitrary atomic address:
    \[
        \begin{prooftree}
            \[
                \[
                    \[
                        \[
                            \justifies
                            \qtp{
                                y : \One
                            }{
                                \PiClose\y
                            }
                            \using\refrule[close]\PiCloseRule
                        \]
                        \quad
                        \[
                            \mathstrut\smash\vdots
                            \justifies
                            \qtp{
                                x : \FormulaF_{ailr}
                            }{
                                \Call\Work\x
                            }
                        \]
                        \justifies
                        \qtp{
                            x : (\One \tfork \FormulaF)_{ail}
                        }{
                            \Fork\x\y{\PiClose\y}{\Call\Work\x}
                        }
                        \using\refrule[fork]\ForkRule
                    \]
                    \justifies
                    \qtp{
                        x : ((\One \tfork \FormulaF) \plinchoice \One)_{ai}
                    }{
                        \Select\ComplexTag\x\dots
                    }
                    \using\refrule[select]\SelectRule
                \]
                \[
                    \justifies
                    \qtp{
                        x : \dots 
                    }{
                        \Select\SimpleTag\x.\PiClose\x
                    }
                    \using\refrule[close]\PiCloseRule,\refrule[select]\SelectRule
                \]
                \justifies
                \qtp{
                    x : ((\One \tfork \FormulaF) \plinchoice \One)_{ai}
                }{
                    \PiChoice{
                        \Select\ComplexTag\x.\Fork\x\y{\PiClose\y}{\Call\Work\x}
                    }{
                        \Select\SimpleTag\x.\PiClose\x
                    }
                }
                \using\refrule\PiChoiceRule
            \]
            \justifies
            \qtp{
                x : \FormulaF_\address
            }{
                \Call\Work\x
            }
            \using\refrule[rec]\RecRule
        \end{prooftree}
    \]

    Note that the only infinite branch in this derivation is unfair because it
    traverses infinitely many choices with rank $1 = \rankof{\Call\Work\x}$. So,
    $\Work$ is well typed.

    Concerning $\Gather$, we obtain the following typing derivation where
    $\FormulaG \eqdef \tnu\X.(\Bot \tjoin X) \plinbranch \Bot$:
    \[
        \begin{prooftree}
            \[
                \[
                    \[
                        \[
                            \mathstrut\smash\vdots
                            \justifies
                            \qtp{
                                x : \FormulaG_{\dual ailr},
                                z : \One
                            }{
                                \Call\Gather{x,z}    
                            }
                        \]
                        \justifies
                        \qtp{
                            x : \FormulaG_{\dual ailr},
                            y : \Bot,
                            z : \One
                        }{
                            \PiWait\y.\Call\Gather{x,z}
                        }
                        \using\refrule[wait]\PiWaitRule
                    \]
                    \justifies
                    \qtp{
                        x : (\Bot \tjoin \FormulaG)_{\dual ail},
                        z : \One
                    }{
                        \Join\x\y.\PiWait\y.\Call\Gather{x,z}
                    }
                    \using\refrule[join]\JoinRule
                \]
                \[
                    \justifies
                    \qtp{
                        x : \Bot,
                        z : \One
                    }{
                        \PiWait\x.\PiClose\z
                    }
                \]
                \justifies
                \qtp{
                    x : ((\Bot \tjoin \FormulaG) \plinbranch \Bot)_{\dual ai},
                    z : \One
                }{
                    \Case\x{
                        \Join\x\y.\PiWait\y.\Call\Gather{x,z}
                    }{
                        \PiWait\x.\PiClose\z
                    }            
                }
                \using\refrule[case]\CaseRule
            \]
            \justifies
            \qtp{
                x : \FormulaG_{\dual a},
                z : \One
            }{
                \Call\Gather{x,z}
            }
            \using\refrule[corec]\CorecRule
        \end{prooftree}
    \]

    Here too there is just one infinite branch, which is fair and supported by
    the $\tnu$-thread $t = (\FormulaG_{\dual a}, ((\Bot \tjoin \FormulaG) \plinbranch
    \Bot)_{\dual ai}, (\Bot \tjoin \FormulaG)_{\dual ail}, \FormulaG_{\dual
    ailr}, \dots)$. Indeed, all the formulas in $\strip{t}$ occur infinitely
    often and $\minf\InfOften{t} = \FormulaG$ which is a $\tnu$-formula.
    Hence, $\Gather$ is well typed and so is the composition
    $\Cut\x{\Call\Work\x}{\Call\Gather{x,z}}$ in the context $z : \One$. We
    conclude that the program is fairly terminating, despite the fact that the
    composition of $\Work$ and $\Gather$ may grow arbitrarily large because
    $\Work$ may spawn an unbounded number of workers.
    \eoe
\end{example}

\begin{example}[Forwarder]
    \label{ex:forwarder}
    In this example we illustrate a deterministic, well-typed process that
    unfolds a least fixed point infinitely many times. In particular, we
    consider once again the formulas $\FormulaF \eqdef \tmu\X.X \plinchoice \One$ and
    $\FormulaG \eqdef \tnu\X.X \plinbranch \Bot$ and the process $\Forwarder$ defined
    by the equation
    \[
        \Forwarder(x,y) =
        \Corec\x.
        \Rec\y.
        \Case\x{
            \Left\y.
            \Call\Forwarder{x,y}
        }{
            \Right\y.
            \PiWait\x.
            \PiClose\y
        }
    \]
    which forwards the sequence of messages received from channel $x$ to channel
    $y$. We derive
    \[
        \begin{prooftree}
            \[
                \[
                    \[
                        \mathstrut\smash\vdots
                        \justifies
                        \qtp{
                            x : \FormulaG_{ail},
                            y : \FormulaF_{bil}
                        }{
                            \Call\Forwarder{x,y}
                        }
                    \]
                    \justifies
                    \qtp{
                        x : \FormulaG_{ail},
                        y : (\FormulaF \plinchoice \One)_{bi}
                    }{
                        \Left\y\dots 
                    }
                    \using\refrule[select]{\SelectRule}
                \]                
                \[
                    \[
                        \justifies
                        \qtp{
                            x : \Bot,
                            y : \One
                        }{
                            \PiWait\x.\PiClose\y
                        }
                        \using\refrule[close]\PiCloseRule,\refrule[wait]\PiWaitRule
                    \]
                    \justifies
                    \qtp{
                        x : \Bot,
                        y : (\FormulaF \plinchoice \One)_{bi}
                    }{
                        \Right\y\dots 
                    }
                    \using\refrule[select]{\SelectRule}
                \]
                \justifies
                \qtp{
                    x : (\FormulaG \plinbranch \Bot)_{ai},
                    y : (\FormulaF \plinchoice \One)_{bi}
                }{
                    \Case\x{
                        \Left\y.
                        \Call\Forwarder{x,y}
                    }{
                        \Right\y.
                        \PiWait\x.
                        \PiClose\y
                    }            
                }
                \using\refrule[case]{\CaseRule}
            \]
            \justifies
            \qtp{x : \FormulaG_a, y : \FormulaF_b}{\Call\Forwarder{x,y}}
            \using\refrule[corec]{\CorecRule}, \refrule[rec]{\RecRule}
        \end{prooftree}
    \]
    and observe that $\rankof{\Call\Forwarder{x,y}} = 0$. This typing derivation
    is valid because the only infinite branch is fair and supported by the
    $\tnu$-thread of $\FormulaG_a$.
    Note that $\FormulaF = \dual\FormulaG$ and that the derivation proves an
    instance of \refrule{\LinkRule}. In general, the axiom is admissible in
    \muMALL \citep{BaeldeDoumaneSaurin16}.
    \eoe
\end{example}

\begin{example}[Slot Machine]
    \label{ex:slot_machine}
    Rank finiteness is not a necessary condition for well typedness. As an
    example, consider the system $\Cut\x{\Call\Player\x}{\Call\Machine{x,y}}$
    where
    \[
        \begin{array}{@{}r@{~}c@{~}l@{}}
            \Player(x) & = &
            \Rec\x.
            \parens{
                \PiChoice{
                    \Select\PlayTag\x.
                    \Case\x{
                        \Call\Player\x
                    }{
                        \Rec\x.\Select\QuitTag\x.\PiClose\x
                    }
                }{
                    \Select\QuitTag\x.\PiClose\x
                }
            }
            \\
            \Machine(x,y) & = &
            \Corec\x.
            \\ & &
            \Case\x{
                \PiChoice{
                    \Select\WinTag\x.\Call\Machine{x,y}
                }{
                    \Select\LoseTag\x.\Call\Machine{x,y}
                }
            }{
                \PiWait\x.\PiClose\y
            }
        \end{array}
    \]
    which models a game between a player and a slot machine. At each round, the
    player decides whether to $\PlayTag$ or to $\QuitTag$. In the first case,
    the slot machine answers with either $\WinTag$ or $\LoseTag$. If the player
    $\WinTag$s, it also $\QuitTag$s. Otherwise, it repeats the same behavior.
    It is possible to show that $\qtp{x : \FormulaF_a}{\Call\Player\x}$ and
    $\qtp{x : \FormulaG_{\dual a}, y : \One}{\Call\Machine{x,y}}$ are derivable
    where $\FormulaF \eqdef \tmu\X.(X \plinbranch X) \plinchoice \One$ and $\FormulaG
    \eqdef \tnu\X.(X \plinchoice X) \plinbranch \Bot$. For
		convenience, we define $\Formula' \eqdef (\Formula \plinbranch \Formula) \plinchoice
		\One$ and we abbreviate $\Player$ to $P$. 
		Note that $\Formula'$ is the unfolding of $\Formula$.
    \[
    \renewcommand{\Player}{P}
    \begin{prooftree}
        \[
            \[
                \[
                    \[
                        \mathstrut\smash\vdots
                        \justifies
                        \qtp{
                            x : \FormulaF_{aill}
                        }{
                            \Call\Player\x
                        }
                    \]
                    \qquad
                    \[
                        \[
                            \[
                                \justifies
                                \qtp{
                                    x : \One
                                }{
                                    \PiClose\x
                                }
                                \using\refrule[close]{\PiCloseRule}
                            \]
                            \justifies
                            \qtp{
                                x : \Formula'_{ailri}
                            }{
                                \Select\QuitTag\x.\PiClose\x
                            }
                            \using\refrule[select]{\SelectRule}
                        \]
                        \justifies
                        \qtp{
                            x : \FormulaF_{ailr}
                        }{
                            \Rec\x\dots
                        }
                        \using\refrule[rec]{\RecRule}
                    \]
                    \justifies
                    \qtp{
                        x : (\FormulaF \plinbranch \FormulaF)_{ail}
                    }{
                        \Case\x{
                            \Call\Player\x
                        }{
                            \Rec\x\dots
                        }    
                    }
                    \using\refrule[case]{\CaseRule}
                \]
                \justifies
                \qtp{
                    x : \Formula'_{ai}
                }{
                    \Select\PlayTag\x.
                    \Case\x{
                        \Call\Player\x
                    }{
                        \Rec\x\dots
                    }
                }
                \using\refrule[select]{\SelectRule}
            \]
            \[
                \[
                    \justifies
                    \qtp{
                        x : \One
                    }{
                        \PiClose\x
                    }
                    \using\refrule[close]{\PiCloseRule}
                \]
                \justifies
                \qtp{
                    x : \Formula'_{ai}
                }{
                    \Select\QuitTag\x.\PiClose\x
                }
                \using\refrule[select]{\SelectRule}
            \]
            \justifies
            \qtp{
                x : \Formula'_{ai}
            }{
                \PiChoice{
                    \Select\PlayTag\x\dots
                }{
                    \Select\QuitTag\x.\PiClose\x
                }
            }
        \]
        \justifies
        \qtp{
            x : \Formula_a
        }{
            \Call\Player\x
        }
        \using\refrule[rec]{\RecRule}
    \end{prooftree}
		\]
    Note that the only infinite branch in the derivation for $\Player$ is unfair since
    $\rankof{\Call\Player\x} = 1$. Hence, $\Player$ is well typed.
    
    Below is the typing derivation showing that $\Machine$ is quasi typed. For
		convenience, we define $\FormulaG' \eqdef \FormulaG \plinchoice \FormulaG$,
		we abbreviate $\Machine$ to $M$ and we show the derivation only for the $\WinTag$
		branch since the $\LoseTag$ is the same.
		\[
    \renewcommand{\Machine}{M}
    \begin{prooftree}
        \[
            \[
                \[
                    \[
                        \mathstrut\smash\vdots
                        \justifies
                        \qtp{
                            x : \FormulaG_{bill},
                            y : \One
                        }{
                            \Call\Machine{x,y}
                        }
                    \]
                    \justifies
                    \qtp{
                        x : \FormulaG'_{bil},
                        y : \One        
                    }{
                        \Select\WinTag\x\dots
                    }
                    \using\refrule[select]\SelectRule
                \]
                \[
                    \vdots
                    \justifies
                    \qtp{
                        \dots     
                    }{
                        \Select\LoseTag\x\dots
                    }
                    \using\refrule[select]\SelectRule
                \]
                \justifies
                \qtp{
                    x : \FormulaG'_{bil},
                    y : \One    
                }{
                    \PiChoice{
                        \Select\WinTag\x\dots
                    }{
                        \Select\LoseTag\x\dots
                    }
                }
                \using\refrule\PiChoiceRule
            \]
            \hspace{-1em}
            \[
                \[
                    \justifies
                    \qtp{
                        y : \One
                    }{
                        \PiClose\y
                    }
                    \using\refrule[close]{\PiCloseRule}
                \]
                \justifies
                \qtp{
                    x : \Bot,
                    y : \One
                }{
                    \PiWait\x.\PiClose\y    
                }
                \using\refrule[wait]{\PiWaitRule}
            \]
            \justifies
            \qtp{
                x : (\FormulaG' \plinbranch \Bot)_{bi},
                y : \One
            }{
                \Case\x{
                    \PiChoice{
                        \Select\WinTag\x\dots
                    }{
                        \Select\LoseTag\x\dots
                    }
                }{
                    \PiWait\x.\PiClose\y
                }            
            }
            \using\refrule[case]{\CaseRule}
        \]
        \justifies
        \qtp{
            x : \FormulaG_b,
            y : \One
        }{
            \Call\Machine{x,y}
        }
        \using\refrule[corec]{\CorecRule}
    \end{prooftree}
		\]
    There are infinitely many branches in the derivation for $\Machine$
    accounting for all the sequences of $\WinTag$ and $\LoseTag$ choices that
    can be made. Since $\rankof{\Call\Machine{x,y}} = \infty$, all these
    branches are fair but also valid.
    So, the system as a whole is well typed.
    \eoe
\end{example}

%% file: ts-ll/soundness.tex
\beginalto
In this section we present the proof of \Cref{thm:ts_ll_sound}.
In general, the proof technique follows again that presented in
\Cref{sec:ts_bin_corr,sec:ts_multi_corr}. However, the different scenario
implies different design choices. Hence, we think that it is worth 
showing a more detailed sketch of the proof of \Cref{thm:ts_ll_sound} by introducing
the main intermediate results and by relating them to \muMALL. 
Then, as in \Cref{sec:ts_bin_corr,sec:ts_multi_corr}, we present all the detailed proofs.


\subsection{Proof Sketch}
\input{ts-ll/proof-sketch}


\subsection{Subject Reduction}
\input{ts-ll/proof-sr}


\subsection{Proximity and Multicuts}
\input{ts-ll/proof-proximity}


\subsection{Additional Properties}
\input{ts-ll/proof-additional}


\subsection{Soundness}
\input{ts-ll/proof-soundness}

%% file: ts-ll/proof-sketch.tex
\beginbass
Here we informally explain the main results that are needed in order to prove 
\Cref{thm:ts_ll_sound}. Roughly, the most important lemmas tell that
typing is preserved by reduction (\emph{subject reduction}) and that 
a well typed process can always reach termination (\emph{weak termination}).

\begin{theorem}[Subject Reduction]
	\label{thm:subj_red_ll}
  If $\piwtp\Ctx{P}$ and $P \red Q$ then $\piwtp\Ctx{Q}$.
\end{theorem}

All reductions in \Cref{fig:red_ll} except those for non-deterministic
choices correspond to cut-elimination steps in a quasi typing derivation. As
an illustration, below is a fragment of derivation tree for two processes
exchanging a pair of $y$ and $z$ on channel $x$.
\[
    \begin{prooftree}
        \[
            \[
                \mathstrut\smash\vdots
                \justifies
                \qtp{\CtxC, y : S}{P}
            \]
            \[
                \mathstrut\smash\vdots
                \justifies
                \qtp{\CtxD, z : T}{Q}
            \]
            \justifies
            \qtp{
                \CtxC, \CtxD, x : S \tfork T
            }{
                \Fork[z]\x\y{P}{Q}
            }
            \using\refrule[fork]\ForkRule
        \]
        \[
            \[
                \mathstrut\smash\vdots
                \justifies
                \qtp{\CtxC', y : \dual{S}, z : \dual{T}}{R}
            \]
            \justifies
            \qtp{\CtxC', x : \dual{S} \tjoin \dual{T}}{\Join[z]\x\y.R}
            \using\refrule[join]\JoinRule
        \]
        \justifies
        \qtp{
            \CtxC, \CtxD, \CtxC'
        }{
            \Cut\x{
                \Fork[z]\x\y{P}{Q}
            }{
                \Join[z]\x\y.R
            }
        }
        \using\refrule\CutRule
    \end{prooftree}
\]

As the process reduces, the quasi typing derivation is rearranged so that
the cut on $x$ is replaced by two cuts on $y$ and $z$. The resulting quasi
typing derivation is shown below.
\[
    \begin{prooftree}
        \[
            \mathstrut\smash\vdots
            \justifies
            \qtp{\CtxC, y : S}{P}
        \]
        \[
            \[
                \mathstrut\smash\vdots
                \justifies
                \qtp{\CtxD, z : T}{Q}
            \]
            \[
                \mathstrut\smash\vdots
                \justifies
                \qtp{\CtxC', y : \dual{S}, z : \dual{T}}{R}
            \]
            \justifies
            \qtp{
                \CtxD, \CtxC', y : \dual{S}
            }{
                \Cut\z{Q}{R}
            }
            \using\refrule\CutRule
        \]
        \justifies
        \qtp{
            \CtxC, \CtxD, \CtxC'
        }{
            \Cut\y{P}{\Cut\z{Q}{R}}
        }
        \using\refrule\CutRule            
    \end{prooftree}
\]

It is also interesting to observe that, when $P \red Q$, the reduct $Q$ is well
typed in the same context as $P$ but its rank may be different. In particular,
the rank of $Q$ can be \emph{greater} than the rank of $P$. Recalling that the
rank of a process estimates the number of choices that the process must perform
to terminate, the fact that the rank of $Q$ increases means that $Q$ \emph{moves
away} from termination instead of getting closer to it (we will see an instance
where this phenomenon occurs in \Cref{ex:parallel_programming}). What really
matters is that a well-typed process is weakly terminating. This is the second
key property ensured by our type system.

\begin{lemma}[Weak Termination]
	\label{lem:weak_termination_ll}
  If $\piwtp{x : \One}{P}$ then $P \wred \PiClose\x$.
\end{lemma}

The proof of \Cref{lem:weak_termination_ll} is a refinement of the cut elimination
property of \muMALL. Essentially, the only new case we have to handle is when a
choice $\PiChoice{P_1}{P_2}$ ``emerges'' towards the bottom of the typing
derivation, meaning that it is no longer guarded by any action. In this case, we
reduce the choice to the $P_i$ with smaller rank, which is guaranteed to lay on
a fair branch of the derivation.
An auxiliary result used in the proof of \Cref{lem:weak_termination_ll} 
is that our type system is a conservative extension of \muMALL.

\begin{lemma}[Conservativity]
	\label{lem:conservativity}
  If $\piwtp{x_1:S_1,\dots,x_n:S_n}{P}$ is derivable then 
  $\vdash S_1,\dots,S_n$ is derivable in \muMALL.
\end{lemma}

Then, the proof of \Cref{thm:ts_ll_sound} is a
simple consequence of \Cref{thm:subj_red_ll} and \Cref{lem:weak_termination_ll}.
The combination of \Cref{thm:fair_termination,thm:ts_ll_sound} also guarantees the
termination of every fair run of the process.

\begin{corollary}[Fair Termination]
	\label{cor:fair_termination_ll}
   If $\piwtp{x : \One}{P}$ then $P$ is fairly terminating.
\end{corollary}

Observe that zero-ranked process do not contain any non-deterministic choice. In
that case, every infinite branch in their typing derivation is fair and our
validity condition coincides with that of \muMALL. As a consequence, we obtain
the following strengthening of \Cref{cor:fair_termination_ll}:

\begin{proposition}
    If $\piwtp{x : \One}{P}$ and $\rankof{P} = 0$ then $P$ is terminating.
\end{proposition}

For regular processes (those consisting of finitely many distinct sub-trees, up
to renaming of bound names) it is possible to adapt the algorithm that
decides the validity of a \muMALL proof so that it decides the validity of a
\piLIN typing derivation.

%% file: ts-ll/proof-sr.tex
\beginbass
We detail the proof of the \emph{subject reduction} theorem (see \Cref{thm:subj_red_ll}).
Notably, the presence of \refrule{rp-struct} in the reduction rules
implies a \emph{subject congruence} lemma, stating that typing is preserved by structural
pre-congruence. The same happened in \Cref{sec:ts_bin_corr,sec:ts_multi_corr}.

\begin{lemma}[Substitution]
\label{lem:substitution}
	If $\qtp{\Ctx, x : S}{P}$ and $y \not\in \dom\Ctx$, then
	$\qtp{\Ctx, y : S}{P\subst\y\x}$.
\end{lemma}
\begin{proof}
	A simple application of the coinduction principle.
\end{proof}


\begin{lemma}[Quasi Subject Congruence]
\label{lem:quasi_subj_cong_ll}
	If\/ $\qtp\Ctx{P}$ and $P \pcong Q$, then $\qtp\Ctx{Q}$.
\end{lemma}
\begin{proof}
	By induction on the derivation of $P \pcong Q$ and by cases on the last rule
	applied. We only discuss the base cases.

	\proofrule{sp-link}
		Then $P = \Link\x\y \pcong \Link\y\x = Q$.
		From \refrule{\LinkRule} we deduce that there exist $\Formula$,
		$\addressA$ and $\addressB$ such that $\Ctx = x :
		\Formula_\addressA, y : \dual{\Formula_\addressB}$. We conclude with an
		application of \refrule{\LinkRule}.

	\proofrule{sp-comm}
		Then $P = \Cut\x{P_1}{P_2} \pcong \Cut\x{P_2}{P_1} = Q$. From
		\refrule{\CutRule} we deduce that there exist $\Ctx_1$,
		$\Ctx_2$, and $S$ such that
		\begin{itemize}
		\item $\Ctx = \Ctx_1, \Ctx_2$
		\item $\qtp{\Ctx_1, x : S}{P_1}$
		\item $\qtp{\Ctx_2, x : \dual{S}}{P_2}$
		\end{itemize}
		We conclude with an application of \refrule{\CutRule}.

	\proofrule{sp-assoc}
		Then $P = \Cut\x{P_1}{\Cut\y{P_2}{P_3}} \pcong
		\Cut\y{\Cut\x{P_1}{P_2}}{P_3} = Q$ and $\x \in \fn{P_2}$.
		From \refrule{\CutRule} we deduce that there exist $\Ctx_1$,
		$\Ctx_2$, $\Ctx_3$, $S$, $T$ such that
		\begin{itemize}
		\item $\Ctx = \Ctx_1, \Ctx_2, \Ctx_3$
		\item $\qtp{\Ctx_1, x : S}{P_1}$
		\item $\qtp{\Ctx_2, x : \dual{S}, y : T}{P_2}$
		\item $\qtp{\Ctx_3, y : \dual{T}}{P_3}$
		\end{itemize}
		We conclude with two applications of \refrule{\CutRule}.
\end{proof}


\begin{lemma}[Quasi Subject Reduction]
	\label{lem:quasi_subj_red_ll}
	If $P \red Q$ then $\qtp\Ctx{P}$ implies $\qtp\Ctx{Q}$.
\end{lemma}
\begin{proof}
	By induction on $P \red Q$ and by cases on the last rule applied.

	\proofrule{rp-link}
		Then $P = \Cut\x{\Link\x\y}{R} \red R\subst\y\x = Q$.
		From \refrule{\CutRule} and \refrule{\LinkRule} we deduce that there
		exist $\Formula_\addressA$, $\Formula_\addressB$ and $\CtxD$ such
		that $\Ctx = y : \dual\Formula_{\dual\addressB}, \CtxD$ and
		$\qtp{\CtxD, x : \dual\Formula_{\dual\addressB}}{R}$
		We conclude by applying \cref{lem:substitution}.

	\proofrule{rp-unit}
		Then $P = \Cut\x{\PiClose\x}{\PiWait{x}.Q} \red Q$.
		From \refrule\CutRule, \refrule[close]{\PiCloseRule} and
		\refrule[wait]{\PiWaitRule} we conclude $\qtp\Ctx{Q}$.

	\proofrule{rp-pair}
		Then $P = \Cut\x{\Fork[z]\x\y{P_1}{P_2}}{\Join[z]\x\y.P_3} \red
		\Cut\y{P_1}{\Cut\z{P_2}{P_3}} = Q$.
		From \refrule{\CutRule}, \refrule[fork]{\ForkRule} and
		\refrule[join]{\JoinRule} we deduce that there exist $\Ctx_1$,
		$\Ctx_2$, $\Ctx_3$, $S$ and $T$ such that
		\begin{itemize}
		\item $\Ctx = \Ctx_1,\Ctx_2,\Ctx_3$
		\item $\qtp{\Ctx_1, y : S}{P_1}$
		\item $\qtp{\Ctx_2, z : T}{P_2}$
		\item $\qtp{\Ctx_3, y : \dual{S}, z : \dual{T}}{P_3}$
		\end{itemize}
		We conclude with two applications of \refrule{\CutRule}.
		
	\proofrule{rp-sum}
		Then $P = \Cut\x{\Select[y]{\InTag_i}\x.R}{\Case[y]\x{P_1}{P_2}} \red
		\Cut\y{R}{P_i} = Q$ for some $i\in\set{1,2}$.
		From \refrule{\CutRule}, \refrule[select]{\SelectRule} and
		\refrule[case]{\CaseRule} we deduce that there exist $\Ctx_1$,
		$\Ctx_2$, $S_1$ and $S_2$ such that
		\begin{itemize}
		\item $\Ctx = \Ctx_1, \CtxD$
		\item $\qtp{\Ctx_1, y : S_i}{R}$
		\item $\qtp{\Ctx_2, y : \dual{S_1}}{P_1}$
		\item $\qtp{\Ctx_2, y : \dual{S_2}}{P_2}$
		\end{itemize}
		We conclude with an application of \refrule{\CutRule}.

	\proofrule{rp-rec}
		Then $P = \Cut\x{\Rec[y]\x.P_1}{\Corec[y]\x.P_2} \red \Cut\y{P_1}{P_2} = Q$.
		From \refrule{\CutRule}, \refrule[rec]{\RecRule} and
		\refrule[corec]{\CorecRule} we deduce that there exist $\Ctx_1$,
		$\Ctx_2$ and $S$ such that
		\begin{itemize}
		\item $\Ctx = \Ctx_1, \Ctx_2$
		\item $\qtp{\Ctx_1, y : S\subst{\tmu\X.S}\X}{P_1}$
		\item $\qtp{\Ctx_2, y : \dual{S}\subst{\tnu\X.\dual{S}}{\X}}{P_2}$
		\end{itemize}
		We conclude with an application of \refrule{\CutRule}.

	\proofrule{rp-choice}
		Then $P = \PiChoice{P_1}{P_2} \red P_i = Q$ for some $i \in \set{1,2}$.
		From \refrule{\PiChoiceRule} we conclude that $\qtp\Ctx{P_i}$.

		\item[Case \refrule{rp-cut}.]
		Then $P = \Cut\x{P'}{R} \red \Cut\x{Q'}{R} = Q$ and $P' \red Q'$.
		From \refrule{\CutRule} we deduce that there exist $\Ctx_1$,
		$\Ctx_2$ and $S$ such that
		\begin{itemize}
		\item $\Ctx = \Ctx_1, \Ctx_2$
		\item $\qtp{\Ctx_1, x : S}{P'}$
		\item $\qtp{\Ctx_2, x : \dual{S}}{R}$
		\end{itemize}
		Using the induction hypothesis on $P' \red Q'$ we deduce
		$\qtp{\Ctx_1, x : S}{Q'}$.
		We conclude with an application of \refrule{\CutRule}.

	\proofrule{rp-struct}
		Then $P \pcong P'$, $P' \red Q'$ and $Q' \pcong Q$ for some $P'$, $Q'$.
		From \Cref{lem:quasi_subj_cong_ll} we deduce $\qtp\Ctx{P'}$.
		Using the induction hypothesis on $P' \red Q'$ we deduce
		$\qtp\Ctx{Q'}$. We conclude by applying \Cref{lem:quasi_subj_cong_ll}
		once more.
\end{proof}

\begin{proof}[Proof of \Cref{thm:subj_red_ll}]
	From the hypothesis $\piwtp\Ctx{P}$ we deduce $\qtp\Ctx{P}$. Using
	\Cref{lem:quasi_subj_red_ll} we deduce $\qtp\Ctx{Q}$.
	Now, let $\gamma$ be an infinite fair branch in the typing derivation for
	$\qtp\Ctx{Q}$. By inspecting the proof of
	\Cref{lem:quasi_subj_red_ll} we observe that $\gamma$ can be
	decomposed as $\gamma = \gamma_1\gamma_2$ where $\gamma_2$ is a branch in
	the typing derivation for $\qtp\Ctx{Q}$. From the fact that $\gamma$ is
	fair we deduce that so is $\gamma_2$.
	From the hypothesis $\piwtp\Ctx{P}$ we deduce that $\gamma_2$ is valid,
	namely there is a $\tnu$-thread $t$ in it. Then $t$ is a $\tnu$-thread also of
	$\gamma$, that is $\gamma$ is also valid.
	We conclude $\piwtp\Ctx{Q}$.
\end{proof}

%% file: ts-ll/proof-proximity.tex
\beginbass
The cut elimination result of \muMALL is proved by introducing a \emph{multicut}
rule that collapses several unguarded cuts in a single rule having a variable
number of premises. The usefulness of working with multicuts is that they
prevent infinite sequences of reductions where two cuts are continuously
permuted in a non-productive way and no real progress is made in the cut
elimination process. In the type system of \piLIN we do not have a typing rule
corresponding to the multicut. At the same time, the troublesome permutations of
cut rules correspond to applications of the associativity law of parallel
composition, namely of the \refrule{sp-assoc} pre-congruence rule. That is, the
introduction of the multicut rule in the cut elimination proof of \muMALL
corresponds to working with \piLIN terms considered equal up to structural
pre-congruence. Nonetheless, since the \piLIN reduction rules require two
processes willing to interact on some channel $x$ to be the children of the same
application of \refrule\CutRule, it is not entirely obvious that \emph{not}
introducing an explicit multicut rule allows us to perform in \piLIN all the
principal reductions that are performed during the cut elimination process in a
\muMALL proof.

Here we show that this is actually the case by proving a so-called
\emph{proximity lemma}, showing that every well-typed process can be rewritten
in a structurally pre-congruent one where any two processes in which the same
channel $x$ occurs free are the children of a cut on $x$.
To this aim, we introduce \emph{process contexts} to refer to unguarded
sub-terms of a process. Process contexts are processes with a single unguarded
\emph{hole} $\Hole$ and are generated by the following grammar:
\[
    \textbf{Process context} \qquad
    \PCtxC, \PCtxD ~~::=~~
    \Hole ~~\mid~~
    \Cut\x\PCtx{P} ~~\mid~~
    \Cut\x{P}\PCtx
\]

As usual we write $\PCtx[P]$ for the process obtained by replacing the
hole in $\PCtx$ with $P$. Note that this substitution is not
capture-avoiding in general and some free names occurring in $P$ may be captured
by binders in $\PCtx$. We write $\bn\cdot$ for the function
inductively defined by
\[
  \bn\Hole = \emptyset
  \text{\qquad and \qquad}
  \bn{\Cut\x\PCtx{P}} = \bn{\Cut\x{P}\PCtx} = \set\x \cup \bn\PCtx
\]

The next lemma shows that a cut on $x$ can always be pushed down towards any
process in which $x$ occurs free.

\begin{lemma}
    \label{lem:push-down}
    If $x\in\fn{P} \setminus \bn\PCtxC$, then $\Cut\x{\PCtxC[P]}{Q} \pcong
    \PCtxD[\Cut\x{P}{Q}]$ for some $\PCtxD$.
  \end{lemma}
\begin{proof}
    By induction on the structure of $\PCtxC$ and by cases on its
    shape.

      \proofcase{Case $\PCtxC = \Hole$}
        We conclude by taking $\PCtxD \eqdef \Hole$ using the
        reflexivity of $\pcong$. 

      \proofcase{Case $\PCtxC = \Cut\y{\PCtxC'}{R}$}
        From the hypothesis $x \in \fn{P} \setminus \bn\PCtxC$ we
        deduce $x \ne y$ and $x \in \fn{P} \setminus \bn{\PCtxC'}$.
        Using the induction hypothesis we deduce that there exists
        $\PCtxD'$ such that $\Cut\x{\PCtxC'[P]}{Q} \pcong
        \PCtxD'[\Cut\x{P}{Q}]$. Take $\PCtxD \eqdef
        \Cut\y{\PCtxD'}{R}$. We conclude
        \[
            \begin{array}{rcll}
                \Cut\x{\PCtxC[P]}{Q}
                & = & \Cut\x{\Cut\y{\PCtxC'[P]}{R}}{Q} & \text{by definition of $\PCtxC$}
                \\
                & \pcong & \Cut\x{Q}{\Cut\y{\PCtxC'[P]}{R}} & \text{by \refrule{sp-comm}}
                \\
                & \pcong & \Cut\y{\Cut\x{Q}{\PCtxC'[P]}}{R} & \text{from $x \in \fn{P} \setminus \bn\PCtxC$} \\
                & & & \text{and \refrule{sp-assoc}}
                \\
                & \pcong & \Cut\y{\Cut\x{\PCtxC'[P]}{Q}}{R} & \text{by \refrule{sp-comm}}
                \\
                & \pcong & \Cut\y{\PCtxD'[\Cut\x{P}{Q}]}{R} & \text{by induction hypothesis}
                \\
                & = & \PCtxD[\Cut\x{P}{Q}] & \text{by definition of $\PCtxD$}
            \end{array}
        \]

      \proofcase{Case $\PCtxC = \Cut\y{R}{\PCtxC'}$}
        Analogous to the previous case.
\end{proof}

The proximity lemma is then proved by applying \Cref{lem:push-down} twice.

\begin{lemma}[Proximity]
    \label{lem:proximity_ll}
    \brk
    If $x\in\fn{P_i}\setminus\bn{\PCtxC_i}$ for $i=1,2$, then
    $\PCtxC[\Cut\x{\PCtxC_1[P_1]}{\PCtxC_2[P_2]} \pcong
    \PCtxD[\Cut\x{P_1}{P_2}]$ for some $\PCtxD$.
\end{lemma}
\begin{proof}
    It suffices to apply \Cref{lem:push-down} twice, thus:
    \[
        \begin{array}{@{}r@{~}c@{~}ll@{}}
            \PCtxC[\Cut\x{\PCtxC_1[P_1]}{\PCtxC_2[P_2]}
            & \pcong & \PCtxC[\PCtxD_1[\Cut\x{P_1}{\PCtxC_2[P_2]}]]
            & \text{using the hypothesis} \\
            & & & \text{and \Cref{lem:push-down}}
            \\
            & \pcong & \PCtxC[\PCtxD_1[\Cut\x{\PCtxC_2[P_2]}{P_1}]]
            & \text{by \refrule{sp-assoc}}
            \\
            & \pcong & \PCtxC[\PCtxD_1[\PCtxD_2[\Cut\x{P_2}{P_1}]]]
            & \text{using the hypothesis} \\
            & & & \text{and \Cref{lem:push-down}}
            \\
            & \pcong & \PCtxC[\PCtxD_1[\PCtxD_2[\Cut\x{P_1}{P_2}]]]
            & \text{by \refrule{sp-assoc}}
        \end{array}
    \]
    and we conclude by taking $\PCtxD \eqdef \PCtxC[\PCtxD_1[\PCtxD_2]]$.
\end{proof}

Notably, \Cref{lem:proximity_ll} corresponds to \Cref{lem:proximity_bin,lem:proximity_multi} that 
we proved in the session based scenarios.

%% file: ts-ll/proof-additional.tex
\beginbass
Most of the soundness proof of the type system relies on the cut elimination
result of \muMALL. Here we gather some auxiliary definitions and properties of
well-typed processes.
First of all, we define a function that makes a ``fair choice'' among two
processes $P$ and $Q$, by selecting the one with smaller rank. If $P$ and $Q$
happen to have the same rank, we choose $P$ by convention.

\begin{definition}
    \label{def:fc}
    The \emph{fair choice} among $P$ and $Q$, written $\fc{P}{Q}$, is defined by
    \[
        \fc{P}{Q} \eqdef
        \begin{cases}
            P & \text{if $\rankof{P} \leq \rankof{Q}$} \\
            Q & \text{otherwise}
        \end{cases}
    \]
\end{definition}

From its definition we have $\rankof{\fc{P}{Q}} = \min\set{ \rankof{P},
\rankof{Q} }$.
Next, we show that in a well-typed process there cannot be an infinite sequence
of choices if we follow the fair ones. To this aim, we define a total function
on processes that computes the length of the longest chain of subsequent fair
choices.

\begin{definition}
    \label{def:depth}
    Let $\depth\cdot$ be the function from processes to $\RankSet$ such that
    \[
        \depth{P} =
        \begin{cases}
            1 + \depth{\fc{P_1}{P_2}} & \text{if $P = \PiChoice{P_1}{P_2}$} \\
            0 & \text{otherwise}
        \end{cases}
    \]
\end{definition}

Note that $\depth{\fc{P}{Q}} < 1 + \depth{\fc{P}{Q}} = \depth{\PiChoice{P}{Q}}$.
We prove that, for well-typed processes, $\depth\cdot$ always yields a natural
number. That is, in a well-typed process there is no infinite chain of fair
choices.

\begin{lemma}
    \label{lem:depth}
    If $\piwtp\Ctx{P}$ then $\depth{P} \in \Nat$.
\end{lemma}
\begin{proof}
    Suppose that $\depth{P} = \infty$. Then the derivation for $\qtp\Ctx{P}$
    has an infinite branch $\gamma = (\qtp\Ctx{P_i})_{i\in\omega}$ solely
    consisting of choices such that $\rankof{P_{i+1}} < \rankof{P_i}$ for every
    $i\in\omega$. Therefore, $\rankof{P_i} = \infty$ for every $i\in\omega$ and
    $\gamma$ is fair.
    From the hypothesis that $P$ is well typed we deduce that $\gamma$ is also
    valid (see \Cref{def:valid_branch}), namely it has a non-stationary
    $\tnu$-thread. This contradicts the fact that the contexts in $\gamma$ are
    all equal to $\Ctx$.
    We conclude that $\depth{P} \in \Nat$.
\end{proof}

Now we define a function $\Resolve\cdot$ on well-typed processes to statically
and fairly resolve all the choices of a process.

\begin{definition}
    \label{def:resolve}
    Let $\Resolve\cdot$ be the function on well-typed processes such that
    $\Resolve{\PiChoice{P}{Q}} = \Resolve{\fc{P}{Q}}$ and extended
    homomorphically to all the other process forms.
\end{definition}

The fact that $\Resolve{P}$ is uniquely defined when $P$ is well typed is a
consequence of \Cref{lem:depth}. Indeed, any branch of $P$ that contains
infinitely many choices also contains infinitely many forms other than choices.
Note that the range of $\Resolve\cdot$ only contains zero-ranked processes.

The next two results prove that $\Resolve{P}$ is well typed if so is $P$.

\begin{lemma}
    \label{lem:resolved_quasi_typed}
    If $\piwtp\Ctx{P}$ then $\qtp\Ctx{\Resolve{P}}$.
\end{lemma}
\begin{proof}
    \newcommand\rrel{\mathcal{R}}
    We apply the coinduction principle to show that every judgment in the set
    \[
        \rrel \eqdef \set{\qtp\Ctx{\Resolve{P}} \mid
        \piwtp\Ctx{P}}
    \]
    is the conclusion of a rule in \Cref{fig:ts_ll} whose premises are
    also in $\rrel$.
    Let $\qtp\Ctx{Q} \in \rrel$. Then $Q = \Resolve{P}$ for some $P$ such
    that $\piwtp\Ctx{P}$.
    From \Cref{lem:depth} we deduce that $\depth{P} \in \Nat$.
    We reason by induction on $\depth{P}$ and by cases on the shape of $P$ to
    show that $\qtp\Ctx{Q}$ is the conclusion of a rule in
    \Cref{fig:ts_ll} whose premises are also in $\rrel$. We only discuss
    a few cases, the others being similar or simpler.
    
      \proofcase{Case $P = \Cut\x{P_1}{P_2}$}
        Then $Q = \Resolve{P} = \Cut\x{\Resolve{P_1}}{\Resolve{P_2}}$.
        From \refrule{\CutRule} we deduce that there exists $\Ctx_1$,
        $\Ctx_2$, $S_1$ and $S_2$ such that $\piwtp{\Ctx_i, x : S_i}{P_i}$
        for $i=1,2$ and $\Ctx = \Ctx_1, \Ctx_2$ and $S_1 =
        \dual{S_2}$.
        Then $\qtp{\Ctx_i, x : S_i}{\Resolve{P_i}} \in \rrel$ by definition
        of $\rrel$ and we conclude observing that $\qtp\Ctx{Q} \in \rrel$ is
        the conclusion of \refrule{\CutRule}.
    
      \proofcase{Case $P = \Choice{P_1}{P_2}$}
        Then $Q = \Resolve{P} = \Resolve{\fc{P_1}{P_2}}$. From
        \refrule{\PiChoiceRule} we deduce $\piwtp\Ctx{\fc{P_1}{P_2}}$. Since
        $\depth{\fc{P_1}{P_2}} < \depth{P} \in \Nat$, we conclude using the
        induction hypothesis.
\end{proof}

\begin{lemma}
    \label{lem:resolved_well_typed}
    If $\piwtp\Ctx{P}$ then $\piwtp\Ctx{\Resolve{P}}$.
\end{lemma}
\begin{proof}
    Using \Cref{lem:resolved_quasi_typed} we deduce $\qtp\Ctx{\Resolve{P}}$.
    Now, consider an infinite branch $\gamma$ in the derivation for
    $\qtp\Ctx{\Resolve{P}}$ and observe that $\gamma$ is necessarily fair,
    since it does not traverse any choice.
    The branch $\gamma$ corresponds to another infinite branch $\gamma'$ in the
    derivation for $\qtp\Ctx{P}$ which always makes fair choices whenever it
    traverses a process of the form $\PiChoice{P_1}{P_2}$. The only differences
    between $\gamma$ and $\gamma'$ are the judgments for the choices in $P$ that
    have been resolved. Nonetheless, all of the contexts that occur in $\gamma'$
    also occur in $\gamma$ because \refrule{\PiChoiceRule} does not affect typing
    contexts.
    If $\gamma'$ traverses infinitely many choices, then $\gamma'$ traverses
    infinitely many processes with strictly decreasing ranks, which must all be
    $\infty$. Therefore, $\gamma'$ is fair.
    Since $P$ is well typed we know that $\gamma'$ is valid. Then, the
    $\tnu$-thread that witnesses the validity of $\gamma'$ corresponds to a
    $\tnu$-thread that witnesses the validity of $\gamma$.
    We conclude that $\Resolve{P}$ is well typed.
\end{proof}

\begin{proof}[Proof of \Cref{lem:conservativity}]
    By \Cref{lem:resolved_well_typed} we deduce
    $\piwtp{x_1:S_1,\dots,x_n:S_n}{\Resolve{P}}$. Since $\rankof{\Resolve{P}} =
    0$, there are no applications of \refrule{\PiChoiceRule} in the derivation for
    $\qtp{x_1:S_1,\dots,x_n:S_n}{\Resolve{P}}$. That is, this derivation
    corresponds to a \muMALL derivation and every infinite branch in it is fair.
    Since $\Resolve{P}$ is well typed, we deduce that every infinite branch in
    this derivation is valid. That is, $\vdash S_1, \dots, S_n$ is derivable in
    \muMALL.
\end{proof}

The key property of zero-ranked processes that are well typed in a context $x :
\One$ is that they are weakly terminating and they eventually reduce to
$\PiClose\x$.

\begin{lemma}
    \label{lem:zero_rank_weak_termination}
    If $\piwtp{x:\One}{P}$ and $\rankof{P} = 0$ then $P \wred \Close\x$.
\end{lemma}
\begin{proof}
    Without loss of generality we may assume that $P$ does not contain links.
    Indeed, the axiom \refrule{\LinkRule} is admissible in
    \muMALL \citep[Proposition 10]{BaeldeDoumaneSaurin16}, so we may assume that
    links have been expanded into equivalent (choice-free) processes. We just
    note that, if the statement holds for link-free processes, then it holds
    also in the case $P$ does contain links, the only difference being that the
    sequence of reductions that are needed to fully eliminate all the cuts
    resulting from an expanded link are replaced by a single occurrence of
    \refrule{rp-link}.

    From the hypothesis that $P$ has rank 0 we deduce that $P$ does not contain
    non-determimnistic choices and the derivation of $\qtp{x:\One}{P}$ does not
    contain applications of the \refrule{\PiChoiceRule} rule. So, every infinite
    branch of this derivation is fair. From the hypothesis that every infinite
    fair branch of this derivation is valid we deduce that every infinite branch
    of this derivation is valid. Then this derivation corresponds to a \muMALL
    proof of $\vdash \One$.
    Observe that every principal reduction rule of \muMALL \cite[Figure
    3.2]{Doumane17} corresponds to a reduction rule of \piLIN
    (see \Cref{fig:red_ll}).
    Also, \Cref{lem:proximity_ll} guarantees that, whenever there are two processes
    in which the same channel name $x$ occurs free we are always able to rewrite
    the process using structural precongruence so that the two sub-processes are
    the children of a cut on $x$.
    Finally, the cut elimination result for \muMALL is proved by reducing
    bottom-most cuts, meaning that principal reductions (also called
    \emph{internal reductions} \citep{Doumane17}) are only applied at the bottom
    of a \muMALL proof.
    Therefore, each step in which a principal reduction is applied in the cut
    elimination result of \muMALL can be mimicked by a reduction of \piLIN.
    From the cut elimination result for \muMALL \cite[Proposition
    3.5]{Doumane17} we deduce that there exists no (fair) infinite sequence of
    principal reductions. That is, there exists a stuck $Q$ such that $P \wred
    Q$. From \Cref{lem:quasi_subj_red_ll} we deduce $\qtp{x:\One}{Q}$.
    Since the only cut-free \muMALL proof of $\vdash \One$ consists of a single
    application of \refrule[close]{\PiCloseRule}, and since this proof corresponds
    to the proof that $Q$ is quasi typed, we conclude that $Q = \Close\x$.
\end{proof}

%% file: ts-ll/proof-soundness.tex
\beginbass
The last auxiliary result for proving \Cref{lem:weak_termination_ll} is to show
that $P$ weakly simulates $\Resolve{P}$, namely that every reduction of a
process $\Resolve{P}$ can be mimicked by one or more reductions of $P$.

\begin{lemma}
    \label{lem:resolved_red}
    If $\piwtp\Ctx{P}$ and $\Resolve{P} \red Q$ then $P \wred R$ for some $R$
    such that $Q = \Resolve{R}$.
\end{lemma}
\begin{proof}
    From \Cref{lem:depth} we deduce $\depth{P} \in \Nat$. We proceed by double
    induction on $\depth{P}$ and on the derivation of $\Resolve{P} \red Q$ and
    by cases on the last rule applied in the derivation of $\qtp\Ctx{P}$.
    Most cases are straightforward since the structure of $\Resolve{P}$ and that
    of $P$ only differ for non-deterministic choices, so we only discuss the
    case \refrule{\PiChoiceRule} in which $P = \Choice{P_1}{P_2}$. Then
    $\qtp\Ctx{\fc{P_1}{P_2}}$ and $\Resolve{P} = \Resolve{\fc{P_1}{P_2}}
    \red Q$. Recall that $\depth{\fc{P_1}{P_2}} < \depth{P} \in \Nat$. Using the
    induction hypothesis we deduce $\fc{P_1}{P_2} \wred R$ for some $R$ such
    that $Q = \Resolve{R}$. We conclude by observing that $P \red \fc{P_1}{P_2}
    \wred R$.
\end{proof}

\Cref{lem:resolved_red} can be easily generalized to arbitrary sequences of
reductions.

\begin{lemma}
    \label{lem:resolved_reds}
    If $\piwtp\Ctx{P}$ and $\Resolve{P} \wred Q$ then $P \wred R$ for some $R$
    such that $Q = \Resolve{R}$.
\end{lemma}
\begin{proof}
    A simple induction on the number of reductions in $\Resolve{P} \wred Q$
    using \Cref{lem:resolved_red}.
\end{proof}

\begin{proof}[Proof of \Cref{lem:weak_termination_ll}]
    Using \Cref{lem:resolved_well_typed} we deduce $\piwtp{x:\One}{\Resolve{P}}$.
    Using \Cref{lem:zero_rank_weak_termination} we deduce $\Resolve{P} \wred
    \Close\x$. Using \Cref{lem:resolved_reds} and
    \Cref{lem:quasi_subj_red_ll} we conclude $P \wred \Close\x$.
\end{proof}

Finally, we can prove \Cref{thm:ts_ll_sound,cor:fair_termination_ll}.

\begin{proof}[Proof of \Cref{thm:ts_ll_sound}]
    By induction on the number of reductions in $P \wred Q$, from the hypothesis
    $\piwtp{x : \One}{P}$ and \Cref{thm:subj_red_ll} we deduce $\piwtp{x :
    \One}{Q}$. We conclude using \Cref{lem:weak_termination_ll}.
\end{proof}

\begin{proof}[Proof of \Cref{cor:fair_termination_ll}]
		\brk
    Immediate consequence of \Cref{thm:ts_ll_sound,thm:fair_termination}.
\end{proof}

%% file: ts-ll/comparison.tex
\beginalto
This last section concludes \Cref{pt:type_systems} so we dedicate it to a comparison
between the two approaches to fair termination that we presented, that is,
the session based one (see \Cref{ch:ft_bin,ch:ft_multi}) and that about linear logic
(in the present chapter).
The most important result is that the two approaches \emph{cannot} be actually compared
since there exist examples that are accepted in the former and rejected in the other and viceversa.
Here we provide two examples of this fact (see \Cref{ex:fwd_comp,ex:slot_comp}).
Then, such result leaves an open problem, that is establishing the meaning of fair subtyping
in the linear logic scenario. Indeed, the type system in \Cref{ssec:ts_ll} does not
use subtyping. 
What makes this research line interesting is how \emph{infinite} behaviors are dealt with 
in the two approaches; the former relies on fair subtyping and the second on validity conditions.
To conclude, in \Cref{ex:delegation_ll} we encode \Cref{ex:invariant_ch} in the linear logic
context and we see how it is rejected by the type system in \Cref{ssec:ts_ll}.

\begin{example}[Forwarder]
	\label{ex:fwd_comp}
	Here we recall the forwarder that we proved to be well typed in \piLIN in \Cref{ex:forwarder}.
	We model the same process using binary sessions and we try to prove that it is well typed
	using the type system in \Cref{ch:ft_bin}.
	\newcommand{\lok}{\Tag[ok]}
	\[
	\Forwarder(x,y) = \x\iact
		\set{\lok.\y\oact\lok.\Call{\Forwarder}{\x,\y},
				 \lstop.\y\oact\lstop.\Wait{x}{\Close\y}
		} 
	\]
	where $\x$ and $\y$ are used according to $S_x$ and $S_y$ defined below
	\[
	\begin{array}{lr}
	S_x = \In\set{\lok.S_x,\lstop.\End[\In]}
	& \qquad
	S_y = \Out\set{\lok.S_y,\lstop.\End[\Out]}	
	\end{array}
	\]
	It is clear that, if we try to provide a typing derivation for $\Forwarder$, then
	we have to apply rule \refrule{tb-tag} that we report here for the sake of clarity.
	\[
	\inferrule
          {\forall i\in I: \wtp[n]{\Ctx, x : S_i}{P_i}}
          {
            \textstyle
            \wtp[n]{\Ctx, x : \Pol\set{\l_i : S_i}_{i \in I}}
            {x\Pol\set{\l_i : P_i}_{i \in I}}
          } 
	\]
	In order to apply such rule the branches must be typed in some
	contexts that differ only with respect to $\x$. 
	Hence, we cannot apply \refrule{tb-tag} by taking into account $S_x$ and $S_y$
	because the behavior of $\y$ changes as well.
	We can try to solve the problem by placing casts in the right positions
	\[
	\Forwarder(x,y) = \x\iact
		\set{\lok.\Cast\y\y\oact\lok.\Call{\Forwarder}{\x,\y},
				 \lstop.\Cast\y\y\oact\lstop.\Wait{x}{\Close\y}
		} 
	\]
	Indeed $\y$ is used according to $\Out\lok.S_y$ and $\Out\lstop.\End[\Out]$
	in the left and right branches, respectively. Moreover,
	the two applications of fair subtyping are correct and we can easily derive
	$S_y \subt[1] \Out\lok.S_y$ and $S_y \subt[1] \Out\lstop.\End[\Out]$.
	Note that the weights are $1$ because the types differ only on the topmost
	output, which means that the contravariant rule \refrule{fsb-tag-out-2} for
	output must be applied.
	Now we can try to provide a typing derivation (we omit the rule names).
	\[
	\begin{prooftree}
		\[
			\[
				\[
					\vdots
					\justifies
					\wtp[n]{\x : S_x , \y : S_y}{\Call{\Forwarder}{\x,\y}}
				\]
				\justifies
				\wtp[n]{\x : S_x , \y : S_y}{\y\oact\lok.\Call{\Forwarder}{\x,\y}}
			\]
			\justifies
			\wtp[n+1]{\x : S_x , \y : \Out\lok.S_y}{\Cast\y\y\oact\lok.\Call{\Forwarder}{\x,\y}}
		\]
		\[
			\[
				\[
					\[
						\justifies
						\wtp[n]{\y : \End[\Out]}{\Close\y}
					\]
					\justifies
					\wtp[n]{\x : \End[\In] : \y : \End[\Out]}{\Wait{\x}{\Close\y}}
				\]
				\justifies
				\wtp[n]{\x : \End[\In], \y : \Out\lstop.\End[\Out]}{\oact\lstop\dots}
			\]
			\justifies
			\wtp[n+1]{\x : \End[\In], \y : S_y}{\Cast\y\dots}
		\]
		\justifies
		\wtp[n+1]{\x : S_x, \y : S_y}{\Call{\Forwarder}{\x,\y}}
	\end{prooftree}
	\]
	where the right branch ends with an application of \refrule{tb-end} which assigns rank $n$.
	Indeed, \refrule{tb-tag} requires that the rank is an upper bound to the rank of the
	branches so we assign it in order to match that of the left one.
	Note that it is not possible to assign a finite rank to the process
	because it requires $n = n+1$, hence the process if \emph{cast unbounded} (\Cref{ssec:boundedness}).
	\eoe
\end{example}

\begin{example}[A Hopeful Player]
	\label{ex:slot_comp}
	\newcommand{\SlotMachineF}{SlotF}
	\newcommand{\SlotMachineU}{SlotU}
	\newcommand{\MachineF}{MF}
	\newcommand{\MachineU}{MU}
	The slot machine was a recurring example in \Cref{pt:type_systems} (see \Cref{ex:non-det,ex:slot_machine}).
	In \Cref{ex:slot_fair_sub} we described the behaviors of a fair and an unfair machine.
	The second one simply never lets the player win. Let $S$, $T$ be
	the types from \Cref{ex:slot_fair_sub} such that
	\[
	\begin{array}{rcl}
		S & = & \In\tplay.(\Out\twin.S + \Out\tlose.S) + \In\tquit.\End[\Out]
		\\
		T & = & \In\tplay.\Out\tlose.T + \In\tquit.\End[\Out]
	\end{array}
	\]  
	Now we can model the two implementations of the machines. For the sake of simplicity,
	we adapt the one in \Cref{ex:non-det} to the binary case. We dub $\SlotMachineF$ the fair
	machine and $\SlotMachineU$ the unfair one.
	\[
	\begin{array}{rcl}
		\SlotMachineF(x) & \peq &
			\x\iact\set{
				\tplay.\x\oact\set{
					\twin.\pinvk\SlotMachineF{x},
					\tlose.\pinvk\SlotMachineF{x}
				},
				\tquit.\pclose{x}
			}
		\\
		\SlotMachineU(x) & \peq &
			\x\iact\set{
				\tplay.\x\oact\tlose.\pinvk\SlotMachineU{x},
				\tquit.\pclose{x}
			}
	\end{array}
	\]
	It is easy to see that $\wtp[0]{\x : S}{\pinvk\SlotMachineF{x}}$ and
	that $\x : S \not\vdash \pinvk\SlotMachineU{x}$ since the $\twin$ branch is missing.
	We can try to rewrite $\SlotMachineU$ in order to be well-typed with $\x : S$.
	First, we can place a cast in the $\tlose$ branch.
	\[
		\SlotMachineU(x) \peq
			\x\iact\set{
				\tplay.\pcast\x\x\oact\tlose.\pinvk\SlotMachineU{x},
				\tquit.\pclose{x}
			}
	\]
	where the cast is correct since we can derive 
	$(\Out\twin.S + \Out\tlose.S) \subt[1] \Out\tlose.S$. The $1$ weight
	is due to the fact that the two types differ only for the topmost output, which requires
	an application of \refrule{fsb-tag-out-2}.
	Let us try to derive a typing judgment.
	\[
	\begin{prooftree}
		\[
			\[
				\[
					\vdots
					\justifies
					\wtp[n]{\x : S}{\pinvk\SlotMachineU{x}}
				\]
				\justifies
				\wtp[n]{\x : \Out\tlose.S}{\x\oact\tlose.\pinvk\SlotMachineU{x}}
			\]
			\justifies
			\wtp[n+1]{\x : \Out\twin.S + \Out\tlose.S}{\pcast\x\x\oact\tlose.\pinvk\SlotMachineU{x}}
		\]
		\[
			\justifies
			\wtp[n+1]{x : \End[\Out]}{\pclose\x}
		\]
		\justifies
		\wtp[n+1]{\x : S}\pinvk\SlotMachineU{x}
	\end{prooftree}
	\]
	Hence, we cannot find a finite rank; the process is \emph{cast unbounded} (see \Cref{ssec:boundedness}).
	Alternatively we can try to place a cast before the first invocation of the process since it is clear
	that $\SlotMachineU$ uses $\x$ according to $T$. Such cast would be invalid
	since in \Cref{ex:slot_fair_sub} we proved that $S \not\subt T$.
	Hence, the only typing judgment that we can derive is $\wtp[0]{\x : T}{\pinvk\SlotMachineU{x}}$.
	
	Now we can apply the same reasoning from the \piLIN point of view.
	Let us model the fair (\Cref{ex:slot_machine}) and the unfair machines.
	We dub $\MachineF$ the fair machine and $\MachineU$ the unfair one.
	\[
	\begin{array}{rcl}
		\MachineF(x,y) & \peq & \Corec\x.
            \\ & &
            \Case\x{
                \PiChoice{
                    \Select\WinTag\x.\Call\MachineF{x,y}
                }{
                    \Select\LoseTag\x.\Call\MachineF{x,y}
                }
            }{
                \PiWait\x.\PiClose\y
            }
		\\
		\MachineU(x,y) & \peq & \Corec\x.
            \Case\x{
                \Select\LoseTag\x.\Call\MachineU{x,y}
            }{
                \PiWait\x.\PiClose\y
            }
	\end{array}
	\]
	In \Cref{ex:slot_machine} we proved that 
	$\qtp{x : \FormulaG_b, y : \One}{\Call\MachineF{x,y}}$ and that $\MachineF$ is well-typed
	where $\FormulaG \eqdef \tnu\X.(X \plinchoice X) \plinbranch \Bot$.
	At the same time we can prove that $\qtp{x : \FormulaG_b, y : \One}{\Call\MachineU{x,y}}$
	\[
    \begin{prooftree}
        \[
            \[
                \[
                    \mathstrut\smash\vdots
                    \justifies
                    \qtp{
                        x : \FormulaG_{bilr},
                        y : \One
                    }{
                        \Call\MachineU{x,y}
                    }
                \]
                \justifies
                \qtp{
                    x : \FormulaG'_{bil},
                    y : \One        
                }{
                    \Select\LoseTag\x.\Call\MachineU{x,y}
                }
                \using\refrule[select]\SelectRule
            \]
            \[
                \[
                    \justifies
                    \qtp{
                        y : \One
                    }{
                        \PiClose\y
                    }
                    \using\refrule[close]{\PiCloseRule}
                \]
                \justifies
                \qtp{
                    x : \Bot,
                    y : \One
                }{
                    \PiWait\x.\PiClose\y    
                }
                \using\refrule[wait]{\PiWaitRule}
            \]
            \justifies
            \qtp{
                x : (\FormulaG' \plinbranch \Bot)_{bi},
                y : \One
            }{
                \Case\x{
                    \Select\LoseTag\x.\Call\MachineU{x,y}
                }{
                    \PiWait\x.\PiClose\y
                }            
            }
            \using\refrule[case]{\CaseRule}
        \]
        \justifies
        \qtp{
            x : \FormulaG_b,
            y : \One
        }{
            \Call\MachineU{x,y}
        }
        \using\refrule[corec]{\CorecRule}
    \end{prooftree}
		\]
		Moreover the process is well-typed since no choices are met.
		This example points out that we can type two different processes with the same type
		while in the session scenario we the correspondence between types and processes is
		tighter.
		
		Now we can focus on the player side and we can try to model a hopeful player
		that plays until it receives $\WinTag$. In that case, it stops playing.
		We first write it in the binary session scenario and then we encode it
		in \piLIN.
		\newcommand{\Play}{Player}
		\[
		\Play(\x) \peq \x\oact\tplay.\x\iact\set{\twin.\pclose\x,\tlose.\pinvk\Play\x}
		\]
		where $\x$ is used according to $T_p = \Out\tplay.\In\set{\twin.\End[\Out], \tlose.T_p}$.
		It can be derived that $\wtp[0]{\x : T_p}{\pinvk\Play\x}$.
		Note that the system consisting of $\MachineF$ and $\Play$ is fairly terminating.
		Indeed $S \compatible T_p$.
		
		Now we model the same process in \piLIN. In particular,
		we slightly modify the player in \Cref{ex:slot_machine}.
		\renewcommand{\Player}{P}
		\[
		\Player(x) =
        \Rec\x.
            \Select\PlayTag\x.
            \Case\x{
                \Call\Player\x
            }{
                \Rec\x.\Select\QuitTag\x.\PiClose\x
            }
		\]
		Let us look at the typing derivation with $\FormulaF \eqdef \tmu\X.(X \plinbranch X) \plinchoice \One$.
		\[
    \begin{prooftree}
        \[
                \[
                    \[
                        \mathstrut\smash\vdots
                        \justifies
                        \qtp{
                            x : \FormulaF_{aill}
                        }{
                            \Call\Player\x
                        }
                    \]
                    \qquad
                    \[
                        \[
                            \[
                                \justifies
                                \qtp{
                                    x : \One
                                }{
                                    \PiClose\x
                                }
                                \using\refrule[close]{\PiCloseRule}
                            \]
                            \justifies
                            \qtp{
                                x : \Formula'_{ailri}
                            }{
                                \Select\QuitTag\x.\PiClose\x
                            }
                            \using\refrule[select]{\SelectRule}
                        \]
                        \justifies
                        \qtp{
                            x : \FormulaF_{ailr}
                        }{
                            \Rec\x.\Select\QuitTag\x.\PiClose\x
                        }
                        \using\refrule[rec]{\RecRule}
                    \]
                    \justifies
                    \qtp{
                        x : (\FormulaF \plinbranch \FormulaF)_{ail}
                    }{
                        \Case\x{
                            \Call\Player\x
                        }{
                            \Rec\x\Select.\QuitTag\x.\PiClose\x
                        }    
                    }
                    \using\refrule[case]{\CaseRule}
                \]
                \justifies
                \qtp{
                    x : \Formula'_{ai}
                }{
                    \Select\PlayTag\x.
                    \Case\x{
                        \Call\Player\x
                    }{
                        \Rec\x.\Select\QuitTag\x.\PiClose\x
                    }
                }
                \using\refrule[select]{\SelectRule}
        \]
        \justifies
        \qtp{
            x : \Formula_a
        }{
            \Call\Player\x
        }
        \using\refrule[rec]{\RecRule}
    \end{prooftree}
		\]
		Note that now $\rankof{\Call\Player\x} = 0$ since there are no choices.
		Hence, the infinite branch is fair because it traverses no choices.
		We can conclude that the $\Call\Player\x$ is not well typed because
		because the infinite, fair branch is not valid.
		We recall that the player in \Cref{ex:slot_machine} was well typed because
		the infinite branch was unfair as it contained infinitely many finitely ranked choices.
		\eoe
\end{example}

\Cref{ex:fwd_comp,ex:slot_comp} show that the type systems in \Cref{ch:ft_bin,ch:ft_multi} cannot be compared
to that in \Cref{ch:ft_ll}. \Cref{ex:fwd_comp} illustrates a process that can be only typed in \piLIN
while \Cref{ex:slot_comp} show a process that can be typed only in the session based calculi.
Moreover \Cref{ex:slot_comp} shows a peculiar difference between the type systems. 
In \Cref{ch:ft_bin,ch:ft_multi} the tight correspondence between types and processes allows us
to use a \emph{fair} type to prove that a \emph{fair} process is well-typed. 
Such correspondence does not hold in \piLIN as we proved that both the fair and the unfair
slot machines can be typed with the same formula.
We conclude this section showing how to encode delegation in \piLIN and trying
to encode \Cref{ex:invariant_ch} to see whether it is accepted or rejected in \piLIN.

\newcommand{\enc}[1]{\lfloor #1 \rfloor}

\begin{definition}[Delegation in \piLIN]
	\label{def:delegation_ll}
	We show delegation how can be encoded in \piLIN. 
	\cite{LindleyM15} provide an encoding from GV (a session typed functional language) to \CP
	(a process calculus based on classical linear logic). 
	Hence, they encode (binary) session types
	to the types of CP, that is, linear logic formulas. Concerning delegation, the encoding
	is provided as
	\[
	\begin{array}{rclrcl}
		\enc{\Out\T.\S} & = & \co{\enc\T} \tfork \enc\S
		& \qquad
		\enc{\In\T.\S} & = & \enc\T \tjoin \enc\S
	\end{array}
	\]
	(see \cite[Fig.9]{LindleyM15}).
	Following the encoding of types we extend the encoding to 
	the processes in \Cref{sec:ts_bin_proc}.
	\[
	\begin{array}{rclrcl}
		\enc{\POutput\x{(\y)}.P} & = & \Fork\x\z{\Link\y\z}{\enc{P}}
		& \qquad
		\enc{\PInput\x{(\y)}} & = & \Join\x\y{\enc{P}}
	\end{array}
	\]
\end{definition}

\begin{example}[Delegation]
	\label{ex:delegation_ll}
	In this example we encode \Cref{ex:invariant_ch} in \piLIN and we see if we obtain a well-typed process or not.
	To encode delegation we refer to \Cref{def:delegation_ll}.
	Let us first recall the process and the main types.
	\[
    	\begin{array}{@{}r@{~}l@{}}
    	& \NewPar\y{\NewPar\x{\Call A {x,y}}{\Call B x}}{\Call B y}
    	\\
    	\Definition{A}{x,y}{& \POutput\x\lmore.\POutput\x\y.\Call B x}
    	\\
    	\Definition{B}{x}{&
        \PRecv\x{
        	\lmore : \PInput\x{(y)}.\Call A {y,x},
        	\lstop : \Wait\x\Done
        }
    	}	
    	\end{array}
    \]
   	where $\x$ and $\y$ are used according to $S_A$ and $T_A$ in $A(x,y)$ and
   	$\x$ is used according to $S_B$ in $B(x)$.
   	\[
	 	\begin{array}{rcl}
	 		S_A & = & \Out\lmore.\Out\T_A.S_B
    	\\
    	S_B & = & \In\lmore.\In\S_A.T_A \branch \In\lstop.\End[\In]
    	\\
    	T_A & = & \Out\lmore.\Out\T_A.S_B \choice \Out\lstop.\End[\Out]
		\end{array}
		\]
		Now we can look at the encoding of processes.
		\[
		\begin{array}{rcl}
			\enc{A(x,y)} & = & \Rec\x.\Select\lmore\x.\Fork\x\z{\Link\y\z}{\enc{B(\x)}}
			\\
			\enc{B(x)} & = & \Corec\x.\Case\x{\Join\x\y\enc{A(y,x)}}{\PiWait\x}
		\end{array}
		\]
		The encoding of types is more involved. Since we do not have subtyping in
		\piLIN, we can encode $S_b$ and $T_A$. Moreover, if we replace $\In{S_A}.T_A$
		with $\In{T_A}.T_A$ in $S_B$, we obtain that $\co{S_B} = T_A$.
		Let $\enc{T_A} = \Formula_A$ and $\enc{S_B} = \Formula_B$.
		\[
		\begin{array}{rlr}
			\Formula_A = & \tmu\Y.(\dual{\Formula_A} \tfork \Formula_B) \plinchoice \One 
				\\ & \hfill \text{by definition of $\enc{T_A}$}
				\\ 			 = & \tmu\Y.(\dual{\Formula_A} \tfork ((\Y \tjoin \Y) \plinbranch \Bot)) \plinchoice \One
				\\ & \hfill \text{by replacing $\Formula_B$}
				\\			 = & \tmu\Y.((\tnu\X.(\Y \tjoin ((\X \tfork \X) \plinchoice \One)) \plinbranch \Bot) 
															\tfork ((\Y \tjoin \Y) \plinbranch \Bot)) \plinchoice \One \qquad
				\\ & \hfill \text{by replacing $\dual{\Formula_A}$}
				\\
			\Formula_B = & \tnu\X.(\Formula_A \tjoin \Formula_A) \plinbranch \Bot
				\\ & \hfill \text{by definition of $\enc{S_B}$}
				\\
				         = & \tnu\X.(\Formula_A \tjoin ((\X \tfork \X) \plinchoice \One)) \plinbranch \Bot
				\\ & \hfill \text{by replacing $\Formula_A$}
				\\
				         = & \tnu\X.((\tmu\Y.(\X \tfork ((\Y \tjoin \Y) \plinbranch \Bot)) \plinchoice \One) 
				                      \tjoin ((\X \tfork \X) \plinchoice \One)) \plinbranch \Bot
				\\ & \hfill \text{by replacing $\Formula_A$}      
		\end{array}
		\]
		From these definition it can be noted that it the encoding is not straightforward since we do not
		know how/when actually unfold the formulas. Indeed we are not able to find the right
		types to prove that the process is quasi-typed.
		
		Furthermore, assuming that we could find such types, it would turn out that $B$ is well-typed
		while $A$ is not as its infinite thread in not a $\nu$ one.
		\eoe
\end{example}

%% file: ts-ll/related.tex
\beginalto
On account of the known encodings of sessions into the linear
$\pi$-calculus \citep{Kobayashi02b,DardhaGiachinoSangiorgi17,ScalasDardhaHuYoshida17},
\piLIN belongs to the family of process calculi providing logical foundations to
sessions and session types. Some representatives of this family are
\piDILL \citep{CairesPfenningToninho16} and its variant equipped with a circular
proof theory \citep{DerakhshanPfenning19,Derakhshan21}, \CP \citep{Wadler14} and
\muCP \citep{LindleyMorris16}, among others.
There are two main aspects distinguishing \piLIN from these calculi. The first
one is that these calculi take sessions as a native feature. This fact can be
appreciated both at the level of processes, where session endpoints are
\emph{linearized} resources that can be used \emph{multiple times} albeit in a
sequential way, and also at the level of types, where the interpretation of the
$\FormulaF \tfork \FormulaG$ and $\FormulaF \tjoin \FormulaG$ formulas is skewed
so as to distinguish the type $\FormulaF$ of the message being sent/received on
a channel from the type $\FormulaG$ of the channel after the exchange has taken
place.
In contrast, \piLIN adopts a more fundamental communication model based on
\emph{linear} channels, and is thus closer to the spirit of the encoding of
linear logic proofs into the $\pi$-calculus proposed by \cite{BellinScott94} 
while retaining the same expressiveness of the
aforementioned calculi. To some extent, \piLIN is also more general than those
calculi, since a formula $\FormulaF \tfork \FormulaG$ may be interpreted as a
protocol that bifurcates into two independent sub-protocols $\FormulaF$ and
$\FormulaG$ (we have seen an instance of this pattern in
\Cref{ex:parallel_programming}). So, \piLIN is natively equipped with the
capability of modeling some multiparty interactions, in addition to all of the
binary ones.
A session-based communication model identical to \piLIN, but whose type system
is based on intuitionistic rather than classical linear logic, has been
presented by \cite{DeYoungCairesPfenningToninho12}. In that work,
the authors advocate the use of explicit continuations with the purpose of
modeling an asynchronous communication semantics and they prove the equivalence
between such model and a buffered session semantics. However, they do not draw a
connection between the proposed calculus and the linear
$\pi$-calculus \citep{KobayashiPierceTurner99} through the encoding of binary
sessions \citep{Kobayashi02b,DardhaGiachinoSangiorgi17} and, in the type system,
the multiplicative connectives are still interpreted asymmetrically.
The second aspect that distinguishes \piLIN from the other calculi is its type
system, which is still deeply rooted into linear logic and yet it ensures fair
termination instead of
progress \citep{CairesPfenningToninho16,Wadler14,QianKavvosBirkedal21},
termination \citep{LindleyMorris16} or strong
progress \citep{DerakhshanPfenning19,Derakhshan21}. Fair termination entails
progress, strong progress and lock freedom \citep{Kobayashi02,Padovani14}, but at
the same time it does not always rule out processes admitting infinite
executions. Simply, infinite executions are deemed unrealistic because they are
unfair.

Another difference between \piLIN and other calculi based on linear logic is
that its operational semantics is completely ordinary, in the sense that it does
not include commuting conversions, reductions under prefixes, or the swapping of
communication prefixes. The cut elimination result of \muMALL, on which the
proof of \Cref{thm:ts_ll_sound} is based, works by reducing cuts from the bottom
of the derivation instead of from the
top \citep{BaeldeDoumaneSaurin16,Doumane17,BaeldeEtAl22}. As a consequence, it is
not necessary to reduce cuts guarded by prefixes or to push cuts deep into the
derivation tree to enable key reductions in \piLIN processes.

The extension of calculi based on linear logic with non-deterministic features
has recently received quite a lot of attention. \cite{RochaCaires21} have proposed 
a session calculus with shared cells
and non-deterministic choices that can model mutable state. Their typing rule
for non-deterministic choices is the same as our own, but in their calculus
choices do not reduce. Rather, they keep track of every possible evolution of a
process to be able to prove a confluence result.
\cite{QianKavvosBirkedal21} introduce \emph{coexponentials}, a new
pair of modalities that enable the modeling of concurrent clients that compete
in order to interact with a shared server that processes requests sequentially.
In this setting, non-determinism arises from the unpredictable order in which
different clients are served. Interestingly, the coexponentials are derived by
resorting to their semantics in terms of least and greatest fixed points. For
this reason, the cut elimination result of \muMALL might be useful to attack the
termination proof in their setting.

%% file: gis-lib/chintro.tex
\begintreble
In this chapter we present a library for supporting generalized inference systems (see \Cref{sec:gis}) in Agda \citep{Agda}.
Summarizing from \Cref{sec:gis},
an \emph{inference system} \citep{Aczel77,LeroyG09,Sangiorgi11}, that is, a set of (meta-)rules stating that 
a consequence can be derived from a set of premises, is a simple, general and widely-used way to express 
and reason about a recursive definition. In most cases such recursive definition is seen as inductive, 
that is,  the denoted set consists of the elements with a finite derivation. 
This  enables \emph{inductive reasoning}, that is, to prove that the elements an inductively 
defined set satisfy a property, it is enough to show that, for each (meta-)rule, the property 
holds for the consequence assuming that it holds for the premises.
In other cases, the recursive definition is seen as coinductive, 
that is, the denoted set  consists of the elements with a possibly infinite derivation. 
This enables \emph{coinductive reasoning}, that is, to prove that all the elements satisfying a 
property belong to the coinductively defined set, it is enough to show that, when the property 
holds for an element, it can be derived from premises for which the property holds as well. 
Recently, a generalization of inference systems has been proposed \citep{AnconaDagninoZucca17,Dagnino19,Dagnino21} 
which handles cases where neither the inductive, nor the purely coinductive intepretation provides the desired meaning. 
This approach is called \emph{flexible coinduction}, and, correspondingly, coinductive reasoning is 
generalized as well by a principle which is called \emph{bounded coinduction}.

The Agda proof assistant \citep{Agda} offers language constructs to inductively/coinductively define predicates, 
and correspondingly  built-in proof principles. 
However, in this way the recursive definition is monolithic, and hard-wired with its chosen interpretation.
\begin{remark}
	In \cite{Ciccone20} we deeply investigated the built-in features of Agda by
	inspecting all the different ways a (co)inductive predicate can be defined.
	We found out that, while pure (co)inductive definitions are easily supported, a predicate mixing 
	both approaches led to complex codes with many duplicated notions due to the lack 
	of modularity (see \Cref{rm:agda_dup}).
\end{remark} 
Our aim, instead, is to provide an Agda library allowing  the user to express a recursive definition 
as an \emph{instance of a parametric type} of inference systems.  
In this way, the user is not committed from the beginning to a given interpretation but, rather, 
gets for free a bunch of properties which have been proved once and for all, including the inductive and 
coinductive intepretation and the corresponding proof principles. 
Moreover, it is possible to define composition operators on inference systems, 
for instance union and restriction.
Finally, flexible coinduction is modularly obtained as well, 
by composing in a certain way the interpretations of two inference systems. 

\emph{Indexed containers} \citep{AltenkirchGHMM15} provide a way to specify possibly recursive definitions 
of predicates independently from their interpretation and are supported in the Agda standard library. 
An Agda implementation of inference systems can be provided by seeing them as indexed containers. 
However, this approach requires to structure definitions in an unusual way. 
Indeed, inference systems are usually presented through a (finite) set of \emph{meta-rules}, 
denoting all the rules which can be obtained by instantiating meta-variables with values satisfying the side condition. 
Hence, we provide a different implementation following this schema, to allow users to write their own inference 
system in an Agda format which closely resembles  that ``on paper''. 
We then prove that the two implementations are equivalent, showing that every  indexed container can be 
encoded in terms of meta-rules and viceversa.

\begin{remark}[Agda Version \& Reference]
	\label{rm:agda_v_link}
	It is important to point out that the material presented in \Cref{ch:gis_lib,ch:agda_prop}
	holds as long as Agda is consistent.
	During the development of the notions we are going to present, Agda received many updates.
	As an example, an inconsistency in mixing \emph{sized types} and \emph{coinductive records}
	has been found.
	The library is available on GitHub (see \cite{GisLib}) and has been tested
	with the Agda 2.6.2.2. In case of future Agda updates, we will try to keep the
	material updated.  
	\eor
\end{remark}

The chapter is structured as follows.
In \Cref{sec:agda_gis_meta} we show the mechanization of the meta theory of (generalized) inference systems (see \Cref{sec:gis}).
In particular, we present the main datatypes to encode inference systems in Agda, how to obtain their interpretations and,
most important, the proof principles. We provide the formalization of the examples that we introduce in \Cref{sec:gis} since
they give an overview of all the features of the library.
Then, in \Cref{sec:agda_gis_lambda} we show a more involved example, that is, a lambda calculus with divergence and
we prove soundness and completeness of its big-step semantics.
At last, in \Cref{sec:agda_gis_container} we investigate the relation between inference systems and indexed containers
\citep{AltenkirchGHMM15} that are supported by the standard library. 

%% file: gis-lib/meta.tex
\beginalto
In this section we describe the main definitions of the library. 
Concerning the meta-theory, the reader can refer to \Cref{sec:gis}.
Notably, we first present an approach mimicking meta-rules. Then we introduce
the notions of \emph{interpretations} and we prove the \emph{principles}.
At last, we formalize the examples we presented in \Cref{sec:gis}.


\subsection{(Meta-)Rules/Inference System}
\input{gis-lib/meta-rules}


\subsection{Interpretations and Principles}
\input{gis-lib/meta-principles}


\subsection{Basic Examples}
\label{ssec:lib_examples}
\input{gis-lib/meta-examples}

%% file: gis-lib/meta-rules.tex
\beginbass
As anticipated, the aim of the Agda library is to allow a user to write meta-rules as ``on paper''. 
To illustrate this format, let us consider, e.g., the $\allpos$ example from \Cref{sec:gis}:
\[
\inferrule{\allpos{\xs}}{\allpos {\x{:}\xs}}
	~ \x > 0
\]
In a meta-rule, we have \emph{meta-variables}, which range over certain sets, in a way possibly restricted by a \emph{side condition}. 
We call \emph{context} the set of the instantiations of meta-variables which satisfy the side-condition, 
hence produce a rule of the inference system.  
In the example, there are two meta-variables, $\x$ and $\xs$, which range over $\N$ and $\FIList{\N}$, 
respectively, with the restriction that $\x$ should be positive. 
Hence the context is $\{\ple{x,l}\in\N\times\FIList{\N} \mid x > 0 \}$, 
see \Cref{ssec:lib_examples} for the Agda version of this meta-rule.
\begin{figure}[t]
\begin{lstlisting}[frame=single]  
record MetaRule {$\ell$c $\ell$p : Level} (U : Set $\ell$u) : Set _ where 
  field 
    Ctx : Set $\ell$c
    Pos : Set $\ell$p 
    prems : Ctx $\rightarrow$ Pos $\rightarrow$ U
    conclu : Ctx $\rightarrow$ U 

  RF[_] : $\forall${$\ell$} $\rightarrow$ (U $\rightarrow$ Set $\ell$) $\rightarrow$ (U $\rightarrow$ Set _)
  RF[_] P u = 
  	$\Sigma$[ c $\in$ Ctx ] (u $\equiv$ conclu c $\times$ ($\forall$ p $\rightarrow$ P (prems c p)))

  RClosed : $\forall${$\ell$} $\rightarrow$ (U $\rightarrow$ Set $\ell$) $\rightarrow$ Set _
  RClosed P = $\forall$ c $\rightarrow$ ($\forall$ p $\rightarrow$ P (prems c p)) $\rightarrow$ P (conclu c)
  
record IS {$\ell$c $\ell$p $\ell$n : Level} (U : Set $\ell$u) : Set _ where
  field
    Names : Set $\ell$n            
    rules : Names $\rightarrow$ MetaRule {$\ell$c} {$\ell$p} U 

  ISF[_] : $\forall${$\ell$} $\rightarrow$ (U $\rightarrow$ Set $\ell$) $\rightarrow$ (U $\rightarrow$ Set _)
  ISF[_] P u = $\Sigma$[ rn $\in$ Names ] RF[ rules rn ] P u

  ISClosed : $\forall${$\ell$} $\rightarrow$ (U $\rightarrow$ Set $\ell$) $\rightarrow$ Set _
  ISClosed P = $\forall$ rn $\rightarrow$ RClosed (rules rn) P  
\end{lstlisting}
\caption{MetaRule datatype}
\label{fig:metarule-dt}
\end{figure} 
Correspondingly, the Agda declaration in \Cref{fig:metarule-dt} defines a meta-rule as a record, parametric on the universe \lstinline{U}. 
The first two components are the context and a set of positions for premises. 
For each element of the context (instantiation of meta-variables satisfying the side condition), 
the last two components produce the premises, one for each position, and the conclusion of the 
rule obtained by this instantiation.

Recall that in Agda the declaration \lstinline{U : Set} introduces the type (set) 
\lstinline{U}, and \lstinline{P : U $\rightarrow$ Set} the dependent type (predicate on \lstinline{U}) \lstinline{P}.  
For each element \lstinline{u} of \lstinline{U}, \lstinline{P u} is the type of the proofs that 
\lstinline{u} satifies \lstinline{P}, hence \lstinline{P u} inhabited means that \lstinline{u} satisfies \lstinline{P}. 
To avoid paradoxes, not every Agda type is in \lstinline{Set}; there is an infinite sequence \lstinline{Set 0},
\lstinline{Set 1}, \ldots, \lstinline{Set $\ell$}, \ldots\ such that \lstinline{Set $\ell$ : Set (suc $\ell$)}, 
where $\ell$ is a \emph{level}, and \lstinline{Set} is an abbreviation for \lstinline{Set 0}.  
The programmer can write a wildcard for a level which can be inferred; to make the Agda code 
reported in the paper lighter, we sometimes use a wildcard even for a level which is explicit in the real code. 

\begin{remark}
	In the Agda code in this section, predicates  \lstinline{P : U $\rightarrow$ Set} encode subsets 
	of the universe, so we speak of subsets and membership, 
	rather than of predicates and satisfaction, to closely follow  the previous formulation. 
	\eor
\end{remark}

The function  \lstinline{RF[_]} encodes the inference operator associated with the meta-rule. 
Given a subset \lstinline{P} of the universe, \lstinline{u} belongs to the resulting subset if we can 
find an instantiation \lstinline{c} of meta-variables satisfying the side condition, producing \lstinline{u} 
as conclusion, and, for each position, a premise in \lstinline{P}. 
Note the use of existential quantification \lstinline{$\Sigma$[ x $\in$ A ] B} where \lstinline{B} depends on \lstinline{x}.  
The predicate \lstinline{RClosed} encodes the property of being closed with respect to the meta-rule.
A subset \lstinline{P} of the universe is closed if, for each instantiation \lstinline{c} of the 
meta-variables satisfying the side-condition,  if all the premises are in \lstinline{P} then the 
conclusion is in \lstinline{P} as well. 
Note the use of universal quantification 
\lstinline{$\forall$  (x : A) $\rightarrow$ B}, where \lstinline{B} depends on \lstinline{x}.
Finally, an inference system is defined in \Cref{fig:metarule-dt} as a record, parametric on the universe \lstinline{U}, 
consisting of a set of meta-rule names and a family of meta-rules. 
The function \lstinline{ISF[_]} and the predicate \lstinline{ISClosed} 
are defined composing those given for a single meta-rule. 
\begin{figure}[t]
\begin{lstlisting}[frame=single]
record FinMetaRule {$\ell$c n} (U : Set $\ell$u) : Set _ where
  field
    Ctx : Set $\ell$c
    comp : Ctx $\rightarrow$ Vec U n $\times$ U

  from : MetaRule {$\ell$c} {zero} U
  from .MetaRule.Ctx = Ctx
  from .MetaRule.Pos = Fin n
  from .MetaRule.prems c i = get (proj$_1$ (comp c)) i
  from .MetaRule.conclu c = proj$_2$ (comp c)
\end{lstlisting}
\caption{Finitary meta-rule}
\label{fig:finmetarule-dt}
\end{figure}

Since in practical cases metarules are very often \emph{finitary}, that is,
premises are a finite set, the library also offers an interface to write a (
finitary) meta-rule (see \Cref{fig:finmetarule-dt}),
by providing, besides the context, 
two components which are the \emph{vector} of premises, with fixed length \lstinline{n}, and the conclusion. 
The injection \lstinline{from} transforms this more concrete format in the generic one for meta-rules, by     
specifying that the set of positions is \lstinline{Fin n} (the set of indexes from $0$ to $n-1$).

%% file: gis-lib/meta-principles.tex
\beginbass
Recall that the inductive interpretation $\Inductive{\mis}$ of an inference system $\mis$ is the set of 
elements of the universe which have a finite proof tree, and finite proof trees are, in turn, 
inductively defined, that is, by a least fixed point operator.  
In Agda, inductive structures are encoded as \emph{datatypes} (see \Cref{fig:coind-dt}), which specify their constructors. 
\begin{figure}[t]
\begin{lstlisting}[frame=single]
data Ind$\llbracket$_$\rrbracket$ {$\ell$c $\ell$p $\ell$n : Level} 
(is : IS {$\ell$c} {$\ell$p} {$\ell$n} U) : U $\rightarrow$  Set _ where
  fold : $\forall$ {u} $\rightarrow$ ISF[ is ] Ind$\llbracket$ is $\rrbracket$ u $\rightarrow$ Ind$\llbracket$ is $\rrbracket$ u
 
record CoInd$\llbracket$_$\rrbracket$ {$\ell$c $\ell$p $\ell$n : Level} 
 (is : IS {$\ell$c} {$\ell$p} {$\ell$n} U) (u : U) : Set _ where
  coinductive
  constructor cofold_
  field
    unfold : ISF[ is ] CoInd$\llbracket$ is $\rrbracket$ u

data SCoInd$\llbracket$_$\rrbracket$ {$\ell$c $\ell$p $\ell$n : Level} 
(is : IS {$\ell$c} {$\ell$p} {$\ell$n} U) : U $\rightarrow$ Size $\rightarrow$ Set _ where
  sfold : $\forall$ {u i} $\rightarrow$ ISF[ is ] ($\lambda$ u $\rightarrow$ Thunk (SCoInd$\llbracket$ is $\rrbracket$ u) i) u 
      $\rightarrow$ SCoInd$\llbracket$ is $\rrbracket$ u i
\end{lstlisting}
\caption{(Co)inductive interpretations - datatype}
\label{fig:coind-dt}
\end{figure}
For each \lstinline{u}, \lstinline{Ind$\llbracket$ is $\rrbracket$ u} is the type of the proofs that \lstinline{u}
satisfies \lstinline{Ind$\llbracket$ is $\rrbracket$},  which are essentially the finite proof trees
\footnote{With some more structure, since the Agda proofs keep trace of the applied meta-rules.} for \lstinline{u}. 
Indeed, the \lstinline{fold} constructor, given a proof that \lstinline{u} can be derived  by applying a 
rule from premises belonging to \lstinline{Ind$\llbracket$ is $\rrbracket$},   
which essentially  consists of a rule with conclusion \lstinline{u} and finite proof trees for its premises, 
builds a finite proof tree for \lstinline{u}. 

The coinductive interpretation $\CoInductive{\mis}$ (see \Cref{fig:coind-dt}), instead, is the set of elements of the universe 
which have a possibly infinite proof tree, and possibly infinite proof trees are, in turn, 
coinductively defined, that is, by a greatest fixed point operator. 
For each \lstinline{u}, \lstinline{CoInd$\llbracket$ is $\rrbracket$ u} is the type of the 
proofs that \lstinline{u} satisfies \lstinline{CoInd$\llbracket$ is $\rrbracket$},  
which are essentially the possibly infinite proof trees for \lstinline{u}, 
and analogously for \lstinline{SCoInd$\llbracket$ is $\rrbracket$}. 

\begin{remark}
	\label{rm:agda-coind}
	In Agda, coinductive structures can be encoded in two different ways: 
	either as \emph{coinductive records} \citep{AbelPTS13}, 
	or as  datatypes  by using the mechanism of \emph{thunks} (suspended computations) 
	together with \emph{sized types} \citep{Abel12,AbelP16,AbelVW17}  to ensure termination. 
\end{remark}  

To allow compatibility with existing code implemented in either way, 
both versions in \Cref{rm:agda-coind} are supported by the library.
In the first version, a possibly infinite proof tree for \lstinline{u} is a record with 
only one field \lstinline{unfold} containing an element of \lstinline{ISF[ is ] CoInd$\llbracket$ is $\rrbracket$ u}, 
that is, a proof that \lstinline{u} can be derived by applying a rule from premises 
belonging to \lstinline{CoInd$\llbracket$ is $\rrbracket$}, 
which  essentially consists of a rule with conclusion \lstinline{u} and possibly infinite proof trees for its premises.  
In the second version, a possibly infinite proof tree is obtained by a \lstinline{data} constructor, 
analogously to a finite one in the inductive interpretation; 
however, since proof trees are encoded as thunks, hence evaluated lazily,
this encoding represents infinite trees as well. 
In other words, coinduction is ``hidden'' in the library type \lstinline{Thunk}, 
which is a coinductive record with only one field \lstinline{force}, 
intuitively representing the suspended computation.

The interpretation of a generalized inference system (see \Cref{fig:fcoind-dt}) can then be encoded following exactly 
the definition in \Cref{sec:gis}: it is the coinductive interpretation of \lstinline{I}, restricted to
rules whose conclusion is in the inductive interpretation of the (standard) inference 
system consisting of both rules \lstinline{I} and corules \lstinline{C}.   
\begin{figure}[t]
\begin{lstlisting}[frame=single]
_$\sqcap$_ : $\forall$ {$\ell$c $\ell$p $\ell$n $\ell$}{U : Set $\ell$u} $\rightarrow$ IS {$\ell$c} {$\ell$p} {$\ell$n} U 
  $\rightarrow$ (U $\rightarrow$ Set $\ell$) $\rightarrow$ IS {$\ell$c $\sqcup$ $\ell$} {_} {_} U
(is $\sqcap$ P) .Names = is .Names
(is $\sqcap$ P) .rules rn = addSideCond (is .rules rn) P

_$\cup$_ : $\forall${$\ell$c $\ell$p $\ell$n $\ell$n'}{U : Set $\ell$} $\rightarrow$ IS {$\ell$c} {$\ell$p} {$\ell$n} U 
  $\rightarrow$ IS {_} {_} {$\ell$n'} U $\rightarrow$ IS {_} {_} {$\ell$n $\sqcup$ $\ell$n'} U
(is1 $\cup$ is2) .Names = (is1 .Names) $\uplus$ (is2 .Names)
(is1 $\cup$ is2) .rules = [ is1 .rules , is2 .rules ]

FCoInd$\llbracket$_,_$\rrbracket$ : $\forall${$\ell$c $\ell$p $\ell$n $\ell$n'} $\rightarrow$ (I : IS {$\ell$c} {$\ell$p} {$\ell$n} U) 
  $\rightarrow$ (C : IS {$\ell$c} {$\ell$p} {$\ell$n'} U) $\rightarrow$ U $\rightarrow$ Set _
FCoInd$\llbracket$ I , C $\rrbracket$ = CoInd$\llbracket$ I $\sqcap$ Ind$\llbracket$ I $\cup$ C $\rrbracket$ $\rrbracket$

SFCoInd$\llbracket$_,_$\rrbracket$ : $\forall${$\ell$c $\ell$p $\ell$n $\ell$n'} $\rightarrow$ (I : IS {$\ell$c} {$\ell$p} {$\ell$n} U) 
  $\rightarrow$ (C : IS {$\ell$c} {$\ell$p} {$\ell$n'} U) $\rightarrow$ U $\rightarrow$ Size $\rightarrow$ Set _
SFCoInd$\llbracket$ I , C $\rrbracket$ = SCoInd$\llbracket$ I $\sqcap$ Ind$\llbracket$ I $\cup$ C $\rrbracket$ $\rrbracket$
\end{lstlisting}
\caption{Interpretation generated by corules - datatype}
\label{fig:fcoind-dt}
\end{figure}
The definition is provided in two flavours where the coinductive interpretation 
is encoded by coinductive records and thunks, respectively, and uses two operators on inference systems, 
restriction $\sqcap$ and union $\cup$. We report the codes in \Cref{fig:fcoind-dt}. 
The former adds  to each rule the side condition that the conclusion should satisfy \lstinline{P}, 
as specified by the function \lstinline{addSideCond} (here omitted). On the other hand, $\cup$ joins two inference
systems.

The library also provides the proofs of relevant properties, e.g., that closed sets coincide with pre-fixed points, 
and consistent sets coincide with post-fixed points. Moreover, it is shown that the two versions of encoding of the
coinductive interpretation (by coinductive records and thunks) are equivalent.    
Finally, the library provides  the induction, coinduction, and bounded coinduction principles (see \Cref{prop:indp,prop:coindp,prop:bcp}). 
We only report the statements in \Cref{fig:principles} and we briefly recall their meaning.
\begin{figure}[t]
\begin{lstlisting}[frame=single]
ind[_] : $\forall${$\ell$c $\ell$p $\ell$n $\ell$} 
    $\rightarrow$ (is : IS {$\ell$c} {$\ell$p} {$\ell$n} U)		-- IS
    $\rightarrow$ (S : U $\rightarrow$ Set $\ell$)			-- specification
    $\rightarrow$ ISClosed is S			-- S is closed
    $\rightarrow$ Ind$\llbracket$ is $\rrbracket$ $\subseteq$ S

coind[_] : $\forall${$\ell$c $\ell$p $\ell$n $\ell$}
    $\rightarrow$ (is : IS {$\ell$c} {$\ell$p} {$\ell$n} U) 
    $\rightarrow$ (S : U $\rightarrow$ Set $\ell$)
    $\rightarrow$ (S $\subseteq$ ISF[ is ] S)		-- S is consistent
    $\rightarrow$ S $\subseteq$ CoInd$\llbracket$ is $\rrbracket$
 
bounded-coind[_,_] : $\forall${$\ell$c $\ell$p $\ell$n $\ell$n' $\ell$} 
    $\rightarrow$ (I : IS {$\ell$c} {$\ell$p} {$\ell$n} U)
    $\rightarrow$ (C : IS {$\ell$c} {$\ell$p} {$\ell$n'} U)
    $\rightarrow$ (S : U $\rightarrow$ Set $\ell$)                   
    $\rightarrow$ S $\subseteq$ Ind$\llbracket$ I $\cup$ C $\rrbracket$		-- S is bounded w.r.t. I $\cup$ C
    $\rightarrow$ S $\subseteq$ ISF[ I ] S		-- S is consistent w.r.t. I
    $\rightarrow$ S $\subseteq$ FCoInd$\llbracket$ I , C $\rrbracket$
\end{lstlisting}
\caption{Proof principles}
\label{fig:principles}
\end{figure}

\begin{itemize}
\item If \lstinline{S} is closed, then each element of the inductively defined 
set \lstinline{Ind$\llbracket$ is $\rrbracket$} satisfies  \lstinline{S}.
\item If \lstinline{S} is consistent, then each element satisfying \lstinline{S} is in the
coinductively defined set \lstinline{CoInd$\llbracket$ is $\rrbracket$}. 
\item If \lstinline{S} is bounded, and  consistent with respect to \lstinline{I},
then each element which satisfies \lstinline{S} belongs to the set 
\lstinline{FCoInd$\llbracket$ I , C $\rrbracket$} defined by flexible coinduction. 
\end{itemize}
 
Another useful theorem is that 
$\FlexCo{\mis}{\mcois}\subseteq\Ind{\mis\cup\mcois}$ (see \Cref{fig:fcoind-to-ind}). 

\begin{figure}[t]
\begin{lstlisting}[frame=single]
fcoind-to-ind : $\forall${$\ell$c $\ell$p $\ell$n $\ell$n'}
    {is : IS {$\ell$c} {$\ell$p} {$\ell$n} U}{cois : IS {$\ell$c} {$\ell$p} {$\ell$n'} U} 
    $\rightarrow$ FCoInd$\llbracket$ is , cois $\rrbracket$ $\subseteq$ Ind$\llbracket$ is $\cup$ cois $\rrbracket$
\end{lstlisting}
	\caption{Extract inductive proof}
	\label{fig:fcoind-to-ind}
\end{figure}

%% file: gis-lib/meta-examples.tex
\beginbass
We continue this section by showing how to use the library to define specific inference systems and prove their properties. 
In particular, we consider the basic examples in \Cref{sec:gis} that allow us to cover
all the cases that we investigated before.

\begin{example}
	Consider the predicate $\member$. We first recall its inference system and we give names to the (meta-)rules.
		\begin{mathpar}
		\inferrule[mem-h]{\mathstrut}{\member(\x , \cons\x\xs)}
		\and
		\inferrule[mem-t]{\member(\x ,\xs)}{\member(\x , \cons\y\xs)}
	\end{mathpar} 
	The universe consists of pairs of elements and possibly infinite lists, 
	implemented by the Agda library \lstinline{Colist} which uses thunks.
	
\begin{lstlisting}
U = A $\times$ Colist A $\infty$
data memberRN : Set where mem$\textrm{-}$h mem$\textrm{-}$t : memberRN

mem$\textrm{-}$h$\textrm{-}$r : FinMetaRule U
mem$\textrm{-}$h$\textrm{-}$r .Ctx = A $\times$ Thunk (Colist A) $\infty$
mem$\textrm{-}$h$\textrm{-}$r .comp (x , xs) =
 	[] ,
 	----------------
 	(x , x :: xs) 

mem$\textrm{-}$t$\textrm{-}$r : FinMetaRule U
mem$\textrm{-}$t$\textrm{-}$r .Ctx = A $\times$ A $\times$ Thunk (Colist A) $\infty$
mem$\textrm{-}$t$\textrm{-}$r .comp (x , y , xs) =
 	((x , xs .force) :: []) ,
 	----------------
 	(x , y :: xs) 

memberIS : IS U
memberIS .Names = memberRN
memberIS .rules mem$\textrm{-}$h = from mem$\textrm{-}$h$\textrm{-}$r
memberIS .rules mem$\textrm{-}$t = from mem$\textrm{-}$t$\textrm{-}$r
\end{lstlisting}

	Here \lstinline{memberRN} are the rule names, and each rule name has an associated element of 
	\lstinline{FinMetaRule U}, which exactly encodes the meta-rule in the inference system at the beginning. 
	Note, in \lstinline{mem$\textrm{-}$t$\textrm{-}$r}, the use of the \lstinline{force} field 
	of \lstinline{Thunk} to actually obtain the tail colist. 
	This inference system is expected to define exactly the pairs \lstinline{(x , xs)} such that 
	\lstinline{x} belongs to \lstinline{xs}, that is, those satisfying the following specification
\begin{lstlisting}	
memSpec : U $\rightarrow$ Set
memSpec (x , xs) = $\Sigma$[ i $\in$ $\N$ ] (Colist.lookup i xs = just x)
\end{lstlisting} 
	where \lstinline{lookup : $\N$ $\rightarrow$ Colist A $\infty$ $\rightarrow$ Maybe A} 
	is the (standard) library function that returns the \lstinline{i}-th element of \lstinline{xs}, if any. 
	As said in \Cref{sec:gis}, to obtain the desired meaning this inference system has to be 
	interpreted inductively, and soundness can be proved by the induction principle, 
	that is, by providing a proof that the specification is closed with respect to the two meta-rules, as shown below. 

\begin{lstlisting}
_member_ : A $\rightarrow$ Colist A $\infty$ $\rightarrow$ Set
x member xs = Ind$\llbracket$ memberIS $\rrbracket$ (x , xs)

memSpecClosed : ISClosed memberIS memSpec
memSpecClosed mem$\textrm{-}$h _ _ = zero , refl
memSpecClosed mem$\textrm{-}$t _ pr =
 	let (i , proof) = pr Fin.zero in (suc i) , proof

memberSound : $\forall$ {x xs} $\rightarrow$ x member xs $\rightarrow$ memSpec (x , xs)
memberSound = ind[memberIS] memSpec memSpecClosed
\end{lstlisting}

	For completeness there is no canonical technique; in this example, it can be proved by induction 
	on the position (the index \lstinline{i} in the specification). 
	For the complete proof see \cite{Ciccone20}.
	\eoe
\end{example}


\begin{example}
	Consider the predicate $\allpos$ from \Cref{sec:gis}. We first recall its inference system and we give names to the (meta-)rules.
	\begin{mathpar}
	\inferrule[allP-$\Lambda$]{\mathstrut}{\allpos\nil}
	\and
	\inferrule[allP-t]{\allpos\xs}{\allpos \cons\x\xs} ~ \x > 0
	\end{mathpar}
	Then the universe consists of possibly infinite lists. 

\begin{lstlisting}
U : Set
U = Colist $\N$ $\infty$
data allPosRN : Set where allP$\textrm{-}\Lambda$ allP$\textrm{-}$t : allPosRN

allP$\textrm{-}\Lambda\textrm{-}$r : FinMetaRule U
allP$\textrm{-}\Lambda\textrm{-}$r .Ctx = $\top$
allP$\textrm{-}\Lambda\textrm{-}$r .comp c =
  [] ,
  -----------------
  [] 

allP$\textrm{-}$t$\textrm{-}$r : FinMetaRule U
allP$\textrm{-}$t$\textrm{-}$r .Ctx = $\Sigma$[ (x , _) $\in$ $\N$ $\times$ Thunk (Colist $\N$) $\infty$ ] x > 0
allP$\textrm{-}$t$\textrm{-}$r .comp ((x , xs) , _) =
  ((xs .force) :: []) ,
  -----------------
  (x :: xs)

allPosIS : IS U
allPosIS .Names = allPosRN
allPosIS .rules allP$\textrm{-}\Lambda$ = from allP$\textrm{-}\Lambda\textrm{-}$r
allPosIS .rules allP$\textrm{-}$t = from allP$\textrm{-}$t$\textrm{-}$r
\end{lstlisting}

	This inference system is expected to define exactly the lists such that all elements are positive, 
	that is, those satisfying the following specification 
	(where for simplicity, we use the predicate $\in$, omitted, directly defined inductively). 
	Notably, we could use $\member$ instead of $\in$.

\begin{lstlisting}
allPosSpec : U $\rightarrow$ Set
allPosSpec xs = $\forall$ {x} $\rightarrow$ x $\in$ xs $\rightarrow$ x > 0
\end{lstlisting}

	As said in \Cref{sec:gis}, to obtain the desired meaning this inference system has 
	to be interpreted coinductively, and completeness can be proved by the coinduction principle, 
	that is, by providing a proof that the specification is consistent with respect to the inference system, as shown below. 

\begin{lstlisting}
allPos : U $\rightarrow$ Set
allPos = CoInd$\llbracket$ allPosIS $\rrbracket$

allPosSpecCons : $\forall$ {xs} 
	$\rightarrow$ allPosSpec xs $\rightarrow$ ISF[ allPosIS ] allPosSpec xs
allPosSpecCons {[]} _ = allP$\textrm{-}\Lambda$ , (tt , (refl , tt , $\lambda$ ()))
allPosSpecCons {(x :: xs)} Sxs = 
  allP$\textrm{-}$t , 
  ((x , xs) , (refl , 
  		(Sxs here , 
  		$\lambda$ {Fin.zero $\rightarrow$ $\lambda$ mem $\rightarrow$ Sxs (there mem)})))

allPosComplete : allPosSpec $\subseteq$ allPos
allPosComplete = coind[ allPosIS ] allPosSpec allPosSpecCons
\end{lstlisting}

	For what concerns the soundness, there is no canonical technique; in this example, 
	when the colist is empty the proof that the specification holds is trivial.
	If the colist is not empty, then the proof proceeds by induction on 
	the position of the element to be proved to be positive.
	For the complete proof see \cite{Ciccone20}.
 	\eoe
\end{example}

\begin{example}
	Consider the predicate $\maxelem$ from \Cref{sec:gis}. We recall once again its meta-(co)rules.
	\begin{mathpar}
		\inferrule[max-h]{\mathstrut}{\maxelem(\x , \cons\x\nil)}
		\and
		\inferrule[max-t]{\maxelem(\x , \xs)}{\maxelem(max(\x , \y) , \cons\y\xs)}
		\and
		\infercorule[co-max-h]{\mathstrut}{\maxelem(\x , \cons\x\xs)}
	\end{mathpar}
	Then the universe consists of pairs of natural numbers and possibly infinite lists.

\begin{lstlisting}
U : Set
U = $\N$ $\times$ Colist $\N$ $\infty$
data maxElemRN : Set where max$\textrm{-}$h max$\textrm{-}$t : maxElemRN
data maxElemCoRN : Set where co$\textrm{-}$max$\textrm{-}$h : maxElemCoRN

max$\textrm{-}$h$\textrm{-}$r : FinMetaRule U
max$\textrm{-}$h$\textrm{-}$r .Ctx = 
  $\Sigma$[ (_ , xs) $\in$ $\N$ $\times$ Thunk (Colist $\N$) $\infty$ ] xs .force $\equiv$ []
max$\textrm{-}$h$\textrm{-}$r .comp ((x , xs) , _) =
  [] ,
  --------------
  x , x :: xs

max$\textrm{-}$t$\textrm{-}$r : FinMetaRule U
max$\textrm{-}$t$\textrm{-}$r .Ctx = 
  $\Sigma$[ (x , y , z , _) $\in$ 
  $\N$ $\times$ $\N$ $\times$ $\N$ $\times$ Thunk (Colist $\N$) $\infty$ ] z $\equiv$ max x y 
max$\textrm{-}$t$\textrm{-}$r .comp ((x , y , z , xs) , _) =
  (x , xs .force) :: [] ,
  --------------
  z , y :: xs

co$\textrm{-}$max$\textrm{-}$h$\textrm{-}$r : FinMetaRule U
co$\textrm{-}$max$\textrm{-}$h$\textrm{-}$r .Ctx = $\N$ $\times$ Thunk (Colist $\N$) $\infty$
co$\textrm{-}$max$\textrm{-}$h$\textrm{-}$r .comp (x , xs) =
  [] ,
  --------------
  (x , x :: xs) 

maxElemIS : IS U
maxElemIS .Names = maxElemRN
maxElemIS .rules max$\textrm{-}$h = from max$\textrm{-}$h$\textrm{-}$r
maxElemIS .rules max$\textrm{-}$t = from max$\textrm{-}$t$\textrm{-}$r

maxElemCoIS : IS U
maxElemCoIS .Names = maxElemCoRN
maxElemCoIS .rules co$\textrm{-}$max$\textrm{-}$h = from co$\textrm{-}$max$\textrm{-}$h$\textrm{-}$r
\end{lstlisting}

	Note that in this example we have defined two inference systems, the rules and the corules. 
	This generalized inference system is expected to define exactly the pairs 
	\lstinline{(x , xs)} such that \lstinline{x} is the maximal element of \lstinline{xs}, 
	that is, those satisfying the following specification, where to be the maximal element 
	\lstinline{x} should belong to \lstinline{xs}, and be greater or equal than any \lstinline{n} in \lstinline{xs}.

\begin{lstlisting}
maxSpec inSpec geqSpec : U $\rightarrow$ Set
inSpec (x , xs) = x $\in$ xs
geqSpec (x , xs) = $\forall${n} $\rightarrow$ n $\in$ xs $\rightarrow$ x $\equiv$ max x n
maxSpec u = inSpec u $\times$ geqSpec u
\end{lstlisting}

	As said in \Cref{sec:gis}, the desired meaning is provided by the interpretation of the generalized inference system.
	
\begin{lstlisting}
_maxElem_ : $\N$ $\rightarrow$ Colist $\N$ $\infty$ $\rightarrow$ Set
x maxElem xs = FCoInd$\llbracket$ maxElemIS , maxElemCoIS $\rrbracket$ (x , xs)
\end{lstlisting}

	and the completeness can be proved by the bounded coinduction principle. 

\begin{lstlisting}
maxElemComplete : $\forall${x xs} $\rightarrow$ maxSpec (x , xs) $\rightarrow$ x maxElem xs
maxElemComplete =
  bounded$\textrm{-}$coind[ maxElemIS , maxElemCoIS ] maxSpec 
    ($\lambda${(x , xs) $\rightarrow$ maxSpecBounded x xs}) 
    $\lambda${(x , xs) $\rightarrow$ maxSpecCons x xs}
\end{lstlisting}

	Notably, we have to prove that the specification is:
	\begin{itemize}
	\item \emph{bounded}, that is, contained in \lstinline{_maxElem$_i$_ }, the inductive interpretation of the 
	standard inference system consisting of both rules and corules, as shown below:
	
\begin{lstlisting}
_maxElem$_i$_ : $\N$ $\rightarrow$ Colist $\N$ $\infty$ $\rightarrow$ Set
x maxElem$_i$ xs = Ind$\llbracket$ maxElemIS $\cup$ maxElemCoIS $\rrbracket$ (x , xs)

maxSpecBounded : $\forall${x xs} $\rightarrow$ inSpec (x , xs) 
  $\rightarrow$ geqSpec (x , xs) $\rightarrow$ x maxElem$_i$ xs
\end{lstlisting}

	\item \emph{consistent} with respect to the inference system consisting of only rules, as shown below:
	
\begin{lstlisting}
maxSpecCons : $\forall${x xs} $\rightarrow$ inSpec (x , xs) $\rightarrow$ 
  geqSpec (x , xs) $\rightarrow$ ISF[ maxElemIS ] maxSpec (x , xs)
\end{lstlisting}

	\end{itemize}
	These proofs are omitted for the sake of brevity. See \cite{Ciccone20} for more details.
	Concerning the soundness there is no canonical technique. 
	The proof can be split for the two components of the specification. 
	It is worth noting that, for the soundness with respect to \lstinline{inSpec}, 
	we  first use \lstinline{fcoind$\textrm{-}$to$\textrm{-}$ind} (see \Cref{fig:fcoind-to-ind}),
	and then define \lstinline{maxElemSound$\textrm{-}$in$\textrm{-}$ind}, omitted, 
	by induction on the inference system consisting of rules and corules.  
	The use of \lstinline{fcoind$\textrm{-}$to$\textrm{-}$ind} in the proof corresponds to the fact 
	that without corules unsound judgments could be derived. 
	
\begin{lstlisting}
maxElemSound$\textrm{-}$in : $\forall$ {x xs} $\rightarrow$ x maxElem xs $\rightarrow$ inSpec (x , xs)
maxElem$\textrm{-}$sound$\textrm{-}$in max = maxElemSound$\textrm{-}$in$\textrm{-}$ind (fcoind$\textrm{-}$to$\textrm{-}$ind max)
\end{lstlisting}

	Soundness with respect to \lstinline{geqSpec} is proved by induction on the position, 
	that is, the proof of membership, of the element that must be proved to be less or equal.  
	In this case, soundness would hold even in the purely coinductive case. 
	\eoe
\end{example}

\begin{remark}[Code Duplication]
	\label{rm:agda_dup}
	Of course, as Agda supports both inductive and coinductive dependent types, 
	one could directly write Agda code for inductive, coinductive and 
	even flexible coinductive definitions of concrete examples. 
	We have explored this possibility in \cite{Ciccone20}. 
	However, in this way, the definition is hard-wired with its semantics, and, for flexible coinduction, 
	one has to manually construct the interpretation by combining in the correct way an inductive and a 
	coinductive type and to prove the bounded coinduction principle for each example. 
	For instance, the definition of \lstinline{maxElem} will look as follows: 
	
\begin{lstlisting}
data _maxElem_ : $\N$ $\to$ CoList $\N$ $\infty$ $\to$ Size $\to$ Set where 
  max$\textrm{-}$h : $\forall$ {x xs i} $\to$force xs $\equiv$ [] $\to$ x maxElem (x :: xs) i 
  max$\textrm{-}$t : $\forall$ {x y xs i} $\to$ Thunk (x maxElem (force xs)) i 
                       $\to$ z $\equiv$ max x y 
                       $\to$ z maxElem$_i$ (y :: xs) 
                       $\to$ z maxElem (y :: xs) i  

data _maxElem$_i$_ : $\N$ $\to$ CoList $\N$ $\infty$ $\to$ Set where 
  imax$\textrm{-}$h : $\forall$ {x xs} $\to$force xs $\equiv$ [] $\to$ x maxElem$_i$ (x :: xs) 
  imax$\textrm{-}$t : $\forall$ {x y xs} $\to$ x maxElem$_i$ (force xs)) $\to$ z $\equiv$ max x y 
                      $\to$ z maxElem$_i$ (y :: xs)  
  co$\textrm{-}$max$\textrm{-}$h : $\forall$ {x xs} $\to$ x maxElem$_i$ (x :: xs) 
\end{lstlisting} 

	Clearly, this approach causes duplication of rules and code, as rules of the coinductive type have to be duplicated in the inductive one, 
	making things rather complex. 
	Our library instead hides all these details, exposing interfaces for interpretations and proof principles, 
	so that the user only has to write code describing rules. 
	\eor
\end{remark}

%% file: gis-lib/lambda.tex
\beginalto
\begin{figure}[t]
\framebox[\textwidth]{
\begin{mathpar}
	\begin{array}{rcll}
		\produzione{\te}{\val \mid\x\mid\AppExp{\te_1}{\te_2}\mid\ldots}{term}\\
		\produzione{\val}{\LambdaExp{\x}{\te}\mid\ldots}{value}\\
		\produzione{\infval}{\val\mid\infty}{result}
	\end{array}
	\and
	\infercorule[coa]{\mathstrut}{\eval{e}{\infty}}\defrule[coa]{}
	\and
	\inferrule[val]{\mathstrut}{\eval{\val}{\val}}\defrule[val]{}
	\and
	\inferrule[app]
		{\eval{\te_1}{\LambdaExp{\x}{\te}}\quad\eval{\te_2}{{\val}}\quad\eval{\lamsubst{\te}{\x}{{\val}}}{\infval}}
		{\eval{\te_1\ \te_2}{\infval}}\defrule[app]{}
	\and
	\inferrule[l-div]
		{\eval{\te_1}{\infty}}
		{\eval{\te_1\ \te_2}{\infty}}\defrule[l-div]{}
	\and
	\inferrule[r-div]
		{\eval{\te_1}{{\val}}\quad\eval{\te_2}{\infty}}
		{\eval{\te_1\ \te_2}{\infty}}\defrule[r-div]{}
\end{mathpar}
}
\caption{$\lambda$-calculus: syntax and big-step semantics}
\label{fig:lambda}
\end{figure}
In \Cref{ssec:lib_examples} we formalized some basic examples to explain how
the library can be used. In this section we describe a more significant example of instantiation: 
an inference system with corules providing a big-step semantics of lambda-calculus including divergence 
among the possible results \cite{AnconaDZ@oopsla17}, reported in \Cref{fig:lambda}.   
In this example, corules play a key role: indeed , considering, e.g., the divergent term 
$\Omega=\AppExp{(\LambdaExp{\x}{\x$ $\x})}{(\LambdaExp{\x}{\x$ $\x})}$, in the standard inductive big-step 
semantics no result can be derived (an infinite proof tree is needed), as for a stuck term; 
in the purely coinductive interpretation, any judgment $\eval{\Omega}{\infval}$ would be obtained \cite{LeroyG09}. 
Since each node of the infinite proof tree for a judgment should also have a finite proof 
tree using the corules, the coaxiom \refrule{coa} forces to obtain only $\infty$ as result, 
see \cite{AnconaDZ@oopsla17} for a more detailed explanation.
\footnote{Other examples of big-step semantic definitions with 
more sophisticated corules are given in \cite{AnconaDZ@ecoop18,AnconaDRZ20}.}

In rule \refrule{app}, $\infval$ is used for the result, so the rule also covers the case
when the evaluation of the body of the lambda abstraction diverges.  
As usual, $\lamsubst{\te}{\x}{\val}$ denotes capture-avoiding substitution. 
Rules \refrule{l-div} and \refrule{r-div} cover the cases when either $\te_1$ or $\te_2$ diverges,
assuming  a left-to-right evaluation strategy.
Terms, values, and results are inductively defined, hence encoded by Agda datatypes. 
As customary in implementations of lambda-calculus, we use the De Bruijn notation: 
notably, \lstinline{Term n} is the set of terms with \lstinline{n} free variables. 

\begin{lstlisting}
data Term (n : $\N$) : Set where
  var : Fin n $\rightarrow$ Term n
  lambda : Term (suc n) $\rightarrow$ Term n
  app : Term n $\rightarrow$ Term n $\rightarrow$ Term n

data Value : Set where 
  lambda : Term 1 $\rightarrow$ Value

term : Value $\rightarrow$ Term 0
term (lambda x) = lambda x

data Value$^\infty$ : Set where
  res : Value $\rightarrow$ Value$^\infty$
  $\infty$ : Value$^\infty$
\end{lstlisting}

The universe consists of big-step judgments (pairs consisting of a term and a result). 
The two inference systems of rules and corules are encoded below.
\begin{lstlisting}
U : Set
U = Term 0 $\times$ Value$^\infty$

data BigStepRN : Set where val app l-div r-div : BigStepRN
data BigStepCoRN : Set where COA : BigStepCoRN

BigStepIS : IS U
BigStepIS .Names = BigStepRN
BigStepIS .rules val = from val$\textrm{-}$r
BigStepIS .rules app = from app$\textrm{-}$r
BigStepIS .rules L$\textrm{-}$DIV = from l$\textrm{-}$div$\textrm{-}$r
BigStepIS .rules R$\textrm{-}$DIV = from r$\textrm{-}$div$\textrm{-}$r

BigStepCoIS : IS U
BigStepCoIS .Names  = BigStepCoRN
BigStepCoIS .rules COA = from coa$\textrm{-}$r
\end{lstlisting}

where \lstinline{BigStepRN} are the rule names, and each rule name has an associated element of \lstinline{FinMetaRule U}. 
For instance, \lstinline{app$\textrm{-}$r} is given below. 
The auxiliary function \lstinline{subst}, omitted, implements capture-avoiding substitution. 

\begin{lstlisting}
app$\textrm{-}$r : FinMetaRule U
app$\textrm{-}$r .Ctx = Term 0 $\times$ Term 1 $\times$ Term 0 $\times$ Value $\times$ Value$^\infty$
app$\textrm{-}$r .comp (t1 , t , t2 , v , v$^\infty$) =
 (t1 , res (lambda t)) :: (t2 , res v) :: 
 (subst t (term v) , v$^\infty$) :: [] ,
  -------------------------
  (app t1 t2 , v$^\infty$) 
\end{lstlisting}

The big-step semantics can be obtained as the interpretation of the generalized inference system,
as shown below. We use the flavour with thunks.
 
\begin{lstlisting}
_$\Downarrow$_ : Term 0 $\rightarrow$ Value$^\infty$ $\rightarrow$ Size $\rightarrow$ Set
(t $\Downarrow$ v$^\infty$) i = SFCoInd$\llbracket$ BigStepIS , BigStepCoIS $\rrbracket$ (t , v$^\infty$) i

_$\Downarrow_\textrm{i}$_ : Term 0 $\rightarrow$ Value$^\infty$ $\rightarrow$ Set
t $\Downarrow_\textrm{i}$ v$^\infty$ = Ind$\llbracket$ BigStepIS $\cup$ BigStepCoIS $\rrbracket$ (t , v$^\infty$)
\end{lstlisting}

The second predicate  (\textrm{i} stands for ``inductive'')  models that a judgment has a finite proof tree in the 
inference system consisting of rules and coaxiom, and will be used in proofs. 

\begin{figure}[t]
\framebox[\textwidth]{
\begin{mathpar}
		\inferrule[$\beta$]{\mathstrut}{\SmallStep{\AppExp{(\LambdaExp{\x}{\te})}{\val}}{\lamsubst{\te}{\x}{\val}}}
		\and
		\inferrule[l-app]{\te_1\ev\te'_1}{\AppExp{\te_1}{\te_2}\ev\AppExp{\te'_1}{\te_2}}
		\and
		\inferrule[r-app]{\te_2\ev\te'_2}{\AppExp{\val}{\te_2}\ev\AppExp{\val}{\te'_2}}
\end{mathpar}
}
\caption{$\lambda$-calculus: small-step semantics}
\label{fig:lambda_ss}
\end{figure}

Small-step semantics, reported in \Cref{fig:lambda_ss}, can also be obtained appropriately instantiating the library. 
In this case, the universe consists of small-step judgments, which are pairs of terms. 
There is only one inference system, where \lstinline{SmallStepRN} are the rule names, 
and each rule name has an associated element of \lstinline{FinMetaRule U}. 

\begin{lstlisting}
U : Set
U = Term 0 $\times$ Term 0

data SmallStepRN : Set where $\beta$ L$\textrm{-}$app R$\textrm{-}$app : SmallStepRN
  
SmallStepIS : IS U
SmallStepIS .Names = SmallStepRN
SmallStepIS .rules $\beta$ = from $\beta\textrm{-}$r
SmallStepIS .rules L$\textrm{-}$app = from l$\textrm{-}$app$\textrm{-}$r
SmallStepIS .rules R$\textrm{-}$app = from r$\textrm{-}$app$\textrm{-}$r
\end{lstlisting}

For instance, $\beta\textrm{-}$r is given below.

\begin{lstlisting}
$\beta\textrm{-}$r : FinMetaRule U
$\beta\textrm{-}$r .Ctx = Term 1 $\times$ Value
$\beta\textrm{-}$r .comp (t , v) =
  [] ,
  -------------------------
  (app (lambda t) (term v) , subst t (term v))
\end{lstlisting}

The one-step relation $\Rightarrow$ is obtained as the inductive interpretation of the (standard) inference system.  
Then, finite computations are modeled by its reflexive and transitive closure $\Rightarrow^\star$, 
defined using \lstinline{Star} in the Agda library,  as shown below.

\begin{lstlisting}
_$\Rightarrow$_ : Term 0 $\rightarrow$ Term 0 $\rightarrow$ Set
t $\Rightarrow$ t' = Ind$\llbracket$ SmallStepIS $\rrbracket$ (t , t')

_$\Rightarrow^\star$_ : Term 0 $\rightarrow$ Term 0 $\rightarrow$ Set
_$\Rightarrow^\star$_ = Star _$\Rightarrow$_
\end{lstlisting}

Infinite computations, instead, are modeled by the relation $\Rightarrow^\infty$, 
coinductively defined by the meta-rule 
\[
\inferrule{\te'\Rightarrow^\infty}{\te\Rightarrow^\infty}{\te\Rightarrow\te'}
\]
that we encoded in Agda by using thunks.  

\begin{lstlisting}
data _$\Rightarrow^\infty$ : Term 0 $\rightarrow$ Size $\rightarrow$ Set where
  step : $\forall$ {t t' i} $\rightarrow$ t $\Rightarrow$ t' $\rightarrow$ Thunk (t' $\Rightarrow^\infty$) i $\rightarrow$ t $\Rightarrow^\infty$ i
\end{lstlisting}

The proof of equivalence between big-step and small-step semantics is structured as follows, 
where $\Spec=\{\Pair{\te}{\val}\mid\te\Rightarrow^\star\val\}\cup\{\Pair{\te}{\infty}\mid\te\Rightarrow^\infty\}$. 
\begin{description}
\item[Soundness]\
\begin{description}

\item[$\eval{\te}{\val}$ implies $\te\Rightarrow^\star\val$] We use \lstinline{fcoind$\texttt{-}$to$\texttt{-}$ind} (see \Cref{fig:fcoind-to-ind}), 
and then reason by induction on the judgment $\te{\Downarrow_\textrm{i}}\val$. 
That is, we show that $\te\Rightarrow^\star\val$ is closed w.r.t.\ the inference system consisting of rules and corules. 
As already pointed out for the \lstinline{maxElem} example, the use of \lstinline{fcoind$\textrm{-}$to$\textrm{-}$ind} 
in the proof corresponds to the fact that, without the coaxiom \refrule{coa}, unsound judgments would be derived, 
e.g., $\eval{\Omega}{\val}$ for $\val\in\Val$. 

\item[$\eval{\te}{\infty}$ implies $\te\Rightarrow^\infty$]  This implication, instead, would hold even in the purely coinductive case.  
It can be proved from \emph{progress} and \emph{subject reduction} properties:
\begin{description} 
\item[Progress] $\eval{\te}{\infty}$ implies that there exists $\te'$ such that $\te\Rightarrow\te'$.
\item[Subject reduction] $\eval{\te}{\infty}$ and $\te\Rightarrow\te'$ implies $\eval{\te'}{\infty}$.
\end{description}
\end{description}
\item[Completeness] By bounded coinduction (see \Cref{prop:bcp}).
\begin{description}
\item[Boundedness]\
\begin{description} 
\item[$\te\Rightarrow^\star\val$ implies $\te{\Downarrow_\textrm{i}}\val$] By induction on the number of steps.
\item[$\te\Rightarrow^\infty$ implies $\te{\Downarrow_\textrm{i}}\infty$] Trivial, since the coaxiom \lstinline{coa} can be applied.
\end{description}
\item[Consistency]  We have to show that, for each $\Pair{\te}{\infval}\in\Spec$,  $\Pair{\te}{\infval}$ 
is the consequence of a big-step rule where the premises are in $\Spec$ as well.
We distinguish two cases.
\begin{description}
\item[$\te\Rightarrow^\star\val$] By induction on the number of steps. If it is $0$, then $\te$ is a value, 
hence we can use rule \refrule{val}. Otherwise, $\te$ is an application, and we can use rule \refrule{app}.
\item[$\te\Rightarrow^\infty$] The term $\te$ is an application $\AppExp{\te_1}{\te_2}.$ We distinguish the following cases:
\begin{itemize}
\item $\te_1$ diverges, hence we can use rule \refrule{l-div}
\item $\te_1$ converges and $\te_2$ diverges, hence we can use rule \refrule{r-div}
\item both $\te_1$ and $\te_2$ converge, hence we can use rule \refrule{app}.
\end{itemize}
Note that in this proof by cases we need to use the \emph{excluded middle} principle, 
which is defined in the standard library, and  postulated  in our proof. 
\end{description}
\end{description}
\end{description}

%% file: gis-lib/container.tex
\beginalto
We conclude this section by investigating the analogies and the differences of
generalized inference systems with respect to \emph{index containers}.
\emph{Indexed containers} \citep{AltenkirchGHMM15} are a rather general notion, 
meant to capture families of datatypes with some form of indexing.
They are part of the Agda standard library. 
We report below the definition, simplified and adapted a little for presentation purpose. 
Notably, we use ad-hoc field names, chosen to reflect the explanation provided below.

\begin{lstlisting}
record Container {$\ell$i $\ell$o} 
 (I : Set $\ell$i) (O : Set $\ell$o) ($\ell$c $\ell$p : Level) : Set _ where
  constructor _ $\lhd$ _/_
  field
    Cons  : (o : O) $\rightarrow$ Set $\ell$c
    Pos : $\forall$ {o} $\rightarrow$ Cons o $\rightarrow$ Set $\ell$p
    input : $\forall$ {o} (c : Cons o) $\rightarrow$ Pos c $\rightarrow$ I

$\llbracket$_$\rrbracket$: $\forall$ {$\ell$i $\ell$o $\ell$c $\ell$p $\ell$} {I : Set $\ell$i} {O : Set $\ell$o} $\rightarrow$ 
      Container I O $\ell$c $\ell$p $\rightarrow$
      (I $\rightarrow$ Set $\ell$) $\rightarrow$ (O $\rightarrow$ Set _)
$\llbracket$ Cons $\lhd$ Pos / input $\rrbracket$ X o = 
      $\Sigma$[ c $\in$ C o ] ((p : P c) $\rightarrow$ X (inp c p))
\end{lstlisting}

To explain the view of an inference system as an indexed container, 
we can think of the latter as describing a family of datatype constructors 
where \lstinline{I} and \lstinline{O} are  input and output sorts,  respectively. 
Then, \lstinline{Cons} specifies, for each output sort \lstinline{o}, 
the set of its constructors; for each constructor for \lstinline{o}, 
\lstinline{Pos} specifies a set of positions to store inputs to the constructor; finally, 
\lstinline{input} specifies the input sort for each position in a constructor.  
The function \lstinline{$\llbracket$_$\rrbracket$} models the ``semantics'' of an indexed container, 
that is, given a family of inputs \lstinline{X} indexed by \lstinline{I}, it returns the family 
of outputs indexed by \lstinline{O} which can be constructed by providing to 
some constructor inputs from \lstinline{P} of correct sorts. 

Then, inference systems can be defined as indexed containers 
where input and output sorts coincide, and are the elements of the universe, as follows. 
 
\begin{lstlisting}
  ISCont : {$\ell$c $\ell$p : Level} $\rightarrow$ (U : Set $\ell$u) $\rightarrow$ Set _
  ISCont {$\ell$c} {$\ell$p} U = Container U U $\ell$c $\ell$p
\end{lstlisting}

In this way, for each \lstinline{u : U}: 
\begin{itemize}
\item \lstinline{Cons u} is the set of (indexes for) all the rules which have consequence \lstinline{u}
\item \lstinline{Pos c} is the set of (indexes for) the premises of the \lstinline{c}-th rule 
\item \lstinline{input c p} is the \lstinline{p}-th premise of the \lstinline{c}-th rule
\end{itemize}

This view comes out quite naturally observing that an inference system is an element 
of $\wp(\wp(\universe)\times\universe)$; equivalently, a function which, for each $\judg\in\universe$,
returns the set of the sets of premises of all the rules with consequence $\judg$.  
In a constructive setting such as Agda, the powerset construction is not available, hence we have to use functions. 
So, for each element \lstinline{u}, we need a type to index all rules with consequence \lstinline{u}, 
and, for each  rule, a type to index its premises, which are exactly the data of an indexed container.  
In other words, this view of inference systems as indexed containers explicitly interprets rules as constructors for proofs. 
Moreover, definitions in \Cref{sec:gis} can be easily obtained as instances of analogous definitions 
for indexed containers, building on the fact that the inference operator associated with an inference 
system turns out to be the semantics \lstinline{$\llbracket$_$\rrbracket$} of the corresponding container.

Whereas this encoding allows reuse of notions and code, a drawback is that information is structured in a 
rather different way from that ``on paper''; notably, we group together rules with the same consequence, 
rather than those obtained as instances of the same ``schema'', that is, meta-rule. 
For this reason we developed the Agda library mimicking meta-rules.
For instance, the inference system for $\allpos$ would be as follows: 

\begin{lstlisting}
allPosCont : ISCont (Colist $\N$ $\infty$) 
allPosCont .Cons [] = $\top$ 
allPosCont .Cons (x :: xs) = x > 0 
allPosCont .Pos {[]} c = $\bot$ 
allPosCont .Pos {x :: xs} c = Fin 1
allPosCont .input {x :: xs} c zero = xs .force 
\end{lstlisting}

However, we can prove that the two notions are equivalent, as shown below. 
To this end, we define a translation \lstinline{C[_]} from inference systems to indexed containers
and a converse translation \lstinline{IS[_]}. 
Note that in the translation \lstinline{C[_]} each meta-rule is transformed in all its instantiations; 
more precisely, for each \lstinline{u:C}, \lstinline{Cons u} gives all the instantiations 
of meta-rules having \lstinline{u} as consequence.
Conversely, in the translation \lstinline{IS[_]}, each rule is transformed in a meta-rule with trivial context.

\begin{lstlisting}
C[_] : $\forall${$\ell$c $\ell$p $\ell$n} $\rightarrow$ IS {$\ell$c} {$\ell$p} {$\ell$n} U $\rightarrow$ Container U U _ $\ell$p
C[ is ] .Cons u = $\Sigma$[ rn $\in$ is .Names ] $\Sigma$[ c $\in$ is .rules rn .Ctx ]  
    u $\equiv$ is .rules rn .conclu c
C[ is ] .Pos (rn , _ , refl) = is .rules rn .Pos
C[ is ] .input (rn , c , refl) p = is .rules rn .prems c p

IS[_] : $\forall${$\ell$c $\ell$p} $\rightarrow$ 
      	Container U U $\ell$c $\ell$p $\rightarrow$ IS {zero} {$\ell$p} {l $\sqcup$ $\ell$c} U 
IS[ C ] .Names = $\Sigma$[ u $\in$ U ] C .Cons u
IS[ C ] .rules (u , c) = 
  record { 
    Ctx = $\top$ ; 
    Pos = C .Pos c ; 
    prems = $\lambda$ _ r $\rightarrow$ C .input c r ; 
    conclu = $\lambda$ _ $\rightarrow$ u }

isf-to-c : $\forall${$\ell$c $\ell$p $\ell$n $\ell$p} {is : IS {$\ell$c} {$\ell$p} {$\ell$n} U}
    {P : U $\rightarrow$ Set $\ell$p} $\rightarrow$ ISF[ is ] P $\subseteq$ $\llbracket$ C[ is ] $\rrbracket$ P
isf-to-c (rn , c , refl , pr) = (rn , c , refl) , pr

c-to-isf : $\forall${l' $\ell$p $\ell$p} {C : Container U U l' $\ell$p}
    {P : U $\rightarrow$ Set $\ell$p} $\rightarrow$ $\llbracket$ C $\rrbracket$ P $\subseteq$ ISF[ IS[ C ] ] P
c-to-isf (c , pr) = (_ , c) , tt , refl , pr
\end{lstlisting}

%% file: agda-prop/chintro.tex
\begintreble
In this chapter we take into account three interesting properties of \emph{binary} session types. 
All the properties under analysis mix safety and liveness. Hence, we use \GISs (\Cref{sec:gis}) as a reference framework
for their definitions and proofs.
Furthermore, for each property we first investigate its safety counterpart and we show how to obtain the desired
one by using corules.
Notably, all the results have been formalized in Agda \citep{CicconePadovani21,FairSubtypingAgda,CicconeP22@lmcs}.
We chose to restrict to the binary case to simplify the mechanizations as much as possible.

The properties that we study are the following:
\begin{itemize}
	\item \emph{Fair Termination} of a session type. A fairly terminating session can always eventually
	reach termination
	\item \emph{Fair Compliance} of two session types. Compliant sessions can always eventually interact to reach
	termination (asymmetric variant of \Cref{def:compatibility})
	\item \emph{Fair Subtyping} between session types. This property is a compliance-preserving subtyping relation
	(see \Cref{def:ssubt})
\end{itemize}

We begin the chapter by mechanizing in \Cref{sec:agda_fairt} the notions presented in \Cref{sec:ft_formally}
and that will be used later.
Then, in \Cref{sec:agda_st} we introduce an alternative definition of (binary) session
types and its formalization.
Then the chapter is split according to the property under analysis 
(\Cref{sec:agda_ft,sec:agda_fc,sec:agda_fs}).
Furthermore, for each property we show its Agda definition and we explain how
the proof principles (see \Cref{sec:gis}) can be used to prove the correctness. 
At last, in \Cref{sec:agda_fc_corr} we
detail the soundness and completeness proof of \emph{fair compliance}.
The formalization of all the results is available on GitHub \citep{FairSubtypingAgda}.

\begin{remark}
	\label{rm:agda_st}
	In this chapter we refer to \emph{binary session types} but we rely on a syntax 
	different from the more conventional one presented
	in \Cref{sec:st}. This choice is mainly motivated by the fact we aimed at
	simplifying the mechanization as much as possible. In \Cref{sec:agda_st} we describe all
	the advantages of using such representation.
\end{remark}

%% file: agda-prop/fair-termination.tex
\beginalto
\begin{figure}
\begin{lstlisting}[frame=single]
open import Relation.Nullary using ($\neg$_)
open import Relation.Unary using (Satisfiable; _$\cup$_)
open import Relation.Binary.Core using (Rel)
open import Relation.Binary.Construct.Closure.ReflexiveTransitive 
		using (Star)
open import Function.Base using (_$\circ$_)
\end{lstlisting}
\caption{Imported modules}
\label{fig:ft_imports}
\end{figure}
In this section we mechanize \emph{fair termination} in its general form, that is,
considering an arbitrary labeled transition system.
In particular, we mechanize the definitions and the notions that we presented in \Cref{sec:ft_formally}.
We will instanciate such property in \Cref{sec:agda_ft,sec:agda_fc} in session based scenarios.
Notably, the entire development is available on GitHub (see \cite{FairTermination}).

\begin{remark}[Imports]
	The codes that we present takes advantage of many features from the standard library.
	\Cref{fig:ft_imports} illustrates the modules that are imported and used in the formalization in this section.
	\lstinline{Satisfiable P} holds when a proof of \lstinline{P} can be exhibited whereas
	\lstinline{Star} is used to obtain the reflexive transitive closure of the reference relation.
	\lstinline{$\neg$ P} is equivalent to \lstinline{P $\rightarrow$ $\bot$}.
	For more details see \cite{FairTermination}.
	\eor
\end{remark}

First, we have to consider an arbitrary labelled transition system. Hence
the Agda module is parametric on
\begin{center}
	\lstinline{(State : Set) (_$\sim>$_ : State $\rightarrow$ State $\rightarrow$ Set)}
\end{center}
that is, a set of states and a transition relation.
Then, we define the reflexive transitive closure of the transition relation and the set 
of predicates over states. The predicates \lstinline{Reduces} and \lstinline{Stuck} hold when
a state can reduce to another one and when no reductions are allowed, respectively.

\begin{lstlisting}
StateProp : Set$_1$
StateProp = State $\rightarrow$ Set

_$\sim>^*$_ : Rel State _
_$\sim>^*$_ = Star _$\sim>$_

Reduces : StateProp
Reduces S = Satisfiable (S $\sim>$_)

Stuck : StateProp
Stuck S = $\neg$ (Reduces S)
\end{lstlisting}

A run is a maximal sequence of states 
\lstinline{S $\sim>$ S$_1$ $\sim>$ S$_2$ $\sim>$ $\dots$}, that is a
sequence of reductions that is either infinite or it ends with a
stuck state.

\begin{lstlisting}
data Run (S : State) : Set
record $\infty$Run (S : State) : Set where
  coinductive
  field force : Run S

data Run S where
  stop : (stuck : Stuck S) $\rightarrow$ Run S
  _::_ : $\forall${S'} (red : S $\sim>$ S') ($\rho$ : $\infty$Run S') $\rightarrow$ Run S
 
RunProp : Set$_1$
RunProp = $\forall${S} $\rightarrow$ Run S $\rightarrow$ Set
\end{lstlisting}

Note that runs are defined using a \emph{coinductive record}. 
\lstinline{RunProp} identifies predicates over runs.
Now we model the fact that each property of states induces a corresponding property of
runs in which there is a state that satisfies such property.
A run is finite if it contains a stuck state.

\begin{lstlisting}
data Eventually (P : StateProp) : RunProp where
  here : $\forall${S} {$\rho$ : Run S} (proof : P S) $\rightarrow$ Eventually P $\rho$
  next : $\forall${S S'} (red : S $\sim>$ S') {$\rho$ : $\infty$Run S'} 
  	(ev : Eventually P ($\rho$ .force)) $\rightarrow$ Eventually P (red :: $\rho$)
 
Finite : RunProp
Finite = Eventually Stuck
\end{lstlisting}

A fairness assumption is a proposition over runs such that every
partial run \lstinline{S $\sim>^*$ S'} can be extended to a fair run. This
condition is called feasibility or machine closure (see \Cref{lem:feasibility}).

\begin{lstlisting}
record FairnessAssumption : Set$_1$ where
  field
    Fair : RunProp
    feasible : $\forall${S S'} (reds : S $\sim>^*$ S') $\rightarrow$ 
    			$\Sigma$[ $\rho$ $\in$ Run S' ] Fair (reds ++ $\rho$)
 
StuckFairness : FairnessAssumption
StuckFairness = record { Fair = Fair' ; feasible = feasible' }
  where
    Fair' : RunProp
    Fair' = Eventually (Stuck $\cup$ NonTerminating)
\end{lstlisting}

where \lstinline{++} is used to concatenate runs and \lstinline{StuckFairness}
denotes our fairness assumption (see \Cref{def:fair_run}). For the sake of 
clarity we omit the definition of \lstinline{feasible'}.
We recall that a run is fair if it contains finitely many weakly terminating
states. This means that the run is either finite or divergent.
Now we can define \emph{fair termination} (see \Cref{def:fair_termination}). 
We recall that a state S is fairly terminating if the fair runs of S are finite.

\begin{lstlisting}
WeaklyTerminating : StateProp
WeaklyTerminating S = $\Sigma$[ $\rho$ $\in$ Run S ] Finite $\rho$

FairlyTerminating : FairnessAssumption $\rightarrow$ StateProp
FairlyTerminating $\phi$ S = $\forall${$\rho$ : Run S} $\rightarrow$ Fair $\phi$ $\rho$ $\rightarrow$ Finite $\rho$

Specification : StateProp
Specification S = $\forall${S'} $\rightarrow$ S $\sim>^*$ S' $\rightarrow$ WeaklyTerminating S'
\end{lstlisting}

\lstinline{Specification} is the alternative characterization of fair termination that
does not use fair runs. A state S satisfies the specification if
any S' that is reachable from S is weakly terminating. 
A state is weakly terminating if it has a finite run.
Now we can state \Cref{thm:fair_termination} (proofs are omitted).
In particular, we prove that the specification is a \emph{necessary} condition for fair
termination, regardless of the fairness assumption being made and that
the specification is a \emph{sufficient} condition for the notion of
fair termination induced by our assumption.

\begin{lstlisting}
ft$\rightarrow$spec : ($\phi$ : FairnessAssumption) $\rightarrow$ $\forall${S} $\rightarrow$ 
	FairlyTerminating $\phi$ S $\rightarrow$ Specification S

spec$\rightarrow$ft : $\forall${S} $\rightarrow$ 
	Specification S $\rightarrow$ FairlyTerminating StuckFairness S
\end{lstlisting}

At last, as a consequence we can prove that our assumption is the fairness assumption that 
induces the largest family of fairly terminating states.

\begin{lstlisting}
ft$\rightarrow$ft : ($\phi$ : FairnessAssumption) $\rightarrow$ 
	$\forall${S} $\rightarrow$ FairlyTerminating $\phi$ S $\rightarrow$ 
	FairlyTerminating StuckFairness S
ft$\rightarrow$ft $\phi$ = spec$\rightarrow$ft $\circ$ ft$\rightarrow$spec $\phi$
\end{lstlisting}

%% file: agda-prop/st.tex
\beginalto
As noted in \Cref{rm:agda_st}, we rely on a different definition of (binary) session types
with respect to that presented in \Cref{sec:st}. 
Although such choice has been originally motivated by the fact that we tried to simplify
as much as possible the Agda mechanization of the properties (\Cref{sec:agda_ft,sec:agda_fc,sec:agda_fs}),
it also has the capability of encoding \emph{dependent} session types, that is,
types that depend on previously exchanged messages. We give more
details later. We first introduce the new formulation of (binary) session types
and then we detail the technical aspects of their formalization.


\subsection{Session Types: An Alternative Formulation}
\input{agda-prop/st-definition}


\subsection{Agda Mechanization of Types}
\label{ssec:agda_st_formalization_types}
\input{agda-prop/st-formalization-types}

\subsection{Agda Mechanization of LTS}
\label{ssec:agda_st_formalization_lts}
\input{agda-prop/st-formalization-lts}

%% file: agda-prop/st-definition.tex
\beginbass
We assume a set $\Message$ of \emph{message tags} that can be exchanged in
communications. This set may include booleans, natural numbers, strings, and so
forth. Hereafter, we assume that $\Message$ contains at least \emph{two}
elements, otherwise branching protocols cannot be described and the theoretical
development that follows becomes trivial. We use $x$, $y$, $z$ to range over the
elements of $\Message$.
We define the set $\SessionType$ of \emph{session types} over $\Message$ using
coinduction, to account for the possibility that session types (and the
protocols they describe) may be infinite.

\begin{definition}[Session Types]
	\label{def:agda_st}
  Session types $T$, $S$ are the possibly infinite trees coinductively generated
  by the productions
  \[
    \begin{array}{rrcl}
      \textbf{Polarity}       & \Pol & \in & \set{{\In}, {\Out}} \\
      \textbf{Session type} & T, S & ::= & \TNil \mid \Pol\set{x:T_x}_{x\in\Message} \\
    \end{array}
  \]
\end{definition}

Hereafter, we write $\co{\Pol}$ for the opposite or dual polarity of $\Pol$, 
that is $\co\In = \Out$ and $\co\Out =
\In$.
Note that input and output session types specify continuations for \emph{all}
possible values in the set $\Message$. The session type $\TNil$, which describes
an \emph{unusable} session channel, can be used as continuation for those values
that \emph{cannot} be received or sent. As we will see shortly, the presence of
$\TNil$ breaks the symmetry between inputs and outputs.

It is convenient to introduce some notation for presenting session types in a
more readable and familiar form.
Given a polarity $\Pol$, a set $X \subseteq \Message$ of values and a family
$T_{x\in X}$ of session types, we let
\[
  \Pol\set{x:T_x}_{x\in X} \eqdef \Pol\left(\set{x:T_x}_{x\in X} \cup \set{x:\TNil}_{x\in\Message\setminus X}\right)
\]
so that we can omit explicit $\TNil$ continuations. As a special case when all
the continuations are $\TNil$, we write $\End[\Pol]$ instead of $\Pol\emptyset$.
Both $\End[\In]$ and $\Win$ describe session channels on which no further
communications may occur, although they differ slightly with respect to the
session types they can be safely combined with.  Describing terminated protocols
as degenerate cases of input/output session types reduces the amount of
constructors needed for their Agda representation (see \Cref{ssec:agda_st_formalization_types}).
Another common case for which we introduce a convenient notation is when the
continuations are the same, regardless of the value being exchanged: in these
cases, we write $\Pol X.T$ instead of $\Pol\set{x:T}_{x\in X}$. For example,
$\Out\Bool.\T$ describes a channel used for sending a boolean and then according
to $T$ and $\In\Nat.\S$ describes a channel used for receiving a natural number
and then according to $S$. We abbreviate $\set\x$ with $\x$ when no confusion
may arise. So we write $\Out\vtrue.\T$ instead of $\Out\set\vtrue.\T$.

Finally, we define a partial operation $+$ on session types such that
\[
  \Pol\set{\x:T_x}_{\x \in X} + \Pol\set{\x:T_x}_{\x \in Y}
  \eqdef
  \Pol\set{\x:T_x}_{\x \in X \union Y}
\]
when $X \cap Y = \emptyset$. For example, $\Out\vtrue.S_1 + \Out\vfalse.S_2$
describes a channel used first for sending a boolean value and then according to
$S_1$ or $S_2$ depending on the boolean value.
It it easy to see that $+$ is commutative and associative and that $\End[\Pol]$
is the unit of $+$ when used for combining session types with polarity $\Pol$. Note
that $T + S$ is undefined if the topmost polarities of $T$ and $S$ differ.
We assume that $+$ binds less tightly than the `$.$' in continuations.

We do not introduce any concrete syntax for specifying infinite session types.
Rather, we specify possibly infinite session types as solutions of equations of
the form $S = \cdots$ where the metavariable $S$ may also occur (guarded) on the
right-hand side of `$=$'. Guardedness guarantees that the session type $S$
satisfying such equation does exist and is unique \citep{Courcelle83}.

\begin{example}
  \label{ex:agda_st_ex}
  The session types $T_1$ and $S_1$ that satisfy the equations
  \[
    T_1 = \Out\vtrue.\Out\Nat.\T_1 + \Out\vfalse.\End[\In]
    \qquad
    S_1 = \Out\vtrue.\Out\NatPlus.\S_1 + \Out\vfalse.\End[\In]
  \]
  both describe a channel used for sending a boolean. If the boolean
  is $\vfalse$, the communication stops immediately ($\End[\In]$). If it
  is $\vtrue$, the channel is used for sending a natural number (a
  strictly positive one in $S_1$) and then according to $T_1$ or
  $S_1$ again. Notice how the structure of the protocol after the
  output of the boolean depends on the \emph{value} of the boolean.

  The session types $T_2$ and $S_2$ that satisfy the equations
  \[
    T_2 = \In\vtrue.\Out\Nat.\T_2 \branch \In\vfalse.\End[\In]
    \qquad
    S_2 = \In\vtrue.\Out\NatPlus.\S_2 \branch \In\vfalse.\End[\In]
  \]
  differ from $T_1$ and $S_1$ in that the channel they describe is
  used initially for \emph{receiving} a boolean.
  \eoe
\end{example}

\begin{figure}[t]
	\framebox[\textwidth]{
	\begin{mathpar}
  	\inferrule[input]{\mathstrut}{
    	\In\x.S + T \lred{\In\x} S
  	}\defrule[input]{}
  	\and
  	\inferrule[output]{\mathstrut}{
    	\Out\x.S + T \lred{\Out\x} S
  	}
  	~ S \ne \TNil \defrule[output]{}
  \end{mathpar}
  }
  \caption{Labeled transition system}
  \label{fig:agda_st_lts}
\end{figure}

We define the operational semantics of session types by means of a
\emph{labeled transition system}. \emph{Labels}, ranged over by
$\actionA$, $\actionB$, $\actionC$, have either the form $\In\x$
(input of message $x$) or the form $\Out\x$ (output of message $x$).
Transitions $T \lred\action S$ are defined in \Cref{fig:agda_st_lts}

There is a fundamental asymmetry between send and receive operations: the act of
sending a message is \emph{active} -- the sender may choose the message to send
-- while the act of receiving a message is \emph{passive} -- the receiver cannot
cherry-pick the message being received.
We model this asymmetry with the side condition $S \ne \TNil$ in
\refrule{output} and the lack thereof in \refrule{input}:
a process that uses a session channel according to
$\Out\set{x:T_x}_{x\in\Message}$ refrains from sending a message $x$ if $T_x =
\TNil$, namely if the channel becomes unusable by sending that particular
message, whereas a process that uses a session channel according to
$\In\set{x:T_x}_{x\in\Message}$ cannot decide which message $x$ it will receive,
but the session channel becomes unusable if an unexpected message arrives. The
technical reason for modeling this asymmetry is that it allows us to capture a
realistic communication semantics. For the time being, these transition rules allow us
to appreciate a little more the difference between $\Win$ and $\End[\In]$. While both
describe a session endpoint on which no further communications may occur, $\Win$
is ``more robust'' than $\End[\In]$ since it has no transitions, whereas $\End[\In]$ is
``more fragile'' than $\Win$ since it performs transitions, all of which lead to
$\TNil$. For this reason, we use $\Win$ to flag successful session termination, 
whereas $\End[\In]$ only means that the protocol has ended.

To describe \emph{sequences} of consecutive transitions performed by
a session type we use another relation $\wlred\actions$ where
$\actions$ and $\actionsB$ range over strings of labels. As usual,
$\es$ denotes the empty string and juxtaposition denotes string
concatenation. The relation $\wlred\actions$ is the least one such
that $T \wlred\es T$ and if $T \lred\action S$ and
$S \wlred\actions R$, then $T \wlred{\action\actions} R$.

\begin{figure}[t]
  \framebox[\textwidth]{
  \begin{mathpar}
  	\inferrule[sync]{\mathstrut}{
    	\session\S\T \red \session{S'}{T'}
  	}
  	~
  	S \lred{\co\action} S', T \lred\action T'
  	\defrule[sync]{}
  \end{mathpar}
  }
  \caption{Reduction of session}
  \label{fig:agda_st_session_red}
\end{figure}

At last, we need to model the evolution of a session as client
and server interact. To this aim, we represent a session as a pair
$\session\R\T$ where $\R$ describes the behavior of the client and $T$ that of
the server.  Sessions reduce according to the rule in \Cref{fig:agda_st_session_red}
where $\co\action$ is the \emph{complementary action} of $\action$ defined by
$\co{\Pol\x} = \co{\Pol}\x$.  We extend $\co{\,\cdot\,}$ to traces in the obvious way
and we write $\wred$ for the reflexive, transitive closure of $\red$. We write
$\session\R\T \red {}$ if $\session\R\T \red \session{R'}{T'}$ for some $R'$ and
$T'$ and $\session\R\T \nred$ if not $\session\R\T\red$.

%% file: agda-prop/st-formalization-types.tex
\beginbass
We postulate the existence of $\Message$, representing the set of
values that can be exchanged in communications.
The actual Agda formalization is parametric on an
arbitrary set $\Message$, the only requirement being that $\Message$
must be equipped with a decidable notion of equality. We will make
some assumptions on the nature of $\Message$ when we present
specific examples.
To begin the formalization, we declare the two data types we use to represent session types. 
Because these types are mutually recursive, we declare them in advance so that 
we can later refer to them from within the definition of each.

\begin{lstlisting}
data SessionType : Set
record $\infty$SessionType : Set

data SessionType where
  nil : SessionType
  inp out : Continuation $\rightarrow$ SessionType
  
record $\infty$SessionType where
  coinductive
  field force : SessionType
\end{lstlisting}

The \lstinline{SessionType} data type provides three constructors corresponding 
to the three forms of a session type (see \Cref{def:agda_st}).
The	\lstinline{$\infty$SessionType} wraps a session type
within a coinductive record, so as to make it possible to represent
\emph{infinite} session types. The record has just one field
\lstinline{force} that can be used to access the wrapped session
type. By opening the record, we make the \lstinline{force} field
publicly accessible without qualifiers. 

\begin{remark}
	Coinductive records are the built-in technique for dealing with infinite
	datatypes. It has been already used in \Cref{ch:gis_lib} and \Cref{sec:agda_ft}
	for implementing the coinductive interpretations of an inference system and
	possibly infinite runs in a labeled transitions system. 
	Although the \emph{sized types} approach leads to syntactically easier datatypes,
	a recent Agda update highlighted an inconsistency when using both approaches together.
	Since \emph{thunks}, on which sized types are based on, are defined as a coinductive
	record, we decided to mainly rely on the former approach for the sake of simplicity.
	\eor
\end{remark}

A key design choice of our formalization is the representation of continuations 
$\set{x:S_x}_{x\in\Message}$, which may have infinitely many branches if $\Message$ is infinite. 
We represent continuations in Agda as \emph{total functions} from $\Message$ to session types, thus:

\begin{lstlisting}
Continuation : Set
Continuation = $\Message$ $\rightarrow$ $\infty$SessionType
\end{lstlisting}

\begin{example}
	\label{ex:agda_st_ex_form}
	Consider once again the session type $T_1$ discussed in \Cref{ex:agda_st_ex}:
	\[
  	T_1 = \Out\vtrue.\Out\Nat.\T_1 + \Out\vfalse.\End[\In]
	\]
	In this case we assume that $\Message$ is the disjoint sum of $\Nat$
	(the set of natural numbers) and $\Bool$ (the set of boolean values)
	with constructors \lstinline{nat} and \lstinline{bool}. 
	It is useful to also define once and for all the \emph{continuation} \lstinline{empty},
	which maps every message to $\TNil$:

\begin{lstlisting}
empty : Continuation
empty _ .force = nil
\end{lstlisting}

	Now, the Agda encoding of $T_1$ is shown below:
	
\begin{lstlisting}
T$_1$ : SessionType
T$_1$ = out f
  where
    f g : Continuation
    f (nat _) .force = nil
    f (bool true) .force = out g
    f (bool false) .force = inp empty
    g (nat _) .force = out f
    g (bool _) .force = nil
\end{lstlisting}

	The continuations \lstinline{f} and
	\lstinline{g} are defined using pattern matching on the
	message argument and using copattern matching to specify the value of the \lstinline{force}
	field of the resulting coinductive record. They represent the two stages of the protocol:
	\lstinline{f} allows sending a boolean (but no natural number)
	and, depending on the boolean, it continues as \lstinline{g} or
	it terminates; \lstinline{g} allows sending a natural number
	(but no boolean) and continues as \lstinline{f}.
	\eoe
\end{example}

\Cref{ex:agda_st_ex_form} illustrates a simple form of \emph{dependency} whereby
the structure of a communication protocol may depend on the content
of previously exchanged messages. The fact that we use Agda to write
continuations means that we can model sophisticated forms of
dependencies that are found only in the most advanced theories of
dependent session types
\citep{ToninhoCairesPfenning11,ToninhoYoshida18,ThiemannVasconcelos20,CicconePadovani20}.

\begin{example}
	Consider the Agda encoding of a session type
	\[
  	\Out{(n:\Nat)}.\underbrace{\Out\Bool\dots\Out\Bool}_n.\End[\Out]
	\]
	describing a channel used for sending a natural number $n$ followed by $n$ boolean values:
	
\begin{lstlisting}
BoolVector : SessionType
BoolVector = out g
  where
    f : $\Nat$ $\rightarrow$ Continuation
    f zero _ .force = nil
    f (suc n) (bool _) .force = out (f n)
    f (suc n) (nat _) .force = nil
    g : Continuation
    g (nat n) .force = out (f n)
    g (bool _) .force = nil
\end{lstlisting}

	We will not discuss further examples of dependent session types. 
	However, note that the possibility of encoding protocols such as \lstinline{BoolVector} 
	has important implications on the scope of our study: it means that the results we have 
	presented and formally proved in Agda hold for a large 
	family of session types that includes dependent ones.
	\eoe
\end{example}

We now provide a few auxiliary predicates on session types and continuations. 
First of all, we say that a session type is \emph{defined} if it is different from $\TNil$:

\begin{lstlisting}
data Defined : SessionType $\rightarrow$ Set where
  inp : $\forall${f} $\rightarrow$ Defined (inp f)
  out : $\forall${f} $\rightarrow$ Defined (out f)
\end{lstlisting}

Concerning continuations, we define the \emph{domain} of a continuation function $f$ 
to be the subset of $\Message$ such that $f~x$ is defined, 
that is \lstinline{dom} $f = \set{x \in \Message \mid f~x \ne \TNil }$.
We say that a continuation is \emph{empty} if so is its domain.
On the contrary, a non-empty continuation is said to have a \emph{witness}. 
We define a \lstinline{Witness} predicate to characterize this condition.

\begin{lstlisting}
dom : Continuation $\rightarrow$ Pred $\Message$ Level.zero
dom f x = Defined (f x .force)

EmptyContinuation : Continuation $\rightarrow$ Set
EmptyContinuation f = Relation.Unary.Empty (dom f)

Witness : Continuation $\rightarrow$ Set
Witness f = Relation.Unary.Satisfiable (dom f)
\end{lstlisting}

where \lstinline{Satisfiable P} and \lstinline{Empty P} mean that
at least an element satisfies \lstinline{P} and no elements satisfy \lstinline{P}, respectively.

\begin{example}
	Consider the \lstinline{empty} continuation defined in \Cref{ex:agda_st_ex_form}.
	We can prove that it is indeed an empty one.
	
\begin{lstlisting}
empty-is-empty : EmptyContinuation empty
empty-is-empty _ ()
\end{lstlisting}
	\eoe
\end{example}

We now define a predicate \lstinline{Win} to characterize the session type $\End[\Out]$ and
\lstinline{End} to characterize $\End[\Pol]$:

\begin{lstlisting}
data Win : SessionType $\rightarrow$ Set where
  out : $\forall${f} $\rightarrow$ EmptyContinuation f $\rightarrow$ Win (out f)

data End : SessionType $\rightarrow$ Set where
  inp : $\forall${f} (U : EmptyContinuation f) $\rightarrow$ End (inp f)
  out : $\forall${f} (U : EmptyContinuation f) $\rightarrow$ End (out f)
\end{lstlisting}

At last, we need a representation of sessions as pairs $\session{R}{S}$ of session types.
To this aim, we introduce the \lstinline{Session} data type as an alias for pairs of session types.

\begin{lstlisting}
Session : Set
Session = SessionType $\times$ SessionType
\end{lstlisting}

%% file: agda-prop/st-formalization-lts.tex
\beginbass
Let us move onto the definition of transitions for session types. 
We begin by defining an \lstinline{Action} data type to represent 
input/output actions, which consist of a polarity and a value.
The \emph{complementary action} of $\action$, denoted by $\co\action$ 
in the previous sections, is computed by the function \lstinline{co-action}.

\begin{lstlisting}
data Action : Set where
  I O : $\Message$ $\rightarrow$ Action

co-action : Action $\rightarrow$ Action
co-action (I x) = O x
co-action (O x) = I x 
\end{lstlisting}

A \emph{transition} is a ternary relation among two session types and an action. 
A value of type \lstinline{Transition S $\action$ T} represents the transition $S \lred\action T$. 
Note that the premise \lstinline{x $\in$ dom f} in the constructor 
\lstinline{out} corresponds to the side condition 
$S \ne \TNil$ of rule \refrule{output}.

\begin{lstlisting}
data Transition : SessionType $\rightarrow$ Action $\rightarrow$ SessionType $\rightarrow$ Set 
  where
    inp : $\forall${f x} $\rightarrow$ Transition (inp f) (I x) (f x .force)
    out : $\forall${f x} $\rightarrow$ 
    	x $\in$ dom f $\rightarrow$ Transition (out f) (O x) (f x .force)
\end{lstlisting}

A session \emph{reduces} when client and server synchronize, by performing 
actions with opposite polarities and referring to the same message.
We formalize the reduction relation in \Cref{fig:agda_st_lts} as the \lstinline{Reduction} 
data type, so that a value of type \lstinline{Reduction ($\session{R}{S}$) ($\session{R'}{S'}$)}
witnesses the reduction $\session{R}{S} \red \session{R'}{S'}$.
At last, the weak reduction relation is called \lstinline{Reductions}
and is defined as the reflexive, transitive closure of \lstinline{Reduction}, 
just like $\wred$ is the reflexive, transitive closure of $\red$. 
We make use of the \lstinline{Star} data type from Agda's standard library to define such closure.

\begin{lstlisting}
data Reduction : Session $\rightarrow$ Session $\rightarrow$ Set where
  sync : $\forall${$\action$ R R' S S'} $\rightarrow$ Transition R (co-action $\action$) R' $\rightarrow$ 
  	Transition S $\action$ S' $\rightarrow$ Reduction (R , S) (R' , S')
  	
Reductions : Session $\rightarrow$ Session $\rightarrow$ Set
Reductions = Star Reduction
\end{lstlisting}

%% file: agda-prop/ft.tex
\beginalto
In this section we characterize fair termination of a session type
according to its general formulation in \Cref{thm:fair_termination}.
We say that a session type is fairly terminating if it preserves the possibility
of reaching $\Win$ or $\End[\In]$ along all of its transitions that do not lead to
$\TNil$.  Fair termination of $S$ does not necessarily imply that there exists an
upper bound to the length of communications that follow the protocol $S$, but it
guarantees the absence of ``infinite loops'' whereby the communication is forced
to continue forever.


\subsection{Definition}
\label{ssec:agda_ft_def}
\input{agda-prop/ft-definition}


\subsection{Agda Formalization}
\label{ssec:agda_ft_form}
\input{agda-prop/ft-formalization}

%% file: agda-prop/ft-definition.tex
\beginbass
To formalize fair termination we need the notion of \emph{trace}, which is a
finite sequence of actions performed on a session channel while preserving
usability of the channel.

\begin{definition}[(Maximal) traces]
  \label{def:traces}
  The \emph{traces} of a session $S$ are defined as $\traces\S \eqdef \set{ \actions \mid
  \exists\T : S \wlred\actions T \ne \TNil }$.  We say that $\actions \in
  \traces\S$ is \emph{maximal} if $\actions\actionsB \in \traces\S$ implies
  $\actionsB = \es$.
\end{definition}

\begin{remark}
	\Cref{def:traces} differs from the $\paths$ of a session type (\Cref{def:path})
	since we require that the involved session type does not reduce to $\TNil$.
	This difference is motivated by the different definition of session types.
	\eor
\end{remark}

\begin{example}
	We have $\traces\TNil = \emptyset$ and $\traces\Win = \traces{\End[\In]} =
	\set\es$. Note that $\Win$ and $\End[\In]$ have the same traces but different
	transitions (hence different behaviors).
	\eoe
\end{example}

A \emph{maximal trace} is a trace that cannot be extended any further. 

\begin{definition}[Fair Termination]
  \label{def:wt}
  We say that $S$ is \emph{fairly terminating} if, for every
  $\actionsA\in\traces\S$, there exists $\actionsB$ such that
  $\actionsA\actionsB \in \traces\S$ and $\actionsA\actionsB$ is maximal.
\end{definition}

\begin{example}
 	$\es$ is a maximal trace of both $\Win$ and $\End[\In]$ but not of
 	$\Out\Bool.\End[\In]$ whereas $\Out\vtrue$ and $\Out\vfalse$ are maximal traces of
 	$\Out\Bool.\End[\In]$.
	\eoe
\end{example}

\begin{example}
  \label{ex:termination}
  All of the session types presented in \Cref{ex:agda_st_ex} are fairly
  terminating. The session type $R = \Out\Bool.R$, which describes a channel
  used for sending an infinite stream of boolean values, is not fairly
  terminating because no trace of $R$ can be extended to a maximal one. Note
  that also $R' \eqdef \Out\vtrue.R + \Out\vfalse.\Win$ is not fairly
  terminating, even though there is a path leading to $\Win$, because fair
  termination must be \emph{preserved} along all possible transitions of the
  session type, whereas $R' \xlred{\Out\vtrue} R$ and $R$ is not fairly
  terminating.
  Finally, $\TNil$ is trivially fairly terminating because it has no trace.
  \eoe
\end{example}

\begin{figure}[t]
	\framebox[\textwidth]{
 		\begin{mathpar}
      \inferrule[t-nil]{\mathstrut}{
        \terminates\TNil
      } \defrule[t-nil]{}
      \and
      \inferrule[t-all]{
        \forall x \in \Message : \terminates{T_x}
      }{
        \terminates{\Pol\set{x:T_x}_{x\in\Message}}
      } \defrule[t-all]{}
      \and
      \infercorule[t-any]{
        \terminates{S}
      }{
        \Pol x.S + T
      }
      ~
      S \ne \TNil \defrule[t-any]{}
    \end{mathpar}
  }
  \caption{Generalized inference system $\gis{\is[T]}{\cois[T]}$ for fair termination}
  \label{fig:ft}
\end{figure}

To find an inference system for fair termination observe that the set
$\FairlyTerminating$ of fairly terminating session types is the largest one that
satifies the following two properties:
\begin{enumerate}
  \item it must be possible to reach either $\Win$ or $\End[\In]$ from every $S \in
\FairlyTerminating \setminus \set\TNil$;
  \item the set $\FairlyTerminating$ must be closed by transitions, namely if $S
\in \FairlyTerminating$ and $S \lred\action T$ then $T \in \FairlyTerminating$.
\end{enumerate}
Neither of these two properties, taken in isolation, suffices to define
$\FairlyTerminating$: the session type $R'$ in \Cref{ex:termination}
enjoys property (1) but is not fairly terminating; the set $\SessionType$ is
obviously the largest one with property (2), but not every session type in it is
fairly terminating.
This suggests the definition of $\FairlyTerminating$ as the largest subset of
$\SessionType$ satisfying (2) and whose elements are \emph{bounded} by property
(1), which is precisely what corules allow us to specify.

\Cref{fig:ft} shows a \GIS $\gis{\is[T]}{\cois[T]}$ for fair termination with
the usual notation for single-lined rules and doubly-lined corules.
The axiom \refrule{t-nil} indicates that $\TNil$ is fairly terminating in a
trivial way (it has no trace), while \refrule{t-all} indicates that fair
termination is closed by all transitions. Note that these two rules, interpreted
coinductively, are satisfied by all session types, hence $\set{ \S \mid
\terminates\S \in \CoInductive{\is[S]} } = \SessionType$.

\begin{theorem}
  \label{thm:termination}
  $T$ is fairly terminating if and only if $\terminates\S \in
  \FlexCo{\is[S]}{\cois[S]}$.
\end{theorem}
\begin{proof}[Proof sketch]
  For the ``if'' part, suppose $\terminates\S \in \FlexCo{\is[S]}{\cois[S]}$ and
  consider a trace $\actions\in\traces\S$. That is, $S \wlred\actions T$ for
  some $T \ne \TNil$. Using \refrule{t-all} we deduce $\terminates\T \in
  \FlexCo{\is[S]}{\cois[S]}$ by means of a simple induction on $\actions$. Now
  $\terminates\T \in \FlexCo{\is[S]}{\cois[S]}$ implies $\terminates\T \in
  \Inductive{\is[S] \cup \cois[S]}$. Another
  induction on the (well-founded) derivation of this judgment, along with the
  witness message $x$ of \refrule{t-any}, allows us to find $\actionsB$ such
  that $\actions\actionsB$ is a maximal trace of $S$.

  For the ``only if'' part, we apply the bounded coinduction principle (see
  \Cref{prop:bcp}). Since we have already argued that the coinductive
  interpretation of the GIS in \Cref{fig:ft} includes all session types, it
  suffices to show that $S$ fairly terminating implies $\terminates\S \in
  \Inductive{\is[S] \cup \cois[S]}$.
  From the assumption that $S$ is fairly terminating we deduce that there exists
  a maximal trace $\actions\in\traces{S}$. An induction on $\actions$ allows us
  to derive $\terminates{S}$ using repeated applications of \refrule{t-any}, one
  for each action in $\actions$, topped by a single application of
  \refrule{t-nil}.
\end{proof}

%% file: agda-prop/ft-formalization.tex
\beginbass
We now use the Agda library for \GIS (see \Cref{ch:gis_lib}) to 
formally define the inference system for fair termination shown in \Cref{fig:ft}. 
First, we encode the universe on which the predicate is defined. In this
case, it consists of session types, that is, $\SessionType$.

\begin{lstlisting}
U : Set
U = SessionType
\end{lstlisting}

Now we define the sets containing the names of the (co)rules. 
Although we try to be as consistent as possible with respect to
the \GIS in \Cref{fig:ft}, we need to split both 
\refrule{t-all} and \refrule{t-any} according to the polarity
of the involved session type. Hence, we obtain three rules
and two corules.

\begin{lstlisting}
data RuleNames : Set where
  nil inp out : RuleNames

data CoRuleNames : Set where
  inp out : CoRuleNames
\end{lstlisting}

We can look at the definitions of the five metarules.
Notably, as we presented in \Cref{ch:gis_lib}, the library offers 
a simpler way for defining a metarule with a finite number of premises
(see \lstinline{FinMetaRule} datatype). We take advantage of such feature
to define the axiom as well as the corules that differ from the rules since they
have a single premise.
We show rules and corules separately.

\begin{lstlisting}
nil-r : FinMetaRule U
nil-r .Ctx = $\top$
nil-r .comp _ =
  [] ,
  ----
  nil

inp-r : MetaRule U
inp-r .Ctx = Continuation
inp-r .Pos _ = $\Message$
inp-r .prems f p = f p .force
inp-r .conclu f = inp f

out-r : MetaRule U
out-r .Ctx = Continuation
out-r .Pos _ = $\Message$
out-r .prems f p = f p .force
out-r .conclu f = out f
\end{lstlisting}

The axiom simply states that $\TNil$ is trivially fairly terminating.
Concerning the context, we use the datatype \lstinline{$\top$} from 
the standard library which can be always instantiated with
the constructor \lstinline{tt}.
The two rules have have the \lstinline{Pos} field set to \lstinline{$\Message$}
since they have a premise for each element in $\Message$. Note also that 
each session type is unfolded in the premises by accessing the \lstinline{force}
filed of the coinductive record \lstinline{$\infty$SessionType}.

\begin{lstlisting}
inp-co-r : FinMetaRule U
inp-co-r .Ctx = $\Sigma$[ (f , x) $\in$ Continuation $\times$ $\Message$ ] x $\in$ dom f
inp-co-r .comp ((f , x) , _) =
  f x .force :: [] ,
  --------------------
  inp f

out-co-r : FinMetaRule U
out-co-r .Ctx = $\Sigma$[ (f , x) $\in$ Continuation $\times$ $\Message$ ] x $\in$ dom f
out-co-r .comp ((f , x) , _) =
  f x .force :: [] ,
  --------------------
  out f
\end{lstlisting}

As mentioned before, the corules differ from the rules because they have a single premise.
This is represented by the existential quantifier in the context which informally
asks that the must be a continuation of the involved session type which is
fairly terminating.
Now we can compose the two inference systems \lstinline{FairTerminationIS} and \lstinline{FairTerminationCOIS}
consisting of the rules and the corules, respectively.
The desired predicate \lstinline{FairTermination} is obtained through the generalized interpretation of the whole
inference system.
\lstinline{FairTerminationI} is the inductive predicate obtained by inductively
interpreting all the rules.

\begin{lstlisting}
FairTerminationIS : IS U
Names FairTerminationIS = RuleNames
rules FairTerminationIS nil = from nil-r
rules FairTerminationIS inp = inp-r
rules FairTerminationIS out = out-r

FairTerminationCOIS : IS U
FairTerminationCOIS .Names = CoRuleNames
FairTerminationCOIS .rules inp = from inp-co-r
FairTerminationCOIS .rules out = from out-co-r

FairTermination : SessionType $\rightarrow$ Set
FairTermination = 
  FCoInd$\llbracket$ FairTerminationIS , FairTerminationCOIS $\rrbracket$

FairTerminationI : SessionType $\rightarrow$ Set
FairTerminationI = 
  Ind$\llbracket$ FairTerminationIS $\cup$ FairTerminationCOIS $\rrbracket$
\end{lstlisting}

where we recall that the function \lstinline{from} turns a \lstinline{FinMetaRule} into
a \lstinline{MetaRule} (see \Cref{fig:finmetarule-dt}).

In order to prove the correctness of \lstinline{FairTermination} we need to encode
the specification, that is, \Cref{def:wt}.

\begin{lstlisting}
FairTerminationS : SessionType $\rightarrow$ Set
FairTerminationS S = $\forall${$\phi$} $\rightarrow$ 
  $\phi$ $\in$ $\llbracket$ S $\rrbracket$ $\rightarrow$ $\exists$[ $\psi$ ] ($\phi$ ++ $\psi$ $\in$ Maximal $\llbracket$ S $\rrbracket$)
\end{lstlisting}

where \lstinline{$\phi$,$\psi$ : Trace} and traces are defined as \lstinline{List Action}.
Hence \lstinline{_++_} is the library function for computing the concatenation of two lists.
\lstinline{$\llbracket$_$\rrbracket$} denotes the set of all the possible traces of
a session type. Finally, \lstinline{Maximal} is a predicate on sets of traces and denotes
all those traces that cannot be extended (see \Cref{def:traces}). 

The correctness of \lstinline{FairTermination} is expressed in terms of soundness
and completeness with respect to \lstinline{FairTerminationS}.
We are not going to detail the proofs. Instead we report the declarations
of the main lemmas.
Concerning the \emph{soundness}, \GISs do not provide a canonical technique to prove it.

\begin{lstlisting}
sound : FairTermination $\subseteq$ FairTerminationS
\end{lstlisting}

where \lstinline{P $\subseteq$ Q} is equivalent to \lstinline{$\forall$$\{$x$\}$ $\rightarrow$ P x $\rightarrow$ Q x}
with \lstinline{P,Q} predicates over some \lstinline{A : Set _} and \lstinline{x : A}.
For what concerns the \emph{completeness}, it is by bounded coinduction (\Cref{prop:bcp}, \Cref{fig:principles});
hence we have to prove that \lstinline{FairTerminationS} is \emph{bounded} and \emph{consistent} with respect
to the \GIS.

\begin{lstlisting}
bounded : FairTerminationS $\subseteq$ FairTerminationI

consistent : 
  FairTerminationS $\subseteq$ ISF[ FairTerminationIS ] FairTerminationS

complete : FairTerminationS $\subseteq$ FairTermination
complete =
  bounded-coind[ FairTerminationIS , FairTerminationCOIS ]
    FairTerminationS bounded consistent
\end{lstlisting}

\begin{remark}
	Assume that we instanciate \lstinline{Specification} from \Cref{sec:agda_fairt} using
	\lstinline{SessionType} (see \Cref{ssec:agda_st_formalization_types}) as set of states and
	\lstinline{Transition} (see \Cref{ssec:agda_st_formalization_lts}) as transition system.
	Such instance of \lstinline{Specification} will not be equivalent to \lstinline{FairTerminationS}
	as, for example, a client sending always \lstinline{true} would be fairly terminating since
	the \lstinline{false} branch would always lead to $\TNil$.
	To make the two specifications equivalent we need to instanciate \lstinline{Specification}
	by using the following transition system
	\[
		\inferrule
			{S \lred\action T}
			{S \xrightarrow{\action}' T}
			~ T \ne \TNil
	\]
	\eor
\end{remark}

%% file: agda-prop/fc.tex
\beginalto
In this section we define and characterize two \emph{compliance} relations for
session types, which formalize the ``successful'' interaction between a client
and a server connected by a session. The notion of ``successful interaction''
that we consider is biased towards client satisfaction, but see
\Cref{rm:asymmetry} for a discussion about alternative notions.


\subsection{Definition}
\label{ssec:agda_fc_def}
\input{agda-prop/fc-definition}


\subsection{Agda Formalization}
\label{ssec:agda_fc_form}
\input{agda-prop/fc-formalization}

%% file: agda-prop/fc-definition.tex
\beginbass
The first compliance relation that we consider requires that, if the interaction
in a session stops, it is because the client ``is satisfied'' and the server
``has not failed'' (recall that a session type can turn into $\TNil$ only if an
unexpected message is received). Formally:

\begin{definition}[Compliance]
  \label{def:comp}
  We say that $R$ is \emph{compliant} with $T$ if
  $\session\R\T \wred \session{\R'}{\T'} \nred$ implies $\R' = \Win$
  and $\T' \neq \TNil$.
\end{definition}

This notion of compliance is an instance of \emph{safety property} in which the
invariant being preserved at any stage of the interaction is that either client
and server are able to synchronize further, or the client is satisfied and the
server has not failed.

The second compliance relation that we consider adds a \emph{liveness}
requirement namely that, no matter how long client and server have been
interacting with each other, it is always possible to reach a configuration in
which the client is satisfied and the server has not failed.

\begin{definition}[Fair Compliance]
  \label{def:fcomp}
  We say that $R$ is \emph{fairly compliant} with $T$ if $\session\R\T \wred
  \session{\R'}{\T'}$ implies $\session{\R'}{\T'} \wred \session\Win{\T''}$ with
  $\T'' \neq \TNil$.
\end{definition}

Notably, fair compliance corresponds to a \emph{successful} form of fair termination.
It is easy to show that fair compliance implies compliance, but there exist
compliant session types that are not fairly compliant, as illustrated in the
following example.

\begin{example}
  \label{ex:fcomp}
  Recall \Cref{ex:agda_st_ex} and consider the session types $R_1$ and $R_2$
  such that
  \[
    R_1 = \In\vtrue.\In\Nat.R_1 \branch \In\vfalse.\Win
    \qquad
    R_2 = \Out\vtrue.(\In0.\Win \branch \In\NatPlus.R_2)
  \]

  Then $R_1$ is fairly compliant with both $T_1$ and $S_1$ and $R_2$ is
  compliant with both $T_2$ and $S_2$.
  Even if $S_1$ exhibits fewer behaviors compared to $T_1$ (it never sends $0$
  to the client), at the beginning of a new iteration it can always send
  $\vfalse$ and steer the interaction along a path that leads $R_1$ to success.
  On the other hand, $R_2$ is fairly compliant with $T_2$ but not with $S_2$. In
  this case, the client insists on sending $\vtrue$ to the server in hope to
  receive $0$, but while this is possible with the server $T_2$, the server
  $S_2$ only sends strictly positive numbers.

  This example also shows that fair termination of both client and server is not
  sufficient, in general, to guarantee fair compliance. Indeed, both $R_2$ and
  $S_2$ are fairly terminating, but they are not fairly compliant. The reason is
  that the sequences of actions leading to $\Win$ on the client side are not
  necessarily the same (complemented) traces that lead to $\End$ on the server
  side. Fair compliance takes into account the synchronizations that can
  actually occur between client and server.
  \eoe
\end{example}

\begin{remark}
  \label{rm:asymmetry}
  With the above notions of compliance we can now better motivate the
  asymmetric modeling of (passive) inputs and (active) outputs in the labeled
  transition system of session types (\Cref{fig:agda_st_lts}). Consider the session
  types $R = \Out\vtrue.\Win + \Out\vfalse.\Out\vfalse.\Win$ and $S =
  \In\vtrue.\End[\In]$. Note that $R$ describes a client that succeeds by either
  sending a single $\vtrue$ value or by sending two $\vfalse$ values in
  sequence, whereas $S$ describes a server that can only receive a single
  $\vtrue$ value. If we add the same side condition $S \ne \TNil$ also for
  \refrule{input} then $R$ would be compliant with $S$. Indeed, the server
  would be unable to perform the $\In\vfalse$-labeled transition, so that the
  only synchronization possible between $R$ and $S$ would be the one in which
  $\vtrue$ is exchanged. In a sense, with the $S \ne \TNil$ side condition in
  \refrule{input} we would be modeling a communication semantics in which
  client and server \emph{negotiate} the message to be exchanged depending on
  their respective capabilities. Without the side condition, the message to be
  exchanged is always chosen by the active part (the sender) and, if the
  passive part (the receiver) is unable to handle it, the receiver fails. The
  chosen asymmetric communication semantics is also key to induce a notion of
  (fair) subtyping that is \emph{covariant} with respect to inputs
  (see \Cref{sec:agda_fc}).
  \eor
\end{remark}

\begin{figure}[t]
	\framebox[\textwidth]{
	\begin{mathpar}
      \inferrule[c-success]{\mathstrut}{
        \compliance\Win\T
      }
      ~T \neq \TNil \defrule[c-success]{}
      \and
      \inferrule[c-inp-out]{
        \forall x \in X : \compliance{S_x}{T_x}
      }{
        \compliance{\In\set{x:S_x}_{x\in\Message}}{\Out\set{x:T_x}_{x\in X}}
      }
      ~X \ne \emptyset \defrule[c-inp-out]{}
      \and
      \infercorule[c-sync]{
        \compliance{S}{T}
      }{
        \compliance{\Pol x.S + S'}{\co{\Pol}x.T + T'}
      } \defrule[c-sync]{}
      \and
      \inferrule[c-out-inp]{
        \forall x \in X : \compliance{S_x}{T_x}
      }{
        \compliance{\Out\set{x:S_x}_{x\in X}}{\In\set{x:T_x}_{x\in\Message}}
      }
      ~X \ne \emptyset \defrule[c-out-inp]{}
  \end{mathpar}
  }
  \caption{Generalized inference system $\gis{\is[C]}{\cois[C]}$ for fair compliance}
  \label{fig:compliance}
\end{figure}

\Cref{fig:compliance} presents the \GIS $\gis{\is[C]}{\cois[C]}$ for fair
compliance. Intuitively, a derivable judgment $\compliance{S}{T}$ means that the
client $S$ is (fairly) compliant with the server $T$.
Rule \refrule{c-success} relates a satisfied client with a non-failed server.
Rules \refrule{c-inp-out} and \refrule{c-out-inp} require that, no matter which
message is exchanged between client and server, the respective continuations are
still fairly compliant. The side condition $X \ne \emptyset$ guarantees progress
by making sure that the sender is capable of sending at least one message.
As we will see, the coinductive interpretation of $\is[C]$, which consists of
these three rules, completely characterizes compliance
(\Cref{def:comp}).
However, these rules do not guarantee that the interaction between client and
server can always reach a successful configuration as required by
\Cref{def:fcomp}. For this, the corule \refrule{c-sync} is essential.
Indeed, a judgment $\compliance{S}{T}$ that is derivable according to the
generalized interpretation of the \GIS $\gis{\is[C]}{\cois[C]}$ must admit a
well-founded derivation tree also in the inference system $\is[C] \cup
\cois[C]$. Since \refrule{c-success} is the only axiom in this inference system,
finding a well-founded derivation tree in $\is[C] \cup \cois[C]$ boils down to
finding a (finite) path of synchronizations from $\session{S}{T}$ to a
successful configuration in which $S$ has reduced to $\Win$ and $T$ has reduced
to a session type other than $\TNil$.
Rule \refrule{c-sync} allows us to find such a path by choosing the appropriate
messages exchanged between client and server.
In general, one can observe a dicotomy between the rules \refrule{c-inp-out} and
\refrule{c-out-inp} having a universal flavor (they have many premises
corresponding to every possible interaction between client and server) and the
corule \refrule{c-sync} having an existential flavor (it has one premise
corresponding to a particular interaction between client and server). This is
consistent with the fact that we use rules to express a safety property (which
is meant to be \emph{invariant} hence preserved by all the possible
interactions) and we use the corule to help us expressing a liveness property.
This pattern in the usage of rules and corules is quite common in \GISs because
of their interpretation and it can also be observed in the \GIS for fair
termination (see \Cref{fig:ft}) and, to some extent, in that for fair subtyping
as well (see \Cref{sec:agda_fs}).

\begin{example}
  \label{ex:fcomp:gis}
  Consider again $R_2 = \Out\vtrue.(\In0.\Win \branch \In\NatPlus.R_2)$ from
  \Cref{ex:fcomp} and $T_2 = \In\vtrue.\Out\Nat.\T_2 \branch
  \In\vfalse.\End[\In]$ from \Cref{ex:agda_st_ex}.
  In order to show that $\compliance{R_2}{T_2} \in \FlexCo{\is[C]}{\cois[C]}$ we
  have to find a possibly infinite derivation for $\compliance{R_2}{T_2}$ using
  the rules in $\is[C]$ as well as finite derivations for all of the judgments
  occurring in this derivation in $\is[C] \cup \cois[C]$.
  
  For the former we have
  \[
    \begin{prooftree}
      \[
        \[
          \justifies
          \compliance\Win{T_2}
          \using\refrule{c-success}
        \]
        \[
          \smash\vdots\mathstrut
          \justifies
          \compliance{R_2}{T_2}
        \]
        \justifies
        \compliance{
          \In0.\Win + \In\NatPlus.R_2
        }{
          \Out\Nat.T_2
        }
        \using\refrule{c-inp-out}
      \]
      \justifies
      \compliance{R_2}{T_2}
      \using\refrule{c-out-inp}
    \end{prooftree}
  \]
  where, in the application of \refrule{c-inp-out}, we have collapsed all of the
  premises corresponding to the $\NatPlus$ messages into a single premise. Thus,
  we have proved $\compliance{R_2}{T_2} \in \CoInductive{\is[C]}$.
  Note the three judgments occurring in the above derivation tree. The finite
  derivation
  \[
    \begin{prooftree}
      \[
        \[
          \justifies
          \compliance\Win{T_2}
          \using\refrule{c-success}
        \]
        \justifies
        \compliance{
          \In0.\Win + \In\NatPlus.R_2
        }{
          \Out\Nat.T_2
        }
        \using\refrule{c-sync}
      \]
      \justifies
      \compliance{R_2}{T_2}
      \using\refrule{c-sync}
    \end{prooftree}
  \]
  shows that $\compliance{R_2}{T_2} \in \Inductive{\is[C] \cup \cois[C]}$. We
  conclude $\compliance{R_2}{T_2} \in \FlexCo{\is[C]}{\cois[C]}$.
  \eoe
\end{example}

Observe that the corule \refrule{c-sync} is at once essential and unsound. For
example in \Cref{ex:fcomp:gis}, without it we would be able to derive the judgment
$\compliance{\R_2}{\S_2}$ despite the fact that $R_2$ is not fair compliant with
$S_2$ (see \Cref{ex:fcomp}). At the same time, if we treated
\refrule{c-sync} as a plain rule, we would be able to derive the judgment
$\compliance{\Out\Nat.\Win}{\In0.\End[\In]}$ despite the reduction
$\session{\Out\Nat.\Win}{\In0.\End[\In]} \red \session\Win\TNil$ since \emph{there
exists} an interaction that leads to the successful configuration
$\session\Win{\End[\In]}$ (if the client sends $0$) but none of the others does.

\begin{theorem}[Compliance]
  \label{thm:compliance}
  For every $R, T \in \SessionType$, the following properties hold:
  \begin{enumerate}
  \item $R$ is compliant with $T$ if and only if $\compliance\R\T \in
    \CoInductive{\is[C]}$;
  \item $R$ is fairly compliant with $T$ if and only if
    $\compliance\R\T \in \FlexCo{\is[C]}{\cois[C]}$.
  \end{enumerate}
\end{theorem}
\begin{proof}[Proof sketch]
  We sketch the proof of item~(2), which is the most interesting one. In
  \Cref{sec:agda_fc_corr} we will describe the full proof formalized in Agda.
  For the ``if'' part, suppose that $\compliance\R\T \in
  \FlexCo{\is[C]}{\cois[C]}$ and consider a reduction $\session\R\T \wred
  \session{R'}{T'}$. An induction on the length of this reduction, along with
  \refrule{c-inp-out} and \refrule{c-out-inp}, allows us to deduce
  $\compliance{R'}{T'} \in \FlexCo{\is[C]}{\cois[C]}$. Then we have
  $\compliance{R'}{T'} \in \Inductive{\is[C] \cup \cois[C]}$ by
  Definition~\ref{def:gis}. An induction on this (well-founded) derivation
  allows us to find a reduction $\session{R'}{T'} \wred \session\Win{T''}$ such
  that $T'' \ne \TNil$.

  For the left to right implication part we apply the bounded coinduction principle
  (\Cref{prop:bcp}). Concerning consistency, we show that whenever
  $R$ is fairly compliant with $T$ we have that $\compliance{R}{T}$ is the
  conclusion of a rule in \Cref{fig:compliance} whose premises are pairs of
  fairly compliant session types. Indeed, from the hypothesis that $R$ is fairly
  compliant with $T$ we deduce that there exists a derivation $\session{R}{T}
  \wred \session\Win{T'}$ for some $T'\ne\TNil$. A case analysis on the shape of
  $R$ and $T$ allows us to deduce that either $R = \Win$ and $T = T' \ne \TNil$,
  in which case the axiom \refrule{c-success} applies, or that $R$ and $T$ must be
  input/output session types with opposite polarities such that the sender has
  at least one non-$\TNil$ continuation and whose reducts are still fairly
  compliant (because fair compliance is preserved by reductions). Then either
  \refrule{c-inp-out} or \refrule{c-out-inp} applies.
  Concerning boundedness, we do an induction on the reduction $\session{R}{T}
  \wred \session\Win{T'}$ to build a well-founded tree made of a suitable number
  of applications of \refrule{c-sync} topped by a single application of
  \refrule{c-success}.
\end{proof}

%% file: agda-prop/fc-formalization.tex
\beginbass
We now use the Agda library for \GIS (see \Cref{ch:gis_lib}) to 
formally define the inference system for fair compliance shown in \Cref{fig:compliance}. 
As we did in \Cref{sec:agda_ft}, the first thing to do is to define 
the universe \lstinline{U} of judgments that we want to 
derive with the inference system. We can equivalently think of  fair compliance as of a binary 
relation on session types or as a predicate over sessions. We take the second point of view, 
as it allows us to write more compact code later on.

\begin{lstlisting}
U : Set
U = Session
\end{lstlisting}

Next, we define two data types to represent the \emph{unique names} with which we 
identify the rules and corules of the \GIS. We use the same labels of \Cref{fig:compliance}
except for the corule \refrule{c-sync} which we split into \emph{two} 
symmetric corules to avoid reasoning on opposite polarities.

\begin{lstlisting}
data RuleNames : Set where
  success inp-out out-inp : RuleNames
  
data CoRuleNames : Set where
  inp-out out-inp : CoRuleNames
\end{lstlisting}

Again, we recall that there are two different ways of defining rules and corules, depending on whether 
these have a finite or a possibly infinite number of premises. 
Clearly, (co)rules with finitely many premises are just a special case of 
those with possibly infinite ones, but the \GIS library provides some syntactic 
sugar to specify (co)rules of the former kind in a slightly easier way (see \Cref{ch:gis_lib}). 
We use a finite rule to specify \refrule{c-success}.

\begin{lstlisting}
success-rule : FinMetaRule U
success-rule .Ctx = $\Sigma$[ Se $\in$ Session ] Success Se
success-rule .comp (Se , _) = [] , Se
\end{lstlisting}

Concerning \refrule{c-out-inp} and \refrule{c-inp-out}, these rules have a possibly 
infinite set of premises if $\Message$ is infinite. Therefore, we specify the rules 
using the most general form allowed by the \GIS Agda library. 

\begin{lstlisting}
out-inp-rule : MetaRule U
out-inp-rule .Ctx = 
	$\Sigma$[ (f , _) $\in$ Continuation $\times$ Continuation ] Witness f
out-inp-rule .Pos ((f , _) , _) = $\Sigma$[ x $\in$ $\Message$ ] x $\in$ dom f
out-inp-rule .prems ((f , g) , _) = 
	$\lambda$ (x , _) $\rightarrow$ f x .force , g x .force
out-inp-rule .conclu ((f , g) , _) = out f , inp g

inp-out-rule : MetaRule U
inp-out-rule .Ctx = 
	$\Sigma$[ (_ , g) $\in$ Continuation $\times$ Continuation ] Witness g
inp-out-rule .Pos ((_ , g) , _) = $\Sigma$[ x $\in$ $\Message$ ] x $\in$ dom g
inp-out-rule .prems ((f , g) , _) = 
	$\lambda$ (x , _) $\rightarrow$ f x .force , g x .force
inp-out-rule .conclu ((f , g) , _) = inp f , out g
\end{lstlisting}

In the above rules, the \lstinline{Pos} field, which is the
\emph{domain} of the function that \emph{generates} the premises,
coincides with that of the continuation
function corresponding to the output session type, since we want to specify a
fair compliance premise for every message that can be sent.
The specification of corules is no different from that of plain rules. 
As we have anticipated, we split \refrule{c-sync} into two corules, each having exactly one premise.

\begin{lstlisting}
out-inp-corule : FinMetaRule U
out-inp-corule .Ctx = 
	$\Sigma$[ (f , _) $\in$ Continuation $\times$ Continuation ] Witness f
out-inp-corule .comp ((f , g) , x , _) = 
	(f x .force , g x .force) :: [] , (out f , inp g)

inp-out-corule : FinMetaRule U
inp-out-corule .Ctx = 
	$\Sigma$[ (_ , g) $\in$ Continuation $\times$ Continuation ] Witness g
inp-out-corule .comp ((f , g) , x , _) = 
	(f x .force , g x .force) :: [] , (inp f , out g)
\end{lstlisting}

We can now define two inference systems, \lstinline{FCompIS} that consists of the plain rules only 
and \lstinline{FCompCOIS} that consists of the corules only. 
These are called $\is[C]$ and $\cois[C]$ in \Cref{ssec:agda_fc_def}.

\begin{lstlisting}
FCompIS : IS U
FCompIS .Names = RuleNames
FCompIS .rules success = from success-rule
FCompIS .rules out-inp = out-inp-rule
FCompIS .rules inp-out = inp-out-rule

FCompCOIS : IS U
FCompCOIS .Names = CoRuleNames
FCompCOIS .rules out-inp = from out-inp-corule
FCompCOIS .rules inp-out = from inp-out-corule
\end{lstlisting}

where the Agda function \lstinline{from} converts a finite rule into its more general
form on-the-fly, so that the internal representation of all rules is uniform.

We obtain the generalized interpretation of $\gis{\is[C]}{\cois[C]}$, 
named \lstinline{FCompG}, through the library function \lstinline{Gen}.
We also define a predicate \lstinline{FCompI} as the inductive interpretation 
of the union of \lstinline{FCompIS} and \lstinline{FCompCOIS}, 
which is useful in the soundness and boundedness proofs of the \GIS.

\begin{lstlisting}
FCompG : Session $\rightarrow$ Set
FCompG = Gen$\llbracket$ FCompIS , FCompCOIS $\rrbracket$

FCompI : Session $\rightarrow$ Set
FCompI = Ind$\llbracket$ FCompIS $\cup$ FCompCOIS $\rrbracket$
\end{lstlisting}

The relation $\compliance{}{}$ defined by the \GIS in \Cref{fig:compliance} 
is now just a curried version of \lstinline{FCompG}.

\begin{lstlisting}
_$\compliance{}{}$_ : SessionType $\rightarrow$ SessionType $\rightarrow$ Set
R $\compliance{}{}$ S = FCompG (R , S)
\end{lstlisting}

We conclude this section without showing correctness results as they will
be detailed separately in \Cref{sec:agda_fc_corr}.

%% file: agda-prop/fs.tex
\beginalto
The notions of compliance given in \Cref{sec:agda_fc} induce
corresponding semantic notions of subtyping that can be used to define a safe
substitution principle for session types \citep{LiskovWing94}. Intuitively, $\S$
is a subtype of $\T$ if any client that successfully interacts with $\S$ does so
with $\T$ as well. 
The reader may look at \Cref{ch:fs} for more details about \emph{fair subtyping}.


\subsection{Definition}
\label{ssec:agda_fs_definition}
\input{agda-prop/fs-definition}


\subsection{Agda Formalization}
\label{ssec:agda_fs_formalization}
\input{agda-prop/fs-formalization}


\subsection{On the Corule}
\label{ssec:agda_fs_conv}
\input{agda-prop/fs-conv}

%% file: agda-prop/fs-definition.tex
\beginbass
As we noted at the beginning, the aim of this chapter is to show how to obtain
a refined, liveness enforcing, property by adding corules to the inference systems
that alone characterize well known safety properties.
Hence, for what concerns subtyping, we refer to the \GIS that we presented in \Cref{ssec:fsub_gis}.
However, we have to take into account that the compliance relation the we illustrated in
\Cref{sec:agda_fc} is \emph{asymmetric}. Thus, we first recall the semantic
definitions of (un)fair subtyping and then we show the asymmetric variant of
the \GIS in \Cref{fig:fsub_gis}. 

\begin{definition}[Subtyping]
  \label{def:sub}
  We say that $S$ is a \emph{subtype} of $T$ if $R$ compliant with
  $S$ implies $R$ compliant with $T$ for every $R$.
\end{definition}

\begin{definition}[Fair Subtyping]
  \label{def:fsub}
  We say that $S$ is a \emph{fair subtype} of $T$ if $R$ fairly compliant with
  $S$ implies $R$ fairly compliant with $T$ for every $R$.
\end{definition}

\begin{figure}[t]
  \framebox[\textwidth]{
    \begin{mathpar}
      \inferrule[s-nil]{\mathstrut}{
        \TNil\subt\T
      } \defrule[s-nil]{}
      \and
      \inferrule[s-end]{\mathstrut}{
        {\End[\Pol]}\subt\T
      }
      ~T \neq \TNil \defrule[s-end]{}
      \\
      \inferrule[s-inp]{
        \forall x \in X : {S_x}\subt{T_x}
      }{
        {\In\set{x:S_x}_{x\in X}}\subt{\In\set{x:T_x}_{x\in X \cup Y}}
      } \defrule[s-inp]{}
      \and
      \inferrule[s-out]{
        \forall x \in X : {S_x}\subt{T_x}
      }{
        {\Out\set{x:S_x}_{x\in X \cup Y}}\subt{\Out\set{x:T_x}_{x\in X}}
      } \defrule[s-out]{}
      \\
      \infercorule{
        \forall\actionsA\in\traces\S\setminus\traces\T:
        \exists\actionsB \leq \actionsA, x \in \Message:
        {S(\actionsB\Out\x)}\subt{T(\actionsB\Out\x)}
      }{
        \S\subt\T
      }
    \end{mathpar}
  }
  \caption{Generalized inference system $\gis{\is[F]}{\cois[F]}$ for fair subtyping}
  \label{fig:subt}
\end{figure}

The \GIS in \Cref{fig:subt} corresponds to that in \Cref{fig:fsub_gis} where the 
two axioms \refrule{s-nil} and \refrule{s-end} model the asymmetry.
The corule is defined exactly as \refrule{fs-converge} in \Cref{fig:fsub_gis}.
The explanation of the \emph{convergence} is given in \Cref{ssec:fsub_gis}.

\begin{example}
  \label{ex:subt}
  Consider once again the session types $T_i$ and $S_i$ from
  \Cref{ex:agda_st_ex}.
  The (infinite) derivation
  \[
    \begin{prooftree}
      \[
        \[
          \vdots
          \justifies
          {T_1}\subt{S_1}
        \]
        \justifies
        {\Out\Nat.T_1}\subt{\Out\NatPlus.S_1}
        \using\refrule{s-out}
      \]
      \quad
      \[
        \justifies
        \End[\In]\subt\End[\In]
        \using\refrule{s-end}
      \]
      \justifies
      {T_1}\subt{S_1}
      \using\refrule{s-out}
    \end{prooftree}
  \]
  proves that ${T_1}\subt{S_1} \in \CoInductive{\is[F]}$ and the (infinite) derivation
  \[
    \begin{prooftree}
      \[
        \[
          \vdots
          \justifies
          {T_2}\subt{S_2}
        \]
        \justifies
        {\Out\Nat.T_2}\subt{\Out\NatPlus.S_2}
        \using\refrule{s-out}
      \]
      \quad
      \[
        \justifies
        \End[\In]\subt\End[\In]
        \using\refrule{s-end}
      \]
      \justifies
      {T_2}\subt{S_2}
      \using\refrule{s-inp}
    \end{prooftree}
  \]
  proves that ${T_2}\subt{S_2} \in \CoInductive{\is[F]}$.
  In order to derive ${\T_i}\subt{\S_i}$ in the \GIS $\gis{\is[F]}{\cois[F]}$ we
  must find a well-founded proof tree in $\is[F] \cup \cois[F]$
  and the only hope to do so is by means of \refrule{fs-converge}, since $T_i$
  and $S_i$ share traces of arbitrary length.
  Observe that every trace $\actionsA$ of $T_1$ that is not a trace
  of $S_1$ has the form $(\Out\vtrue\Out{p_k})^k\Out\vtrue\Out0\dots$
  where $p_k \in \NatPlus$. Thus, it suffices to take
  $\actionsB = \es$ and $x = 0$, noted that
  $T_1(\Out0) = S_1(\Out0) = \End$, to derive
  \[
    \begin{prooftree}
      \[
        \justifies
        \End[\In]\subt\End[\In]
        \using\refrule{fs-converge}
      \]
      \justifies
      {T_1}\subt{S_1}
      \using\refrule{fs-converge}
    \end{prooftree}
  \]
  On the other hand, traces $\actionsA \in \traces{T_2} \setminus
  \traces{S_2} = (\In\vtrue\Out{p_k})^k\In\vtrue\Out0\dots$ where
  $p_k \in \NatPlus$.  All the prefixes of such traces that are followed by an
  output and are shared by both $T_2$ and $S_2$ have the form
  $(\In\vtrue\Out{p_k})^k\In\vtrue$ where $p_k \in \NatPlus$, and
  $T_2(\actionsB\Out{p}) = T_2$ and $S_2(\actionsB\Out{p}) = S_2$ for all such
  prefixes and $p \in \NatPlus$. It follows that we are unable to derive
  ${T_2}\subt{S_2}$ with a well-founded proof tree in $\is[F] \cup \cois[F]$.
  This is consistent with the fact that, in \Cref{ex:fcomp}, we have
  found a client $R_2$ that is fairly compliant with $T_2$ but not with $S_2$.
  Intuitively, $R_2$ insists on poking the server waiting to receive $0$. This
  may happen with $T_2$, but not with $S_2$.
  In the case of $T_1$ and $S_1$ no such client can exist, since the server may
  decide to interrupt the interaction at any time by sending a $\vfalse$ message
  to the client.
  \eoe
\end{example}

\begin{remark}
  Part of the reason why rule \refrule{fs-converge} is so contrived and hard to
  understand is that the property it enforces is fundamentally \emph{non-local}
  and therefore difficult to express in terms of immediate subtrees of a session
  type as we saw for the purely coinductive formulation of fair subtyping (see \Cref{sec:fair_sub}).
  To better illustrate the point, consider the following alternative set of
  corules meant to replace \refrule{fs-converge}:
  
  \begin{mathpar}
    \infercorule[co-inc]{
      \mathstrut
    }{
      \S\subt\T
    } \defrule[co-inc]{}
    ~ \traces\S \subseteq \traces\T
    \and
    \infercorule[co-inp]{
      \forall x \in \Message {S_x}\subt{T_x}
    }{
      {\In\set{x:S_x}_{x\in\Message}}\subt{\In\set{x:T_x}_{x\in\Message}}
    } \defrule[co-inp]{}
    \and
    \infercorule[co-out]{
      {S}\subt{T}
    }{
      {\Out x.S + S'}\subt{\Out x.T + T'}
    } \defrule[co-out]{}
  \end{mathpar}

  It is easy to see that these rules provide a sound approximation of
  \refrule{s-converge}, but they are not complete. Indeed, consider the session
  types $S = \In\vtrue.S \branch \In\vfalse.(\Out\vtrue.\End[\In] +
  \Out\vfalse.\End[\In])$ and $T = \In\vtrue.T \branch \In\vfalse.\Out\vtrue.\End[\In]$.
  We have $\S\subt\T$ and yet $\S\isubt\T$ cannot be proved with the above corules: 
  it is not possible to prove $\S\isubt\T$ using \refrule{co-inc} because $\traces\S \not\subseteq \traces\T$. 
  If, on the other hand, we insist on visiting both branches of the topmost input as required by \refrule{co-inp}, 
  we end up requiring a proof of $\S\isubt\T$ in order to derive $\S\isubt\T$.
  \eor
\end{remark}

\begin{theorem}
  \label{thm:sub}
  For every $S,T \in \SessionType$ the following properties hold:
  \begin{enumerate}
  \item $S$ is a subtype of $T$ if and only if
    $\S\subt\T \in \CoInductive{\is[F]}$;
  \item $S$ is a fair subtype of $T$ if and only if
    $\S\subt\T \in \FlexCo{\is[F]}{\cois[F]}$.
  \end{enumerate}
\end{theorem}
\begin{proof}[Proof sketch]
  As usual we focus on item~(2), which is the most interesting property.
	We have already presented the correctness proofs for the purely coinductive
	formulation of fair subtyping in \Cref{ssec:fsub_sound,ssec:fsub_complete}. 
	So we just sketch the proofs for the \GIS one.
  For the ``if'' part, we consider an arbitrary $R$ that fairly complies with
  $S$ and show that it fairly complies with $T$ as well. More specifically, we
  consider a reduction $\session{R}{T} \wred \session{R'}{T'}$ and show that it
  can be extended so as to achieve client satisfaction.
  The first step is to ``unzip'' this reduction into $R \wlred{\co\actions} R'$
  and $T \wlred\actions T'$ for some string $\actions$ of actions. Then, we show
  by induction on $\actions$ that there exists $S'$ such that ${S'}\subt{T'} \in
  \FlexCo{\is[F]}{\cois[F]}$ and $S \wlred\actions S'$, using the hypothesis
  $\S\subt\T \in \CoInductive{\is[F]}$ and the hypothesis that $R$ complies
  with $S$. This means that $R$ and $S$ may synchronize just like $R$ and $T$,
  obtaining a reduction $\session{R}{S} \wred \session{R'}{S'}$. At this point
  the existence of the reduction $\session{R'}{T'} \wred \session\Win{T''}$ is
  proved using the arguments in the discussion of rule \refrule{fs-converge}
  given earlier.

  For the ``only if'' part we use once again the bounded coinduction principle.
  In particular, we use the hypothesis that $S$ is a fair subtype of $T$ to show
  that $S$ and $T$ must have one of the forms in the conclusions of the rules in
  \Cref{fig:subt}. This proof is done by cases on the shape of $S$,
  constructing a canonical client of $S$ that must succeed with $T$ as well.
  Then, the coinduction principle allows us to conclude that $\S\subt\T \in
  \CoInductive{\is[F]}$.
  The fact that $\S\subt\T \in \Inductive{\is[F] \cup \cois[F]}$ also holds is
  by far the most intricate step of the proof. First of all, we establish that
  $\Inductive{\is[F] \cup \cois[F]} = \Inductive{\cois[F]}$. That is, we
  establish that rule \refrule{fs-converge} subsumes all the rules in $\is[F]$
  when they are inductively interpreted. Then, we provide a characterization of
  the \emph{negation} of \refrule{fs-converge}, which we call divergence. At this
  point we proceed by contradiction: under the hypothesis that $\S\subt\T \in
  \CoInductive{\is[F]}$ and that $S$ and $T$ ``diverge'', we are able to
  corecursively define a \emph{discriminating} (see \Cref{def:discriminator}) client $R$ that fairly complies
  with $S$ but not with $T$. This contradicts the hypothesis that $S$ is a fair
  subtype of $T$ and proves, albeit in a non-constructive way, that $\S\subt\T
  \in \Inductive{\cois[F]}$ as requested.
\end{proof}

\begin{remark}
  \label{rem:duality}
  Most session type theories adopt a symmetric form of session type
  compatibility whereby client and server are required to terminate
  the interaction at the same time.
  It is easy to define a notion of \emph{symmetric compliance} 
  by turning $T' \ne \TNil$ into $T' = \End[\In]$ in \Cref{def:comp}. 
  The subtyping relation induced
  by symmetric compliance has essentially the same characterization of
  \Cref{def:sub}, except that the axiom \refrule{s-end} is replaced by
  the more familiar \refrule{fs-end} \citep{GayHole05}.
  On the other hand, the analogous change in \Cref{def:fcomp} has much
  deeper consequences: the requirement that client and server must end the
  interaction at the same time induces a large family of session types that are
  syntactically very different, but semantically equivalent. For example, the
  session types $S$ and $T$ such that $S = \In\Nat.S$ and $T = \Out\Bool.T$,
  which describe completely unrelated protocols, would be equivalent for the
  simple reason that no client successfully interacts with them (they are not
  fairly terminating, since they do not contain any occurrence of $\End[]$).
  \eor
\end{remark}

%% file: agda-prop/fs-formalization.tex
\beginbass
Now we show the mechanization of the \GIS in \Cref{fig:subt}.
We still start defining the right universe and then we move to 
the definition of the (co)rules. At last, we show the key
lemmas that are needed to prove the correctness of the predicate.
The universe is clearly made of pairs of session types.

\begin{lstlisting}
U : Set
U = SessionType $\times$ SessionType
\end{lstlisting}

Then we have to define names of the rules and the corules.
For the sake of clarity, we try to be consistent with the names in \Cref{fig:subt}.

\begin{lstlisting}
data FSubIS-RN : Set where
  s-nil s-end : FSubIS-RN
  s-inp s-out : FSubIS-RN

data FSubCOIS-RN : Set where
  converge : FSubCOIS-RN
\end{lstlisting}

where the corules involve only \refrule{fs-converge}.
We start looking at the definitions of the axioms. 
In this case we can use the finitary formulation by using \lstinline{FinMetaRule}.

\begin{lstlisting}
s-nil-r : FinMetaRule U
s-nil-r .Ctx = SessionType
s-nil-r .comp T =
  [] ,
  ------------------
  (nil , T)

s-end-r : FinMetaRule U
s-end-r .Ctx = 
  $\Sigma$[ (S , T) $\in$ SessionType $\times$ SessionType ] End S $\times$ Defined T
s-end-r .comp ((S , T) , _) =
  [] ,
  ------------------
  (S , T)
\end{lstlisting}

Note that the definitions are consistent with \refrule{s-nil} and \refrule{s-end}-
Indeed \lstinline{s-end-r} requires that the first session type has the form $\End[\Pol]$
and that the second is different from $\TNil$ by using the predicates \lstinline{End} and
\lstinline{Defined}, respectively (see \Cref{sec:agda_st} for more details).
Next, we show the rules for modeling covariance of input and contravariance of output.

\begin{lstlisting}
s-inp-r : MetaRule U
s-inp-r .Ctx = 
  $\Sigma$[ (f , g) $\in$ Continuation $\times$ Continuation ] dom f $\subseteq$ dom g
s-inp-r .Pos ((f , _) , _) = $\Sigma$[ t $\in$ $\Message$ ] t $\in$ dom f
s-inp-r .prems ((f , g) , _) (t , _) = f t .force , g t .force
s-inp-r .conclu ((f , g) , _) = inp f , inp g

s-out-r : MetaRule U
s-out-r .Ctx = $\Sigma$[ (f , g) $\in$ Continuation $\times$ Continuation ] 
		  dom g $\subseteq$ dom f $\times$ Witness g
s-out-r .Pos ((_ , g) , _) = $\Sigma$[ t $\in$ $\Message$ ] t $\in$ dom g
s-out-r .prems ((f , g) , _) (t , _) = f t .force , g t .force
s-out-r .conclu ((f , g) , _) = out f , out g
\end{lstlisting}

We can point out a couple of things. First, both rules have a premise for each
element in the domain of the first/second \lstinline{Continuation}. Then, 
co/contra-variance is encoded through the domain inclusion as side condition.
We recall that \lstinline{Witness g} means that the domain of \lstinline{g} is
not empty.
Now we can move to the \refrule{fs-converge} corule.

\begin{lstlisting}
converge-r : FinMetaRule U
converge-r .Ctx = 
  $\Sigma$[ (S , T) $\in$ SessionType $\times$ SessionType ] S $\downarrow$ T
converge-r .comp ((S , T) , _) =
  [] ,
  ------------------
  (S , T)
\end{lstlisting}

where \lstinline{S $\downarrow$ T} is the predicate obtained by inductively encoding
the corule \refrule{fs-converge} alone.
Although this approach might seem inconsistent with the other mechanizations
for it does not use the actual corule, it can be proved that it is equivalent in this
example. We will give more details in \Cref{ssec:agda_fs_conv}.

For the sake of simplicity, we omit its definition.
As usual, we proceed by defining the inference systems.

\begin{lstlisting}
FSubIS : IS U
FSubIS .Names = FSubIS-RN
FSubIS .rules s-nil = from s-nil-r
FSubIS .rules s-end = from s-end-r
FSubIS .rules s-inp = s-inp-r
FSubIS .rules s-out = s-out-r

FSubCOIS : IS U
FSubCOIS .Names = FSubCOIS-RN
FSubCOIS .rules converge = from converge-r
\end{lstlisting}

At last, we can encode the desired predicate by the generalized interpretation.
We also consider the inductive interpretation of the whole inference system which
will be used later.

\begin{lstlisting}
_$\subt$F_ : SessionType $\rightarrow$ SessionType $\rightarrow$ Set
S $\subt$F T = FCoInd$\llbracket$ FSubIS , FSubCOIS $\rrbracket$ (S , T)

_$\subt$F$_i$_ : SessionType $\rightarrow$ SessionType $\rightarrow$ Set
S $\subt$F$_i$ T = Ind$\llbracket$ FSubIS $\cup$ FSubCOIS $\rrbracket$ (S , T)
\end{lstlisting}

Concerning the specification against which we prove the correctness of \lstinline{_$\subt$F_}
it is important to highlight that it can be defined both by using \emph{ground} notions
and by \lstinline{_$\compliance{}{}$_} (fair compliance) defined in \Cref{ssec:agda_fc_form}. In the mechanization we proved that they 
are equivalent and we used them differently according to our needs. Here,
show only that using the compliance predicate.

\begin{lstlisting}
FSSpec : U $\rightarrow$ Set
FSSpec (S , T) = $\forall${R} $\rightarrow$ R $\compliance{}{}$ S $\rightarrow$ R $\compliance{}{}$ T
\end{lstlisting}

To see the advantage of relying on such specification based on \lstinline{_$\compliance{}{}$_},
we just need to think that the conclusion \lstinline{R $\compliance{}{}$ T} is defined using a
\GIS. Hence, for what concerns the soundness, we can formulate an ad-hoc specification and 
use the bounded coinduction principle (\Cref{prop:bcp}). We omit some details.

\begin{lstlisting}
SpecAux : U $\rightarrow$ Set
SpecAux (R , T) = $\Sigma$[ S $\in$ SessionType ] S $\subt$F T $\times$ R $\compliance{}{}$ S 

spec-aux-sound : SpecAux $\subseteq$ $\lambda$ (R , S) $\rightarrow$ R $\compliance{}{}$ S
spec-aux-sound = 
  bounded-coind[ FCompIS , FCompCOIS ] 
    SpecAux spec-bounded spec-cons

sound : $\forall${S T} $\rightarrow$ S $\subt$F T $\rightarrow$ FSSpec (S , T)
sound {S} fs fc = spec-aux-sound (S , fs , fc)
\end{lstlisting}

where \lstinline{spec-bounded} and \lstinline{spec-cons} are boundedness and consistency proofs
of \lstinline{SpecAux} with respect to the \GIS in \Cref{fig:compliance}.
For what concerns the completeness, the proof is clearly by bounded coinduction (\Cref{prop:bcp}).

\begin{lstlisting}
bounded : $\forall${S T} $\rightarrow$ FSSpec (S , T) $\rightarrow$ S $\subt$F$_i$ T

consistent : FSSpec $\subseteq$ ISF[ FSubIS ] FSSpec

complete : $\forall${S T} $\rightarrow$ FSSpec (S , T) $\rightarrow$ S $\subt$F T
complete = 
  bounded-coind[ FSubIS , FSubCOIS ] FSSpec bounded consistent
\end{lstlisting}

%% file: agda-prop/fs-conv.tex
\beginbass
We dedicate this last part of the section answering a couple of questions
about the corule \refrule{fs-converge} and its mechanization.
First, the reader might be confused about whether \refrule{fs-converge} is
actually a (meta)rule for it contains an existential quantifier.
Hence, \refrule{fs-converge} is not a proper metarule written in that form.
Indeed, in the premise both a universal and an existential quantifier are mixed.
The issue can be solved by turning such premise in \emph{Skolem normal form}, that is,
without existential quantifiers.
The drawback of this approach is that quantifiers are replaced by functions making
the defined predicate hard to understand.
Although for the sake of clarity we could present \refrule{fs-converge}
in a \emph{wrong format}, in the Agda mechanization we should have followed the
Skolemization approach because the
\lstinline{MetaRule} datatype requires a precise formulation of all the components. 
For the sake of simplicity, we decided to opt for the more convenient approach that
we presented in \Cref{ssec:agda_fs_formalization} and that we comment here.

The second question that we try to answer is whether the approach that we adopted in 
\Cref{ssec:agda_fs_formalization} for defining \refrule{fs-converge}, that is by using
a coaxiom with a side condition, is equivalent to defining the actual corule.
Notably, the side condition \lstinline{S $\downarrow$ T} is obtained by
considering \refrule{fs-converge} as a stand-alone predicate. 
Let us show such rule forgetting about subtyping.
We use the notation with message tags that we adopted in \Cref{sec:st}.

\[
  \inferrule{
    \forall\actionsA\in\traces\S\setminus\traces\T:
    \exists\actionsB\prefix\actionsA, \l:
    S(\actionsB\Out\l) \converge T(\actionsB\Out\l)
  }{
    S \converge T
  }
\]

The Agda predicate \lstinline{_$\downarrow$_} is defined analogously.
Now we can prove that the predicate obtained by inductively interpreting such rule
is equivalent to the inductive interpretation of the whole inference system in \Cref{fig:subt}.

\begin{lemma}
	 $S \isubt T$ if and only if $S \converge T$.
\end{lemma}
\begin{proof}
  The ``if'' part is trivial since the sole rule defining
  $\converge$ is the same as \refrule{fs-converge}.  We prove the
  ``only if'' part by induction on the derivation of $S \isubt T$
  and by cases on the last rule applied.
	
	\proofcase{Case \refrule{s-nil}}
	Then $S = \TNil$ and $T \ne \TNil$. We conclude $S \converge T$
  observing that $\emptyset \setminus \traces\T = \emptyset$.
  
  \proofcase{Case \refrule{s-end}}
  Then $S = \End[\Pol]$ for some $\Pol$ and $T \ne \TNil$. 
  Notably, $\traces{\End[\Pol]} = \emptyset$ by \Cref{def:traces}.
  Hence, we conclude $S \converge T$
  observing that $\emptyset \setminus \traces\T = \emptyset$.
 
  \proofcase{Case \refrule{s-inp}}
  Then $S = \Branch{\l_i : S_i}_{i\in I}$ and
  $T = \Branch{\l_j : T_j}_{j\in J}$ and $I \subseteq J$ and
  $S_i \isubt T_i$ for every $i \in I$.
  Using the induction hypothesis we deduce that $S_i \converge T_i$
  for every $i\in I$.
  We conclude $S \converge T$ observing that
  $\actionsA \in \traces\S \setminus \traces\T$ implies
  $\actionsA = \In\l_k\actionsB$ for some $k\in I$ and
  $\actionsB \in \traces{S_k} \setminus \traces{T_k}$.

  \proofcase{Case \refrule{s-out}}
  Then $S = \Choice{\l_i : S_i}_{i\in I}$ and
  $T = \Choice{\l_j : T_j}_{j\in J}$ and $J \subseteq I$ and
  $S_j \isubt T_j$ for every $j\in J$.
  Using the induction hypothesis we deduce that $S_j \converge T_j$
  for every $j\in J$.
  In order to conclude $S \converge T$ we have to show that, for
  every $\actionsA \in \traces\S \setminus \traces\T$, we are able to
  find $\actionsB \prefix \actionsA$ and $\l \in \Message$ such that
  $S(\actionsB\Out\l) \converge T(\actionsB\Out\l)$.
  We distinguish two sub-cases:
  \begin{itemize}
  \item Sub-case $\actions = \Out\l_j\actions'$ where
    $\actions' \in \traces{S_j} \setminus \traces{T_j}$ for some
    $j\in J$.
    From $S_j \converge T_j$ we deduce that there exist
    $\actionsB' \prefix \actionsA'$ and $\l \in \Message$ such that
    $S_j(\actionsB'\Out\l) \converge T_j(\actionsB'\Out\l)$.
    We conclude by taking $\actionsB \eqdef \Out\l_j\actionsB'$
    observing that $S(\actionsB\Out\l) = S_j(\actionsB'\Out\l)$ and
    $T(\actionsB\Out\l) = T_j(\actionsB'\Out\l)$.
  \item Sub-case $\actions = \Out\l_i\actions'$ where
    $\actions' \in \traces{S_i}$ for some $i\in I \setminus J$.
    We conclude by taking $\actionsB \eqdef \es$ and $\l \eqdef \l_j$
    for some $j\in J$ and observing that $S(\actionsB\Out\l) = S_j$
    and $T(\actionsB\Out\l) = T_j$.
  \end{itemize}

  \proofcase{Case \refrule{fs-converge}}
  Then, for every $\actionsA \in \traces\S \setminus \traces\T$, there
  exist $\actionsB \prefix \actions$ and $\l\in\Message$ such that
  $S(\actionsB\Out\l) \isubt T(\actionsB\Out\l)$.
  We conclude immediately using the induction hypothesis to deduce
  that for each such $\actionsB$ and $\l$ we have
  $S(\actionsB\Out\l) \converge T(\actionsB\Out\l)$.
\end{proof}

In the Agda mechanization it happens that we take the inductive interpretation of the \GIS with
the corule replace by the coaxion.
Hence, in order to prove the equivalence between the two approaches, we need to prove
that the inductive interpretation of the \GIS in Agda is equivalent to the $\converge$ predicate.

\begin{lstlisting}
$\subt$F$_i$$\rightarrow\downarrow$ : $\forall${S T} $\rightarrow$ S $\subt$F$_i$ T $\rightarrow$ S $\downarrow$ T
\end{lstlisting}

We omit the detailed proof (it can be found in \cite{FairSubtypingAgda}). 
The other direction is trivial since we can directly apply the coaxiom providing the convergence proof.

%% file: agda-prop/fc-corr.tex
\beginalto
In this section we detail the mechanized correctness proof of \emph{fair compliance}
(see point 2. of \Cref{thm:compliance}).
As usual, correctness is expressed in terms of \emph{soundness} and \emph{completeness}
of \lstinline{_$\compliance{}{}$_} with respect to some specification.
Such specification is defined in \Cref{def:fcomp}. 
Before looking at such proofs, we formalize \Cref{def:fcomp}.
To formalize fair compliance, we define a \lstinline{Success} predicate that 
characterizes those configurations $\session{R}{S}$ in which the client has succeeded 
($R = \Win$) and the server has not failed ($S \neq \TNil$).

\begin{lstlisting}
data Success : Session $\rightarrow$ Set where
  success : $\forall${R S} $\rightarrow$ Win R $\rightarrow$ Defined S $\rightarrow$ Success (R , S)
\end{lstlisting}

We can weaken \lstinline{Success} to \lstinline{MaySucceed}, 
to characterize those configurations \emph{that can be extended} so as to become successful ones.
For this purpose we make use of the \lstinline{Satisfiable} predicate and of the 
intersection \lstinline{$\cap$} of two sets from Agda's standard library.

\begin{lstlisting}
MaySucceed : Session $\rightarrow$ Set
MaySucceed Se = 
	Relation.Unary.Satisfiable (Reductions Se $\cap$ Success)
\end{lstlisting}

In words, \lstinline{Se} may succeed if there exists \lstinline{Se'}
such that \lstinline{Se $\wred$ Se'} and \lstinline{Se'} is a successful configuration. 
We can now formulate fair compliance as the property of those sessions that may succeed 
no matter how they reduce (see \Cref{def:fcomp}). This is the specification against 
which we prove soundness and completeness of the \GIS for fair compliance (\Cref{fig:compliance}).

\begin{lstlisting}
FCompS : Session $\rightarrow$ Set
FCompS Se = $\forall${Se'} $\rightarrow$ Reductions Se Se' $\rightarrow$ MaySucceed Se'
\end{lstlisting}

\begin{remark}
	As we did in \Cref{sec:agda_ft}, assume that we want to instaciate \lstinline{Specification}
	from \Cref{sec:agda_fairt} by using \lstinline{Session} (see \Cref{ssec:agda_st_formalization_types})
	as set of states and \lstinline{Reduction} (see \Cref{ssec:agda_st_formalization_lts}) as transition system.
	Such instance of \lstinline{Specification} would not be equivalent to \lstinline{FCompS} as fari compliance
	corresponds to a successful form of fair termination.
	In order to make them equivalent we would need to parametrize \lstinline{WeaklyTerminating} on a predicate
	that the reducts have to satisfy. In this case we would say that a state is weakly terminating
	if it reduces to another one that is in normal form and that satisfies the input predicate.
	\eor
\end{remark}


\subsection{Soundness}
\input{agda-prop/fc-corr-soundness}


\subsection{Completeness}
\input{agda-prop/fc-corr-completeness}

%% file: agda-prop/fc-corr-soundness.tex
\beginbass
\GISs provide no canonical way for proving the soundness of the generalized 
interpretation of an inference system, so we have to handcraft the proof. 
We start by proving that the inductive interpretation of the inference system 
with the corules implies the existence of a reduction leading to a successful configuration.

\begin{lstlisting}
FCompI$\rightarrow$MS : $\forall${Se} $\rightarrow$ FCompI Se $\rightarrow$ MaySucceed Se
FCompI$\rightarrow$MS (fold (inj$_1$ success , (_ , succ) , refl , _)) = 
  _ , $\epsilon$ , succ
FCompI$\rightarrow$MS (fold (inj$_1$ out-inp , (_ , _ , fx) , refl , pr)) =
  let _ , reds , succ = FCompI$\rightarrow$MS (pr (_ , fx)) in
  _ , sync (out fx) inp $\triangleleft$ reds , succ
FCompI$\rightarrow$MS (fold (inj$_1$ inp-out , (_ , _ , gx) , refl , pr)) =
  let _ , reds , succ = FCompI$\rightarrow$MS (pr (_ , gx)) in
  _ , sync inp (out gx) $\triangleleft$ reds , succ
FCompI$\rightarrow$MS (fold (inj$_2$ out-inp , (_ , _ , fx) , refl , pr)) =
  let _ , reds , succ = FCompI$\rightarrow$MS (pr Data.Fin.zero) in
  _ , sync (out fx) inp $\triangleleft$ reds , succ
FCompI$\rightarrow$MS (fold (inj$_2$ inp-out , (_ , _ , gx) , refl , pr)) =
  let _ , reds , succ = FCompI$\rightarrow$MS (pr Data.Fin.zero) in
  _ , sync inp (out gx) $\triangleleft$ reds , succ 
\end{lstlisting}

where \lstinline{$\epsilon$} and \lstinline{red $\triangleleft$ reds} 
are the constructors of \lstinline{Star}. While the former represents 
the base case (when there are no reductions), the second
represents a chain of reductions starting with the single reduction 
\lstinline{red} followed by the reductions \lstinline{reds}.
There are two things worth noting here. First, in the union of the two inference 
systems each rule is identified by a name of the form \lstinline{inj$_1$ n} 
or \lstinline{inj$_2$ n} where \lstinline{n} is either the name of a rule 
or of a corule, respectively.
Also, we use the function \lstinline{pr} to access the premises of a 
(co)rule by their position. For the plain rules \lstinline{out-inp} 
and \lstinline{inp-out} the position is the witness \lstinline{fx}
or \lstinline{gx} that the value exchanged in the synchronization belongs to the 
domain of the continuation function of the sender. For the corules 
\lstinline{out-inp} and \lstinline{inp-out}, 
we use the position \lstinline{Data.Fin.Zero} 
to access the first and only premise in their list of premises.

The next auxiliary result establishes a ``subject reduction'' property for 
fair compliance: if $\compliance{R}{S}$ and $\session{R}{S} \wred \session{R'}{S'}$, 
then $\compliance{R'}{S'}$. Note that this property is trivial to prove when we 
consider the specification of fair compliance (see \Cref{def:fcomp}), 
but here we are referring to the predicate defined by the \GIS. 
The proof consists of a simple induction on the reduction $\session{R}{S} \wred \session{R'}{S'}$.

\begin{lstlisting}
sr : $\forall${Se Se'} $\rightarrow$ FCompG Se $\rightarrow$ Reductions Se Se' $\rightarrow$ FCompG Se'
sr fc $\epsilon$ = fc
sr fc (_ $\triangleleft$ _) with fc .CoInd$\llbracket$_$\rrbracket$.unfold
sr _ (sync (out fx) inp $\triangleleft$ _) | 
  success , ((_ , success (out e) _) , _) , refl , _ = 
  $\bot$-elim (e _ fx)
sr _ (sync inp (out gx) $\triangleleft$ reds) | inp-out , _ , refl , pr = 
  sr (pr (_ , gx)) reds
sr _ (sync (out fx) inp $\triangleleft$ reds) | out-inp , _ , refl , pr = 
  sr (pr (_ , fx)) reds
\end{lstlisting}

Note that we have an absurd case, in which a successful configuration apparently reduces,
which we rule out using false elimination (\lstinline{$\bot$-elim}).
The soundness proof is a simple combination of the above auxiliary results. We
use the library function \lstinline{fcoind-to-ind} (see \Cref{fig:fcoind-to-ind})
to extract an inductive derivation of \lstinline{FCompI Se} from a
derivation of \lstinline{FCompG Se} in the \GIS for fair
compliance.

\begin{lstlisting}
sound : $\forall${Se} $\rightarrow$ FCompG Se $\rightarrow$ FCompS Se
sound fc reds = FCompI$\rightarrow$MS (fcoind-to-ind (sr fc reds))
\end{lstlisting}

%% file: agda-prop/fc-corr-completeness.tex
\beginbass
For the completeness result we appeal to the bounded coinduction principle of 
\GISs (\Cref{prop:bcp}, \Cref{fig:principles}), which requires us to prove boundedness and 
consistency of \lstinline{FCompS}. Concerning boundedness, 
we start by computing a proof of \lstinline{FCompI Se}
for every session \lstinline{Se} that may reduce a successful 
configuration, by induction on the reduction.

\begin{lstlisting}
MS$\rightarrow$FCompI : $\forall${Se} $\rightarrow$ MaySucceed Se $\rightarrow$ FCompI Se
MS$\rightarrow$FCompI (_ , reds , succ) = aux reds succ
  where
    aux : $\forall${Se Se'} $\rightarrow$ Reductions Se Se' $\rightarrow$ 
      	Success Se' $\rightarrow$ FCompI Se
    aux $\epsilon$ succ = apply-ind (inj$_1$ success) (_ , succ) $\lambda$ ()
    aux (sync (out fx) inp $\triangleleft$ red) succ =
      apply-ind (inj$_2$ out-inp) 
        (_ , _ , fx) $\lambda${Data.Fin.zero $\rightarrow$ aux red succ}
    aux (sync inp (out gx) $\triangleleft$ red) succ =
      apply-ind (inj$_2$ inp-out) 
        (_ , _ , gx) $\lambda${Data.Fin.zero $\rightarrow$ aux red succ}
\end{lstlisting}

where \lstinline{apply-ind} is the function in the \GIS library that applies 
a rule for an inductively defined predicate.
Then, boundedness follows by observing that $R$ fairly compliant with $S$ 
implies the existence of a successful configuration reachable from $\session{R}{S}$.

\begin{lstlisting}
bounded : $\forall${Se} $\rightarrow$ FCompS Se $\rightarrow$ FCompI Se
bounded fc = MS$\rightarrow$FCompI (fc $\epsilon$)
\end{lstlisting}

Showing that \lstinline{FCompS} is consistent means showing that every 
configuration \lstinline{Se} that satisfies \lstinline{FCompS} is 
found in the conclusion of a rule in the inference system \lstinline{FCompIS}
whose premises are all configurations that in turn satisfy \lstinline{FCompS}. 
This follows by a straightforward case analysis on the \emph{first} reduction 
of \lstinline{Se} that leads to a successful configuration.

\begin{lstlisting}
consistent : $\forall${Se} $\rightarrow$ FCompS Se $\rightarrow$ ISF[ FCompIS ] FCompS Se
consistent fc with fc $\epsilon$
... | _ , $\epsilon$ , succ = success , (_ , succ) , refl , $\lambda$ ()
... | _ , sync (out fx) inp $\triangleleft$ _ , _ =
out-inp , (_ , _ , fx) , refl , 
  $\lambda$ (_ , gx) reds $\rightarrow$ fc (sync (out gx) inp $\triangleleft$ reds)
... | _ , sync inp (out gx) $\triangleleft$ _ , _ =
inp-out , (_ , _ , gx) , refl , 
  $\lambda$ (_ , fx) reds $\rightarrow$ fc (sync inp (out fx) $\triangleleft$ reds)
\end{lstlisting}

We obtain the completeness proof using the bounded coinduction principle, 
\ie the library function
\lstinline{bounded-coind} (see \Cref{fig:principles})
which applies the principle to the boundedness and
consistency proofs.

\begin{lstlisting}
complete : $\forall${Se} $\rightarrow$ FCompS Se $\rightarrow$ FCompG Se
complete = 
  bounded-coind[ FCompIS , FCompCOIS ] FCompS bounded consistent
\end{lstlisting}

%% file: agda-prop/related.tex
\beginalto
We have shown that generalized inference systems are an effective
framework for defining sound and complete proof systems of (some)
combined safety and liveness properties of (dependent) session types
(\Cref{def:wt,def:fcomp}), as well as of a liveness-preserving
subtyping relation (\Cref{def:fsub}).

\paragraph{Properties of Session Types.}
We think that this achievement is more than a coincidence.  One of
the fundamental results in model checking states that every property
can be expressed as the conjunction of a safety property and a
liveness
property \citep{AlpernSchneider85,AlpernScheider87,BaierKatoen08}.
The connections between safety and liveness on one side and
coinduction and induction on the other make \GISs appropriate for
characterizing combined safety and liveness properties.

\cite{Murgia19} studies a wide range of compliance relations
for processes and session types, showing that many of them are fixed
points of a functional operator, but not necessarily the least or
the greatest ones. In particular, he shows that 
\emph{progress compliance}, which is akin to our compliance (\Cref{def:comp}), is
a greatest fixed point and that \emph{should-testing compliance},
which is akin to our fair compliance (\Cref{def:fcomp}), is an
intermediate fixed point. These results are consistent with
\Cref{thm:compliance}.

\paragraph{Dependent Session Types.}
The first theories of dependent session types are those of
\cite{ToninhoCairesPfenning11} and \cite{GriffithGunter13}. These
works augment session types with binders, thus allowing for the
specification of message predicates.
\cite{ToninhoYoshida18} present a full calculus combining functions
and processes in which the structure of both types and processes may
depend on the content of messages. 
\cite{ThiemannVasconcelos20} propose a full model of functions and
processes enabling a simplified form of dependency whereby the
structure of types and processes may depend on labels and possibly
natural numbers. They introduce a conditional context extension
operator that prevents dependencies on linear values.

\cite{Zhou19} describes the theory and implementation of a
refinement session type system where the type of messages can be
refined by predicates that specify their properties and
relationships.

\paragraph{Formalizations of Session Type Systems.}
\cite{Thiemann19} gives the first mechanized proof of a calculus of
functions and sessions. His type system distinguishes between types
and session types, but only non-dependent pairs are considered.
\cite{HughesBradyVanderbauwhede19} describe an Idris EDSL where
dependent types enable reasoning on value dependencies between
exchanged messages.
\cite{ZalakainDardha20} give another Agda formalization of the
linear $\pi$-calculus. They focus exclusively on the process layer
and only consider channel types, using typing with
leftovers \citep{Allais17}.
\cite{RouvoetEtAl20} describe a technique inspired by separation
logic to specify and verify in Agda interpreters using linear
resources. Among the case studies they discuss is a linearly-typed
lambda calculus with primitives for session communications.

%% file: finale.tex
\begintreble
Let us briefly summarize the two main contents of this thesis.
First, we introduced \emph{fair termination} as a desirable property in
systems  based on communication as it entails many well known properties
that are studied in the literature. 
However, providing a type system for enforcing fair termination is not straightforward
as there exist some peculiar scenarios in which the liveness is compromised. 
Then, due to the increasing interest of research communities in proof assistants we
decided to focus on some mechanization aspects. Notably, generalized inference
systems have been a transversal topic throughout the thesis as most of the time we relied on them
for characterizing the definitions that we needed.

\paragraph{Type Systems}
On the one hand, in \Cref{ch:ft_bin,ch:ft_multi} we presented a type system for
binary/multiparty sessions which involved \emph{fair subtyping}. Then we showed
that fair subtyping must be used carefully, that is, finitely many times when it has a strictly
positive weight. The main advantage of the type systems under analysis is that 
they allow for compositional reasoning.
On the other hand, in \Cref{ch:ft_ll} we presented a linear logic inspired type system for
the linear $\pi$-calculus. In this case there is no subtyping relation. A peculiar 
aspect was that we could type both a fair and an unfair process (see \Cref{ex:slot_comp})
by using the same type/formula. Differently with respect to the type systems in
\Cref{ch:ft_bin,ch:ft_multi}, the enforcement of fair termination 
is based on a global \emph{validity condition}.
At last, we showed that the type systems cannot be compared which leads to interesting
research directions.
Apart from the technique being used for developing the type systems, we hope
that the reader has been convinced that fair termination is a fundamental
property for communication-based calculi.

\paragraph{Agda Mechanizations}
In \Cref{ch:gis_lib} we introduce a library for supporting
generalized inference systems in Agda. Differently, to the built-in techniques, 
the library allows to provide \emph{modular}
definitions that can be composed without duplicating notions.
We then took advantage of such library for characterizing
the properties of session types in \Cref{ch:agda_prop}. 
The main aim was to show that well known properties
could be turned into \emph{fair} ones by adding corules and without changing
the main inference systems (see \Cref{sec:agda_fs}). 
The application of foundational aspects to session types tried to provide
a useful guideline for those researchers involved in the formalization of
session type based calculi. Indeed, we also relied on a peculiar definition
of session types that aims at simplifying the definition of the predicates on them.
We hope to have provided some helpful suggestions for realizing the dream according
to which one day all the proofs will be certified.

\section*{Future Work}

We conclude the thesis by inspecting some possible research directions that we can 
take in the future. We are grateful to all the anonymous reviewers of the articles
that we have submitted during these years as they suggested many interesting
topics that we mention in this section.

\paragraph{Fair Termination of Sessions}
An open question that may have a relevant practical impact is whether the
type system in \Cref{ch:ft_bin,ch:ft_multi} remain \emph{sound} 
in a setting where communications are
\emph{asynchronous}. We expect the answer to be positive, as is the case for
other synchronous multiparty session types systems \citep{ScalasYoshida19}, but
we have not worked out the details yet. 
Then, a natural development of these type systems is their application to a
real programming environment. We envision two approaches that can be followed to
this aim. A bottom-up approach may apply our static analysis technique to a
program (in our process calculus) that is extracted from actual code and that
captures the code's communication semantics. We expect that suitable annotations
may be necessary to identify those branching parts of the code that represent
non-deterministic choices in the program. Most typically, these branches will
correspond to finite loops or to queries made to the human user of the program
that have several different continuations. A top-down approach may provide
programmers with a \emph{generative tool} that, starting from a global
specification in the form of a \emph{global type} \citep{HondaYoshidaCarbone16},
produces template code that is ``well-typed by design'' and that the programmer
subsequently instantiates to a specific application.
Scribble \citep{YoshidaHuNeykovaNg13,AnconaEtAl16} is an example of such a tool.
Interestingly, the usual notion of global type projectability is not sufficient
to entail that the session map resulting from a projection is coherent. However,
coherence would be guaranteed by requiring that the projected global type is
fairly terminating.
Finally, we plan to investigate the adaptation of the type system for ensuring
the fair termination in the popular actor-based model. This is a drastically
different setting in which the order of messages is not as controllable as in
the case of sessions. As a consequence, type based analyses require radically
different formalisms such as mailbox types \citep{deLiguoroP18}, for which the
study of fair subtyping and of type systems enforcing fair termination is
unexplored. 

\paragraph{Validity in \piLIN}
One drawback of the type system proposed in \Cref{ch:ft_ll} is that establishing whether a
quasi-typed process is also well typed requires a global check on the whole
typing derivation. This does not happen in the type systems for sessions
in \Cref{ch:ft_bin,ch:ft_multi}.
The need for a global check seems to arise commonly in infinitary proof
systems \citep{DaxHofmannLange06,Doumane17,BaeldeDoumaneSaurin16,BaeldeEtAl22},
so an obvious aspect to investigate is whether the analysis can be localized. A
possible source of inspiration for devising a local type system for \piLIN might
come from the work of \cite{DerakhshanPfenning19}. They
propose a compositional technique for dealing with infinitary typing derivations
in a session calculus, although their type system is limited to the additive
fragment of linear logic.

\paragraph{Fair Subtyping and Linear Logic}
In \Cref{sec:comparison} we showed that the type systems in \Cref{ch:ft_bin,ch:ft_multi}
and the one in \Cref{ch:ft_ll} cannot be compared. 
In particular, \Cref{ex:slot_comp} points out a fundamental difference between the two
approaches, that is, the presence or absence of fair subtyping.
Given the rich
literature exploring the connections between linear logic and session types,
\piLIN and its type system might provide the right framework for investigating
the logical foundations of fair subtyping.

\paragraph{Inference Systems in Agda}
For future work we plan to extend the library that we 
presented in \Cref{ch:gis_lib} in several directions. 
The first one is to support other interpretations of inference systems, 
such as the \emph{regular} one \citep{Dagnino20}, which is basically 
coinductive but allows only proof trees with finitely many distinct subtrees.
To this end, useful starting points are works on regular terms and streams 
\citep{Spadotti16,UustaluV17} and on finite sets \citep{FirsovU15} in dependent type theories.
The challenging part is the finiteness constraint, which is not trivial in a type-theoretic setting. 
A second direction is to implement other proof techniques for (flexible) coinduction, 
as parametrized coinduction \citep{HurNDV13} and up-to techniques \citep{Pous16,Danielsson18}. 
Finally, another direction could be the development of a full framework for composition 
of inference systems, along  the lines of seminal work on module systems \citep{Bracha92}. 
On the more practical side, a further development is to transform the methodology in 
an automatic translation. That is, a user should be allowed to write an inference system 
(with corules) in a natural syntax, and the corresponding Agda types should be 
generated automatically, either by  an external tool, or, more interestingly, 
using \emph{reflection}, recently added in Agda.

%% file: appendix.tex

\chapter{Supplement to Chapter 4}
\input{appendix/ts-bin}

%% file: appendix/ts-bin.tex

\section{Algorithms}

\beginalto
In this section we discuss the ingredients to obtain a type checking algorithm 
for the type system presented in \Cref{sec:ts_bin_ts}.
In \Cref{ssec:app_min_rank} we show how to compute the \emph{minimum rank of a process} while
in \Cref{ssec:app_algo} we present a variant of the rules in \Cref{fig:ts_bin} that allows
us to obtain a type checking algorithm.
Notably, such algorithm has been implemented in \cite{FairCheck}. The tool
is available on GitHub and contains the codes of most of the examples from 
this thesis as well as from \cite{CicconePadovani22}.


\subsection{Minimum Rank of a Process}
\label{ssec:app_min_rank}
\input{appendix/ts-bin-rank}


\subsection{Type Checking Algorithm}
\label{ssec:app_algo}
\input{appendix/ts-bin-algo}

%% file: appendix/ts-bin-rank.tex
\renewcommand{\ple}{\sqsubseteq}

\beginbass
In this section we develop a function to compute the \emph{minimum rank} of a process, 
namely the least quantity that is necessary in order to find a typing derivation for $P$. 
In the following we assume that bound names and casts are annotated with session types as well
as with the \emph{weight} of the subtyping relation ($\Cast{\x , m} P$ means that the subtyping
being applied has weight $m$).

\begin{remark}
	The weight of a fair subtyping application is the least solution of the equations
	in \Cref{def:fsub_wg}. We did not carry out an analysis about the decidability, so we require
	that such information is given by the programmer by annotating the process. 
	If the solution of \Cref{def:fsub_wg} can be found in a finite amount of time, then
	the weight can be inferred.
	\eor
\end{remark}

Moreover, we assume that a non deterministic choice $P \pchoice_k Q$ is labeled with a number
$m \in \set{1,2}$ which identifies the path leading to successful termination. 
The function is defined below.

\begin{definition}[Minimum Rank of a Process]
	\label{def:process_rank}
	The \emph{minimum rank} of a process $P$, written $\Rank[]{P}$, is the least
	upper bound to the number of casts that $P$ may need to perform and of
	sessions that $P$ may need to create in order to terminate. Formally, let
	$\Rank{P}$ be the function inductively defined by the following equations,
	where $\aset$ is a set of process names:
	\[
		\begin{array}{r@{~}c@{~}ll}
			\Rank\Done = \Rank{\Close\x} & = & 0
			\\
			\Rank{\Wait\x P} & = & \Rank P
			\\
			\Rank{\Call{A}{\seqof\x}} & = & 0 & \text{if $A \in \aset$}
			\\
			\Rank{\Call{A}{\seqof\x}} & = & \Rank[\aset \cup \set{A}]{P}
			& \text{if $A \not\in \aset$ and $\Definition A {\seqof\x} P$}
			\\
			\Rank{\PInput\x{(\y)}.P} = \Rank{\POutput\x\y.P} & = & \Rank{P}
			\\
			\Rank{\Cast{\x , m} P} & = & m + \Rank P
			\\
			\Rank{\NewPar\x P Q} & = & 1 + \Rank P + \Rank Q
			\\
			\Rank{x\p\set{\l_i : P_i}_{i \in I}} & = & \bigsqcup_{i\in I} \Rank{P_i}
			\\
			\ifextension
			\Rank{P_1 \pchoice_k P_2} & = & \Rank{P_k}
			& \text{if $k\in\set{1,2}$}
			\fi
		\end{array}
	\]
	We write $\Rank[] P$ for $\Rank[\emptyset] P$.
\end{definition}

Now we have to show that this function allows us to compute the minimum rank that is necessary in a typing derivation.
The first step is to provide a characterization of those processes that play a primary role in computing $\Rank[]\cdot$. 
We do so introducing an order relation $\ple$ on processes.

\begin{definition}[Termination Path]
	\label{def:ple}
	Let $\ple$ be the least preorder such that
	\begin{mathpar}
		\inferrule{\mathstrut}{
			P_k \ple P_1 \choice_k P_2
		} ~k \in \set{1,2}
		\and
		\inferrule{\mathstrut}{P \ple \POutput\x\y.P}
		\and
		\inferrule{\mathstrut}{P \ple \PInput\x{(\y)}.P}
		\and
		\inferrule{\mathstrut}{
			P_k \ple x\Pol\set{\l_i:P_i}_{i\in I}
		} ~k\in I
		\and
		\inferrule{
			\Definition{A}{\seqof\x}{P}
		}{
			P \ple \Call{A}{\seqof\x}
		}
	\end{mathpar}
	We say that $P$ is on a \emph{termination path} of $Q$ if $P \ple Q$.
\end{definition}

Intuitively, $P \ple Q$ means that the rank of $Q$ may be affected by the rank of $P$, 
because $P$ occurs along a path that leads $Q$ to termination. 
Hereafter, we often omit the arguments of a process invocation involved in a 
process order relation and write, for example, $A \ple P$ and $P \ple A$ instead of 
$\Call{A}{\seqof\x} \ple P$ and $P \ple \Call{A}{\seqof{x}}$.
The notation $A \ple P \ple A$, shortcut for $A \ple P$ and $P \ple A$, means that $P$ 
is found in between two invocations of the same definition $A$. These ``loops'' in termination 
paths are the dangerous places in which no casts (with strictly positive weight) 
can be performed and no sessions can be created.

\begin{definition}[Safe Program]
	\label{def:safe_program}
	We say that a program $\set{\Definition{A_i}{\seqof{x_i}}{P_i}}_{i\in I}$ is \emph{safe} 
	if $A_i \ple P \ple A_i$ implies that $P$ is not a cast or a session for every $P$ and $i\in I$.
\end{definition}

Note that it is straightforward to define an algorithm that checks whether a program is safe, 
since the number of processes is finite and so is the number of process names.

Now we show that, in a safe program, the rank of a process invocation corresponds to that of its unfolding. 
This requires some auxiliary results that allow us to express the relationship between the ranks of a process 
depending on the set $\aset$ or process names used. First, we show that the larger the set of 
process names, the smaller the rank. Intuitively, this is because in \Cref{def:process_rank} every invocation 
of a process name that occurs in $\aset$ results in a null rank.

\begin{lemma}
	\label{lem:rank-contr}
	If $\aset \subseteq \bset$, then $\Rank[\bset]{P} \leq \Rank[\aset]{P}$.
\end{lemma}
\begin{proof}
	By induction on the definition of rank and by cases on the shape of $P$. We
	only discuss the base case in which $P = \Call A {\seqof\x}$, distinguishing
	three sub-cases: if $A \in \aset$, then we conclude $\Rank[\bset] P = 0 =
	\Rank[\aset] P$; if $A \in \bset \setminus \aset$, then we conclude
	$\Rank[\bset] P = 0 \leq \Rank[\aset] P$; if $A \not\in \bset$, then we
	conclude $\Rank[\bset] P = \Rank[\bset \cup \set{A}] Q \leq \Rank[\aset \cup
	\set{A}] Q = \Rank[\aset] P$ using the induction hypothesis and $\Definition
	A {\seqof\x} Q$.
\end{proof}

Next we show that the rank of a process $P$ does not depend on the presence or
absence of $A$ in the set $\aset$ if $A \not\ple P$ if there is no invocation to
$A$ along any termination path of $P$.

\begin{lemma}
	\label{lem:rank_not_ple}
	If $A \not\ple P$, then $\Rank{P} = \Rank[\aset \cup \set{A}]{P}$.
\end{lemma}
\begin{proof}
	By induction on the definition of $\Rank P$ and by cases on the shape of
	$P$.

	\proofcase{Cases $P = \Done$ and $P = \Close\x$}
	We conclude $\Rank{P} = \Rank[\aset \cup \set{A}]{P} = 0$.

	\proofcase{Case $P = \Call{B}{\seqof\x}$ where $\Definition{B}{\seqof\x}Q$}
	From the hypothesis $A \not \ple P$ we deduce $B \neq A$. We distinguish two
	sub-cases.
	If $B \in \aset$, then we conclude
	\[
		\begin{array}{rcll}
			\Rank{P} & = & \Rank{\Call{B}{\seqof\x}} & \text{by definition of $P$}
			\\
			& = & 0 & \text{by definition of $\Rank[]\cdot$}
			\\
			& = & \Rank[\aset \cup \set{A}]{\Call{B}{\seqof\x}} & \text{by definition of $\Rank[]\cdot$}
			\\
			& = & \Rank[\aset \cup \set{A}]{P} & \text{by definition of $P$}
		\end{array}
	\]

	If $B \not\in \aset$, then from the hypothesis $A \not\ple P$ we deduce $A \not\ple Q$ and we conclude
	\[
		\begin{array}{rcll}
			\Rank{P} & = & \Rank{\Call{B}{\seqof\x}} & \text{by definition of $P$}
			\\
			& = & \Rank[\aset \cup \set{B}]{Q} & \text{by definition of $\Rank[]\cdot$}
			\\
			& = & \Rank[\aset \cup \set{A,B}]{Q}
			& \text{using the induction hypothesis}
			\\
			& = & \Rank[\aset \cup \set{A}]{\Call{B}{\seqof\x}}
			& \text{by definition of $\Rank[]\cdot$}
			\\
			& = & \Rank[\aset \cup \set{A}]{P} & \text{by definition of $P$}
		\end{array}
	\]

	\proofcase{Case $P = P_1 \pchoice_k P_2$ where $k\in\set{1,2}$}
	From the hypothesis $A \not\ple P$ we deduce $A \not\ple P_k$. We conclude
	\[
		\begin{array}{rcll}
			\Rank{P} & = & \Rank{P_1 \pchoice_k P_2} & \text{by definition of $P$}
			\\
			& = & \Rank{P_k} & \text{by definition of $\Rank[]\cdot$}
			\\
			& = & \Rank[\aset \cup \set{A}]{P_k} & \text{using the induction hypothesis}
			\\
			& = & \Rank[\aset \cup \set{A}]{P_1 \pchoice_k P_2} & \text{by definition of $\Rank[]\cdot$}
			\\
			& = & \Rank[\aset \cup \set{A}]{P} & \text{by definition of $P$}
		\end{array}
	\]

	\proofcase{Case $P = \POutput\x\y.Q$}
	From the hypothesis $A \not\ple P$ we deduce $A \not\ple Q$. We conclude
	\[
		\begin{array}{rcll}
			\Rank{P} & = & \Rank{\POutput\x\y.Q} & \text{by definition of $P$}
			\\
			& = & \Rank{Q} & \text{by definition of $\Rank[]\cdot$}
			\\
			& = & \Rank[\aset \cup \set{A}]{Q} & \text{using the induction hypothesis}
			\\
			& = & \Rank[\aset \cup \set{A}]{\POutput\x\y.Q} & \text{by definition of $\Rank[]\cdot$}
			\\
			& = & \Rank[\aset \cup \set{A}]{P} & \text{by definition of $P$}
		\end{array}
	\]
	
	\proofcase{Case $P = \PInput\x{(\y)}.Q$}
	Analogous to the previous case.

	\proofcase{Case $P = \Cast{\x,m} Q$}
	From $A \not\ple P$ we deduce $A \not\ple Q$. We conclude
	\[
		\begin{array}{rcll}
			\Rank{P} & = & \Rank{\Cast\x Q} & \text{by definition of $P$}
			\\
			& = & m + \Rank{Q} & \text{by definition of $\Rank[]\cdot$}
			\\
			& = & m + \Rank[\aset \cup \set{A}]{Q} & \text{using the induction hypothesis}
			\\
			& = & \Rank[\aset \cup \set{A}]{\Cast\x Q} & \text{by definition of $\Rank[]\cdot$}
			\\
			& = & \Rank[\aset \cup \set{A}]{P} & \text{by definition of $P$}
		\end{array}
	\]

	\proofcase{Case $P = \NewPar\x{P_1}{P_2}$}
	From $A \not\ple P$ we deduce $A \not\ple P_i$ for $i=1,2$. We conclude
	\[
		\begin{array}{rcll}
			\Rank{P} & = & \Rank{\NewPar\x{P_1}{P_2}} & \text{by definition of $P$}
			\\
			& = & 1 + \Rank{P_1} + \Rank{P_2} & \text{by definition of $\Rank[]\cdot$}
			\\
			& = & 1 + \Rank[\aset \cup \set{A}]{P_1} + \Rank[\aset \cup \set{A}]{P_2} & \text{using the induction hypothesis}
			\\
			& = & \Rank[\aset \cup \set{A}]{\NewPar\x{P_1}{P_2}} & \text{by definition of $\Rank[]\cdot$}
			\\
			& = & \Rank[\aset \cup \set{A}]{P} & \text{by definition of $P$}
		\end{array}
	\]

	\proofcase{Case $P = x\Pol\set{\l_i : P_i}_{i \in I}$}
	From the hypothesis $A \not\ple P$ we deduce $A \not\ple P_i$ for every
	$i\in I$. We conclude
	\[
		\begin{array}[b]{rcll}
			\Rank{P} & = & \Rank{x\p\set{\l_i:P_i}_{i\in I}} & \text{by definition of $P$}
			\\
			& = & \bigsqcup_{i\in I} \Rank{P_i} & \text{by definition of $\Rank[]\cdot$}
			\\
			& = & \bigsqcup_{i\in I} \Rank[\aset \cup \set{A}]{P_i}
			& \text{using the induction hypothesis}
			\\
			& = & \Rank[\aset \cup \set{A}]{x\p\set{\l_i:P_i}_{i\in I}}
			& \text{by definition of $\Rank[]\cdot$}
			\\
			& = & \Rank[\aset \cup \set{A}]{P} & \text{by definition of $P$}
		\end{array}
	\]
\end{proof}

Finally, we can show that the rank of a process $P$ such that $A \ple P \ple A$
cannot exceed the rank of $A$. Recall that $A \ple P$ means that there is an
invocation to $A$ along a termination path of $P$ and that $P \ple A$ means that
$P$ occurs along a termination path of $A$.

\begin{lemma}
	\label{lem:rank_ple}
	In a safe program, if\/ $A \ple P \ple A$ and $\Definition{A}{\seqof\x}{Q}$,
	then $\Rank{P} \leq \Rank[\aset \cup \set{A}]{P} \psup \Rank[]{\Call A
	{\seqof\x}}$.
\end{lemma}
\begin{proof}
	\newcommand{\RankA}{\Rank[]{\Call A {\seqof\x}}}
	By induction on $\Rank P$ and by cases on the shape of $P$. Note that $P$
	cannot be a cast or a session because of the hypothesis that the program is
	safe.

	\proofcase{Case $P = \Call{B}{\seqof\x}$ and $B \in \aset$}
	We conclude
	\[
		\begin{array}{rcll}
			\Rank{P} & = & \Rank{\Call{B}{\seqof\x}} & \text{by definition of $P$}
			\\
			& = & 0 & \text{by definition of $\Rank[]\cdot$}
			\\
			& = & \Rank[\aset \cup \set{A}]{\Call{B}{\seqof\x}} & \text{by definition of $\Rank[]\cdot$}
			\\
			& \leq & \Rank[\aset \cup \set{A}]{\Call{B}{\seqof\x}} \psup \RankA
			& \text{property of $\psup$}
			\\
			& = & \Rank[\aset \cup \set{A}]{P} \psup \RankA
			& \text{by definition of $P$}
		\end{array}
	\]

	\proofcase{Case $P = \Call{A}{\seqof\y}$ and $A \not\in \aset$} We may assume, without loss of generality, 
	that $\seqof\y = \seqof\x$ since the rank of a process invocation does \emph{not} depend on channel names.
	We conclude
	\[
		\begin{array}{rcll}
			\Rank{P} & = & \Rank{\Call{A}{\seqof\x}} & \text{by definition of $P$}
			\\
			& = & \Rank[\aset \cup \set{A}]{Q} & \text{by definition of $\Rank[]\cdot$}
			\\
			& \leq & \Rank[\set{A}]{Q} & \text{by \Cref{lem:rank-contr}}
			\\
			& = & \RankA & \text{by definition of $\Rank[]\cdot$}
			\\
			& = & 0 \psup \RankA & \text{property of $\psup$}
			\\
			& = & \Rank[\aset \cup \set{A}]{\Call{A}{\seqof\x}} \psup \RankA & \text{by definition of $\Rank[]\cdot$}
			\\
			& = & \Rank[\aset \cup \set{A}]{P} \psup \RankA
			& \text{by definition of $P$}
		\end{array}
	\]

	\proofcase{Case $P = \Call{B}{\seqof\x}$ and $B \not\in \aset \cup \set{A}$ where $\Definition{B}{\seqof\x}R$}
	From the hypotheses $A \ple P \ple A$ and $B \not\in \aset \cup \set{A}$ we deduce $A \ple R \ple A$.
	\[
		\begin{array}{rcll}
			\Rank{P} & = & \Rank{\Call{B}{\seqof\x}} & \text{by definition of $P$}
			\\
			& = & \Rank[\aset \cup \set{B}]{R} & \text{by definition of $\Rank[]\cdot$}
			\\
			& \leq & \Rank[\aset \cup \set{A,B}]{R} \psup \RankA
			& \text{using the induction hypothesis}
			\\
			& = & \Rank[\aset \cup \set{A}]{\Call{B}{\seqof\x}} \psup \RankA
			& \text{by definition of $\Rank[]\cdot$} \\
			& & & \text{and the hypothesis $B \not\in \aset \cup \set{A}$}
			\\
			& = & \Rank[\aset \cup \set{A}]{P} \psup \RankA
			& \text{by definition of $P$}
		\end{array}
	\]

	\proofcase{Case $P = P_1 \pchoice_k P_2$ where $k\in\set{1,2}$}
	From the hypotheses $A \ple P \ple A$ we deduce $A \ple P_k \ple A$. We conclude
	\[
		\begin{array}{rcll}
			\Rank{P} & = & \Rank{P_1 \choice_k P_2} & \text{by definition of $P$}
			\\
			& = & \Rank{P_k} & \text{by definition of $\Rank[]\cdot$}
			\\
			& \leq & \Rank[\aset \cup \set{A}]{P_k} \psup \RankA
			& \text{using the induction hypothesis}
			\\
			& = & \Rank[\aset \cup \set{A}]{P_1 \pchoice_k P_2} \psup \RankA
			& \text{by definition of $\Rank[]\cdot$}
			\\
			& = & \Rank[\aset \cup \set{A}]{P} \psup \RankA
			& \text{by definition of $P$}
		\end{array}
	\]

	\proofcase{Case $P = \POutput\x\y.Q$}
	From the hypotheses $A \ple P \ple A$ we deduce $A \ple Q \ple A$. We conclude
	\[
		\begin{array}{rcll}
			\Rank{P} & = & \Rank{\POutput\x\y.Q} & \text{by definition of $P$}
			\\
			& = & \Rank{Q} & \text{by definition of $\Rank[]\cdot$}
			\\
			& \leq & \Rank[\aset \cup \set{A}]{Q} \psup \RankA
			& \text{using the induction hypothesis}
			\\
			& = & \Rank[\aset \cup \set{A}]{\POutput\x\y.Q} \psup \RankA & \text{by definition of $\Rank[]\cdot$}
			\\
			& = & \Rank[\aset \cup \set{A}]{P} \psup \RankA & \text{by definition of $P$}
		\end{array}
	\]

	\proofcase{Case $P = \PInput\x{(\y)}.Q$}
	Analogous to the previous case.
	
	\proofcase{Case $P = x\Pol\set{\l_i : P_i}_{i \in I}$}
	For every $i\in I$ we distinguish two possibilities, depending on whether $A \not\ple P_i$ or $A \ple P_i$. 
	In the first case we deduce $\Rank{P_i} = \Rank[\aset \cup \set{A}]{P_i}$ using \Cref{lem:rank_not_ple}. 
	In the second case we deduce $\Rank{P_i} \leq \Rank[\aset \cup \set{A}]{P_i} \psup \RankA$ using the induction hypothesis. 
	Either way, we have $\Rank{P_i} \leq \Rank[\aset \cup \set{A}]{P_i} \psup \RankA$ for every $i\in I$. We conclude
	\[
		\begin{array}[b]{@{}rcll@{}}
			\Rank{P} & = & \Rank{x\p\set{\l_i:P_i}_{i\in I}} & \text{by definition of $P$}
			\\
			& = & \bigsqcup_{i\in I} \Rank{P_i} & \text{by definition of $\Rank[]\cdot$}
			\\
			& \leq & \bigsqcup_{i\in I} (\Rank[\aset \cup \set{A}]{P_i} \psup \RankA)
			& \text{using \Cref{lem:rank_not_ple}} \\
			& & & \text{or the induction hypothesis}
			\\
			& = & (\bigsqcup_{i\in I} \Rank[\aset \cup \set{A}]{P_i}) \psup \RankA
			& \text{distributivity of $\psup$}
			\\
			& = & \Rank[\aset \cup \set{A}]{x\p\set{\l_i:P_i}_{i\in I}} \psup \RankA
			& \text{by definition of $\Rank[]\cdot$}
			\\
			& = & \Rank[\aset \cup \set{A}]{P} \psup \RankA & \text{by definition of $P$}
		\end{array}
	\]
\end{proof}

Next is the key lemma stating that the given notion of rank
(\Cref{def:process_rank}) behaves well, in the sense that it is preserved by
unfolding a process invocation.

\begin{lemma}
	\label{lem:rank-empty}
	In a safe program such that $\Definition{A}{\seqof\x}P$ we have $\Rank[]P =
	\Rank[]{\Call{A}{\seqof\x}}$.
\end{lemma}
\begin{proof}
	Clearly $P \ple A$ since $P$ is the body of the definition of process $A$.
	We distinguish two sub-cases that cover all possibilities.
	If $A \not\ple P$, then using \Cref{lem:rank_not_ple} we deduce $\Rank[]P =
	\Rank[\emptyset]P = \Rank[\set A]P = \Rank[]{\Call{A}{\seqof\x}}$.
	If $A \ple P$, then using \Cref{lem:rank_not_ple} we deduce $\Rank[] P =
	\Rank[\emptyset] P \leq \Rank[\set A] P \psup \Rank[]{\Call A {\seqof\x}} =
	\Rank[]{\Call A {\seqof\x}} \psup \Rank[]{\Call A {\seqof\x}} =
	\Rank[]{\Call A {\seqof\x}}$.
	Using \Cref{lem:rank-contr} we conclude $\Rank[]{\Call{A}{\seqof\x}} = \Rank[\set{A}]{P} \leq \Rank[]P$.
\end{proof}

In order to show that $\Rank[]P$ is indeed the minimum rank of $P$ that allows us to find a typing derivation for $P$, 
provided there is one, the first thing to do is to prove that a well-typed program is also safe. So from now on, 
when we reason about well-typed processes, we may assume that they belong to a safe program.

\begin{lemma}
	\label{lem:well_typed_safe}
	A well-typed program is safe.
\end{lemma}
\begin{proof}
	First of all observe that $\wtp[m]\CtxC{P}$ and $\wtp[n]\CtxD{Q}$ and $P \ple Q$ imply $m \leq n$. 
	This follows by considering the base cases of $P \ple Q$ and looking at the typing rules in which $Q$ 
	is in the conclusion and $P$ is one of the premises.
	Then, $A \ple P \ple A$ implies that $P$ occurs in a typing judgment having exactly the 
	same rank annotation as that of $A$. It follows that $P$ cannot be a cast or a session, 
	for these forms strictly increase the rank annotation.
\end{proof}

Next we show that $\Rank[]P$ is no greater than any rank that may appear is a
typing derivation for $P$.

\begin{lemma}
	\label{lem:minimum_rank}
	If $\wtp[n]\Ctx{P}$, then $\Rank[]P \leq n$.
\end{lemma}
\begin{proof}
	We prove that $\wtp[n]\Ctx{P}$ implies $\Rank[\aset]P \leq n$ by a straightforward induction 
	on the definition of $\Rank[\aset]P$. The conclusion $\Rank[]P \leq n$ 
	is then just the particular case when $\aset = \emptyset$.
\end{proof}

Finally, we show that if a program is well typed under \emph{some} assignment then it 
is also well typed under the assignment that uses the minimum ranks. 
With \Cref{lem:minimum_rank}, this result justifies the definition of $\Rank[]P$ as \emph{minimum rank} of $P$.

\begin{theorem}
	\label{thm:rank_typing}
	If $\set{\Definition{A_i}{\seqof{x_i}}{P_i}}_{i\in I}$ is well typed under
	the global assignment $\set{\tass{A_i}{\seqof{S_i}}{n_i}}_{i\in I}$, then it
	is well typed also under the global assignment
	$\set{\tass{A_i}{\seqof{S_i}}{\Rank[]{P_i}}}_{i\in I}$.
\end{theorem}
\begin{proof}
	\newcommand{\relr}{\mathcal{R}}
	First we use the coinduction principle to show that every judgment in $\relr
	\eqdef \set{\wtp[m]\Ctx{P} \mid \wtpc[n]\Ctx{P}, \Rank[]P \leq m}$
	is the conclusion of a rule in \Cref{fig:ts_bin} whose premises are also in
	$\relr$. This allows us to deduce that $\wtpc[n]\Ctx{P}$ implies $\wtpc[\RankE{P}]\Ctx{P}$.
	Suppose $\wtp[m]\Ctx{P} \in \relr$. Then $\wtpc[n]\Ctx{P}$ and
	$\Rank[]P \leq m$.
	We reason by cases on the last rule used in the derivation of
	$\wtpc[n]\Ctx{P}$. We only discuss two representative cases.

	\proofrule{tb-call}
	Then $P = \Call{A_k}{\seqof\x}$ and $\Definition{A_k}{\seqof\x}{Q}$ and
	$\wtpc[n']\Ctx{Q}$ for some $n' \leq n$ and some $k\in I$.
	From the definition of rank we deduce $\Rank[]P = \Rank[]Q = \Rank[]{Q_k}$
	since the rank of a process does not depend on its free names.
	From \Cref{lem:minimum_rank} we deduce $\Rank[]{Q_k} \leq n_k$.
	We conclude by observing that $m \geq n_k$ and that $\wtp[m]\Ctx{Q}$ is
	the conclusion of \refrule{tb-call}.

	\proofrule{tb-tag}
	Then $P = x\Pol\set{\l_i:Q_i}_{i\in I}$ and $\Ctx = \CtxD, x :
	\Pol\set{\l_i:S_i}_{i\in K}$ and $\wtpc[n]{\CtxD, x : S_i}{Q_i}$ for every
	$i\in I$.
	From the definition of rank we have $\Rank[]P = \bigsqcup_{i\in I}
	\Rank[]{Q_i}$.
	From $m \geq \Rank[]P$ we deduce $m \geq \Rank[]{Q_i}$ for every $i\in I$.
	Then $\wtp[m]{\CtxD, x : S_i}{Q_i} \in \relr$ for every $i\in I$ by
	definition of $\relr$ and we conclude by observing that $\wtp[m]\Ctx{P}$
	is the conclusion of \refrule{tb-label}.

	To show that $\wtpi[n]\Ctx{P}$ implies $\wtpi[\RankE{P}]\Ctx{P}$ it
	suffices a straightforward induction on the derivation of
	$\wtpi[n]\Ctx{P}$.
	By the bounded coinduction principle this is enough to conclude.
\end{proof}

%% file: appendix/ts-bin-algo.tex
\beginbass
In this section we show how to obtain an alternative version of the typing rules
from which it is easy to derive a type checking algorithm, provided that bound
names and casts are explicitly annotated with session types and the weight of the
subtyping being applied. As we pointed out in \Cref{ssec:app_min_rank}, we assume that
non deterministic choices are labeled with the branch which leads to successful termination.
There are three
aspects that make the type system presented in \Cref{fig:ts_bin} not strictly
algorithmic:
\begin{enumerate}
    \item\label{item:alg_coind} the fact that typing derivations are potentially
    infinite
    \item\label{item:alg_ind} the need for building finite derivations using the
        corules \refrule{cob-choice} and \refrule{cob-tag}, which overlap with
        \refrule{tb-choice} and \refrule{tb-tag} respectively
    \item\label{item:alg_rank} the rank annotation to be used in each typing
    judgment
\end{enumerate}

Concerning \Cref{item:alg_rank}, in \Cref{ssec:app_min_rank} we have seen how the
rank annotation can be computed for any process in a safe program. So here we
focus on \Cref{item:alg_coind,item:alg_ind}.

\begin{figure}[t]
    \framebox[\textwidth]{
      \begin{mathpar}
        \displaystyle
          \inferrule[a-done]{\mathstrut}{
            \wtp\EmptyCtx\Done
          } \defrule[a-done]{}
          \and
          \inferrule[a-call]{
              \mathstrut
          }{
              \wtp{\seqof{x : S}}{\Call A {\seqof x}}
          }
          ~ \tass{A}{\seqof{S}}{n} \defrule[a-call]{}
          \and
          \inferrule[a-choice]{
            \wtp\Ctx{P}
            \\
            \wtp\Ctx{Q}
          }{
            \wtp\Ctx{P \pchoice Q}
          } \defrule[a-choice]{}
          \and
          \inferrule[a-close]{\mathstrut}
          {
            \wtp{x : \End[\Out]}{\Close\x}
          } \defrule[a-close]{}
          \and
          \inferrule[a-wait]
          {
            \wtp\Ctx{P}
          }{
            \wtp{\Ctx, x : \End[\In]}{\Wait\x P}
          } \defrule[a-wait]{}
          \and
          \inferrule[a-channel-in]{
            \wtp{\Ctx, x : S, y : T}{P}
          }{
            \wtp{\Ctx, x : \In\T.S}{\PInput\x{(y)}.P}
          } \defrule[a-channel-in]{}
          \and
          \inferrule[a-tag]
          {
            \forall i \in I : \wtp{\Ctx, x : S_i}{P_i}
          }{
            \textstyle
            \wtp{
              \Ctx, x : \Pol\set{\l_i : S_i}_{i \in I}
            }{
              x\Pol\set{\l_i : P_i}_{i \in I}
            }
          } \defrule[a-tag]{}
          \and
          \inferrule[a-channel-out]{
            \wtp{\Ctx, x : S}{P}
          }{
            \wtp{\Ctx, x : \Out\T.S, y : T}{\POutput\x\y.P}
          } \defrule[a-channel-out]{}
          \and
          \inferrule[a-par]{
            \wtp{\CtxC, x : S}{P}
            \\
            \wtp{\CtxD, x : T}{Q}
          }{
            \wtp{
              \CtxC, \CtxD
            }{
              \NewPar{x}{P}{Q}
            }
          }
          ~
          S \compatible T \defrule[a-par]{}
          \and
          \inferrule[a-cast]{
            \wtp{\Ctx, x : T}{P}
          }{
            \wtp{\Ctx, x : S}{\Cast{x} P}
          }
          ~ S \subt T \defrule[a-cast]{}
      \end{mathpar}
    }
    \caption{Algorithmic typing rules for processes}
    \label{fig:as}
\end{figure}

\Cref{fig:as} presents an (inductively interpreted) set of typing rules that are
a ``stripped down'' version of those given in \Cref{fig:ts_bin}. There are two main
differences between these rules and those given in the main body of the paper:
first, there is no rank annotation in typing judgments; second, \refrule{a-call}
is an \emph{axiom}, unlike \refrule{tb-call}. The remaining structure of the
rules and the kind of constraints they impose is exactly the same as before.
Henceforth, we write $\wtpa\Ctx{P}$ if $\wtp\Ctx{P}$ is (inductively)
derivable using the typing rules in \Cref{fig:as}.

\begin{lemma}
    \label{lem:coind_alg}
    Let $\set{ \Definition{A_i}{\seqof{x_i}}{P_i}}_{i\in I}$ be a safe program
    and let $\set{ \tass{A_i}{\seqof{S_i}}{n_i} }_{i\in I}$ be a global assignment.
    The following properties are equivalent:
    \begin{enumerate}
        \item $\wtpc[n_i]{\seqof{x_i : S_i}}{P_i}$ for every $i\in I$;
        \item $\wtpa{\seqof{x_i : S_i}}{P_i}$ for every $i\in I$.
    \end{enumerate}
\end{lemma}
\begin{proof}
    \newcommand{\relr}{\mathcal{R}}
    $1 \Rightarrow 2$. Just observe that a (finite) derivation for
    $\wtpa\Ctx{P}$ can be obtained from a (possibly infinite) derivation for
    $\wtpc\Ctx{P}$ by truncating each application of \refrule{tb-call} in the
    latter derivation to an application of \refrule{a-call} with the same
    conclusion.

    $2 \Rightarrow 1$. Let $\relr \eqdef \set{ \wtp[n]\Ctx{P} \mid
    \wtpa\Ctx{P}, \Rank[]P \leq n }$. Using the coinduction principle it
    suffices to show that each judgment found in $\relr$ is the conclusion of a
    rule in \Cref{fig:ts_bin} whose premises are also in $\relr$.
    Let $\wtp[n]\Ctx{P} \in \relr$, meaning that $\wtpa\Ctx{P}$ and
    $\Rank[]P \leq n$.
    We reason by cases on the rule used to derive $\wtpa\Ctx{P}$. We only discuss a few cases.

    \proofrule{a-done}
    We conclude observing that $P$ is the conclusion of \refrule{tb-done}.

    \proofrule{a-call}
    Then $P = \Call{A}{\seqof\x}$ for some $\Definition{A}{\seqof\x}{Q}$.
    Note that $n \geq \Rank[]P = \Rank[]Q$.
    From the hypothesis we know that $\wtpa\Ctx{Q}$ and we conclude by observing that $\wtp[n]\Ctx{P}$ 
    is the conclusion of \refrule{tb-call} and that $\wtp[n]\Ctx{Q} \in \relr$ by definition of $\relr$.

    \proofrule{a-par}
    Then $P = \NewPar\x{P_1}{P_2}$ and $\Ctx = \Ctx_1, \Ctx_2$ and
    $\wtpa{\Ctx_i, x : S_i}{P_i}$ for $i=1,2$ and $S_1 \compatible S_2$.
    Note that $n \geq \Rank[]P = 1 + \Rank[]{P_1} + \Rank[]{P_2}$. Hence, there
    exist $n_1$ and $n_2$ such that $n = 1 + n_1 + n_2$ and $\Rank[]{P_i} \leq
    n_i$ for $i=1,2$.
    We conclude by observing that $\wtp[n]\Ctx{P}$ is the conclusion of
    \refrule{tb-par} and that $\wtp[n_i]{\Ctx_i, x : S_i}{P_i} \in \relr$ by
    definition of $\relr$.

    \proofrule{a-tag}
    Then $P = x\Pol\set{\l_i:P_i}_{i\in I}$ and $\Ctx = \Ctx', x : \Pol\set{\l_i:S_i}_{i\in I}$ 
    and $\wtpa{\Ctx', x : S_i}{P_i}$ for every $i\in I$.
    Note that $n \geq \Rank[]P = \bigsqcup_{i\in I} \Rank[]{P_i}$, 
    hence $n \geq \Rank[]{P_i}$ for every $i\in I$.
    We conclude by observing that $\wtp[n]\Ctx{P}$ is the conclusion of \refrule{tb-tag} 
    and that $\wtp[n]{\Ctx', x : S_i}{P_i} \in \relr$ by definition of $\relr$.
\end{proof}

\begin{figure}[t]
    \framebox[\textwidth]{
      \begin{mathpar}
        \displaystyle
            \inferrule{\mathstrut}{
                \pbounded\Done
            }
            \and
            \inferrule{\mathstrut}{
                \pbounded{\Close\x}
            }
            \and
            \inferrule{
                \pbounded[\aset \cup \set{A}]P
            }{
                \pbounded{\Call{A}{\seqof\x}}
            }
            ~ A \not\in \aset, \Definition{A}{\seqof\x}{P}
            \and
            \inferrule{
                \pbounded P_k
            }{
                \pbounded{P_1 \pchoice_k P_2}
            }
            ~ k \in \set{1,2}
            \and
            \inferrule{
                \pbounded P
            }{
                \pbounded{\PInput\x{(\y)}.P}
            }
            \and
            \inferrule{
                \pbounded P
            }{
                \pbounded{\POutput\x\y.P}
            }
            \and
            \inferrule{
                \pbounded{P_k}
            }{
                \pbounded{x\Pol\set{\l_i:P_i}_{i\in I}}
            }
            ~ k\in I
            \and
            \inferrule{
                \forall i \in \set{1,2} : \pbounded{P_i}
            }{
                \pbounded{\NewPar\x{P_1}{P_2}}
            }
            \and
            \inferrule{
                \pbounded P
            }{
                \pbounded{\Cast\x P}
            }
    \end{mathpar}
    }
    \caption{Algorithmic rules for action boundedness}
    \label{fig:pbounded}
\end{figure}

To filter out those judgments derivable in the algorithmic type system for which
there is no finite derivation using the original type system with the corules
\refrule{cob-choice} and \refrule{cob-label}, we separately define the (inductive)
inference system shown in \Cref{fig:pbounded} for action boundedness. Note that
this inference system can be trivially turned into an algorithm by checking
whether, for a process of the form $x\Pol\set{\l_i:P_i}_{i\in I}$, there is at least
one branch for which $\pbounded{P_i}$ is derivable.

\begin{lemma}
    \label{lem:ind_alg}
    If $\wtpc[n]\Ctx{P}$, then $\wtpi[n]\Ctx{P}$ if and only if
    $\pbounded[\emptyset]P$.
\end{lemma}
\begin{proof}
    For the ``if'' part we prove that $\pbounded P$ implies
    $\wtpi[A]\Ctx{P}$ by induction on the derivation of $\pbounded P$.
    For the ``only if'', we first prove that if $\wtpi[A]\Ctx{P}$ and none
    of the process names occurring in the \refrule{tb-call} applications of this
    derivation is in $\aset$, then $\pbounded P$.
    Then, the result follows by considering the \emph{smallest} derivation
    $\wtpi[A]\Ctx{P}$, in which no process definition is expanded twice.
\end{proof}

\begin{theorem}
    Let $\set{\Definition{A_i}{\seqof{x_i}}{P_i}}_{i\in I}$ be a safe program
    and let $\set{\tass{A_i}{\seqof{S_i}}{n_i} }_{i\in I}$ be a global assignment.
    The following properties are equivalent:
    \begin{enumerate}
        \item $\wtp[n_i]{\seqof{x_i : S_i}}{P_i}$ for every $i\in I$;
        \item $\wtpa{\seqof{x_i : S_i}}{P_i}$ for every $i\in I$ and $\pbounded[\emptyset]Q$ 
        is derivable for every $Q$ occurring in the derivations.
    \end{enumerate}
\end{theorem}
\begin{proof}
    Consequence of \Cref{lem:coind_alg,lem:ind_alg}.
\end{proof}